\documentclass[twoside,a4paper,11pt]{book}
\usepackage[hmargin=3cm,vmargin=3.4cm]{geometry}

\usepackage[utf8]{inputenc}
\usepackage[T1]{fontenc}
\usepackage{float}
\usepackage{CJKutf8}
\AtBeginDvi{\input{zhwinfonts}}
\usepackage{ifpdf}
\usepackage{fancyhdr} %encabezados guais
\usepackage[nottoc]{tocbibind} %para incluir la bibliografía en el índice
\usepackage{graphicx}
\usepackage{tensor}
\usepackage{epsfig}\parskip 5pt plus 1pt
\usepackage{multirow}
\usepackage{booktabs}  %for nicer tables
\usepackage{epsfig}
\usepackage{epstopdf}
\usepackage{tikz}
\usepackage{braket}
\usepackage{pgfplots}
\usepackage{bbm}
\usepackage{pifont}
\usepackage{enumerate}
\usepackage{caption}
\usepackage{amsfonts}
\usepackage{mathrsfs}
\usepackage[hang,center]{subfigure}    
\usepackage{mathtools}
\usepackage{amsbsy}
\usepackage{rotate}
\usepackage{bbm}
\usepackage{color}
\usepackage{verbatim}  % this is so we can use \begin{comment}

\usepackage{wasysym}
\usepackage{harmony}
\usepackage{abc}
\usepackage{mathtools}

\usepackage{slashed}

\usepackage{tabularx} %para poder hacer tablas con footnotes que funcionen

\definecolor{MyGrey}{rgb}{0,0,0} %defining the color 'MyDarkGreen'
\definecolor{MyDarkBlue}{rgb}{0.23,0.21,0.69} %defining the color 'MyDarkBlue'
\definecolor{MyLightBlue}{rgb}{0.22,0.51,0.86}

\usepackage[colorlinks=true,linkcolor=MyDarkBlue,citecolor=MyDarkBlue,urlcolor=MyLightBlue,bookmarksnumbered=true,bookmarksopen]{hyperref}

\usepackage{cite} % The cite package must be loaded after hyperref package

\usepackage{breakurl}

\usepackage[times]{quotchap}

\def\be{\begin{equation}}
\def\ee{\end{equation}}

\title{Higher-curvature gravity, black holes and holography}
\author{Pablo Antonio Cano Molina-Ni\~nirola}
\makeindex
\usepackage{amssymb,amsmath}

\newtheorem{theorem}{Theorem}
\newtheorem{conjecture}{Conjecture}
\newtheorem{corollary}{Corollary}

\newcommand{\arxiv}[1]{{\tt
\href{http://www.arXiv.org/abs/#1}{arXiv:#1}}}

\newcommand{\rdd}[1]{{\color{black}#1}}

\newcommand{\rh}{{r_h}}

\newcommand{\quotient}[1]{_{\hskip-2pt\lower1pt\hbox{$/$}\lower2pt\hbox{\hskip-1pt$#1$}}}

\def\place#1#2#3{\vbox to0pt{\kern-\parskip\kern-7pt
                             \kern-#2truein\hbox{\kern#1truein #3}
                             \vss}\nointerlineskip}

\newcommand{\Tr}{\mathrm{Tr}}

\newcommand{\symm}[1]{_{\hskip-3pt\lower3pt\hbox{$\left\{#1\right\}$}}}

\newcommand{\cicystop}{~\lower8pt\hbox{.}}

\def\place#1#2#3{\vbox to0pt{\kern-\parskip\kern-7pt
                             \kern-#2truein\hbox{\kern#1truein #3}
                             \vss}\nointerlineskip}

\newcommand{\beq}{\begin{equation}}
\newcommand{\eeq}{\end{equation}}
\newcommand{\bea}{\begin{eqnarray}}
\newcommand{\eea}{\end{eqnarray}}
\newcommand{\bean}{\begin{eqnarray*}}
\newcommand{\eean}{\end{eqnarray*}}

\newcommand{\req}[1]{(\ref{#1})} %{Eq.\thinspace(\ref{#1})}  
\newcommand{\labell}[1]{\label{#1}}

\newcommand{\ba}{\begin{eqnarray}}
\newcommand{\ea}{\end{eqnarray}}

\newcommand{\beqa}{\begin{eqnarray}}
\newcommand{\eeqa}{\end{eqnarray}}
\newcommand{\beqar}{\begin{eqnarray*}}
\newcommand{\eeqar}{\end{eqnarray*}}

\newcommand{\ssc}{\scriptscriptstyle}
\newcommand{\eg}{{\it e.g.,}\ }
\newcommand{\ie}{{\it i.e.,}\ }

\newcommand{\E}{\mathcal{E}}

\newcommand{\fin}{f_\infty}

\newcommand{\ct}{C_{T}} %{C_\mt{T}}

\newcommand{\cs}{c_{\ssc S}}
\newcommand{\cse}{c_{\ssc S,E}}
\newcommand{\ctt}{C_{\ssc T}}
\newcommand{\ctte}{C_{\ssc T}^{{\rm E}}}

\DeclareMathOperator{\tr}{Tr}

\begin{document}
\allowdisplaybreaks

\setlength{\belowdisplayskip}{12pt} \setlength{\belowdisplayshortskip}{12pt}
\setlength{\abovedisplayskip}{12pt} \setlength{\abovedisplayshortskip}{12pt}

\pagestyle{headings}
\pagestyle{empty}

\newcommand{\HRule}{\rule{\linewidth}{1mm}}
\setlength{\parindent}{1cm}
\setlength{\parskip}{1mm}
\noindent

\noindent
\HRule
\begin{center}
\huge{\textbf{Higher-Curvature Gravity,\\ Black Holes and Holography}}
 \vspace{0.2cm}
\end{center}
\HRule

\vspace{0.5cm}

\begin{center}

\large{	   
Memoria de la Tesis Doctoral realizada por \\[3mm]
\textbf{\large{Pablo Antonio Cano Molina-Ni\~nirola}} \\[3mm]
presentada ante el Departamento de F\'isica Te\'orica \\[1mm]                  
de la Universidad Aut\'onoma de Madrid \\[1mm]
para optar al T\'itulo de Doctor en F\'isica Te\'orica \\[1mm]
}

\vspace{1cm}
Tesis Doctoral dirigida por el \textbf{\large{Prof. D. Tom\'as Ort\'in}}$^{1}$ \\
y por el \textbf{\large{Dr. D. Pablo Bueno}}$^{2}$ \\[2mm]
%\vspace{1cm}
$^{1}$Profesor de Investigaci\'on del Instituto de F\'isica Te\'orica UAM/CSIC \\
$^{2}$Investigador postdoctoral en Instituto Balseiro \\

\end{center}

\vspace{0.5cm}

\begin{center}
{\Large {Departamento de F\'isica Te\'orica\\[1mm] Universidad Aut\'onoma de Madrid }}\\
\vspace{0.3cm}
{\Large {Instituto de F\'isica Te\'orica\\[2mm] UAM/CSIC}}\\
\end{center}
%\vspace{1.2cm}

\begin{figure}[ht]
\centering
\hspace{-0.6cm}
%\includegraphics[scale=0.9]{uam.jpg}
% \hspace{0.6cm}
\includegraphics[height=0.1\textheight]{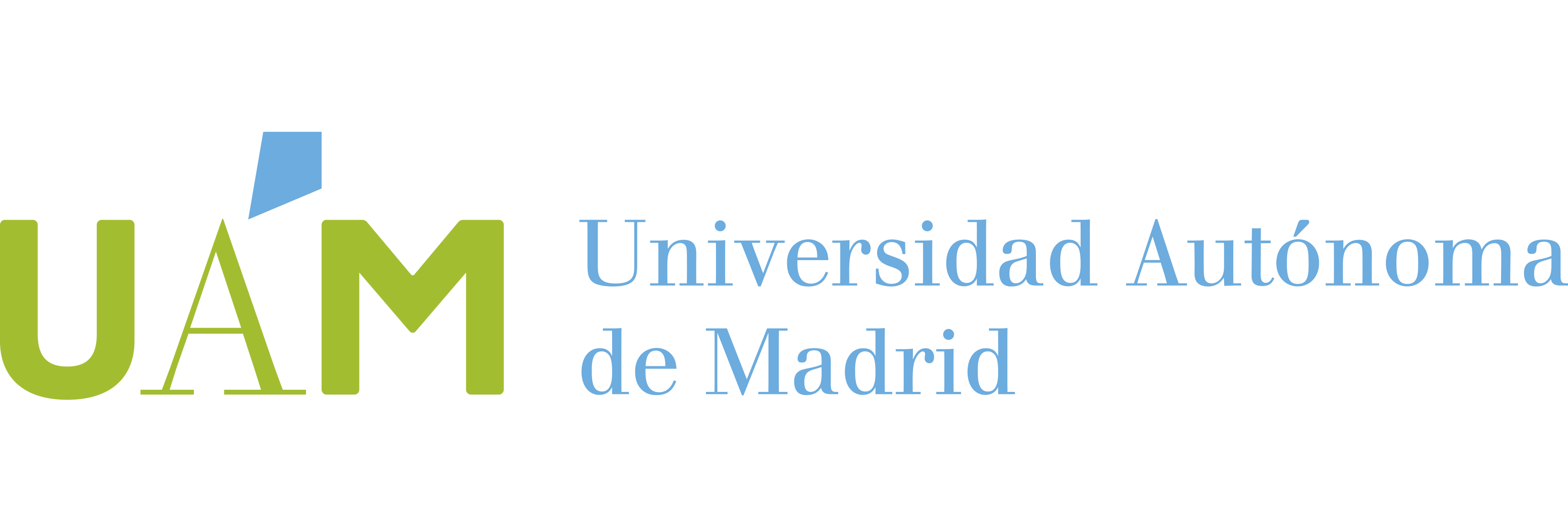}
\qquad
\includegraphics[height=0.1\textheight]{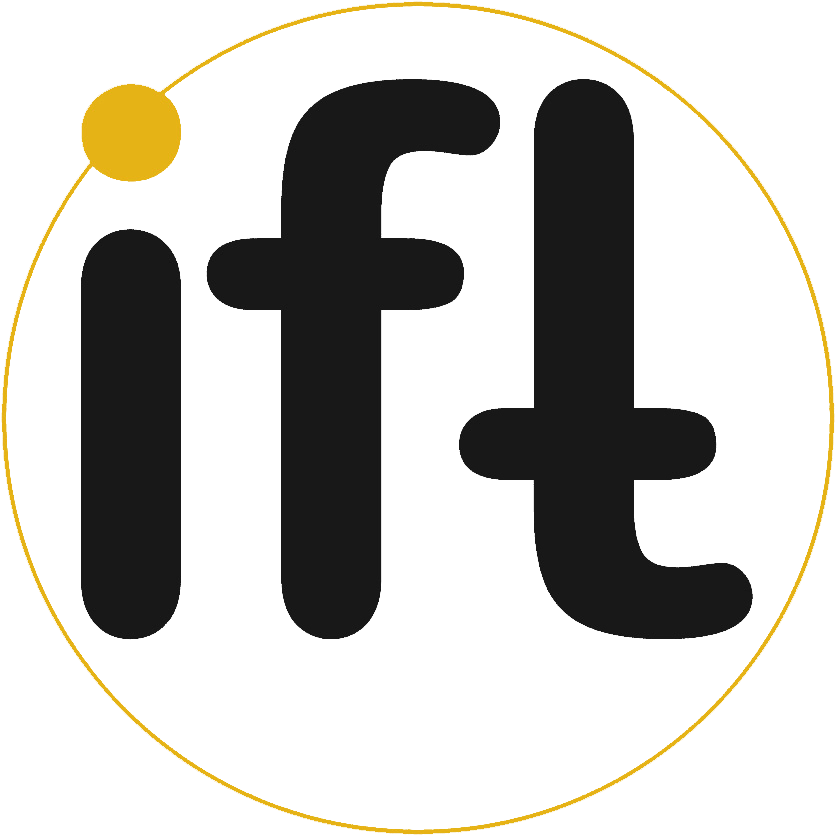}
\end{figure}

\vspace{0.2cm}

\begin{center}
{\large Madrid, septiembre de 2019}
\end{center}

\newpage

%\newpage
%\thispagestyle{empty}
%\phantom{lala}

%\newpage
%\thispagestyle{empty}
%\phantom{lala}

\chapter*{}
\begin{samepage}

\begin{flushright}
\textit{There are more things in heaven and earth, Horatio, \\ than are dreamt of in your philosophy.}\\
\small
William Shakespeare (Hamlet)
\normalsize
\end{flushright}
\vspace{0.2cm}

\Large
\emph{Agradecimientos}
\normalsize

\vspace{0.5cm}
\thispagestyle{empty}

Esta tesis no habría sido posible si durante estos cuatro años no hubiera contado con la ayuda de gente extraordinaria. Ha sido toda una suerte poder realizar mi doctorado en el grupo de Tom\'as Ort\'in, quien ha sido un verdadero mentor para m\'i, y no s\'olo en el campo de la supergravedad y la ciencia en general, sino en todos los ámbitos (echaré de menos las discusiones sobre política). Gracias, Tomás, por todo lo que me has ense\~nado. Por otra parte, Pablo Bueno me introdujo en el mundillo de las higher-order gravities y la holograf\'ia, y juntos hemos formado un equipo imparable. Ha sido un placer trabajar contigo y te doy las gracias por haberme ayudado a progresar. En el mismo sentido, debo expresar mi gratitud hacia Patrick Meessen por su incondicional apoyo.

Pero a parte de tener buenos mentores, he sido afortunado al encontrarme con grandes amigos y colaboradores.  Esta es la lista de excelentes científicos con los que he tenido el placer de trabajar (adem\'as de los mencionados arriba):  Gustavo Arciniega, Samuele Chimento, Jos\'e D. Edelstein, Frederik Goelen, Robie A. Hennigar, Thomas Hertog, Luisa G. Jaime, \'Oscar Lasso, Robert B. Mann, Hugo Marrochio, Vincent S. Min, Javier Moreno, \'Angel Murcia, Pedro F. Ram\'irez, Alejandro Ruip\'erez, Camilla Santoli, Bert Vercnocke y Manus R. Visser. Pero tampoco me puedo olvidar de los que simplemente me han dado su amistad: Iván M. Soler,  Nieves López, Verónica P\'erez, Tamara Vega, Carlos S. Shahbazi, Diego Medrano, Luca Romano... y muchos más.  Gracias a todos, pues mi estancia en Madrid habría estado vacía sin vosotros.

Tambi\'en quiero hacer un agradecimiento especial a los que me han acogido cuando he salido de Madrid. En Lovaina, gracias a Thomas Hertog y de nuevo a Pablo B. por invitarme y gracias a todos los miembros del ITF por hacerme sentir como en casa. 
El duro invierno en el Perimeter Institute habría sido más difícil de llevar sin la calidad de la gente que allí encontré, y sobre todo quiero acordarme de Robie Hennigar y Hugo Marrochio, con quienes hice buenas migas y un artículo. También estoy agradecido a Rob Myers, quien hizo las veces de tutor durante mi estancia, y a Robb Mann por su hospitalidad y su generosidad. Igualmente, fue un placer poder visitar Santiago de Compostela gracias a la invitación de José Edelstein. 

Finalmente, doy las gracias a mis padres, Antonio y Cecilia, quienes me dieron una buena educación y me inculcaron la pasión por la ciencia. Esta tesis se la dedico a ellos. Extiendo este agradecimiento a mis hermanas Cecilia y Rosa, y a toda mi familia en general. Y por supuesto, agradezco a mi novia Mar\'ia Dolores el apoyo que me ha dado y la paciencia que ha tenido conmigo durante estos años.

Esta tesis también ha sido posible gracias a la financiación que he recibido por parte de la Fundación ``la Caixa'' a través de una beca ``la Caixa - Severo Ochoa'' asociada al Instituto de Física Teórica UAM-CSIC.

\vspace{\fill}

\begin{minipage}[b]{\textwidth}
%\begin{flushright}
\hskip7cm Madrid, 17 de junio de 2019\\ 
%\end{flushright}
\end{minipage}

\end{samepage}

%\newpage 
%\thispagestyle{empty}
%\phantom{lala}

\chapter*{List of Publications\label{publicationlist}}

\thispagestyle{empty}
The following articles, some of which are unrelated to the content of this thesis, were published by the candidate during the realization of this work:
\\~\\
\begin{enumerate}
\item
 P.~Bueno, P.~A.~Cano, J.~Moreno and \'A.~Murcia,
  \textbf{``All higher-curvature gravities as Generalized quasi-topological gravities,''}
  \rdd{JHEP {\bf 1911} (2019) 062}.\\
  \arxiv{1906.00987} [hep-th].
  
\item 
  P.~A.~Cano and A.~Ruip\'erez,
  \textbf{``Leading higher-derivative corrections to Kerr geometry,''}
   \rdd{JHEP {\bf 1905} (2019) 189}. 
  \arxiv{1901.01315} [gr-qc].

\item 
  G.~Arciniega, P.~Bueno, P.~A.~Cano, J.~D.~Edelstein, R.~A.~Hennigar and L.~G.~Jaime,
 \textbf{ ``Geometric Inflation,''}
  \arxiv{1812.11187} [hep-th].

\item 
  P.~A.~Cano, P.~F.~Ramirez and A.~Ruiperez,
  \textbf{``The small black hole illusion,''}
  \arxiv{1808.10449} [hep-th].

\item 
  P.~A.~Cano, S.~Chimento, P.~Meessen, T.~Ortin, P.~F.~Ramirez and A.~Ruiperez,
  \textbf{``Beyond the near-horizon limit: Stringy corrections to Heterotic Black Holes,''}
  \rdd{JHEP {\bf 1902} (2019) 192}.
  \arxiv{1808.03651} [hep-th].

\item 
  P.~Bueno, P.~A.~Cano, R.~A.~Hennigar and R.~B.~Mann,
  \textbf{``Universality of squashed-sphere partition functions,''}
   \rdd{Phys.\ Rev.\ Lett.\  {\bf 122} (2019) no.7,  071602}.\\
  \arxiv{1808.02052} [hep-th].

\item  
  P.~Bueno, P.~A.~Cano, R.~A.~Hennigar and R.~B.~Mann,
  \textbf{``NUTs and bolts beyond Lovelock,''}
  \rdd{JHEP {\bf 1810} (2018) 095}.
  \arxiv{1808.01671} [hep-th].

\item 
  P.~A.~Cano, S.~Chimento, T.~Ortin and A.~Ruiperez,
  \textbf{``Regular Stringy Black Holes?,''}
  \rdd{Phys.\ Rev.\ D {\bf 99}, no. 4, 046014 (2019)}.
  \arxiv{1806.08377} [hep-th].

\item 
  P.~A.~Cano and T.~Ort\'in,
  \textbf{``The structure of all the supersymmetric solutions of ungauged $\mathcal{N} = (1,0),d=6$ supergravity,''}
  \rdd{Class.\ Quant.\ Grav.\  {\bf 36} (2019) no.12,  125007.}
  \arxiv{1804.04945} [hep-th].

\item 
  P.~A.~Cano, R.~A.~Hennigar and H.~Marrochio,
  \textbf{``Complexity Growth Rate in Lovelock Gravity,''}
  \rdd{Phys.\ Rev.\ Lett.\  {\bf 121}, 121602 (2018)}.
  \arxiv{1803.02795} [hep-th]. 
   
\item 
  P.~A.~Cano, P.~Meessen, T.~Ort\'in and P.~F.~Ram\'irez,
  \textbf{``$\alpha'$-corrected black holes in String Theory,''}
  \rdd{JHEP {\bf 1805} (2018) 110}.
  \arxiv{1803.01919} [hep-th].
 
\item 
  P.~A.~Cano,
  \textbf{``Lovelock action with nonsmooth boundaries,''}
  \rdd{Phys.\ Rev.\ D {\bf 97} (2018) 104048}.
  \arxiv{1803.00172} [gr-qc].
 
\item 
  P.~Bueno, P.~A.~Cano and A.~Ruip\'erez,
  \textbf{``Holographic studies of Einsteinian cubic gravity,''}
  \rdd{JHEP {\bf 1803} (2018) 150}.
  \arxiv{1802.00018} [hep-th].

\item 
  P.~Bueno, P.~A.~Cano, F.~Goelen, T.~Hertog and B.~Vercnocke,
    \textbf{``Echoes of Kerr-like wormholes,''}
  \rdd{Phys.\ Rev.\ D {\bf 97} (2018) no.2,  024040}.
  \arxiv{1711.00391} [gr-qc].
 
\item 
  P.~A.~Cano and T.~Ort\'in,
  \textbf{``Non-perturbative decay of Non-Abelian hair,''}
  \rdd{JHEP {\bf 1712} (2017) 091}.
  \arxiv{1710.05052} [hep-th].

\item 
  P.~Bueno and P.~A.~Cano,
  \textbf{``Universally stable black holes,''}
  \rdd{Int.\ J.\ Mod.\ Phys.\ D {\bf 26} (2017) no.12,  1743024}.
 \textit{This essay received an Honorable Mention in the 2017 Essay Competition of the Gravity Research Foundation.}

\item 
  P.~Bueno and P.~A.~Cano,
   \textbf{``Universal black hole stability in four dimensions,''}
   \rdd{Phys.\ Rev.\ D {\bf 96} (2017) no.2,  024034}.
  \arxiv{1704.02967} [hep-th].

\item 
  P.~A.~Cano, P.~Meessen, T.~Ort\'in and P.~F.~Ram\'irez,
   \textbf{``Non-Abelian black holes in string theory,''}
      \rdd{JHEP {\bf 1712} (2017) 092}.
  \arxiv{1704.01134} [hep-th].

\item 
  P.~A.~Cano, T.~Ort\'in and P.~F.~Ram\'irez,
  \textbf{``A gravitating Yang-Mills instanton,''}
  \rdd{JHEP {\bf 1707} (2017) 011}.
  \arxiv{1704.00504} [hep-th].

\item 
  P.~Bueno and P.~A.~Cano,
  \textbf{``On black holes in higher-derivative gravities,''}
  \rdd{Class.\ Quant.\ Grav.\  {\bf 34} (2017) no.17,  175008}.
  \arxiv{1703.04625} [hep-th].

\item 
Pablo~Bueno, Pablo~A.~Cano, Vincent S. Min and Manus R. Visser,
 \textbf{``Aspects of general higher-order gravities,''}
  \rdd{Phys. Rev. D {\bf 95} (2017) 044010}.
\arxiv{1610.08519} [hep-th].

\item 
Pablo~Bueno and Pablo~A.~Cano,
 \textbf{``Four-dimensional black holes in Einsteinian cubic gravity,''}
 \rdd{Phys. Rev. D {\bf 94} (2016) 124051 }.
\arxiv{1610.08019} [hep-th].

\item 
Pablo~Bueno and Pablo~A.~Cano,
 \textbf{``Einsteinian cubic gravity,''}
  \rdd{Phys. Rev. D {\bf 94} (2016), 104005}.
\arxiv{1607.06463} [hep-th].

\item 
  P.~A.~Cano, T.~Ort\'in and C.~Santoli,
  \textbf{``Non-Abelian black string solutions of $ \mathcal{N} = (2,0), d = 6$ supergravity,''}
  \rdd{JHEP {\bf 1612} (2016) 112}.
  \arxiv{1607.02595} [hep-th]].

\item 
Pablo~Bueno, Pablo~A.~Cano, \'Oscar~Lasso~A. and Pedro~F.~Ram\'irez,
 \textbf{``f(Lovelock) theories of gravity,''}
\rdd{JHEP \textbf{1604} (2016) 028}.
\arxiv{1602.07310} [hep-th].

\end{enumerate}

\chapter*{Abstract\label{abstract}}

\thispagestyle{empty}

Higher-curvature theories of gravity are extensions of General Relativity (GR) that arise in effective descriptions of quantum gravity theories, such as String Theory. While at low energies the behaviour of the gravitational field in higher-curvature gravities is almost indistinguishable from the one predicted by GR, the differences can be dramatic in extreme gravity scenarios, such as in the case of black holes (BHs). It is therefore an exciting task to study how black hole geometries are modified by higher-curvature corrections, with the hope that some problematic characteristics of BHs observed in GR could be improved, providing hints on the effects of an underlying UV-complete theory of gravity.  However, there are some difficulties associated with higher-derivative theories, such as the existence of instabilities, propagation of ghost modes, or simply the extreme complexity of the differential equations governing the dynamics of the gravitational field. In this thesis we identify a new family of higher-curvature gravities that avoid some of these problems. Known as \emph{Generalized quasi-topological gravities} (GQGs), such theories represent extensions of GR that are free of instabilities and  ghosts at the linear level, and whose equations of motion for static, spherically symmetric spaces acquire a sufficiently simple form so as to allow for the non-perturbative study of black hole solutions. The simplest non-trivial member of this family in four dimensions --- and also  the first one to be discovered --- is known as \emph{Einsteinian cubic gravity}, and it will have a starring role in this thesis.  Besides the intrinsic interesting properties of GQGs, we argue that they capture the most general higher-derivative correction to GR when field redefinitions are included into the game.
Then, we use these theories to study the non-perturbative corrections to the Schwarzschild black hole in four dimensions and we focus our attention on the modified thermodynamic relations. The most impressive prediction of these theories is that the Hawking temperature of static, neutral black holes vanishes in the zero-mass limit instead of diverging --- which is the result predicted by GR. As a consequence, small black holes become thermodynamically stable and their evaporation process takes an infinite time. 
In addition, higher-curvature gravities find very rewarding applications in the Anti-de Sitter/Conformal Field Theory (AdS/CFT) correspondence, a duality that relates a classical theory of gravity in AdS space to a quantum field theory that lives in the boundary of AdS. In this context, holographic higher-curvature gravities are useful toy models that we can use, for instance, to extract general lessons about the dynamics of CFTs or to question the generality of the predictions of holographic Einstein gravity. In this thesis we explore the holographic applications of four-dimensional Einsteinian cubic gravity, which provides a toy model for a non-supersymmetric holographic CFT in three dimensions. In addition, we construct new Euclidean-AdS-Taub-NUT solutions, which are dual to conformal field theories placed on squashed spheres. Using these results, we derive a universal expression for the expansion of the free energy of three-dimensional CFTs on squashed spheres up to cubic order in the deformation parameter.

\cleardoublepage

\thispagestyle{empty}
\pagenumbering{gobble}
\tableofcontents
\clearpage

%
%\newpage
%\thispagestyle{empty}
%\mbox{}

%anadir esto a la version de pagina pequena
%\newpage
%\thispagestyle{empty}
%\phantom{lala}

%%%%%%%%%%%%%%%%%%%%%%%%%%

\newpage

%----------------------------------------------------------------------%

\pagestyle{fancy}
\fancyhead{}
\fancyfoot{}

\fancyhead[LE,RO] {\itshape\nouppercase\leftmark}
\fancyfoot[C]{\thepage}
\renewcommand{\headrulewidth}{0.3pt}
\renewcommand{\footrulewidth}{0pt}

%----------------------------------------------------------------------%

\pagenumbering{arabic}

\chapter{Introduction}
%%%%%%%%%%%%%%%%%%%%%%%%%%%%%%%%%%

\section{Higher-curvature gravity}

General Relativity (GR) describes gravity as spacetime curvature. Einstein's field equations, that rule the dynamics of the gravitational field, can be derived from the Einstein-Hilbert (EH) action
\begin{equation}\label{eq:EH}
S=\frac{1}{16\pi G}\int d^{4}x\sqrt{|g|}R\, ,
\end{equation}
which is essentially the \textit{simplest} non-trivial covariant action one can write for the metric tensor. This beautiful theory has passed a large number of experimental tests and it is broadly accepted as the correct description of the gravitational interaction. Amongst the most impressive confirmations of GR predictions we must mention the recent detection of gravitational waves coming from black hole and neutron star binaries \cite{Abbott:2016blz,Abbott:2016nmj,TheLIGOScientific:2016src,TheLIGOScientific:2017qsa,Abbott:2017vtc,Abbott:2017oio} and the first image of a black hole captured by the Event Horizon Telescope \cite{Akiyama:2019cqa}. 

However, these experiments are only able to probe gravity in situations of relatively small curvature, and there are good reasons to think that GR will be modified when the curvature of the spacetime exceeds a certain value. One of these reasons is that GR seems to be incompatible with quantum mechanics. In fact, the problem of reconciling the principles of quantum mechanics with gravitation is a long standing one. The first attempts to apply the standard quantization procedures to the Einstein-Hilbert action determined that this theory is non-renormalizable \cite{tHooft:1974toh,Deser:1974cz,Deser:1974xq}. As a consequence of this result, it is usually accepted that GR should be regarded as an effective theory, since in that case renormalizability is not necessary.\footnote{However, GR could be non-perturbatively renormalizable, and this is the perspective behind the asymptotic safety proposal --- see \cite{Niedermaier:2006wt} and references therein.} An astounding example of this is provided by Fermi's theory of beta decay, which successfully describes the observed phenomena despite being non-renormalizable. It was later understood that Fermi's interaction arises from electroweak interactions mediated by the W$^{\pm}$ bosons in the Standard Model. In the regime where energies are much smaller than the mass of the gauge bosons, Fermi's theory describes the phenomena of beta decay with great accuracy, but deviations are found when the electroweak scale is reached. 
In a similar way, General Relativity is expected to be an effective description of an underlying UV-complete theory of gravity, and, consequently, it will be modified at some energy scale, such as, for instance (but not necessarily), the Planck scale. 

From this point of view, the Einstein-Hilbert action \req{eq:EH} would just be the leading term in an effective theory which is expected to contain additional subleading terms. According to the prescriptions of Effective Field Theory, one should add to the action all the terms that are consistent with the symmetries of the theory \cite{Weinberg:1995mt}. 
In the case of gravity, we take the metric tensor $g_{\mu\nu}$ as the fundamental field, and our guiding principle is diffeomorphism invariance.\footnote{See \textit{e.g.} \cite{Li:2010cg,Horava:2009uw,deRham:2010kj,Olmo:2011uz} for other possible extensions of GR.} This means that the Lagrangian must be a scalar quantity formed from the Riemann tensor and its covariant derivatives --- which are the only tensors one can construct out of the metric.  In this sense, the Einstein-Hilbert Lagrangian is the simplest possible choice --- besides the addition of a cosmological constant --- and the only one that contains two derivatives of the metric. The additional terms we can include take the form of higher-derivative operators, which provide a modification of GR in the UV. The lowest-order terms of that type we can add are quadratic in the curvature, so that the leading-order corrections to GR take the form
\begin{equation}\label{eq:HDG}
S=\frac{1}{16\pi G}\int d^{4}x\sqrt{|g|}\left[R+\frac{1}{M_{\rm \ssc new}^2}\left(\alpha_1R^2+\alpha_2R_{\mu\nu}R^{\mu\nu}+\alpha_3R_{\mu\nu\rho\sigma}R^{\mu\nu\rho\sigma}\right)+\ldots\right]\, .
\end{equation}
Here we are assuming that the expansion parameter is some mass scale $M_{\rm \ssc new}$ --- the scale of new physics --- which should not necessarily coincide with Planck's mass. Terms of higher order in the derivative expansion are suppressed with higher powers of $M_{\rm \ssc new}$ and their effect is increasingly irrelevant in the IR. Therefore, higher-derivative gravities\footnote{Throughout the text we use equivalently the denominations ``higher-derivative gravity'', ``higher-curvature gravity'' and ``higher-order gravity''.}  arise naturally in the framework of Effective Field Theory, but we expect that a UV-complete theory of gravity predicts which precise terms appear in this expansion. Nowadays, there are several frameworks that attempt to unify gravity and quantum mechanics, but perhaps the most prominent one is String Theory (ST) \cite{Schwarz:1982jn,Green:1987sp,Green:1987mn,Polchinski:1998rq,Polchinski:1998rr}. Although the underlying principles of ST depart largely from a field theory formulation of gravity, it is a definite prediction of this theory that, at low enough energies, gravity is described by the Einstein-Hilbert action improved with higher-derivative terms \cite{Gross:1986mw,Gross:1986iv,Grisaru:1986vi,Grisaru:1986px,Bergshoeff:1989de,Green:2003an,Frolov:2001xr}. Therefore, higher-derivative gravities stand as valuable effective field theories that we can use in order to learn about quantum gravity effects.

From a different perspective, it is possible to consider higher-curvature gravities as classical alternatives to GR \cite{Stelle:1977ry}. As a matter of fact, the gravitational action becomes renormalizable when it is supplemented with quadratic curvature terms \cite{Stelle:1976gc}, and higher-order terms are even super-renormalizable.  Thus, if our starting classical theory is a higher-derivative gravity rather than General Relativity, we get a renormalizable theory, and in this sense higher-derivative gravities could be considered as quantum gravity candidates.  However, renormalizability comes at a price. Indeed, one peculiarity of gravitational theories is that higher-derivative operators, almost inevitably,\footnote{The only exception to this is provided by Lovelock-Lanczos theories, that we review in Section \ref{sec:hogexamples}.} yield equations of motion of higher-order in derivatives of the metric. In fact, the Riemann tensor contains second derivatives of the metric and it is a well-known fact that non-degenerate Lagrangians containing second (or higher) derivatives of a field give rise to instabilities. Ostrogradsky's theorem \cite{Ostrogradsky:1850fid,Woodard:2015zca} shows that for this type of theories the Hamiltonian is unbounded from below, with the corresponding appearance of instabilities and negative energy modes. This is precisely what happens in the gravitational action \req{eq:HDG}. 
When the propagator of this theory is analyzed, one finds that additional degrees of freedom that are not present in GR appear. Along with a massless spin-2 graviton, the spectrum of \req{eq:HDG} contains a scalar mode and a massive graviton of mass $m_{g}^2=-M_{\rm \ssc new}^2/(\alpha_2+4\alpha_3)$ \cite{Stelle:1977ry}. Independently of the sign of the higher-derivative couplings, the propagator of the massive graviton always comes with the ``wrong'' sign, so that it carries negative energy. When one tries to quantize the theory, this mode produces non-normalizable states and therefore it becomes a \emph{ghost} \cite{Alvarez-Gaume:2015rwa}. 
This inconsistency implies that higher-derivative gravities cannot be considered complete theories so easily. However, it should be noted that not all the higher-derivative gravities contain ghost modes in their spectrum --- see \eg \cite{Lovelock1,Lovelock2,Woodard:2006nt,Sotiriou:2008rp,Biswas:2011ar}. Thus, at least some of these theories can be considered as self-consistent classical alternatives to GR. 

Independently of the origin of the higher-derivative corrections, it is interesting to look at the effects of these terms on gravity, which sometimes lead to very intriguing conclusions. 
Higher-curvature corrections modify the dynamics of the gravitational field in extreme situations, and one of the greatest hopes is that they will be able to improve the behaviour of gravity precisely where GR fails, providing signs of an underlying UV-complete theory. In this respect, there are two instances where gravity is especially violent. One of them is the Big-Bang singularity, which a quantum theory of gravity should be able to resolve. There is evidence that higher-curvature corrections are indeed able to improve the Big-Bang singularity on their own, replacing it by an infinite period of cosmic inflation. This is accomplished \eg by Starobinsky's model of inflation \cite{Starobinsky:1980te}, which introduces an $R^2$ term in the gravitational action.  Furthermore, we have recently learned that inflation can be realized by a different mechanism known as \emph{Geometric inflation}, which seems to be a general phenomenon of higher-derivative gravity \cite{Arciniega:2018fxj,Cisterna:2018tgx,Arciniega:2018tnn} --- see also \cite{Hohm:2019jgu}.  The other instance that comes to mind when one thinks of extreme gravity is a black hole. Here, one hopes that a UV-complete theory of gravity should be able to resolve the singularity in the black hole interior. Higher-derivative corrections are able to improve the divergence\cite{Boulware:1985wk,PabloPablo2}, and in some exceptional cases --- that usually involve the introduction of specific matter fields --- can resolve the singularity \cite{Biswas:2011ar,Olmo:2015axa,Menchon:2017qed,Cano:2018aod,delaCruz-Dombriz:2018aal}, producing a \emph{regular black hole} \cite{Dymnikova:1992ux,Borde:1996df,Frolov:2016pav}. However, the corrections can also significantly modify the properties of a black hole at the level of the horizon if its mass is small enough. Assuming that the higher-derivative operators are weighted by an energy scale $M_{\rm \ssc new}$, the deviations from GR will become important when $M\sim M_{\rm \ssc P}^2/M_{\rm \ssc new}$.\footnote{This is the scale at which the horizon radius becomes of the size of $1/M_{\rm \ssc new}$} The behaviour of black holes of smaller masses generically departs greatly from the GR prediction. For example, it is known that the divergence of the Hawking temperature in the limit $M\rightarrow 0$ of higher-dimensional $(D\ge5)$ black holes can be cured by higher-derivative interactions \cite{Myers:1988ze,Cai:2001dz} --- we review these results in Sec.~\ref{sec:introbhshdg}. One of the main results of this thesis is the proof that a similar conclusion holds for four-dimensional black holes.

Finally, due to their relation with String Theory, higher-derivative gravities play a prominent role in the context of the AdS/CFT correspondence \cite{Maldacena,Gubser,Witten}. We will review this correspondence in Section~\ref{sec:holoint}, but in short, it states that there is an equivalence between a classical theory of gravity in anti-de Sitter (AdS) space and a strongly coupled, large $N$ Conformal Field Theory (CFT) which lives on the boundary of AdS. According to the original form of the duality derived from String Theory, higher-curvature corrections in the bulk geometry are dual to finite $N$ and finite coupling corrections in the CFT, so that these terms allow us to probe CFTs in more realistic regimes. However, the AdS/CFT duality is nowadays understood as a general principle, and in this context, higher-curvature terms have proven to be valuable tools that can be used, for instance, to establish statements about general CFTs, whose validity is usually difficult to determine from field theory considerations --- see \eg \cite{Myers:2010tj,Myers:2010xs,Mezei:2014zla,Bueno1,Bueno2}.  In this thesis we will provide a new example of this type of application of higher-derivative gravity --- see Chapter~\ref{Chap:8}.

%Next, we review in more detail some of the topics we have mentioned.  

\subsection{Higher-curvature corrections from String Theory}\label{sec:stringyintro}
In order to further motivate the study of higher-curvature gravities, it is interesting to understand how gravity emerges in String Theory.\footnote{See \textit{e.g.}  \cite{Schwarz:1982jn,Green:1987sp,Green:1987mn,Polchinski:1998rq,Polchinski:1998rr,Becker:2007zj} for an introduction to String Theory.}
According to superstring theories, the basic constituents of matter are fundamental strings that live in some target space which is in principle undetermined. The dynamics of these strings is ruled by a non-linear sigma model defined on the worldsheet swept out by the string, where the variables are the embedding functions of the string on the spacetime $X^{\mu}$ \cite{Polyakov:1981rd}. The worldsheet action contains couplings between these variables and background fields defined in the embedding spacetime: these include the metric $\hat g_{\mu\nu}$, the Kalb-Ramond 2-form $\hat B_{\mu\nu}$ and the dilaton $\hat\phi$. 
%Conformal invariance and supersymmetry imply a large number of restrictions on the possible theories that we can define, and it leads to the construction of the different superstring theories: type I, type IIA, type IIB, heterotic $\mathrm{SO}(32)$ and heterotic $\mathrm{E}_8\times\mathrm{E}_8$. 
Consistency of the worldsheet theory at the quantum level imposes important restrictions on these background fields and on the embedding spacetime itself.  When the theory is quantized, negative norm states appear, but this can be avoided if the target space has $D=10$ dimensions --- in that case many of the negative norm states become spurious (they acquire vanishing norm) and the theory is \emph{critical}. On the other hand, in order to preserve conformal invariance --- which is present at the classical level --- one must impose the vanishing of the beta functions, which leads to several constraints on the spacetime geometry as well as on the rest of the background fields \cite{Callan:1985ia,Sen:1985eb,Callan:1986jb}. At leading order in the worldsheet loop expansion, these constraints coincide with the equations of motion of ten-dimensional supergravity theories, whose action always contains the common Neveu-Schwarz sector \cite{Ortin:2015hya}
\begin{equation}\label{eq:sugra}
S_{\rm sugra}=\frac{g_{s}^2}{16\pi G^{(10)}}\int d^{10}x\sqrt{|\hat{g}|}e^{-2\hat{\phi}}\left[\hat{R}+4\left(\partial\hat{\phi}\right)^2-\frac{1}{2 \cdot 3!} \hat{H}^2+\ldots\right]\, .
\end{equation}
Here, $\hat H=d\hat{B}$ is the field strength of the Kalb-Ramond 2-form and $\hat{R}$ is the Ricci scalar of the ten-dimensional metric $\hat g_{\mu\nu}$. In addition, $g_{s}$ is the string coupling constant, related to the asymptotic value of the dilaton according to $g_s=\langle e^{\phi}\rangle$ in solutions that asymptote a vacuum, while the ten-dimensional Newton's constant reads
\begin{equation}
G^{(10)}=8\pi^6 g_s^2\ell_s^8\, ,
\end{equation}
where $\ell_s$ is the string length.  Therefore, String Theory is telling us that Einstein's field equations (with specific matter couplings) arise as a consistency condition in order for the fundamental strings to behave correctly. 
An equivalent way to derive the action \req{eq:sugra} entails the computation of scattering amplitudes in String Theory for massless modes, such as gravitons. Then, the effective action is constructed in a way that it reproduces the S-matrix in the $\alpha'\rightarrow 0$ limit (where $\alpha'=\ell_s^2$) \cite{Scherk:1971xy,Neveu:1971mu,Scherk:1974mc,Gross:1986mw}.

Now, in order to produce four-dimensional (or other lower-dimensional) theories, one has to compactify the action \req{eq:sugra} assuming a decomposition of the spacetime of the form $\mathcal{M}_{10}=\mathcal{M}_{4}\times \mathcal{Y}_{6}$, where $\mathcal{Y}_{6}$ is a compact six-dimensional space.\footnote{An alternative to compactification is provided by Randall-Sundrum braneworld scenarios \cite{Randall:1999ee,Randall:1999vf}.} In this way, one obtains four-dimensional Einstein gravity coupled to different types of matter fields --- the possibilities for the matter sector are almost endless due to the freedom in choosing $\mathcal{Y}_{6}$, but  the gravitational sector is always ruled by Einstein's field equations $G_{\mu\nu}=8\pi G^{(4)} T_{\mu\nu}$.  Let us also note that the four-dimensional Newton's constant reads
\begin{equation}
G^{(4)}=\frac{8\pi^6 g_s^2\ell_s^8}{V\left(\mathcal{Y}_{6}\right)}\, ,
\end{equation}
where $V\left(\mathcal{Y}_{6}\right)$ is the volume of the compact dimensions.  Thus, if the typical length scale of $\mathcal{Y}_{6}$ is of order, say $\ell_{c}$, then the four-dimensional Planck's length is of the order of $\ell_{\rm \ssc P}\sim g_s \ell_s^4/\ell_c^3$.  

Higher loop contributions to the beta functions provide corrections to the constraints on the background fields, and as a result, the effective action \req{eq:sugra} receives additional contributions in the form of higher-derivative terms \cite{Grisaru:1986vi,Grisaru:1986px} --- alternatively, those terms can be computed by studying higher-point scattering amplitudes \cite{Gross:1986mw,Gross:1986iv}.
In general, the low-energy  effective actions predicted by String Theory are a double series in the string coupling constant $g_{s}$  and in the parameter $\alpha'=\ell_{s}^2$,
\begin{equation}\label{eq:effstring}
S_{\rm eff}=\sum_{n=0}^{\infty}\sum_{k=0}^{\infty} (\alpha')^n g_s^{2k}S_{n,k}
\end{equation}
The leading term in this expansion is the two-derivative supergravity action \req{eq:sugra}, while the rest of the terms contain higher-derivative interactions for the metric and for the rest of the fields. Thus, when the action is compactified and truncated down to $D=4$ (or to other lower dimension) one finds that the dynamics of gravity is not given by the Einstein-Hilbert action anymore, but by a higher-derivative gravity. In particular, $\alpha'$ corrections modify the gravitational interaction at a distance $\ell_s$, so that gravity behaves differently when the size of the strings becomes relevant. It is important to note that this scale is not necessarily the same as Planck's one; in fact, according to our previous estimation we have $\ell_s/\ell_{\rm \ssc P}= g_{s}^{-1} \ell_{c}^3/\ell_{s}^3$, so that the possibility of corrections below Planck's scale should not be discarded. 

The computation of the higher-derivative terms in the expansion \req{eq:effstring} is extremely involved and only the few first ones have been computed for the different superstring theories. We show here a couple of examples.

\subsubsection{Quadratic terms in Heterotic String Theory}
The effective action of the Heterotic Superstring receives higher-curvature corrections at first and higher orders in the $\alpha'$ expansion.  At first order, the ten-dimensional action is given by\footnote{Here we are truncating the Yang-Mills fields for simplicity.} \cite{Bergshoeff:1989de}
%\footnote{With respect to the conventions of \cite{Cano:2018qev, Chimento:2018kop}, here we are using mostly plus signature $g_{\mu\nu}\rightarrow - g_{\mu\nu}$ and the definition of the Riemann tensor differs by a minus sign, i.e. $R_{\mu\nu\rho}{}^\sigma\rightarrow -R_{\mu\nu\rho}{}^\sigma$.}
\begin{equation}\label{eq:actionheterotic}
S_{\rm Het}=\frac{g_{s}^2}{16\pi G^{(10)}}\int d^{10}x \sqrt{|\hat g|} \, e^{-2\hat \phi}\left[\hat R+4\left(\partial\hat\phi\right)^2-\frac{1}{2 \cdot 3!}\hat  H^2 -\frac{\alpha'}{8}\tensor{\hat R}{_{(-)\mu\nu}^{a}_{b}} \tensor{\hat R}{_{(-)}^{\mu\nu b}_{a}}\right]+\dots\, ,
\end{equation}
where $\hat R_{(-)}{}^a{}_b$ is the curvature of the torsionful spin-connection $\hat \Omega_{(-)}{}^a{}_b=\hat \omega^a{}_b-\frac{1}{2}\hat H_{\mu}{}^a{}_b dx^\mu$. In addition, the Kalb-Ramond 3-form $H$ receives corrections due to the modified Bianchi identity
\begin{equation}\label{eq:Bianchiid}
d\hat H=\frac{\alpha'}{4}\hat R_{(-)}{}^a{}_b\wedge \hat R_{(-)}{}^b{}_a+\ldots\ ,
\end{equation}
%The curvature-squared term is needed in order to supersymmetrize the action at first order in $\alpha'$, which otherwise would not be supersymmetric due to the presence of the Chern-Simons terms in the definition of the 3-form field strength $H$ (see \cite{Bergshoeff:1989de} for more details). 
This theory can be compactified and truncated in several ways in order to produce lower-dimensional dynamics. 
For instance, if we compactify (\ref{eq:actionheterotic}) on a six torus and truncate all the Kaluza-Klein degrees of freedom we get the following action \cite{Cano:2019ore}
\begin{equation}
\begin{aligned}
S=\frac{1}{16\pi G^{(4)}}\int d^{4}x \sqrt{|g|} \left[R-\frac{1}{2}\left(\partial\phi^1\right)^2-\frac{1}{2} \left(\partial\phi^2\right)^2 -\frac{\alpha'}{8}\phi^1\mathcal X_4+\frac{\alpha'}{8}\phi^2\,R_{\mu\nu\rho\sigma}\tilde R^{\mu\nu\rho\sigma}\right]\ ,
\end{aligned}
\end{equation}
where  we also used field redefinitions in order to upgrade the Riemann squared term to the Gauss-Bonnet (GB) density $\mathcal{X}_{4}=R^{\mu\nu\rho\sigma}R_{\mu\nu\rho\sigma}-4R_{\mu\nu}R^{\mu\nu}+R^2$ \cite{Metsaev:1987zx}.  The other quadratic term is the Pontryagin density, where $\tilde R_{\mu\nu\rho\sigma}=\frac{1}{2}\epsilon_{\mu\nu\alpha\beta}\tensor{R}{^{\mu\nu}_{\rho\sigma}}$ is the dual Riemann tensor. We note that the two scalars --- the axion and the dilaton --- cannot be truncated, since the equations imply that they have a non-trivial profile whenever the curvature is non-vanishing. Since the GB and Pontryagin densities are topological in four dimensions, they only modify Einstein's field equations thanks to the scalar couplings. However, in higher dimensions the GB term is not topological and it alone gives the leading $\alpha'$ correction to Einstein gravity \cite{Boulware:1985wk}. 

\subsubsection{Quartic terms in Type IIB String Theory}

In the case of  type IIB Superstring Theory, the leading $\alpha'$ corrections to the usual supergravity action appear at order  $(\alpha')^3$ and they are quartic in the curvature \cite{Gross:1986iv,Grisaru:1986px}. Schematically we have 
\begin{equation}\label{wtf}
   S_{\text{IIB}}=S_{\text{IIB}}^{(0)}+{\alpha'}^3 S_{\text{IIB}}^{(1)}+\ldots,
\end{equation}
where $S_{\text{IIB}}^{(0)}$ is the  two-derivative supergravity action \cite{Howe:1983sra}, and the dots stand for subleading corrections in  $\alpha'$. When the theory is considered in $\mathcal{A}_5\times\mathbb{S}^5$ where $\mathcal{A}_5$ is a negatively curved Einstein manifold, it is consistent to truncate all fields except for the metric and it is possible to write an effective action for the five-dimensional metric \cite{Myers:2008yi,Buchel:2008ae,Galante:2013wta}. This is given by  \cite{Gubser:1998nz,Buchel:2004di}
\begin{equation}\label{acads5}
 S_{\text{IIB}_{{\mathcal{A}}_5 \times \mathbb{S}^5}}[g_{\mu\nu}]=\frac{1}{16\pi G}\int d^5 x\sqrt{|g|}\left[ R+\frac{12}{\ell^2}+\frac{\zeta(3)}{8}{\alpha'}^3 W^4 \right] \, ,
\end{equation}
where $W^4$ is a particular combination of contractions of four Weyl tensors given by
\begin{equation}
    W^4=\left(W_{\mu \nu \rho \sigma}W^{\lambda \nu \rho \gamma}+\frac{1}{2}W_{\mu \sigma\mu \rho }W^{\lambda \gamma\nu \rho}\right)W\indices{^{\mu}^{\tau}_{\eta}_{\lambda}}W\indices{_{\gamma}_{\tau}^{\eta}^{\sigma}}.
\end{equation}
In different compactification schemes, such as $\mathcal{M}_{10}=\mathcal{M}_{4}\times \mathcal{Y}_{6}$, where $\mathcal{Y}_{6}$ is a non-compact Calabi-Yau manifold, type IIB theories can also produce quadratic curvature terms  \cite{Alvarez-Gaume:2015rwa}.

\subsection{Equations of motion, boundary terms and conserved charges}\label{sec:equationsintro}
We have just seen that, even within String Theory, the higher-derivative terms that one obtains depend on the type of strings (Heterotic, IIB, etc.) as well as on the compactification chosen. In addition, the scale at which these corrections appear does not necessarily coincide with the Planck scale. Thus, we will remain agnostic about the kind of terms we include in our gravitational action and about the energy scale of new physics. We take higher-derivative gravity as our starting point and our goal will be to determine some of the most relevant consequences of these theories. Hopefully, some of them will ultimately capture the UV effects in gravity predicted by String Theory, or by any other putative quantum theory of gravity. 

Our guiding principle for the construction of a metric gravitational theory is diffeomorphism invariance.  The most general diffeomorphism-invariant action that one can consider takes the form 
\begin{equation}\label{eq:actiongen}
S=\int d^{D}x\sqrt{|g|}\mathcal{L}\left(g^{\alpha\beta}, R_{\mu\nu\rho\sigma}, \nabla_{\alpha} R_{\mu\nu\rho\sigma},\nabla_{\beta}\nabla_{\alpha} R_{\mu\nu\rho\sigma}, \ldots\right)\, ,
\end{equation}
where the Lagrangian must be a scalar function formed from correct contractions of indices of the curvature tensor and its derivatives.\footnote{Additionally, one could include terms containing the Levi-Civita symbol $\epsilon_{\mu_1\ldots\mu_D}$, which generically break parity invariance. We will discard those terms under the assumption that gravity is parity invariant. In case they are kept, they are irrelevant for spherically symmetric solutions, but in turn they have interesting consequences for rotating black holes \cite{Cardoso:2018ptl,Cano:2019ore}.} %If the Lagrangian is formed from terms that contain up to $p$ covariant derivatives of the Riemann tensor, \ie $\nabla_{\alpha_1}\ldots\nabla_{\alpha_p}R_{\mu\nu\rho\sigma}$, the equations of motion generically will be of order $2p+4$. %Incrementing the order of the equations leads to the appearance of new degrees of freedom as well as to aggravating the problem of instabilities that are common to these theories. 
In addition, we will usually assume that the Lagrangian can be expanded as a polynomial in terms containing an increasing number of derivatives. These terms are weighted by inverse powers of an energy scale $M_{\rm \ssc new}$, or equivalently by the powers of the associated length scale $L=1/M_{\rm \ssc new}$. Thus, schematically we expand the Lagrangian as
\begin{equation}\label{eq:HDGintro}
\mathcal{L}=\frac{1}{16\pi G}\left[-2\Lambda+R+\sum_{n,p} \lambda_{n,p,i}L^{2n+p-2}(\nabla^p\mathcal{R}^n)_{i}\right]\, ,
\end{equation}
where $\lambda_{n,p,i}$ are dimensionless constants and $(\nabla^p\mathcal{R}^n)_{i}$ is a monomial formed out of $p$ covariant derivative operators and $n$ curvatures, while the subscript $i$ denotes a specific way of contracting the indices. One of these terms contains in total $2n+p$ derivatives and gives rise, generically, to equations of motion of order $2p+4$. For this reason, we will mainly focus on the terms with $p=0$, which in general produce fourth-order equations of motion.\footnote{However, increasing the order of the equations of motion is not always problematic: there are Lagrangians that contain an infinite number of derivatives and that give rise to well-behaved theories \cite{Biswas:2011ar,Biswas:2013cha}.}  We will refer to these theories as $\mathcal{L}$(Riemann)  or $\mathcal{L}\left(g^{\alpha\beta}, R_{\mu\nu\rho\sigma}\right)$ gravities. However, some of these theories produce second-order equations of motion: this is obviously the case of the Einstein-Hilbert Lagrangian, and more generally of the family of Lovelock gravities, that we will review in the next section.

\subsubsection{Variational problem}
Let us now study the variational problem defined by the action \req{eq:actiongen}, which we assume to be defined in a region $\mathcal{M}$ with boundary $\partial\mathcal{M}$. For simplicity, we restrict ourselves to Lagrangians of the form $\mathcal{L}\left(g^{\alpha\beta}, R_{\mu\nu\rho\sigma}\right)$, \ie without covariant derivatives of the Riemann tensor. Before imposing any boundary conditions, the variation of the action with respect to the metric yields
\begin{equation}
\delta S=\int_{\mathcal{M}}d^Dx\sqrt{|g|}\E_{\mu\nu}\delta g^{\mu\nu}+\epsilon \int_{\partial\mathcal{M}}d^{D-1}x\sqrt{|h|}n_{\mu}\delta v^{\mu}\, ,
\label{varintro}
\end{equation}
where 
\begin{equation}
\mathcal{E}_{\mu\nu}\equiv P_{\mu}\,^{\alpha\beta\sigma}R_{\nu \alpha\beta\sigma}-\frac{1}{2}g_{\mu\nu}\mathcal{L}+2\nabla^{\alpha}\nabla^{\beta}P_{\mu \alpha\nu\beta}\, , %=\frac{1}{2}T_{ab}\, .
\label{fieldequations}
\end{equation}
and
\begin{align}
\delta v^{\mu}=&2\tensor{P}{_{\alpha}^{\beta \mu \sigma}}\nabla^{\alpha}\delta g_{\beta\sigma}\, , \\
&\notag \\
P^{\mu\nu\rho\sigma}\equiv &\left[\frac{\partial \mathcal{L}}{\partial R_{\mu\nu\rho\sigma}}\right]_{g^{\alpha\beta}}\, .\label{Ptensorintro}
\end{align}
In addition, $n^{\mu}$ is the unit normal\footnote{It is future-directed if it is timelike, or outward if it is spacelike.} to $\partial \mathcal{M}$, normalized as $n^{\mu}n_{\mu}\equiv \epsilon=\pm 1$, and  $h_{\mu\nu}=g_{\mu\nu}-\epsilon n_{\mu}n_{\nu}$ is the induced metric on the boundary. Ignoring momentarily the surface term in \req{varintro}, we see that the equations of motion read $\mathcal{E}_{\mu\nu}=0$ --- or more generally $\mathcal{E}_{\mu\nu}=\frac{1}{2}T_{\mu\nu}$ if we couple the theory to a matter Lagrangian with energy-momentum tensor $T_{\mu\nu}$. Using the fact that the tensor $P^{\mu\nu\rho\sigma}$ inherits all the symmetries of Riemann tensor,  one can check that the equations of motion are identically divergence-free (off-shell), 
\begin{equation}
\nabla^{\mu}\E_{\mu\nu}\equiv 0\, ,
\end{equation}
as required for any diffeomorphism-invariant theory. Let us now focus our attention on the surface term in \req{varintro}.  In order to specify a solution of the equations of motion we need to impose some boundary conditions, and these must be taken into account when computing the variation of the action. Then, one hopes that when these conditions are imposed on \req{varintro}, the surface term vanishes, 
so that the action is stationary for solutions of the equations of motion. If this happens we say that the variational problem is \emph{well posed}. We can also say that an action is well posed if it is ``differentiable'', in the sense that $\delta S\propto \delta g^{\mu\nu}$ for variations that satisfy the boundary conditions. Let us note that a well-posed action is essential in order to define a semiclassical partition function for gravity --- see Section \ref{sec:thermointro}. 

However, it is a well-known fact that the gravitational actions, defined as in \req{eq:actiongen}, are not well posed. In order to see why, it is convenient to review the situation in GR. In that case, the equations of motion are of second order, and in a problem with Dirichlet boundary conditions, we only need to specify the value of the induced metric on the boundary, $h_{ab}$.\footnote{The letters $a, b=0,1,\ldots D-2$ denote intrinsic boundary indices.} However, when the condition $\delta h_{ab}=0$ is imposed in \req{varintro} (specified for the GR case), one finds that the surface contribution does not vanish. Hence, the Einstein-Hilbert action is not well posed. This problem can be solved by introducing appropriate surface terms to the action, and in the case of the Einstein-Hilbert Lagrangian $\mathcal{L}_{\rm EH}=\left[R-2\Lambda \right]/(16\pi G)$, this issue is solved by the addition of the Gibbons-Hawking-York term \cite{York:1972sj,Gibbons:1976ue},
\begin{equation}
S_{\rm GHY}=\frac{\epsilon}{8\pi G}\int_{\partial\mathcal{M}}d^{D-1}x\sqrt{|h|}K\, ,
\end{equation}
where $K=K_{\mu\nu}g^{\mu\nu}$ is the trace of the second fundamental form of the boundary, $K_{\mu\nu}=h_{\mu}^{\ \alpha} \nabla_{\alpha} n_{\nu}$. When this term is included, the Dirichlet variation of the action reads
\begin{equation}
\delta(S_{\rm EH}+S_{\rm GHY})\Big|_{\delta h_{ab}=0}=\frac{1}{16 \pi G}\int_{\mathcal{M}}d^{D-1}x\sqrt{|g|}\left[R_{\mu\nu}-\frac{1}{2}g_{\mu\nu}\mathcal{L}_{\rm EH}\right]\delta g^{\mu\nu}\, ,
\end{equation} 
and now the action is stationary whenever the metric satisfies Einstein's field equations. 
 
For higher-order gravities one finds again that the bulk action \req{eq:actiongen} is not well posed, but solving this issue is considerably more involved than in GR. 
One of the main difficulties arises from the fact that these theories generally possess fourth-order equations of motion. This implies that the boundary-value problem is not fully determined by the induced metric on $\partial\mathcal{M}$, and one needs to specify additional data in the boundary, such as the value of some derivatives of the metric. Furthermore, even if we know the variables we have to fix on the boundary, determining what surface term needs to be added in order to yield a differentiable action for such variations is a far from trivial task. 
With the goal of providing a canonical formulation for arbitrary $\mathcal{L}$(Riemann$)$ gravities, an interesting proposal for constructing satisfactory boundary terms for such class of theories was presented in \cite{Deruelle:2009zk} --- see also \cite{Teimouri:2016ulk}.
The procedure used in that work involves the introduction of auxiliary fields which account for the variables that one needs to fix on the boundary. 
However, explicit surface terms in a purely metric formulation have been constructed for some particular theories. In this respect, we must mention the case of Lovelock gravities \cite{Lovelock1,Lovelock2}, which are the only higher-curvature theories with second-order equations of motion. The Dirichlet problem here is similar to the one in Einstein gravity as it only requires fixing the induced metric on the boundary, and it is possible to construct an appropriate surface term analogous to the Gibbons-Hawking-York one \cite{Teitelboim:1987zz,Myers:1987yn}. We show this term in Eq.~\req{bdrylove}. 
Some other examples for which differentiable actions have been constructed are: quadratic gravities (perturbatively in the couplings) \cite{Smolic:2013gz}, $f(R)$ \cite{Madsen:1989rz,Dyer:2008hb,Guarnizo:2010xr} and, more generally, $f($Lovelock$)$ gravities \cite{Love}. In these cases, it is also necessary to fix the value of some of the densities on the boundary --- \eg $\delta R\big|_{\partial\mathcal{M}}=0$ for $f(R)$ --- which is related to the fact that these theories propagate additional scalar modes.

\subsubsection{Gravitational energy}
Another aspect of gravitational theories that will be relevant for us is the issue of conserved charges, and, in particular, the definition of energy. Due to the Equivalence Principle, it is not possible to define a local notion of gravitational energy, because the gravitational field can always be removed locally  --- although non-covariant or quasi-local definitions of energy are possible \cite{Komar:1962xha,Chang:1998wj}.  However, it is possible to provide a satisfactory definition of total energy if the spacetime is asymptotically maximally symmetric --- in particular, asymptotically flat or asymptotically anti-de Sitter.\footnote{For asymptotically de Sitter spaces the definition of energy presents ambiguities due to the presence of a horizon \cite{Abbott:1981ff}.}

In the case of Einstein gravity, the total energy can be computed using the Hamiltonian formalism developed by Arnowitt, Deser and Misner (ADM) \cite{Arnowitt:1960es,Arnowitt:1960zzc,Arnowitt:1961zz,Arnowitt:1962hi}. A careful evaluation of the gravitational Hamiltonian for asymptotically flat spaces yields the celebrated ADM mass formula,
\begin{equation}
M_{\rm ADM}=\frac{1}{16 \pi G}\int_{\mathbb{S}^{D-2}_{\infty}} d\Sigma^{j}\Big(\partial_{i} h_{ij}-\partial_{j}h_{ii}\Big)\, ,
\label{MADM}
\end{equation}
where the integral is taken over a sphere at infinity, and where $h_{\mu\nu}=g_{\mu\nu}-\eta_{\mu\nu}$ is the metric perturbation expressed in asymptotically Cartesian coordinates ($i,j$ denote the spatial indices). 
A different approach that leads to global conservation laws of energy and momentum was proposed by Abbott and Deser (AD) \cite{Abbott:1981ff}.  The main idea behind the Abbott-Deser approach is to use the asymptotic symmetries of the spacetime to derive an asymptotic law of conservation of energy-momentum. When the spacetime is asymptotically maximally symmetric one has the maximal number of asymptotic Killing vectors and for each one of them there is a conserved charge. However,  we only require the existence of one timelike asymptotic Killing vector in order to define the global energy.  When applied to the asymptotically flat case, the AD result coincides with the ADM formula \req{MADM}.
An important advantage of the Abbott-Deser method is that it is easily generalizable to higher-derivative gravities. In particular, Refs. \cite{Deser:2002jk,Senturk:2012yi} constructed the appropriate generalization for the case of $\mathcal{L}$(Riemann) gravities --- see also the recent review \cite{Adami:2017phg}. It turns out that, with few modifications, the AD result applies as well in the case of higher-order gravities, providing that the gravitational Lagrangian has a well-defined Einstein gravity limit $\mathcal{L}=\left[R-2\Lambda +\ldots\right]/(16\pi G)$. In fact, in the asymptotically flat case the energy is identified exactly in the same way as in GR, and one can apply the formula \req{MADM} without any modification to higher-derivative gravities. In the asymptotically anti-de Sitter (or de Sitter) case, the only difference is that we must replace Newton's constant $G$ --- which is the one that appears in the Lagrangian --- by the effective Newton's constant $G_{\rm eff}$ in the corresponding mass formula. The effective Newton's constant is the one that determines the coupling between matter and gravity, and in general will be different from $G$ --- see Chapter~\ref{Chap:1} and \eg Refs.~\cite{Tekin1,Aspects} for a precise definition. 

Since in this thesis we will mostly deal with static and spherically symmetric spaces, let us conclude this section by identifying the total energy for a general $D$-dimensional metric of that type, which can be written as
\begin{equation}
ds^2=g_{tt} dt^2+g_{rr} dr^2+r^2d\Omega_{(D-2)}^2\, .
\end{equation}
According to the generalized Abbott-Deser prescription \cite{Deser:2002jk,Senturk:2012yi}, the total energy or mass of this spacetime can be obtained by looking at the coefficient of the term $1/r^{D-3}$ in the asymptotic expansion of $1/g_{rr}$. The precise identification reads
\begin{equation}\label{eq:Msphere}
\frac{1}{g_{rr}}=-\mathcal{K} r^2+1-\frac{16\pi G_{\rm eff} M}{(D-2)\Omega_{(D-2)}}\frac{1}{r^{D-3}}+\ldots\, ,
\end{equation}
where $\mathcal{K}$ is the curvature of the asymptotic solution --- Minkowski $(\mathcal{K}=0)$, anti-de Sitter $(\mathcal{K}<0)$, or de Sitter $(\mathcal{K}>0)$ space ---  and $\Omega_{(D-2)}$ is the area of the $D-2$ sphere,
\begin{equation}
\Omega_{(D-2)}=\frac{2\pi^{\frac{D-1}{2}}}{\Gamma\left(\frac{D-1}{2}\right)}\, .
\end{equation}
 
\subsection{Some examples of higher-order gravities}\label{sec:hogexamples}

In general, even within the family of $\mathcal{L}$(Riemann) theories, the number of independent higher-derivative operators in the action \req{eq:HDGintro} grows very fast with the order in curvature, making a general analysis of these terms inaccessible. In fact, the problem of finding all the basic invariants at a given order in the derivative expansion is a far from trivial one \cite{0264-9381-9-5-003,MartinGarcia:2008qz}.
In addition, studying the properties of any of these theories is usually a very hard task. Even the problem of finding static, spherically symmetric solutions becomes so challenging that only a few solutions of that kind are known for some particular theories --- we review some of them in Section~\ref{introspherebhs}. Furthermore, as we mentioned earlier, higher-order gravities usually propagate ghost-like modes along with the Einstein graviton, which are a source of potential instabilities. 
These reasons motivate the search for particular higher-derivative gravities that circumvent some of these problems. The idea is that these theories serve as useful models that can be used to explore possible general features of higher-derivative gravities.
Some of them overcome the problem of Ostrogradski's instability and are, therefore, viable models for alternative classical theories of gravity.  Other theories have simple enough field equations so as to allow us to study black hole solutions and analyze the differences with respect to those of Einstein gravity. Let us comment here three well-known examples of interesting higher-order gravities. 

\subsubsection{$f(R)$ theories}
One of the most popular modified gravity models is $f(R)$ gravity \cite{Sotiriou:2008rp}, whose Lagrangian is a certain function of the Ricci scalar $R$,
\begin{equation}
S_f=\frac{1}{16\pi G}\int_Md^Dx\sqrt{|g|}f(R)\, .
\label{f(R)}
\end{equation}
%The field equations for this theory are
%\begin{equation}
%f'(R)R_{\mu\nu}-\frac{1}{2}f(R)g_{\mu\nu}+\big(g_{\mu\nu}\Box-\nabla_{\mu}\nabla_{\nu}\big)f'(R)=0.
%\label{fReq}
%\end{equation}
The main feature of these theories is that they introduce a scalar degree of freedom not present in GR. Assuming that $f$ is a convex function, it is possible to define its Legendre transform $V$, and in that case one can check that the following action
\begin{equation}\label{eq:fRphi}
S_{\varphi}=\frac{1}{16\pi G}\int d^Dx\sqrt{|g|}\Big\{\varphi R-V(\varphi)\Big\}\, ,
\end{equation}
is indeed equivalent to \req{f(R)}. Thus, $f(R)$ gravities are equivalent to Brans-Dicke theories with a scalar potential, in which a scalar field is non-minimally coupled to gravity. Formulated in this way, it is straightforward to derive a generalized Gibbons-Hawking-York boundary term that makes the action \req{eq:fRphi} well posed \cite{Madsen:1989rz,Dyer:2008hb,Guarnizo:2010xr} 
\begin{equation}
S_{\varphi, {\rm bdry}}=\frac{\epsilon}{8\pi G}\int_{\partial\mathcal{M}}d^{D-1}x\sqrt{|h|}\varphi K\, .
\end{equation}
In this case, the scalar must be fixed on the boundary: $\delta\varphi\big|_{\partial\mathcal{M}}=0$. When expressed in terms of the metric, this is equivalent to demanding that the value of Ricci scalar $R$ must be fixed by boundary conditions, since $\varphi=f'(R)$. 

An interesting property of $f(R)$ gravities is that they can avoid the issue of Ostrogradsky instability \cite{Woodard:2006nt}, since all the higher-derivative terms can be absorbed in the definition of $\varphi$.  This makes $f(R)$ models very useful for cosmology, where they can be used to explain inflation in the early universe \cite{Starobinsky:1980te} as well as late time cosmic acceleration, see \eg \cite{Nojiri:2003ft,Woodard:2006nt,Nojiri:2010wj}. However, since these models are equivalent to a scalar-tensor theory, they do not really provide a modification of the dynamics of gravity, understood as a spin-2 field. The vacuum equations of motion of $f(R)$ gravity can always be solved by Einstein metrics, \ie those with $R_{\mu\nu}\propto g_{\mu\nu}$. Hence, all vacuum solutions of Einstein gravity are also solutions of $f(R)$ gravity. In particular, black holes do not receive any correction and are still given by the Schwarzschild \cite{Schwarzschild:1916uq} and Kerr \cite{Kerr:1963ud} metrics. Although this might be in part an advantage, these theories do not tell us anything about how black holes are modified by higher-curvature corrections. 

\subsubsection{Lovelock gravity}
Another generalization of Einstein gravity is the Lanczos-Lovelock theory (or Lovelock gravity in short) \cite{Lovelock1, Lovelock2,Padmanabhan:2013xyr}, which is the most general higher-derivative gravity that possesses second-order equations of motion. The Lagrangian of Lovelock gravity is a sum of dimensionally continued Euler densities (ED) $\mathcal{X}_{2n}$,
which are given by\footnote{The alternate Kronecker symbol is defined by $\delta^{\mu_1\mu_2...\mu_r}_{\nu_1\nu_2...\nu_r}=r!\delta^{[\mu_1}_{\nu_1}\delta^{\mu_2}_{\nu_2}... \delta^{\mu_r]}_{\nu_r}.$}
\begin{equation}
\mathcal{X}_{2n}=\frac{1}{2^n}\delta^{\mu_1...\mu_{2p}}_{\nu_1...\nu_{2n}}R^{\nu_1\nu_2}_{\mu_1\mu_2}... R^{\nu_{2n-1}\nu_{2n}}_{\mu_{2n-1}\mu_{2n}}\, .
\label{ED}
\end{equation}
These densities have a different behaviour depending on the dimension. When $D<2n$, the Euler density $\mathcal{X}_{2n}$ vanishes identically, as follows from the antisymmetrization in \req{ED}. In the critical dimension $D=2n$, the corresponding density becomes topological, which is the origin of its name. In particular, if $\mathcal{M}$ is a compact $2n$-dimensional Euclidean manifold, the Chern-Gauss-Bonnet theorem states that 
\begin{equation}
\chi(\mathcal{M})=\frac{1}{n!(4\pi)^n}\int_\mathcal{M}d^{2n}x\sqrt{g}\mathcal{X}_{2n}\, 
\end{equation}
is the Euler characteristic of $\mathcal{M}$.
From the point of view of the gravitational action, this means that $\mathcal{X}_{2n}$ does not contribute to the equations of motion. Finally, when $D>2n$, the density $\mathcal{X}_{2n}$ becomes dynamical. Thus, the most general $D$-dimensional Lovelock action reads\footnote{$\lfloor D/2 \rfloor$ is the floor function of $D/2$, so it is $D/2$ for even $D$ and $(D-1)/2$ for odd $D$.}
\begin{equation}
S_{LL}=\frac{1}{16\pi G}\int d^Dx\sqrt{-g}\sum_{n=0}^{\lfloor D/2 \rfloor}\lambda_n L^{2n-2}\mathcal{X}_{2n}\, ,
\label{LLaction}
\end{equation}
where $L$ is a length scale, $\lambda_n$ are dimensionless constants, and by convention, $\mathcal{X}_{0}=1$.
Hence, in $D=4$ the most general Lovelock theory is Einstein gravity $(\mathcal{X}_2=R)$ plus a cosmological constant; in $D=5,6$ the Gauss-Bonnet term becomes non-trivial
\begin{equation}\label{eq:introGB}
\mathcal{X}_4=R^2-4R_{\mu\nu}R^{\mu\nu}+R_{\mu\nu\alpha\beta}R^{\mu\nu\alpha\beta}.
\end{equation}
In $D=7, 8$ we can also add the following Euler density $\mathcal{X}_6$ and so on. Taking the variation with respect to the metric in \req{LLaction} we obtain the equations of motion
\begin{equation}
\sum_{n=0}^{\lfloor D/2 \rfloor}\lambda_n L^{2n-2}\E_{\mu\nu}^{(n)}=0\, ,\,\,\,\, \text{where}\,\,\, \E_{\mu\nu}^{(n)}=-\frac{1}{2^{n+1}}g_{\alpha \mu}\delta^{\alpha \mu_1... \mu_{2n}}_{\nu \nu_1... \nu_{2n}} R^{\nu_1\nu_2}_{\mu_1\mu_2}... R^{\nu_{2n-1}\nu_{2n}}_{\mu_{2n-1}\mu_{2n}}\, ,
\label{LLeq}
\end{equation}
which are algebraic in the curvature, hence of second order in derivatives of the metric. For this reason, Lovelock gravity propagates the same degrees of freedom as GR and it provides a natural generalization of Einstein gravity in higher dimensions. In addition, it is possible to add a generalized Gibbons-Hawking-York term that makes the action \req{LLaction} well posed. This term was found by Myers  \cite{Myers:1987yn} and Teitelboim and Zanelli \cite{Teitelboim:1987zz}, and for the $n$-th order Lovelock action it reads
\begin{equation}\label{bdrylove}
S_{\rm bdry}^{(n)}=\int_{\partial\mathcal{M}}d\Sigma \mathcal{Q}_{n}\, ,
\end{equation}
where $\mathcal{Q}_{n}$ is given by 
\begin{equation}
\mathcal{Q}_{n}=2n\int_0^1dt\, \delta^{a_1\dots a_{2n-1}}_{b_1\dots b_{2n-1}}K^{b_1}_{a_1}\left[\frac{1}{2}\mathcal{R}^{b_2 b_3}_{a_2a_3}-\epsilon t^2K^{b_2}_{a_2}K^{b_3}_{a_3}\right]\cdots \left[\frac{1}{2}\mathcal{R}^{b_{2n-2}b_{2n-1}}_{a_{2n-2}a_{2n-1}}-\epsilon t^2K^{b_{2n-2}}_{a_{2n-2}}K^{b_{2n-1}}_{a_{2n-1}}\right]\, ,
\label{Qterm1}
\end{equation}
where $\mathcal{R}^{b_2 b_3}_{a_2a_3}$ is the intrinsic curvature of the boundary, $K^{a}_{b}$ is the extrinsic curvature and $\epsilon=n^2=\pm 1$ is the sign of the normal to the boundary. Also, $d\Sigma=d^{D-1}x\sqrt{|h|}$ is the volume element on $\partial\mathcal{M}$ and the orientation is such that, as a 1-form, $n=n_{\mu}dx^{\mu}$ points outside of $\mathcal{M}$

Unlike those of $f(R)$ gravities, Lovelock's equations of motion are not solved by Einstein metrics (except in $D\le 4$), and the higher-order Euler densities introduce corrections to GR black hole solutions. In particular, it is possible to solve the equations of motion to find explicit spherically symmetric black hole solutions  \cite{Wheeler:1985nh,Wheeler:1985qd,Boulware:1985wk,Cai:2001dz,Dehghani:2009zzb,deBoer:2009gx,Camanho:2011rj} --- we will review these solutions in Section~\ref{introspherebhs}. 
Besides, some of the Lovelock's invariants appear explicitly in String Theory when considering effective actions --- we saw before that this is the case for the Gauss-Bonnet density \cite{Boulware:1985wk,Metsaev:1987zx}. This makes the study of Lovelock gravities very interesting from the point of view of the AdS/CFT correspondence \cite{Camanho:2009vw,deBoer:2009pn,Buchel:2009sk,deBoer:2009gx,Camanho:2009hu,Camanho:2010ru,Camanho:2013pda,Grozdanov:2014kva,Grozdanov:2016fkt,Andrade:2016rln,Konoplya:2017zwo}, that we will review in Section~\ref{sec:holoint}.

\subsubsection{Quasi-topological gravity}
One of the disadvantages of Lovelock theories is that the number of densities one can include in the action is highly restricted by the spacetime dimension. Thus, in $D=4$ Lovelock gravity reduces to GR, while in $D=5$ it only allows for the introduction of the quadratic Gauss-Bonnet term.  In order to explore the effect of higher-order terms it is necessary to consider more general theories. Since Lovelock gravities are the most general theories with second order equations of motion \cite{Lovelock1,Lovelock2}, any other model will lose this characteristic.
However, the idea is to find some theories that at least are able to mimic some of the nice properties of Lovelock theories, such as absence of additional degrees of freedom propagated in the vacuum, or the existence of ``simple''\footnote{We will be more explicit in Section~\ref{introspherebhs} about what we mean by simple.} black hole solutions.  This is the philosophy behind the construction of Quasi-topological gravity \cite{Quasi,Quasi2}. This is a cubic curvature interaction that in general dimensions $D\ge5$ is given by 
\begin{equation}\label{eq:QuasiTopointro}
\begin{aligned}
\mathcal{Z}_{D} =
&\tensor{R}{_{\mu}^{\rho}_{\nu}^{\sigma}}\tensor{R}{_{\rho}^{\alpha}_{\sigma}^{\beta}}\tensor{R}{_{\alpha}^{\mu}_{\beta}^{\nu}}
+ \frac{1}{(2D-3)(D-4)} \Big ( 
- 3(D-2) \tensor{R}{_{\mu \nu\rho\sigma}}\tensor{R}{^{\mu \nu\rho}_{\alpha}}R^{\sigma \alpha} \\
&+ \frac{3 (3D-8)}{8}  \tensor{R}{_{\mu\nu\rho\sigma}}\tensor{R}{^{\mu\nu\rho\sigma}}R
+3D \tensor{R}{_{\mu\nu\rho\sigma}}\tensor{R}{^{\mu\rho}}\tensor{R}{^{\nu\sigma}} \\
&+ 6(D-2) \tensor{R}{_{\mu}^{\nu}}\tensor{R}{_{\nu}^{\rho}}\tensor{R}{_{\rho}^{\mu}}
- \frac{3(3D-4)}{2}  R_{\mu\nu }R^{\mu\nu }R
+ \frac{3D}{8} R^3 \Big) \, .
\end{aligned}
\end{equation}
This density was originally identified in its five-dimensional version in \cite{Quasi}, by the condition that the trace of the field equations is of second order. As a consequence, it was found that the field equations in the presence of spherical symmetry are of second order, facilitating the obtention of explicit black hole solutions and giving rise to a Birkhoff theorem. Independently, the quasi-topological interaction was derived in \cite{Quasi2}, where black hole solutions where further analyzed. In addition, it was noted that the linearized equations of motion of this theory around maximally symmetric vacua become of second order, and identical to Einstein's gravity ones. Hence, Quasi-topological gravity only propagates a massless spin-2 graviton, just as GR.  These properties make this theory a very useful toy model for holographic applications \cite{Myers:2010jv}. 
It is possible to construct higher-order generalizations of Quasi-topological gravity possessing similar properties, and, so far, the quartic \cite{Dehghani:2011vu} and quintic \cite{Cisterna:2017umf} versions have been identified. Thus, these theories provide a way to probe higher-curvature effects --- specially, on black holes --- beyond Lovelock gravity. However, they still have an important drawback: Quasi-topological gravities only exist in $D\ge5$ and they cannot be used to learn anything about higher-curvature corrections in our own universe. One of the goals of this thesis is to present a new family of theories that generalizes the quasi-topological family and that is non-trivial in $D=4$.

\section{Black holes}\label{sec:introbhshdg}
One of the most remarkable predictions of GR is that of the existence of black holes: regions of spacetime where gravity is so strong that nothing, not even light, can escape. Nowadays we have astonishing experimental evidence of the existence of these objects thanks to the gravitational wave detectors LIGO/Virgo \cite{Abbott:2016blz,Abbott:2016nmj,TheLIGOScientific:2016src,TheLIGOScientific:2017qsa,Abbott:2017vtc,Abbott:2017oio} and the Event Horizon Telescope \cite{Akiyama:2019cqa}, and in the next years the experimental data will be accurate enough to allow for precision tests on black holes \cite{Giddings:2014ova,Berti:2015itd,Johannsen:2015hib,Cardoso:2016rao,Yunes:2016jcc,Barack:2018yly,Berti:2018cxi}. 
Yet, from the theoretical point of view, black holes remain one of the most exciting and active areas of study, since they pose intriguing questions about the interaction between gravity and quantum mechanics. But, before going into details, it will be useful to remind very briefly some basic concepts about black holes.

%From the classical perspective, black holes are the simplest macroscopic objects, since any black hole can be described \emph{exactly} by a reduced set of parameters such as the mass, spin and electric charge --- a fact known as the ``no-hair theorem''. But, when quantum mechanics are taking into consideration, one is led to the conclusion that black holes have a temperature and an entropy, and that they act as thermodynamic system. This result suggests that black holes are ensembles composed from a vast number of microstates, and it is believed that a correct theory of quantum gravity should provide an explanation for the black hole entropy as a microstate counting. 

\subsubsection{Basic definitions}
 A black hole in an asymptotically flat spacetime $(\mathcal{M}, g_{\mu\nu})$ is defined as a region
\begin{equation}
\mathcal{B}=\mathcal{M}-I^{-}(\mathcal{I}^{+})\, ,
\end{equation}
where $\mathcal{I}^{+}$ is the future null infinity and $I^{-}$ the chronological past. In words, this expresses that the future of $\mathcal{B}$ is not contained in the asymptotic region. The event horizon of the black hole is a null hypersurface corresponding to the common boundary between $\mathcal{B}$ and the past of $\mathcal{I}^{+}$.  It is important to remark that the event horizon is a global characteristic of the spacetime which requires knowing all its future history, hence it lacks a local significance. 
We will be interested in \emph{stationary} black holes: a spacetime containing a black hole is said to be stationary if it possesses a Killing vector field $\xi^{\mu}$ that is asymptotically timelike. Furthermore, the spacetime is \emph{static} if this vector is hypersurface orthogonal, \ie if the associated one-form satisfies $\xi \wedge d\xi=0$. One important class of stationary black holes corresponds to axisymmetric spacetimes, which additionally to $\xi^{\mu}$ contain another Killing vector $\phi^{\mu}$ that generates rotations at infinity. 

Under very general circumstances, the ``rigidity theorems'' \cite{1973blho.conf...57C,Hawking:1973uf,Wald:1984rg,Friedrich:1998wq} state that the event horizon of a stationary black hole is always a Killing horizon, meaning that there is a Killing vector $k^{\mu}$ that becomes null on the horizon.\footnote{This vector is normal to the horizon and in this sense we say that it \emph{generates} the horizon.}  Hawking \cite{Hawking:1973uf} showed that this is the case for any stationary black hole in GR, while Carter \cite{1973blho.conf...57C} provided a geometrical proof for axisymmetric black holes that does not rely on equations of motion. If $k^{\mu}$ is the Killing vector that generates the horizon, then it is easy to see that there must exist a quantity $\kappa$ such that 
\begin{equation}
k^{\alpha}\nabla_{\alpha} k^{\mu}=\kappa k^{\mu}\, ,
\end{equation}
when evaluated at the horizon. The proportionality factor $\kappa$ is known as surface gravity and it possible to show that it takes a constant value on the horizon  \cite{1973blho.conf...57C,Bardeen:1973gs,Racz:1995nh,Wald:1999vt}. Again, this can proven directly for black holes in GR \cite{Bardeen:1973gs}, or for axisymmetric black holes \cite{1973blho.conf...57C}. 

We can see that the value of $\kappa$ depends on the normalization of $k^{\mu}$, so in order to give it a physical interpretation we must choose the Killing vector appropriately. In the case of static black holes, the horizon is generated by the asymptotically timelike Killing vector, and we should normalize it so that $k^{\mu}k_{\mu}=-1$ at infinity. %Defined in this way, $\kappa$ really measures the tension $\kappa m$ in the fishing line of an asymptotic observer that is holding an object of mass $m$ near the horizon with a fishing pole. 
We distinguish two different situations depending on the value of the surface gravity. When $\kappa\neq 0$, it is possible to show that $k^{\mu}$ generates a bifurcate Killing horizon in the maximally extended spacetime, of which the event horizon of the black hole is a branch \cite{Racz:1995nh}. The two null hypersurfaces that compose the bifurcate Killing horizon intersect at a codimension two spacelike surface --- the bifurcation surface --- which is placed at the points where $k^{\mu}$ vanishes. On the other hand, when $\kappa=0$ the Killing horizon is degenerate, and the corresponding black hole is said to be \emph{extremal}. 

\subsubsection{The Schwarzschild black hole}
The best way to illustrate the concept of a black hole is to study one particular example, and perhaps the most relevant one is given by Schwarzschild's black hole. 
Schwarzschild's metric\cite{Schwarzschild:1916uq} was the first exact solution of vacuum Einstein field equations to be discovered, and it still stands as one of the most studied ones. It describes a static and spherically symmetric gravitational field, and in standard ``Schwarzschild coordinates'', it is given by  
\begin{equation}\label{eq:SCHW}
ds^2=-\left(1-\frac{2GM}{r}\right)dt^2+\frac{dr^2}{\left(1-\frac{2GM}{r}\right)}+r^2\left(d\theta^2+d\phi^2\sin^2\theta\right)\, ,
\end{equation}
where $M$ is the total mass (or energy) of the spacetime, as can be computed from the ADM \cite{Arnowitt:1960es,Arnowitt:1960zzc,Arnowitt:1961zz} or Abbott-Deser \cite{Abbott:1981ff} prescriptions. 
In addition, Schwarzschild's metric is the unique spherically symmetric solution of vacuum Einstein equations, by virtue of Birkhoff's theorem. The spacetime described by the metric \req{eq:SCHW} is asymptotically flat (for $r\rightarrow\infty$) and it contains a black hole whose event horizon is placed at the Schwarzschild radius $r_s=2GM$. However, this is not evident in the present form of the solution, since the metric \req{eq:SCHW} seems to be singular at $r=r_s$. In order to unveil the global structure of the spacetime, an analytic continuation of the solution beyond $r_s$ is required. The usual approach consists in introducing the advanced and retarded Eddington-Finkelstein coordinates \cite{PhysRevLett.14.57}
\begin{equation}
u=t-r_*\, ,\quad v=t+r_*\, , 
\end{equation}
which are constant on outgoing and ingoing null radial geodesics, respectively, and where $r_*$ is the tortoise coordinate, defined as
\begin{equation}
r_*=\int dr g_{rr}=r+2GM\log\left|\frac{r}{2GM}-1\right|\, .
\end{equation}
Replacing the coordinate $t$ in terms of one of the variables $v$ or $u$ already allows one to extend the metric to the black hole interior. For instance, written in terms of $v$, Schwarzschild's solution reads
\begin{equation}\label{eq:SCHW2}
ds^2=-\left(1-\frac{2GM}{r}\right)dv^2+2dvdr+r^2\left(d\theta^2+d\phi^2\sin^2\theta\right)\, ,
\end{equation}
and now it is clear that $r=2GM$ is a null hypersurface. Besides, it is a Killing horizon for the vector $k=\partial_{v}$, and it is a straightforward computation to show that the surface gravity reads
\begin{equation}\label{eq:kappaSCHW}
\kappa=\frac{1}{4GM}\, .
\end{equation}
However, the metric \req{eq:SCHW2} is still not the maximally extended spacetime. This is revealed by introducing the Kruskal-Szekeres coordinates \cite{Kruskal:1959vx,Szekeres:1960gm}
\begin{equation}\label{eq:UVKS}
U=e^{-\frac{u}{4GM}}\, ,\qquad V= e^{\frac{v}{4GM}}\, ,
\end{equation}
in terms of which the metric takes the form
\begin{equation}
ds^2=-32\frac{(GM)^3}{r}e^{-\frac{r}{2GM}}dUdV+r^2\left(d\theta^2+d\phi^2\sin^2\theta\right)\, ,
\end{equation}
where now the radius depends on the coordinates $U$ and $V$ according to the relation
\begin{equation}
UV=e^{\frac{r}{2GM}}\left(\frac{r}{2GM}-1\right)\, .
\end{equation}
The analytic extension comes from the fact that now we allow the coordinates $U$, $V$ to take values in all the real line, and not only positive values as the relations \req{eq:UVKS} would imply.
When the metric is expressed in this form, we see that it is regular everywhere except at $r=0$, which is a real singularity. Furthermore, in terms of the coordinates $U$, $V$, the horizon is placed at $UV=0$, and we see that it is bifurcate since it contains the branches $V=0$ and $U=0$, with the bifurcation surface placed at $U=V=0$. We can also check that this is a bifurcate Killing horizon for the Killing vector $k=(V\partial_V-U\partial_U)/(4GM)$. Therefore, we distinguish four regions: $U>0$, $V>0$ is the exterior region originally covered by the Schwarzschild coordinates; $U<0$, $V>0$ corresponds to the black hole interior, that contains a future singularity when $UV=-1$; $U<0$, $V>0$ is the interior of a ``white hole'', which contains a past singularity; and $U<0$, $V<0$ is another exterior region.

Besides serving as a simple example of a black hole solution, Schwarzschild's metric will be very relevant for the purposes of this thesis, since one of our goals will be to find generalizations of this solution in higher-order gravity. 
Generically, the introduction of higher-derivative interactions implies that Einstein metrics no longer solve the gravitational field equations. As a consequence, the Schwarzschild and Kerr \cite{Kerr:1963ud} metrics, that describe static and rotating black holes in four-dimensional GR (or Tangherlini \cite{Tangherlini:1963bw} and Myers-Perry \cite{Myers:1986un} metrics in the higher-dimensional case), are not solutions of the modified theories. Therefore, it is interesting to solve the equations of motion of higher-derivative gravity in order to search for corrected black hole geometries and to study the deviations with respect to the GR predictions. 
We are particularly interested in the thermodynamic properties of black holes, that we review now.

\subsection{Black hole thermodynamics}\label{sec:thermointro}
One of the most important breakthroughs of the last fifty years was the realization that black holes can be identified with a thermodynamic system. By studying the behaviour of quantum fields placed in the background of a black hole, Hawking \cite{Hawking:1974sw} showed that any black hole emits thermal radiation at a temperature
\begin{equation}\label{eq:Hawking}
T=\frac{\kappa}{2\pi}\, ,
\end{equation}
where $\kappa$ is the surface gravity of the event horizon. Correspondingly, black holes must have an entropy, which in the case of GR is given by Bekenstein-Hawking formula, 
\begin{equation}\label{eq:BekHawk}
S_{\rm \ssc BH}=\frac{A}{4G}\, .
\end{equation}
which states that the entropy of a black hole is proportional to the area of its horizon \cite{Bekenstein:1973ur,Bekenstein:1974ax}. Due to these intriguing properties, which have a quantum-mechanical origin, black holes are an excellent laboratory to test the interaction between quantum mechanics and gravity.
For instance, any prospective quantum theory of gravity should provide an explanation for the black hole entropy from a counting of microscopic states. In this respect, it is worth emphasizing that String Theory is able to recover Bekenstein-Hawking result, at least in some special situations \cite{Strominger:1996sh,Callan:1996dv,David:2002wn}. 
However, we would like to remark that the formulas  \req{eq:Hawking} and \req{eq:BekHawk} have a different status. The identification of the temperature with the surface gravity is a completely general result because it only depends on the very concept of event horizon, but the entropy formula is theory-dependent and has a different form if we consider theories other than Einstein gravity, as we will see later on. This discrepancy has to be taken into account in order to perform a precise matching of microscopic and macroscopic black hole entropies in String Theory \cite{Sen:2005iz,Sen:2007qy,Prester:2008iu,Cano:2018qev,Cano:2018brq}.

Black hole solutions are characterized by different conserved charges, such as the mass $M$, the angular momentum $J$ or the electric charge $Q$,\footnote{In GR, these black holes are given by Kerr \cite{Kerr:1963ud}, Reissner-Nordstrom \cite{1916AnP...355..106R,1918KNAB...20.1238N} and Kerr-Newmann \cite{Newman:1965my} solutions.}  and these play a role as thermodynamic variables. 
It turns out that these quantities satisfy the \emph{four laws of black hole mechanics} \cite{Bardeen:1973gs}, in analogy to the laws of thermodynamics. At the classical level, these laws involve several relations that hold for black hole solutions of the gravitational theory. But the fact that Hawking's temperature represents a real phenomena implies that black holes are, indeed, thermodynamic systems on their own. Let us briefly review these laws.

The zeroth law establishes that the surface gravity is constant on the horizon of a stationary black hole. This result can be derived in several ways \cite{1973blho.conf...57C,Bardeen:1973gs,Racz:1995nh,Wald:1999vt} and it implies that a stationary black hole is in thermal equilibrium --- there are no gradients of temperature.  On the other hand, the first law of black hole mechanics states that the following equality holds for perturbations of a black hole solution,
\begin{equation}
\delta M=\frac{\kappa}{8\pi}\delta A+\Omega \delta J+\ldots \, ,
\end{equation}
where $M$ is the total spacetime's energy as computed from the Hamiltonian, $\Omega$ is the angular velocity of the horizon, $J$ is the total angular momentum, and the ellipsis denote additional variables on which the black hole solution may depend. 
This result, together with Hawking's formula \req{eq:Hawking} and the standard thermodynamic relation $T^{-1}=\frac{\partial S}{\partial M}$ can be used to derive Bekenstein-Hawking formula \req{eq:BekHawk}. In a way, the first law is equivalent to Einstein's field equations \cite{Jacobson:1995ab}, a fact that has been sometimes interpreted as a sign of gravity being an \emph{emerging} phenomena. 

The second law of thermodynamics is replaced, at the classical level, by the area increase theorem \cite{Hawking:1971tu}, that establishes that the area of the event horizon never decreases, hence implying $\Delta S_{\rm \ssc BH}\ge 0$ as time evolves. However, this law is violated by an isolated black hole due to Hawking radiation (which is a quantum effect): the conservation of energy implies that the black hole loses mass as it radiates, hence decreasing its area. In order to obtain a consistent second law one needs to take into account the entropy of the surrounding matter as well, and one has to define the generalized entropy \cite{Bekenstein:1974ax} $S_{\rm \ssc gen}=S_{\rm \ssc BH}+S_{\rm \ssc out}$. Then, the entropy of the Hawking quanta emitted by the black hole compensates the entropy loss due to the decrease of the area, and the total entropy increases $\Delta S_{\rm \ssc gen}\ge 0$. The generalized second law of black hole mechanics states that this happens for all physical processes. 

Finally, there is a third law of black hole mechanics that asserts that, starting from a black hole with a non-degenerate horizon, it is not possible to form an extremal black hole --- one with vanishing surface gravity --- by performing a finite number of operations \cite{PhysRevLett.57.397}. This is equivalent to the third law of thermodynamics in the sense that it states that it is not possible to reach the absolute zero of temperature   

\subsubsection{Black hole thermodynamics in higher-derivative gravity}
Let us now explore how the previous discussion extends to black holes in higher-derivative gravity. 
We already mentioned that Hawking's result for the temperature of black holes \req{eq:Hawking} is universal, in the sense that it relies on the spacetime geometry but not on the underlying dynamics of the gravitational field. Thus, a stationary black hole in a higher-derivative gravity also possesses a temperature proportional to its surface gravity, $T=\kappa/(2\pi)$. In addition, it is possible to prove that the surface gravity is constant on the horizon without making use of the equations of motion \cite{1973blho.conf...57C,Racz:1995nh,Wald:1999vt}, so the zeroth-law of black hole mechanics is automatically satisfied. However, the relationship between different quantities such as temperature, mass and area will be generically modified by higher-curvature corrections. As a consequence, it will no longer be true that $\delta M=\frac{\kappa}{2\pi}\frac{\delta A}{4G}+\ldots$, so if we want to preserve the first law of black hole mechanics we must conclude that the entropy is not given by the area law anymore $S_{\rm \ssc BH}\neq A/(4G)$. In fact, it is not evident that a first law should exist with the entropy being a universally defined quantity. It is an extraordinary fact of higher-derivative gravities that the first law of black hole mechanics does extend naturally.  As shown by Wald \cite{Wald:1993nt}, the first law is a consequence of diff. invariance, hence it is not a characteristic feature of GR, but of any covariant theory of gravity. 
According to Wald's result, the area law in the Bekenstein-Hawking entropy formula \req{eq:BekHawk} should be replaced by the integral, on the horizon, of the Noether current associated with diffeomorphism invariance.  More precisely, Wald's entropy formula reads \cite{Wald:1993nt,Iyer:1994ys,Jacobson:1993vj} 
\begin{equation}\label{eq:Waldintro}
S_{\rm \ssc W}=-2\pi \int_{\mathcal{H}} d^{D-2}x\sqrt{h} \frac{\delta \mathcal{L}}{\delta R_{\mu\nu\rho\sigma}}\epsilon_{\mu\nu}\epsilon_{\rho\sigma}\, ,
\end{equation}
where the integral is taken over the bifurcation surface of the horizon and $\frac{\delta \mathcal{L}}{\delta R_{\mu\nu\rho\sigma}}$ is the Euler-Lagrange derivative of the gravitational Lagrangian as if the Riemann tensor were an independent variable, this is,
\begin{equation}
\frac{\delta \mathcal{L}}{\delta R_{\mu\nu\rho\sigma}}=\frac{\partial \mathcal{L}}{\partial R_{\mu\nu\rho\sigma}}-\nabla_{\alpha}\left(\frac{\partial \mathcal{L}}{\partial \nabla_{\alpha}R_{\mu\nu\rho\sigma}}\right)+\ldots
\end{equation}
In addition, $h$ is the determinant of the induced metric on the horizon and $\epsilon_{\mu\nu}$ is the binormal of the horizon, normalized as $\epsilon_{\mu\nu}\epsilon^{\mu\nu}=-2$. 
Wald showed that the entropy defined in this way satisfies the first law,
\begin{equation}
\delta M=\frac{\kappa}{2\pi}\delta S_{\rm \ssc W}+\Omega \delta J+\ldots\, .
\end{equation}

On the other hand, the (ordinary) second law is not guaranteed to hold even classically, because for isolated black holes it might happen $S_{\rm \ssc W}(M_1,J_1,\ldots)+S_{\rm \ssc W}(M_2,J_2,\ldots)>S_{\rm \ssc W}(M_1+M_2,J_1+J_2,\ldots)$ --- see \eg \cite{Chatterjee:2013daa} for an example of this --- which implies that if the two initial black holes merge, the total black hole entropy decreases. However, it has been proposed that the generalized second law still holds if one replaces Wald's entropy by a related quantity, which coincides with Dong's formula for holographic entanglement entropy \cite{Dong:2013qoa}.  According to the results in \cite{Wall:2011hj,Sarkar:2013swa,Wall:2015raa}, the quantity defined as $S_{\rm\ssc gen}=S_{\rm \ssc Dong}+S_{\rm\ssc out}$ is monotonically increasing, $\Delta S_{\rm \ssc gen}>0$, at least for linear perturbations of the metric.  Since Dong's formula is equivalent to Wald's one for stationary black holes, this result tells us that the total entropy does increase, and a version of the generalized second law of black hole mechanics holds. 

\subsubsection{Euclidean path integrals and partition functions}
A useful approach in order to formalize the study of black hole thermodynamics consists in considering the Euclidean version of black hole solutions. In order to illustrate how this works, let us perform a Wick rotation in Schwarzschild's metric \req{eq:SCHW} by introducing the Euclidean time as $\tau=it$, so that we get a solution of Euclidean GR, 
\begin{equation}\label{eq:SCHWE}
ds^2_{E}=\left(1-\frac{2GM}{r}\right)d\tau^2+\frac{dr^2}{\left(1-\frac{2GM}{r}\right)}+r^2\left(d\theta^2+d\phi^2\sin^2\theta\right)\, .
\end{equation}
The properties of the Euclidean solution are drastically different from those of the Lorentzian one. In particular, the solution cannot be extended beyond $r=2GM$, which instead of a horizon is an ending point for the coordinate $r$. Near that point, the Euclidean metric takes the form 
\begin{equation}\label{eq:SCHWE}
ds^2_{E}\approx\frac{\rho^2 d\tau^2}{(4G M)^2}+d\rho^2+(2GM)^2\left(d\theta^2+d\phi^2\sin^2\theta\right)\,\,\, \text{when}\,\,\, \rho\rightarrow 0\, ,
\end{equation}
where we have introduced the coordinate $\rho=\sqrt{8GM(r-2GM)}$. Now, we see that the line element $\frac{\rho^2 d\tau^2}{(4G M)^2}+d\rho^2$ will have a conical singularity at $\rho=0$ unless the coordinate $\tau/(4GM)$ has period $2\pi$, in whose case that line element simply represents a plane in polar coordinates. Thus, the absence of a conical singularity imposes that the Euclidean time must have a periodicity 
\begin{equation}
\beta=8\pi G M\, .
\end{equation}
It is a well-known fact that the periodicity of the Euclidean time represents the inverse of the temperature, $\beta=1/T$, for QFTs in a thermal state, so this is telling us that Schwarzschild's black hole has a temperature $T=1/(8\pi G M)$. The identification of the temperature by demanding regularity of the Euclidean metric works for any other black hole solution and is equivalent to Hawking's result \req{eq:Hawking}. 

Now, once the black hole temperature has been identified, the rest of thermodynamic quantities can be obtained by computing the Euclidean path integral for gravity \cite{Gibbons:1976ue}, which we can write formally as 
\begin{equation}
\mathcal{Z}=\int\mathcal{D}[g_{\mu\nu}]e^{-S_{\rm E}}\, ,
\end{equation}
where $S_{\rm E}$ is the Euclidean action for gravity. The Euclidean version of a black hole solution yields a solitonic contribution to the Euclidean path integral, and in the semiclassical approximation it can be estimated by
\begin{equation}\label{eq:Zonshell}
\mathcal{Z}\sim e^{-S_{\rm E}\big|_{\rm on-shell}}\, .
\end{equation}
However, in order for the saddle-point approximation that is implied in the derivation of this formula to make sense, one must make sure that the action is extremized by solutions of the equations of motion. In other words, the action must be well posed, and as we saw in Section~\ref{sec:equationsintro} this means that one has to add appropriate surface terms to the bulk action. 
Even if this is accomplished, the resulting action usually produces divergences when evaluated on-shell and one needs to regularize it. In that case one has to introduce counterterms, which contribute to the action but have no effect on the variational problem. For instance, in the case of Einstein gravity with asymptotically flat solutions, a well-posed and regularized Euclidean action reads
\begin{equation}
S_{E}=-\frac{1}{16\pi G}\int_{\mathcal{M}} d^4x\sqrt{g}R-\frac{1}{8\pi G}\int_{\partial\mathcal{M}} d^3x\sqrt{h}\left(K-K_0\right)\, ,
\end{equation}
where $K_0$ is the extrinsic curvature of the same boundary $\partial\mathcal{M}$ when embedded in flat space. Other regularization schemes are possible in the asymptotically AdS case \cite{Emparan:1999pm}. 

From \req{eq:Zonshell} we can see that the free energy of the system is simply $F=T S_{\rm E}\big|_{\rm on-shell}$, and we can compute the entropy and the mass by using the relations
\begin{equation}
S=-\left(\frac{\partial F}{\partial T}\right)_{M}\, ,\quad M=F+TS\, .
\end{equation}
If there are additional thermodynamic variables we can obtain their respective potentials by taking the corresponding derivatives of the free energy. This method has the advantage that it can be applied without any formal modification to any other gravity theory besides Einstein gravity. However, it requires knowing an appropriate boundary term and counterterms for the corresponding action, something that is not always available. 

In passing, let us also mention that there is an extended thermodynamic approach known as \emph{black hole chemistry}, in which the cosmological constant is treated as a pressure and whose conjugate thermodynamic variable is a volume \cite{Kubiznak:2014zwa,Kubiznak:2016qmn}. In that case, the Euclidean action computes the Gibbs free energy $G=T S_{\rm E}\big|_{\rm on-shell}$, so that $M$ is identified with an enthalpy, $M=G+TS$. 

\subsection{Evaporation of black holes and the information paradox}
As we have just seen, black holes can be described as thermodynamic systems, but let us stop here a moment to take a closer look at the thermodynamic properties of the prototypical case of a Schwarzschild's black hole. Its temperature and entropy read
\begin{equation}
T=\frac{1}{8\pi G M}\, ,\qquad  S_{\rm \ssc BH}=4\pi G M^2\, ,
\end{equation}
while, in turn, $M$ represents the energy. There is something really unusual about these relations because they imply that, as the black hole loses mass in the form of thermal radiation, its temperature rises. This implies that the heat capacity of the black hole is negative, which characterizes the black hole as an unstable thermodynamic system. The instability is revealed if we consider the time evolution of a black hole placed in a thermal bath at some temperature $T_0$. If the initial temperature of the black hole is below the temperature of the environment, $T<T_0$, then it will absorb more radiation than it emits and as a result it will gain mass, hence decreasing its temperature even more. On the contrary, when $T>T_0$ the full balance for the black hole is to emit radiation, and as the mass decreases the temperature rises. Eventually, the temperature diverges and the evaporation process ends with a catastrophic explosion, after which the black hole has disappeared \cite{Hawking:1974rv}. This could be, in particular, the fate of black holes in our own universe \cite{PhysRevD.13.198}.

The evaporation of black holes due to Hawking radiation gives rise to the famous \emph{information paradox}, which poses a serious conflict between gravity and quantum mechanics --- see the reviews \cite{Mathur:2009hf,Harlow:2014yka,Chen:2014jwq} and references therein. Imagine a spacetime that contains a black hole, and that the full system is initially described by a pure state. It is natural to split the system as the black hole interior region plus the exterior region, which is the only part of the wavefunction that an external observer can measure. Now, during the Hawking process a Schwinger pair is created in the surroundings of the black hole horizon. One of these quanta falls into the black hole and the other one escapes to infinity, and since the pair is entangled, this process creates entanglement between the exterior and interior regions. 
In addition, the quanta emitted at some latter stage cannot be entangled with those emitted at early times --- a process that would reduce the amount of entanglement --- because those are necessarily entangled to their partner in the interior of the black hole, and a tripartite entangled state is not allowed by the ``monogamy theorem''.  As a consequence, the entanglement between the interior of the black hole and the exterior region grows as the black hole evaporates. If, eventually, the black hole disappears after the evaporation process, the Hawking quanta are in an entangled state, but since there is nothing to which they are entangled with, they must be described by a density matrix. Hence, our pure state has evolved into a mixed state, and we conclude that unitarity has been violated.

Other alternatives are possible, but an inconsistency seems to be unavoidable. For instance, one can assume that Hawking evaporation stops when the black hole reaches some minimum mass, in whose case it becomes a \emph{remnant}. Such object would be entangled with all the Hawking quanta emitted previously, and since the original black hole can have an arbitrarily large entropy, the remnant would need to have an infinite number of internal states to get entangled with an arbitrarily large number of quanta. The problem lies in remnants having a finite mass and volume, in which case they will produce divergences in the partition function yielding a spontaneous production of these objects at an alarming rate \cite{Preskill:1992tc,Harvey:1992xk,Giddings:1994qt} --- see \cite{Susskind:1995da} for a different argument against remnants.

Some proposed solutions to the information problem involve the invocation of \emph{firewalls} \cite{Almheiri:2012rt,Almheiri:2013hfa}, black hole complementarity \cite{Susskind:1993if}, or the fuzzball proposal \cite{Mathur:2005zp,Skenderis:2008qn}, but it is not our intention to discuss these possibilities here. Instead, one of the goals of this thesis is to show that part of the conclusions above rely on the particular properties of black holes within General Relativity. We will see that the thermodynamic properties and evaporation process of black holes can be dramatically affected if one considers modifications of gravity in the UV.

\subsection{Spherically symmetric black hole solutions}\label{introspherebhs}
Although we have described a general framework to study black hole thermodynamics, the construction of explicit black hole solutions in higher-order gravity is a very different (and complicated) issue. For simplicity, we focus our attention on static and spherically symmetric (SSS) black holes, which is the first case one would attempt to solve. The most general ansatz for a SSS metric can be written as follows
\begin{equation}\label{eq:Nfintro}
ds^2_{N,f}=-N(r)^2f(r)dt^2+\frac{dr^2}{f(r)}+r^2d\Omega_{(D-2)}^2\, ,
\end{equation}
where $N(r)$ and $f(r)$ are two independent functions and where $d\Omega_{(D-2)}^2$ is the metric of the $(D-2)$-sphere. In general, the functions $N(r)$ and $f(r)$ are determined by a system of differential equations that can be chosen as the components $\mathcal{E}_{tt}=0$ and $\mathcal{E}_{rr}=0$ of the equations of motion.\footnote{Using the Bianchi identities of the equations of motion $\nabla^{\mu}\mathcal{E}_{\mu\nu}=0$, that any diff. invariant theory satisfies, it is easy to prove that the rest of components are proportional to $\mathcal{E}_{tt}$, $\mathcal{E}_{rr}$ and their derivatives.} In the case of GR with a cosmological constant, $\mathcal{L}=(R-2\Lambda)/(16\pi G)$, these equations turn out to be of first order and their only solution is given by $N(r)=N_0$ (customarily taken to $N_0=1$) and 
\begin{equation}\label{SAdSintro}
f(r)=1-\frac{16\pi GM}{(D-2)\Omega_{(D-2)} r^{D-3}}-\frac{2\Lambda r^2}{(D-1)(D-2)}\, , 
\end{equation}
which corresponds to the Schwarzschild-Tangherlini black hole. Here $M$ is an integration constant which represents the ADM mass  \cite{Arnowitt:1960es,Arnowitt:1960zzc,Arnowitt:1961zz} of the black hole. On the contrary, for a general $\mathcal{L}$(Riemann) gravity the equations $\mathcal{E}_{tt}=0$ and $\mathcal{E}_{rr}=0$ form a system of non-linear, coupled differential equations of fourth- and third-order, respectively, for the functions $N(r)$ and $f(r)$. The resolution of these equations, even numerically, poses a challenging problem, and even the existence and unicity of solutions is not guaranteed. Consequently, there are only few instances of higher-order gravities in which one can solve the equations of motion for the metric \req{eq:Nfintro} in a more or less explicit way and study black hole solutions.
We must warn at this point that the literature on black hole solutions in modified gravity theories is extensive, but we are especially interested in theories and solutions that satisfy a number of conditions that we consider to correspond to natural situations. 
\begin{enumerate} 
\item We wish to study black hole solutions of higher-order gravity in the vacuum, \ie without introducing additional fields. 
\item We would like to consider higher-order gravities that are smooth deformations of GR, in the sense that they are of the form \req{eq:HDGintro}, with the higher-derivative terms being controlled by free parameters that we can set to zero independently. 
\item The corresponding black hole solutions should be smooth non-trivial deformations of GR solutions as well, so that when the higher-order couplings are taken to zero one recovers \req{SAdSintro}. 
\end{enumerate}
The first condition discards, for instance, scalar-tensor theories with non-minimal couplings of the scalars to higher-curvature terms, such as Einstein-dilaton-Gauss-Bonnet gravity \cite{EdGB,Kanti:1995vq,Torii:1996yi,Alexeev:1996vs} and other related models \cite{Sotiriou:2014pfa,Doneva:2017bvd,Silva:2017uqg,Antoniou:2017acq}. While we consider these theories interesting, we shall not study those cases for the sake of concreteness --- we focus exclusively on metric theories. The second item removes theories that are pure higher-derivative gravities and lack an Einstein-gravity limit, such as conformal gravity $\mathcal{L}=\alpha C_{\mu\nu\rho\sigma}C^{\mu\nu\rho\sigma}$  \cite{Riegert:1984zz,Klemm:1998kf} --- see \eg \cite{Banados:1993ur,Cai:2006pq,Oliva:2010zd,Lu:2013hx} for similar examples. 
Additional theories that do not satisfy the second item are those that involve fine-tuned terms across different orders in curvature, as the case of perfect-square Lagrangians such as $\mathcal{L}=-(R-4\Lambda)^2/(8\Lambda)$ \cite{Cai:2009ac,Love}.
The third condition discards solutions that do not exist in the Einstein gravity limit. As an example of this, the works \cite{Lu:2015psa,Lu:2015cqa} showed that quadratic gravity in four-dimensions allows for ``non-Schwarzschild'' black holes besides the Schwarzschild one. These non-Schwarzschild solutions have exotic properties and they do not survive in the limit in which the higher-derivative couplings are set to zero. On the other hand, there are theories whose equations are solved by Einstein metrics, such as the aforementioned quadratic gravity in $D=4$, $f(R)$ gravity \cite{delaCruzDombriz:2009et}, or more generally, any theory whose Lagrangian only depends on Ricci curvature \cite{Li:2017ncu}. These theories do not modify GR vacuum black hole solutions and the third item states that we must also discard them because we are interested in investigating the effects of non-trivial corrections.  
Additionally, we would like to obtain exact black hole solutions of a given theory. Obtaining perturbative solutions is an accessible problem --- especially in the spherically symmetric case --- but the non-perturbative effects provided by exact solutions are far most interesting, as we will see.  

When these criteria are taken into account, few examples of black hole solutions are left, and all of them in $D\ge5$. Before the results of this thesis were presented, the only theories for which exact black hole solutions deforming the Schwarzschild geometry had been constructed were Lovelock gravities \cite{Wheeler:1985nh,Wheeler:1985qd,Boulware:1985wk,Cai:2001dz,Dehghani:2009zzb,deBoer:2009gx}, Quasi-topological gravity \cite{Quasi2,Quasi} and its quartic \cite{Dehghani:2011vu} and quintic \cite{Cisterna:2017umf} generalizations. Let us review these cases now. 

\subsubsection{Black holes in Lovelock gravity}
Previously, we introduced Lovelock gravity as the most general higher-curvature gravity that possesses second order equations of motion. Its action, given by \req{LLaction}, is composed as a linear combination of dimensionally extended Euler densities, and we saw that in $D=4$ the only non-trivial terms are the Einstein-Hilbert one plus a cosmological constant. Thus, Lovelock theory only introduces corrections to GR in higher dimensions.  As a first example, let us study the black hole solutions in five-dimensional Lovelock gravity, in which case the Gauss-Bonnet density $\mathcal{X}_{4}$ --- given by \req{eq:introGB}--- becomes non-trivial. For simplicity, let us also momentarily set to zero the cosmological constant, so that we consider the action 
\begin{equation}\label{eq:GBactionintro}
S_{GB}=\frac{1}{16\pi G}\int d^{5}x\sqrt{|g|}\left[R+ \frac{\lambda}{2}L^{2}\mathcal{X}_{4}\right]\, ,
\end{equation}
where $\lambda$ is a dimensionless parameter and $L$ a length scale. The black hole solutions of this theory were first analyzed in the works \cite{Wheeler:1985nh,Wheeler:1985qd,Boulware:1985wk}. A certain combination of the equations of motion evaluated on \req{eq:Nfintro} implies that $N'(r)=0$, so that this function is a constant. In order to ensure that the solution is asymptotically flat, where the coordinate $t$ is identified with the time of an asymptotic observer, we set $N(r)=1$. On the other hand, it is found that the differential equation satisfied by the remaining variable $f(r)$ can be integrated, yielding an algebraic equation 
\begin{equation}\label{eq:polGB}
-r^2(f(r)-1)+\lambda L^2 \left(f(r)-1\right)^2=\omega^2\, ,
\end{equation}
where $\omega$ is an integration constant with units of length. The two roots of this equation provide two different solutions
\begin{equation}\label{eq:GBsolintro}
f(r)=1+\frac{r^2}{2 \lambda L^2}\left[1\pm\sqrt{1+\frac{4 \lambda L^2 \omega ^2}{r^4}}\right]\, ,
\end{equation}
but only one of these solutions --- the one with the ``$-$'' sign --- is asymptotically flat and has a well-defined Einstein gravity limit $\lambda\rightarrow 0$. The other solution, corresponding to the ``$+$'' sign, is asymptotically de Sitter $(\lambda<0)$ or anti-de Sitter $(\lambda>0)$ and it is singular in the limit $\lambda\rightarrow 0$. Let us therefore consider the former case. 
First of all, by expanding the solution for large $r$, and using the usual ADM prescription for the mass of a spacetime \cite{Arnowitt:1960es,Arnowitt:1960zzc,Arnowitt:1961zz} --- which extends to the case of higher-order gravity as well  \cite{Deser:2002jk,Senturk:2012yi} --- we can see that $\omega^2$ is related to the mass according to 
\begin{equation}
\omega^2=\frac{8GM}{3\pi}\, .
\end{equation}
In addition, one can easily check that in the limit $\lambda\rightarrow 0$ one recovers the five-dimensional Schwarzschild-Tangherlini solution \req{SAdSintro}, so \req{eq:GBsolintro} is a smooth deformation of an Einstein gravity solution. 
However, the properties of the Gauss-Bonnet black hole are quite different from those of Einstein gravity black holes. First we must determine the radius of the horizon, which is fixed by the condition $f(r_h)=0$. By looking at \req{eq:polGB} we derive immediately the following relation between $r_h$ and the parameter $\omega$:
\begin{equation}
r_h^2=\omega^2-\lambda L^2\, .
\end{equation}
We see that something very unusual happens for $\lambda>0$: the black hole becomes of zero size for a finite value of the mass, namely $M_{\rm min}=3\pi \lambda L^2/(8 G)$. For smaller masses, the solution does not represent a black hole anymore, but a naked singularity. In the case of negative $\lambda$, the solution becomes singular for $M=3\pi |\lambda| L^2/(8 G)$, but now this happens at a finite radius. Then, we can analyze the black hole's temperature, which can be computed according to the simple relation $T=f'(r_h)/(4\pi)$, and as a function of $\omega$ it reads
\begin{equation}
T=\frac{\sqrt{\omega ^2-\lambda L^2}}{2\pi\left(\lambda  L^2+\omega ^2\right)}\, ,
\end{equation}
For large masses, $\omega^2\rightarrow\infty$, we observe that the temperature reduces to the Einstein gravity value, $T\propto M^{-1/2}$. However, the behaviour for small masses is completely different, and it is especially interesting for $\lambda>0$. In that case we get that the temperature vanishes when $\omega ^2=\lambda L^2$, \ie when the minimum mass $M=M_{\rm min}$ is reached, so the black holes of the minimal mass are extremal. This also implies that there exists a maximum temperature and that black holes of small masses have positive specific heat. This behaviour differs dramatically from the GR prediction, where one finds that the temperature of small black holes diverges, and it has intriguing consequences for the evaporation process of black holes \cite{Myers:1988ze}. In fact, the final stage of an evaporating Gauss-Bonnet black hole is one of these extremal black holes of mass $M_{\rm min}$. In addition, since the temperature vanishes as the mass approaches the minimum value, the extremal limit is never reached in a finite time. This is an excellent example of how the introduction of higher-derivative terms can cure some of the divergences present in GR.

Let us now study in a less detailed way the black hole solutions of general $D$-dimensional Lovelock gravity, as introduced in \req{LLaction}. For convenience, let us we rewrite that action as follows
\begin{equation}\label{LoveF}
S_{LL}=\frac{1}{16\pi G}\int d^{D}x\sqrt{|g|}\left\{-2\Lambda+R+ \sum_{n=2}^{\lfloor \frac{(D-1)}{2}\rfloor}\lambda_n \frac{(D-1-2n)!}{(D-3)!}(-1)^{n}L^{2n-2}\mathcal{X}_{2n}\right\}\, ,
\end{equation}
where $\lambda_n$ are dimensionless coupling constants and we have included an arbitrary cosmological constant term. For completeness, we will not restrict ourselves to spherically symmetric solutions only, and we will study as well black holes with planar and hyperbolic horizons --- these appear naturally when considering asymptotically AdS solutions. Thus, we assume a metric ansatz of the form
\begin{equation}\label{eq:Nfintro2}
ds^2=-N(r)^2f(r)dt^2+\frac{dr^2}{f(r)}+r^2d\Sigma_{k,(D-2)}^2\, ,
\end{equation}
where $d\Sigma_{k,(D-2)}^2$ is the metric of a maximally symmetric space of curvature $k=-1,0,1$, corresponding, respectively, to spherical, planar and hyperbolic geometries.

As in the case of Gauss-Bonnet gravity, the equations of motion imply that $N'(r)=0$, so that this function is a constant $N(r)=N_k$ --- for $k=1$ we set $N_k=1$ but other choices are more natural for $k\neq1$. On the other hand, the differential equation for $f(r)$ can be integrated yielding an algebraic equation. It is convenient to introduce the function $g(r)= L^2(f(r)-k)/r^2$, in terms of which the equation reads
\begin{equation}\label{BHeqLove}
h\left(g\right)=\frac{\omega^{D-3}L^2}{r^{D-1}}\, ,
\end{equation}
where $\omega$ is an integration constant and  $h(x)$ is the polynomial function
\begin{equation}
h(x)=-\frac{2\Lambda L^2}{(D-1)(D-2)}-x+\sum_{n=2}^{\lfloor \frac{D-1}{2}\rfloor}\lambda_n x^n\, .
\end{equation}
In this way, the black hole solutions of Lovelock gravity are obtained by solving the roots of  \req{BHeqLove} ---  see Ref.~\cite{Camanho:2011rj} for a detailed discussion.  The integration constant $\omega$ can be related to the mass $M$ of the black hole according to $\omega^{D-3}=\frac{16\pi G M}{(D-2)V_{k}}$ where $V_{k}$ is the volume of the transverse space. Nevertheless, $M$ only represents the mass in the spherically symmetric case, $k=1$. In the non-compact cases $k=0,-1$ we should interpret $M/V_{k}$ as a mass density --- this makes special sense in the holographic context, as we will see. 
Finally, it is possible to study the thermodynamic properties of Lovelock black holes as we just did in the Gauss-Bonnet case.  
In particular, the temperature and entropy --- the latter computed from Wald's formula \req{eq:Waldintro}, or equivalently from the Jacobson-Myers' one~\cite{Jacobson:1993xs} --- of these black holes read
\begin{align}
T&=-\frac{N_k}{4\pi L}\left[\frac{r_h(D-1) h\left(-k L^2/r_h^2\right)}{L h'\left(-k L^2/r_h^2\right)}+\frac{2kL}{r_h}\right]\, ,\\
\label{ENT}
S&=\frac{r_h^{D-2}V_{k}}{4 G}\left[1-\sum_{n=2}^{\lfloor \frac{D-1}{2}\rfloor}\lambda_n\left(-\frac{k L^2}{r_h^2}\right)^{n-1}\frac{n(D-2)}{(D-2n)}\right]\, .
\end{align}
Analyzing these formulas, one finds that the existence of stable black holes or black holes with degenerate horizons are quite common features of Lovelock gravities \cite{Myers:1988ze}.

\subsubsection{Black holes in Quasi-topological gravity}
As we have seen, the equations of motion of Lovelock gravity for SSS metrics take a relatively simple form and it is possible to solve them explicitly. In general, for  other higher-order gravities the equations of motion are much more involved and a similar resolution is not possible. In addition, we have found that the corresponding black hole solutions of Lovelock gravity have very interesting properties that depart greatly from the GR case.  
However, Lovelock theories have the disadvantage of being highly constrained by the spacetime dimension. Thus, in $D=4$ they introduce no corrections, and in $D=5$ the only non-trivial term beyond GR is given by the Gauss-Bonnet density. In order to overcome this restriction, a new theory known as Quasi-topological gravity was introduced, independently, in Refs.~\cite{Quasi,Quasi2}.  Although this theory does not share the unique property of Lovelock gravity of possessing second-order equations of motion, it does mimic some aspects of these theories. We already mentioned that Quasi-topological gravity has the nice property of satisfying second-order linearized equations on constant-curvature backgrounds, which guarantees that it propagates the same degrees of freedom as GR. In addition, the spherically symmetric black hole solutions of Quasi-topological gravity can be constructed explicitly, in a very similar fashion as in Lovelock gravity. 
Let us review here the construction of five-dimensional black holes in this theory as presented in \cite{Quasi2}. We consider the action
\begin{equation}\label{5dsg}
S=\frac{1}{16\pi G}\int d^5x\sqrt{|g|}\left[\frac{12}{L^2}+R+\frac{\lambda L^2}{2} \mathcal{X}_4+\frac{7\mu L^4}{4}  \mathcal{Z}_5\right]\, ,
\end{equation}
where we chose the cosmological constant to be negative and given by $\Lambda=-6/L^2$, where the scale $L$ coincides with the AdS$_5$ radius when $\mu=\lambda=0$. On the other hand, we include the Gauss-Bonnet term $\mathcal{X}_4$ as well as the five-dimensional Quasi-topological gravity term, $\mathcal{Z}_5$, as given by \req{eq:QuasiTopointro}. 
%\begin{equation}
%\begin{aligned}
%\mathcal{Z}_5&=\tensor{R}{_{a}^{c}_{b}^{d}}\tensor{R}{_{c}^{e}_{d}^{f}}\tensor{R}{_{e}^{a}_{f}^{b}}+\frac{1}{56}\Big(-72\tensor{R}{_{abcd}}\tensor{R}{^{abc}_{e}}R^{de}+21\tensor{R}{_{abcd}}\tensor{R}{^{abcd}}R+120\tensor{R}{_{abcd}}\tensor{R}{^{ac}}\tensor{R}{^{bd}}\\
%&+144 R^{b}_{a} R_{b}^{c} R_{c}^{a}- 132R_{ab}R^{ab}R+15 R^3\Big)\, ,
%\end{aligned}
%\end{equation}
We consider a metric ansatz with different possible choices of horizon topologies, as before
\begin{equation}\label{eq:Nfintro3}
ds^2=-N(r)^2f(r)dt^2+\frac{dr^2}{f(r)}+r^2d\Sigma_{k,(3)}^2\, ,
\end{equation}
where $N(r)$ and $f(r)$ are unknown functions. As in the case of Lovelock gravity, one finds that the equations of motion set $N(r)$ to a constant $N_k$, while the equation for $f(r)$ can be reduced to an algebraic one. Introducing $f(r)=k+r^2g(r)/L^2$ the equation satisfied by $g(r)$ reads
\begin{equation}\label{eq:QTsol}
1-g+\lambda g^2+\mu g^3=\frac{\omega^4}{r^4}\, ,
\end{equation}
where $\omega$ is an integration constant.  Thus, the situation is very similar to the Lovelock case, but now we have a cubic term in the left-hand-side of this equation --- such term would not be allowed in five-dimensional Lovelock gravity.  Equation \req{eq:QTsol} can have several solutions and a different character depending on the values of the couplings $\lambda$ and $\mu$, and we refer to the original work \cite{Quasi2} for a detailed analysis. In any case, let us emphasize that asymptotically, this equation becomes
\begin{equation}\label{eq:QTsol2}
1-g_{\infty}+\lambda g_{\infty}^2+\mu g_{\infty}^3=0\, ,
\end{equation}
The roots of this polynomial determine the possible AdS vacua of the theory \req{5dsg}, which have a radius $\tilde L^2=L^2/g_{\infty}$. Only one of the possible vacua is smoothly connected to the Einstein gravity one when $\lambda, \mu\rightarrow 0$, and this is the one that is taken as physical. 

It is possible to repeat the construction of black holes with the quasi-topological interaction \req{eq:QuasiTopointro} in higher dimensions $D\ge 6$. However, in those cases the contribution of this term to the equations of motion for metrics of the form \req{eq:Nfintro2} is the same as the one from the cubic Lovelock density $\mathcal{X}_6$, so that it does not provide new modifications.\footnote{In particular, for $D=6$ the term $\mathcal{Z}_6$ does not contribute to the equations of motion for SSS metrics, and this is the origin of the name ``Quasi-topological''.} On the other hand, it is possible to obtain higher-order generalizations of the cubic density $\mathcal{Z}_{D}$. In particular, Refs.~\cite{Dehghani:2011vu,Cisterna:2017umf} constructed, respectively, the quartic and quintic versions of Quasi-topological gravity. These terms share the same properties of the cubic interaction that we mentioned here: they possess second-order linearized equations on constant-curvature backgrounds,\footnote{Actually, this has only been explicitly proven for the quintic theory \cite{Cisterna:2017umf} thanks to the efficient linearization method presented in \cite{Aspects}, that we review in  Chapter \ref{Chap:1}.} and they allow for black hole solutions of the form \req{eq:Nfintro3} with $N(r)=$ constant and $g(r)$ satisfying an equation similar to \req{eq:QTsol} (with the polynomial in the left-hand-side containing higher powers of $g$).  Most likely, it will be possible to extend the construction of this type of theories to arbitrary orders in curvature.

However, there is an important drawback: none of these quasi-topological theories exist in four dimensions. At the same time, this means that the study of non-trivial modifications of the Schwarzschild solution in $D=4$ still remains as an open subject. 
Let us note here that all the black hole solutions that we have reviewed satisfy the property of having $N(r)=$constant, or in other words, they have $g_{tt}g_{rr}=$ const. when expressed in Schwarzschild-like coordinates. In addition, all the theories in which these types of solutions have been constructed possess second-order linearized equations, and this suggests that there might be a relation between both properties. We will show in Chapter~\ref{Chap:2} that there is indeed a connection between these properties, and furthermore, we will see that the theories that satisfy the condition $g_{tt}g_{rr}=$ const. are very appealing to study SSS black hole solutions. This will lead us to the construction of a generalization of Quasi-topological gravity that is non-trivial in $D=4$. The new theories will allow us to provide the first examples of exact, non-trivial modifications of the four-dimensional Schwarzschild solution in higher-order gravity.

\section{Holography}\label{sec:holoint}
Earlier in this introduction, we already met String Theory as a promising candidate for a quantum theory of gravity. One of the most important characteristics of ST is the existence of an extensive net of \emph{dualities} that relate the different types of string theories, or different ways of realizing them. When two theories are dual, they are equivalent in a mathematical sense, but physically they may describe dramatically different systems. The dualities discovered in ST have had a profound impact in other areas of physics --- and even in mathematics \cite{Kontsevich:1994na,Kontsevich:1994dn} ---  and they have led to connections that no one would have expected. In this respect, one of the most impressive dualities predicted by ST is known as the Anti-de Sitter/Conformal Field Theory (AdS/CFT) correspondence \cite{Maldacena,Gubser,Witten}. On general grounds, this correspondence relates String Theory placed on a $D$-dimensional negatively curved space --- Anti-de Sitter space --- to a conformal field theory on $(D-1)$-dimensional flat space. There are two striking facts about this duality. First, it relates theories of two different dimensionalities, and in this sense the AdS/CFT correspondence is a particular case of the \emph{holographic principle} \cite{tHooft:1993dmi,Susskind:1994vu,Bousso:2002ju}. From this point of view, it is usually interpreted that the CFT lives on the boundary of the AdS space, and that the physics in the \emph{bulk} are a hologram of the physics in the boundary. Second, it maps a theory of quantum gravity such as String Theory, to a quantum field theory without gravity. Since quantum field theories are much better understood than String Theory, it has been claimed that the AdS/CFT correspondence can be used to define ST. In addition, it is also believed that this duality solves the problem of information loss in black holes: since the process of Hawking evaporation will be described by unitary evolution in the CFT side, by construction no unitarity loss is possible --- see \cite{Mathur:2009hf} for a criticism though. 

However, the AdS/CFT duality can also be applied in the opposite direction: we can use a theory of gravity in order to learn about conformal field theories. The success of this approach relies on the fact that, in an appropriate limit, the gravitational theory becomes classical, and the calculations in the bulk are often simpler than in the CFT side. In order to illustrate this, let us review the original form of the conjecture.

\subsubsection{Type IIB String Theory/Super-Yang-Mills correspondence}
There are many explicit forms of the AdS/CFT correspondence that arise from String Theory, but the most well-known is the one due to Maldacena \cite{Maldacena}. Explicitly, it states the equivalence between $d=4$, $\mathcal{N}=4$ Super-Yang-Mills (SYM) theory with gauge group $SU(N)$ and coupling constant $g_{\rm YM}$, and type IIB Superstring Theory on AdS$_5\times\mathbb{S}^5$, with radius of curvature $L$ and $N$ units of $F_{(5)}$ flux on $\mathbb{S}^5$. Both theories contain two parameters: the SYM theory depends on $N$ and on $g_{\rm YM}$, while String Theory depends on $\alpha'=\sqrt{\ell_s}$ and the string coupling $g_s$. According to the AdS/CFT duality, these parameters are related in the following way
\begin{equation}
g_{\rm YM}^2=2\pi g_s\, ,\qquad 2 g_{\rm YM}^2 N=\frac{L^4}{\alpha'^2}\, .
\end{equation}
Then, the fundamental identity of the duality relates the partition functions of both theories
\begin{equation}
\mathcal{Z}_{\rm CFT}=\mathcal{Z}_{\rm ST}\, .
\end{equation}
We emphasize that the correspondence is conjectured to hold for any value of the parameters and that in general it relates two seemingly different quantum theories. However, there is an interesting limit which is usually expressed in terms of the 't Hooft coupling $\lambda=g_{\rm YM}^2 N$. First, when we fix $\lambda$ to a constant value and we take $N\rightarrow\infty$, the string coupling constant goes to zero $g_s\rightarrow0$, and in this case String Theory becomes classical. If in addition we take $\lambda$ to be very large, we get $\sqrt{\alpha'}/L\rightarrow 0$, and in this situation, String Theory is accurately described by a classical supergravity theory. 
The fact that the large $N$ and large 't Hooft coupling limit of a CFT is described by a classical gravity theory in the bulk is a generic result that extends to other forms of the AdS/CFT correspondence. The main advantage is that now the partition function of the bulk theory can be computed at tree level using the corresponding classical action,\footnote{We show the identification in Euclidean signature.}
\begin{equation}
\mathcal{Z}_{\rm ST}\sim e^{-S_{\rm grav}}\, .
\end{equation}
Thus, identifying the fields in both sides one can compute correlation functions of the CFT  by taking the appropriate functional derivatives in the bulk action. More generally, the relation between quantities in both sides of the duality is known as the \emph{holographic dictionary}.

\subsubsection{Finite $N$ and $\lambda$ corrections }
 As we saw in Section~\ref{sec:stringyintro}, the stringy effective actions are a double series in the parameters $g_s$ and $\alpha'$, and if these are very small the full action can be approximated by the leading term, which corresponds to a supergravity theory. 
This is the reason why the strongly coupled and large $N$ regime of a CFT corresponds  to a classical (super)gravity theory in the bulk. However, strictly speaking that approximation only captures the behavior in the limit $N,\lambda \rightarrow\infty$. 
If one wishes to study a CFT with large but finite values of $N$ and $\lambda$, the dual bulk theory must have small, but non-zero values of $g_s$ and $\sqrt{\alpha'}/L$. This means that one should consider the subleading terms in the stringy effective actions \req{eq:effstring}, which in general contain higher-derivative operators.  As an example, let us consider the quartic action \req{acads5} which arises in type IIB Superstring Theory on AdS$_5\times\mathbb{S}^5$. When the origin of the quartic terms is analyzed one finds that the pre-factor is related in the following way with the SYM theory parameters \cite{Myers:2008yi}
\begin{equation}
\zeta(3)\frac{\alpha'^3}{L^6}\rightarrow\frac{\zeta(3)}{\lambda^{3/2}}+\frac{\lambda^{1/2}}{48 N^2}\, .
\end{equation}
Thus, when we apply the AdS/CFT correspondence to a classical gravity theory supplemented with higher-derivative corrections we are capturing finite $N$ and $\lambda$ effects in the dual CFT. However, the precise terms that have to be included as well as the relation with $N$ and $\lambda$ are dictated by String Theory in every particular case.

\subsection{Higher-curvature gravities as holographic toy models}\label{sec:holotoymodels}
Even though the AdS/CFT correspondence has its origin in String Theory, the holographic relationship between classical gravity and conformal field theories is nowadays understood as a general principle, sometimes called \emph{gauge/gravity duality}. In this sense, ST establishes what theories are related and provides a precise dictionary between the parameters and fields in one and the other side of the correspondence. However, it is also a valid strategy to consider a given classical theory of gravity with a negative cosmological constant as a holographic toy model.  In support of this statement, imagine that we are only interested in the gravitational sector and we take a bulk theory whose only field is the metric. Such theory might probably appear as the truncation of many other theories that contain additional fields and for which an explicit holographic correspondence with a CFT exists. Thus, when we study the holographic dictionary of this gravitational theory we are not exploring a unique CFT, but a whole family of CFTs that belong to the same \emph{universality class}. 

In this respect, when one studies the holographic dual of Einstein gravity, only a restricted set of CFTs is explored. A very successful approach that has usually been applied in order to broaden the spectrum of CFTs that one can analyze consists in considering higher-derivative gravities as bulk theories. This allows us to explore CFTs that belong to different universality classes from the one defined by GR \cite{Buchel:2008vz,Hofman:2008ar,Hofman:2009ug}. For instance, these theories can have $a\neq c$ in $d=4$ \cite{Nojiri:1999mh,Blau:1999vz} or a more general 3-point function of the stress-energy tensor from the one predicted by holographic Einstein gravity \cite{Camanho:2009vw,Buchel:2009sk,Camanho:2009hu,Camanho:2010ru,Camanho:2013pda,Myers:2010jv} --- we review these cases below. In relation to this, higher-order gravities can be used in order to obtain relationships between quantities of CFTs valid for general theories. The idea is that, if a certain property holds for all holographic CFTs dual to higher-order gravities, then it might actually hold for arbitrary CFTs ---  see \cite{Myers:2010tj,Myers:2010xs,Mezei:2014zla,Bueno1,Bueno2} for successful examples of this type of applications. In other cases, higher-order gravities have served just the opposite purpose, namely providing counterexamples of previously conjectured relationships. One interesting example is given by the Kovtun-Son-Starinets (KSS) bound for the shear viscosity over entropy density ratio \cite{Kovtun:2004de}. The KSS conjecture states that, for any fluid or plasma in nature, the ratio between these quantities is not smaller than $1/(4\pi)$ (in natural units), which is the value predicted by Einstein gravity holography. However, holographic hydrodynamics computations with higher-curvature gravities provide evidence that this bound might be violated in some cases \cite{Buchel:2004di,Kats:2007mq,Brigante:2007nu,Myers:2008yi,Cai:2008ph,Ge:2008ni}.  
Thus, holographic higher-order gravities should be regarded as useful toy models that allow us learn about new phenomena in CFTs, and to perform many computations of physical quantities otherwise practically inaccessible in a field theory approach--- see \eg \cite{HoloRen,Hung:2014npa,deBoer:2011wk,Bianchi:2016xvf} for additional examples.

There are many aspects of a CFT that can be studied looking only at the gravitational sector of the bulk theory. In what follows we review in more detail some entries of the holographic dictionary and discuss the relevance of higher-curvature corrections in each case. 
 
\subsubsection{Correlators of the stress-energy tensor}
According to the holographic dictionary, the metric perturbation on the boundary of AdS, $h_{ab}$, couples to the stress-energy tensor of the CFT, $T^{ab}$. Thus, the gravitational sector of a bulk theory determines the different correlators of the dual CFT stress-energy tensor. These can be computed by studying perturbations on a pure AdS geometry and evaluating the corresponding gravitational action. For instance, if we study a perturbation of AdS space in the Poincar\'e patch of the form
\begin{equation}
ds^2=\frac{L^2dr^2}{r^2}+\frac{r^2}{L^2}dx^{a}dx^{b}\left(\eta_{ab}+h_{ab}(x)\right)\, .
\end{equation}
we can compute the 2-point function as
\begin{align}\label{2pointintro}
\langle T_{ab}(x)T_{cd}(x')\rangle_{\ssc\text{CFT}} =-\frac{\delta S_{\rm grav}}{\delta h^{ab}(x)\delta h^{cd}(x')}\bigg|_{h_{ab}=0}\, .
\end{align}
A very interesting property of conformal invariance is that the form of these correlators is enormously constrained. Thus, in the case of the 2-point function above, it takes the following form for any CFT
\begin{equation}
\braket{ T_{ab}(x)T_{cd}(x')} =\frac{\ctt}{|x-x'|^{2d}}\mathcal{I}_{ab,cd}(x-x')\, ,
\end{equation}
where
\begin{equation}
\mathcal{I}_{ab,cd}(x)\equiv\frac{1}{2}\left(I_{ac}(x)I_{bd}(x)+I_{ad}(x)I_{bc}(x)\right)-\frac{1}{d}\eta_{ab}\eta_{cd}\, ,\quad \text{and} \quad I_{ab}(x)\equiv\eta_{ab}-2\frac{x_{a}x_{b}}{x^2}\ ,
\end{equation}
is a fixed tensorial structure and the only theory-dependent quantity in each case is the \emph{central charge} $\ctt$. Likewise, in dimension $d\ge4$ the 3-point function of $T_{ab}$ only depends on three constants \cite{Erdmenger:1996yc,Osborn:1993cr}, and one combination of them is related to $\ctt$, hence implying that only two additional parameters are needed in order to fix the 3-point function of the CFT. In $d=3$ only one additional constant besides $\ctt$ is required, while in $d=2$ the 3-point function is completely determined by the central charge $\ctt$. A direct holographic computation of the 3-point function using an expression analogous to \req{2pointintro} is possible \cite{Arutyunov:1999nw}, but in many cases it involves an extremely lengthy computation. An alternative way of obtaining the parameters of the 3-point function entails examining energy fluxes in the boundary of AdS after a local perturbation was created by the insertion of the stress-energy tensor of the form $\epsilon_{ab}T^{ab}$ \cite{Hofman:2008ar}.
The energy flux that escapes at null infinity in the direction of the unit vector $\vec{n}$ is then given by
\begin{equation}\label{introenergyflux}
\langle\E(\vec{n})\rangle=\frac{E}{\Omega_{(d-2)}}\left[1+t_2\left(\frac{\epsilon^{*}_{ab}\epsilon_{ac}n^bn^c}{\epsilon^{*}_{ab}\epsilon^{*}_{ab}}-\frac{1}{d-1}\right)+t_4\left(\frac{\left|\epsilon^{*}_{ab}n^an^b\right|^2}{\epsilon^{*}_{ab}\epsilon^{*}_{ab}}-\frac{2}{d^2-1}\right)\right]\, .
\end{equation}
and the two coefficients $t_2$ and $t_4$, together with the central charge $\ctt$, can be used to determine the full 3-point function ---see the appendix of \cite{Buchel:2009sk}. Note that the structure multiplied by $t_2$ vanishes in $d=3$, and in that case only the coefficient $t_4$ is relevant, in agreement with the number of independent terms in the 3-point function in that dimension.
Now, the holographic computation of energy fluxes is a much more amenable task than the direct computation of 3-point correlators \cite{Camanho:2009vw,Buchel:2009sk}. 

Let us then consider Einstein gravity as a holographic toy model,
\begin{equation}\label{eq:EGholointro}
S=\frac{1}{16\pi G}\int d^{d+1}x\sqrt{|g|}\left[\frac{d(d-1)}{L^2}+R\right]\, ,
\end{equation}
where the negative cosmological constant $\Lambda=-\frac{d(d-1)}{2L^2}$ is defined in a way that $L$ represents the ``radius'' of AdS.
When the holographic dictionary of this theory is studied, one finds that the central charge of the stress tensor 2-point function reads
\begin{equation}\label{cteintro}
\ctte=\frac{\Gamma[d+2] L^{d-1}}{8\pi^{\frac{d+2}{2}}(d-1)\Gamma\left[\frac{d}{2} \right]G}\, ,
\end{equation}
while the coefficients $t_2$ and $t_4$ vanish. Thus, as a holographic toy model, Einstein gravity only explores CFTs with a particular form of the stress-energy tensor 3-point function. If one wants to study holographic CFTs with a more general 3-point function, then one must consider adding higher-curvature terms in the bulk action. The simplest non-trivial extension of GR that serves to this purpose is Gauss-Bonnet gravity \cite{Camanho:2009vw,Buchel:2009sk}, or more generally, Lovelock gravities \cite{Camanho:2009hu,Camanho:2010ru,Camanho:2013pda}. We already found these theories in Section~\ref{sec:hogexamples}, where we saw that they are the most general higher-derivative gravities possessing second-order equations of motion. In addition, we also saw that  some of them appear naturally in string effective actions \cite{Boulware:1985wk,Metsaev:1987zx}. For these reasons, Lovelock gravities provide interesting holographic toy models. 
In the previous references \cite{Camanho:2009vw,Buchel:2009sk,Camanho:2009hu,Camanho:2010ru,Camanho:2013pda}, it was determined that Lovelock gravities in general give rise to a non-vanishing $t_2$ in the energy flux \req{introenergyflux}, so that they explore CFTs inequivalent to those captured by EG holography. However, all CFTs dual to Lovelock gravities have $t_4=0$, so they still do not capture CFTs with a general stress tensor 3-point function  --- in particular, $t_4=0$ is a property that holds for any supersymmetric CFT. 
This was one of reasons to introduce Quasi-topological gravity in Ref.~\cite{Quasi2}, a theory that we reviewed in Section~\ref{sec:hogexamples}. Several holographic aspects of Quasi-topological gravity were studied in \cite{Myers:2010jv}, and we reproduce here the results found for the flux parameters. We consider the five-dimensional version of Quasi-topological gravity together with a Gauss-Bonnet term,
\begin{equation}\label{5dsg-2}
S_{\rm\ssc QT}=\frac{1}{16\pi G}\int d^5x\sqrt{|g|}\left[\frac{12}{L^2}+R+\frac{\lambda L^2}{2} \mathcal{X}_4+\frac{7\mu L^4}{4}  \mathcal{Z}_5\right]\, .
\end{equation}
First one must determine the AdS solutions of this theory. Unlike the case of EG in \req{eq:EGholointro}, the length scale $L$ does not coincide with the radius of AdS anymore. Instead, it is customary to denote the radius of AdS as $\tilde L=L/\sqrt{f_{\infty}}$, where $f_{\infty}$ is a constant. Then, the equations of motion imply that $f_{\infty}$ must be a root of the following polynomial
\begin{equation}\label{eq:QTsol2-2}
1-f_{\infty}+\lambda f_{\infty}^2+\mu f_{\infty}^3=0\, ,
\end{equation}
This equation may have several solutions with $f_{\infty}>0$, but we must choose the one that is connected to the EG vacuum, \ie the one that satisfies $\lim_{\mu,\lambda\rightarrow0}f_{\infty}=1$. Once the AdS radius is determined, one can proceed to compute the 2-point function and the energy fluxes --- see the details in \cite{Myers:2010jv} --- and the result reads
\begin{equation}\label{cteintro2}
\ctt=\frac{5 \tilde L^{3}}{\pi^{3}G}\left(1-2\lambda f_{\infty} -3\mu f_{\infty}^2\right)\, ,
\end{equation}
\begin{align}
\label{eq:t2gqtg5D}
t_2=\frac{24f_{\infty} \left(\lambda-87 \mu f_{\infty}\right)}{1-2\lambda f_{\infty}  -3\mu f_{\infty}^2}\, ,\qquad 
t_4=\frac{3780f_{\infty}^2  \mu}{1-2\lambda f_{\infty} -3\mu f_{\infty}^2}\ .
\end{align}
Thus, the Quasi-topological interaction yields a non-zero value of $t_4$, and hence provides a holographic toy model of a non-supersymmetric CFT$_4$. In addition, the three parameters $\ctt$, $t_2$, $t_4$ are now independent and therefore the theory \req{eq:QTsol2-2} can be used to explore CFTs with arbitrary 2- and 3-point functions. However, there are causality and unitarity constraints that $t_2$ and $t_4$ must satisfy in order for the dual theory to be well-behaved.  These constraints restrict the range of values that the couplings $\lambda$ and $\mu$ can take. 

A similar analysis can probably be performed for higher-dimensional Quasi-topological gravity (or for its higher-curvature generalizations \cite{Dehghani:2011vu,Cisterna:2017umf}). However, since Quasi-topological and Lovelock gravities are trivial in $D=4$, they cannot be used to study three-dimensional CFTs with a general stress tensor 3-point function --- which requires a non-vanishing $t_4$. 

\subsubsection{Trace anomaly}
Conformal symmetry at the classical level implies that the stress-energy tensor of a CFT is traceless $\tensor{T}{^{a}_{a}}=0$. At the quantum level, this identity should transform into $\langle\tensor{T}{^{a}_{a}}\rangle=0$ but this not always satisfied due to the presence of anomalies.  In the case of even-dimensional CFTs a \emph{trace anomaly} appears when we place the theory in a curved background.  
In $d=4$ the anomaly is controlled by the two central charges $a$ and $c$ and it takes the form
\begin{equation}
\langle\tensor{T}{^{a}_{a}}\rangle=\frac{c}{16\pi^2}I_4-\frac{a}{16\pi^2}\mathcal{X}_4\, ,
\end{equation}
where $I_4=W_{abcd}W^{abcd}$ is the Weyl squared invariant and $\mathcal{X}_4$ is the Gauss-Bonnet density, both of them evaluated on the background geometry. These charges satisfy the relationship (for $d=4$)\cite{Hofman:2008ar,Buchel:2009sk,Myers:2010jv}
\begin{equation}\label{eq:cma}
\frac{c-a}{c}=\frac{1}{6}t_2+\frac{4}{45}t_4\, ,
\end{equation}
so that for holographic CFTs with an Einstein gravity dual we will have $a=c$ --- the precise result of a holographic computation yields \cite{Henningson:1998ey,Henningson:1998gx,Schwimmer:2008yh} $a=c=\pi L^3/(8G)$. However, as happened for the three-point function charges, the degeneracy between $a$ and $c$ is broken when we consider higher-derivative gravities in the bulk --- and this is specially evident looking at \req{eq:cma}. Following with the example of Quasi-topological gravity in \req{5dsg-2}, the charges $a$ and $c$ read \cite{Myers:2010jv} 
\begin{equation}
a=\frac{\pi \tilde L^3}{8G}\left(1-6\lambda f_{\infty}+9\mu f_{\infty}^2\right)\, ,\quad c=\frac{\pi \tilde L^3}{8G}\left(1-2\lambda f_{\infty}-3\mu f_{\infty}^2\right)\, ,
\end{equation}
which now are different. 
Higher-order gravities have played a central role in establishing the monotonicity theorems of Refs.~\cite{Myers:2010tj,Myers:2010xs} precisely because they allow distinction between $a$ and $c$ charges.

\subsubsection{Thermodynamic phase space}
According to the AdS/CFT correspondence, the dual description of a black hole bulk geometry is a CFT in a thermal state. This is directly related to the fact that black holes are thermodynamic systems, and indeed, the thermal partition function of the CFT is computed by evaluating the Euclidean gravitational action on black hole solutions, according to $-\log \mathcal{Z}_{\rm\ssc CFT}=S_E$. As we discussed at the end of Section~\ref{sec:thermointro}, the gravitational bulk action needs to be supplemented with appropriate boundary terms and counterterms that make it well-posed and that regularize it. We reviewed boundary terms for some theories in Section~\ref{sec:hogexamples},
and as for counterterms, there is a natural way to construct them for asymptotically AdS spaces \cite{Emparan:1999pm}. The method involves the introduction of intrinsic curvatures in the boundary term, and for instance, for Einstein gravity we get
\begin{equation}\label{EGBdryintro}
\begin{aligned}
S_E^{\rm E}&=-\frac{1}{16\pi G}\int_{\mathcal{M}}d^{d+1}x\sqrt{g}\left[\frac{d(d-1)}{L^2}+R\right]\\
&-\frac{1}{8 \pi G}\int_{\partial \mathcal{M}}d^{d}x\sqrt{h}\Bigg[K-\frac{d-1}{L}-\frac{L}{2(d-2)}\mathcal{R}+\ldots\Bigg]\, ,
\end{aligned}
\end{equation}
where $\mathcal{R}$ is the Ricci scalar of the induced metric on $\partial \mathcal{M}$ and we have added the counterterms needed for  $d\le 4$. 

An interesting property of AdS space is that it can be foliated in different ways, and this gives rise to CFTs placed in various geometries, each one corresponding to black holes with a different horizon topology. For instance, if we wish to study CFTs in flat spacetime we must search for static black hole solutions of the form
\begin{equation}
ds^2= \frac{r^2}{L^2}\left(-N(r)^2 f(r)dt^2+dx^idx^i\right)+\frac{L^2 dr^2}{r^2 f(r)}\, ,
\end{equation}
which have planar horizons --- they are known as black branes. The integration constants must be fixed appropriately so that asymptotically the solution behaves as $f(r)\rightarrow r^2f_{\infty}/L^2$, $N(r)^2\rightarrow 1/f_{\infty}$, and in that case the boundary of AdS is $d$-dimensional Minkowski space. The CFT dual to one of these black branes is in a thermal state analogous to the quark-gluon plasma and it always satisfies a relation of the form
\begin{equation}\label{eq.thermobb}
s=\cs T^{d-1}\, ,
\end{equation}
where $s$ is the entropy density, $T$ is the temperature and $\cs$ is a constant known as the thermal entropy charge. For instance, in the case of Einstein gravity the black brane solutions above are given by $N(r)=1$, $f(r)=1-\omega^{d}/r^{d}$, where $\omega$ is an integration constant. The thermal entropy charge for CFTs dual to EG reads
\begin{equation}
\cs^{\rm\ssc E}=\frac{4^{d-2}\pi^{d-1}L^{d-1}}{d^{d-1}G}\, .
\end{equation}
As we can see, this constant is essentially indistinguishable from $\ctte$ in \req{cteintro}, because it is proportional again to $L^{d-1}/G$. In fact, all central charges in holographic EG are proportional to this ratio, which is the only dimensionless quantity that can be formed out of $L$ and $G$. However, the addition of higher-curvature terms breaks the degeneracy of charges and allows to distinguish between $\ctt$ and $\cs$ --- see \eg \cite{Quasi}.

On the other hand, the usual spherically symmetric black hole solutions are dual to a thermal state of a CFT placed on $\mathbb{R}\times\mathbb{S}^{d-1}$. In this case, the Euclidean version of the CFT has two scales: the period of the Euclidean time $\beta$ and the size of the sphere, and for this reason the thermodynamic quantities do not satisfy a power-law relation as in \req{eq.thermobb}. An interesting phenomena is that a phase transition takes place for certain value of the temperature $T=1/\beta$; this the CFT equivalent of the Hawking-Page phase transition in the bulk \cite{Hawking:1982dh,Witten:1998zw}. Let us review this phenomena in the case of five-dimensional holographic Einstein gravity. There are two Euclidean bulk geometries whose boundary is a CFT on $\mathbb{S}_{\beta}\times\mathbb{S}^{3}$; one of them is a black hole solution of the form
\begin{equation}
ds^2=f(r)d\tau^2+\frac{dr^2}{f(r)}+r^2d\Omega_{(3)}\, ,\quad \text{where}\quad f(r)=1+\frac{r^2}{L^2}\left(1-\frac{\omega^4}{r^4}\right)
\end{equation}
and the other one is pure (thermal) AdS given by the metric above with $\omega=0$. In both cases the coordinate $\tau$ has period $\beta=1/T$ and the boundary $r\rightarrow\infty$ is $\mathbb{S}_{\beta}\times\mathbb{S}^{3}$, but in the case of the black hole we must impose regularity of the Euclidean geometry at the point $r_h$ where $f(r_h)=0$. The absence of a conical singularity imposes the condition $f'(r_h)=4\pi T$, which in turn provides the following relations between $T$, $\omega$ and $r_h$:
\begin{equation}
T=\frac{1}{2\pi}\left[\frac{2r_h}{L^2}+\frac{1}{r_h}\right]\, ,\quad \omega^4=r_h^4+L^2 r_h^2\, .
\end{equation}
Finally, for both solutions one may compute the Euclidean action \req{EGBdryintro}, which yields
\begin{eqnarray}
S_E^{\rm E, \,\, \text{Radiation}}&=&\frac{3\pi \beta L^2}{32 G}\, ,\\
S_E^{\rm E, \,\, \text{BH}}&=&\frac{\pi \beta }{8 G L^2}\left[\frac{3L^4}{4}+ L^2r_h^2-r_h^4\right]\, .
\end{eqnarray}
Then, the one with smallest value dominates the partition function. We see that for large $r_h$, the action for black holes becomes negative and is the one that dominates. However, when $r_h=L$, which corresponds to a temperature $T_{\rm HP}= \frac{3}{2\pi L}$, both phases contribute equally to the partition function, and a phase transition from black holes to radiation takes place. This indicates as well the existence of a phase transition in the CFT: for $T>T_{\rm HP}$ its thermal state is dual to a large black hole, while for $T<T_{\rm HP}$ it must be described by pure (thermal) AdS. 
The introduction of higher-derivative bulk corrections in this context generically imply that the temperature of the phase transition as well as the latent heat will change. However, they can also introduce other phenomena such as the appearance of additional phases \cite{Cai:2001dz,Cho:2002hq}. In some cases the new phases can dominate the partition function and produce a large/small BH transition instead of a large BH/radiation one \cite{Mir:2019ecg,Mir:2019rik}. Furthermore, some higher-order gravities allow for the construction of more sophisticated phase transitions that generalize the Hawking-Page one \cite{Camanho:2013uda}. 

Finally, it is possible to construct black holes with hyperbolic horizons, whose asymptotic boundary is the hyperbolic cylinder $\mathbb{R}\times\mathbb{H}^{d-1}$. According to the Casini-Huerta-Myers map \cite{CHM}, the thermal entropy of a CFT placed in this geometry is equal to the entanglement entropy across a spherical region $\mathbb{S}^{d-2}$ of the same CFT placed on flat space. Thus, this map allows us to compute the entanglement entropy of a holographic CFT as the entropy of a certain hyperbolic black hole. More generally, it is possible to compute holographic R\'enyi entropies \cite{renyi1961,renyi1} from the thermodynamic entropy of hyperbolic black holes \cite{HoloRen}. 
R\'enyi entropies provide a measure of the entanglement of a quantum system \cite{Klebanov:2011uf,Laflorencie:2015eck} and they represent a generalization of the usual entanglement entropy. In order to define R\'enyi entropies, imagine that we take a constant time slice of the CFT's spacetime and we split it as the sum of a subregion $V$ and its complement $\bar V$. Then we assume that this split also produces a bi-partition of the Hilbert space as the sum of the degrees of freedom that live in $V$ plus the ones that live in its complement. We define the reduced density matrix of the subsystem $V$ as the partial trace $\rho_V=\tr_{\bar V} \rho$ and then R\'enyi entropies are defined as
\begin{equation}\label{rrintro}
S_q(V)=\frac{1}{1-q}\log \Tr \rho_V^q \, , \quad q\geq 0\, , q\neq 1\, .
\end{equation}
The limit $q\rightarrow 1$ corresponds to the entanglement entropy, defined as the von Neumann entropy of $\rho_V$. In the holographic context, these entropies are computed according to \cite{HoloRen}
\begin{equation}\label{sqqintro}
S_q=\frac{q}{(1-q)T_0}\int_{T_0/q}^{T_0} S_{\rm \ssc thermal}(T) dT\, ,
\end{equation}
where $S_{\rm \ssc thermal}(T)$ is the corresponding thermal entropy on $\mathbb{R}\times \mathbb{H}^{d-1}$ at temperature $T$, and  $T_0=1/(2\pi R)$, where $R$ is the radius of curvature of $\mathbb{H}^{d-1}$. The thermal entropy in the case of Einstein gravity is simply given by Bekenstein-Hawking formula \req{eq:BekHawk}, but this result is generalized to higher-order gravities by using Wald's formula \req{eq:Waldintro}. An interesting application in the latter case is that, since the degeneracy of central charges if broken, it is possible to study the dependence of R\'enyi entropies on some of these charges \cite{HoloRen} --- see \cite{Galante:2013wta,Belin:2013dva,Dey:2016pei,Puletti:2017gym} for additional examples.

As a general remark, let us note that if we want to generalize the results from holographic Einstein gravity to higher-order gravities, these must possess accessible black hole solutions. Of particular interest is the case in which the computations can be performed non-perturbatively in the higher-order couplings, and for this reason Lovelock and Quasi-topological theories are especially appealing in this context for $d\ge4$. However, the situation in $d=3$ is more precarious because these theories become trivial in $D=4$.

\subsubsection{Hydrodynamics and KSS bound}
When considering perturbations around thermal equilibrium, a CFT behaves as a plasma that can be described under the hydrodynamic approximation. We have just seen that the bulk geometry dual to a CFT in a thermal state corresponds, generically, to a black hole --- in particular to a black brane if the CFT is defined on flat space --- and therefore, holographic plasmas can be studied by analyzing perturbations over black hole solutions. 
One quantity that has attracted much attention in the context of holographic hydrodynamics is the shear viscosity, $\eta$, which measures the velocity gradient between layers of fluid that move between two plates. This quantity scales with temperature according to $\eta\propto T^{d-1}$, and for this reason it is interesting to look at the ratio between the shear viscosity and the entropy density, $\eta/s$, which is independent of $T$. At weak coupling, this ratio is expected to be large since it must diverge in the limit of free fields. Thus, $\eta/s$ gets smaller at strong coupling and in this regime it can be studied using the AdS/CFT correspondence --- see \cite{Son:2002sd,Policastro:2002se,Iqbal:2008by,Myers:2009ij,Paulos:2009yk} for different holographic methods of computing $\eta$.

When the ratio $\eta/s$ is computed for holographic Einstein gravity in any number of dimensions, the following answer is obtained \cite{Son:2002sd,Policastro:2002se}
\begin{equation}\label{etasEGintro}
\frac{\eta}{s}\Big|_{\rm E}=\frac{1}{4\pi}\, .
\end{equation}
This result is universal in the sense that it holds for all CFTs whose dual bulk gravity sector is GR, and the fact it corresponds to the the strong coupling limit of these CFTs suggests that this might be the minimal universal value for $\eta/s$. 
Indeed, the Kovtun-Son-Starinets (KSS) bound \cite{Kovtun:2004de} claims that any fluid or plasma in nature satisfies
\begin{equation}
\frac{\eta}{s}\ge\frac{1}{4\pi}\, ,
\end{equation}
a bound which is saturated for Einstein gravity holography. However, when higher-curvature terms are included in the gravitational action, the relation \req{etasEGintro} receives corrections \cite{Buchel:2004di,Kats:2007mq,Brigante:2007nu,Myers:2008yi,Cai:2008ph,Ge:2008ni}. For instance, in the context of $\mathcal{N}=4$ SYM theory, whose dual gravity theory contains the quartic terms in \req{acads5}, it was found that the leading perturbative corrections to \req{etasEGintro} are given by \cite{Myers:2008yi}
\begin{equation}
\frac{\eta}{s}\Big|_{\rm SYM}=\frac{1}{4\pi}\left(1+\frac{15\zeta(3)}{\lambda^{3/2}}+\frac{5}{16}\frac{\lambda^{1/2}}{N^2}+\ldots\right)\, .
\end{equation}
In this instance the corrections respect the KSS bound, but there has been an intensive search for toy models of holographic higher-order gravities that can violate the bound --- see \cite{Cremonini:2011iq} for a review. For example, in the case of Quasi-topological gravity, given by \req{5dsg-2}, the following value of the shear viscosity to entropy density ratio was found \cite{Myers:2010jv}
\begin{equation}
\frac{\eta}{s}\Big|_{\rm QT}=\frac{1}{4\pi}\left[1-4\lambda-36\mu(9-64\lambda +128\lambda^2+48\mu)\right]\, ,
\end{equation}
and it is obvious that for some values of the parameters the KSS bound is violated. However, in order for this result to be meaningful it is necessary to make sure that the dual CFT is not pathological: it must respect unitarity, causality and positivity of energy. A thorough study of these constraints in Quasi-topological gravity \cite{Myers:2010jv} led to the conclusion that the KSS bound can be lowered to $\eta/s\sim 0.4/(4\pi)$. A similar analysis in \cite{Camanho:2010ru} for the case of general Lovelock gravity determined that $\eta/s$ can be made arbitrarily close to zero if we take $d\rightarrow\infty$. 
However, it was later realized that higher-derivative terms generically lead to causality violations when three-point graviton scattering is involved \cite{Camanho:2014apa}. This can only be fixed by introducing an infinite tower of higher-spin fields which appear at the same energy scale as the higher-derivative terms. Although the effect of such additional degrees of freedom on the holographic setup is unclear, this result seems to compromise the validity of the previous estimations. On the bright side, it should be noted that holographic higher-curvature models have been successful in other applications where they have provided correct intuitions about CFTs \cite{Myers:2010tj,Myers:2010xs,Mezei:2014zla,Bueno1,Bueno2}, where no inconsistencies were found. 
In addition, there are theories whose three-point function structure agrees with the one of Einstein gravity, and therefore are not directly affected by the results in \cite{Camanho:2014apa}. An example of this is precisely provided by the quartic terms that appear in the effective action of type IIB ST \req{acads5}. 
Even if the validity of previous estimations of $\eta/s$ is not clear, holographic higher-order gravities at least suggest that some non-trivial bound, lower than the KSS one, does exist for general $d$  --- see \eg \cite{Fouxon:2008pz}.

\subsubsection{Squashed holography}
Placing a conformal field theory on a curved background is a fruitful strategy in order to gain relevant information about the dynamics of the theory, \eg we have already seen the case of the trace anomaly. We are interested now in studying the change in the partition function of the CFT when we perturb or deform the background metric. 
One class of manifolds that has attracted some attention in this context is that of deformed spheres \cite{Bobev:2016sap,Bobev:2017asb,Hertog:2017ymy,Fischetti:2017sut}, and we will consider in particular a class of odd-dimensional squashed spheres given by the metric
\begin{equation}\label{squaintro}
ds^2_{\mathbb{S}_{\varepsilon}^d}=\frac{ds^2_{\mathbb{CP}^k}}{(d+1)}+(1+\varepsilon)\left(d\psi+\frac{A_{\mathbb{CP}^k}}{(d+1)}\right)^2\, ,
\end{equation}
where $ds^2_{\mathbb{CP}^k}$ is the Einstein metric on the projective space $\mathbb{CP}^k$ and $J=dA_{\mathbb{CP}^k}$ is the K\"ahler form on $\mathbb{CP}^k$ (where $k\equiv (d-1)/2$).  The deformation is controlled by the squashing parameter $\varepsilon$: for $\varepsilon=0$ we recover the round sphere $\mathbb{S}^{d}$, while a non-zero $\varepsilon$ yields a squashed sphere with isometry group SU$(\frac{d+1}{2})\times$U$(1)$. 
Ref.~\cite{Bobev:2017asb} studied several aspects of the free energy of CFTs, defined as $\mathcal{F}=- \log |\mathcal{Z}|$, when they are placed on one of these squashed spheres. Among other results, it was found that $\varepsilon=0$ is a local maximum of $\mathcal{F}$, while the second derivative is universally controlled by $\ctt$ --- the explicit coefficient was computed in $d=3,5$. However, the subleading terms $\mathcal{O}(\varepsilon^3)$ in the expansion of the free energy did not appear to have a simple expression in terms of other charges. 

On the other hand, this type of squashed spheres can also be studied holographically: the dual bulk geometries are given by Taub-NUT metrics
\cite{Hawking:1998ct,Dowker:1998pi,Chamblin:1998pz,Emparan:1999pm,Hartnoll:2005yc,Bobev:2016sap}.
Let us review here the $d=3$  case for holographic Einstein gravity. A solution of the equations of motion of \req{eq:EGholointro} is given by a metric of the form
\begin{equation}
	ds^2=V(r)(d\tau+2 n \cos\theta d\phi)^2+\frac{dr^2}{V(r)}+(r^2-n^2)\left(d\theta^2+\sin^2\theta d\phi^2\right)\, ,
	\label{taubintro}
\end{equation}
where $n$ is an arbitrary constant and $V(r)$ reads
\begin{equation}\label{boltEiintro}
V(r)=\frac{(r-r_h)\left[(6n^2 r r_h-3 n^4+ L^2 (n^2-r r_h)-r r_h (r^2+r r_h+ r_h^2)) \right]}{L^2(n^2-r^2)r_h}\, .
\end{equation}
Here $r_h$ is an integration constant that we choose so that $V(r_h)=0$, but we still need to impose several regularity conditions. First, 
absence of Misner strings \cite{Misner:1963fr} fixes the periodicity of the Euclidean time $\tau$ to 
$\beta_{\tau}=8\pi n$. 
On the other hand, in order to avoid a conical singularity at $r=r_h$, we must demand $V'(r_h)=1/(2n)$. This is an equation for $r_h$ that has two types of solutions: one possible choice is $r_h=n$, and the other one is 
\begin{equation}\label{boltrbintro}
r_h=\frac{L^2 }{12 n} \left[ 1\pm \sqrt{ 1-\frac{48n^2}{L^2} +\frac{144 n^4}{L^4}}\right]\, .
\end{equation}
In the former case, the solution is called a NUT, while in the second one we have a Bolt. For both types of solutions, the asymptotic boundary of AdS for $r\rightarrow\infty$ is given by 
\begin{equation}
	\frac{^{(3)}ds^2}{r^2}=\frac{4n^2}{L^2}(d\psi+\cos\theta d\phi)^2+d\theta^2+\sin^2\theta d\phi^2+\mathcal{O}(r^{-2})\, ,
\end{equation}
where $\psi=\tau/(2n)$ has period $4\pi$. This metric coincides with the squashed sphere one \req{squaintro} if we take into account that $\mathbb{CP}^1\equiv \mathbb{S}^2$ and if we identify
\begin{equation}
\frac{n^2}{L^2} =\frac{(1+\varepsilon)}{4}\, .
\end{equation}
Thus, both Taub-NUT and Bolt solutions are bulk duals of a CFT on a squashed sphere. 
For each type of solution, the on-shell Euclidean action computes the corresponding contribution to the holographic free energy, and  the dominating phase is the one with a lower value. For small values of $\varepsilon$ it is found that the NUT solution dominates, and it yields the following value for the holographic free energy
\begin{equation}
\mathcal{F}^{\rm E}_{\mathbb{S}_{\varepsilon}^{3}}=\frac{\pi L^2}{2G} (1-\varepsilon^2)\, .
\end{equation}
We check that $\varepsilon=0$ is indeed a maximum of the free energy, and furthermore, we have $\mathcal{F}_{\mathbb{S}^3_{\varepsilon}}''(0)=-\frac{\pi^4}{3}\ctte$, where $\ctte$ is the central charge for Einstein gravity in \req{cteintro}. Thus, the holographic result agrees with the field-theory expectations \cite{Bobev:2017asb}, but it gives us no additional information. In particular, one expects that the free energy contains subleading terms in the $\varepsilon$ expansion, but this cannot be explored with holographic Einstein gravity because the expansion stops at quadratic order. A possible way to go beyond this result is to consider higher-curvature corrections in the bulk. We will follow this approach in Chapter~\ref{Chap:8}.

 \section{Summary}
This thesis is based on the results found in references \cite{Aspects,PabloPablo,PabloPablo2,PabloPablo3,PabloPablo4,ECGholo,NewTaub2,Bueno:2018yzo,Bueno:2019ltp}. Some parts of the text have been directly extracted from these papers, others have been adapted with minor modifications, and some parts have been written from scratch --- especially chapters \ref{Chap:4} and \ref{Chap:5}. In any case, the conventions have been adapted in order to provide a (more or less) uniform notation in all the chapters. In addition, there are some new results and discussions with respect to those references. 

The thesis is structured in three parts that deal with different topics but which are interrelated among them. In the first part, which consists of chapters~\ref{Chap:1}, \ref{Chap:2} and \ref{Chap:3}, we study some general aspects of higher-derivative gravities and the purpose is mainly to identify and classify certain theories that have interesting properties. The second part contains chapters~\ref{Chap:4} and \ref{Chap:5} and is devoted to the study of asymptotically flat black holes in a family of four-dimensional higher-order gravities that were previously identified in the first part. Finally, the third part is composed of chapters~\ref{Chap:6}, \ref{Chap:7} and \ref{Chap:8}, and it studies asymptotically AdS solutions of this type of theories, as well as a number of applications in the context of the AdS/CFT correspondence. We provide here a more detailed summary of the main achievements in every chapter.

\subsubsection{Chapter~\ref{Chap:1} (based on \cite{Aspects,PabloPablo})}
We study the linearized equations of general $\mathcal{L}$(Riemann$)$ theories on maximally symmetric backgrounds and we provide an efficient algorithm to compute them for particular theories. This algorithm has been afterwards used by other authors, \eg \cite{Cisterna:2017umf,Hennigar:2017ego}. We demonstrate that  the theories of the form $\mathcal{L}$(Riemann$)$ in general propagate a massive, ghost graviton and a scalar mode along with the massless graviton, but we show that the additional modes can be absent in some special cases and we provide a classification of theories according to their spectrum.  Among them, there are non-trivial theories whose linearized equations coincide with those of Einstein gravity and we call them \emph{Einstein-like}. The most interesting achievement of the chapter is the identification of an even more special theory, named \emph{Einsteinian cubic gravity}, which is given by the following cubic interaction 
 \begin{align}\label{ECGintro2}
\mathcal{P}&=12 \tensor{R}{_{\mu}^{\rho}_{\nu}^{\sigma}}\tensor{R}{_{\rho}^{\alpha}_{\sigma}^{\beta}}\tensor{R}{_{\alpha}^{\mu}_{\beta}^{\nu}}+\tensor{R}{_{\mu\nu}^{\rho\sigma}}\tensor{R}{_{\rho\sigma}^{\alpha\beta}}\tensor{R}{_{\alpha\beta}^{\mu\nu}}-12R_{\mu\nu\rho\sigma}R^{\mu\rho}R^{\nu\sigma}+8\tensor{R}{_{\mu}^{\nu}}\tensor{R}{_{\nu}^{\rho}}\tensor{R}{_{\rho}^{\mu}}\, .
\end{align} 
This density satisfies the following properties
\begin{enumerate}
\item In any number of dimensions $D$, the linearized equations of $\mathcal{P}$ around maximally symmetric backgrounds are of second order and proportional to the linearized Einstein tensor. Therefore, only a massless graviton is propagated on those backgrounds.
\item Up to cubic order in curvature, $\mathcal{P}$ is the only higher-order term with this property besides the Lovelock densities $\mathcal{X}_4$ and $\mathcal{X}_6$.
\item However, unlike the Lovelock densities, $\mathcal{P}$ is not trivial nor topological in $D=4$. 
\end{enumerate}
Furthermore, the ECG density satisfies additional properties that are described in the next chapter. Due to these interesting features, ECG has triggered numerous follow up research by other groups worldwide and has inspired the construction of similar theories, see \eg \cite{Dey:2016pei,Hennigar:2016gkm,Hennigar:2017ego,Ahmed:2017jod,Feng:2017tev,Hennigar:2018hza,Li:2017ncu,Poshteh:2018wqy,Arciniega:2018fxj,Cisterna:2018tgx,Li:2019auk,Erices:2019mkd,Mehdizadeh:2019qvc,Mir:2019rik,Mir:2019ecg,Emond:2019crr}

\subsubsection{Chapter~\ref{Chap:2} (based on \cite{PabloPablo3})}
In this chapter we study the equations of motion of higher-order gravity for static, spherically symmetric (SSS) spaces, and we establish several results for theories that satisfy the seemingly anecdotic condition of possessing solutions of the form
\begin{equation}\label{Fmetricintro}
ds^2=-f(r)dt^2+\frac{dr^2}{f(r)}+r^2d\Omega_{(D-2)}^2\, ,
\end{equation}
\ie with $g_{tt}g_{rr}=-1$. The main result is the systematic characterization of a new family of theories known as \emph{Generalized quasi-topological gravities} (GQGs), which were first proposed at cubic order in the curvature in \cite{Hennigar:2017ego}. These theories are particularly interesting in order to study spherically symmetric black hole solutions because they share the following properties. 
\begin{enumerate}
\item Their linearized equations are of the Einstein-like type and they only propagate a traceless and massless graviton on maximally symmetric vacua. 
\item They allow for vacuum solutions of the form \req{Fmetricintro}, where  $f(r)$ satisfies a differential equation of order $\le 2$. 
\item For a fixed mass $M$, there is at most a discrete number of black hole solutions of the form \req{Fmetricintro}. If there are several black holes, there is one (and only one) of them which is a smooth deformation of the Schwarzschild solution when $M\rightarrow\infty$.
\item The thermodynamic properties of these black holes can be determined analytically by solving a system of algebraic equations without free parameters.
\item The exterior gravitational field of a spherically symmetric matter distribution is again given by a metric of the form \req{Fmetricintro}.
\end{enumerate}
We then show that the ECG term \req{ECGintro2} is the simplest member of the GQG class in $D=4$ and we construct additional theories of this family. 

\subsubsection{Chapter~\ref{Chap:3} (based on \cite{Bueno:2019ltp})}
Here we examine the effect of field redefinitions on the gravitational action, with the ambitious goal of demonstrating that GQG theories provide a basis to construct the most general effective higher-derivative Lagrangian for gravity. We show that all the terms of the form $\mathcal{L}(g^{\mu\nu},R_{\mu\nu\rho\sigma})$ (without covariant derivatives of the Riemann tensor) of a given order can be mapped to a GQG using metric redefinitions, provided that there exists one non-trivial GQG at that order. There is conclusive evidence  that non-trivial GQGs exists at any order and in any dimension, so this result virtually proves that all the densities of the form $\mathcal{L}(g^{\mu\nu},R_{\mu\nu\rho\sigma})$ can be mapped to a GQG. 
Furthermore, we show that densities with two covariant derivatives of the Riemann tensor can also be mapped to GQGs and that they are irrelevant for the purposes of studying spherically symmetric solutions. As a conclusion, we propose the following conjectures
\begin{enumerate}
\item Any higher-order Lagrangian can be mapped, order by order in the derivative expansion, to a sum of Generalized quasi-topological gravity densities by performing redefinitions of the metric.
\item The Generalized quasi-topological gravities of the form $\mathcal{L}(g^{\mu\nu},R_{\mu\nu\rho\sigma})$ capture 
the physics of spherically symmetric black holes --- in particular, their thermodynamic properties --- in the most general higher-derivative effective theory for gravity. 
\end{enumerate}

\subsubsection{Chapter~\ref{Chap:4} (based on \cite{PabloPablo2})}
We study the spherically symmetric, asymptotically flat black hole solutions of Einsteinian cubic gravity in $D=4$, with Lagrangian $\mathcal{L}=R-\frac{\mu L^4}{8}\mathcal{P}$. These are the first examples of non-perturbative black hole solutions deforming the Schwarzschild geometry in a four-dimensional higher-order gravity, and they are given by a metric of the form \req{Fmetricintro}. The equations of motion that we have to solve are reduced to a single second-order differential equation for the function $f(r)$ and we show that it admits a unique black hole solution whose profile can be found numerically. However, the most impressive characteristic of ECG is that the thermodynamic properties of this black hole can be obtained analytically and non-perturbatively. A particularly interesting relation is the one between temperature and radius, which reads
\begin{equation}\label{eq:TECGintro}
T^{\rm ECG}=\frac{r_h}{2\pi \left(r_h^2+ \sqrt{r_h^4+3 \mu  L^4}\right)}\, .
\end{equation}
In the limit of large $r_h$ this relation reduces to the one in Einstein gravity, $T^{\rm E}=1/(2\pi r_h)$, but for $r_h\rightarrow 0$ the temperature vanishes instead of diverging. As a consequence, we show that there exists a maximum temperature which separates two types of black holes. Large black holes have negative specific heat and are thermodynamically unstable, just as Schwarzschild's black hole. On the other hand, below certain mass the specific heat becomes positive and small black holes are stable. 

\subsubsection{Chapter~\ref{Chap:5} (based on \cite{PabloPablo4})}
We extend the results of previous chapter for the family of all the Generalized quasi-topological gravities of the form $\mathcal{L}(g^{\mu\nu},R_{\mu\nu\rho\sigma})$ in $D=4$. We show that at a given order in curvature all the different GQG densities contribute equally to the equations of motion for the metric \req{Fmetricintro}, and this implies that we only need to include one ``representative'' GQG at every order in curvature. Thus, we consider an action of the form
\begin{equation}\label{lalintro}
S= \frac{1}{16\pi G}\int d^4x\sqrt{|g|}\left[R+\sum_{n=2}^{\infty}L^{2n-2}\lambda_{n}\mathcal{R}_{(n)}\right]\, ,
\end{equation}
where $\mathcal{R}_{(n)}$ are particular non-trivial GQG densities. A explicit computation of the equations of motion for these densities up to order ten allows us to derive the general form of the equation satisfied by the metric function $f(r)$ in \req{Fmetricintro} for arbitrary $n$. In this way, we can study black hole solutions with corrections at all orders in curvature. 
We observe that there is at most a discrete number of black hole solutions of this equation, but only one of them represents a smooth deformation of Schwarzschild's geometry. Again the profile of the solutions can be found numerically, but the thermodynamic properties of these black holes can be studied analytically. We will show that the relations between the mass $M$, the temperature $T$ and the entropy $S$ are exactly given by the following parametric equations in terms of a variable $x$
\begin{equation}\label{eq:MTSintro}
\begin{aligned}
M&=\frac{Lh(x)}{G}\left[\frac{h'(x)}{3h(x)-x h'(x)}\right]^{3/2}\, ,\\
T&=\frac{x}{2\pi L}\left[\frac{h'(x)}{3h(x)-x h'(x)}\right]^{1/2}\, ,\\
S&=\frac{\pi L^2}{G}\left[\frac{h'^2(x)}{3h(x)-x h'(x)}+\int dx \frac{h''(x)}{x} \right]\, ,\\
\end{aligned}
\end{equation}
where $h(x)$ is the function
\begin{equation}\label{eq:hfuncintro}
h(x):=x-\sum_{n=3}^{\infty}\lambda_{n} x^n\, .
\end{equation}
A simple computation shows that these relations satisfy the first law of thermodynamics $dM=TdS$ and that the usual formulas for Einstein gravity $T^{\rm E}=1/(8\pi G M)$, $S^{\rm E}=4\pi G M^2$ are recovered when $\lambda_{n}=0$ $\forall n$.\footnote{In the case of asymptotically AdS or dS black holes the same relations apply, but now $h(x):=\lambda_0+x-\sum_{n=3}^{\infty}\lambda_{n} x^n$, where the constant term is related to the cosmological constant according to $\lambda_0=\Lambda L^2/3$. For $\lambda_{n\ge 3}=0$ we recover the thermodynamic relations for Schwarzschild-A(dS) black holes. }
Studying the thermodynamic relations above, we determine that the existence of small, stable black holes is a common phenomena to all the theories in \req{lalintro}. In addition, we observe that the limit of zero mass is approached universally: regardless of the maximum order of corrections involved, small black holes always satisfy $T\propto M^{1/3}$, $S\propto M^{2/3}$. Finally, we study the evaporation process of these black holes and we conclude that they have infinite lifetimes, with a last stage corresponding to a sort of cold remnant. This is just opposite to the case of Schwarzschild black holes, which evaporate completely after a finite time and whose temperature diverges in the last stages. 

Using the results of Chapter~\ref{Chap:3} we will argue that these intriguing results might apply not only to Generalized quasi-topological gravities, but probably to any higher-curvature gravity that corrects GR. In particular, this would imply that for any higher-order Lagrangian there are some coefficients $\lambda_n$ for which the relations \req{eq:MTSintro} describe exactly the thermodynamics of spherically symmetric black holes. 

\subsubsection{Chapter~\ref{Chap:6} (based on \cite{ECGholo})}
In this chapter we study several holographic applications of Einsteinian cubic gravity in $D=4$. We determine various entries of the holographic dictionary non-perturbatively in the ECG coupling; these include the central charge of the stress-tensor 2-point function $\ctt$, the thermal entropy charge $\cs$, the charge $a^*$ controlling the universal contribution to the entanglement entropy in a circle region $\mathbb{S}^1$ and the coefficient of the stress-tensor 3-point function $t_4$ --- they are listed in Table~\ref{dictionintro}. Since $t_4$ is non-vanishing, ECG provides a holographic toy model of a non-supersymmetric CFT$_3$ analogous to the one defined by Quasi-topological gravity \cite{Myers:2010jv} in $d=4$. 
Then, we describe asymptotically AdS black holes in ECG with spherical, planar and hyperbolic horizon geometries which allow us to explore the thermodynamic phase space of the dual CFT --- again, we are able to obtain non-perturbative expressions in the ECG coupling. In order to compute gravitational on-shell actions, we propose a generalized Gibbons-Hawking-York boundary term that, we argue, is valid for theories with Einstein-like linearized equations as long as we consider asymptotically AdS solutions. This boundary term is simply proportional to the GHY term and counterterms of GR, and the proportionality constant has a very nice interpretation in the context of holography: it is the aforementioned entanglement entropy charge $a^*$. More precisely, according to our proposal, a well-posed and regularized Euclidean action for higher-order gravities with Einstein-like equations would read
\begin{equation}\label{SEcomplete2intro}
\begin{aligned}
S_E=&-\int_{\mathcal{M}}d^Dx\sqrt{g}\mathcal{L}(g^{\alpha\beta},R_{\mu\nu\rho\sigma})\\
&-\frac{2a^*}{\Omega_{(d-1)}\tilde{L}^{d-1}} \int_{\partial\mathcal{M}}d^{D-1}x\sqrt{|h|} \left[ K-\frac{d-1}{\tilde{L}}-\frac{\tilde L}{2(d-2)}\mathcal{R}+\cdots \right]\, ,
\end{aligned}
\end{equation}
where $\tilde L$ is the length scale of the asymptotic AdS vacuum, and $a^*$ is related to the gravitational Lagrangian evaluated on AdS according to \cite{Imbimbo:1999bj,Schwimmer:2008yh,Myers:2010tj,Myers:2010xs}
\begin{equation}\label{astarintro}
a^*=-\frac{\pi^{d/2}\tilde{L}^{d+1}}{d \Gamma(d/2)}\mathcal{L}
|_{{\rm AdS}}\, .
\end{equation}
We show that this boundary term passes numerous tests and provides correct answers for ECG as well as for other theories in which other prescriptions exist. 
Finally, we also examine the shear viscosity to entropy density ratio in ECG and we observe that it does not violate the KSS bound, which is only saturated in the Einstein gravity limit. However, we obtain new phenomena: the dependence of $\eta/s$ on the ECG coupling is non-perturbative (in the sense that it does not allow for a perturbative Taylor expansion) and the ratio diverges to $+\infty$ for a finite value of the coupling. 

\begin{table}[t] 
\centering
	\begin{tabular}{|c|c|c|c|c|}
		\hline
		&   $\ctt$  & $\ctt\cdot t_4$   & $\cs $   & $a^* $\\
		\hline\hline
		& & & &\\[-10pt]
		Einstein &  $\displaystyle\frac{3\tilde{L}^{2}}{\pi^3G}$  & $0$ & $\displaystyle\frac{4\pi^2\tilde{L}^{2}}{9G}$ &  $\displaystyle\frac{\tilde{L}^{2}}{4G}$ \\ [6pt] \hline
		& & & &\\[-10pt]
		ECG & $\displaystyle(1-3\mu f_{\infty}^2)\ctte$   & $\displaystyle-1260 \mu f_{\infty}^2 \ctte$ & $\displaystyle\left(1-\frac{27}{4}\mu \right)f_{\infty}^2 \cse $ & $\displaystyle\left(1+3\mu f_{\infty}^2\right)a^{*{\rm \ssc E}}$ \\ [10pt]\hline
		%QTG $(d=4)$	& $(1-3\mu f_{\infty}^2)\ctte $   & $3780 \mu f_{\infty}^2 \ctte$ & $f_{\infty}^3\cse $ & $\left(1+9\mu f_{\infty}^2\right)a^{*{\rm \ssc E}}$ \\
		%\hline
	\end{tabular}
	\caption{CFT charges for holographic Einstein gravity and Einsteinian cubic gravity in $d=3$. }
	\label{dictionintro}
\end{table}

\subsubsection{Chapter~\ref{Chap:7} (based on \cite{NewTaub2})}
In this chapter we show that some Generalized quasi-topological gravities allow for Taub-NUT solutions of the form 
\begin{equation}\label{FFnutintro}
ds^2=V_{\mathcal{B}}(r) (d\tau+ n A_{\mathcal{B}})^2+\frac{dr^2}{V_{\mathcal{B}}(r)}+(r^2-n^2)d\sigma_{\mathcal{B}}^2\, ,
\end{equation}
for various $(D-2)$-dimensional K\"ahler-Einstein base spaces $\mathcal{B}$ with metric $g_{\mathcal{B}}$ and  K\"ahler form $J=dA_{\mathcal{B}}$. We use these theories to construct Euclidean-AdS-Taub-NUT solutions, which in the holographic context can be used to probe CFTs on squashed spheres when the base space is $\mathcal{B}=\mathbb{CP}^{(D-2)/2}$. 
In $D=4$, we find that ECG allows for this type of solutions and we construct Taub-NUT and Bolt geometries both with spherical and toroidal base spaces. Similarly to the case of black hole solutions, the equations of motion of ECG are reduced to a second order equation for $V_{\mathcal{B}}$ which can be solved numerically. This analysis is extended to a quartic GQG in $D=4$ and to quartic Quasi-topological and Generalized quasi-topological gravities in $D=6$, where we focus on the base space $\mathbb{CP}^{2}$. In all cases, the thermodynamic properties of the solutions can be determined analytically, and in particular we compute the Euclidean on-shell action using the boundary term introduced in the preceding chapter. Then, we provide a thorough analysis of the phase space of these solutions, determining the existence of new types of phase transitions with respect to the Einstein gravity case. 
In addition, we find that the equations can be solved analytically in the critical limit of ECG and we obtain special NUT-charged solutions that represent different types of AdS wormholes and non-isotropic bouncing cosmologies.

\subsubsection{Chapter~\ref{Chap:8} (based on\cite{Bueno:2019ltp})}
In the final chapter of this thesis we apply the knowledge we have gathered to derive new results about the free energy of CFTs on squashed spheres. We first propose a direct formula that computes the holographic free energy of CFTs on squashed spheres in terms of the gravitational Lagrangian. This formula reads
\begin{equation}\label{fee0eintro}
\mathcal{F}_{\mathbb{S}_{\varepsilon}^{d}}=(-1)^{\frac{(d-1)}{2}}\frac{\pi^{\frac{(d+2)}{2}} }{\Gamma\left[\frac{d+2}{2}\right]}\frac{\mathcal{L}\left[f_{\infty}/(1+\varepsilon)\right] L^{d+1}}{[f_{\infty}/(1+\varepsilon)]^{\frac{(d+1)}{2}}}\, ,
\end{equation}
where $\mathcal{L}(x)$ is the Lagrangian evaluated on an auxiliary AdS$_{(d+1)}$ space with radius $L/\sqrt{x}$.  Our formula reproduces the result obtained from evaluation of the Euclidean action on Taub-NUT geometries for all the cases that have been studied in the literature --- including Einstein and Lovelock gravities in arbitrary dimension ---  and we argue that it applies at least to all the GQGs of the type identified in Chapter~\ref{Chap:7}, which presumably constitute an infinite set.  Besides, we show that the expansion of \req{fee0eintro} near $\varepsilon=0$ has a universal character that must hold for all Einstein-like theories. Given that the formula covers an infinite number of holographic toy models, it can be used to search for possible general characteristics valid for any CFT (holographic or not). First, we show that our formula yields the following relation between the second derivative of the free energy and the central charge $\ctt$:
\begin{equation}\label{genD2intro}
\mathcal{F}_{\mathbb{S}_{\varepsilon}^{d}}''(0)=\frac{(-1)^{\frac{(d-1)}{2}}\pi^{d+1} (d-1)^2 }{  2\,  d!}\ctt\, .
\end{equation}
This value generalizes the results found in \cite{Bobev:2017asb} for $d=3, 5$ for arbitrary $d$, and it must hold for arbitrary CFTs.  Second, focusing in the $d=3$ case and using the holographic dictionary derived for ECG, we observe that our formula \req{fee0eintro} predicts the following subleading contribution to the free energy
\begin{equation}\label{3conjintro}
\mathcal{F}_{\mathbb{S}_{\varepsilon}^{3}}=\mathcal{F}_{\mathbb{S}^{3}}-\frac{\pi^4\ctt}{6}\varepsilon^2\left[1-\frac{t_4}{630}\varepsilon + \mathcal{O}(\varepsilon^2) \right]\, ,
\end{equation} 
which implies that the $\mathcal{O}(\varepsilon^3)$ term is controlled by $\ctt t_4$. 
Furthermore, we prove that the same result holds for infinitely many holographic higher-order gravities besides ECG, and we crosscheck this prediction with numeric computations for free fields finding excellent agreement. Based on these evidences, we conjecture that the small-squashing expansion of the free energy up to order $\varepsilon^3$ is universally given by \req{3conjintro} for arbitrary three-dimensional CFTs. 

\subsubsection*{Note on conventions}
We use natural units $c=\hbar=1$ throughout the thesis, but we will leave $G$ explicit. We define Planck's length as $G=\ell_{\rm \ssc P}^{D-2}$, while $M_{\rm \ssc P}=1/\ell_{\rm\ssc P}$. $D$ stands for the number of spacetime dimensions, while in the context of holography $d\equiv D-1$ is the dimension of the boundary theory. We use signature $(-,+,+,\dots)$, greek indices for bulk tensors, $\mu,\nu,\dots=0,1,\dots,D$, Latin indices from the beginning of the alphabet for boundary tensors, $a, b,\dots=0,1,\dots,d$ and $i,j,\dots=1,\dots,d$ for spatial indices on the boundary.  Riemann's tensor is defined by the relation
\begin{equation}
u_{\mu}R^{\mu}_{\ \ \nu\rho\sigma}=-\left[\nabla_{\rho},\nabla_{\sigma}\right] u_{\nu}\, .
\end{equation}

\part{General aspects of higher-derivative gravities}\label{Part:I}

\chapter{The weak-field limit of $\mathcal{L}($Riemann$)$ theories}\label{Chap:1}

One of the first aspects one can study about higher-curvature gravity theories is their constant curvature (or maximally symmetric) solutions.  We will refer to one of these solutions as the vacuum of theory. In Einstein gravity, there is a unique vacuum that is fully determined by the cosmological constant $\Lambda$, and it is either de Sitter (dS), anti-de Sitter (AdS) or Minkowski spacetime depending on the value of $\Lambda$. The situation is different in the presence of higher-derivative terms since there might be several vacua and their curvature is not proportional to the cosmological constant anymore. As we show in this chapter, the problem of finding these solutions in higher-order gravity is always reduced to solving an algebraic equation for the background curvature.

Next, one can perform linear perturbation theory over the previous backgrounds. This analysis, that can be carried out in full generality, will reveal the degrees of freedom that these theories propagate on the vacuum.  One of the main goals of the present chapter is to present a systematic method that allows computation of the linearized equations of any higher-curvature theory in a much simpler and faster way than previous methods.  For simplicity, we perform the analysis for $\mathcal{L}($Riemann$)$ theories, whose Lagrangian is constructed out of Riemann invariants without introducing covariant derivatives of the curvature. This is a very general type of theory to which most of this thesis is devoted. Using the proposed method, we will linearize explicitly all the $\mathcal{L}($Riemann$)$ theories up to quartic order in curvature. However, the linearized equations of these theories are so constrained that we will be able to study their properties without making reference to specific Lagrangians. 

In any higher-derivative theory, the existence of new degrees of freedom, including negative energy modes, is expected. We show in this chapter that, for $\mathcal{L}($Riemann$)$ gravities, the linearized spectrum contains, along with the massless spin-2 graviton, a massive, ghost-like spin-2 mode (that we will call a massive graviton) and a scalar mode. We find, however, that there are some special cases in which these modes do not propagate in the vacuum, and we will provide a classification of theories according to their spectrum. There are, in particular, non-trivial theories whose linearized equations of motion are of second order, and whose only degree of freedom on the vacuum is the massless graviton. These theories are called \emph{Einstein-like} and they will play a central role in this work --- especially when we come to holography \ref{Part:III}.
As a refinement of the formers, we shall describe \emph{Einsteinian} theories as those possessing Einstein-like spectrum in any dimension, and this will lead us to the discovery of \emph{Einsteinian cubic gravity}, a theory with even more special properties that we will find in the forthcoming chapters.

As an application of the results of this chapter, we will compute the polarization modes of gravitational waves in $\mathcal{L}($Riemann$)$ theories, as well as the Newtonian limit of these theories, obtaining the metric produced by a point-like mass in any dimension.

%%%%%%%%%%%%%%%%%%%%%%%%%%%%%%%%%%%%%%%%%%%%%%%%%%%%%%%%%%%%%%%
%%%%%%%%%%%%%%%%%%%%%%%%%%%%%%%%%%%%%%%%%%%%%%%%%%%%%%%%%%%%%%%
%%%%%%%%%%%%%%%%%%%%%%%%%%%%%%%%%%%%%%%%%%%%%%%%%%%%%%%%%%%%%%%%%%%%%%%%%%%%%%%%%%%%%%%%%%%%%%%%%%%%%%%%%%%%%%%%%%%%%%%%%%%%%%
\section{Linearized equations of $\mathcal{L}$(Riemann) theories}\label{section2}\label{chap:2sec:1}

$\mathcal{L}$(Riemann$)$ theories of gravity are a very general class of higher-curvature theories, which are given by an action of the form

\begin{equation}\label{Smassc2}
S=\int_{\mathcal{M}}d^{D}x\sqrt{|g|}\mathcal{L}(g^{\mu\nu}, R_{\mu\nu\rho\sigma})\, ,
\end{equation}
where the Lagrangian is some invariant formed from products and contractions of the Riemann tensor and the metric.\footnote{In some dimensions it also possible to construct parity-violating Lagrangians by allowing contractions between the Riemann tensor and the Levi-Civita tensor $\epsilon_{\mu_{1}\ldots \mu_{D}}$, but we will not consider this case here. See \eg \cite{Gruzinov:2006ie,Alexander:2009tp,Endlich:2017tqa}.}
The field equations computed from the action \req{Smassc2} are of fourth order in derivatives of the metric and they read \cite{Padmanabhan:2011ex}
\begin{equation}
\mathcal{E}_{\mu\nu}\equiv P_{\mu}\,^{\sigma\rho\lambda}R_{\nu\sigma \rho\lambda}-\frac{1}{2}g_{\mu\nu}\mathcal{L}+2\nabla^{\alpha}\nabla^{\beta}P_{\mu\alpha\nu\beta}=\frac{1}{2}T_{\mu\nu}\, ,
\label{fieldequationsc2}
\end{equation}
where we defined the object
\begin{equation}
P^{\mu\nu\rho\sigma}\equiv \frac{\partial \mathcal{L}}{\partial R_{\mu \nu\rho\sigma}}\, .\label{Ptensor2}  
\end{equation}
We also included an energy-momentum tensor in the right-hand-side of the equation,
 
\be
T_{\mu\nu}\equiv -\frac{2}{\sqrt{|g|}}\frac{ \delta S_{\rm matter}}{\delta g^{\mu\nu}}\, ,
\ee
in case we couple our theory to some matter with action $S_{\rm matter}$.

The aim of this section is to study the linearized equations of general $\mathcal{L}$(Riemann$)$ theories on maximally symmetric backgrounds (msb) in arbitrary dimensions. In order to do that, we use a new method to obtain the linearized equations that was originally introduced in \cite{PabloPablo,Aspects}, which is more efficient and simpler than previous linearization methods. The new method allows us to obtain the linearized equations of any $\mathcal{L}$(Riemann$)$ theory without the need of knowing the explicit form of the Lagrangian. 
In fact, as we show in the next subsection, it is possible to linearize the equations \req{fieldequationsc2} up to the identification of four constants $a,b,c$ and $e$ that are the only theory-dependent parameters. We propose a recipe that can be used to obtain those constants from the corresponding Lagrangian following some simple steps that we detail. 
Then, we rewrite the linearized equations in terms of three physical parameters which can be easily obtained from $a,b,c$ and $e$. These are the effective gravitational constant $G_{\rm eff}$, and the masses of the two extra modes that appear in the linearized spectrum of generic $\mathcal{L}$(Riemann) theories, $m_g^2$ and $m_s^2$. As we show, both in (anti-)de Sitter and Minkowski backgrounds, the usual massless graviton is generically accompanied by a massive ghost-like graviton of mass $m_g$ and a scalar mode of mass $m_s$. In Subsection \ref{physmo} we obtain the matter-coupled wave equations satisfied by these modes. We close the section by constructing a quadratic effective action from which the linearized equations can be obtained from the variation of the metric perturbation. 

\subsection{Linearization procedure}\labell{lll}
%Let us start giving a detailed account of the fast-linearization method for general $\mathcal{L}($Riemann$)$ theories presented in \cite{PabloPablo, Aspects}.

%\subsubsection*{First-order variations on a general background metric}
Let $\bar{g}_{\mu\nu}$ be a solution of the full non-linear equations \req{fieldequationsc2} and let us consider a perturbed metric of the form

\begin{equation}
g_{\mu\nu}=\bar{g}_{\mu\nu}+h_{\mu\nu}\, ,
\end{equation}
where $h_{\mu\nu}$ is the perturbation. Our goal is to expand the field equations \req{fieldequationsc2} to linear order in $h_{\mu\nu}$. We will soon assume that $\bar{g}_{\mu\nu}$ is maximally symmetric, but let us for the moment assume it is any solution of  \req{fieldequationsc2}. In order to linearize the equations, it is useful to define the tensor 
\begin{eqnarray}\label{P-def}
 C_{\sigma\rho\lambda\eta}^{\mu\gamma\sigma\nu}&\equiv &g_{\sigma\alpha} g_{\rho\beta} g_{\lambda\chi} g_{\eta\xi} \frac{\partial P^{\mu\gamma\sigma\nu}}{\partial R_{\alpha\beta\chi\xi}}\, ,
%=\frac{\partial }{\partial R^{\sigma\rho\lambda\eta}}\frac{\partial\mathcal{L}}{\partial R_{\mu\alpha\beta\nu}}\Big|_{B},
\end{eqnarray}
where $P^{\mu\nu\rho\sigma}$ was defined in \req{Ptensor2}. Now, using the identity \cite{Padmanabhan:2011ex}

\begin{equation}\label{idus}
\left[\frac{\partial \mathcal{L}}{\partial g^{\mu\nu}}\right]_{R_{\rho\sigma\gamma\delta}}=2P_{\mu}^{\ \rho\sigma\gamma}R_{\nu\rho\sigma\gamma}\, ,
\end{equation}
it is possible to prove that the variations of $\mathcal{L}$ and $P^{\mu\alpha\beta\nu}$ read respectively\footnote{Observe that we choose $\{R_{\mu\nu\rho\sigma},g^{\gamma\delta}\}$ to be the fundamental variables in $\mathcal{L}$. As explained in \cite{Padmanabhan:2011ex,Padmanabhan:2013xyr}, all expressions obtained using these variables are consistent with alternative elections such as $\{R^{\mu}_{\ \nu\rho\sigma},g^{\alpha\beta}\}$ or $\{R_{\mu\nu}^{\rho\sigma}\}$. In particular, using the identities analogous to \req{idus} obtained in \cite{Padmanabhan:2011ex} for the different elections of variables it is possible to show that \req{deltas} and \req{deltas2} are correct independently of such election. For example, if we choose $\{R_{\mu\nu}^{\rho\sigma}\}$, \req{deltas} and \req{deltas2} can be written as
\begin{equation}\labell{deltas22}
\delta\mathcal{L}=\bar P_{\mu \nu}^{\rho\lambda}\delta R_{ \rho\lambda}^{\mu\nu}\, , \quad 
\delta P^{\mu\alpha\beta\nu}=2\delta g^{\lambda[\mu}\bar P_{\lambda}^{\ \alpha]\beta\nu}+\bar C^{\mu\alpha\beta\nu}_{\lambda\rho\sigma\tau}\bar g^{\lambda\eta}  \bar g^{\rho \gamma}\delta R_{\eta \gamma}^{ \kappa \upsilon}\, .
\end{equation}
}
\begin{eqnarray}\labell{deltas}
\delta\mathcal{L}&=&2\delta g^{\mu\nu} \bar P_{\mu}^{\ \sigma\rho\lambda}\bar R_{\nu\sigma\rho\lambda}+\bar P^{\mu \sigma\rho\lambda}\delta R_{ \mu \sigma\rho\lambda}\, ,\\ \labell{deltas2}
\delta P^{\mu\alpha\beta\nu}&=&2\delta g^{\lambda[\mu}\bar P_{\lambda}^{\ \alpha]\beta\nu}+2\delta g^{\rho\eta}\bar C^{\mu\alpha\beta\nu}_{\lambda\eta\sigma\tau}\bar R^{\lambda \ \ \sigma\tau}_{\ \ \rho}+\bar C^{\mu\alpha\beta\nu}_{\lambda\rho\sigma\tau}\bar g^{\lambda\eta}  \bar g^{\rho \gamma}\bar g^{\sigma \kappa}\bar g^{\tau \upsilon} \delta R_{\eta \gamma \sigma \tau}\, ,
\end{eqnarray}
where the bars mean evaluation on the background metric $\bar g_{\mu\nu}$.  Now, if we plug these expressions in \req{fieldequationsc2} and we take into account the expressions for $\delta g^{\mu\nu}$ and $\delta R_{ \mu \sigma\rho\lambda}$ as functions of the metric perturbation,
\be
\delta g^{\mu\nu}=-h^{\mu\nu}=-\bar{g}^{\mu\alpha}\bar{g}^{\nu\beta}h_{\alpha\beta}\, ,\quad \delta \tensor{R}{_{\mu\sigma\rho\lambda}}=h_{\alpha[\mu}\tensor{\bar{R}}{^{\alpha}_{\sigma]\rho\lambda}}+\bar{\nabla}_{[\rho|}\bar{\nabla}_{\sigma}h_{|\lambda]\mu}-\bar{\nabla}_{[\rho|}\bar{\nabla}_{\mu}h_{|\lambda]\sigma}
\ee
we obtain the linearized equations. However, on an arbitrary background, the tensors $\bar{C}$, $\bar{P}$, and also the curvature, can take any form. Thus, we cannot obtain any useful information unless we specify a particular background.

\subsubsection*{Maximally symmetric background}
In order to make progress, from now on we will assume that the unperturbed spacetime $(\bar{\mathcal{M}},\bar{g}_{\mu\nu})$ is a maximally symmetric background (msb), or equivalently, a constant curvature space. This implies that $\bar g_{\mu\nu}$ satisfies
\begin{equation}\label{msb}
\bar R_{\mu\nu\alpha\beta}=2\mathcal{K} \bar g_{\mu[\alpha}\bar g_{\beta]\nu}\, ,
\end{equation}
for some constant $\mathcal{K}$ that characterizes the solution. This constant is not arbitrary, but is determined from the field equations as we show below. But before that, let us note that, on a maximally symmetric space, the form of the tensors $\bar P^{\mu\alpha\beta\nu}$ and $\bar C_{\sigma\rho\lambda\eta}^{\mu\alpha\beta\nu}$  is highly constrained.
Obviously, their explicit expressions still depend on the particular Lagrangian $\mathcal{L}$ considered, but when these objects are evaluated on a msb, the resulting expressions can only contain terms involving combinations of $\bar g_{\mu\nu}$, $\bar g^{\mu\nu}$ and $\delta_{\ \mu}^{\nu}$, because these objets are formed only from the metric and the Riemann tensor, and the latter is again given in terms of the metric for these backgrounds \req{msb}. In addition, it is clear from \req{Ptensor2} and \req{P-def}, that $P^{\mu\nu\rho\sigma}$ and $C_{\mu\nu\rho\sigma}^{\alpha\beta\gamma\delta}$ inherit the symmetries of the Riemann tensors appearing in their definitions. 
These conditions force $\bar P^{\mu\alpha\beta\nu}$ to be given by
\begin{equation}
\bar P^{\mu\alpha\beta\nu}=2e \bar g^{\mu[\beta}\bar g^{\nu]\alpha}\, ,
\label{e-def}
\end{equation}
where the value of the constant $e$ depends on the theory.  Similarly, $\bar C_{\sigma\rho\lambda\eta}^{\mu\alpha\beta\nu}$ is fully determined by three tensorial structures, namely
 \begin{equation}
\begin{aligned}
\bar C^{\sigma\rho\lambda\eta}_{\mu\alpha\beta\nu}&=a\left[\delta^{[\sigma}_{\mu}\delta^{\rho]}_{\alpha}\delta^{[\lambda}_{\beta}\delta^{\eta]}_{\nu}+\delta^{[\lambda}_{\mu}\delta^{\eta]}_{\alpha}\delta^{[\sigma}_{\beta}\delta^{\rho]}_{\nu}\right]+b\left[\bar g_{\mu\beta}\bar g_{\alpha\nu}-\bar g_{\mu\nu}\bar g_{\alpha\beta}\right]\left[\bar g^{\sigma\lambda}\bar g^{\rho\eta}-\bar g^{\sigma\eta}\bar g^{\rho\lambda}\right]\\
&+4c\,\delta^{[\sigma}_{\ (\tau}\bar g^{\rho][\lambda}\delta^{\eta]}_{\ \epsilon)}\delta^{\tau}_{\ [\mu}\bar g_{\alpha][\beta}\delta^{\epsilon}_{\ \nu]}\, ,
\end{aligned}
\label{abc-def}
\end{equation}
where the only theory-dependent quantities are in turn the constants $a$, $b$ and $c$.

%This equation allows us to express $\bar{\mathcal{L}}$ in terms of $e$ and $\mathcal{K}$. 
\subsubsection*{Background embedding equation}
Let us now determine the equation for the curvature scale $\mathcal{K}$. Imposing the constant curvature  metric $\bar g_{\mu\nu}$ to solve the field equations \req{fieldequationsc2} with $T_{\mu\nu}=0$, one finds
\begin{equation}
\bar{\mathcal{L}}(\mathcal{K})=4e(D-1)\mathcal{K}\, ,
\label{embedding}
\end{equation}
where we used \req{e-def}.
This is a relation between the background scale $\mathcal{K}$ defined in \req{msb} and all the possible couplings appearing in the higher-order Lagrangian $\mathcal{L}($Riemann$)$. 
%We also have to take into account that the background metric $\bar g_{\mu\nu}$ must be a solution of the field equations. With our definitions, the equation for a maximally symmetric background is written as
Another equation relating $e$ and $\mathcal{K}$ can be obtained using \req{msb} and \req{e-def}. This reads in turn
\begin{equation}
\frac{d\bar{\mathcal{L}}(\mathcal{K})}{d \mathcal{K}}=\bar P^{\mu\nu\rho\sigma}2\bar g_{\mu[\rho} \bar g_{\sigma]\nu}=2eD(D-1)\, ,
\end{equation}
which, along with \req{embedding} produces the nice expression
\begin{equation}
\label{Lambda-eq}
\bar{\mathcal{L}}(\mathcal{K})-\frac{2\mathcal{K}}{D}\frac{d\bar{\mathcal{L}}(\mathcal{K})}{d \mathcal{K}}=0\, .
\end{equation}
This is the algebraic equation that needs to be solved in order to determine the possible vacua of the theory, \ie the allowed values of $\mathcal{K}$ as functions of the scales and couplings appearing in $\mathcal{L}($Riemann$)$.
For instance, in the case of Einstein gravity with a cosmological constant, $\mathcal{L}=R-2\Lambda$, one finds $\mathcal{K}=\tfrac{2\Lambda}{(D-1)(D-2)}$.\footnote{Another example: for Gauss-Bonnet with a negative cosmological constant $\mathcal{L}=R+(D-1)(D-2)/L^2+L^2\lambda_{\rm GB}/((D-3)(D-4))\mathcal{X}_4$, one finds the well-known relation $-L^2\mathcal{K}=(1\pm\sqrt{1-4\lambda_{\rm GB}})/(2\lambda_{\rm GB})$, see \eg \cite{Buchel:2009sk}.} For this reason, we will in general define the ``effective'' cosmological constant $\Lambda_{\rm eff}$ according to
\be\label{eq:lambdaeff}
\mathcal{K}=\frac{2\Lambda_{\rm eff}}{(D-1)(D-2)}\, .
\ee

Equation~\req{Lambda-eq} can also be derived directly from the \emph{reduced} gravitational action, which is obtained by evaluating the full action on an arbitrary constant curvature metric. The reduced action reads
\begin{equation}
S(\mathcal{K})=V_0 |\mathcal{K}|^{-D/2}\bar{\mathcal{L}}(\mathcal{K})\, ,
\end{equation}
where $V_0$ is the regularized ``spacetime volume'' for some reference value of $\mathcal{K}$. Extremizing this expression with respect to $\mathcal{K}$ yields

\begin{equation}
\frac{\delta S(\mathcal{K})}{\delta \mathcal{K}}=V_0 |\mathcal{K}|^{-D/2}\left(\frac{d \bar{\mathcal{L}}(\mathcal{K})}{d\mathcal{K}}-\frac{D}{2\mathcal{K}}\bar{\mathcal{L}}(\mathcal{K})\right)=0\, ,
\end{equation}
where we recognize again Eq.~\req{Lambda-eq}.  This way of deriving the equation is completely general and it allows us to see that Eq.~\req{Lambda-eq} is actually valid for any higher-derivative gravity at all, and not only for $\mathcal{L}($Riemann$)$ theories.

%%%%%%%%%%%%%%%%%%%%%%%%%%%%%%%%%%%%%%%%%%%%%%%%%%%%%%%%%%%%%%%
\subsubsection*{Linearization procedure}
With the information from the previous items, we are ready to linearize \req{fieldequationsc2}.
The result of a long computation in which we make use of \req{Ptensor2}-\req{abc-def} reads
\begin{align}\notag
\frac{1}{2}\mathcal{E}_{\mu\nu}^{L}=&+\left[e-2\mathcal{K}(a(D-1)+c)+(2a+c)\bar \Box\right]G_{\mu\nu}^{ L}+\left[a+2b+c\right]\left[\bar g_{\mu\nu}\bar\Box-\bar\nabla_{\mu}\bar\nabla_{\nu}\right]R^{ L}\\  &-\mathcal{K}\left[a(D-3)-2b(D-1)-c \right] \bar g_{\mu\nu}R^{ L}\, , \label{lineareqs}
\end{align}
where the linearized Einstein tensor and the linearized Ricci scalar read, respectively\footnote{More precisely, $G_{\mu\nu}^L$ is the linearized \emph{cosmological} Einstein tensor, that can be written as
\begin{align}
G_{\mu\nu}^L&=R_{\mu\nu}^L-\frac{1}{2}\bar g_{\mu\nu}R^L-(D-1)\mathcal{K} h_{\mu\nu}\, , 
\end{align}
where the linearized Ricci tensor reads
\begin{align}
R_{\mu\nu}^L&=\bar\nabla_{(\mu|}\bar\nabla_{\sigma}h^{\sigma}_{\ |\nu)}-\frac{1}{2}\bar\Box h_{\mu\nu}-\frac{1}{2}\bar\nabla_{\mu}\bar\nabla_{\nu}h+D \mathcal{K} h_{\mu\nu}-\mathcal{K} h \bar g_{\mu\nu}\, .
\end{align}
In these expressions we use the standard notation $h\equiv \bar g^{\mu\nu} h_{\mu\nu}$ and indices are raised and lowered with $\bar g^{\mu\nu}$ and $\bar g_{\mu\nu}$ respectively. 
}
\begin{align}\label{eq:GL}
G_{\mu\nu}^L&=-\frac{1}{2}\bar\Box h_{\mu\nu}+\bar\nabla_{(\mu|}\bar\nabla_{\sigma}h^{\sigma}_{\ |\nu)}-\frac{1}{2}\bar\nabla_{\mu}\bar\nabla_{\nu}h+\frac{1}{2}\bar g_{\mu\nu}\left(\bar \Box h-\bar\nabla^{\alpha}\bar\nabla^{\beta}h_{\alpha\beta}\right)\\ \nonumber
&+ \mathcal{K} h_{\mu\nu}+\frac{D-3}{2}\mathcal{K} h \bar g_{\mu\nu}\, ,\\ \label{ricciscalar}
R^{L}&=\bar\nabla^{\mu}\bar\nabla^{\nu}h_{\mu\nu}-\bar \Box h-(D-1)\mathcal{K} h\, .
\end{align}
Also, we have introduced $\bar\Box=\bar{g}^{\mu\nu}\bar{\nabla}_{\mu}\bar{\nabla}_{\nu}$. The above equations are quartic in derivatives of the perturbation for generic higher-derivative theories, as expected. The problem is hence reduced to the evaluation of $a$, $b$, $c$ and $e$ for a given theory, something that can be done using \req{Ptensor2}, \req{P-def}, (\ref{e-def}) and (\ref{abc-def}). However, this is a very tedious procedure in general, which involves the computation of first and second derivatives of $\mathcal{L}($Riemann$)$ with respect to the Riemann tensor.

We present here a method (originally reported in \cite{Aspects, PabloPablo})  that allows for an important simplification of this problem.  The procedure has several steps which we explain now. %We will omit the bars from the metrics $g_{\mu\nu}$ for the rest of the subsection in order to avoid confusion.

\begin{enumerate}
\item Consider an auxiliary symmetric tensor $k_{\mu\nu}$ satisfying the following properties.
\begin{equation}
\label{k-def}
k^{\mu}_{\ \mu}=\chi\, , \quad k^{\mu}_{\ \alpha}k^{\alpha}_{\ \nu}=k^{\mu}_{\ \nu}\, .
\end{equation} 
Note that both imply that $\chi$ is an integer constant corresponding to the rank of $\tensor{k}{_{\mu}^{\nu}}$. However, it does not really matter what $\chi$ is and what $k$ is; we only need the properties above in order to make the computations that we specify below.

\item Define the following ``Riemann tensor''\footnote{The associated ``Ricci tensor'' and ``Ricci scalar'' are: $
\tilde R_{\mu\nu}=\mathcal{K}(D-1)g_{\mu\nu}+\alpha(\chi-1)k_{\mu\nu}$ and $\tilde R=\mathcal{K} D(D-1)+\alpha \chi(\chi-1)$ respectively.
}
\begin{equation}
\label{Riemalpha}
\tilde R_{\mu\nu\sigma\rho}(\mathcal{K}, \alpha)\equiv 2\mathcal{K}  g_{\mu[\sigma} g_{\rho]\nu}+2\alpha k_{\mu[\sigma}k_{\rho]\nu}\, , 
\end{equation}
where  $\alpha$ and $\mathcal{K}$ are two parameters. Observe that $\tilde R_{\mu\nu\sigma\rho}(\mathcal{K}, \alpha)$  does not correspond --- or more precisely, it does not need to correspond ---  to the Riemann tensor of any actual metric in general, even though it respects the symmetries of a true Riemann tensor. An exception occurs when $\alpha=0$, as $\tilde R_{\mu\nu\sigma\rho}(\mathcal{K},0)$ becomes the Riemann tensor of a msb of curvature $\mathcal{K}$ associated to a metric $g_{\mu\nu}=\bar g_{\mu\nu}$ as defined in \req{msb}. 

\item Evaluate the higher-derivative Lagrangian on $\tilde R_{\mu\nu\sigma\rho}(\mathcal{K}, \alpha)$, \ie replace all Riemann tensors appearing in $\mathcal{L}(g^{\alpha\beta}, R_{\mu\nu\rho\sigma})$ by the object defined in \req{Riemalpha}. This gives rise to a function of $\mathcal{K}$ and $\alpha$,
%Then, we can evaluate the Lagrangian $\mathcal{L}(R_{\mu\nu\rho\sigma})$ on this deformed background, so we obtain a function of $\alpha$ and $\mathcal{K}$:
\begin{equation}
\mathcal{L}(\mathcal{K},\alpha)\equiv \mathcal{L}\left(g^{\alpha\beta}, R_{\mu\nu\rho\sigma}=\tilde R_{\mu\nu\rho\sigma}(\mathcal{K},\alpha)\right)\, .
\label{Ldefinition}
\end{equation}

\item The values of  $a,b,c$ and $e$ can be obtained from the expressions
\begin{align}\label{tete}
\frac{\partial \mathcal{L}}{\partial\alpha}\Big|_{\alpha=0}&=2e\,\chi(\chi-1)\,, \\ \label{teta}
\frac{\partial^2 \mathcal{L}}{\partial \alpha^2}\Big|_{\alpha=0}&=4\chi(\chi-1)\left(a+b\, \chi(\chi-1)+c(\chi-1)\right)\, ,
\end{align}
as can be proven using the chain rule along with equations \req{Ptensor2}, (\ref{P-def}), (\ref{e-def}) and (\ref{abc-def}).
Interestingly, since $a$, $b$, $c$ and $e$ do not depend on $\chi$ and they appear multiplied by factors involving different combinations of this parameter, we can identify them unambiguously for any theory by simple inspection. Once $\mathcal{L}(\mathcal{K},\alpha)$ and its derivatives are computed, we just need to compare the resulting expressions with the RHS of \req{tete} and \req{teta} to obtain $a$, $b$, $c$ and $e$\footnote{Observe that we only need $\mathcal{L}(\mathcal{K},\alpha)$ up to $\alpha^2$ order, \ie from
$
\mathcal{L}(\mathcal{K},\alpha)=\mathcal{L}(\mathcal{K})+\left[2\chi(\chi-1)e\right]\, \alpha+ \left[ 2\chi(\chi-1)(a+b\,\chi(\chi-1)+c(\chi-1))\right]\, \alpha^2+\mathcal{O}(\alpha^3)
$ we can read off the values of all the relevant constants.}.

\item Replace the values of $a$, $b$, $c$ and $e$ in the general expression \req{lineareqs}.

\end{enumerate}

This procedure is obviously simpler than computing $\bar{P}^{\mu\nu\rho\sigma}$ and $\bar C^{\mu\nu\alpha\beta}_{\lambda\eta\sigma\tau}$ explicitly using their definitions \req{Ptensor2} and (\ref{P-def}). Indeed, the most difficult step is the evaluation of $\mathcal{L}(\mathcal{K},\alpha)$, which simply involves contractions of $g_{\mu\nu}$ and $k_{\mu\nu}$ for any theory. The function $\mathcal{L}(\mathcal{K},\alpha)$ is a sort of ``prepotential'' containing all the information needed for the linearization of a given higher-derivative theory of the form \req{Smassc2} on a msb. 

We will apply this method in various sections of the chapter --- \eg see Sec.~\ref{quartic} for the linearization of general quartic theories and Sec.~\ref{fscalars} for theories constructed as functions of curvature invariants. Appendix \ref{lineapp} contains a detailed application of our linearization procedure to quadratic theories and to a particular Born-Infeld-like theory.

Let us mention that in \cite{Tekin1,Tekin2,Senturk:2012yi} a more refined method than the naive brute-force linearization of the full non-linear equations was also introduced for general $\mathcal{L}($Riemann$)$ theories. This incorporates decompositions similar to the ones in \req{e-def} and \req{abc-def}, but still requires the somewhat tedious explicit evaluation of  $\bar P^{\mu\nu\rho\sigma}$ and $\bar C_{\mu\nu\rho\sigma}^{\alpha\beta\gamma\delta}$ for each theory considered.

We close this subsection by mentioning that our linearization method reproduces all the particular cases previously studied in the literature. These include: quadratic gravities \cite{Smolic:2013gz,Lu,Tekin1,Tekin2,Deser:2011xc}, Quasi-topological gravity \cite{Quasi2,Quasi}, $f(R)$ \cite{Bueno2} and general $f($Lovelock$)$ theories \cite{Love}.

\subsubsection*{Equivalent quadratic theory}\label{quadra}
The linearized equations \req{lineareqs} of any higher-order gravity of the form \req{Smassc2} characterized by some parameters $a$, $b$, $c$ and $e$, can always be mapped to those of a quadratic theory of the form
\begin{equation}\labell{quadratic0}
\mathcal{L}_{\rm quadratic}=\lambda (R-2\Lambda)+\alpha R^2+\beta R_{\mu\nu}R^{\mu\nu}+\gamma \mathcal{X}_4\, ,
\end{equation}
where $\mathcal{X}_4=R_{\mu\nu\rho\sigma}R^{\mu\nu\rho\sigma}-4R_{\mu\nu}R^{\mu\nu}+R^2$ is the dimensionally-extended four-dimensional Euler density, also known as Gauss-Bonnet term. Indeed, the parameters $\lambda$, $\alpha$, $\beta$ and $\gamma$ of the equivalent quadratic theory can be obtained in terms of $a$, $b$, $c$ and $e$ through 
\begin{equation}\labell{quadratic}
\lambda=2e-4\mathcal{K}\left[a+bD(D-1)+c(D-1)\right]\, ,\quad
\alpha=2b-a\, , \quad \beta=4a+2c\, , \quad \gamma=a\, .
\end{equation}
Similarly, the cosmological constant $\Lambda$ can be trivially related to the parameters appearing in \req{Smassc2} through
$ \Lambda=-\mathcal{L}(R_{\mu\nu\rho\sigma}=0)/(2\lambda)$.

Notice that the mapping from \req{Smassc2} to \req{quadratic0} is surjective but not injective, \ie all $\mathcal{L}($Riemann$)$ theories are mapped to some quadratic theory, but (infinitely) many of them are mapped to the same one. Observe also that the existence of this mapping is a consequence of the fact that the linearized equations of any theory come from its action expanded at quadratic order in $h_{\mu\nu}$ --- see Subsection \ref{quadact}. This means that the most general quadratic theory, namely \req{quadratic0} already contains all the possible kinds of terms produced in the action at order $\mathcal{O}(h^2)$ of any $\mathcal{L}($Riemann$)$ theory. Observe however that the fact that the parameters $a$, $b$, $c$ and $e$ for a given theory can be related to those appearing in \req{quadratic0} does not immediately help in identifying the values of those parameters for a given theory. The mapping was explicitly performed for general cubic theories in \cite{Tekin2}.

%%%%%%%%%%%%%%%%%%%%%%%%%%%%%%%%%%%%%%%%%%%%%%%%%%%%%%%%%%%%%%%
%%%%%%%%%%%%%%%%%%%%%%%%%%%%%%%%%%%%%%%%%%%%%%%%%%%%%%%%%%%%%%%
\subsection{Physical modes}\label{physmo}
As we just reviewed, $\mathcal{E}_{\mu\nu}^{L}$ depends on four constants $a$, $b$, $c$ and $e$ as well as on the background curvature $\mathcal{K}$. For a given theory, the four constants can be computed using the procedure explained in Subsection \ref{lll}, from which one can obtain the full linearized equations through \req{lineareqs}. In this subsection we will explore how \req{lineareqs} can be further simplified using the gauge freedom of the metric perturbation and used to characterize the additional physical modes propagated by the metric in a general theory of the form \req{Smassc2}. 

Let us start with the following observation. If we parametrize $a$, $b$ and $c$ in terms of three new constants $m_g^2$, $m_s^2$ and $\kappa_{\rm eff}$  as

\begin{equation}
\label{parameters23}
\begin{aligned}
a&=\left[4 e \kappa_{\rm eff}-1\right]/\left[8\mathcal{K}(D-3)\kappa_{\rm eff}\right]\, ,\\
b&=\left[(4 e \kappa_{\rm eff}-1)(D-1)m_s^2m_g^2+2(3-2D+2(D-1)De \kappa_{\rm eff})m_g^2\mathcal{K} \right. \\  &\left.\quad +(D-3)\mathcal{K}(D m_s^2+4(D-1)\mathcal{K})\right]/\left[16\mathcal{K}(D-3)\kappa_{\rm eff}m_g^2(D-1)(m_s^2+D\mathcal{K})\right]\, ,\\
c&=-\left[(4e \kappa_{\rm eff}-1)m_g^2+(D-3)\mathcal{K}\right]/\left[4\mathcal{K}(D-3)\kappa_{\rm eff }m_g^2\right]\, ,
\end{aligned}
\end{equation}
it is possible to rewrite \req{lineareqs} in terms of four different parameters, namely, $\kappa_{\mbox{{\scriptsize eff}}}$, $m_s^2$, $m_g^2$ and $\mathcal{K}$. Indeed, one finds
\begin{align}\label{lineareq2s}
\mathcal{E}_{\mu\nu}^{L}=\frac{1}{2\kappa_{\rm eff}m_g^2}&\left\{ \left[m_g^2+2\mathcal{K}-\bar \Box \right]G_{\mu\nu}^{ L}+\left[\frac{(D-2)(m_g^2+m_s^2+2\mathcal{K})}{2(m_s^2+D\mathcal{K})}\right]\mathcal{K} \bar g_{\mu\nu} R^{ L}\right.\\ \notag
&\left.+\left[\frac{(D-2)(m_g^2-m_s^2-2(D-1)\mathcal{K})}{2(D-1)(m_s^2+D \mathcal{K})} \right]\left[\bar g_{\mu\nu}\bar\Box-\bar\nabla_{\mu}\bar\nabla_{\nu}\right]R^{ L} \right\}
\, ,
\end{align}
so the dependence on $e$ disappears, while that on $\kappa_{\rm eff}$ gets factorized out from all terms. While \req{lineareqs} is more useful when computing the linearized equations of a particular theory --- because we know a simple procedure to obtain $a$, $b$, $c$ and $e$ --- \req{lineareq2s} is more illuminating from a physical point of view. Indeed, as we will see in a moment, $\kappa_{\rm eff}$ will be the effective gravitational constant, related to the effective Newton's constant according to\footnote{During this chapter we will use $\kappa_{\rm eff}$ rather than $G_{\rm eff}$ for the sake of conciseness.} 
\be
\kappa_{\rm eff}\equiv 8\pi G_{\rm eff}
\ee
while $m_g^2$ and $m_s^2$ will correspond, respectively, to the squared-masses of additional spin-2 and scalar modes. 

It is straightforward to invert the relations \req{parameters23} to obtain the values of such physical quantities in terms of $a$, $b$, $c$ and $e$. One finds
\begin{eqnarray}\label{kafka}
\kappa_{\rm eff}&=&\frac{1}{4e-8\mathcal{K}(D-3)a}\, ,\\ \label{kafka1}
m_s^2&=&\frac{e(D-2)-4\mathcal{K}(a+bD(D-1)+c(D-1))}{2a+Dc+4b(D-1)}\, ,\\ \label{kafka2}
m_g^2&=&\frac{-e+2\mathcal{K}(D-3)a}{2a+c}\, .
\end{eqnarray}
Let us stress that if we consider a theory consisting of a linear combination of invariants --- like the one in \req{quarticaction} below --- the values of $a$, $b$, $c$ and $e$ of that theory can be simply computed as the analogous linear combination of the parameters for each of those terms. However, that is not the case for $\kappa_{\rm eff}$, $m_s^2$ and $m_g^2$, since they are not linear combinations of $a$, $b$, $c$ and $e$. Hence, in order to determine these quantities for a given linear combination of invariants, the natural procedure should be obtaining the total values of $a$, $b$, $c$ and $e$ first, and then using \req{kafka}-\req{kafka2} to compute the corresponding values of $\kappa_{\rm eff}$, $m_s^2$ and $m_g^2$. For example, for a general quadratic theory of the form
\begin{equation}
\begin{aligned}
S=\int_{\mathcal{M}}d^Dx\sqrt{|g|}\Bigg[&\frac{1}{2\kappa}(-2\Lambda+R)+ \kappa^{\frac{(4-D)}{D-2}}(\alpha_1R^2+\alpha_2 R_{\mu\nu}R^{\mu\nu}+\alpha_3 R_{\mu\nu\rho\sigma}R^{\mu\nu\rho\sigma})\Bigg]\, ,
\end{aligned}
\label{quarticaction}
\end{equation}
the values of $\kappa_{\rm eff}$, $m_g$ and $m_s$ read, respectively
  \begin{align}\label{quadrit1}
 \kappa_{\text{eff}} &= \frac{ \kappa }{  1+ 4 \mathcal{K} \kappa^{\frac{2}{D-2}} (\alpha_1 D(D-1)+\alpha_2 (D-1)-2 \alpha_3 (D-4))    }\, ,
     \\\label{quadrit2}
 m_s^2 &= \frac{(D-2) + 4 (D-4)   \mathcal{K} \kappa^{\frac{2}{D-2}}   \left(  \alpha_1 D (D-1)   + \alpha_2 (D-1)  +  2
   \alpha _3\right)}{2 \kappa^{\frac{2}{D-2}}
   \left(4 \alpha _1 (D-1)+\alpha _2 D  +   4 \alpha _3\right)} \, ,  \\\label{quadrit3}
  m_g^2 &=  \frac{-1  - 4 \mathcal{K} \kappa^{\frac{2}{D-2}} \left (   \alpha_1 D (D-1)    + \alpha_2 (D-1) 
    -2      \alpha _3 (D-4)  \right)
   }{2 \kappa^{\frac{2}{D-2}} \left(\alpha _2+4 \alpha _3\right)  } \, ,
    \end{align}
which we obtained using \req{kafka}-\req{kafka2} and the values of $a$, $b$, $c$ and $e$ which appear in Table \ref{tabla2}.
During the remainder of this section, we will write all expressions in terms of $\kappa_{\rm eff}$, $m_s^2$ and $m_g^2$, which will make the presentation clearer. Nonetheless, all equations can be converted back to the language of $a$, $b$, $c$ and $e$ using the above relations. 

The discussion proceeds slightly differently depending on whether we consider AdS/dS or Minkowski as the background spacetime, so we will consider the two cases separately. Let us start with the first.

 \subsubsection{(Anti-)de Sitter background}\label{AdS space}
 When studying the physical modes propagated by the metric perturbation on an AdS/dS background, it is customary and very convenient to work in the transverse gauge, in which\footnote{The metric decomposition performed in this section is similar to the one considered in \cite{Smolic:2013gz}.}
 \begin{equation}\label{trans}
 \bar \nabla_{\mu} h^{\mu\nu}=\bar \nabla^{\nu}h\, .
 \end{equation} 
Imposing this condition, many terms in (\ref{lineareq2s}) cancel out. Let us now expand the metric perturbation into its trace and traceless parts, which we denote by $h$ and $h_{\langle\mu\nu\rangle}$ respectively\footnote{In this section, we denote the trace and traceless parts of rank-2 tensors $P_{\mu\nu}$ linear in $h_{\mu\nu}$ as $P\equiv \bar g^{\mu\nu}P_{\mu\nu}$ and $P_{\langle\mu\nu\rangle}\equiv P_{\mu\nu}-\frac{1}{D}\bar g_{\mu\nu}P$ respectively. In the case of the equations of motion, one can use the same notation, \ie $\mathcal{E}^L\equiv \bar g^{\mu\nu}\mathcal{E}^L_{\mu\nu}$, $T^L\equiv \bar{g}^{\mu\nu}T^L_{\mu\nu}$ --- and similarly for the traceless part --- because $\bar{\mathcal{E}}_{\mu\nu}=\bar T_{\mu\nu}=0$. Observe however that $R^L=(g^{\mu\nu}R_{\mu\nu})^L$ is not the trace of $R^L_{\mu\nu}$, but rather $R^L=\bar g^{\mu\nu}R_{\mu\nu}^L-h^{\mu\nu}\bar{R}_{\mu\nu}=\bar g^{\mu\nu}R_{\mu\nu}^L-(D-1)h\mathcal{K}$.},
 \begin{equation}
 h_{\mu\nu}=h_{\langle\mu\nu\rangle}+\frac{1}{D}\bar g_{\mu\nu}h\,.
 \end{equation}
Doing the same with the field equations (\ref{lineareq2s}), we find
\begin{align}
\label{traceless}
\mathcal{E}^L_{\langle\mu\nu\rangle}=&+\frac{1}{2}T^L_{\langle\mu\nu\rangle}=\frac{1}{4m_g^2\kappa_{\rm eff}}\bigg\{ \left[\bar\Box -2\mathcal{K}\right]\left[\bar\Box-2\mathcal{K}-m_g^2\right]h_{\langle\mu\nu\rangle}-\bar\nabla_{\langle\nu}\bar\nabla_{\mu\rangle}\bar\Box h\\ \notag&+\left[\frac{m_g^2(m_s^2+2(D-1)\mathcal{K})+\mathcal{K}((4-3D)m_s^2-4(D-1)^2\mathcal{K})}{(m_s^2+D\mathcal{K})} \right]\bar\nabla_{\langle\nu}\bar\nabla_{\mu\rangle}h\bigg\}
\, ,\\
\label{trace}
\mathcal{E}^L=&+\frac{1}{2}T^L=-\left[\frac{(D-1)(D-2)\mathcal{K}(m_g^2-(D-2)\mathcal{K})}{4\kappa_{\rm eff}m_g^2(m_s^2+D\mathcal{K})}\right]\left[\bar\Box -m_s^2\right]h\, ,
\end{align}
where, for the sake of generality, we are including an energy-momentum tensor $T^L_{\mu\nu}$ as a matter source.
The second is the equation of motion of a free scalar field of mass $m_s$, while the first is an inhomogeneous equation for $h_{\langle \mu\nu\rangle}$ as it involves also $h$.
In order to obtain an independent equation for the traceless part, we define another traceless tensor:
\begin{equation}\label{tttt}
t_{\mu\nu}\equiv h_{\langle\mu\nu\rangle}-\frac{\bar\nabla_{\langle \mu}\bar\nabla_{\nu\rangle}h}{(m_s^2+D\mathcal{K})} \, ,
\end{equation} 
where we have implicitly assumed that $m_s^2\neq -D\mathcal{K}$. After some manipulations, it can be seen that 
$t_{\mu\nu}$ satisfies the equation
 \begin{equation}
 \frac{1}{2\kappa_{\mbox{{\scriptsize eff}}}m_g^2}(\bar\Box-2\mathcal{K})(\bar\Box-2\mathcal{K}-m_g^2)t_{\mu\nu}=T^{L, \mbox{\scriptsize eff}}_{\langle\mu\nu\rangle}\, ,
 \end{equation}
where we have defined the effective energy-momentum tensor
\begin{equation}
T^{L, \mbox{\scriptsize eff}}_{\langle\mu\nu\rangle}\equiv T^L_{\langle\mu\nu\rangle}+\frac{\left[\bar\Box+(D-4)\mathcal{K}-m_g^2\right]\bar\nabla_{\langle\mu}\bar\nabla_{\nu\rangle}T^L}{\mathcal{K}(D-1)(D-2)(m_g^2-(D-2)\mathcal{K})}\,.
\end{equation}
Now, observe that the object
\begin{equation}
t^{(m)}_{\mu\nu}\equiv -\frac{1}{m_g^2}(\bar\Box-2\mathcal{K}-m_g^2)t_{\mu\nu}\, ,
\label{massless}
\end{equation}
satisfies the equation of the usual massless graviton, namely
\begin{equation}
-(\bar\Box-2\mathcal{K})t^{(m)}_{\mu\nu}=2\kappa_{\rm eff }T_{\langle\mu\nu\rangle}^{L, \rm eff}\, ,
\label{masslesseq}
\end{equation}
but with a non-standard coupling to matter. %(TASK: PROVE IT IS ALWAYS TRANSVERSE). 
On the other hand, using (\ref{massless}) and (\ref{masslesseq}), it is easy to see that the tensor 
\begin{equation}\label{MaS}
t^{(M)}_{\mu\nu}\equiv t_{\mu\nu}-t^{(m)}_{\mu\nu}=\frac{1}{m_g^2}(\bar \Box-2\mathcal{K})t_{\mu\nu}\,,
\end{equation}
satisfies instead
\begin{equation}\label{MaSeq}
(\bar\Box-2\mathcal{K}-m_g^2)t^{(M)}_{\mu\nu}=2\kappa_{\mbox{{\scriptsize eff}}}T_{\langle\mu\nu\rangle}^{L, \mbox{\scriptsize eff}}\, .
\end{equation}
Hence, we identify $t^{(M)}_{\mu\nu}$ with a massive traceless spin-2 field with mass $m_g$. Observe that the coupling to matter of this mode has the wrong sign, which reflects its ghost-like behavior. Note that, apart from being a ghost, this mode is also tachyonic whenever $m_g^2<0$. The same occurs for the scalar when $m_s^2<0$.

In sum, using definitions \req{tttt}, \req{massless} and \req{MaS}, we can decompose the metric perturbation $h_{\mu\nu}$ as
\begin{equation}\labell{decoo}
h_{\mu\nu}=t_{\mu\nu}^{(m)}+t_{\mu\nu}^{(M)}+\frac{\bar\nabla_{\langle\mu}\bar\nabla_{\nu\rangle}h}{(m_s^2+D\mathcal{K})}+\frac{1}{D}\bar g_{\mu\nu} h\, ,
\end{equation}
where $h$, $t_{\mu\nu}^{(M)}$ and $t_{\mu\nu}^{(m)}$ satisfy \req{trace}, \req{MaSeq} and \req{masslesseq}, and represent respectively: a scalar mode of mass $m_s$, a ghost-like spin-2 mode of mass $m_g$ --- which we will often refer to as ``massive graviton'' throughout the text --- and a massless graviton.

\begin{comment}
The physical constants that appear in these expressions are given by
\begin{equation}
\label{parameters}
\boxed{
\begin{aligned}
 \frac{1}{2\kappa_{\mbox{{\scriptsize eff}}}}&=2e-4\mathcal{K}(D-3)a,\\
m_s^2&=\frac{e(D-2)-4\mathcal{K}(a+bD(D-1)+c(D-1))}{2a+Dc+4b(D-1)},\\
m_g^2&=\frac{-e+2\mathcal{K}(D-3)a}{2a+c},
\end{aligned}}
\end{equation}
\end{comment}

\subsubsection{Minkowski background}\label{Flat space}
If we set $\mathcal{K}=0$ in \req{trace}, this equation would lead us to conclude that $T=0$. This inconsistency is a reflection of the fact that the transverse gauge cannot be used in flat spacetime. The usual choice is in this case the so-called \emph{de Donder gauge}, given by
\begin{equation}
\partial_{\mu}h^{\mu\nu}=\frac{1}{2}\partial^{\nu}h\, .
\label{Donder}
\end{equation}
In this gauge, the linearized field equations (\ref{lineareq2s}) in a Minkowski background can be written as

\begin{equation}\label{flat1}
\mathcal{E}_{\mu\nu}^L=-\frac{1}{4\kappa_{\mbox{{\scriptsize eff}}}}\bar\Box \hat h_{\mu\nu}=\frac{1}{2}T^L_{\mu\nu}\, ,
\end{equation}
where we have defined

\begin{equation}
\label{hath}
\begin{aligned}
 \hat h_{\mu\nu}&\equiv h_{\mu\nu}-\frac{1}{2}\eta_{\mu\nu}h-\frac{1}{m_g^2}\left[\bar\Box h_{\mu\nu}-\frac{1}{2}\partial_{\mu}\partial_{\nu}h\right]+\left[\frac{m_g^2(D-2)+m_s^2}{2(D-1)m_g^2 m_s^2}\right]\left[\eta_{\mu\nu}\bar\Box-\partial_{\mu}\partial_{\nu}\right]h\,.
\end{aligned}
\end{equation}
Using the gauge condition (\ref{Donder}) it is easy to see that $\hat h_{\mu\nu}$ is transverse, \ie

\begin{equation}
\partial_{\mu}\hat h^{\mu\nu}=0\, .
\end{equation}
Naturally, $\hat h_{\mu\nu}$ is the usual spin-2 massless graviton, as it satisfies the linearized Einstein equation \req{flat1}. However, there are more degrees of freedom (dof). In particular, we find that the metric can be decomposed as 

\begin{equation}
\begin{aligned}
\label{flatdecomp}
h_{\mu\nu}&=\hat h_{\mu\nu}-\frac{1}{D-2}\eta_{\mu\nu}\hat h+\frac{1}{D-1}(m_g^{-2}-m_s^{-2})\partial_{\langle \mu}\partial_{\nu\rangle}\hat h\\
&+t_{\mu\nu}+\frac{2}{D(D-2)}\eta_{\mu\nu}\phi+\frac{1}{(D-1)m_s^2}\partial_{\langle \mu}\partial_{\nu\rangle}\phi\,,
\end{aligned}
\end{equation}
where $t_{\mu\nu}$ is traceless and $\phi$ is a scalar field. These objects satisfy the equations
\begin{eqnarray}
\label{flat2}
-(\bar\Box-m_s^2)\phi&=&2\kappa_{\mbox{{\scriptsize eff}}}T^L\, ,\\
\label{flat3}
(\bar\Box-m_g^2)t_{\mu\nu}&=&2\kappa_{\mbox{{\scriptsize eff}}}\left[T^L_{\langle \mu\nu\rangle}+\frac{1}{(D-1)m_g^2}\partial_{\langle \mu}\partial_{\nu\rangle}T^L\right]\, .
\end{eqnarray}
Hence, even though we have proceeded in a different way as compared to the $\mathcal{K}\neq 0$ case, we have found the same physical modes: we have a massless spin-2 graviton $\hat h_{\mu\nu}$, a massive one $t_{\mu\nu}$ and a scalar $\phi$, the masses of the last two being the same as the ones we found for $t^{(M)}_{\mu\nu}$ and $h$ in the (A)dS case. Note however that even though the degrees of freedom and the masses are the same, the metric decomposition as well as the coupling of the fields to matter are different --- compare \req{trace} and \req{MaSeq} with \req{flat2} and \req{flat3}, and \req{decoo} with \req{flatdecomp}. This can be understood as a consequence of the fact that the gauge which is convenient for (A)dS \req{trans} differs from the de Donder one \req{Donder} utilized for Minkowski.

%Indeed, the transverse gauge $\bar\nabla_{\mu}h^{\mu\nu}=\bar\nabla^{\nu}h$ used in the $\mathcal{K}\neq 0$ case cannot be used in the flat case, since it leads to an inconsistency (it imposes $T=0$).

\subsection{Quadratic action}
\label{quadact}
As pointed out in Sec.~\ref{quadra}, the linearized equations \req{lineareq2s} come from terms of order $\mathcal{O}(h^2)$ in the action, which means that the structure of the linearized equations for the most general $\mathcal{L}($Riemann$)$ is already captured by the most general quadratic theory. Expanding the action of a higher-order gravity to $\mathcal{O}(h^2)$ is not trivial in general.
However, we can use the expression for the linearized equations \req{lineareq2s} to find an action that yields these equations when varied with respect to $h_{\mu\nu}$. The easiest possibility is
\begin{equation}
S_2=-\frac{1}{2}\int_{\mathcal{M}} d^Dx\,h^{\mu\nu}\mathcal{E}^L_{\mu\nu}\, .
\end{equation}
Using \req{lineareq2s} and integrating by parts several times we find the effective action

\begin{equation}\label{equiv}
S_2=\int_{\mathcal{M}} \frac{d^Dx}{4\kappa_{\rm eff}} \left[\frac{(D-2)\left[m_g^2+(D-2)(m_s^2+(D-1)\mathcal{K})\right]}{2(D-1)m_g^2(m_s^2+D\mathcal{K})}({R^L})^2-\left[h^{\mu\nu}+\frac{2{G^{L}}^{\mu\nu}}{m_g^2}\right]G^L_{\mu\nu}\right]\, .
\end{equation}
As pointed out in \cite{Tekin1}, where an analogous action was found, \req{equiv} is manifestly invariant under ``gauge'' transformations $h_{\mu\nu}\rightarrow h_{\mu\nu}+\bar \nabla_{\mu}\xi_{\nu}+\bar \nabla_{\nu}\xi_{\mu}$ as follows from the invariance of the linearized Einstein tensor and Ricci scalar under such transformations.

%%%%%%%%%%%%%%%%%%%%%%%%%%%%%%%%%%%%%%%%%%%%%%%%%%%%%%%%%%%%%%%
%%%%%%%%%%%%%%%%%%%%%%%%%%%%%%%%%%%%%%%%%%%%%%%%%%%%%%%%%%%%%%%
\section{Classification of theories}\label{Classification}
In this section we will classify all gravity theories of the form \req{Smassc2} according to the properties of their physical modes. Depending on the values of the parameters $a$, $b$, $c$ and $e$, we will distinguish six different situations. One of them corresponds to the general case we already studied, $0< |m_g|^2<+\infty$, $0\leq |m_s|^2<+\infty$, and then there are five special cases: 1) theories without massive gravitons, \ie those for which the additional spin-2 mode is absent but the spin-0 one is dynamical; 2) theories without dynamical scalar, \ie those for which the additional graviton is dynamical but the spin-0 mode is absent; 3) theories with two massless gravitons and a massive scalar, \ie those for which the extra graviton is massless --- a property which to some extent cures its problematic behavior; 4) generalized \emph{critical} gravities \ie those which belong to the previous category and, in addition, have no additional spin-0 mode; 5) and finally, Einstein-like theories, \ie theories for which the only mode is the usual massless graviton.\footnote{In principle, one could also impose more exotic conditions like $\kappa_{\rm eff}=0$, which would remove all propagating modes, see \eg \cite{Fan:2016zfs}.} A summary of the different cases can be found in Table \ref{tablaa} and various examples of particular theories belonging to each class are provided in Appendix \ref{Classificationexamples}. Let us note in passing that boundary conditions can be sometimes used to remove spurious modes from the spectrum of certain higher-order gravities --- see \cite{Maldacena:2011mk}. We shall not discuss this issue here.  
Finally, let us also mention that related analyses were previously performed in the absence of matter in \cite{PabloPablo,Tekin1,Tekin2}.
\begin{table}[hpt] 
\begin{center}
\hspace*{-1cm}
\begin{tabular}{|c|c|c|c|}
\hline
 &$m_g^2=0$&$0<m_g^2 <+\infty$&$m_g^2=+\infty$\\ \hline
 $0\leq m_s^2<+\infty$& Massless gravitons + scalar & General case & No massive graviton\\
\hline
$m_s^2=+\infty$& Critical  & No dynamical scalar & Einstein-like  \\
\hline
\end{tabular}
\hspace*{-1cm}
\caption{Classification of theories according to their spectrum on a msb.}
\label{tablaa}
\end{center}
\end{table}

\subsection{Theories without massive graviton}\label{noghost}
The ghost-like massive spin-2 mode $t_{\mu\nu}^{(M)}$ found in the previous section can be removed from the linearized spectrum of the theory by imposing $m_g^2=+\infty$.  In terms of the parameters characterizing a given higher-derivative theory as described in Sec.~\ref{section2}, such condition will be satisfied whenever

\begin{equation}\label{2ac}
2a+c=0\, .
\end{equation}
When this condition holds, the linearized equations \req{lineareq2s} become
\begin{equation}
\begin{aligned}
\mathcal{E}_{\mu\nu}^{L}=\frac{1}{2\kappa_{\rm eff}}&\left\{ G_{\mu\nu}^{ L}+\left[\frac{(D-2)}{2(D-1)(m_s^2+D\mathcal{K})}\right]\left[(D-1)\mathcal{K} \bar g_{\mu\nu} +\bar g_{\mu\nu}\bar\Box-\bar\nabla_{\mu}\bar\nabla_{\nu}\right]R^{ L}\right\}\, .
\end{aligned}
\label{lineareq33s}
\end{equation}
Observe that \req{2ac} has the effect of making the $\bar \Box G_{\mu\nu}^{ L}$ term --- responsible for the appearance of the extra spin-2 graviton --- disappear. As a consequence, even though \req{lineareq33s} still contains quartic derivatives of $h_{\mu\nu}$, the equations do become second-order when we choose the transverse gauge $\bar \nabla^{\mu}h_{\mu\nu}=\bar \nabla_{\nu}h$, as it can be immediately checked from \req{lineareq33s} using \req{ricciscalar} --- or alternatively from \req{traceless} taking the limit $m_g^2\rightarrow +\infty$ there.

On AdS/dS backgrounds --- the extension to Minkowski is straightforward --- \req{2ac} imposes $t_{\mu\nu}^{(M)}=0$, so the metric decomposition becomes now 
%This condition is satisfied, for example, by $f( R )$ theories. Le us simplify the previous formulas in this case. Now we have the following metric decomposition
\begin{equation}
h_{\mu\nu}=t_{\mu\nu}^{(m)}+\frac{\bar\nabla_{\langle\mu}\bar\nabla_{\nu\rangle}h}{(m_s^2+D\mathcal{K})} +\frac{1}{D}\bar g_{\mu\nu} h\, ,
\label{hdecomp22}
\end{equation}
%where $\kappa_{\mbox{{\scriptsize eff}}}$ is given in (\ref{parameters}), and $K$ can be written now in a simpler way:
%\begin{equation}
%\frac{1}{4K}=2\mathcal{K} (D-1)^2(2b-a).
%\end{equation}
where $h$ and $t_{\mu\nu}^{(m)}$ still satisfy \req{trace} and \req{masslesseq} respectively. Observe that using \req{trace} and \req{hdecomp22} along with the transverse gauge condition \req{trans}, it is possible to show that $t_{\mu\nu}^{(m)}$ is transverse in the vacuum,  
%The field equations are
%\begin{equation}
%\begin{aligned}
%-\frac{1}{4K}\left[\bar\Box -m_s^2\right]h&=\frac{1}{2}T,\\
%-(\bar\Box-2\mathcal{K})t^{(m)}_{\mu\nu}&=2\kappa_{\mbox{{\scriptsize eff}}}\left(T_{\langle\mu\nu\rangle}-\frac{1}{\mathcal{K} (D-1)(D-2)}\bar\nabla_{\langle\mu}\bar\nabla_{\nu\rangle}T\right),\\
%\end{aligned}
%\label{tmeq2}
%\end{equation}
%with $m_s^2$ given also in (\ref{parameters}). 
%From the scalar equation and the metric decomposition (\ref{hdecomp2}), and by using the gauge condition $\bar\nabla^{\mu}h_{\mu\nu}=\bar\nabla_{\nu} h$, it can be shown that $t_{\mu\nu}^{(m)}$ is transverse in vacuum:
\begin{equation}\label{tratra}
\bar\nabla^{\mu}t_{\mu\nu}^{(m)}=0\, .
\end{equation}
Notice also that after imposing \req{trans} we still have some gauge freedom, because a gauge transformation $h_{\mu\nu}\rightarrow h_{\mu\nu}+2\bar\nabla_{(\mu}\xi_{\nu)}$ for any vector $\xi_{\mu}$ satisfying $\bar\nabla^{\mu}\bar\nabla_{(\mu}\xi_{\nu)}=\bar\nabla_{\nu}\bar\nabla_{\mu}\xi^{\mu}$ preserves \req{trans}. This allows us to impose additional conditions on $h_{\mu\nu}$. In particular, we can choose
\begin{equation}
t_{0\mu}^{(m)}=t_{\mu 0}^{(m)}=0\,,
\end{equation}
so that only the spatial components $t_{ij}^{(m)}$, $i,j=1,...,D-1$ are non-zero. Then this tensor has $D(D-1)/2$ components, but we have also 
\begin{equation}
\bar\nabla^{i}t_{ij}^{(m)}=0, \quad \bar g^{ij}t_{ij}^{(m)}=0\,,
\end{equation}
which follow from \req{tratra} and the tracelessness of $t_{\mu\nu}^{(m)}$ respectively. These are $(D-1)+1=D$ constraints, so the number of polarizations of $t^{(m)}_{\mu\nu}$ is $D(D-3)/2$, just like for the usual Einstein graviton. Of course, the trace $h$ provides an additional degree of freedom, so these theories propagate $(D-1)(D-2)/2$ physical degrees of freedom in the vacuum.
%\subsubsection*{Examples}\commentt{tbw}
%As we have said, a paradigmatic example of theories without massive gravitons but with a scalar field is $f( R )$ gravity, and more generally, $f($Lovelock$)$. If we restrict to $D=4$ and to quadratic order in curvature, the only theory with a scalar is $R^2$:
%\begin{equation}
%S=\frac{1}{2\kappa}\int_{\mathcal{M}}d^4x\sqrt{|g|}\left\{-2\Lambda+R+\frac{\alpha}{\Lambda}R^2\right\}.
%\end{equation}
%The mass of the scalar is given by $m_s^2=\frac{\Lambda}{6\alpha}$, thus it is never massless. Indeed, if we include the term $R$ in the action, there is no quadratic gravity with a massless scalar, because only pure quadratic theories are scale-invariant. Therefore, if we include the EH term $R$ in the action, one should include cubic terms in curvature to obtain a theory with a massless scalar. 

\subsection{Theories without dynamical scalar}\label{noscalar}
The condition for the absence of the scalar mode is naturally given by $m_s^2=+\infty$. In terms of the parameters $a$, $b$, $c$ and $e$, this reads

\begin{equation}\label{wos}
2a+Dc+4b(D-1)=0\, .
\end{equation}

The linearized equations of motion \req{lineareq2s} become in that case
\begin{equation}
\begin{aligned}
\mathcal{E}_{\mu\nu}^{L}=\frac{1}{2\kappa_{\rm eff}m_g^2}&\left\{\left[m_g^2+2\mathcal{K}-\bar\Box\right] G_{\mu\nu}^{ L}+\frac{(D-2)}{2(D-1)}\left[(D-1)\mathcal{K} \bar g_{\mu\nu} -\bar g_{\mu\nu}\bar\Box+\bar\nabla_{\mu}\bar\nabla_{\nu}\right]R^{ L}\right\}\, .\\
\end{aligned}
\label{lineareq22s}
\end{equation}
The metric decomposition simplifies to
\begin{equation}
h_{\mu\nu}=t_{\mu\nu}^{(m)}+t_{\mu\nu}^{(M)}+\frac{1}{D}\bar g_{\mu\nu} h\, ,
\end{equation}
where the trace of the metric perturbation is simply determined by the matter stress-tensor through the expression   
\begin{equation}
h=\frac{2\kappa_{\rm eff} m_g^2}{(D-1)(D-2)\mathcal{K} (m_g^2-(D-2)\mathcal{K})}T^L\,.
\end{equation}
The massless and massive gravitons satisfy the same equations as in the general case, \ie \req{masslesseq} and \req{MaSeq} respectively.

\subsection{Theories with two massless gravitons}\label{Criticaal gravity}
As we saw, $t^{(M)}_{\mu\nu}$ is a ghost. In order to remove this instability, the simplest solution is to consider theories in which it is absent. Another possibility is to set $m_g=0$, namely, impose its mass to be zero like for the usual graviton. The condition to be satisfied is in this case
\begin{equation}\labell{tetis}
-e+2\mathcal{K}(D-3)a=0\, .
\end{equation}
From \req{kafka} we learn that \req{tetis} also imposes the effective Einstein constant to diverge, $\kappa_{\rm eff}=+\infty$. This inconsistency is artificial and comes from a wrong identification of $\kappa_{\rm eff}$ in this case. In fact, 
 the effective gravitational constant must be defined now as
\begin{equation}\label{kak}
\hat\kappa_{\rm{eff}}\equiv m_g^2\kappa_{\rm{eff}}=-\frac{1}{4(2a+c)}\, ,
\end{equation}
which remains finite when we impose \req{tetis}.
Then, the equation for the trace reads
\begin{equation}
\left[\frac{(D-1)(D-2)^2\mathcal{K}^2}{2\hat\kappa_{\rm eff}(m_s^2+D\mathcal{K})}\right]\left[\bar\Box -m_s^2\right]h=T^L\, .
\end{equation}
On the other hand, we cannot decompose the traceless perturbation $t_{\mu\nu}$ into two independent fields. Instead, it fulfills the equation
\begin{equation}
\frac{1}{2\hat\kappa_{\rm{eff}}}(\bar\Box-2\mathcal{K})^2 t_{\mu\nu}=T_{\langle\mu\nu\rangle}^{L,\rm{eff}}\, ,
\end{equation}
with a metric decomposition given now by
\begin{equation}
h_{\mu\nu}=t_{\mu\nu}+\frac{\bar\nabla_{\langle\mu}\bar\nabla_{\nu\rangle}h}{(m_s^2+D\mathcal{K})}+\frac{1}{D}\bar g_{\mu\nu} h\, .
\end{equation}

\subsection{Critical gravities}\label{Critical gravity}
Critical gravities \cite{Lu} are theories in which the extra graviton is massless and, in addition, the scalar mode is absent, \ie it satisfies $m_s^2=+\infty$. As shown in \cite{Lu} for the quadratic case in $D=4$, the energies of both $t^{(m)}_{\mu\nu}$ and $t^{(M)}_{\mu\nu}$ become zero for this class of theories. We can easily check this statement from the quadratic action \req{equiv}. Specifying for the critical gravity case, it reads
\begin{equation}
S_2=\int_{\mathcal{M}} \frac{d^Dx}{4\hat\kappa_{\rm eff}} \left[\frac{(D-2)^2}{2(D-1)}({R^L})^2-2{G^{L}}^{\mu\nu}G^L_{\mu\nu}\right]\, .
\end{equation}
 Now, in the vacuum the field equations imply that $h=0$, so that $R^L=0$,  and $(\Box-2\mathcal{K})^2 h_{\langle\mu\nu\rangle}=0$. There are solutions, corresponding to the usual massless graviton, which are annihilated by $(\Box-2\mathcal{K})$, and they have $G^L_{\mu\nu}=0$. Therefore, for these solutions the Lagrangian as well as its derivatives vanish on-shell. In particular, the Hamiltonian vanishes, since it is constructed from the Lagrangian and its first derivatives, so the gravitons have zero energy. However, there are additional logarithmic modes which are not annihilated by  $(\Box-2\mathcal{K})$, but by the full operator $(\Box-2\mathcal{K})^2$ instead, and these modes do carry positive energy \cite{Lu}.

The conditions to be imposed for this class of theories are \req{tetis} and \req{wos} as well as the redefinition of the Einstein constant in \req{kak}.
Then, the traceless part of the metric satisfies
\begin{equation}
\frac{1}{2\hat\kappa_{\mbox{{\scriptsize eff}}}}\left[ (\bar\Box-2\mathcal{K})^2h_{\langle\mu\nu\rangle} -\bar\nabla_{\langle\nu}\bar\nabla_{\mu\rangle}\bar \square h\right]=T^L_{\langle\mu\nu\rangle}\,,
\end{equation}
while the trace is determined by matter,
\begin{equation}
h=-\frac{2\hat\kappa_{\mbox{{\scriptsize eff}}}}{(D-1)(D-2)^2\mathcal{K}^2}T^L\, .
\end{equation}

\subsection{Einstein-like theories}\label{eee}
When both the massive graviton and the scalar mode are absent, we are left with a theory whose only propagating degree of freedom is a massless graviton. The conditions $m_g^2=m_s^2=+\infty$ can be expressed as

\begin{equation}\label{eq:ELcond}
2a+c=4b+c=0\, .
\end{equation}
The linearized equations of motion drastically simplify and become identical to those of Einstein gravity with an effective Einstein constant, 

\begin{equation}
\begin{aligned}
\mathcal{E}_{\mu\nu}^{L}=\frac{1}{2\kappa_{\rm eff}}G_{\mu\nu}^{ L}=\frac{1}{2}T^L_{\mu\nu}\, .\\
\end{aligned}
\label{lineareq223s}
\end{equation}
The metric decomposition is very simple now,

\begin{equation}
h_{\mu\nu}=t_{\mu\nu}^{(m)}+\frac{1}{D}\bar g_{\mu\nu} h\, , 
\end{equation}
with $t_{\mu\nu}^{(m)}$ satisfying \req{masslesseq}, and $h$ being again completely determined by matter,

\begin{equation}
h=\frac{2\kappa_{\mbox{{\scriptsize eff}}}}{\mathcal{K} (D-1)(D-2)}T^L\, .
\end{equation}
Hence, according to the discussion in \ref{noghost}, the only propagating mode is the transverse and traceless part of the metric perturbation, which carries $D(D-3)/2$ dof, like in Einstein gravity. 
The only theory-dependent parameter in the linearized equations of Einstein-like theories is the effective Einstein constant $\kappa_{\rm eff}$. Once we know that our theory is Einstein-like, we can derive a direct relation that allows us to obtain this constant directly from the on-shell Lagrangian $\mathcal{L}(\mathcal{K})$. In order to see this, let us note the following relations,
\begin{align}
\frac{d\mathcal{L}(\mathcal{K})}{d\mathcal{K}}&=2e D(D-1)\,, \\ 
\frac{d^2\mathcal{L}(\mathcal{K})}{d\mathcal{K}^2}&=4D(D-1)\left(a+b D(D-1)+c (D-1)\right)\, ,
\end{align}
analogous to \req{teta}. Using these equations together with the Einstein-like conditions \req{eq:ELcond}, we determine the four parameters $a, b, c, e$ in terms of these derivatives of the Lagrangian. Then, we only need to insert the values of these constants in \req{kafka}, and we obtain

\begin{equation}\label{eq:kappaEL}
\begin{aligned}
\frac{1}{\kappa_{\rm eff}}&=\frac{2}{D(D-1)}\left(\frac{d\mathcal{L}(\mathcal{K})}{d\mathcal{K}}-\frac{2\mathcal{K}}{(D-2)}\frac{d^2\mathcal{L}(\mathcal{K})}{d\mathcal{K}^2}\right)\\
&=\frac{2}{(D-1)(D-2)}\frac{d}{d\mathcal{K}}\left(\mathcal{L}(\mathcal{K})-\frac{2\mathcal{K}}{D}\frac{d\mathcal{L}(\mathcal{K})}{d\mathcal{K}}\right).
\end{aligned}
\end{equation}
We get that the inverse of $\kappa_{\rm eff}$ appears to be proportional to the slope of the function that determines the vacuum equation \req{Lambda-eq}. This result will be very useful to us in other chapters.

Let us stress at this point that throughout the text we use the labels \emph{Einstein-like} and \emph{Einsteinian} with different meanings. By Einstein-like theories we mean theories for which the extra modes are absent and the only dynamical field at the linearized level is the usual massless graviton of general relativity. By Einsteinian we refer to those Einstein-like theories which are defined in a dimension-independent way --- see Sec.~\ref{sec:ECG}. 

\begin{comment}
\subsubsection*{Example}
There is no non-trivial quadratic theory in four dimensions which possess the same spectrum as GR. Indeed, if one wants a modification of Einstein gravity which preserves the dof, one have to include cubic curvature terms. For example, we will show that the following theory has the same spectrum as GR in every dimension and that it is non-trivial in $D=4$:
\begin{equation}
S=\frac{1}{16\pi G}\int_{\mathcal{M}}d^Dx\sqrt{|g|}\left\{-2\Lambda+R+\beta\mathcal{X}\right\},
\end{equation}
where 
\begin{equation}
\mathcal{X}=12 R_{a\ b}^{\ c \ d}R_{c\ d}^{\ e \ f}R_{e\ f}^{\ a \ b}+R_{ab}^{cd}R_{cd}^{ef}R_{ef}^{ab}-12R_{abcd}R^{ac}R^{bd}+8R_{a}^{b}R_{b}^{c}R_{c}^{a}
\end{equation}
\end{comment}

\section{Explicit linearization of some theories}\label{quartic}
So far, our analysis was completely general and systematic, without making reference to any concrete theory.  In this section we apply the linearization method explained in Sec.~\ref{section2} to some relevant cases that will be useful to us.

\subsection{Linearization of all theories up to quartic order}
Up to quartic order in curvature, the most general $D$-dimensional theory of the form \req{Smassc2} can be written as

\begin{align}
S=\frac{1}{2\kappa}\int_{\mathcal{M}}d^Dx\sqrt{|g|}\bigg\{&-2\Lambda+R+\sum_{i=1}^{3}\alpha_i \mathcal{L}_i^{(2)}
+\sum_{i=1}^{8}\beta_i\mathcal{L}_i^{(3)}+\sum_{i=1}^{26}\gamma_i\mathcal{L}_i^{(4)}\bigg\}.
\label{quarticaction}
\end{align}
Here, $\mathcal{L}_i^{(2)}$, $\mathcal{L}_i^{(3)}$ and $\mathcal{L}_i^{(4)}$ represent, respectively, the quadratic, cubic and quartic curvature invariants enumerated in Table \ref{tabla2}, $\alpha_i$, $\beta_i$ and $\gamma_i$ are dimensionful constants and $\kappa=8\pi G$ is again Einstein's constant. Also, $\Lambda$ is the cosmological constant. In general dimensions, there are three independent quadratic invariants, eight cubic and twenty-six quartic \cite{0264-9381-9-5-003}. However, not all of these invariants are linearly independent as we consider small enough $D$. For example, in $D=4$ there are only two quadratic, six cubic and thirteen quartic invariants. 

Using the procedure explained in Sec.~\ref{section2} we have linearized the quartic action \req{quarticaction}, \ie we have computed the quantity $\mathcal{L}(\mathcal{K},\alpha)$ defined in (\ref{Ldefinition}) at order $\mathcal{O}(\alpha^2)$ for every term in the action and obtained the values of $a$, $b$, $c$ and $e$ from there.%\footnote{Note that from this table one can find easily $\mathcal{L}(\mathcal{K},\alpha)$ for any other higher-order invariant formed as a product of invariants appearing on the table. The corresponding on-shell Lagrangian is simply the product. This way, one can easily linearize this kind of higher-order terms.}.
 The results are shown in Table \ref{tabla2}. Finally, the parameters $a$, $b$, $c$ and $e$ of the full theory \req{quarticaction} can be found by adding linearly the contribution of each term, with the corresponding coefficients in front in each case, namely
 \begin{align}
 e&=\frac{1}{2\kappa}\left(e[R]+\sum_{i=1}^{3}\alpha_i\, e\left[\mathcal{L}_i^{(2)}\right] +\sum_{i=1}^{8}\beta_i \, e\left[\mathcal{L}_i^{(3)}\right]+\sum_{i=1}^{26}\gamma_i \, e\left[\mathcal{L}_i^{(4)}\right]\right)  \, ,
 \end{align}
 where \eg $e[R]=1/2$ is the value of $e$ corresponding to the Einstein-Hilbert term $R$, and so on.  Completely analogous expressions hold for $a$, $b$ and $c$.
 
Table \ref{tabla2} along with the results in Sec.~\ref{Classification} can be easily used to classify the different theories in \req{quarticaction} according to their spectrum.

\begingroup
\everymath{\footnotesize}
\footnotesize

\begin{table}[!htp] 
\begin{center}
\hspace*{-1cm}
\scalebox{0.9}{
\begin{tabular}{|c|c|c|c|c|c|c|}
\hline
Label&Term&$e$&$a$&$b$&$c$\\ \hline
\hline
$\mathcal{L}^{(1)}_{1}$&$R$&$\frac{1}{2}$&0&0&0\\
\hline \hline
$\mathcal{L}^{(2)}_{1}$&$R^2$&$D(D-1)\mathcal{K}$&$0$&$\frac{1}{2}$&$0$ \\ \hline
$\mathcal{L}^{(2)}_{2}$&$R_{\mu\nu}R^{\mu\nu}$&$(D-1)\mathcal{K}$&$0$&$0$&$\frac{1}{2}$\\ \hline
$\mathcal{L}^{(2)}_{3}$&$R_{\mu\nu\rho\sigma}R^{\mu\nu\rho\sigma}$&$2\mathcal{K}$&$1$&0&0\\ \hline
\hline
$\mathcal{L}^{(3)}_{1}$&$R_{\mu\ \nu}^{\ \rho \ \sigma}R_{\rho\ \sigma}^{\ \delta \ \gamma}R_{\delta\ \gamma}^{\ \mu \ \nu}$&$\frac{3}{2}(D-2)\mathcal{K}^2$&$-\frac{3}{2}\mathcal{K}$&$0$&$\frac{3}{2}\mathcal{K}$\\ \hline
$\mathcal{L}^{(3)}_{2}$&$R_{\mu\nu }^{\ \ \rho\sigma }R_{\rho\sigma }^{\ \ \delta\gamma }R_{\delta\gamma }^{\ \ \mu\nu }$&$6\mathcal{K}^2$&$6\mathcal{K}$&$0$&$0$\\ \hline
$\mathcal{L}^{(3)}_{3}$&$R_{\mu\nu\rho\sigma }R^{\mu\nu\rho }_{\ \ \ \delta}R^{\sigma \delta}$&$3(D-1)\mathcal{K}^2$&$(D-1)\mathcal{K}$&$0$&$2\mathcal{K}$\\ \hline
$\mathcal{L}^{(3)}_{4}$&$R_{\mu\nu\rho\sigma }R^{\mu\nu\rho\sigma }R$&$3D(D-1)\mathcal{K}^2$&$D(D-1)\mathcal{K}$&$2\mathcal{K}$&$0$\\ \hline
$\mathcal{L}^{(3)}_{5}$&$R_{\mu\nu\rho\sigma }R^{\mu\rho}R^{\nu\sigma}$&$\frac{3}{2}(D-1)^2\mathcal{K}^2$&$0$&$\frac{1}{2}\mathcal{K}$&$\frac{1}{2}(2D-3)\mathcal{K}$\\ \hline
$\mathcal{L}^{(3)}_{6}$&$R_{\mu}^{\ \nu}R_{\nu}^{\ \rho}R_{\rho}^{\ \mu}$&$\frac{3}{2}(D-1)^2\mathcal{K}^2$&$0$&$0$&$\frac{3}{2}(D-1)\mathcal{K}$\\ \hline
$\mathcal{L}^{(3)}_{7}$&$R_{\mu\nu }R^{\mu\nu }R$&$\frac{3}{2}D(D-1)^2\mathcal{K}^2$&$0$&$(D-1)\mathcal{K}$&$\frac{1}{2}D(D-1)\mathcal{K}$\\ \hline
$\mathcal{L}^{(3)}_{8}$&$R^3$&$\frac{3}{2}D^2(D-1)^2\mathcal{K}^2$&$0$&$\frac{3}{2}D(D-1)\mathcal{K}$&$0$\\
\hline \hline
$\mathcal{L}^{(4)}_{1}$&$R^{\mu\nu\rho\sigma }R_{\mu \ \rho}^{\ \delta\ \gamma}R_{\delta\ \nu}^{\ \chi\ \xi}R_{\gamma \chi \sigma \xi}$&$2(3D-5)\mathcal{K}^3$&$2(D-4)\mathcal{K}^2$&$0$&$7\mathcal{K}^2$\\ \hline
$\mathcal{L}^{(4)}_{2}$&$R^{\mu\nu\rho\sigma }R_{\mu\ \rho}^{\ \delta\ \gamma}R_{\delta\ \gamma}^{\ \chi\ \xi}R_{\nu \chi \sigma \xi}$&$2(D^2-3D+4)\mathcal{K}^3$&$6\mathcal{K}^2$&$\mathcal{K}^2$&$2(D-3)\mathcal{K}^2$\\ \hline
$\mathcal{L}^{(4)}_{3}$&$R^{\mu\nu\rho\sigma }R_{\mu\nu }^{\ \ \delta\gamma }R_{\rho\ \delta}^{\ \chi \ \xi}R_{\sigma \chi \gamma \xi}$&$4(D-2)\mathcal{K}^3$&$(D-7)\mathcal{K}^2$&$0$&$5\mathcal{K}^2$\\ \hline
$\mathcal{L}^{(4)}_{4}$&$R^{\mu\nu\rho\sigma }R_{\mu\nu }^{\ \ \delta\gamma }R_{\rho\delta}^{\ \ \chi\xi}R_{\sigma \gamma \chi \xi}$&$8\mathcal{K}^3$&$12\mathcal{K}^2$&$0$&$0$\\ \hline
$\mathcal{L}^{(4)}_{5}$&$R^{\mu\nu\rho\sigma }R_{\mu\nu }^{\ \ \delta\gamma }R_{\delta\gamma }^{\ \ \chi\xi}R_{\rho\sigma \chi\xi}$&$16\mathcal{K}^3$&$24\mathcal{K}^2$&$0$&$0$\\ \hline
$\mathcal{L}^{(4)}_{6}$&$R^{\mu\nu\rho\sigma }R_{\mu\nu\rho }^{\ \ \ \ \delta}R_{\gamma \xi \chi \sigma}R^{\gamma \xi \chi}_{\ \ \ \ \delta}$&$8(D-1)\mathcal{K}^3$&$4(D-1)\mathcal{K}^2$&$0$&$8\mathcal{K}^2$\\ \hline
$\mathcal{L}^{(4)}_{7}$&$(R_{\mu\nu\rho\sigma }R^{\mu\nu\rho\sigma })^2$&$8D(D-1)\mathcal{K}^3$&$4D(D-1)\mathcal{K}^2$&$8\mathcal{K}^2$&$0$\\ \hline
$\mathcal{L}^{(4)}_{8}$&$R^{\mu\nu }R^{\rho\sigma \delta\gamma }R_{\rho\ \delta\mu}^{\ \xi}R_{\sigma \xi \gamma \nu}$&$2(D-1)(D-2)\mathcal{K}^3$&$-\frac{3}{2}(D-1)\mathcal{K}^2$&$\frac{1}{2}\mathcal{K}^2$&$\frac{1}{2}(5D-9)\mathcal{K}^2$\\ \hline
$\mathcal{L}^{(4)}_{9}$&$R^{\mu\nu }R^{\rho\sigma \delta\gamma }R_{\rho\sigma \ \mu}^{\ \ \ \xi}R_{\delta\gamma \xi\nu}$&$8(D-1)\mathcal{K}^3$&$6(D-1)\mathcal{K}^2$&$0$&$6\mathcal{K}^2$\\ \hline
$\mathcal{L}^{(4)}_{10}$&$R^{\mu\nu }R_{\mu\ \nu}^{\ \rho\ \sigma}R_{\delta\gamma \xi \rho}R^{\delta\gamma \xi}_{\ \ \ \sigma}$&$4(D-1)^2\mathcal{K}^3$&$(D-1)^2\mathcal{K}^2$&$2\mathcal{K}^2$&$(3D-5)\mathcal{K}^2$\\ \hline
$\mathcal{L}^{(4)}_{11}$&$RR_{\mu\ \nu}^{\ \rho \ \sigma}R_{\rho\ \sigma}^{\ \delta \ \gamma}R_{\delta\ \gamma}^{\ \mu \ \nu}$&$2D(D-1)(D-2)\mathcal{K}^3$&$-\frac{3}{2}D(D-1)\mathcal{K}^2$&$\frac{3}{2}(D-2)\mathcal{K}^2$&$\frac{3}{2}D(D-1)\mathcal{K}^2$\\ \hline
$\mathcal{L}^{(4)}_{12}$&$RR_{\mu\nu }^{\ \ \rho\sigma }R_{\rho\sigma }^{\ \ \delta\gamma }R_{\delta\gamma }^{\ \ \mu\nu }$&$8D(D-1)\mathcal{K}^3$&$6D(D-1)\mathcal{K}^2$&$6\mathcal{K}^2$&$0$\\ \hline
$\mathcal{L}^{(4)}_{13}$&$R^{\mu\nu }R^{\rho\sigma }R^{\delta\ \gamma}_{\ \mu\ \rho}R_{\delta \nu \gamma \sigma}$&$4(D-1)^2\mathcal{K}^3$&$(D-1)^2\mathcal{K}^2$&$\frac{1}{2}\mathcal{K}^2$&$\frac{1}{2}(9D-10)\mathcal{K}^2$\\ \hline
$\mathcal{L}^{(4)}_{14}$&$R^{\mu\nu }R^{\rho\sigma }R^{\delta\ \gamma}_{\ \mu\ \nu}R_{\delta \rho \gamma \sigma}$&$2(D-1)^3\mathcal{K}^3$&$0$&$\frac{1}{2}(3D-4)\mathcal{K}^2$&$\frac{1}{2}(3D^2-8D+6)\mathcal{K}^2$\\ \hline
$\mathcal{L}^{(4)}_{15}$&$R^{\mu\nu }R^{\rho\sigma }R^{\delta\gamma }_{\ \ \mu\rho}R_{\delta\gamma \nu \sigma}$&$4(D-1)^2\mathcal{K}^3$&$(D-1)^2\mathcal{K}^2$&$\mathcal{K}^2$&$(4D-5)\mathcal{K}^2$\\ \hline
$\mathcal{L}^{(4)}_{16}$&$R^{\mu\nu }R_{\nu}^{\ \rho}R^{\sigma \delta\gamma }_{\ \ \ \mu}R_{\sigma\delta\gamma \rho}$&$4(D-1)^2\mathcal{K}^3$&$(D-1)^2\mathcal{K}^2$&$0$&$5(D-1)\mathcal{K}^2$\\ \hline
$\mathcal{L}^{(4)}_{17}$&$R_{\delta\gamma }R^{\delta\gamma }R_{\mu\nu\rho\sigma }R^{\mu\nu\rho\sigma }$&$4D(D-1)^2\mathcal{K}^3$&$D(D-1)^2\mathcal{K}^2$&$4(D-1)\mathcal{K}^2$&$D(D-1)\mathcal{K}^2$\\ \hline
$\mathcal{L}^{(4)}_{18}$&$RR_{\mu\nu\rho\sigma }R^{\mu\nu\rho }_{\ \ \ \delta}R^{\sigma\delta}$&$4D(D-1)^2\mathcal{K}^3$&$D(D-1)^2\mathcal{K}^2$&$3(D-1)\mathcal{K}^2$&$2D(D-1)\mathcal{K}^2$\\ \hline
$\mathcal{L}^{(4)}_{19}$&$R^2R_{\mu\nu\rho\sigma }R^{\mu\nu\rho\sigma }$&$4D^2(D-1)^2\mathcal{K}^3$&$D^2(D-1)^2\mathcal{K}^2$&$5D(D-1)\mathcal{K}^2$&$0$\\ \hline
$\mathcal{L}^{(4)}_{20}$&$R^{\mu\nu }R_{\mu\rho\nu\sigma}R^{\delta\rho}R_{\delta}^{\ \sigma}$&$2(D-1)^3\mathcal{K}^3$&$0$&$(D-1)\mathcal{K}^2$&$(D-1)(2D-3)\mathcal{K}^2$\\ \hline
$\mathcal{L}^{(4)}_{21}$&$RR_{\mu\nu\rho\sigma }R^{\mu\rho}R^{\nu\sigma}$&$2D(D-1)^3\mathcal{K}^3$&$0$&$\frac{1}{2}(D-1)(4D-3)\mathcal{K}^2$&$\frac{1}{2}D(D-1)(2D-3)\mathcal{K}^2$\\ \hline
$\mathcal{L}^{(4)}_{22}$&$R_{\mu}^{\ \nu}R_{\nu}^{\ \rho}R_{\rho}^{\ \sigma}R_{\sigma}^{\ \mu}$&$2(D-1)^3\mathcal{K}^3$&$0$&$0$&$3(D-1)^2\mathcal{K}^2$\\ \hline
$\mathcal{L}^{(4)}_{23}$&$(R_{\mu\nu }R^{\mu\nu })^2$&$2D(D-1)^3\mathcal{K}^3$&$0$&$2(D-1)^2\mathcal{K}^2$&$D(D-1)^2\mathcal{K}^2$\\ \hline
$\mathcal{L}^{(4)}_{24}$&$RR_{\mu}^{\ \nu}R_{\nu}^{\ \rho}R_{\rho}^{\ \mu}$&$2D(D-1)^3\mathcal{K}^3$&$0$&$\frac{3}{2}(D-1)^2\mathcal{K}^2$&$\frac{3}{2}D(D-1)^2\mathcal{K}^2$\\ \hline
$\mathcal{L}^{(4)}_{25}$&$R^2R_{\mu\nu }R^{\mu\nu }$&$2D^2(D-1)^3\mathcal{K}^3$&$0$&$\frac{5}{2}D(D-1)^2\mathcal{K}^2$&$\frac{1}{2}D^2(D-1)^2\mathcal{K}^2$\\ \hline
$\mathcal{L}^{(4)}_{26}$&$R^4$&$2D^3(D-1)^3\mathcal{K}^3$&$0$&$3D^2(D-1)^2\mathcal{K}^2$&$0$\\ 
\hline
\end{tabular}
}
\hspace*{-1cm}
\caption{Parameters $e$, $a$, $b$, $c$ of the linearized equations for all Riemann curvature invariants up to fourth order. We have cross-checked all the terms independently for $D=3,4,5$ using Mathematica.}
\label{tabla2}
\end{center}
\end{table}
\endgroup

%%%%%%%%%%%%%%%%%%%%
%%%%%%%%%%%%%%%%%%%%%%%
%%%%%%%%%%%%%%%%%%%%
%%%%%%%%%%%%%%%%%%%%%%%
\subsection{$f($scalars$)$ theories}\label{fscalars}
As we can see from Table~\ref{tabla2}, the number of Riemann invariants grows very rapidly with the order in curvature, so working out the most general case becomes unpractical.  However, we can construct at least a subset of higher-order curvature terms by considering densities formed from products of lower-order invariants. Therefore, it would be useful to have a way of determining the contribution to the linearized equations of these ``product densities'' by knowing the contribution of their factors.
More generally, one could consider theories whose Lagrangian is an arbitrary function --- not necessarily a polynomial --- of Riemann invariants, such as $f(R)$ gravity.
Thus, in this subsection we will linearize the equations of motion of a theory of the form

\begin{equation}
\mathcal{L}=f(\mathcal{R}_1,\dots,\mathcal{R}_m)\, ,
\label{scalarfunction0}
\end{equation}  
where the $\mathcal{R}_i$ are arbitrary curvature scalars.
For a theory of this form, using the objects

\begin{equation}
 P_i^{\mu\alpha\beta\nu}\equiv \frac{\partial\mathcal{R}_i}{\partial R_{\mu\alpha\beta\nu}}\, , \quad 
 C_{i\  \sigma\rho\lambda\eta}^{\mu\gamma\sigma\nu}\equiv g_{\sigma\alpha} g_{\rho\beta} g_{\lambda\chi} g_{\eta\xi} \frac{\partial P^{\mu\gamma\sigma\nu}_i}{\partial R_{\alpha\beta\chi\xi}}\, ,
\end{equation}
we get the following result for the tensors defined in \req{Ptensor2} and \req{P-def} evaluated on the background,

\begin{eqnarray}
\bar{P}^{\mu\alpha\beta\nu}=\partial_i f(\bar{\mathcal{R}})\bar{P}_i^{\mu\alpha\beta\nu}\, , \quad
\bar C_{\sigma\rho\lambda\eta}^{\mu\alpha\beta\nu}=\partial_i f(\bar{\mathcal{R}})\bar C_{i \ \sigma\rho\lambda\eta}^{\mu\alpha\beta\nu}+\partial_i \partial_jf(\bar{\mathcal{R}})\bar{P}_i^{\mu\alpha\beta\nu}\bar{P}_{j\ \sigma\rho\lambda\eta}\, ,
\end{eqnarray}
where $\partial_i$ denotes derivative with respect to $\mathcal{R}_i$, and $\bar{\mathcal{R}}$ means that we evaluate all the scalars on the background. Using these expressions it is possible to obtain the values of the parameters $a, b, c$ and $e$ defined in (\ref{abc-def}) and (\ref{e-def}) for the theory \req{scalarfunction0}. The result is

\begin{equation}
\begin{aligned}
a=\partial_i f(\bar{\mathcal{R}}) a_i\, , \quad
b=\partial_i f(\bar{\mathcal{R}}) b_i+\partial_i \partial_jf(\bar{\mathcal{R}})e_ie_j\, , \quad
c=\partial_i f(\bar{\mathcal{R}}) c_i\,  \quad
e=\partial_i f(\bar{\mathcal{R}}) e_i\, .
\label{transfrules0}
\end{aligned}
\end{equation}
Hence, once we have computed the parameters $a_i, b_i, c_i, e_i$ for the set of scalars $\mathcal{R}_i$, we can easily find the corresponding parameters for any other Lagrangian $\mathcal{L}=f(\mathcal{R}_1,\dots,\mathcal{R}_m)$. Plugging the values \req{transfrules0} in \req{lineareqs}, we obtain the linearized equations.

\subsubsection{Theories without massive graviton}
In Sec.~\ref{Classification} we classified general $\mathcal{L}($Riemann$)$ theories according to their spectrum on a msb. One of the cases under consideration was that corresponding to theories for which $m_g^2=+\infty$, \ie those containing a single massless graviton plus an additional spin-0 mode. In terms of the parameters defined in the first section, this condition is $2a+c=0$.
% Observe that this equation depends on the curvature of the background $\mathcal{K}$, since $a=a(\mathcal{K})$, $c=c(\mathcal{K})$. Hence, the condition for the absence of massive gravitons must be satisfied independently of $\mathcal{K}$, which imposes a condition at every order in curvature.
% What we want to remark is that the equation $2a+c=0$ should be satisfied independently of $\mathcal{K}$, and this imposes a condition in every order in curvature. Otherwise, the theory would be unstable: in some vacuum we would not have massive gravitons but in other we would have them. 
Assume now that for certain scalars $\mathcal{R}_i$ the condition $2a_i+c_i=0$ is satisfied for all $i$, so that a theory consisting of a linear combination of  $\mathcal{R}_i$ would be free of massive gravitons. From \req{transfrules0} we learn that in fact, this property is shared by any theory of the form $\mathcal{L}=f(\mathcal{R}_1,\dots,\mathcal{R}_m)$ since in that case we find
\begin{equation}\labell{21a}
2a+c=\partial_i f(\bar{\mathcal{R}})(2a_i+c_i)=0\, .
\end{equation}
Therefore, theories constructed as general functions of scalars whose linear combinations do not produce massive gravitons are also free of those modes. This is a straightforward way of understanding why $f(R)$, or more generally $f($Lovelock$)$ theories --- see Appendix \ref{Classificationexamples} --- inherit the property of Lovelock gravities \cite{Lovelock1,Lovelock2} of not propagating the massive graviton \cite{Bueno2,Love}.

\noindent
Furthermore, note   that the condition for the absence of  scalar mode reads in turn
\begin{equation}\labell{21a1}
2a+D c+4b(D-1)= \partial_i f(\bar{\mathcal{R}})(2a_i+D c_i+4b_i(D-1)) +4(D-1)\partial_i\partial_j f(\bar{\mathcal{R}})e_i e_j =0\, .
\end{equation}
This expression is more complicated than \req{21a} since the expression for $b$ in \req{transfrules0} contains a term involving the $e_i$.  This is not surprising: $f(R)$ does propagate the additional scalar mode even though Einstein gravity does not.

%this theory do not have massive gravitons either.  The conclusion is that, if several Lagrangian do not produce massive gravitons, then, a Lagrangian which is a function if the previous do not produce them either. For example, in GR the Lagrangian is $R$, the Ricci scalar, and the only degree of freedom is a massless graviton, so, in particular, there is no massive graviton. Therefore, any theory whose Lagrangian is a function of $R$  --- this is, $f( R )$ gravity ---   does not contain massive gravitons in its spectrum. By the same reasoning, this corollary explains why $f($Lovelock$)$ is free of massive gravitons. However, not all theories which are free of massive gravitons can be built in this way.
%%%%%%%%%%%%%%%%%%%%
%%%%%%%%%%%%%%%%%%%%%%%

%\commentt{

%$f($Riemann$)$ gravities are a wide kind of theories with the only common property that the Lagrangian is a scalar formed with the Riemann tensor and the metric. It turns out that, in any given dimension $D$, there always exists a minimum set of polynomial Riemann scalars , $\{\mathcal{R}_1,\mathcal{R}_2,..,\mathcal{R}_m\}$, such that any scalar can be expressed as $\mathcal{L}=f(\mathcal{R}_1,..,\mathcal{R}_m)$ 
%where $f$ is a differentiable function of $m$ variables. This way, these $\mathcal{R}_i$ are the building blocks of any $f($Riemann$)$ theory. For example, in $D=4$, the minimum set of scalars with the property above contains $m=17$ scalars (although not all of them are independent).}

\section{Einsteinian cubic gravity}\label{sec:ECG}
We have seen in Sec.~\ref{eee} that there is a special class of theories --- that we named \emph{Einstein-like}  theories --- whose linearized equations on constant curvature backgrounds coincide with the linearized Einstein's equations, up to the identification of the effective Newton's constant. Thus, at linear level these theories behave as Einstein gravity and their only 
degree of freedom is a massless graviton. Since Einstein-like theories will play a prominent role in this thesis, in this section we take a close look at them. We will introduce a refined set of Einstein-like theories that we will call \emph{Einsteinian} theories, whose main property is that they are defined in a dimension-independent way. 

We recall that the theories that only propagate a massless graviton on the vacuum are those for which the additional modes are infinitely heavy, \ie those for which $m_g^2=m_s^2=+\infty$. In terms of the parameters of the linearized equations, these conditions translate into $2a+c=4b+c=0$. Let us then explicitly construct these theories at leading order in the curvature expansion. Obviously, Einstein-gravity belongs to this class, and the following terms that we can add to the gravitational action are quadratic in the curvature:
\begin{equation}
\mathcal{L}^{(2)}=\alpha_{1} R^2+\alpha_{2} R_{\mu\nu}R^{\mu\nu}+\alpha_{3} R_{\mu\nu\rho\sigma}R^{\mu\nu\rho\sigma}\, .
\end{equation}
Now, from Table~\ref{tabla2} we can read the value of the constants $a, b, c$ associated to this Lagrangian: $a=\alpha_3$, $b=\alpha_1/2$, $c=\alpha_2/2$. Then the conditions $2a+c=4b+c=0$ impose $\alpha_1=\alpha_3=-\alpha_2/4$, and the only quadratic theory whose spectrum coincides with that of Einstein gravity is $\mathcal{L}^{(2)}=\alpha_3\mathcal{X}_4$, where $\mathcal{X}_4=R^2-4R_{\mu\nu}R^{\mu\nu}+R_{\mu\nu\rho\sigma}R^{\mu\nu\rho\sigma}$ is the Gauss-Bonnet density. This term belongs to the class of Lovelock theories, which are the most general higher-curvature theories whose fully non-linear equations of motion are of second-order.  However, $\mathcal{X}_4$ is topological in four dimensions, hence there are no dynamical Einstein-like theories at quadratic order in $D=4$.

Let us then work out the cubic case. The most general cubic Lagrangian can be written as
\begin{align}\label{eq:cubicLag}
\mathcal{L}^{(3)}&=\beta_1\tensor{R}{_{\mu}^{\rho}_{\nu}^{\sigma}}\tensor{R}{_{\rho}^{\alpha}_{\sigma}^{\beta}}\tensor{R}{_{\alpha}^{\mu}_{\beta}^{\nu}}+\beta_2 \tensor{R}{_{\mu\nu}^{\rho\sigma}}\tensor{R}{_{\rho\sigma}^{\alpha\beta}}\tensor{R}{_{\alpha\beta}^{\mu\nu}}+\beta_3 \tensor{R}{_{\mu \nu\rho\sigma}}\tensor{R}{^{\mu \nu\rho}_{\alpha}}R^{\sigma \alpha}\\
&+\beta_4\tensor{R}{_{\mu\nu\rho\sigma}}\tensor{R}{^{\mu\nu\rho\sigma}}R+\beta_5\tensor{R}{_{\mu\nu\rho\sigma}}\tensor{R}{^{\mu\rho}}\tensor{R}{^{\nu\sigma}}+\beta_6\tensor{R}{_{\mu}^{\nu}}\tensor{R}{_{\nu}^{\rho}}\tensor{R}{_{\rho}^{\mu}}+\beta_7R_{\mu\nu }R^{\mu\nu }R+\beta_8R^3\, .\nonumber
\end{align}
From Table~\ref{tabla2} we read the parameters of the linearized equations and we can write the two Einstein-like constraints:

\begin{align}\label{eq:cubicEL1}
\mathcal{K}^{-1}(2a+c)=&-\frac{3}{2} \beta _1+12 \beta _2+2 D \beta _3+2 D(D-1) \beta
   _4+\frac{1}{2} (2D-3) \beta _5\\ \nonumber
   &+\frac{3}{2}(D-1) \beta _6+\frac{1}{2}D(D-1)\beta _7=0\, ,\\ \label{eq:cubicEL2}
\mathcal{K}^{-1}(4b+c)= & \frac{3}{2} \beta _1+2 \beta _3+8 \beta _4+\frac{1}{2} (2D+1) \beta _5\\ \nonumber
&+\frac{3}{2}(D-1) \beta _6+\frac{1}{2} \left(D^2+7 D-8\right) \beta _7+6 D(D-1)\beta _8=0\, .
\end{align}

These constraints would leave us with a six-parameter family of cubic theories whose spectrum is identical to that of Einstein gravity. However, not all of them are dynamical depending on the dimension. In $D=4$ there are two cubic combinations that are identically zero, and they belong trivially to the Einstein-like class. Thus, in four dimensions there are only four non-trivial, linearly independent Einstein-like cubic Lagrangians. In $D=5$ we find five non-trivial theories of this kind, and in $D>6$ there are six of them. The case $D=6$ is special because all the six independent Einstein-like terms are non-vanishing, but there is a topological combination that corresponds to the six-dimensional Euler density $\mathcal{X}_6$ --- see Eq.~\req{eq:cubicLove} below. 

Observe however that in most cases, these theories have dimension-dependent couplings $\beta_i$, \ie the relative coefficients of the different curvature invariants change with the spacetime dimension. In other words, they are actually \emph{different} theories in different dimensions. This is the case, for example, of Quasi-topological gravity \cite{Quasi2,Quasi} and of certain $f($Lovelock$)$ theories \cite{Love,Karasu:2016ifk}. On the other hand, there are certain theories that satisfy the Einstein-like condition in all dimensions, while keeping the form of the Lagrangian independent from $D$. The most prominent example of this type of theories is Lovelock gravity \cite{Lovelock1,Lovelock2}, that is actually the most general higher-curvature theory possessing second-order field equations --- not only at linear level. The Lovelock Lagrangian is constructed as a linear combination of dimensionally extended Euler densities, given by

\begin{equation}
\mathcal{X}_{2k}  \equiv  \frac{1}{2^k} \delta_{\alpha_1 \beta_1 \dots \alpha_k \beta_k}^{\mu_1 \nu_1 \dots \mu_k \nu_k} {R_{\mu_1 \nu_1}}^{\alpha_1 \beta_1} \cdots {R_{\mu_k \nu_k}}^{\alpha_k \beta_k}\, ,
\end{equation}
where the generalized Kronecker symbol is defined as $\delta^{\mu_1 \nu_1 \dots   \mu_k \nu_k}_{\alpha_1 \beta_1 \dots \alpha_k \beta_k} \equiv (2k)! \delta^{[\mu_1}_{\alpha_1} \delta^{\nu_1}_{\beta_1} \cdots \delta^{\mu_k}_{\alpha_k} \delta^{\nu_k]}_{\beta_k}$. As we can see, the form of these densities does not have an explicit dependence on the spacetime dimension $D$. Thus, these are examples of any-dimensional Einstein-like theories. Our aim here is to search for more theories of this kind, that we will call \emph{Einsteinian} theories, on account of their similarities to Einstein gravity. 

As we have seen, the number of Einstein-like theories is still quite large --- we have up to six of them at cubic order ---  so imposing the additional \emph{Einsteinian} condition seems to be a nice way to reduce the number of theories, providing us with a subset of theories with possibly more interesting properties. 
One advantage of these theories is that their spectrum is Einstein-like not only on constant curvature backgrounds, but also on other vacua that are product of constant curvature spaces. These are the type of vacua that appear in String Theory, where one usually has $\mathcal{M}^{D}=\mathcal{M}_{\rm nc}^{D-n}\times \mathcal{M}_{\rm c}^{n}$, where $\mathcal{M}_{\rm c}^{n}$ is some compact manifold of dimension $n$. For instance, let us consider a $D$-dimensional theory and let us assume that it possesses vacua of the form AdS$_{D-n}\times\mathbb{S}^{n}$, possibly for several values of $n$. If the gravitational sector of the $D$-dimensional theory is of the Einsteinian type, then we can be sure that the spectrum of the gravitational sector in the lower-dimensional AdS$_{D-n}$ factor will only contain a massless graviton, for any value of $n$. The reason is that when one compactifies the theory, the resulting effective action on the non-compact dimensions will involve the same gravitational term. Thus, this term must belong to the Einsteinian class if we want the lower-dimensional theory to have also Einstein-like spectrum.  This is exactly what happens with the Einstein-Hilbert term in general String Theory compactifications.\footnote{For example, the $10$-dimensional type-IIA String Theory  effective action reduces to a class of $D=4$, $\mathcal{N}=2$ Supergravity theories when $6$ of the dimensions are compact on a Calabi-Yau threefold --- see \eg \cite{Mohaupt:2000mj}. In the type-IIA action, the leading contribution from the metric is the $10$-dimensional Einstein-Hilbert term $R^{(10)}$. Under compactification, this produces $R^{(4)}$ --- plus additional terms involving other fields. }

Let us find the Einsteinian theories at cubic order in the curvature. When we demand the parameters $\beta_i$ to solve \req{eq:cubicEL1} and \req{eq:cubicEL2} simultaneously for any $D$, the number of constraints is six instead of two. This is so because both $2a+c$ and $4b+c$ are polynomials of order $D^2$, so we need to impose that the coefficients of the terms proportional to $D^0$, $D$ and $D^2$ in \req{eq:cubicEL1} and \req{eq:cubicEL2} vanish independently. This leaves us with a two-parameter family of theories. In particular, we find
\begin{align}
&\beta_1=12\lambda-8\beta,\, \beta_2=\lambda+4\beta,\, \beta_3=-24\beta, \, \beta_4=3\beta,\\ \notag
&\beta_5=-12\lambda+24\beta,\, \beta_6=8\lambda+16\beta, \, \beta_7=-12\beta\, ,\, \beta_8=\beta\, ,
\end{align}
where $\lambda$ and $\beta$ are the free parameters. Now we have the freedom to choose a basis of two cubic invariants satisfying the above constraints. The first element is somewhat canonical, and corresponds to the dimensionally-extended Euler density $\mathcal{X}_6$, which one finds for $\lambda=0$, $\beta=1$:
\begin{equation}\label{eq:cubicLove}
\begin{aligned}
\mathcal{X}_6=&-8\tensor{R}{_{\mu}^{\rho}_{\nu}^{\sigma}}\tensor{R}{_{\rho}^{\alpha}_{\sigma}^{\beta}}\tensor{R}{_{\alpha}^{\mu}_{\beta}^{\nu}}+4\tensor{R}{_{\mu\nu}^{\rho\sigma}}\tensor{R}{_{\rho\sigma}^{\alpha\beta}}\tensor{R}{_{\alpha\beta}^{\mu\nu}}-24\tensor{R}{_{\mu \nu\rho\sigma}}\tensor{R}{^{\mu \nu\rho}_{\alpha}}R^{\sigma \alpha}+3\tensor{R}{_{\mu\nu\rho\sigma}}\tensor{R}{^{\mu\nu\rho\sigma}}R\\
&+24\tensor{R}{_{\mu\nu\rho\sigma}}\tensor{R}{^{\mu\rho}}\tensor{R}{^{\nu\sigma}}+
16\tensor{R}{_{\mu}^{\nu}}\tensor{R}{_{\nu}^{\rho}}\tensor{R}{_{\rho}^{\mu}}-12R_{\mu\nu }R^{\mu\nu }R+R^3,
\end{aligned}
\end{equation}
 Any other choice produces another invariant. A particularly simple one corresponds to setting $\lambda=1$, $\beta=0$, for which we get the following cubic term:
 \begin{align}\label{ECG}
\mathcal{P}&=12 \tensor{R}{_{\mu}^{\rho}_{\nu}^{\sigma}}\tensor{R}{_{\rho}^{\alpha}_{\sigma}^{\beta}}\tensor{R}{_{\alpha}^{\mu}_{\beta}^{\nu}}+\tensor{R}{_{\mu\nu}^{\rho\sigma}}\tensor{R}{_{\rho\sigma}^{\alpha\beta}}\tensor{R}{_{\alpha\beta}^{\mu\nu}}-12R_{\mu\nu\rho\sigma}R^{\mu\rho}R^{\nu\sigma}+8\tensor{R}{_{\mu}^{\nu}}\tensor{R}{_{\nu}^{\rho}}\tensor{R}{_{\rho}^{\mu}}\, .
\end{align} 
From now, we will refer to the term $\mathcal{P}$ as the \emph{Einsteinian cubic gravity} (ECG) density. 
Hence, we find that, up to cubic order in the curvature, the most general theory that possesses Einstein-like spectrum in any dimension is
 \begin{equation}
\begin{aligned}
S=\frac{1}{16\pi G}\int_{\mathcal{M}}d^{D}x\sqrt{|g|}\left\{-2\Lambda+R+\alpha \mathcal{X}_4 + \beta \mathcal{X}_6 + \lambda \mathcal{P} \right\}\, .
\end{aligned}
\label{quartiii}
\end{equation}
It is quite remarkable that, besides the Einstein-Hilbert and Lovelock terms, there is only one additional contribution $\mathcal{P}$.  Let us stop here a moment to write down some properties of this theory. First, from \req{Lambda-eq} we can write the equation for the curvature of the vacuum, $\mathcal{K}$: 
\be\label{eq:vaceqECG}
\mathcal{K}+2(D-3)(D-4)\alpha \mathcal{K}^2 +2(D-3)(D-6)\left[(D-4)(D-5)\beta-4\lambda\right]\mathcal{K}^3=\frac{2\Lambda}{(D-1)(D-2)}\, .
\ee
Depending on the values of the parameters this cubic equation for $\mathcal{K}$ has one or three real solutions,\footnote{Or two or none if the cubic couplings vanish.} but it is customary to choose the one that is smoothly connected to the Einstein gravity vacuum when the couplings vanish. 
The linearized equations of \req{quartiii} around any of the vacua are almost identical to those of Einstein gravity, 
 \begin{equation}
\begin{aligned}
\mathcal{E}_{\mu\nu}^{L}&=\frac{1}{16\pi G_{\rm eff}}G_{\mu\nu}^{ L}\, ,
\end{aligned}
\label{lineareq44s}
\end{equation}
the only signature of the higher-derivative theories being an effective Newton's constant given by 
\be\label{eq:GeffECG}
\frac{G}{G_{\rm eff}}=1+4(D-3)(D-4)\alpha \mathcal{K} +6(D-3)(D-6)\left[(D-4)(D-5)\beta-4\lambda\right]\mathcal{K}^2\, .
\ee
  
We observe that, while the quadratic and cubic Lovelock densities are non-dynamical in $D=4$ (the former is topological and the latter is trivial), the new density $\mathcal{P}$ is non-trivial, since it contributes to the vacuum equation and to the effective Newton's constant. In this sense, up to cubic order in curvature $\mathcal{P}$ is the only Einsteinian density that is non-trivial in four dimensions besides the Einstein-Hilbert term itself.

%%%%%%%%%%%%%%%%%%%
%%%%%%%%%%%%%%%%%%%%%
\subsubsection{Einsteinian quartic gravities}\label{EQG1}
%%%%%%%%%%%%%%%%%

We can go on and try to find all Einsteinian theories at higher orders in curvature. With the results from Sec.~\ref{quartic} we can perform the analysis up to quartic order in the curvature. Imposing the absence of massive graviton and scalar for the quartic terms in the theory \req{quarticaction}, one is left with two constraints for the coupling values, $F_g^{(4)}(\gamma_i,D)=F_s^{(4)}(\gamma_i,D)=0$ --- see Appendix \ref{appEQG} for the explicit expressions.  Imposing each constraint to be satisfied independently of the dimension multiplies the number of constraints and gives rise to the Einsteinian theories. As we have seen, this is because \eg in the cubic case $F^{(3)}_{g,s}(\beta_i,D)$ is a polynomial of degree $2$ in $D$, so we need to impose the coefficients of the $D^0$, $D^1$ and $D^2$ terms to vanish independently. More generally, at $n$-th order in curvature, the corresponding constraints are polynomials of degree $2n-4$ in $D$, and hence we will find $2n-3$ constraints coming from the absence of the massive graviton, and the same number from imposing the absence of scalar, which makes $2(2n-3)$ in total.
 At the quartic level this means $10$ constraints. Since in general dimensions there are up to $26$ independent invariants at this order in curvature \cite{0264-9381-9-5-003} --- see Table \ref{tabla2}, this means that there exists a $16$-parameter family of Einsteinian quartic gravities (EQGs). If we choose the 16 parameters to be $\{\gamma_1,\gamma_2,\gamma_3,\gamma_4,\gamma_5,\gamma_6,\gamma_7,\gamma_8, \gamma_9,\gamma_{10},\gamma_{12}, \gamma_{13},\gamma_{14},\gamma_{18},\gamma_{20},\gamma_{26}\}$, the rest of couplings are given in terms of these as
 \begin{align}\label{ggf}  
 \gamma_{11}=&+\frac{1}{3} (12 \gamma_{12} -4 \gamma_1 +  12 \gamma_2 - 8 \gamma_3 + 36 \gamma_4 + 72 \gamma_5 + 16 \gamma_6 + 16 \gamma_7 - 
   3 \gamma_8 + 12 \gamma_9)\, ,\\ \notag
   \gamma_{15}=&+\frac{1}{2} (-10  \gamma_1 - 4  \gamma_{10} -  \gamma_{13}+  \gamma_{14} + 16  \gamma_2 - 14  \gamma_3 + 48  \gamma_4 + 96  \gamma_5 + 
   16  \gamma_6 - 4  \gamma_8 + 12  \gamma_9)\, ,\\ \notag
   \gamma_{16}=&+\frac{1}{10} (36 \gamma_1 + 10 \gamma_{10} - 24 \gamma_{12} - 5 \gamma_{13} - 5 \gamma_{14} - 74 \gamma_2 - 2 \gamma_{20} + 
   1140 \gamma_{26} + 57 \gamma_3 - 210 \gamma_4 \\ \notag &- 420 \gamma_5 - 84 \gamma_6 - 20 \gamma_7 + 17 \gamma_8 - 72 \gamma_9)\, , \\ \notag
    \gamma_{17}=&- \gamma_{18 }- 120  \gamma_{26}\, , \\ \notag
     \gamma_{19}=&+6\gamma_{26}\, , \\ \notag
     \gamma_{21}=&+8 \gamma_1 - 12 \gamma_{12} - 3 \gamma_{14} + 2 \gamma_{18} - 18 \gamma_2 - 2 \gamma_{20} + 900 \gamma_{26} + 13 \gamma_3 - 
 54 \gamma_4 - 108 \gamma_5 - 20 \gamma_6\\ \notag & - 20 \gamma_7 + 3 \gamma_8 - 12 \gamma_9\, ,\\ \notag
 \gamma_{22}=&+\frac{1}{10} (16 \gamma_1 - 24 \gamma_{12} - 10 \gamma_{14} - 14 \gamma_{2} - 2 \gamma_{20} + 1140 \gamma_{26} + 17 \gamma_3 - 
   50 \gamma_4 - 100 \gamma_5 - 4 \gamma_6 \\ \notag&- 20 \gamma_7 + 2 \gamma_8 + 8 \gamma_9)\, ,\\ \notag
   \gamma_{23}=&+\frac{1}{20} (-154 \gamma_1 + 216 \gamma_{12} + 60 \gamma_{14} - 40 \gamma_{18} + 306 \gamma_{2} + 38 \gamma_{20} - 
   22260 \gamma_{26} - 233 \gamma_3\\ \notag& + 930 \gamma_4 + 1860 \gamma_5 + 316 \gamma_6 + 340 \gamma_7 - 48 \gamma_8 + 
   168 \gamma_9)\, ,\\ \notag
   \gamma_{24}=&+\frac{1}{30} (-6 \gamma_1 + 24 \gamma_{12} + 54 \gamma_2 + 2 \gamma_{20} + 9060 \gamma_{26} - 27 \gamma_3 + 150 \gamma_4 + 
   300 \gamma_5 + 84 \gamma_6 \\   \notag&+ 60 \gamma_7 - 12 \gamma_8 + 72 \gamma_9)\, ,\\ \label{ggf}
   \gamma_{25}=&-24\gamma_{26}\, .
 \end{align}
 Plugging these back in the original quartic action, we obtain the family of $16$  Einsteinian quartic gravities. All of these theories will be linearly independent for large enough $D$, but in lower dimensions many of them will be trivial.
In four dimensions, it can be seen that only $13$ of the $26$ invariants in Table \ref{tabla2} are linearly independent \cite{0264-9381-9-5-003}. We can use this fact to easily construct three Einsteinian quartic gravities that are non-trivial in four dimensions. In particular, we can set $\gamma_{1}=\gamma_2=\gamma_3=\gamma_4=\gamma_6=\gamma_8=\gamma_9=\gamma_{10}=\gamma_{12}=\gamma_{13}=\gamma_{14}=\gamma_{18}=\gamma_{20}=0$ --- the choice being non-unique --- that guarantees that the rest of the densities are independent. Then, \req{ggf} becomes now
\begin{equation}
 \begin{aligned}\label{ggfe}
 \gamma_{11}=&+8/3(9\gamma_5 + 2\gamma_7 )\, , &  \gamma_{15}=&+48  \gamma_5 \, , \\  
   \gamma_{16}=&+114 \gamma_{26} -42\gamma_5-2\gamma_7\, , &
    \gamma_{17}=&-120  \gamma_{26}\, , \\  
     \gamma_{19}=&+6\gamma_{26}\, , &
     \gamma_{21}=&+4 (225 \gamma_{26} -27\gamma_5-5\gamma_7)\, ,\\  
 \gamma_{22}=&+2(57\gamma_{26}-5\gamma_5-\gamma_7)\, , &
   \gamma_{23}=&-1113\gamma_{26}+93\gamma_5+17\gamma_7\, ,\\  
   \gamma_{24}=&+2(151\gamma_{26}+5\gamma_5+\gamma_7)\, , &
   \gamma_{25}=&-24\gamma_{26}\, .
 \end{aligned}
 \end{equation}
where we only have three free parameters $\{ \gamma_5,\gamma_7,\gamma_{26}\}$. Using these relations we have constructed the following invariants 
\begin{equation}
\begin{aligned}\label{QQ}  
\mathcal{Q}_1\equiv& +3R^{\mu\nu\rho\sigma}R_{\mu\nu}^{\gamma\delta}R_{\gamma\delta}^{\alpha\beta}R_{\rho\sigma\alpha\beta}-15 (R_{\mu\nu\rho\sigma}R^{\mu\nu\rho\sigma})^2-8 R R_{\mu \ \ \nu }^{\ \ \rho \ \ \sigma} R_{\rho \ \ \sigma }^{\ \ \gamma \ \ \delta}R_{\gamma \ \ \delta }^{\ \ \mu\ \ \nu}\\   &+144 R^{\mu\nu}R^{\rho\sigma}R^{\gamma\delta}_{\mu\rho}R_{\gamma\delta\nu\sigma}-96 R^{\mu\nu}R_{\nu}^{\rho}R^{\alpha\beta\gamma}_{\ \ \ \ \mu}R_{\alpha\beta\gamma\rho}-24R R_{\mu\nu\rho\sigma}R^{\mu\rho}R^{\nu\sigma}\\   &+24(R_{\mu\nu}R^{\mu\nu})^2
\, , \\
\mathcal{Q}_2\equiv& +3(R_{\mu\nu\rho\sigma}R^{\mu\nu\rho\sigma})^2+16R R_{\mu \ \ \nu }^{\ \ \rho \ \ \sigma} R_{\rho \ \ \sigma }^{\ \ \gamma \ \ \delta}R_{\gamma \ \ \delta }^{\ \ \mu\ \ \nu}-6 R^{\mu\nu}R_{\nu}^{\rho}R^{\alpha\beta\gamma}_{\ \ \ \ \mu}R_{\alpha\beta\gamma\rho}\\   &-60 R R_{\mu\nu\rho\sigma} R^{\mu\rho}R^{\nu\sigma}-6R_{\mu}^{\nu}R_{\nu}^{\rho}R_{\rho}^{\sigma}R_{\sigma}^{\mu}+51 (R_{\mu\nu}R^{\mu\nu})^2+6 R R_{\mu}^{\nu}R_{\nu}^{\rho} R_{\rho}^{\mu}\, ,\\
\mathcal{Q}_3\equiv&+R^4+57(R_{\mu\nu\rho\sigma}R^{\mu\nu\rho\sigma})^2-120 R_{\gamma\delta}R^{\gamma\delta}R_{\mu\nu\rho\sigma}R^{\mu\nu\rho\sigma}+6R^2R_{\mu\nu\rho\sigma}R^{\mu\nu\rho\sigma}\\   
&-240 R R_{\mu\nu\rho\sigma}R^{\mu\rho}R^{\nu\sigma} -144(R_{\mu\nu}R^{\mu\nu})^2+416 R R_\mu^\nu R_\nu^\rho R_\rho^\mu-24 R^2 R_{\mu\nu}R^{\mu\nu}\\   & +304RR_{\mu\ \nu}^{\ \rho \ \sigma}R_{\rho\ \sigma}^{\ \delta \ \gamma}R_{\delta\ \gamma}^{\ \mu \ \nu}
\, .
\end{aligned}
\end{equation}
Just like its cubic cousin $\mathcal{P}$ defined in \req{ECG}, $\mathcal{Q}_1$, $\mathcal{Q}_2$ and $\mathcal{Q}_3$ --- or any linear combination of them --- only propagate the usual massless graviton when linearized on a msb, not only in $D=4$, but in any number of dimensions.\footnote{We have cross-checked the linearized equations of $\mathcal{P}$ and $\mathcal{Q}_i$, $i=1,2,3$ for $D=4,5,6$ using the Mathematica package xAct \cite{xact}.}

It is important to note that these three are not necessarily the only EQG theories in $D=4$. As we explained, there are 13 independent cubic invariants in that case, which means that there are $11$ independent four-dimensional quartic Einstein-like invariants --- because we have to impose two conditions on the couplings in that case, namely $m_g^2=m_s^2=+\infty$. In order to determine all the possible theories, one should construct the $16$ independent $D$-dimensional EQGs using \req{ggf} and then analyze how many of them are independent when $D=4$. Given that EQGs are particular cases of Einstein-like theories, we conclude that there could actually be up to $8$ additional EQG invariants that are non-trivial in $D=4$.

\section{Solutions to the linearized equations}\label{sec:sollin}
After having determined the general form of the linearized equations of any $\mathcal{L}$(Riemann) theory, in this section we  solve them in few interesting situations.  We first investigate the Newtonian limit of these theories by computing the metric of a gravitating point particle in arbitrary dimensions, finding in all cases that the divergence of the Newtonian potential at $r=0$ is smoothened with respect to the EG case. We also study the polarization modes of gravitational waves in these theories as well as the gravitational radiation from sources.

For simplicity, we will focus in the case of Minkowski vacuum, which was analyzed in \ref{Flat space}. It is easy to see that, in flat backgrounds, only terms up to quadratic order in the curvature contribute to the linearized equations. Thus, we are essentially considering the theory

\begin{equation}\label{eq:QuadAction}
\begin{aligned}
S_2=\frac{1}{16\pi G}\int_{\mathcal{M}}d^Dx\sqrt{|g|}\Bigg[&R+ \alpha_1R^2+\alpha_2 R_{\mu\nu}R^{\mu\nu}+\alpha_3 \mathcal{X}_{4}\Bigg]\, ,
\end{aligned}
\end{equation}
where $\mathcal{X}_{4}=R^2-4R_{\mu\nu}R^{\mu\nu}+R_{\mu\nu\rho\sigma}R^{\mu\nu\rho\sigma}$ is the Gauss-Bonnet density. For this theory, $m_g^2$ and $m_s^2$ read, respectively

\begin{align}\label{eq:MassQuadAction}
m_g^2&=  -\frac{1}{2\alpha _2} \, , \quad  m_s^2 = \frac{(D-2)}{2\left(4 \alpha _1 (D-1)+\alpha _2 D \right)}  \, .
\end{align}
while $G_{\rm eff}=G$. The same study that we perform in this section for flat space can be naturally carried out for an (A)dS background using the expressions in Sec.~\ref{AdS space}. Our results are also applicable in that case provided we consider distances shorter than the (A)dS scale $r<<|\mathcal{K}|^{-1/2}$ and that  $m_g^2>> |\mathcal{K}|$.\footnote{We must also assume this because the limit $m_g^2\rightarrow 0$ produces a qualitatively different theory, as we explained in Sec.~\ref{Criticaal gravity}, and the effects become relevant when $m_g^2< |\mathcal{K}|$} The main difference in the (A)dS case is that terms of all orders contribute to $m_{g}^2$ and $m_{s}^2$, so they are not simply given by \req{eq:MassQuadAction}, and we also have $G_{\rm eff}\neq G$.  Thus, we will let these parameters to be arbitrary, with the only assumption that $m_s^2\ge 0$ and $m_g^2>0$ so that we avoid the presence of tachyons.

\subsection{Generalized Newton potential}\labell{GeneralizedNewton}
In this section we use the results of Sec.~\ref{section2} to compute the Newton potential $U_D(r)$ and the \emph{Parametrized Post-Newtonian} (PPN) parameter $\gamma$ for a general theory of the form \req{Smassc2} in general dimensions. We start reviewing the four-dimensional case and then we extend our results to arbitrary $D$, pointing out interesting differences with respect to the $D=4$ case. 

\subsubsection{Four dimensions}
The analysis performed in Sec.~\ref{Flat space} tells us that in order to obtain a solution of the linearized equations in a flat background we must solve equations \req{flat1}, \req{flat2} and \req{flat3}, and then reconstruct the metric perturbation \req{flatdecomp}. If we denote by $H_{\mu\nu}(m)$ a field that satisfies the Klein-Gordon equation with source

\begin{equation}\labell{kgg}
\left(\bar \Box-m^2\right) H_{\mu\nu}(m)=-4\pi T_{\mu\nu}(x)\, ,
\end{equation}
and by $H(m)$ its trace, the solutions to \req{flat1}, \req{flat2} and \req{flat3} can be written as

\begin{eqnarray}
\label{solh}
\hat h_{\mu\nu}&=&4G_{\rm eff}H_{\mu\nu}(0)\,, \,\,
\quad \phi=4G_{\rm eff}H(m_s)\, , \,\,\\
\nonumber
&&\\
t_{\mu\nu}&=&-4G_{\rm eff}\left[H_{\langle\mu\nu\rangle}(m_g)+\frac{1}{3m_g^2}\partial_{\langle\mu}\partial_{\nu\rangle}H(m_g)\right]\, .
\end{eqnarray}
where we introduced the effective Newton's constant $\kappa_{\rm eff}=8\pi G_{\rm eff}$. Inserting this into the metric perturbation (\ref{flatdecomp}) we get
\begin{equation}
\begin{aligned}
h_{\mu\nu}&=4G_{\rm eff}\Bigg[H_{\mu\nu}(0)-H_{\mu\nu}(m_g)+\eta_{\mu\nu}\left(-\frac{1}{2}H(0)+\frac{1}{4}H(m_g)+\frac{1}{4}H(m_s)\right)\\
&+\frac{1}{3}\partial_{\langle\mu}\partial_{\nu\rangle}\Big( (m_g^{-2}-m_s^{-2})H(0)+m_s^{-2}H(m_s)-m_g^{-2}H(m_g)\Big)\Bigg]\, .
\end{aligned}
\end{equation}
This is the metric perturbation in the de Donder gauge, but we can make an infinitesimal gauge transformation in order to simplify the expression above. In particular, let us introduce the ``Newtonian gauge'', by performing the following transformation

\begin{equation}
h^N_{\mu\nu}\equiv h_{\mu\nu}-\partial_{(\mu}\xi_{\nu)}\, ,
\end{equation}
where

\begin{equation}
\xi_{\nu}\equiv \frac{1}{3}\partial_{\nu}\left( (m_g^{-2}-m_s^{-2})H(0)+m_s^{-2}H(m_s)-m_g^{-2}H(m_g)\right)\, .
\end{equation}
After some simplifications we obtain

\begin{equation}
\label{newtonianmetric2}
h^N_{\mu\nu}=G_{\rm eff}\Bigg[4H_{\mu\nu}(0)-4H_{\mu\nu}(m_g)+\eta_{\mu\nu}\left(-2H(0)+\frac{4}{3}H(m_g)+\frac{2}{3}H(m_s)\right)\Bigg]\, .
\end{equation}
Now if we restrict ourselves to static configurations, \req{kgg} reduces to the so-called screened Poisson equation, $\left(\triangle-m^2\right) H_{\mu\nu}(\vec{x};m)=-4\pi T_{\mu\nu}(\vec{x})$, whose general solution reads
\begin{equation}
H_{\mu\nu}(\vec{x};m)=\int d^{3}\vec{x}'\frac{T_{\mu\nu}(\vec{x}')}{|\vec{x}-\vec{x}'|}e^{-m|\vec{x}-\vec{x}'|}\, .
\label{staticsol}
\end{equation}
This can be seen as a superposition of functions $1/|\vec{x}-\vec{x}'|$ weighted by the source $T_{\mu\nu}(\vec{x}^{\prime})$ and with an exponential screening controlled by the mass $m$.
Using this we can rewrite \req{newtonianmetric2} as
\begin{equation}
h^N_{\mu\nu}(x)=G_{\rm eff}\int d^3\vec{x}' T_{\alpha\beta}(\vec{x}')\Pi^{\alpha\beta}_{\ \ \mu\nu}(\vec{x}-\vec{x}')\, ,
\end{equation}
where the static propagator reads
\begin{align}
\Pi^{\alpha\beta}_{\ \ \mu\nu}(\vec{x}-\vec{x}')=\frac{1}{|x-x'|}&\left[ 4\delta^{\alpha}_{\ (\mu}\delta^{\beta}_{\ \nu)}\Big(1-e^{-m_g|\vec{x}-\vec{x}'|}\Big)\right. \\ \notag &\left.-2\eta^{\alpha\beta}\eta_{\mu\nu}\Big(1-\frac{2}{3}e^{-m_g|\vec{x}-\vec{x}'|}-\frac{1}{3}e^{-m_s|\vec{x}-\vec{x}'|}\Big)\right]\, .
\end{align}
Now, let us apply the previous expressions to the case of a solid and static sphere of radius $R$ and mass $M$ on a flat background. For this distribution of matter, the only non-vanishing component of the stress-tensor reads
\begin{equation}
T_{00}(r)=\rho( r )=\rho_0\, \theta(R-r)\, ,\quad \text{with} \quad  \rho_0\equiv \frac{M}{4\pi R^3/3}\, ,
\end{equation} 
%Therefore, the trace is $T=-\rho$. 
where $\theta (x)$ is the Heaviside step function. For this configuration the result for $H_{00}(r;m)= -H(r;m)$ in the outer region $r>R$ obtained from \req{staticsol} reads
\begin{equation}
H(r;m)=-f(mR)\frac{M}{r}e^{-mr}\, ,
\end{equation}
where $f(mR)$ is a form factor given by
\begin{equation}
f(mR)=\frac{3}{(mR)^3}\Big[mR\cosh(mR)-\sinh(mR)\Big]\, ,
\end{equation}
which behaves as $f(mR)\approx \frac{3}{2}\frac{1}{(mR)^2}e^{mR}$ if $mR>>1$ and as $f(mR)\approx 1$ in the point-like limit, \ie when $mR<<1$. Finally, inserting these results into the metric $h_{\mu\nu}^N$ in (\ref{newtonianmetric2}) and this in $g^N_{\mu\nu}=\eta_{\mu\nu}+h_{\mu\nu}^N$ we obtain
\begin{equation}
\label{Nmetric}
ds^2_N=-(1+2U( r ))dt^2+(1-2\gamma( r )U( r ))\delta_{ij}dx^idx^j\,,
\end{equation}
where $U( r )$ and $\gamma(r)$ are given by
\begin{align}\label{eq:NeWpotform}
U( r )&=-\frac{G_{{\rm eff}}M}{r}\left[1-\frac{4}{3}f(m_gR)e^{-m_g r}+\frac{1}{3}f(m_sR)e^{-m_s r}\right]\, , \\ 
\gamma( r )&=\frac{3-2f(m_gR)e^{-m_g r}-f(m_sR)e^{-m_s r}}{3-4f(m_gR)e^{-m_g r}+f(m_sR)e^{-m_s r}}\, .
\end{align}
Evaluating these expressions in the point-like limit of the sphere $f(mR)=1$ we finally obtain the generalized Newtonian potential and the PPN parameter $\gamma$
\begin{eqnarray}
\label{NewtonPotential}
U( r )=-\frac{G_{\rm eff}M}{r}\left[1-\frac{4}{3}e^{-m_g r}+\frac{1}{3}e^{-m_s r}\right]\, ,\quad
\gamma( r )=\frac{3-2e^{-m_g r}-e^{-m_s r}}{3-4e^{-m_g r}+e^{-m_s r}}\, .
\label{gamma}
\end{eqnarray}
Let us make some comments about these results. First, observe that the usual Newton potential gets corrected by two Yukawa-like terms controlled by the masses of the two extra modes which can be computed for a given theory through \req{kafka1} and \req{kafka2}. 
The above expression for $U(r)$ has been obtained before using different methods --- see \eg \cite{Stelle:1977ry,Stelle:1976gc,Prue}\footnote{See \eg \cite{Modesto:2014eta,Giacchini:2016xns} for results corresponding to higher-order gravities involving covariant derivatives of the Riemann tensor.}. Note that while the contribution from the scalar has the usual sign for a Yukawa potential, the massive graviton one comes with the opposite sign, which is another manifestation of its ghost nature. Observe also that the whole contribution from the higher-derivative terms appears through $m_g$ and $m_s$, the coefficients $-4/3$ and $1/3$ in front of the exponentials being common to all theories.
\begin{table}[!hpt] 
\begin{center}
\begin{tabular}{|c|c|c|}
\hline
 &$U(r)/G_{\rm eff}$ &$\gamma$ \\
 \hline
 $m_s=m_g=+\infty$ & $-M/r$ & $1$ \\
 \hline
  $m_s =+\infty$, $|m_g r|\ll 1$ &$+M/(3r) $& $-1$ \\
 \hline
  $m_s =0$, $m_g=+\infty$ &$-4M/(3r)$ & $1/2$\\
% \hline
 %  $m_s r\ll1$, $m_g r\ll1$ &$-M(4m_g-m_s)/3+M(4m_g^2-m_s^2)r/6$ & $(2m_g+m_s)/(4m_g-m_s)$\\
 \hline
 $m\equiv m_g=m_s$ &$-M(1-e^{-mr})/r$ & $1$\\
 \hline
\end{tabular}
\caption{Newton's potential and $\gamma(r)$ for various values of the masses of the extra modes.}
\label{tabla12}
\end{center}
\end{table}
 In Table \ref{tabla12} we present the values of $U(r)$ and $\gamma$ for different limiting values of $m_s$ and $m_g$. Naturally, when $m_g,m_s\gg 1$ one is left with the Einsteinian values of the Newton potential and $\gamma$, and the same happens if we go sufficiently far away from $M$ for arbitrary values of the extra mode masses. It is also interesting that the only cases for which the potential is divergent as $r\rightarrow 0$ are those for which at least one of the extra modes is absent, \ie when either $m_s=+\infty$, or $m_g=+\infty$ or both $m_g=m_s=+\infty$.

Indeed, $U(r)$ does not diverge as $r\rightarrow 0$ in the general case. In fact, one finds
\begin{equation}\labell{near04}
U( r )=-G_{\rm eff}M \left[\frac{(4m_g-m_s)}{3}-\frac{(4m_g^2-m_s^2)r}{6}+\mathcal{O}(r^2) \right]\, ,
\end{equation}
which is a negative (positive) constant at $r=0$ when $m_g>m_s/4$ ($m_g<m_s/4$). The potential grows linearly with $r$ at first order for $m_g>m_s/2$ and in that case it is monotonous in the whole range of $r$. When $m_g<m_s/2$ instead, $U(r)$ decreases linearly near $r=0$ and it has a minimum at some intermediate value of $r$. Plots of $U(r)/G_{\rm eff}$ for various values of the masses satisfying the different situations can be found in Fig. \ref{fig1}. 
 \begin{figure}[!hpt]
        \centering
                \includegraphics[width=0.65\textwidth]{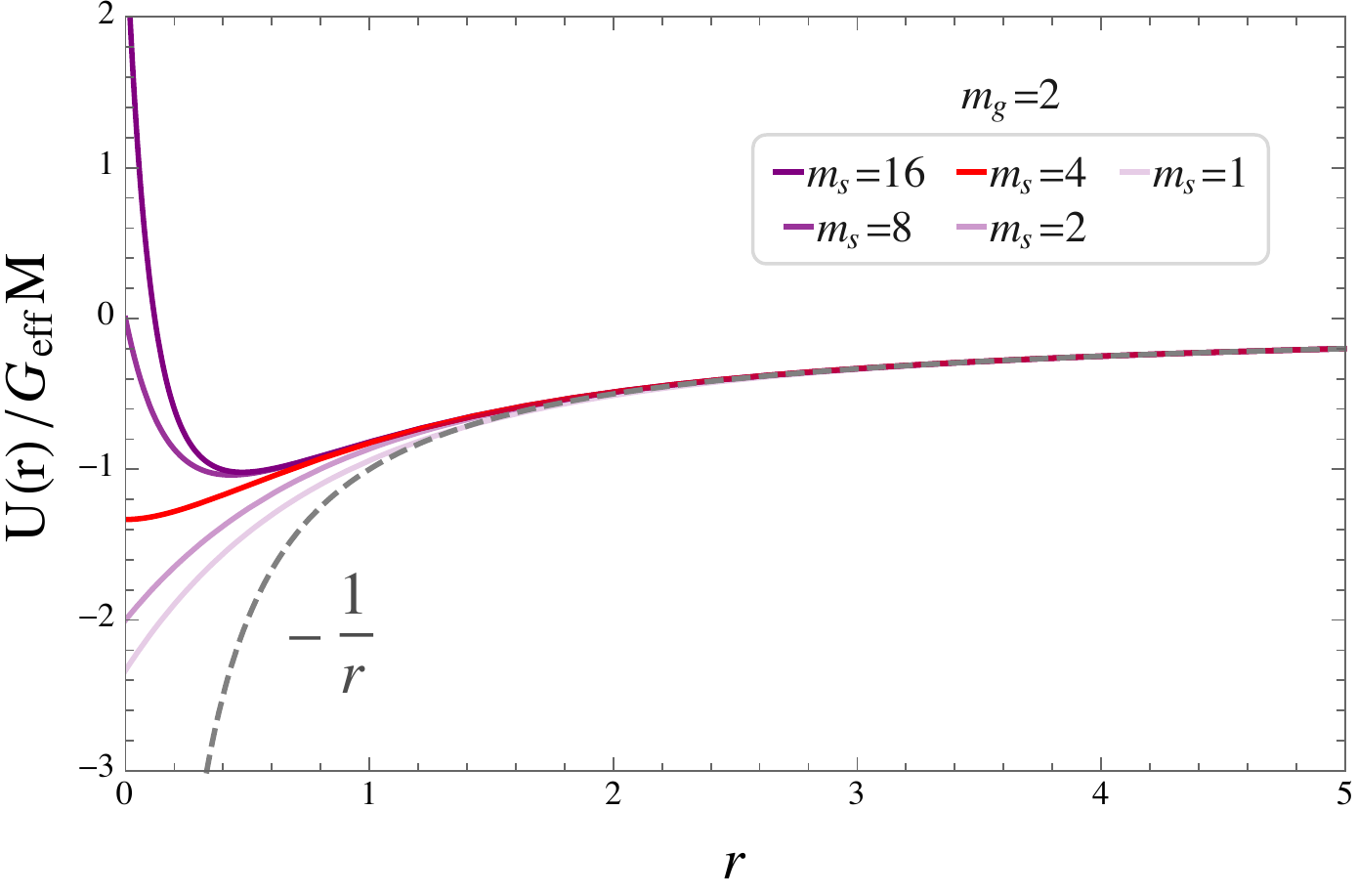}
                \caption{$U(r)/(G_{\rm eff}M)$ for $m_g=2$ and $m_s=16,8,2,1$ (purple curves), and $m_s=4$ (red) and the usual Newton potential (dashed gray).}
\labell{fig1}
\end{figure}

\subsubsection{Higher dimensions}
The analysis of the previous section can be extended to general dimensions $D\geq 4$. The metric perturbation in the Newtonian gauge can be seen to be given by 
\begin{align}\notag
h^N_{\mu\nu}=4G_{\rm{eff}}\Big[&H_{\mu\nu}(0)-H_{\mu\nu}(m_g)\\ &+\frac{\eta_{\mu\nu}}{(D-1)(D-2)}\left(-(D-1)H(0)+(D-2)H(m_g)+H(m_s)\right)\Big]\, ,
\end{align}
where again $H_{\mu\nu}(m)$ is a solution of \req{kgg}. In the static case, we can write the solution explicitly as
\begin{equation}
H_{\mu\nu}(\vec{x};m)=2\left(\frac{m}{2\pi}\right)^{\frac{D-3}{2}}\int d^{D-1}\vec{x}'
\frac{T_{\mu\nu}(\vec{x}')}{|\vec{x}-\vec{x}'|^{\frac{D-3}{2}}}
 K_{\frac{D-3}{2}}(m|\vec{x}-\vec{x}'|) \, ,
\end{equation}
where $K_{\ell}(x)$ is the modified Bessel function of the second kind. 
%In the massless case, this expression is simplified to
%\begin{equation}
%H_{\mu\nu}(\vec{x};m=0)=\frac{\Gamma({\frac{D-3}{2}})}{\pi^{\frac{D-3}{2}}}\int d^{D-1}\vec{x}'\frac{T_{\mu\nu}(\vec{x}')}{|\vec{x}-\vec{x}'|^{D-3}}.
%\end{equation}
Now, specializing to a static point-like particle of mass $M$, we can obtain the $D$-dimensional version of (\ref{Nmetric}). The Newtonian potential and the gamma parameter read, respectively,
\begin{align}\notag
U_D( r )&=-\mu(D) \frac{G_{\rm{eff}} M}{r^{D-3}}\left[1+\nu(D) r^{\frac{D-3}{2}}\left[-m_g^{\frac{D-3}{2}}K_{\frac{D-3}{2}}(m_g r)+\frac{m_s^{\frac{D-3}{2}}}{(D-2)^2}K_{\frac{D-3}{2}}(m_s r) \right]\right]\, , \\
\label{gagaa}
\gamma_D( r )&=\frac{1-\frac{2}{(D-1)\Gamma({\frac{D-3}{2}})}\Big[(D-2)\left(\frac{m_g r}{2}\right)^{\frac{D-3}{2}}K_{\frac{D-3}{2}}(m_g r)
+\left(\frac{m_s r}{2}\right)^{\frac{D-3}{2}}K_{\frac{D-3}{2}}(m_s r)\Big]}{D-3-\frac{2}{(D-1)\Gamma({\frac{D-3}{2}})}\Big[(D-2)^2\left(\frac{m_g r}{2}\right)^{\frac{D-3}{2}}K_{\frac{D-3}{2}}(m_g r)
-\left(\frac{m_s r}{2}\right)^{\frac{D-3}{2}}K_{\frac{D-3}{2}}(m_s r)\Big]}\, ,
\end{align}
with
\begin{equation}
\mu(D) \equiv  \frac{8 \pi}{(D-2) \Omega_{D-2}} \, , \quad \text{and} \quad \nu(D) \equiv\frac{(D-2)^2}{\Gamma\left[{\frac{D+1}{2}}\right]2^{\frac{D-1}{2}}}\, ,
\end{equation}
and where $\Omega_{D-2} \equiv 2 \pi^{\frac{D-1}{2}} /\Gamma[\frac{D-1}{2}]$ is the volume of the $(D-2)$-dimensional unit sphere. When $2\ell$ is odd, \ie for even $D$, the Bessel functions $K_{\ell}(x)$ can be written explicitly in terms of elementary functions as
%\begin{equation}
%K_{1/2}(x)=\sqrt{\frac{\pi}{2x}}e^{-x},\quad K_{3/2}(x)=\sqrt{\frac{\pi}{2x}}e^{-x}(1+1/x),\quad \ldots
%\end{equation}
\begin{equation}
K_{\frac{D-3}{2}}(x)=e^{-x}\sqrt{\frac{\pi}{2x}}\sum_{j=1}^{\frac{D-2}{2}}\frac{(D-3-j)!}{(j-1)!(\frac{D-2}{2}-j)! (2x)^{\frac{D-2}{2}-j}}\,, \quad \text{(even $D$)} 
\end{equation}
which allows for a simplification of \req{gagaa} in those cases, and from which it is easy to reproduce the $D=4$ results \req{NewtonPotential} presented in the previous section.
%This allows us to simplify the previous expressions when $D$ is even. For $D=4$ the result reduces to the one previously found, while, for example, in $D=6$ we obtain
%\begin{equation}
%U_6( r )=-\frac{3G_{\rm{eff}} M}{4\pi r^{3}}\Bigg[1-\frac{16}{15}(1+m_g r)e^{-m_g r}+\frac{1}{15}(1+m_s r)e^{-m_s r}\Bigg],
%\end{equation}
%
%\begin{equation}
%\gamma_6( r )=\frac{5-4(1+m_g r)e^{-m_g r}-(1+m_s r)e^{-m_s r}}{15-16(1+m_g r)e^{-m_g r}+(1+m_s r)e^{-m_s r}}.
%\end{equation}
From \req{gagaa} we infer that the usual four-dimensional Yukawa potential for a force-mediating particle of mass $m$ generalizes to higher dimensions as
\begin{equation}
U_{D,\rm Yukawa}(r)\sim \left( \frac{m}{r}\right)^{\frac{D-3}{2}} K_{\frac{D-3}{2}}(m r)\, .
\end{equation}
Going back to higher-order gravities, observe that close to the origin, the generalized Newton potential $U_D(r)$ behaves for $D>5$ as
\begin{equation}\labell{diviD}
U_D(r\rightarrow 0)\sim -\frac{G_{\rm eff}M\left[(D-2)^2m_g^2-m_s^2\right]}{r^{D-5}}+\dots\, ,
\end{equation}
up to a positive dimension-dependent constant for generic values of $m_g$ and $m_s$. For $D=4$ we find a constant term \req{near04}, while for $D=5$ one finds a logarithmic divergence instead
\begin{equation}
U_5(r\rightarrow 0)=\frac{G_{\rm eff}M}{12\pi}(9m_g^2-m_s^2)\log r+\mathcal{O}(r^0)\, .
\end{equation}
This means that for generic values of the extra mode masses, $U_D(r)$ is divergent at $r=0$ in all dimensions higher than four. In Fig. \ref{fig2} we plot $U_5(r)$, which can be explicitly written as
 \begin{equation}
 U_5 (r) = - \frac{G_{\rm eff} M}{6 \pi r^2} \left[8 - 9 m_g r K_1( m_g r) + m_s r K_1( m_s r)\right]\, .
 \end{equation}
 \begin{figure}[!hpt]
        \centering
                \includegraphics[width=0.65\textwidth]{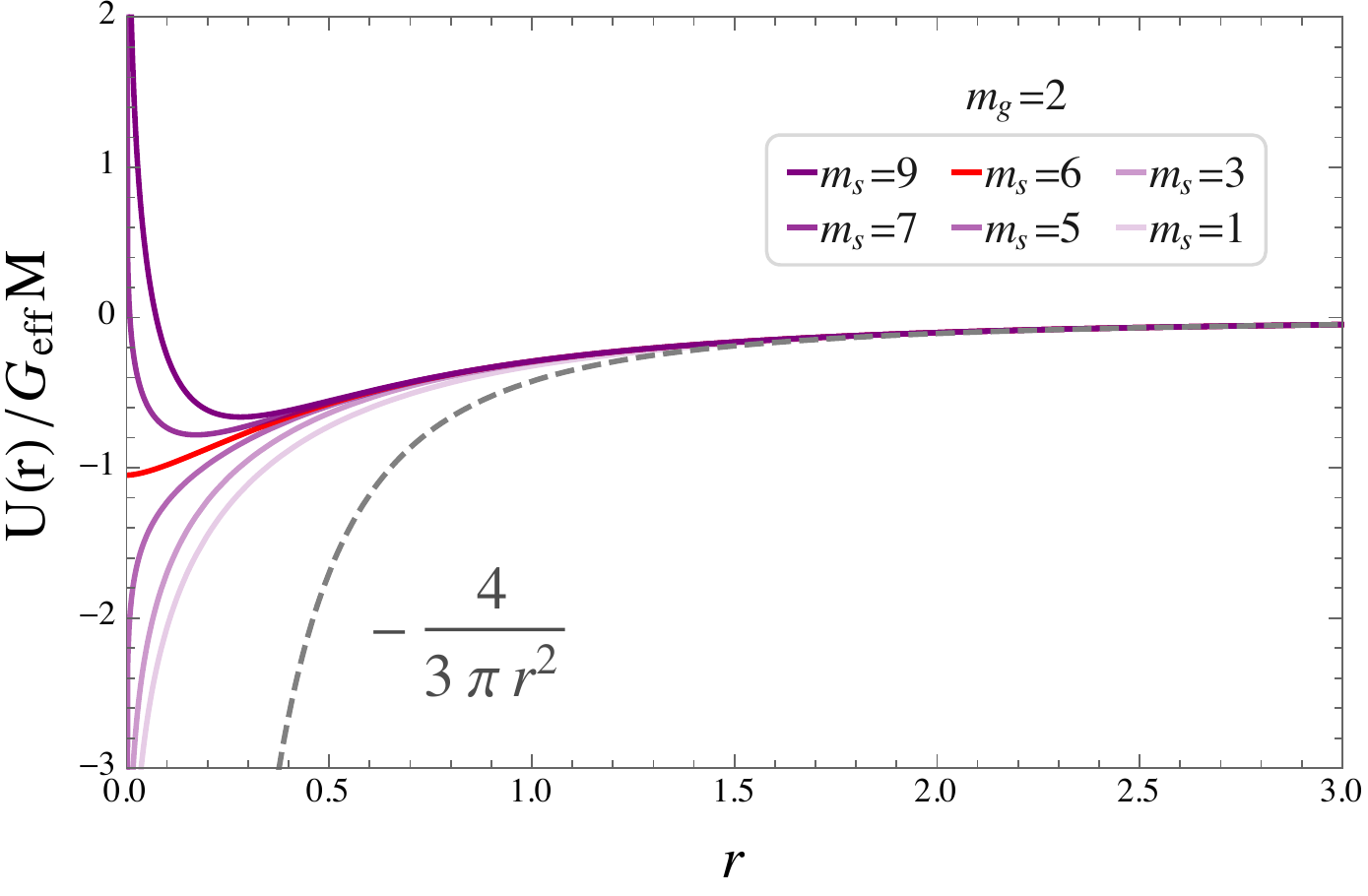}
                \caption{ $U(r)/(G_{\rm eff}M)$ in $D=5$ for $m_g=2$ and $m_s=1,3,5,7,9$ (purple curves), and $m_s=6$ (red) and the usual Newton potential in five dimensions (dashed gray).}
                \labell{fig2}
\end{figure}
As expected, most curves in Fig. \ref{fig2} diverge at the origin. There is an exception (and only one) though, which corresponds to the case $m_g=m_s/3$, for which the potential is finite everywhere. The value $m_g=\frac{m_s}{(D-2)}$ is special in general dimensions, as it determines the transition between two kinds of potentials. In particular, when $m_g>\frac{m_s}{(D-2)}$, $U_D(r)$ is monotonous in the whole range of $r$ and diverges to $-\infty$ at the origin, while for $m_g<\frac{m_s}{(D-2)}$ it has a minimum at some finite value of $r$ and $U_D(r\rightarrow 0)\rightarrow +\infty$ instead --- see Fig. \ref{fig2} for an illustration of these features in the five-dimensional case. For the particular value $m_g=\frac{m_s}{(D-2)}$, the potential is also finite at the origin for $D=6$, but not for $D\geq 7$.

In Table \ref{tabla23} we present some particular cases for $U_D(r)$ and $\gamma_D$\footnote{We use the following two limits of the modified Bessel functions:
\begin{equation}
 \lim_{x\rightarrow \infty} x^\ell K_\ell (x) 
 %= \lim_{x\rightarrow \infty} \left[ \sqrt{\frac{\pi}{2x}} x^{\ell } e^{-x} + \mathcal O(x^{-3/2}) \right]
 = 0 \, ,
 \quad
 \text{and}
 \quad
 \lim_{x\rightarrow 0} x^\ell K_\ell (x) = 2^{\ell-1} \Gamma (\ell)\,.
 %+\mathcal O\left(x^2\right) +x^{2 \ell} \left(2^{-\ell-1} \Gamma (-\ell)+O\left(x^2\right)\right)
\end{equation}} corresponding to different limiting values of $m_g$ and $m_s$.
\begin{table}[!hpt] 
\begin{center}
\begin{tabular}{|c|c|c|}
\hline
 &$U_D(r)/(\mu(D)G_{\rm eff} M)$ &$\gamma_D$ \\
  \hline
 $m_g=m_s=+\infty$ & $-1/r^{D-3}$ & $1/(D-3)$ \\
  \hline
  $m_s =+\infty$, $|m_g r|\ll 1 $ &$+1/ \left[(D-3)(D-1)r^{D-3} \right] $& $-1$ \\
 \hline
  $m_s =0$, $m_g=+\infty$ &$-(D-2)^2 / \left[(D-3)(D-1)r^{D-3} \right]$ & $1/(D-2) $\\
 \hline
 $m\equiv m_g=m_s$ &$-  \left[1-     \frac{(D-3) \Omega_{D-2}}{(2\pi)^{(D-1)/2}} (m r)^{\frac{D-3}{2}}K_{\frac{D-3}{2}}(m r)    \right]/r^{D-3}$ & $1/(D-3)$\\
  \hline
\end{tabular}
\caption{Newton's potential and $\gamma (r)$ in higher dimensions $D\geq 4$ for various values of the masses of the extra modes.}
\label{tabla23}
\end{center}
\end{table}
 Once again, when $m_g,m_s \gg 1$, one is left with the Einsteinian values of the corresponding Newton potentials and $\gamma_D$, and the same happens at sufficiently large distances from $M$ for general values of the extra mode masses. Just like in four dimensions, when the masses of the extra modes are equal, $m_s=m_g$, the gamma parameter coincides with that of Einstein gravity, $\gamma_D=1/(D-3)$. Note also that when one of the modes is absent, the divergence of $U_D(r)$ at $r=0$ becomes stronger than in the generic case \req{diviD} --- namely, of order $1/r^{D-3}$ instead of $1/r^{D-5}$.

%%%%%%%%%%%%%%%%%%%%%%%%%%%
%%%%%%%%%%%%%%%%%%%%%%%%%%%
\subsection{Gravitational waves}\label{gww}
In this section we study some aspects of gravitational waves in $\mathcal{L}($Riemann$)$ theories in flat space. We first obtain plane-wave solutions of the linearized equations in arbitrary dimensions and we determine the number of polarizations of each mode. Then, focusing on four dimensions, we compute the gravitational wave radiation from sources in the far-field and non-relativistic approximations.

\subsubsection{Polarization of gravitational waves}
In the de Donder gauge \req{Donder}, the relevant components of the metric perturbation decomposed as in \req{flatdecomp} satisfy equations \req{flat1}, \req{flat2} and \req{flat3}. In the vacuum, these reduce to
\begin{equation}\labell{eqi}
\bar{\Box} \hat h_{\mu\nu}=0\, ,\quad (\bar{\Box} -m_g^2)t_{\mu\nu}=0\, ,\quad (\bar{\Box} -m_s^2)\phi=0\, .
\end{equation}
Using the tracelessness of $t_{\mu\nu}$, the gauge condition \req{Donder} and equations \req{eqi} along with \req{flatdecomp}, one can show that $\partial^{\mu}t_{\mu\nu}=0$. However, the gauge redundancy has not been completely exploited, as we still have the freedom to make gauge transformations $h_{\mu\nu}\rightarrow h_{\mu\nu}+2\partial_{(\mu}\xi_{\nu)}$ where $\xi_{\mu}$ satisfies $\bar{\Box} \xi_{\mu}=0$.  Let us first note that the transformation rule for the ``massless graviton'' $\hat h_{\mu\nu}$ can be derived from \req{hath}, and it reads
\be
\hat{h}_{\mu\nu}\rightarrow \hat{h}_{\mu\nu}+2\partial_{(\mu}\xi_{\nu)}-\eta_{\mu\nu}\partial_{\alpha}\xi^{\alpha}+\frac{D-2}{2(D-1)}\left(m_{g}^{-2}-m_{s}^{-2}\right)\partial_{\mu}\partial_{\nu}\partial_{\alpha}\xi^{\alpha}
\ee
Then, since both $\hat h_{\mu\nu}$ and the vector $\xi_{\mu}$ are harmonic, we can use the latter to impose $D$ additional conditions on the former. In particular, we can set $\hat h=0$ and $\hat h_{t i}=0$,\footnote{Here $i=1,2,\ldots D-1$ are the spatial indices and the Minkowski metric is $(\eta_{\mu\nu})=\operatorname{diag}(-1,1,\ldots,1)$.} which is called the \emph{traceless-transverse gauge} ($TT$). Observe that we cannot impose similar conditions on $t_{\mu\nu}$ because, being massive, it is not harmonic. Hence, no degrees of freedom in $t_{\mu\nu}$ can be removed with such a gauge transformation and, as a consequence, the massive particles conserve all their polarizations. 

Let us now look for plane-wave solutions of frequency $\omega$,
\begin{equation}\labell{planew}
\hat h^{TT}_{\mu\nu}=A_{\mu\nu}e^{-ik_{\mu} x^{\mu}}\, ,\quad t_{\mu\nu}=B_{\mu\nu}e^{-ip_{\mu} x^{\mu}}\, ,\quad \phi=C e^{-iq_{\mu} x^{\mu}}\, ,
\end{equation}
where $k_{\mu}=(\omega, k_i)$, $p_{\mu}=(\omega, p_i)$, $q_{\mu}=(\omega, q_i)$, and $A_{\mu\mu}$, $B_{\mu\nu}$ and $C$ are constant. 
Equations \req{eqi} produce the following dispersion relations
\begin{equation}
k^2=\omega^2\, , \quad p^2=\omega^2-m_g^2\, , \quad q^2=\omega^2-m_s^2\, ,
\end{equation}
where $k^2=k_i k^{i}$, etc. 
Note that, in order for the massive modes to propagate, the frequency must be greater than the corresponding mass, \ie $\omega^2>m_g^2$ and $\omega^2>m_s^2$ respectively. Otherwise,  the wave will be damped. Now, since we are working in the $TT$ gauge, the polarization tensor $A_{\mu\nu}$  satisfies the following constraints
\begin{equation}
A_{t\mu}=0\, ,\quad k^iA_{ij}=0\, , \quad \tensor{A}{^{i}_{i}}=0\, .
\end{equation}
These conditions leave us only with components transverse to the time direction and transverse to $k$, with the additional requirement of tracelessness. Hence, the number of polarizations corresponds to the number of linearly independent $(D-2)\times (D-2)$ symmetric, traceless matrices; this is, $(D-2)(D-1)/2-1=D(D-3)/2$ polarization modes. In $D=4$ these are the two ``$+$'' and ``$\times$'' polarizations. On the other hand, the polarization tensor $B_{\mu\nu}$ of the massive graviton only satisfies the constraints
\begin{equation}
p^{\mu}B_{\mu\nu}=0\,, \quad \eta^{\mu\nu}B_{\mu\nu}=0\, .
\end{equation}
These imply that the time components are then given by
\begin{equation}
B_{tt}=\tensor{B}{^{i}_{i}}\, ,\quad B_{ti}=\frac{p_j}{\omega}B_{ij}\, .
\end{equation}
and the spatial part of the polarization tensor, $B_{ij}$, only has to satisfy one constraint,
\begin{equation}
p^ip^jB_{ij}=\omega^2 \tensor{B}{^{i}_{i}}\, .
\end{equation}
Thus, there are $(D-2)(D+1)/2$ polarization modes, which include the $D(D-3)/2$ transverse polarizations of the massless graviton plus $D-1$ non-transverse ones. In four dimensions, this reduces to 5 polarizations, which is the correct number for a massive spin-2 particle. 

Finally, we can see in \req{flatdecomp} that we can remove the term $\partial_{\mu}\partial_{\nu}\phi$ by means of another gauge transformation, so that we can write the full metric perturbation for plane waves as 
\begin{equation}
h^{TT}_{\mu\nu}=A_{\mu\nu}e^{-ik_{\mu} x^{\mu}}+B_{\mu\nu}e^{-ip_{\mu} x^{\mu}}+\frac{C}{D(D-1)}\eta_{\mu\nu}e^{-iq_{\mu} x^{\mu}}.
\end{equation}
Since the polarization tensor associated to the scalar mode is pure trace, it is linearly independent from $A_{\mu\nu}$ and $B_{\mu\nu}$.

In sum, gravitational waves in higher-order gravity can propagate up to six different polarizations in four dimensions --- one for the scalar and five for the massive and massless gravitons --- and up to $D(D-1)/2$ polarizations in arbitrary dimension $D$. However, it is important to note that the massive modes do not propagate at lower frequencies, so the accesible polarizations depend on the frequency.

\subsubsection{Gravitational radiation from sources}
Let us now consider a source $T_{\mu\nu}(t,\vec{x})$ concentrated in a region whose diameter is much smaller than the distance $r$ to the observer and which moves at a non-relativistic characteristic speed. Under such approximations
\begin{equation}
|\vec{x}-\vec{x}'|\approx r\, , \quad \left|\frac{d\vec{x}}{dt}\right|\ll 1\, ,
\end{equation}
where $\vec{x}'$ represents the observer's position, the solutions in (\ref{solh}) can be further simplified. In particular, for the massless graviton $\hat h_{\mu\nu}$ one finds
\begin{equation}
\hat h_{\mu\nu}=\frac{4 G_{\rm{eff}}}{r}\int d^3\vec{x}' T_{\mu\nu}(t-r,\vec{x}')\, .
\end{equation}
Our interest here is in the radiative contributions of the solutions, \ie the ones that change with time. For gravitational waves, the time components $\hat h_{\mu 0}$ are determined by the purely space-like ones, so we only need to compute those. The spatial components are radiative in general, and the leading contribution is given by the well-known quadrupole formula
\begin{equation}
\int d^3\vec{x}' T_{ij}(t-r,\vec{x}')=\frac{1}{2}\ddot q_{ij}(t-r)\,,
\end{equation}
where $q_{ij}$ is the quadrupole moment of the source
\begin{equation}
q_{ij}(t-r)=\int d^3\vec{x} x^ix^j\rho(t-r, \vec{x})\,,
\label{quadrupole}
\end{equation}
$\rho$ is the energy density and each dot denotes a time derivative. Therefore, the radiative part of $\hat h_{\mu\nu}$ is given by
\begin{equation}
\hat h_{ij}=\frac{2 G_{\rm{eff}}}{r}\ddot q_{ij}(t-r)\,.
\label{hradiation}
\end{equation}
Obviously, in the case of Einstein gravity --- or for Einstein-like theories --- this is the end of the story. However, in general $\mathcal{L}$(Riemann) theories, we also have to take into account the additional modes.  For the scalar $\phi$ one finds
\begin{equation}
\phi=\frac{4 G_{\rm{eff}}}{r}\int d^3\vec{x}' T(t-r,\vec{x}')-4 G_{\rm{eff}}m_s\int_{r}^{\infty}dt' \frac{J_1(m_s \sqrt{t'^2-r^2})}{\sqrt{t'^2-r^2}}\int d^3\vec{x}' T(t-t',\vec{x}')\, ,
\end{equation}
where $J_1(x)$ is a Bessel function of the first kind.
The integration of the trace yields
\begin{equation}
\int d^3\vec{x}' T(t-r,\vec{x}')=\int d^3\vec{x}' \left(-T_{00}(t-r,\vec{x}')+T_{ii}(t-r,\vec{x}')\right)=-M_0-E(t-r)+\frac{1}{2}\ddot q_{ii}(t-r)\, ,
\end{equation}
where $M_0$ is the rest mass and $E$ is the kinetic energy of the source.
Since the rest mass is constant, it does not source any radiation, and the radiative part of the field is
\begin{equation}
\begin{aligned}
\phi&=\frac{4 G_{\rm{eff}}}{r}\left(\frac{1}{2}\ddot q_{ii}(t-r)-E(t-r)\right)\\
&-4 G_{\rm{eff}}m_s\int_{r}^{\infty}dt' \frac{J_1(m_s \sqrt{t'^2-r^2})}{\sqrt{t'^2-r^2}}\left(\frac{1}{2}\ddot q_{ii}(t-r)-E(t-r)\right)\,.
\end{aligned}
\end{equation}
However, it is important to note that this field does not always radiate. Indeed, if one considers the source to be a set of non-relativistic gravitationally interacting bodies or a pressureless perfect fluid (dust) then one gets $\frac{1}{2}\ddot q_{ii}-E=\text{constant}$.\footnote{The energy-momentum tensor of a pressureless fluid has the form $T_{\mu\nu}=\rho u^{\mu}u^{\nu}$, where $\rho$ is the energy density and $u^{\mu}$ is  the 4-velocity field, satisfying $u^{\mu}u_{\mu}=-1$. Therefore $T=-\rho$ and its integral yields the rest mass of the system. The same argument works for a set of point-like particles. Also, an explicit computation in that case shows that, in the Newtonian approximation, $\frac{1}{2}\ddot q_{ii}-E=E+V$, where $V$ is the gravitational potential energy of the system, and the previous quantity is a constant of motion.} Thus, in situations of interest for gravitational wave emission, such as binary systems, there is no scalar radiation --- at least when the system can be described in the Newtonian approximation. 

Finally, we have to determine the radiative part of $t_{\mu\nu}$. From (\ref{solh}) we can express this field as
\begin{equation}
 t_{\mu\nu}=-H_{\langle\mu\nu\rangle}-\frac{1}{3m_g^2}\partial_{\langle\mu}\partial_{\nu\rangle}H\,,
 \label{tdecomp}
\end{equation}
where $H_{\mu\nu}$ satisfies 
\be
(\bar\Box-m_g^2)H_{\mu\nu}=-4\pi T_{\mu\nu}\, .
\ee
Thus, the purely spacelike components of $H_{\mu\nu}$ for far sources are given by 
\begin{equation}\label{eq:Hrad}
H_{ij}=-\frac{2 G_{\rm{eff}}}{r}\ddot q_{ij}(t-r)+2G_{\rm{eff}}m_g\int_{r}^{\infty}dt'\frac{J_1(m_g \sqrt{t'^2-r^2})}{\sqrt{t'^2-r^2}}\ddot q_{ij}(t-t')\,.
\end{equation}
Moreover, in the vacuum we get $0=\partial_{\mu}t^{\mu\nu}=\partial_{\mu}H^{\mu\nu}$, so this allows us to characterize all the components of $H_{\mu\nu}$ and $t_{\mu\nu}$\footnote{For example, for a plane wave solution we have $p^{\mu}H_{\mu\nu}=0$, so we obtain the time-like components in terms of the purely space-like ones: $H_{0i}=p^j H_{ij}/\omega$, $H_{00}=p^ip^j H_{ij}/\omega^2$. In the general case, the relations that we obtain are not algebraic but differential.}.

Note that the perturbation at a distance $r$ depends on the radiation emitted at all times previous to $t-r$ and not only on the radiation emitted at the time $t-r$. This is related to the fact that the massive graviton and the scalar do not propagate at the speed of light. Indeed, according to the dispersion relation $\omega=\sqrt{m^2_{g,s}+k^2}$, a wave packet with a central frequency $\omega$ will travel at a velocity
\begin{equation}\label{velo}
v_{g,s}=\sqrt{1-\frac{m_{g,s}^2}{\omega^2}}\,.
\end{equation}

By using (\ref{flatdecomp}), (\ref{tdecomp}) and the solutions for $\hat h_{ij}$, $\phi$ and $H_{ij}$ that we have just found, the full metric perturbation can be computed. Assuming that there is no scalar radiation ---  as we argued, this is the natural case --- the metric perturbation can be written as\footnote{We are taking into account in Eq.~\req{flatdecomp} that the traces $H$ and $\hat h$ are not radiative.}
\begin{equation}
h_{ij}=2G_{\rm{eff}}\int_{0}^{\infty}d\tau \frac{J_1(\tau)}{\sqrt{ r^2+\tau^2/m_g^2}}\ddot q_{ij}\left(t-\sqrt{r^2+\tau^2/m_g^2}\right),
\label{quadrupoleformula}
\end{equation}
where we performed a change of variables in the integration in \req{eq:Hrad}. Using this expression, it is easy to see that when $m_g\rightarrow \infty$ we recover the prediction of Einstein gravity $h_{ij}=\frac{2 G_{\rm{eff}}}{r}\ddot q_{ij}(t-r)$. However, the behaviour is quite different depending on the frequency of the source. In order to see this, let us consider a harmonic source, whose quadrupole moment takes the form $q_{ij}(t)=a_{ij} e^{-i\omega t}+c_{ij}$, where $a_{ij}$ is the polarization tensor and $c_{ij}$ is some possible constant term. For this kind of time-dependence, the integral above can be computed and the metric perturbation takes the following form
\begin{eqnarray}
 h_{ij}&=&-\frac{2 G_{\rm{eff}}\omega^2}{r} a_{ij} e^{-i\omega (t-r)}\left(1-e^{i(\sqrt{\omega^2-m_g^2}-\omega)r}\right)\, .
\end{eqnarray}
Thus, there are two regimes. When $\omega^2<m_g^2$ the contribution of the massive graviton is exponentially damped and only the massless graviton propagates at long distances. When $\omega^2\ge m_g^2$, the massive graviton starts propagating and its contribution has the same weight as the massless graviton but with a difference phase, due to the different dispersion relation. As a consequence, the wave is modulated by a factor of modulus $|A(r)|^2=4\sin^2(\pi r/\lambda)$, where the wavelength of the modulation $\lambda$ is
\begin{equation}
\lambda=\frac{2\pi}{m_g^2}\left(\omega+\sqrt{\omega^2-m_g^2}\right)\, .
\end{equation}
Thus, when the massive graviton radiates, the profile of the wave is significantly different from the GR prediction.

%Now, in a realistic case we will not have a perfectly harmonic motion, but it can be expanded as Fourier series with a central frequency $\omega_0$, corresponding to the orbital frequency. Then, even though $\omega_0^2<m_g^2$, there exists certain harmonic $n\omega_0$ such that $n^2\omega_0^2>m_g^2$, and all the frequencies higher than this do propagate. This way, what we obtain is the prediction of Einstein gravity plus a correction of frequencies  $m_g$ and higher. 

\section{Discussion}
In this chapter we have studied the linearized equations of general $\mathcal{L}($Riemann$)$ theories on constant curvature backgrounds. We found that these theories contain additional degrees of freedom --- including a massive ghost graviton --- but one of the most important results was the characterization of a special set of theories whose linearized equations are of second order and that only propagate a massless graviton. These were named  ``Einstein-like'' theories, while Lagrangians satisfying this condition on arbitrary dimensions were called ``Einsteinian''. The full non-linear equations of some of these theories are actually of second-order --- this is the case for Lovelock gravity. However, most of them have fourth-order equations that somehow reduce to second order at linear level, and we should investigate if possible pathologies appear at non-linear level.  We argue here that Einstein-like theories could actually evade some of these problems.

It is well-known that non-degenerate higher-derivative Lagrangians (containing two or more derivatives of the fields) give rise to instabilities \cite{Ostrogradsky:1850fid,Woodard:2015zca}. The result proven by Ostrogradsky states that for this type of Lagrangians the Hamiltonian is unbounded from below, which is the origin of a potentially disastrous instability. At the linear level in higher-order gravity, this instability is manifest in the presence of the ghost-like massive graviton, which propagates negative energy. In Einstein-like theories this mode is not present at the linear level, and consequently the linear theory is free from the Ostrogradsky instability; but if the fully-non linear equations are of higher-order in derivatives, then this instability must appear at a certain point. However, we may argue that the possible instability will not be dangerous for Einstein-like theories providing that we consider spacetimes that tend asymptotically to the maximally symmetric vacuum. For instance, let us consider some Einstein-like theory whose vacuum is AdS. The linear perturbations on AdS are of second order and there is no instability. But now we could consider any other solution of the equations that is not pure AdS, but \emph{asymptotically} AdS, \eg a spacetime containing a black hole. Then we could study the linear perturbations on that solution and, since the full equations of motion are of fourth-order, we expect that there are additional degrees of freedom besides the massless graviton. Some of these modes would carry negative energy in agreement with Ostrogradsky's theorem. However, in Einstein-like theories these modes cannot escape to infinity because they become infinitely heavy as they approach the asymptotically AdS region. Thus, the negative energy modes are confined in a finite region inside the bulk. The fact that there cannot be negative energy fluxes to infinity prevents possible instabilities, or at least some of them. Thus, as long as we consider configurations with the appropriate boundary conditions, part of the possible instabilities are avoided. 
Of course, this is just a qualitative argument and in order to understand possible dynamical pathologies of these theories it would be necessary to study perturbation theory in backgrounds other than constant curvature solutions, or even to perform a Hamiltonian analysis of these theories.

There is another issue related to the previous discussion. Intuitively, the Einstein-like condition is imposed by demanding that the additional modes are infinitely heavy on the vacuum, so that they are infinitely suppressed. In practice, this means that we tune the parameters of the theory so that the coefficients of the higher-derivative kinetic terms vanish at leading order in the perturbative expansion. It is a common lore that this type of situations appears when the vacuum has too many symmetries and gives rise to a strong coupling problem --- see \textit{e.g.} \cite{ArkaniHamed:2002sp}. However, we do not observe this type of behaviour. Indeed, let us assume that we take the masses of the additional modes $m_g^2$, $m_s^2$ to be arbitrarily large (and positive) but not infinite. Then, we can already see these modes at the linear level and the higher-derivative kinetic terms are in fact multiplied by a very small coefficient. However, the modes are not strongly coupled as we can see for example in \req{MaSeq}. Indeed, the coupling constant between these modes and matter is always $\kappa_{\rm eff}$, which is independent from $m_g^2$ and $m_s^2$. Even so, a more careful analysis would be convenient in order to determine whether these theories suffer from some strong coupling problem.

Let us finish by mentioning that higher-derivative gravities of the Einstein-like type find many interesting applications in the AdS/CFT correspondence \cite{Camanho:2009vw,Buchel:2009sk,Camanho:2009hu,Camanho:2010ru,Camanho:2013pda,Myers:2010jv}. %Since  the behaviour of the metric perturbation near the boundary of AdS in these theories is the same as in Einstein gravity, the holographic dictionary can be applied straightforwardly. 
In the cases studied so far, these theories provide finite and consistent answers, even when the computations are beyond the linear regime --- see \textit{e.g.} \cite{Myers:2010jv}. This indicates that these theories could be well-behaved after all.

%%%%%%%%%%%%%%%%%%%%%%%%
%%%%%%%%%%%%%%%%%%%%%%%%%
\chapter{Single-function spherically symmetric solutions}\label{Chap:2}

In the previous chapter we studied the weak-field limit of a broad class of higher-order gravities, and we even obtained some solutions of the linearized equations. However, we have not studied exact solutions of these theories yet, besides constant curvature spaces. In this chapter we analyze the field equations of general higher-derivative gravity theories for spherically symmetric and static (SSS) spacetimes. We will focus on a seemingly anecdotic property of Einstein gravity SSS solutions: in Schwarzschild coordinates the metric satisfies $g_{tt}g_{rr}=-1$ (see \req{Fmetric} below).  We will usually say that the metrics satisfying this property are characterized by a ``single function''. This property is satisfied in any dimension and also holds, for instance, for charged black holes in Einstein-Maxwell theory. In the case of higher-derivative gravity, this property does not hold in general, \ie typically the solutions have $g_{tt}g_{rr}\neq \text{const}.$ However, almost all exact black hole solutions in higher-derivative gravity that are known so far satisfy this property. This is for instance the case of black holes in Lovelock gravity \cite{Wheeler:1985nh,Wheeler:1985qd,Boulware:1985wk,Cai:2001dz,Dehghani:2009zzb,deBoer:2009gx} and in Quasi-topological gravity \cite{Quasi2,Quasi, Dehghani:2011vu,Cisterna:2017umf}. Remarkably, all of these theories also happen to belong to the class of Einstein-like theories defined in the previous chapter, \ie their linearized equations on maximally symmetric vacua are of second order. This makes us think that there could be something special about the condition $g_{tt}g_{rr}=-1$ and, consequently, in this chapter we study in full generality the properties of the theories that satisfy it. 

We first show that there is indeed a connection between possessing single-function solutions and having second-order linearized equations, but the key point consists in assuming that those solutions represent the exterior gravitational field of a spherically symmetric body.\footnote{For instance, the Schwarzschild solution is a vacuum solution of Einstein's equations, but it is also the exterior solution of any spherically symmetric gravitating body. However, an analogous situation is not true for all theories.} Then, we provide a sufficient and necessary condition for characterizing all theories of this type and we show that their field equations are dramatically simplified with respect to the general case, as they allow for a partial integration. We will see in the next chapters that this partial integrability is enough to determine the black hole thermodynamics in an exact way, which is one of the most appealing properties of these theories.  We will check that the aforementioned Lovelock and Quasi-topological theories belong to this class and we will find more examples. In particular, we will show that Einsteinian cubic gravity, introduced in the previous chapter  (see Sec.~\ref{sec:ECG}), is a new member of this class of theories in $D=4$ --- the first theory of this type identified in four dimensions. We will also review the results of \cite{Hennigar:2017ego} where all the cubic theories of this type were constructed in any dimension. The whole family of theories has been given the name of \emph{Generalized quasi-topological gravity} (GQG) and now we know that these theories exist in any dimension \cite{Hennigar:2017ego,Ahmed:2017jod} and at any order in curvature \cite{PabloPablo4}.  Due to their appealing properties,  they will provide us with a rich playground to probe higher-curvature effects on black holes and in holography in the next chapters. 

\section{Generalizations of Schwarzschild's solution}\label{sec:GoSs}
Unless otherwise stated, in this chapter we will consider general higher-derivative theories of the form
\begin{equation}\label{higherd}
S=\int d^Dx \sqrt{|g|}\, \mathcal{L}(g^{\mu\nu},R_{\mu\nu\rho\sigma},\nabla_{\alpha}R_{\mu\nu\rho\sigma},\ldots)\, .
\end{equation}
Throughout the chapter we will be mostly interested in extensions of Einstein gravity, \ie we will assume that the above action reduces to the Einstein-Hilbert one when the curvature is small. In that case, the above Lagrangian can be written in the form
\begin{equation}\label{higherd}
\mathcal{L}(g^{\mu\nu},R_{\mu\nu\rho\sigma},\nabla_{\alpha}R_{\mu\nu\rho\sigma},\ldots)=\frac{1}{16\pi G}\left[-2\Lambda+R+\text{higher-derivative terms}\right]\, ,
\end{equation}
where the higher-derivative terms are assumed to be arbitrary linear combinations of monomials of the Riemann tensor and its covariant derivatives. 
In particular, our interest will be on static and spherically symmetric solutions of the previous theories. In the case of Einstein gravity, those solutions are given by the well-known Schwarzschild-Tangherlini-(A)dS metric, that takes the form\footnote{We will sometimes refer to \req{Fmetric} with \req{SAdS} as `Schwarzschild-(A)dS' solution, by which we will be referring to the three possible asymptotic behaviors: Anti-de Sitter ($\Lambda<0$), de Sitter ($\Lambda>0$) or flat ($\Lambda=0$).}
\begin{equation}
 ds^2_f=-f( r )dt^2+\frac{dr^2}{f( r )}+r^2 d\Omega_{(D-2)}^2\, .
 \label{Fmetric}
 \end{equation}
with 
\begin{equation}\label{SAdS}
f(r)=1-\frac{16\pi GM}{(D-2)\Omega_{(D-2)}r^{D-3}}-\frac{2\Lambda r^2}{(D-1)(D-2)}\, , \quad \text{where} \quad \Omega_{(D-2)}=\frac{2 \pi^{\frac{D-1}{2}}}{\Gamma[\frac{D-1}{2}]}
\end{equation}
is the area of the $(D-2)$-dimensional unit sphere and $M$ is the ADM mass \cite{Arnowitt:1960es,Arnowitt:1960zzc,Arnowitt:1961zz}.

Usually, the introduction of higher-derivative interactions in the action will modify this solution, but we are interested in modifications that preserve the form of the metric \req{Fmetric}. This is, the corrections will appear through the function $f(r)$ --- that will not be given by \req{SAdS} anymore --- but the metric will still satisfy $g_{tt}g_{rr}=-1$.\footnote{Metrics of this form have Ricci tensors with vanishing radial null-null components, see \eg \cite{Salgado:2003ub,Jacobson:2007tj}.} In this sense, this type of solutions can be called \emph{Schwarzschild-like}, because they conserve this property of Schwarzschild solution.
Our discussion will be focused on the spherically symmetric case, but our results are equally applicable when we replace the metric of the sphere $d\Omega_{(D-2)}^2$ in \req{Fmetric} by any other constant curvature metric, as in the case of planar or hyperbolic black holes. 

On general grounds, finding static and spherically symmetric black hole solutions for $D$-dimensional theories of the form \req{higherd} is a challenging task. In particular, if the Lagrangian contains up to $n$ derivatives of the Riemann tensor, the equations of motion generally involve $2n+4$ derivatives of the metric, which, in the case of a general static and spherically symmetric ansatz, --- see \req{Nf} below --- usually translate into a system of coupled differential equations of such order --- see \eg \cite{Lu:2015psa,Lu:2015cqa}. As a matter of fact, some simple analytic solutions of the form \req{Fmetric}, \ie
characterized by the condition $g_{tt}g_{rr}=-1$, have been in fact constructed for certain higher-derivative theories. However, these solutions fall, very often, within one of the following three categories:
\begin{itemize}
\item[i)] They are the ``same'' solution as in Einstein gravity, \ie they correspond to embeddings of Einstein gravity solutions in some higher-derivative theory. This is what happens, for instance, in $f(R)$ gravity \cite{delaCruzDombriz:2009et}. More generally, the Lagrangians that only contain Ricci curvature always allows for Einstein metrics as solutions, and they do not introduce corrections to vacuum Einstein gravity solutions --- see \eg \cite{Li:2017ncu}. 

\item[ii)] They are solutions to pure higher-derivative gravities, so that the action does not include the Einstein-Hilbert term --- and hence they lack an Einstein gravity limit. For instance, pure Weyl-squared gravity, $\mathcal{L}=\alpha C_{\mu\nu\rho\sigma}C^{\mu\nu\rho\sigma}$, in $D=4$ allows for solutions of the form \req{Fmetric}  that are different from Schwarzschild-(A)dS \cite{Riegert:1984zz,Klemm:1998kf}, while $\mathcal{L}=-2\Lambda+R+\alpha C_{\mu\nu\rho\sigma}C^{\mu\nu\rho\sigma}$ does not \cite{Lu:2012xu}. Examples of this kind involving Weyl-cubed terms in $D=6$ can be found \eg in \cite{Oliva:2010zd,Lu:2013hx}. Similar comments apply to pure Lovelock gravity solutions, like those constructed \eg in \cite{Banados:1993ur,Cai:2006pq}.
\item[iii)] They involve the fine-tuning of some of the higher-derivative couplings --- and hence, again, they lack an Einstein gravity limit. A simple example corresponds to perfect-square (or other powers greater than $2$) actions, such as $\mathcal{L}=-(R-4\Lambda)^2/(8\Lambda)$, which of course admits as solution any metric satisfying $R=4\Lambda$.  Examples belonging to this class have been constructed, \eg in \cite{Cai:2009ac,Love}.
\end{itemize}
We will not consider cases ii) and iii) because of their lack of an Einstein gravity limit. With regards to case i), while perfectly valid, it is not interesting to us, because our aim is precisely to study non-trivial modifications of black hole solutions. In addition, we will show that the solutions in the case i) are in some sense  ``unnatural'', since, although they are vacuum solutions they cannot correspond to the exterior gravitational field of generic spherically symmetric distributions of mass --- see next subsection and Sec.~\ref{unnatural} for details.

On the bright side, genuine (analytic or semianalytic) single-function extensions of \req{SAdS} have been constructed in $D\geq 5$ for Lovelock gravities \cite{Wheeler:1985nh,Wheeler:1985qd,Boulware:1985wk,Cai:2001dz,Dehghani:2009zzb,deBoer:2009gx}, Quasi-topological gravity \cite{Quasi2,Quasi} and its quartic \cite{Dehghani:2011vu} and quintic \cite{Cisterna:2017umf} generalizations. The $D=4$ case turns out to be a considerably harder nut to crack, given that all the Lovelock and Quasi-topological densities (except for the Einstein-Hilbert term) are either topological or trivial in that case. 
The main goal of this chapter is to identify and characterize some of the properties which make all these theories special as well as to provide explanations for some previously conjectured relations. As a result, we will be able to identify new theories of this kind, that will include Einsteinian cubic gravity in four dimensions --- a theory that we presented in Chapter \ref{Chap:1} --- as well as the recently proposed \emph{Generalized quasi-topological gravity}\cite{Hennigar:2017ego,Ahmed:2017jod}.\footnote{Chronologically, the discovery of black hole solutions in Einsteinian cubic gravity \cite{PabloPablo,Hennigar:2016gkm,PabloPablo2} and in Generalized quasi-topological gravity \cite{Hennigar:2017ego} happened before we learned about the general results presented in this chapter \cite{PabloPablo3}. The presentation in this thesis follows instead the ``logical'' order: we first obtain the general result and we use it afterwards to construct the new theories.} 
Let us now state the main results of this chapter. 

\subsection{Main results}

Most of the material presented in Sections \ref{bhl} and \ref{bho} can be encapsulated in two main ``theorems'' and a corollary
%\footnote{Note that we have refrained form using the word ``Theorem'' to refer to our ``Theorems''. While we think our proofs are completely robust, we have considered that some additional level of rigor could perhaps be added  }
 (which we prove) and two conjectures (in favor of which we provide strong evidence). In order to formulate them, let us start with some definitions. First, let us note that the most general static and spherically symmetric ansatz can be written as 
 \begin{equation}\label{Nf}
ds^2_{N,f}=-N^2( r )f( r )dt^2+\frac{dr^2}{f( r )}+r^2 d\Omega_{(D-2)}^2\, ,
\end{equation}
 where we need two independent functions $N(r)$ and $f(r)$. Then, let $L_{N,f}$ be the effective Lagrangian resulting from the evaluation of $\sqrt{|g|}\mathcal{L}$ in \req{Nf}, this is
\begin{equation}\label{LNf}
L_{N,f}(r,f(r), N( r ),f'(r), N'( r ), \ldots)\equiv N( r )r^{D-2}\mathcal{L}\big|_{g_{\mu\nu}=g_{\mu\nu}^{N,f}}\, .
\end{equation}
%where $\Omega_{(D-2)}$ is the area of $S^{(D-2)}$.
Analogously, we will denote by $L_f$ the expression resulting from setting $N=1$ in $L_{N,f}$, which of course corresponds to the effective Lagrangian for $f(r)$ resulting from the evaluation of $\sqrt{|g|}\mathcal{L}$ in the single-function ansatz (\ref{Fmetric}). With these definitions at hand, we are ready to enumerate our results:

\begin{theorem}\label{ti1}
Let us consider a higher-derivative gravity of the form $\mathcal{L}(g^{\mu\nu},R_{\mu\nu\rho\sigma})$. If the exterior gravitational field of a spherically symmetric mass distribution is given by a metric of the form (\ref{Fmetric}), \ie characterized by a single function, then the theory only propagates a traceless and massless graviton on the vacuum.\footnote{Following the conventions of Chapter~\ref{Chap:1}, here the vacuum is a maximally symmetric solution of the equations of motion. If the theory possesses several vacua, this result holds for all of them.}
 \end{theorem}
%\\
%This theorem is telling us that, in order to obtain theories whose solutions are characterized by a single function and that satisfy the reasonable requirement of coinciding with an exterior solution of a mass distribution, we must search among the Einstein-like theories.\\

\begin{theorem} \label{theo}Let us consider a higher-derivative gravity Lagrangian $\mathcal{L}(g^{\mu\nu},R_{\mu\nu\rho\sigma}, \nabla_{\alpha}R_{\mu\nu\rho\sigma},\ldots)$ of the form \req{higherd} involving terms with up to $n$ covariant derivatives of the Riemann tensor. If the Euler-Lagrange equation of $L_f$ vanishes identically, \ie if 
\begin{equation}\label{EELL}
\frac{\delta L_f}{\delta f}\equiv\frac{\partial L_f}{\partial f}-\frac{d}{dr}\frac{\partial L_f}{\partial f'}+\frac{d^2}{dr^2}\frac{\partial L_f}{\partial f''}-\ldots= 0\quad \forall \, f(r)\, ,
\end{equation}
 then: 
 \begin{enumerate}
 	\item{The theory allows for vacuum solutions of the form (\ref{Fmetric}), where the equation of $f(r)$ can be integrated once, yielding a differential equation of order $\le 2n+2$ and where the integration constant is the ADM mass.}
 	\item{At least in the $ \mathcal{L}(g^{\mu\nu},R_{\mu\nu\rho\sigma})$ case, the theory only propagates a traceless and massless graviton on the vacuum.}
 \end{enumerate}
\end{theorem}

\begin{corollary}\label{col} For all theories fulfilling the hypothesis of Theorem \ref{theo}, the exterior gravitational field of a spherically symmetric matter distribution is again given by a metric of the form (\ref{Fmetric}), where $f(r)$ satisfies the same differential equation as in the vacuum.
\end{corollary}

  \begin{conjecture}\label{conj} For all theories fulfilling the hypothesis of Theorem \ref{theo}:
  	
  	\begin{enumerate}
  		\item{For a fixed mass $M$, there is at most a discrete number of black hole solutions of the form (\ref{Fmetric}), each one characterized by different thermodynamic relations $T(M)$, $S(M)$. If there are several black holes, only one of them is a smooth deformation of the Schwarzschild solution.}
  		\item{The thermodynamic properties of these black holes can be determined analytically by solving a system of algebraic equations without free parameters.}
  	\end{enumerate}
  \end{conjecture}

 \begin{conjecture}\label{conji2} Given a $\mathcal{L}(g^{\mu\nu},R_{\mu\nu\rho\sigma})$ theory fulfilling the hypothesis of Theorem \ref{theo}, if $L_{N,f}$ can be written as
 	\begin{equation}\label{siss}
 	L_{N,f}=N L_f+ N' F_1+N'' F_2\, ,
 	\end{equation} 
 	where $F_{1,2}$ are functions of $f(r)$ and its derivatives, then the equation determining $f(r)$ is algebraic.
 \end{conjecture}
 
Some comments are in order. 
\begin{itemize}
\item Theorem \ref{ti1}, which we prove in Sec.~\ref{bhl}, explains the previously noticed \cite{Quasi,PabloPablo,Hennigar:2016gkm,PabloPablo2,Aspects,Cisterna:2017umf,Hennigar:2017ego} (but so far unexplained) fact that certain higher-order gravities admitting simple black hole solutions of the form \req{Fmetric} have the interesting property of sharing the linearized spectrum of Einstein gravity. Strictly speaking, the hypothesis of this result is that the theory has solutions of the form \req{Fmetric} describing the exterior field of spherically symmetric mass distributions, so it does not say anything about black holes yet. However, if the theory has solutions of the form \req{Fmetric} outside a spherically symmetric mass distribution, it is clear that there are also vacuum solutions of that form. The converse is not true. The apparent contradiction of Theorem \ref{ti1} with the fact that certain theories which propagate extra modes at the linearized level do admit vacuum solutions of the form \req{Fmetric} is not such. The reason is that, in those cases, the vacuum solutions cannot correspond to the exterior gravitational field of generic spherical distributions, which makes them somewhat ``unnatural''. We illustrate this very explicitly for the well-known Schwarzschild-(A)dS solution in $f(R)$ gravity in Sec.~\ref{unnatural}. 

\item The main result of this chapter is Theorem \ref{theo}, which we prove in Section \ref{bho}. It yields a sufficient condition for a theory to possess solutions of the form \req{Fmetric} and shows additionally that such theories have an Einstein-like spectrum. The hypothesis of Theorem \ref{theo} defines a broad family of theories, and, following the nomenclature of \cite{Hennigar:2017ego}, we will say that they belong to the \emph{Generalized quasi-topological gravity} (GQG) class. 
Interestingly, the proof of the first item of Theorem \ref{theo}, provides a very efficient method for identifying these theories as well as for obtaining the differential equation determining the metric function $f(r)$ in each case. We present this method in the form of a simple recipe in Section \ref{recipe}, and afterwards we construct several examples of theories of this class. 
 	
\item Note that, strictly speaking, there is no connection between the theories of Theorems \ref{ti1} and \ref{theo}, since they satisfy slightly different properties. Intuitively, however, one would expect that the theories of both theorems are the same. In Corollary~\ref{col}, we establish the equivalence in one direction by showing that the theories that satisfy the hypothesis of Theorem \ref{theo} allow for exterior solutions of the form \req{Fmetric}, hence they correspond to theories of Theorem \ref{ti1}. 

\item Note that the reason for considering the subclass of theories $\mathcal{L}(g^{\mu\nu},R_{\mu\nu\rho\sigma})$ (which do not include terms involving covariant derivatives of the Riemann tensor), in Theorem \ref{ti1} and in item 2 of Theorem \ref{theo} is that the spectrum and Newtonian limit of these theories has been exhaustively classified  (see Chapter \ref{Chap:1}), which we use to prove our results. We are not aware of an analogous general classification in the general higher-derivative case. However, in light of some related recent works \cite{Modesto:2014eta,Giacchini:2016xns}, we strongly believe those results apply in the general case as well.

\item Observe that Theorem \ref{theo} does not really make reference to whether the solutions described by \req{Fmetric} in each case correspond, in particular, to black holes. However, the great amount of evidence accumulated so far \cite{Wheeler:1985nh,Wheeler:1985qd,Boulware:1985wk,Cai:2001dz,Dehghani:2009zzb,Quasi,Quasi2,deBoer:2009gx,Dehghani:2011vu,PabloPablo,Hennigar:2016gkm,PabloPablo2,Cisterna:2017umf,Hennigar:2017ego}, along with the results that we will expose in the next chapters, provide strong support for the validity of Conjecture \ref{conj}.

\item  For a $\mathcal{L}(g^{\mu\nu},R_{\mu\nu\rho\sigma})$ theory satisfying the hypothesis of Theorem \ref{ti1}, the order of the differential equation determining $f(r)$ is usually $2$ --- see \eg \cite{Hennigar:2016gkm,PabloPablo2,Hennigar:2017ego}. However, in some well-known cases
	\cite{Wheeler:1985nh,Wheeler:1985qd,Boulware:1985wk,Cai:2001dz,Dehghani:2009zzb,deBoer:2009gx,Quasi,Quasi2,Dehghani:2011vu,Cisterna:2017umf}, the equation is algebraic instead, which represents a considerable simplification. Conjecture \ref{conji2}, which we motivate in Sections \ref{quadratic} and \ref{cubic}, provides a straightforward guiding principle for identifying such class of theories from a given larger set.

\item Finally, note that, even though we focus on the spherically symmetric case, the solutions which can be constructed using our method can be straightforwardly generalized to the hyperbolic and flat transverse geometry cases. 
	%In particular, the recipe explained in section \ref{recipe} can equally be applied to those cases. We illustrate this in section \ref{planar}, where we obtain new five-dimensional asymptotically AdS planar black hole solutions of the recently constructed Generalized quasi-topological gravity \cite{Hennigar:2017ego} --- we also comment on this theory in section \ref{cubic}. This also serves to further support Conjecture \ref{conj} and is of course motivated by holography, where planar black hole solutions to various higher-order gravities have proven to be remarkably useful for different purposes --- see	\eg \cite{Brigante:2007nu,deBoer:2009gx,Buchel:2009sk,Myers:2010jv,Myers:2010tj}. 

\end{itemize}

 \section{$g_{tt}g_{rr}=-1$ and absence of massive modes}\label{bhl}
  It has been previously observed that certain higher-order gravities admitting simple black hole solutions possess particularly simple linearized spectra. This was emphasized by Myers and Robinson in \cite{Quasi}, where they observed that Quasi-topological gravity \cite{Quasi2,Quasi} satisfies the following two unusual properties: first, it admits black holes characterized by a single function $f(r)$, \ie solutions of the form \req{Fmetric}; and second, its linearized spectrum coincides with the Einstein gravity one, namely, the only dynamical mode propagated by the metric perturbation in a maximally symmetric spacetime is a transverse and traceless graviton. These two apparently unrelated properties are also known to hold for general Lovelock theories \cite{Lovelock1,Lovelock2,Wheeler:1985nh,Wheeler:1985qd,Boulware:1985wk,Cai:2001dz,Dehghani:2009zzb,deBoer:2009gx,Love},  for quartic \cite{Dehghani:2011vu} and quintic Quasi-topological gravity \cite{Cisterna:2017umf}.  Hence, it is natural to wonder how generally this connection between the linearized regime and the genuinely non-linear one holds for general higher-order gravities. As we will see, this relation also holds for Einsteinian cubic gravity in four dimensions \cite{PabloPablo,Hennigar:2016gkm,PabloPablo2} and for the recently constructed Generalized quasi-topological gravity \cite{Hennigar:2017ego,Ahmed:2017jod}.

It is important to note that not all higher-order theories that share the linearized spectrum of Einstein gravity admit single-function black hole solutions. Examples of such theories include for instance certain $f($Lovelock$)$ theories \cite{Love,Aspects,Karasu:2016ifk}. Furthermore, there are theories which propagate extra modes at the linearized level and yet they posses (vacuum) solutions of the form \req{Fmetric}. This is the case, \eg of $f(R)$ gravity \cite{delaCruzDombriz:2009et}. Naturally, these observations clearly show that the connection between both properties cannot be a double implication.

In this section we will prove that, in fact, given a general $ \mathcal{L}(g^{\mu\nu},R_{\mu\nu\rho\sigma})$ theory, if the exterior gravitational field of a spherically symmetric body is given by a metric of the form \req{Fmetric}, then the theory only propagates a massless graviton at the linearized level. In other words, only theories sharing the linearized spectrum of Einstein gravity are susceptible of admitting single-function solutions corresponding to the exterior field of a spherically symmetric body.  Note in addition that if a theory has solutions of that kind in the presence of a spherically symmetric body, it will also have single-function vacuum solutions. However, the converse is not true, as we illustrate in Sec.~\ref{unnatural}.

The idea behind the proof is that, if the exterior gravitational field of a spherically symmetric body is given by a metric of the form \req{Fmetric}, this must be true, in particular, in the weak-field limit.  Now, in the previous chapter we studied in full generality the linear approximation of arbitrary $\mathcal{L}(g^{\mu\nu},R_{\mu\nu\rho\sigma})$ theories. In particular, in Sec.~\ref{GeneralizedNewton} we showed that the Newtonian metric of a spherically symmetric body in $D$ dimensions is given by 
 \begin{equation}
 	ds_N^2=-(1+2U(\rho))dt^2+(1-2V(\rho))(d\rho^2+\rho^2 d\Omega_{(D-2)}^2)\,,
 	\label{Newtonian}
 \end{equation}
where $U(\rho)$ is the generalized Newton potential and $V(\rho)=\gamma(\rho) U(\rho)$ where $\gamma(\rho)$ is one of the so-called parametrized Post-Newtonian parameters. In four dimensions, and in the point-like limit of the mass distribution, these functions are explicitly given by

\begin{equation}
 	U ( \rho )=-\frac{G_{\rm{eff}}M}{\rho}\left[1-\frac{4}{3}e^{-m_g \rho}+\frac{1}{3}e^{-m_s \rho}\right],\quad  V ( \rho)=-\frac{G_{\rm{eff}}M}{\rho}\left[1-\frac{2}{3}e^{-m_g \rho}-\frac{1}{3}e^{-m_s \rho}\right],
 	\label{potentials}
 \end{equation}
while the higher-dimensional generalizations of these expressions are written in \req{gagaa}. For a general matter distribution, the difference would be the appearance of form factors in front of the exponential terms, as in \req{eq:NeWpotform}, but this detail is not important for our discussion. In the previous expressions, $m_g$ and $m_s$ are, respectively, the masses of the additional spin-2 and spin-0 modes propagated by the metric perturbation for a generic $ \mathcal{L}(g^{\mu\nu},R_{\mu\nu\rho\sigma})$ theory. These can be easily computed for a given theory using the method developed in Chapter \ref{Chap:1}.

The Newtonian metric \req{Newtonian} is written in isotropic coordinates. In order to express it in Schwarzschild coordinates, we perform the change of variable $r^2=\rho^2(1-2V( \rho))$.  Keeping only terms linear in $U$ and $V$, we obtain\footnote{Note that $V(\rho)=V(r)+\mathcal{O}(V^2)$ and the same is true for $U$.}
 \begin{equation}\label{ttec3}
 	ds_N^2=-(1+2 U( r ))dt^2+(1+2r V'( r ))dr^2+r^2 d\Omega_{(D-2)}^2\,.
 \end{equation}
 Now we make use of our hypothesis and  assume that this is indeed the linearized limit of a full non-linear solution of the form (\ref{Fmetric}). This means that $g_{tt}g_{rr}=-1$, which, when  applied to \req{ttec3}, imposes the following condition on $U$ and $V$,
 \begin{equation}
 	U( r )+ r V'( r )=0\, .
 	\label{condition1}
 \end{equation}
From the expressions in (\ref{potentials}), it follows that this condition holds only when the exponential terms are removed, and this is equivalent to $m_g^2=m_s^2=+\infty$. In other words, the single-function condition on the non-linear solution
implies the absence of the massive graviton and the scalar field at the linear level, so that the only mode that is propagated on the vacuum is the massless graviton. The argument extends straightforwardly to general dimensions using \req{gagaa}. Note also that the argument is independent of the presence of form factors  when the mass distribution is not point-like. Essentially, the condition \req{condition1} is only satisfied by the term $\sim 1/r$ (in general dimension $\sim 1/r^{D-3}$) in the potentials $U$ and $V$. The exponential terms, associated to the presence of massive modes, never fulfill \req{condition1}, hence the massive modes must be absent. 

%In general dimensions $D>4$ the specific expressions for $U$ and $V$ are different, but the conclusion is the same; (\ref{condition1}) only holds whenever the scalar and the massive graviton are removed.
This simple argument shows that if a higher-order gravity allows for a solution of the form (\ref{Fmetric}) representing the exterior field of a spherical body, the theory only propagates a massless graviton on the vacuum. Thus, the theory would belong to the Einstein-like class that we introduced in Sec.~\ref{Classification}. 

There is a last subtlety that we must overcome. Note that, strictly speaking, the Newtonian metric (\ref{Newtonian}) only applies in the asymptotically flat case. In that situation, only terms up to quadratic order in curvature contribute to the masses $m_g$ and $m_s$, so our argument does not immediately go through beyond the quadratic level. In order to extend it to terms of arbitrary order in curvature, we must generalize this argument to include asymptotically (A)dS solutions, since in that case all terms do contribute to the masses of the modes. For a background of curvature $\Lambda$, the Newtonian solution (\ref{Newtonian}) with the potentials \req{potentials} is also a good approximation as long as $\rho<<|\Lambda|^{-1/2}$. Then, we may consider a matter distribution whose size is much smaller than $|\Lambda|^{-1/2}$ and we might zoom in the region near the distribution in which the expressions \req{potentials} apply.  Hence, the same analysis as in the flat case can be applied and the same conclusions are reached, namely, $m_g^2=m_s^2=+\infty$, where now the masses contain information about terms at every order in curvature. Alternatively, one can compute explicitly the Newtonian potential in the (A)dS case and repeat the analysis, but the conclusions remain unchanged.

As we have seen, the results in this section rely on the analysis performed in Chapter~\ref{Chap:1} for the Newtonian metric of a general $\mathcal{L}(g^{\mu\nu},R_{\mu\nu\rho\sigma})$ theory. Hence, these do not include the more general higher-derivative case  $\mathcal{L}(g^{\mu\nu},R_{\mu\nu\rho\sigma},\nabla_{\alpha}R_{\mu\nu\rho\sigma},\ldots)$. However, in view of some results available in the literature for the Newtonian potential in some of these theories \cite{Modesto:2014eta,Giacchini:2016xns}, we are confident that Theorem \ref{ti1} extends as well to that case. The Newtonian metric in the general case seems to have the same structure as in \req{Newtonian}, with exponential terms including the masses of all the additional modes. Then, we would expect that the condition $g_{tt}g_{rr}=-1$ would only hold if all the masses are infinite, similarly to what we found in our analysis. This suggests that the Theorem \ref{ti1} would also hold in the general higher-derivative case. A rigorous analysis would require a systematic classification of the spectrum of these theories.

%Let us estress again that the reason for considering a theory of the form $ \mathcal{L}(g^{ab},R_{abcd})$, instead of the more general \req{higherd}, is that the results used above for the Newtonian metric were obtained in \cite{PabloPablo,Aspects} only for the former subclass of theories.

%the spectrum and the Newtonian limit of these theories has been exhaustively studied \cite{PabloPablo,Aspects}, while a complete classification is still lacking for the general case $\mathcal{L}(g^{ab},R_{abcd},\nabla_{e}R_{abcd},\ldots)$. However, in view of the expression for the Newtonian potential in quadratic theories with an arbitrary number of derivatives \cite{Modesto:2014eta,Giacchini:2016xns}, we think it is very likely that the result applies also in the general case.

%Of course, our result can be rephrased as the fact that theories which include extra modes in their linearized spectrum do not admit static and spherically symmetric solutions of the form \req{Fmetric} representing the gravitational field of a spherical mass distribution.
% In the next subsection we illustrate this point in the simple case of $f(R)$ gravity.

 \subsection{Counterexample: spherically symmetric solutions in $f(R)$ gravity}\label{unnatural}
 %{\bf Counter-example: $f( R )$}\\
Naively, the result found in the previous subsection seems to be incompatible with the fact that 
certain theories which propagate extra modes at the linearized level do also admit single-function black hole solutions of the form \req{Fmetric}. This apparent contradiction is not such. The reason is that, as we have stressed, our result holds only whenever \req{Fmetric} describes the gravitational field of a spherical mass distribution.

%, \ie when the corresponding solution is the \emph{natural} static and spherically symmetric black hole of the corresponding theory generalizing Einstein's gravity Schwarzschild-(A)dS solution in a non-trivial way, and reducing to it when the corresponding higher-order couplings are set to zero. 

In order to illustrate this point, let us consider the case of $f( R )$ gravity. It is well-known that this theory allows for the Schwarzschild-(A)dS solution in the absence of matter --- \eg \cite{delaCruzDombriz:2009et}. Hence, it possesses black hole solutions characterized by a single function. However, we also know that this theory propagates a scalar mode along with the massless graviton on the vacuum. Hence, our result implies that, even though the Schwarzschild-(A)dS metric is a vacuum solution of $f( R )$ gravity, it does not describe the external field of a generic spherically symmetric mass distribution for this theory. Let us verify this statement explicitly.  The $f( R )$ field equations coupled to matter read
 \begin{equation}\label{fr}
 	f'( R ) R_{\mu\nu}-\frac{1}{2}f( R )g_{\mu\nu}+\left(g_{\mu\nu}\Box-\nabla_{\mu}\nabla_{\nu}\right) f'( R )=\kappa T_{\mu\nu}\, ,
 \end{equation}
 where $\kappa$ is proportional to Newton's constant. Now, let us consider a static and spherically symmetric configuration with an energy-momentum tensor $T_{\mu\nu}$ such that $T_{\mu\nu}( r )=0$ if $r> r_0$, for certain $r_0$. %We allow the matter distribution to be discontinue at $r_0$. 
Further, let us assume this situation to be compatible with an exterior metric of constant scalar curvature, \ie satisfying  $R=\bar R$, where the constant $\bar R$ would be obtained from the algebraic equation $2\bar R f'( \bar R )-Df( \bar R )=0$.
If that was the case, \req{fr} would imply $R_{\mu\nu}\propto g_{\mu\nu}$, and we would obtain the Schwarzschild-(A)dS metric in the exterior region. 

However, we will show that no constant-$R$ solution compatible with the above assumptions can exist in the outside region. 
The trace of the field equations reads
 \begin{equation}
 	(D-1)\Box f'( R )+R f'( R )-\frac{D}{2}f( R )=\kappa T\, .
 	\label{Requation}
 \end{equation}
 This can be thought of as an equation for $R$ (or $f'( R )$).
 Since we are considering a spherically symmetric situation, we can assume $R=R( r )$, which reduces \req{Requation} to an ordinary second-order differential equation for $R( r )$. A solution to this equation is then determined by specifying the values of $R$ and $dR/dr$ at some $r$.  
  Now, in the transition point $r_0$ we must demand continuity and differentiability (otherwise there is no solution). Taking into account our assumptions for the exterior solution, this fixes the following boundary conditions for the internal one:
 \begin{equation}\label{bdc}
 R(r_0)=\bar R\, , \quad \frac{dR}{dr}(r_0)=0\, .
\end{equation} 
  The solution $R(r)$ for $r<r_0$ is then completely specified.  However, let us now consider a $(D-2)$-sphere of radius $r_s>r_0$ and unit normal $n^{\mu}$ at some time slice. Then, using the spherical symmetry of the problem and Stokes' theorem, it is straightforward to prove the following equalities
 \begin{equation}
 	\Omega_{(D-2)} r_s^{D-2} \frac{d f'( R )}{dr}(r_s)=\oint_{S_{r_s}} d^{D-2}S\, n^{\mu}\nabla_{\mu}f'( R )=\int_{r<r_s} d^{D-1}x\sqrt{|g|}\Box f'( R )\, .
 \end{equation}
 Finally, taking into account that $f'( R )$ is constant for $r>r_0$ and using equation (\ref{Requation}), which holds in the $r<r_0$ region, we get
 \begin{equation}
 	\Omega_{(D-2)} r_s^{D-2} \frac{d f'( R )}{dr}(r_s)=\frac{1}{D-1}\int_{r<r_0} d^{D-1}x\sqrt{|g|}\left(\kappa T-R f'( R )+\frac{D}{2}f( R )\right).
	\label{contradiction}
 \end{equation}
 Now it is immediate to see that there is a problem here. Indeed, while the left-hand side (lhs) is zero, the right-hand side (rhs) is non-vanishing in general.
 The reason is that the interior solution is already completely specified by the boundary conditions \req{bdc} but those conditions do not depend on the particular form of the matter distribution, while the integral in the rhs does. Hence, that integral will be in general non-vanishing, which leads to a contradiction when compared to the lhs. This implies that no constant-$R$ solutions can describe the gravitational field in the outer region of a spherically symmetric matter distribution 
 for general $f(R)$ theories.\footnote{The fact that $f(R)$ theories do not allow for constant-$R$ solutions outside a source has been in fact known since long ago, see \textit{e.g.} \cite{Michel:1973iu,Mignemi:1991wa}.}
 
 Observe that in the Einstein gravity case, \ie when $f(R)=R-2\Lambda$, the contradiction in \req{contradiction} disappears, as the rhs vanishes in that case by virtue of Einstein's equation. This is naturally related to the absence of terms involving covariant derivatives of $f'(R)$ --- such as $\Box f'(R)$ --- in the equations of motion, which are in fact ultimately responsible for the appearance of the extra spin-0 mode in the linearized spectrum in the general $f(R)$ case.  
 
The reason why black hole metrics in $f(R)$ do not coincide with the exterior field of a spherically symmetric body is due to the existence of a scalar degree of freedom in these theories, besides the massless graviton. In vacuum black hole solutions this mode is not active and consequently the solution is given by the Schwarzschild metric. However, in the moment we introduce some matter the scalar is excited, yielding a different solution. In this way, the gravitational field of a black hole is qualitatively different from the one of a mass distribution. One would need to explain in this theory how the scalar field associated to a matter distribution disappears during the gravitational colapse.

 \section{Sufficient condition for single-function solutions}\label{bho}
In the previous section we showed that only theories sharing the linearized spectrum of Einstein gravity are susceptible of admitting single-function solutions of the form \req{Fmetric} --- providing that those solutions describe the exterior gravitational field of spherical distributions of matter. In this section we prove Theorem \ref{theo}, which provides in turn a sufficient condition for identifying such theories. The proof of this result gives rise to a simple method for constructing higher-derivative gravities satisfying the hypothesis of Theorem \ref{theo}, as well as for obtaining the equation that determines $f(r)$ in each case. 
Here we will consider the general action \req{higherd}, \ie with respect to the previous section we also allow for an arbitrary dependence on terms involving covariant derivatives of the Riemann tensor.
Then, as explained before, if the Lagrangian contains up to $n$ derivatives of the Riemann tensor, the field equations generally involve $2n+4$ derivatives of the metric.  

\subsubsection{Proof of part 1}
Let us first convince ourselves that the equations of motion for a static and spherically symmetric metric of the form \req{Nf} can be studied by considering the reduced action functional $S[N,f]$, obtained by evaluating the action on that ansatz, \ie
\begin{equation}
	S[N,f]=\Omega_{(D-2)}\int dt \int dr L_{N,f}\, ,
\end{equation}
where $L_{N,f}$ was defined in \req{LNf}.
Using the chain rule, we see that the variations of $S[N,f]$ with respect to $N$ and $f$ are related to the $tt$ and $rr$ components of the corresponding field equations $\mathcal{E}_{\mu\nu}\equiv \frac{1}{\sqrt{ |g|}}\frac{\delta S}{\delta g^{\mu\nu}}$ according to 
 \begin{equation}\label{eq:chain}
 \frac{1}{\Omega_{(D-2)}r^{D-2}}\frac{\delta S[N,f]}{\delta N}=\frac{2\mathcal{E}_{tt}}{ f N^2}\, , \quad \frac{1}{\Omega_{(D-2)}r^{D-2}}\frac{\delta S[N,f]}{\delta f}=\frac{\mathcal{E}_{tt}}{N f^2}+N \mathcal{E}_{rr}\, .
 \end{equation}
 Hence, imposing the Euler-Lagrange equations of $N$ and $f$ to hold is equivalent to imposing $\mathcal{E}_{tt}=\mathcal{E}_{rr}=0$. Finally, the Bianchi identity $\nabla^{\mu} \mathcal{E}_{\mu\nu}=0$ automatically makes the angular components vanish whenever  $\mathcal{E}_{tt}=\mathcal{E}_{rr}=0$.
 
 Observe now that the constant rescaling $N\rightarrow N\alpha$, for an arbitrary $\alpha$, is equivalent to the time rescaling $t\rightarrow t\, \alpha$ in \req{Nf}, which leads to the following identities
\begin{equation}
	S[\alpha N, f]=\Omega_{(D-2)}\int dt \int dr L_{\alpha N,f}=\Omega_{(D-2)}\int d(\alpha t) \int dr L_{N,f}
	%=\alpha\Omega_{(D-2)}\int dt \int dr L_{N,f}
	=\alpha S[N,f]\, .
\end{equation}
This implies that both $S[N,f]$ and $L_{N,f}$ are homogeneous of degree 1 in $N$. Moreover, we are assuming that the theory is of the form \req{higherd}, so that the higher-derivative terms appear as polynomials of the Riemann tensor and its derivatives. Thus, the Lagrangian is formed from products, quotients and derivatives of the metric components, and the only homogeneous monomials of degree 1 which can be formed in this way with $N$ and its derivatives are of the form $N^{i_1}N'^{i_2}N''^{i_3}\cdots(N^{(n+2)})^{i_{n+3}}$, where $i_k$ are integers such that $i_1+\ldots +i_{n+3}=1$. Also, we must have $i_k\ge0$ for $k>1$ because the derivatives cannot appear in the denominator. Taking this into account, we observe that the Lagrangian can always be expanded in the following way,\footnote{$N^{(i)}$ stands for the $i$-th derivative of $N$ with respect to $r$, and so on.}
\begin{equation}
	L_{N,f}=N L_{f}+\sum_{i=1}^{n+2}N^{(i)}F_i+\mathcal{O}(N'^2/N)\, ,
	\label{homogeneus}
\end{equation}
where 	$L_f(r, f, f', f'',\ldots)\equiv L_{N=1,f}$ is the effective Lagrangian resulting from the evaluation of the gravitational Lagrangian in the single-function ansatz (\ref{Fmetric}), and the $F_i=F_i(r,f, f', f'',\ldots, f^{(n+2)})$ are functions of $f$ and its derivatives. Finally, $\mathcal{O}(N'^2/N)$ denotes all the terms which are at least quadratic  in derivatives of $N$, 
\begin{equation}\label{ooo}
	\mathcal{O}(N'^2/N)\equiv \sum_{i,j=1}^{n+2}\frac{N^{(i)}N^{(j)}}{N}F_{ij}+\sum_{i,j,k=1}^{n+2}\frac{N^{(i)}N^{(j)}N^{(k)}}{N^2}F_{ijk}+\ldots\, ,
\end{equation}
where, again, $F_{ij}$, $F_{ijk}$, etc., only depend on $f$ and its derivatives.

The analysis so far is completely general. Let us now make use of the hypothesis of Theorem \ref{theo}: we assume that \req{EELL} holds, \ie that the Euler-Lagrange equation of $f(r)$ for the Lagrangian $L_f$ vanishes identically. Of course, this is equivalent to the assumption that $L_f$ is a total derivative, this is, that there exists a function $F_0(r, f, f',\ldots, f^{(n+1)})$ such that
\begin{equation}
	L_f=F_0'\,,
\end{equation}
where again the prime denotes a total derivative with respect to $r$. Using this result and the expansion (\ref{homogeneus}), we can express the reduced action functional as
\begin{equation}\label{acc}
	S[N,f]=\Omega_{(D-2)}\int dt \int dr \left[ N \left(F_0+\sum_{i=1}^{n+1}(-1)^iF_ i^{(i-1)}\right)'+\mathcal{O}(N'^2/N)\right]\, ,
\end{equation}
where we have integrated by parts several times. Now we are ready to compute the Euler-Lagrange equations for $N$ and $f$. First, it is immediate to see that the equation $\delta_f S=0$ is trivially satisfied whenever $N'=0$. Hence, we can just set $N$ to a constant value, which we choose to be one. On the other hand, variation with respect to $N$ and evaluation at $N=1$ yields
\begin{equation}\label{eq:deltaN}
	\delta_N S=\Omega_{(D-2)}\int dt \int dr  \delta N \left(F_0+\sum_{i=1}^{n+1}(-1)^iF_ i^{(i-1)}\right)'=0\, .
\end{equation}
Thus, $f(r)$ satisfied the differential equation 

\begin{equation}
\left(F_0+\sum_{i=1}^{n+1}(-1)^iF_ i^{(i-1)}\right)'=0\, ,
\end{equation}
that can be integrated once, yielding 
\begin{equation}
	F_0+\sum_{i=1}^{n+1}(-1)^iF_ i^{(i-1)}=C\, ,
	\label{fequation}
\end{equation}
for some integration constant $C$. This is the differential equation which determines $f(r)$ in each case. In order to determine its order, let us consider the Bianchi identity, $\nabla_{\mu}\mathcal{E}^{\mu\nu}=0$, in the case of the metric \req{Fmetric}. The $\nu=r$  component reads
\begin{equation}
	\frac{ d \mathcal{E}^{rr}}{dr}+\left(\frac{2}{r}-\frac{1}{2}f^{-1}f'\right)\mathcal{E}^{rr}+\frac{1}{2}f f'\mathcal{E}^{tt}-\frac{f}{r}g_{ij}\mathcal{E}^{ij}\, ,
\end{equation}
where $i,j$ are the angular components. Since all the components of $\mathcal{E}^{\mu\nu}$ contain derivatives up to order $2n+4$ and this identity relates the derivative of $\mathcal{E}^{rr}$ to the rest of components (without derivatives), we must conclude that in fact $\mathcal{E}^{rr}$ contains derivatives up to order $2n+3$.
Now, in \req{fequation} we have integrated the equation once, so the order of the equation is reduced yet another order. Therefore, \req{fequation} is in general of order $2n+2$, which means two orders less than the equations determining $N(r)$ and $f(r)$ in the general case. Naturally, in $ \mathcal{L}(g^{\mu\nu},R_{\mu\nu\rho\sigma})$ theories, for which $n=0$,  (\ref{fequation}) becomes a differential equation of order 2 or less --- see \req{fff} below. 

Finally, in order to complete the proof of the first part of Theorem \ref{theo}, we are going to see that the integration constant $C$ in \req{fequation} is proportional to the ADM mass.  Let us first review how this mass is computed in higher-derivative gravity. In the asymptotically flat case, it turns out that the ADM mass is given by the same prescription as in Einstein gravity \cite{Arnowitt:1960es,Arnowitt:1960zzc,Arnowitt:1961zz,Deser:2002jk}, and for the metric \req{Fmetric} it reduces to the well-known expression
\begin{equation}
M=\frac{(D-2)\Omega_{(D-2)}}{16\pi G}\lim_{r\rightarrow\infty} r^{D-3}\left(\frac{1}{f(r)}-1\right)\, .
\label{massC}
\end{equation}
Hence, $M$ is identified, up to a constant, with the coefficient of the term $1/r^{D-3}$ in $f(r)$ when $r\rightarrow \infty$. Now, the same relation holds in the asymptotically AdS (or dS) case, with the difference that we must trade $G$ by the effective Newton's constant $G_{\rm eff}$ \cite{Deser:2002jk,Senturk:2012yi}, that was introduced in the previous chapter. Thus, the ADM mass is always identified by looking at the asymptotic expansion of $f(r)$, that must take the form

\begin{equation}\label{eq:asymptoticf}
f(r)=1-\frac{2\Lambda_{\rm eff}}{(D-1)(D-2)} r^2-\frac{16 \pi G_{\rm eff} M}{(D-2) \Omega_{(D-2)} r^{D-3}}+\mathcal{O}(r^{2-D})\, ,
\end{equation}
where $\Lambda_{\rm eff}$ (that we defined in \req{eq:lambdaeff}) is the effective cosmological constant. Then, we aim to solve Eq.~\req{fequation} asymptotically in order to identify $C$. We will make use of the second statement of this theorem, but note that this is not tautological because the identification of $C$ is irrelevant to prove the second part. According to this statement, the linearized equations of the theory are Einstein-like and therefore they are given by 
\begin{equation}
\mathcal{E}^{L}_{\mu\nu}=\frac{1}{16\pi G_{\rm eff}} G_{\mu\nu}^{L}\, ,
\end{equation}
where the linearized Einstein tensor is given in \req{eq:GL}. Then, assuming that the solution tends asymptotically to the maximally symmetric vacuum,\footnote{This is, $f(r)$ behaves as $f(r)=1-\frac{2\Lambda_{\rm eff}}{(D-1)(D-2)} r^2+h(r)$ where $h(r)\rightarrow 0$ when $r\rightarrow\infty$} we will have $\mathcal{E}_{\mu\nu}\rightarrow\mathcal{E}^{L}_{\mu\nu}$ when $r\rightarrow\infty$. Using then \req{eq:chain} we get

\begin{equation}
	\delta_N S\Big|_{N=1}=\frac{\Omega_{(D-2)}}{8\pi G_{\rm eff}}\int dt \int dr  \delta N  \frac{r^{D-2}}{f(r)}G_{tt}^{L} =0\, .
\end{equation}
Now we can evaluate the linearized Einstein tensor asymptotically, which yields
\be
\frac{r^{D-2}}{f(r)}G_{tt}^{L} =\frac{d}{dr}\left[-\frac{1}{(D-1)}\Lambda_{\rm eff} r^{D-1}-\frac{(D-2)}{2}(f-1)r^{D-3}\right]\, ,
\ee
and comparing with \req{eq:deltaN}, we identify the left-hand side of Eq.~\req{fequation}. Thus, the asymptotic form of that equation is 
\begin{equation}
\frac{1}{16\pi G_{\rm eff}}\left[-\frac{2}{(D-1)}\Lambda_{\rm eff} r^{D-1}-(D-2)(f-1)r^{D-3}+\ldots\right]=C\, ,
\end{equation}
and we see that $f(r)$ is given by \req{eq:asymptoticf}, where we identify
\begin{equation}
C=\frac{M}{\Omega_{(D-2)}}\, .
\label{Cmass}
\end{equation}
Hence, $C$ is universally identified with the ADM mass of the spacetime in all cases. Strictly speaking, we derived this relation only for the theories of the form $\mathcal{L}(g^{\mu\nu},R_{\mu\nu\rho\sigma})$, because, among other issues, the definition of energy  has been carried out in full generality only for those theories \cite{Deser:2002jk,Senturk:2012yi}. However, there is no reason to expect that the formula \req{Cmass} will not extend to the general case $\mathcal{L}(g^{\mu\nu},R_{\mu\nu\rho\sigma},\nabla_{\alpha}R_{\mu\nu\rho\sigma},\ldots)$.

In sum, $f(r)$ satisfies a differential equation of order $\le 2n+2$ that can always be written as
\begin{equation}
F_0-F_1+F_2'-\ldots=\frac{M}{\Omega_{(D-2)}}\, .
\end{equation}

\subsubsection{Proof of part 2}
We provide here a direct proof of part 2. The idea is simply to impose the hypothesis $\delta L_{f}/\delta f=0$ at the level of the linearized equations. Observe first that this condition is equivalent to 
\be\label{eq:condeq}
\mathcal{E}_{tt}+f^2 \mathcal{E}_{rr}=0\, ,
\ee
\ie we demand that this combination of equations vanishes identically for any metric of the form \req{Fmetric}. It is more convenient to perform a change of variable so that we rewrite metric \req{Fmetric} as 
\be
ds^2=-f(r)du^2-2 du dr+r^2d\Omega_{(D-2)}^2\, .
\ee
where we introduced the coordinate $u=t+r_{*}$, where $dr_{*}=dr/f(r)$. In this coordinate system, the condition \req{eq:condeq} is simply
\be
\tilde{\mathcal{E}}_{rr}=0, 
\ee
where the tilde is used to distinguish the two different coordinate systems. We will apply this condition at the linear level on the maximally symmetric vacuum. Thus, we assume that
\be
f(r)=1-\frac{2\Lambda_{\rm eff}}{(D-1)(D-2)} r^2+h(r)\, ,
\ee
where $h(r)$ is treated as a perturbation, and we use it to compute the component $\tilde{\mathcal{E}}^{L}_{rr}$ of the linearized equations using \req{lineareqs}. The result is
\begin{align}
\tilde{\mathcal{E}}^{L}_{rr}=&(4 b+c) \left[(D-2) \left(\frac{3 (D-3) h}{r^4}-\frac{2 (D-4) h'}{r^3}+\frac{(D-7)h''}{2 r^2}+\frac{h^{(3)}}{r}\right)+\frac{h^{(4)}}{2}\right]\\ \nonumber
   &+(2 a+c) \left[(D-2)\left(\frac{(D-3) h}{r^4}-\frac{(D-4) h'}{r^3}+\frac{(D-5) h''}{2r^2}+\frac{h^{(3)}}{r}\right)+\frac{h^{(4)}}{2}\right]\, .
\end{align}
Now, since this must vanish identically for any function $h(r)$ we conclude that the parameters of the linearized equations must satisfy $2a+c=4b+c=0$, but these are precisely the conditions for Einstein-like theories that we determined in Sec.~\ref{Classification}. QED.

\subsection{Single-function solutions with a matter source}\label{sec:Fmatter}
Here we provide a sketch of the proof of Corollary \ref{col} by trying to construct the solution explicitly. The idea is that, unlike the case of $f(R)$, that we encounter in Sec.~\ref{unnatural}, there is no obstruction in constructing single-function solutions when there is a spherically symmetric matter distribution, and we describe how one could in principle construct these solutions for general theories. A more solid analysis should be performed on a case-by-case basis, but that would be far beyond the aim of the present work. 

Let us start by adding some minimally coupled matter to the gravity action \req{higherd}, $S\rightarrow S + S_{\ssc \rm matter}$, where $S_{\ssc \rm matter}= \int d^Dx \sqrt{|g|}\, L_{\ssc \rm matter} $. The field equations would read now $\mathcal{E}_{\mu\nu}=\frac{1}{2}T_{\mu\nu}$, where the matter stress-energy tensor is defined as usual,
$
T_{\mu\nu}=-\frac{2}{\sqrt{|g|}}\frac{\delta S_{\ssc \rm matter}}{\delta g^{\mu\nu}}.
$
Using the reduced gravitational action as we did in the proof of Theorem \ref{theo}, the equations for the metric functions $f$ and $N$  can be written as
\begin{align}
\label{eq:sphmat1}
&\left(F_0+\sum_{i=1}^{n+1}(-1)^iF_ i^{(i-1)}\right)'+\mathcal{O}(N'^2/N)=r^{D-2} f N^2T^{tt}\, ,\\
\label{eq:sphmat2}
&\frac{\delta S[N,f]}{\delta f}=\Omega_{(D-2)}\frac{r^{D-2}}{2}\left(N^3T^{tt}+\frac{N}{f^2}T^{rr}\right)\, .
\end{align}
In addition to these equations, one would need to specify the equation of state of matter. 
We are interested in a compact, spherically symmetric and static source of radius $r_0$, so that $T_{\mu\nu}( r )=0$ if $r>r_0$. 
Let us first consider the exterior region $r>r_0$. In that case, as we have just seen, $N(r)=1$ solves Eq.~\req{eq:sphmat2} and we only need to solve the other one. Integrating Eq.~\req{eq:sphmat1} from some radius $r>r_0$ up to $r=0$, we get
\begin{equation}
F_0(r)+\sum_{i=1}^{n+1}(-1)^iF_ i^{(i-1)}(r)=\frac{1}{\Omega_{(D-2)}}\int_{r<r_0}d^{D-1}x\sqrt{|g|}\left[f N T^{tt}-\frac{\mathcal{O}(N'/N^2)}{N r^{D-2}}\right]\, ,
\label{FequationT}
\end{equation}
where the rhs is an integral over the matter distribution, and thus, it is independent of $r$. Therefore, \req{fequation} holds outside the source but now the constant $C$ is not arbitrary, but determined by the mass distribution. Interestingly, since $C$ is related to the ADM mass according to \req{Cmass}, this result provides us with a formula for the ADM mass expressed as an integral over the matter distribution. Observe that \req{FequationT} is analogous to the expression \req{contradiction} that we found in the case of $f(R)$ gravity. The crucial difference is that the existence of single-function solutions in the case of $f(R)$ required the right-hand side of \req{contradiction} to be zero, which was an inconsistent condition. Here, however, we can find single-function exterior solutions for arbitrary values of the right-hand-side of \req{FequationT}. Thus, we find no obstruction for the construction of exterior solutions of the form \req{Fmetric} for the theories that satisfy the hypothesis of Theorem \ref{theo}. 

In order to complete the demonstration one would need to prove that these single-function solutions can be smoothly glued to an interior solution. This is the part that is highly theory-dependent and we cannot provide a rigorous proof without further assumptions on the form of the field equations. However, we can provide evidence that this construction is possible by counting the number of integration constants that are required in the problem. Let us for simplicity focus on the case $\mathcal{L}(g^{\mu\nu},R_{\mu\nu\rho\sigma})$, so that the field equations are generically of fourth order. 
We first consider the exterior solution. The equation \req{FequationT} for $f(r)$ is of second order, so that for every value of the rhs there is a two-parameter family of solutions.  Typically, imposing the solution to have the correct asymptotic behaviour fixes one integration constant, leaving one free parameter.\footnote{We will see this explicitly in Chapters~\ref{Chap:4} and \ref{Chap:5}. See also Refs.~\cite{PabloPablo2,Hennigar:2017ego,Ahmed:2017jod}.} Now we turn our attention to the interior solution. First, note that the component $tt$ of the equations is of fourth-order, but as we saw, the $rr$ component is of third order. Deriving the last one and extracting $f^{(4)}$ it is possible to rewrite the $tt$ component using at most third-order derivatives of $f$. Thus, at the end we have a third-order equation for $f$ and a fourth-order equation for $N$, and in order to fix a solution we need to set seven conditions. Now, at $r=r_0$ the interior solution must be smoothly glued with the exterior one. This means, in particular, that we must demand continuity and differentiability of the function $N(r)$. Since it satisfies a fourth-order equation and in the exterior region we have $N(r)=1$, we must impose  $N(r_0)=1$, $N'( r_0)=N''(r_0)=N^{(3)}(r_0)=0,$ so that we fix four conditions. On the other hand, we demand regularity of the solution at the core, and this implies that $f(0)=1$, $f'(0)=0$.  Since we already have six conditions, the solution contains at this stage only one free parameter. Finally, we also demand continuity and differentiability of $f(r)$ at $r=r_0$, so this fixes the boundary conditions for the exterior solution. Then, the free parameter is chosen so that the exterior solution has the correct asymptotic behaviour. In sum, we see that it would be in principle possible to construct solutions with a spherically symmetric matter source with exterior metric given by \req{Fmetric}. In the case of $\mathcal{L}($Riemann) gravity, we need to impose seven conditions, and this coincides with the number of integration constants of the system of differential equations, so there might be unicity of solutions. However, unicity is not expected on general grounds if we increase the order of the equations. 

\section{Construction of theories}\label{sec:construction}
The central result of this chapter is Theorem \ref{theo}, which characterizes a very interesting class of theories for which the problem of finding spherically symmetric solutions is drastically simplified and whose linearized equations are of second order.
The proof of Theorem \ref{theo} provides us a with a useful procedure to characterize these theories and to find their spherically symmetric solutions. We detail this procedure next, and afterwards we apply it to find the $D$-dimensional quadratic and cubic gravities of this type. Finally, focusing on the four-dimensional case, we will see that it is possible to find these theories at any order in curvature. 

%which serves as an illustration of the method, and provides further evidence for Conjectures \ref{conj} and \ref{conji2} .

\subsection{A recipe}\label{recipe}
The results obtained in the previous section provide a very simple and efficient method for identifying higher-derivative gravities with simple black hole solutions of the form \req{Fmetric}, and for characterizing those solutions. Our method is a refinement of an often utilized procedure, \eg in \cite{Palais:1979rca,Deser:2003up,Quasi,PabloPablo2,Dehghani:2011vu,Cisterna:2017umf,Hennigar:2017ego}, consisting in evaluating $L_{N,f}$ and performing a repeated integration by parts in the aim of bringing it to the form \req{acc} for a particular combination of couplings. 

Let us now present our method in the form of a recipe ready to be applied to any higher-derivative gravity. From a computational perspective, our procedure is considerably faster than the one just described. It involves a on-shell evaluation of the higher-derivative action, computing the Euler-Lagrange equation of a one-dimensional Lagrangian, writing an expression as a total derivative of another function (which is guaranteed to be possible), and computing some derivatives. Here is the recipe:

\begin{enumerate}
	\item Evaluate the gravity Lagrangian on the single-function ansatz \req{Fmetric}, namely, $L_f(r,f',f'',\ldots)\equiv r^{D-2}\mathcal{L}|_{g^{\mu\nu}=g^{\mu\nu}_f}$.
	\item Compute the Euler-Lagrange equation of $f(r)$ for the effective Lagrangian $L_f(r,f',f'',\ldots)$.
	\item Fix the higher-derivative couplings in a way such that this equation is identically satisfied, \ie impose $\delta L_f/\delta f=0$ for all $f(r)$.
	\item Find $F_0$, namely, the function of $f(r)$ satisfying $L_f=F_0'$. 
	\item Substitute the general ansatz \req{Nf} in the corresponding gravity Lagrangian: $L_{N,f}(r,f, N,f', N', \ldots)$ $\equiv N( r )r^{D-2}\mathcal{L}\big|_{g^{\mu\nu}=g^{\mu\nu}_{N,f}}$. The result should take the form \req{homogeneus}, where now $L_f=F_0'$. 
	\item Identify the functions $F_i$ by inspection. 
	\item Plug $F_0$ and the corresponding derivatives of the $F_i$ in \req{fequation}. This is the equation that determines $f(r)$.
\end{enumerate}
The first three steps select, from all the possible theories considered originally, the ones which allow for single-function solutions of the form \req{Fmetric}. The last four allow one to determine the differential (or algebraic) equation which needs to be solved in order to determine $f(r)$ for the corresponding theory.  Alternatively, once the theory is identified, the equation for $f(r)$ can be explicitly written as

\begin{equation}
\Omega_{(D-2)}\int dr\frac{\delta L_{N,f}}{\delta N}\bigg|_{N=1}=M\, ,
\end{equation}
where it should be possible to perform the integration formally, since $\frac{\delta L_{N,f}}{\delta N}\big|_{N=1}$ must be a total derivative, and $M$ is the mass of the solution. 

\subsection{Quadratic gravities}\label{quadratic}
In order to illustrate this method, let us apply it to the $D$-dimensional quadratic theory
\begin{equation}
\mathcal{L}_{\rm \ssc quadratic}=\frac{1}{16\pi G}\left[-2\Lambda+R+ \alpha_1 R^2+\alpha_2 R_{\mu\nu}R^{\mu\nu}+\alpha_3 R_{\mu\nu\rho\sigma}R^{\mu\nu\rho\sigma} \right]\, .
\end{equation}
Evaluating the Lagrangian on the single-function metric ansatz, we obtain the effective Lagrangian
\begin{equation}
\begin{aligned}
L_f=&\frac{1}{16\pi G}\Big[-2\Lambda r^{D-2}+(D-2)(D-3) r^{D-4} (f-1)+2(D-2)r^{D-3}f'+r^{D-2}f''\\
&+\alpha_1 r^{D-6}\left((D-2)(D-3)(f-1)+2(D-2)rf'+r^2f''\right)^2\\
&+\alpha_2r^{D-6}\left((D-2)((D-3)(f-1)+rf')+((D-2)rf'+r^2f'')^2/2\right)\\
&+\alpha_3r^{D-6}\left(2(D-2)(D-3)(f-1)^2+2(D-2)r^2f'^2+r^4 f''^4\right)\Big].
\end{aligned}
\end{equation}
From this, it is straightforward to compute the Euler-Lagrange derivative, which yields
\begin{equation}
\begin{aligned}
\frac{\delta L_f}{\delta f}=&\frac{(D-2)}{16\pi G}\left[ (3\alpha_1+\alpha_2+\alpha_3)\left(4(D-3)(f-1)-2r^2f''\right)\right.\\
&\left.+(2\alpha_1+\alpha_2+2\alpha_3)\left((D-4)r^2 f''+2r^2f^{(3)}+r^4 f^{(4)}\right)\right]\,.
\end{aligned}
\end{equation}
 Then, applying the third step of the recipe, we find that
imposing $\delta L_f/ \delta f= 0$ $ \forall f(r)$ fixes $\alpha_1=\alpha_3=-\alpha_2/4=\alpha$, which, unsurprisingly, leads to the usual Gauss-Bonnet combination $\mathcal{X}_4= R^2-4 R_{\mu\nu}R^{\mu\nu}+ R_{\mu\nu\rho\sigma}R^{\mu\nu\rho\sigma}$. Thus, the only quadratic Lagrangian satisfying the hypothesis of Theorem \ref{theo} is
\begin{equation}
\mathcal{L}_{\rm \ssc quadratic}=\frac{1}{16\pi G}\left[-2\Lambda+R+ \alpha \mathcal{X}_{4} \right]\, .
\end{equation}
Having fixed these couplings, we can compute $F_0$ for this theory, which turns out to read
\begin{equation}
\begin{aligned}
16&\pi GF_0=(D-2)r^{D-3}(1-2r^2\Lambda/((D-2)(D-1))-f)\\
&+f'(2(D-3)(D-2) r^{D-4}(f-1)\alpha-r^{D-2}) +(D-4)(D-3)(D-2)r^{D-5} (f-1)^2\alpha\, .
\end{aligned}
\end{equation}
The next step is to evaluate the Lagrangian in the general metric ansatz with two functions. Amusingly, the effective Lagrangian $L_{N,f}$, which in general takes the form \req{homogeneus}, does not contain any $\mathcal{O}(N'^2/N)$ term in this case, and is simply given by \req{siss},
where
\begin{align}\notag
16\pi GF_1&=r^{D-5}(-3f' r^3+2(D-3)(D-2)(5f-3)f' r \alpha\\
&-2(D-2)f (r^2-2(D-4)(D-3)(f-1)\alpha))\, ,\\ \nonumber
&&\\
16\pi GF_2&=-2r^{D-4}f(r^2+2(D-3)(D-2)\alpha(1-f))\, .
\end{align}
This is all we need to determine the equation of $f(r)$, \req{Fmetric}, which in this case (and for any $\mathcal{L}($Riemann) theory satisfying the hypothesis of Theorem \ref{theo}) reads: 
\begin{equation}\label{fff}
 F_0-F_1+F_2'=\frac{M}{\Omega_{(D-2)}}\, ,
\end{equation}
where we have taken into account \req{Cmass}. Explicitly, one finds 
\begin{equation}\label{GB}
r^{D-3}\left((D-2)(1-f)-\frac{2r^2\Lambda}{(D-1)}\right)+\alpha (D-4)(D-3)(D-2) r^{D-5}(f-1)^2=\frac{16\pi G M}{\Omega_{(D-2)}}\,,
\end{equation}
This equation can be solved for $f(r)$ to yield
\begin{equation}
f(r)=1+\frac{r^2}{2(D-3)(D-4)\alpha}\left[1\mp \sqrt{1+\frac{8\alpha \Lambda (D-3)(D-4)}{(D-2)(D-1)}+\frac{64 \alpha\pi G M(D-3)(D-4)}{(D-2)\Omega_{(D-2)}r^{D-1}}} \right]\,.
\end{equation}
There are two different solutions, but note that only the one with the $-$ sign reduces to the Schwarzschild solution when $\alpha\rightarrow 0$, so we should choose that one. This is the well-known $D$-dimensional static and spherically-symmetric Gauss-Bonnet black hole.\footnote{Observe that for $D=4$ and $D=3$, it reduces to the usual Schwarzschild-(A)dS solution.}

It is easy to check explicitly that $M$ is in fact the total mass in any dimension and for any asymptotic behavior. Indeed, we can see that, when $r\rightarrow\infty$, the function $f(r)$ takes the form \req{eq:asymptoticf}, provided that we identify
\begin{eqnarray}
\Lambda_{\rm eff}&=&\frac{(D-1)(D-2)}{4(D-3)(D-4)\alpha}\left[-1+\sqrt{1+\frac{8\alpha \Lambda (D-3)(D-4)}{(D-2)(D-1)}}\right]\, ,\\
G_{\rm eff}&=&\frac{G}{\sqrt{1+\frac{8\alpha \Lambda (D-3)(D-4)}{(D-2)(D-1)}}}\, . 
\end{eqnarray}
Using \req{eq:vaceqECG} and \req{eq:GeffECG} one can check that these are, in fact, the effective cosmological constant and effective Newton's constant of the theory, so that $M$ is truly the ADM mass according to the prescription of \cite{Deser:2002jk,Senturk:2012yi}.

In addition, we note that the equation \req{GB} is algebraic, instead of a second-order differential equation. This is expected in the present case, because the proof of Theorem \ref{theo}  tells us that this equation is reduced at least in two orders with respect to the general order of equations of motion. Of course, Gauss-Bonnet gravity possesses second-order equations, so the equation for $f$ must be algebraic.  The same occurs for general Lovelock theories \cite{Wheeler:1985nh,Wheeler:1985qd,Boulware:1985wk,Cai:2001dz,Dehghani:2009zzb,deBoer:2009gx}. Remarkably, there are also theories with general fourth-order equations of motion for which the equation \req{fequation} becomes algebraic too. This is the case of Quasi-topological theories \cite{Quasi,Quasi2,Dehghani:2011vu,Cisterna:2017umf}.  Interestingly, the algebraicity of \req{fff} (which is a result of the non-trivial cancellation of the different terms in $F_0$, $F_1$ and $F_2'$ involving derivatives of $f(r)$)  appears to be related to the absence of $\mathcal{O}(N'^2/N)$ terms in $L_{N,f}$. This is, very likely, a general feature which we encapsule in Conjecture \ref{conji2}. 

However, as we are going to see, in most cases the equation for $f$ is indeed differential \cite{PabloPablo2,Hennigar:2017ego}. As for Conjecture \ref{conj}, this is trivial when the equation is algebraic, because we know the exact form of the solution and we can determine the thermodynamics of black holes straight away. It is far from trivial in the differential case, but we will see that it works by examining many examples in the next chapters. 
%this is not the general case, but not in other cases like Einsteinian cubic gravity \cite{PabloPablo2} or the recently constructed Generalized quasi-topological gravity \cite{Hennigar:2017ego} --- see below.

\subsection{Six-derivative gravities}\label{cubic}
As happened when we studied the linear spectrum of higher-curvature theories in Sec.~\ref{sec:ECG}, the behaviour of quadratic theories is independent of the dimension. In particular, the Gauss-Bonnet invariant $\mathcal{X}_4$ was the only quadratic term having Einstein-like linearized equations, and it is also the only term satisfying the hypothesis of Theorem \ref{theo}. However, as we move to higher-orders in the curvature, the analysis is different depending on the dimension. Thus, we consider here the case of six-derivative terms focusing first on the case $D=4$, and commenting later the situation for arbitrary $D$. 

At the level of six derivatives, there are two non-trivial terms containing derivatives of the Riemann tensor, besides eight cubic curvature invariants. Thus, the most general six-derivative Lagrangian reads
\begin{align}\label{eq:sixderiv}
\mathcal{L}^{(3)}&=\beta_1\tensor{R}{_{\mu}^{\rho}_{\nu}^{\sigma}}\tensor{R}{_{\rho}^{\alpha}_{\sigma}^{\beta}}\tensor{R}{_{\alpha}^{\mu}_{\beta}^{\nu}}+\beta_2 \tensor{R}{_{\mu\nu}^{\rho\sigma}}\tensor{R}{_{\rho\sigma}^{\alpha\beta}}\tensor{R}{_{\alpha\beta}^{\mu\nu}}+\beta_3 \tensor{R}{_{\mu \nu\rho\sigma}}\tensor{R}{^{\mu \nu\rho}_{\alpha}}R^{\sigma \alpha}\\ \nonumber
&+\beta_4\tensor{R}{_{\mu\nu\rho\sigma}}\tensor{R}{^{\mu\nu\rho\sigma}}R+\beta_5\tensor{R}{_{\mu\nu\rho\sigma}}\tensor{R}{^{\mu\rho}}\tensor{R}{^{\nu\sigma}}+\beta_6\tensor{R}{_{\mu}^{\nu}}\tensor{R}{_{\nu}^{\rho}}\tensor{R}{_{\rho}^{\mu}}+\beta_7R_{\mu\nu }R^{\mu\nu }R\\ \nonumber
&+\beta_8R^3+\beta_9 \nabla_{\sigma}R_{\mu\nu} \nabla^{\sigma}R^{\mu\nu}+\beta_{10}\nabla_{\mu}R\nabla^{\mu}R\, .
\end{align}
Apparently, one could add other terms such as $\nabla_{\alpha}\tensor{R}{_{\mu\nu\rho\sigma}}\nabla^{\alpha}\tensor{R}{^{\mu\nu\rho\sigma}}$, but they can be reduced to a combination of the terms that already appear in \req{eq:sixderiv}. 

Next we apply the recipe \ref{recipe} to this Lagrangian in order to identify the theories satisfying Theorem \ref{theo}. The analysis of the resulting field equations and their solutions will be the object of other chapters. 

\subsubsection{Four dimensions}

Let us first consider the case $D=4$, which will be specially important for us. When we evaluate $L_f=r^2\mathcal{L}^{(3)}\big|_{g^{\mu\nu}_f}$ and we demand $\delta L_f/\delta f\equiv 0$, we find that the terms with covariant derivatives of the curvature must vanish, $\beta_9=\beta_{10}=0$, while the cubic couplings must satisfy four constraints, that can be expressed as
\begin{equation}\label{eq:constrgqg4}
\begin{aligned}
\beta_4&=\frac{3 \beta _1}{56}-\frac{9 \beta _2}{14}-\frac{\beta _3}{4}\, ,\\
\beta_5&=-\frac{3 \beta _1}{7}-\frac{48 \beta _2}{7}-2 \beta _3\, ,\\
\beta_7&=\frac{3 \beta _1}{14}+\frac{24 \beta _2}{7}+\frac{\beta _3}{2}-\frac{3 \beta _6}{4}\, ,\\
\beta_8&=-\frac{3 \beta _1}{56}-\frac{5 \beta _2}{14}+\frac{\beta _6}{8}\, .
\end{aligned}
\end{equation}
The solution of this system of equations leaves us, in principle, with four independent cubic densities. However, we must take into account that not all of the cubic terms in \req{eq:sixderiv} are independent  in $D=4$. Indeed, there are two combinations that vanish identically, and these can be chosen as the cubic Euler density
\begin{equation}\label{eq:cubicLove2}
\begin{aligned}
\mathcal{X}_6=&-8\tensor{R}{_{\mu}^{\rho}_{\nu}^{\sigma}}\tensor{R}{_{\rho}^{\alpha}_{\sigma}^{\beta}}\tensor{R}{_{\alpha}^{\mu}_{\beta}^{\nu}}+4\tensor{R}{_{\mu\nu}^{\rho\sigma}}\tensor{R}{_{\rho\sigma}^{\alpha\beta}}\tensor{R}{_{\alpha\beta}^{\mu\nu}}-24\tensor{R}{_{\mu \nu\rho\sigma}}\tensor{R}{^{\mu \nu\rho}_{\alpha}}R^{\sigma \alpha}+3\tensor{R}{_{\mu\nu\rho\sigma}}\tensor{R}{^{\mu\nu\rho\sigma}}R\\
&+24\tensor{R}{_{\mu\nu\rho\sigma}}\tensor{R}{^{\mu\rho}}\tensor{R}{^{\nu\sigma}}+
16\tensor{R}{_{\mu}^{\nu}}\tensor{R}{_{\nu}^{\rho}}\tensor{R}{_{\rho}^{\mu}}-12R_{\mu\nu }R^{\mu\nu }R+R^3,
\end{aligned}
\end{equation}
and 

\begin{equation}
\begin{aligned}
\tensor{R}{_{[\mu\nu}^{\mu\nu}}\tensor{R}{_{\rho\sigma}^{\rho\sigma}}\tensor{R}{_{\alpha]}^{\alpha}}=&
-\frac{2}{15}\tensor{R}{_{\mu \nu\rho\sigma}}\tensor{R}{^{\mu \nu\rho}_{\alpha}}R^{\sigma \alpha}+\frac{1}{30} R_{\mu\nu\rho\sigma }R^{\mu\nu\rho\sigma }R+\frac{4}{15}R_{\mu\nu\rho\sigma }R^{\mu\rho}R^{\nu\sigma}\\
&+\frac{4}{15}\tensor{R}{_{\mu}^{\nu}}\tensor{R}{_{\nu}^{\rho}}\tensor{R}{_{\rho}^{\mu}}-\frac{4}{15}R_{\mu\nu }R^{\mu\nu }R+\frac{1}{30}R^3\, ,
\end{aligned}
\end{equation}
which is obviously zero due to the antisymmetrization. Since these terms vanish identically, they satisfy trivially the hypothesis of Theorem \ref{theo}, and one can check that they in fact solve the equations in \req{eq:constrgqg4}. Thus, there are only two independent and non-trivial solutions of these equations. The two independent terms can be chosen as
\begin{align}\label{eq:GQG4D}
\hskip-0.5cm\mathcal{P}&=12 \tensor{R}{_{\mu}^{\rho}_{\nu}^{\sigma}}\tensor{R}{_{\rho}^{\alpha}_{\sigma}^{\beta}}\tensor{R}{_{\alpha}^{\mu}_{\beta}^{\nu}}+\tensor{R}{_{\mu\nu}^{\rho\sigma}}\tensor{R}{_{\rho\sigma}^{\alpha\beta}}\tensor{R}{_{\alpha\beta}^{\mu\nu}}-12R_{\mu\nu\rho\sigma}R^{\mu\rho}R^{\nu\sigma}+8\tensor{R}{_{\mu}^{\nu}}\tensor{R}{_{\nu}^{\rho}}\tensor{R}{_{\rho}^{\mu}}\, ,\\
\mathcal{C}&=\tensor{R}{_{\mu \nu\rho\sigma}}\tensor{R}{^{\mu \nu\rho}_{\alpha}}R^{\sigma \alpha}-\frac{1}{4}\tensor{R}{_{\mu\nu\rho\sigma}}\tensor{R}{^{\mu\nu\rho\sigma}}R-2\tensor{R}{_{\mu\nu\rho\sigma}}\tensor{R}{^{\mu\rho}}\tensor{R}{^{\nu\sigma}}+\frac{1}{2}R_{\mu\nu }R^{\mu\nu }R\, .
\end{align}
Quite surprisingly, we find again the Einsteinian cubic gravity term $\mathcal{P}$ that we described in the previous chapter (Sec.~\ref{sec:ECG}). Let us recall that this term was identified by the condition of possessing Einstein-like linearized equations in arbitrary dimensions. What we see is that, in addition, in $D=4$ it also possesses solutions of the form \req{Fmetric}, since it satisfies the hypothesis of Theorem \ref{theo}. On the other hand, $\mathcal{C}$ is a new density that was firstly identified in \cite{Hennigar:2017ego}. 
Thus, the most general theory up to six-derivative terms possessing the interesting properties of Theorem \ref{theo} in four dimensions reads

\begin{equation}
S=\frac{1}{16\pi G}\int d^4x\sqrt{|g|}\left[-2\Lambda+R+\mu \mathcal{P}+\lambda \mathcal{C}\right]
\end{equation}

%\begin{equation}
%\label{fequation}
%\begin{aligned}
%-\frac{1}{3}\Lambda r^3-(f-1)r+\mu \bigg[2f'^3+6\frac{f'^2}{r}-12f(f-1)\frac{f'}{r^2}-6ff''\left(f'-\frac{2(f-1)}{r}\right)\bigg]=2M\, ,
%\end{aligned}
%\end{equation}

It turns out that, although $\mathcal{C}$ is non-trivial, it makes no contribution to the equations of motion for spherically symmetric and static spacetimes. Thus, the only term that modifies the function $f(r)$ in the metric \req{Fmetric} is the Einsteinian cubic gravity density $\mathcal{P}$.

\subsubsection{Higher dimensions}
Repeating the previous analysis for $D>4$ one finds again that $\beta_{9}=\beta_{10}=0$, but now there are five constraints on the cubic couplings. Thus, in general there are three independent theories of this type.  One of them is always the cubic Euler density $\mathcal{X}_{6}$ that we wrote in \req{eq:cubicLove2}. However, this term is vanishing for $D<6$ and it is topological in $D=6$, so it only makes a non-trivial contribution to the equations of motion for $D\ge7$. 
The second density of this type is the cubic Quasi-topological gravity term $\mathcal{Z}_D$ \cite{Quasi,Quasi2}, that is given by
\begin{equation}\label{eq:QuasiTopo}
\begin{aligned}
\mathcal{Z}_{D} =
&\tensor{R}{_{\mu}^{\rho}_{\nu}^{\sigma}}\tensor{R}{_{\rho}^{\alpha}_{\sigma}^{\beta}}\tensor{R}{_{\alpha}^{\mu}_{\beta}^{\nu}}
+ \frac{1}{(2D-3)(D-4)} \Big ( 
- 3(D-2) \tensor{R}{_{\mu \nu\rho\sigma}}\tensor{R}{^{\mu \nu\rho}_{\alpha}}R^{\sigma \alpha} \\
&+ \frac{3 (3D-8)}{8}  \tensor{R}{_{\mu\nu\rho\sigma}}\tensor{R}{^{\mu\nu\rho\sigma}}R
+3D \tensor{R}{_{\mu\nu\rho\sigma}}\tensor{R}{^{\mu\rho}}\tensor{R}{^{\nu\sigma}} \\
&+ 6(D-2) \tensor{R}{_{\mu}^{\nu}}\tensor{R}{_{\nu}^{\rho}}\tensor{R}{_{\rho}^{\mu}}
- \frac{3(3D-4)}{2}  R_{\mu\nu }R^{\mu\nu }R
+ \frac{3D}{8} R^3 \Big) \, .
\end{aligned}
\end{equation}
Although this term gives rise to fourth-order equations of motion, they reduce to second-order for spherically symmetric and static metrics. In particular, this theory has solutions of the form \req{Fmetric}, where the equation for $f(r)$ \req{fequation} is algebraic.  Indeed, for $D\ge 6$ this term contributes to the equation of $f(r)$ in the same way as the Lovelock term $\mathcal{X}_6$.\footnote{In particular, in $D=6$ it makes no contribution to the equations of motion in the presence of spherical symmetry, and this was the origin of the name  ``quasi'' topological \cite{Quasi}} Thus, the primary interest of Quasi-topological gravity is in $D=5$, where $\mathcal{X}_6$ is trivial, but $\mathcal{Z}_5$ is not. Note also that the case $D=4$ of \req{eq:QuasiTopo} does not exist. In fact, there are no theories with analogous properties in four dimensions. 

Finally, there is a third independent density satisfying the hypothesis of Theorem \ref{theo}, and it can be expressed explicitly as
\begin{equation}
\begin{aligned}
\mathcal{S}_D=&14\tensor{R}{_{\mu}^{\rho}_{\nu}^{\sigma}}\tensor{R}{_{\rho}^{\alpha}_{\sigma}^{\beta}}\tensor{R}{_{\alpha}^{\mu}_{\beta}^{\nu}}
 +2\tensor{R}{_{\mu \nu\rho\sigma}}\tensor{R}{^{\mu \nu\rho}_{\alpha}}R^{\sigma \alpha}-\frac{(38-29 D+4 D^2)}{4(D-2)(2D-1)}\tensor{R}{_{\mu\nu\rho\sigma}}\tensor{R}{^{\mu\nu\rho\sigma}}R\\
&-\frac{2(-30+9D+4D^2)}{(D-2)(2D-1)}\tensor{R}{_{\mu\nu\rho\sigma}}\tensor{R}{^{\mu\rho}}\tensor{R}{^{\nu\sigma}}
-\frac{4(66-35D+2D^2))}{3(D-2)(2D-1)}\tensor{R}{_{\mu}^{\nu}}\tensor{R}{_{\nu}^{\rho}}\tensor{R}{_{\rho}^{\mu}}\\
&+\frac{(34-21D+4D^2)}{(D-2)(2D-1)}R_{\mu\nu }R^{\mu\nu }R-\frac{(30-13D+4D^2)}{12(D-2)(2D-1)} R^3\, .
\end{aligned}
\label{SD}
\end{equation}
The term $\mathcal{S}_D$ is known as the Generalized quasi-topological gravity density and it was introduced in  \cite{Hennigar:2017ego}. It has some important differences with respect to $\mathcal{X}_6$ and $\mathcal{Z}_{D}$. To begin with, in this case the equation \req{fequation} that satisfies the metric function $f(r)$ is of second order instead of algebraic --- see \cite{Hennigar:2017ego} for details. In addition, it is non-trivial in all dimensions and its contribution to the equation of $f(r)$ is (obviously) different from the contribution of $\mathcal{X}_6$ and $\mathcal{Z}_{D}$. 
Finally, when restricted to four dimensions, one finds that \cite{Hennigar:2017ego} 
\begin{equation}
\mathcal{S}_4=\mathcal{P}-\frac{1}{4}\mathcal{X}_6+4\mathcal{C}\, ,
\end{equation}
where $\mathcal{P}$ and $\mathcal{C}$ are the terms that we found in the four-dimensional case, Eq.~\req{eq:GQG4D}. Since $\mathcal{C}$ makes no contribution to the equations of the metric \req{Fmetric}, $\mathcal{S}_D$ provides a $D$-dimensional generalization of four-dimensional Einsteinian cubic gravity, in the sense of possessing solutions of the form \req{Fmetric}.

Thus, we have found the most general $D$-dimensional six-derivative theory that satisfies the hypothesis of Theorem \ref{theo}:
\begin{equation}\label{gQuasi2}
S_{\rm GQG}=\frac{1}{16\pi G}\int d^Dx\sqrt{|g|}\left[-2\Lambda+R+\alpha \mathcal{X}_4+ \gamma \mathcal{X}_6+\xi \mathcal{Z}_D+\mu \mathcal{S}_D\right]\, .
\end{equation}

We also recall that in the absence of $\mathcal{S}_D$ (or, equivalently, $\mathcal{P}$ in $D=4$) the equation that determines $f(r)$ is algebraic, and no $\mathcal{O}(N'^2/N)$ terms of the form \req{ooo} appear in $L_{N,f}$. This is no longer the case when $\mathcal{S}_D$ is included. These observations strongly support Conjecture \ref{conji2}.
Finally, let us also mention that for all the different theories contained in \req{gQuasi2}, Conjecture \ref{conj} holds.

\subsection{$D=4$ case at all orders in curvature}\label{sec:D4all}
It is possible to keep searching for this type of theories at higher orders in curvature and in arbitrary dimensions. But performing the analysis in full generality becomes more and more challenging as we increase the curvature order.   In the case of quartic terms we can still perform a complete analysis thanks to the full list of quartic invariants provided in Table~\ref{tabla2}.
In order to simplify matters, let us comment the result in $D=4$ (the case of arbitrary $D$ was studied in \cite{Ahmed:2017jod}).
One finds that several quartic Generalized quasi-topological terms exist, but remarkably enough, it turns out that all of them contribute in the same way to the equation of $f(r)$ \req{fequation}. More precisely, if $\mathcal{Q}$ and $\tilde{\mathcal{Q}}$ are two of these terms and $\mathcal{E}(r,f,f',f'')$, $\tilde{\mathcal{E}}(r,f,f',f'')$ are their respective contributions to \req{fequation}, one finds that they are proportional $\mathcal{E}\propto \tilde{\mathcal{E}}$. Thus, it is always possible to choose a basis for these terms such that there is only one of them that contributes non-trivially to the equation \req{fequation}.\footnote{Note that this does not imply that the rest of terms are trivial; their equations of motion are only trivial for spherically symmetric and static metrics.} This is analogous to the cubic case, where the term $\mathcal{P}$ contributes to \req{fequation} but $\mathcal{C}$ does not. For instance, the following quartic term is an appropriate choice
\begin{equation}
	\begin{aligned}
		\mathcal{Q}=&44R^{\mu\nu\rho\sigma}\tensor{R}{_{\mu\nu}^{\alpha\beta}}\tensor{R}{_{\rho}^{ \gamma}_{\alpha}^{\delta}}R_{\sigma \gamma \beta \delta}-5 R^{\mu\nu\rho\sigma }\tensor{R}{_{\mu\nu}^{\alpha\beta}}\tensor{R}{_{\rho\alpha}^{\gamma\delta}}R_{\sigma \beta \gamma \delta}+5 R^{\mu\nu\rho\sigma }\tensor{R}{_{\mu\nu\rho }^{\alpha}}R_{\beta \gamma \delta \sigma}\tensor{R}{^{\beta \gamma \delta}_{\alpha}}\\
		&+24 R^{\mu\nu }R^{\rho\sigma \alpha\beta }\tensor{R}{_{\rho}^{\gamma}_{\alpha \mu}}R_{\sigma \gamma \beta \nu}\, ,
	\end{aligned}
\end{equation}

Repeating the analysis at higher orders (in a non-exhaustive way, because we lack a complete set of invariants at arbitrary order), one concludes that this situation seems to be general: at every order in curvature, there is a unique way of modifying the equation of $f(r)$ \req{fequation}. Thus, for the purpose of studying the solutions of these theories, it suffices for us to find \emph{one} non-trivial term at every order in curvature satisfying $\delta L_{f}/\delta f=0$. 

A practical approach consists in constructing higher-order Lagrangians as polynomials of a reduced set of curvature invariants. Let us consider, in particular, the following set of invariants $\{R, Q_1, Q_2, C_1, C_2\}$, where

\begin{align}
Q_1\equiv R_{\mu\nu}R^{\mu\nu}\, , \hskip0.2cm  Q_2 \equiv R_{\mu\nu\rho\sigma}R^{\mu\nu\rho\sigma}\, , \hskip0.2cm C_1\equiv \tensor{R}{_{\mu}^{\rho}_{\nu}^{\sigma}}\tensor{R}{_{\rho}^{\alpha}_{\sigma}^{\beta}}\tensor{R}{_{\alpha}^{\mu}_{\beta}^{\nu}} \, , \hskip0.2cm C_2\equiv \tensor{R}{_{\mu\nu}^{\rho\sigma}}\tensor{R}{_{\rho\sigma}^{\alpha\beta}}\tensor{R}{_{\alpha\beta}^{\mu\nu}}\, .
\end{align}
Using the Ricci scalar, the two quadratic invariants, $Q_{1,2}$, and the two cubic  ones, $C_{1,2}$, we can construct $3$ independent invariants at quadratic order: $\{R^2,Q_1,Q_2 \}$; $5$ at cubic order $\{R^3, R Q_1, R Q_2,C_1,C_2\}$; $8$ at quartic order: $\{R^4, R^2 Q_1, R^2 Q_2, R C_1,R C_2, Q_1^2,Q_2^2, Q_1 Q_2 \}$; $12$ at quintic order, $\{R^5, R^3 Q_1, R^3 Q_2, R^2 C_1,R^2 C_2, R Q_1^2, R Q_2^2,R Q_1 Q_2, Q_1C_1,Q_1 C_2,Q_2 C_1,\\ Q_2 C_2 \}$; 19 at sextic order; 25 at septic order; $36$ at octic order, $45$ at nonic order, and so on. Observe that this number grows considerably slower than the total number of independent invariants at each order in curvature \cite{0264-9381-9-5-003}. As it turns out, these terms suffice to construct theories of the Generalized quasi-topological type at arbitrary orders in the curvature. Here we have the following explicit set of invariants up to $n=10$: 

  \begin{align} \label{r3}
 \mathcal{R}_{(3)}=&+\frac{1}{16}\left(R^3-32C_1+2 C_2-3 R Q_2\right) \, , \\
  \mathcal{R}_{(4)}=&-\frac{1}{384}\left(4R^4+108 Q_1^2+15 {Q_2}^2- 128R C_1+8R C_2-24R^2 Q_1 -84 Q_1 Q_2 \right)\, , \\ 
  \mathcal{R}_{(5)}=&+\frac{1}{3456}\big(5R^5+132 R Q_1^2+18 R Q_2^2-272 R^2 C_1+10 R^2 C_2-30 R^3 Q_1\\ \notag&-102 R Q_1 Q_2+552 Q_1 C_1-156 Q_2 C_1\big)\, ,\\
    \mathcal{R}_{(6)}=&-\frac{1}{24576}\big(4R^6+180 R^2 Q_1^2+33 R^2Q_2^2-384 R^3 C_1+8 R^3 C_2-24R^4Q_1\\ \notag&-156R^2 Q_1 Q_2+768 R C_1 Q_1-192 Q_1^3-12Q_2^3+1728C_1^2+64 C_1 C_2+144 Q_1^2Q_2\big)\, , 
    \end{align}\begin{align}
      \mathcal{R}_{(7)}=&+\frac{1}{2949120}\big(84R^7-504 R^5Q_1+168 R^4 C_2-5760 R^4 C_1+293 R^3 Q_2^2\\ \notag &-1676 R^3 Q_1 Q_2+2180 R^3 Q_1^2+11776 R^2 C_1 Q_1-4800 R C_1 C_2+51904 R C_1^2\\ \notag &+208 Q_2^2 C_2-4064 Q_2^2C_1-832 Q_1Q_2C_2+16640 Q_1Q_2C_1+832Q_1^2C_2-17024 Q_1^2C_1\big)\, ,
\\
      \label{r8}
           \mathcal{R}_{(8)}=&-\frac{1}{2793996288}\big(18130 R^8-108780 R^6 Q_1+36260 R^5 C_2-1023592 R^5 C_1\\ \notag&-19437 R^4 Q_2^2-31032 R^4 Q_1 Q_2+139812 R^4 Q_1^2+6515280 R^3 Q_1 C_1\\ \notag&-1881680 R^2 C_1 C_2+12416172 R Q_2^2 C_1-21222000 R Q_1 Q_2 C_1-7220688 R Q_1^2 C_1\\ \notag&-1073478 Q_2^4+6549648 Q_1 Q_2^3-13534416 Q_1^2 Q_2^2+9893184 Q_1^3 Q_2-642448 Q_2 C_1 C_2\\ \notag&+56702496 Q_2 C_1^2-870240 Q_1^4+1284896 Q_1 C_1 C_2-5812928 Q_1 C_1^2\big) \, , 
           \\
           \label{r9}
            \mathcal{R}_{(9)}=&+\frac{1}{99090432}\big(1820 R^9-29400 Q_1 R^7+6300 Q_2 R^7 -64896 C_1 R^6 \\ \notag&+ 4760 C_2 R^6+187596 Q1^2 R^5 - 90156 Q_1 Q_2 R^5 + 6999 Q_2^2 R^5+1285632 C_1 Q_1 R^4\\ \notag& - 43680 C_2 Q_1 R^4 - 640128 C_1 Q_2 R^4 + 12600 C_2 Q_2 R^4 -1208768 C_1^2 R^3 \\ \notag&- 55872 C_1 C_2 R^3 + 2240 C_2^2 R^3 - 
 767856 Q_1^3 R^3 + 855996 Q_1^2 Q_2 R^3 - 338064 Q_1 Q_2^2 R^3\\ \notag&+ 51015 Q_2^3 R^3 -5208960 C_1 Q_1^2 R^2 + 100512 C_2 Q_1^2 R^2 + 3737856 C_1 Q_1 Q_2 R^2 \\ \notag&- 66912 C_2 Q_1 Q_2 R^2 - 332448 C_1 Q_2^2 R^2 + 8328 C_2 Q_2^2 R^2 -705792 C_1^2 Q_1 R \\ \notag&+ 137472 C_1 C_2 Q_1 R + 1596672 Q_1^4 R + 5192256 C_1^2 Q_2 R - 131136 C_1 C_2 Q_2 R\\ \notag& - 2717568 Q_1^3 Q_2 R + 1580544 Q_1^2 Q_2^2 R - 340704 Q_1 Q_2^3 R + 15120 Q_2^4 R\\ \notag&+23224320 C_1^3 - 591360 C_1^2 C_2 - 53760 C_1 C_2^2 + 5117952 C_1 Q_1^3 + 111360 C_2 Q_1^3 \\ \notag&- 6398592 C_1 Q_1^2 Q_2 - 58560 C_2 Q_1^2 Q_2 + 2888448 C_1 Q_1 Q_2^2 - 24960 C_2 Q_1 Q_2^2 \\ \notag&- 484320 C_1 Q_2^3 + 13200 C_2 Q_2^3\big)\, , 
 \end{align}\begin{align}
\label{r10}
   \mathcal{R}_{(10)}=&-\frac{1}{452984832}\big(2100 R^{10}-6300 R^8 Q_2-12600 R^8 Q_1+8400 R^7 C_2\\ \notag&+3840 R^7 C_1 -46435 R^6 Q_2^2+210940 R^6 Q_1 Q_2-160 540 R^6 Q_1^2-12600 R^5 Q_2C_2\\ \notag&-266880 R^5 C_1 Q_2 -25200 R^5 Q_1 C_2-270080 R^5 Q_1 C_1+11625 R^4 Q_2^3\\ \notag& +448260 R^4 Q_1 Q_2^2-1856940 R^4 Q_1^2 Q_2+1827840 R^4 Q_1^3 +8400 R^4 C_2^2\\ \notag&-291520 R^4 C_1 C_2-194368 R^4 C_1^2-87670 R^3 C_2 Q_2^2+2265984 R^3 Q_2^2 C_1\\ \notag&+325480 R^3 Q_1 Q_2 C_2-6631296 R^3 Q_1 Q_2 C_1-300280 R^3 Q_1^2 C_2+8883456 R^3 Q_1^2 C_1\\ \notag&-201348R^2 Q_2^4+1731744 R^2 Q_1 Q_2^3-6767712 R^2 Q_2^2 Q_1^2+12733056 R^2 Q_1^3 Q_2\\ \notag&-1037760 R^2 Q_2 C_1 C_2+34126272 R^2 Q_2 C_1^2-9027648 R^2 Q_1^4+2309120 R^2 Q_1 C_1 C_2\\ \notag&-46106624 R^2 Q_1 C_1^2+51600 R Q_2^3 C_2-3286368 R Q_2^3 C_1+62400 R Q_1 Q_2^2 C_2\\ \notag&+13847808 R Q_1 Q_2^2 C_1-868800 R Q_1^2 Q_2 C_2-9561216 R Q_1^2 Q_2 C_1+1075200 R Q_1^3 C_2\\ \notag&-9977856 R Q_1^3 C_1-598400 R C_1 C_2^2+2492800 R C_1^2 C_2+113305600 R C_1^3\\ \notag&+126000 Q_2^5-705600 Q_1 Q_2^4-604800 Q_1^2 Q_2^3+10483200 Q_1^3 Q_2^2+10400 Q_2^2 C_2^2\\ \notag&-1041600 Q_2^2C_1 C_2-3990272 Q_2^2C_1^2-22176000Q_1^4 Q_2-41600Q_2 Q_1 C_2^2\\ \notag&+5260800Q_1 Q_2 C_1 C_2+11138048 Q_1 Q_2 C_1^2+14515200 Q_1^5+41600 Q_1^2 C_2^2\\ \notag&-6355200 Q_1^2 C_1 C_2+19489792Q_1^2 C_1^2\big)\, .
  \end{align}
  
We have checked that all of these terms contribute in a non-trivial way to the equation \req{fequation}, hence they capture the most general correction to that equation at every order --- we will study the black hole solutions of these theories in Chapter \ref{Chap:5}. Using the linearization method introduced in the previous chapter and the results in Sec.~\ref{quartic}, one can also check explicitly that all of these terms give rise to Einstein-like linearized equations.

\chapter{Effective Field Theory and Field redefinitions}\label{Chap:3}
In chapters \ref{Chap:1} and \ref{Chap:2} we found some special classes of higher-curvature theories with very appealing properties. In particular, in the previous chapter we described the general properties of a broad family of theories known as \emph{Generalized quasi-topological gravity} (GQG).  In these theories, the problem of finding static, spherically symmetric solutions is drastically simplified, and they have the additional property of possessing second-order linearized equations on maximally symmetric backgrounds. These theories will play a central role in the rest of this thesis and they will prove to be useful models that will allow us to test non-perturbative effects of higher-curvature gravity. 

From a more profound perspective, one would like to consider higher-derivative gravity as an effective description of an underlying UV-complete theory of gravity.  The Effective Field Theory (EFT) approach requires the introduction of all possible terms that are compatible with the symmetries of the theory, and in the case of gravity this means that one should include all diffeomorphism-invariant higher-derivative operators at every order. Since the theories described in Chapter \ref{Chap:2} involve very precise combinations of curvature invariants, one could say that they are \emph{fine-tuned} models, hence not valid for EFT.  

However, another ingredient of EFT is the possibility to perform field redefinitions. Classically, the transformed field will, in general, have different properties from the original one. But from a more fundamental perspective, both fields provide equivalent effective descriptions of the same underlying theory. This is manifest, for instance, in the path integral formalism of Quantum Field Theory, where the fields are just integration variables. Performing a field redefinition is equivalent to a change of variables, and this should leave scattering amplitudes and other observables unaffected. Thus, even though the classical fields will be different, there are certain properties that are expected to be invariant under the redefinition of fields. In the case of gravity, the thermodynamic properties of black holes --- such as temperature and entropy --- are invariant under such transformations \cite{Jacobson:1993vj}.  Thus, if our aim is to study black hole thermodynamics, we may work equivalently in the original frame or in the transformed one.

One may wonder whether the theories defined in Chapter \ref{Chap:2} are still fine-tuned when one takes into account the freedom to perform field redefinitions. In fact, in this chapter we will try to answer the following question: can we map any higher-derivative gravity to one of these theories by using field redefinitions? In that case, these theories would actually capture the most general Effective Field Theory of gravity, and they could be used, for instance, in order to learn about black hole thermodynamics in arbitrary higher-derivative theories. As we are going to see, the answer to the previous question is, most likely, positive.

%We first review the issue of redefinitions of the metric and how they change the gravitational action, understood as an effective theory. Then, we show explicitly that, up the eight-derivative level, the most general gravitational action can always be expressed as a sum of Generalized quasi-topological terms --- if one redefines the metric in an appropriate way. We provide as well other results that will help us to argue that this can always be done. 

\section{Field redefinitions in higher-derivative gravity}
Let us consider the most general metric-covariant theory of gravity\footnote{We also assume that parity is preserved so that we do not have to include terms containing the Levi-Civita symbol. Nevertheless, all the results of this chapter also apply if we include those terms.}
\begin{equation}\label{eq:generalhdg}
S=\int d^Dx\sqrt{|g|}\mathcal{L}\left(g^{\mu\nu}, R_{\mu\nu\rho\sigma}, \nabla_{\alpha}R_{\mu\nu\rho\sigma},  \nabla_{\beta}\nabla_{\alpha}R_{\mu\nu\rho\sigma},\ldots\right)\, .
\end{equation}
We will soon assume that the Lagrangian can be expanded as a series in higher-derivative terms, but before that, let us study the effects of field redefinitions on the previous theory. In particular, we want to determine how the action \req{eq:generalhdg} transforms when we redefine the metric $g_{\mu\nu}$ in terms of other metric $\tilde g_{\mu\nu}$ according to, 
\begin{equation}\label{eq:metricredef}
g_{\mu\nu}=\tilde g_{\mu\nu}+\tilde Q_{\mu\nu}\, ,
\end{equation}
where $\tilde Q_{\mu\nu}$ is a symmetric tensor constructed from $\tilde g_{\mu\nu}$. Ideally, we would like the field redefinition to be algebraic, so that the relation between $g_{\mu\nu}$ and $\tilde g_{\mu\nu}$ is functional. However, the only tensor that we can construct out of the metric without introducing derivatives is (proportional to) the metric itself. Hence, $\tilde Q_{\mu\nu}$ can be generically formed from the curvature and the redefinition \req{eq:metricredef} is differential.  
The action $\tilde S$ for the new metric $\tilde g_{\mu\nu}$ is simply obtained by substituting the redefinition in the original action
\begin{equation}
\tilde S[\tilde g_{\mu\nu}]=S[\tilde g_{\mu\nu}+\tilde Q_{\mu\nu}]
\end{equation}

However, there are some subtleties related to the fact that the field redefinition \req{eq:metricredef} involves derivatives of the metric, because it is not equivalent to extremize the action with respect to $g_{\mu\nu}$ or with respect to $\tilde g_{\mu\nu}$. Essentially, one finds that, if $g_{\mu\nu}^{\rm sol}$ is a solution of the original theory, then the relation \req{eq:metricredef} always produces a solution $\tilde g_{\mu\nu}^{\rm sol}$ of the transformed theory when we invert it. However, the converse is not true: there are some solutions of  the equations of motion of $\tilde g_{\mu\nu}$ that do not produce a solution of the original theory when we apply the map \req{eq:metricredef}. The reason is that the presence of derivatives in the field redefinition increases the number of derivatives in the equations of motion derived from $\tilde S$ and this introduces some spurious solutions that we must discard. This issue is further discussed in Appendix \ref{App:2}. Once this point is taken into account, both theories $S$  and $\tilde S$ are equivalent. 

Note that when we keep only the meaningful solutions --- those that are related by \req{eq:metricredef} ---  the on-shell action is the same,
\begin{equation}
\tilde{S}\left[\tilde{g}_{\mu\nu}^{\rm sol}\right]=S\left[g_{\mu\nu}^{\rm sol}\right]\, .
\end{equation}
Since black hole thermodynamics can be determined, in the Euclidean path integral approach, by evaluating the on-shell action \cite{Gibbons:1976ue}, this simple observation proves that black hole thermodynamics is the same in both frames. 
Alternatively, using directly Wald's entropy formula \cite{Wald:1993nt}, it can be proven that the black hole entropy is the same in both frames \cite{Jacobson:1993vj}.\footnote{In order to prove rigorously this statement, it is necessary to assume some mild conditions on $\tilde Q_{\mu\nu}$, namely, it should fall off at infinity fast enough. All the redefinitions that we will consider are well-behaved in this sense.}

Let us now determine how the redefinition of the metric \req{eq:metricredef} changes the action. For that, we assume that the redefinition is perturbative, \ie we treat $\tilde Q_{\mu\nu}$ as a perturbation and we work at linear order. This is enough for our purposes, since, following the EFT approach, we will also expand the action in a perturbative series of higher-derivative terms. Observe that in this case the relation \req{eq:metricredef} can be inverted as
\begin{equation}\label{eq:metricredefinv}
\tilde g_{\mu\nu}=g_{\mu\nu}-Q_{\mu\nu}+\mathcal{O}(Q^2)\, .
\end{equation}
where $Q_{\mu\nu}$ has the same expression as $\tilde Q_{\mu\nu}$ but replacing $\tilde g_{\mu\nu}\rightarrow g_{\mu\nu}$. 
Let us introduce the equations of motion of the original theory as 
\begin{equation}
\E_{\mu\nu}=\frac{1}{\sqrt{|g|}}\frac{\delta S}{\delta g^{\mu\nu}}\, .
\end{equation}
Then, at linear order in $\tilde Q_{\mu\nu}$, the transformed action $\tilde S$ reads
\begin{equation}\label{eq:tildeaction}
\tilde S=\int d^Dx\sqrt{|\tilde g|}\left[\tilde{\mathcal{L}}-\tilde{\E}_{\mu\nu}\tilde Q^{\mu\nu}+\mathcal{O}(Q^2)\right]\, .
\end{equation}
where the tildes denote evaluation on $\tilde g_{\mu\nu}$. Thus, the redefinition introduces a term in the action proportional to the equations of motion of the original theory \cite{Chemissany:2007he}. Let us be more explicit about the form of the Lagrangian by expanding it as a sum over all possible higher-derivative terms
\begin{equation}\label{eq:faction}
S=\frac{1}{16\pi G}\int d^Dx\sqrt{|g|}\Big\{R+\sum_{n=2}^{\infty} L^{2(n-1)}\mathcal{L}^{(n)}\Big\}\, ,
\end{equation}
where $L$ is a length scale and $\mathcal{L}^{(n)}$ represents the most general Lagrangian containing $2n$ derivatives. 
For instance, we already saw in Sec.~\ref{sec:construction} that the four- and six-derivative Lagrangians are given by
\begin{align}\label{eq:sixderiv2}
\mathcal{L}^{(2)}&=\alpha_1 R^2+\alpha_2 R_{\mu\nu}R^{\mu\nu}+\alpha_3 R_{\mu\nu\rho\sigma}R^{\mu\nu\rho\sigma}\, ,\\
\mathcal{L}^{(3)}&=\beta_1\tensor{R}{_{\mu}^{\rho}_{\nu}^{\sigma}}\tensor{R}{_{\rho}^{\alpha}_{\sigma}^{\beta}}\tensor{R}{_{\alpha}^{\mu}_{\beta}^{\nu}}+\beta_2 \tensor{R}{_{\mu\nu}^{\rho\sigma}}\tensor{R}{_{\rho\sigma}^{\alpha\beta}}\tensor{R}{_{\alpha\beta}^{\mu\nu}}+\beta_3 \tensor{R}{_{\mu \nu\rho\sigma}}\tensor{R}{^{\mu \nu\rho}_{\alpha}}R^{\sigma \alpha}\\ \nonumber
&+\beta_4\tensor{R}{_{\mu\nu\rho\sigma}}\tensor{R}{^{\mu\nu\rho\sigma}}R+\beta_5\tensor{R}{_{\mu\nu\rho\sigma}}\tensor{R}{^{\mu\rho}}\tensor{R}{^{\nu\sigma}}+\beta_6\tensor{R}{_{\mu}^{\nu}}\tensor{R}{_{\nu}^{\rho}}\tensor{R}{_{\rho}^{\mu}}+\beta_7R_{\mu\nu }R^{\mu\nu }R\\ \nonumber
&+\beta_8R^3+\beta_9 \nabla_{\sigma}R_{\mu\nu} \nabla^{\sigma}R^{\mu\nu}+\beta_{10}\nabla_{\mu}R\nabla^{\mu}R\, .
\end{align}
The number of terms grows very rapidly, and the eight derivative Lagrangian already contains 92 terms \cite{0264-9381-9-5-003}.\footnote{Ref.~\cite{0264-9381-9-5-003} provides the number of \emph{linearly independent} invariants, but presumably many of them will differ by total derivative terms, which are irrelevant for the action. So the number of relevant terms is presumably much smaller --- yet quite large.}
Let us further choose $\tilde Q_{\mu\nu}^{(k)}$ to be a symmetric tensor containing $2k$ derivatives of the metric and let us perform the following field redefinition
\begin{equation}\label{eq:metricredefk}
g_{\mu\nu}=\tilde g_{\mu\nu}+L^{2k}\tilde Q_{\mu\nu}^{(k)}\, .
\end{equation}
Then, the transformed action \req{eq:tildeaction} reads
\begin{equation}\label{eq:faction2}
\tilde S=\frac{1}{16\pi G}\int d^Dx\sqrt{|\tilde g|}\left\{\tilde R+\sum_{n=1}^{k} L^{2(n-1)}\tilde{\mathcal{L}}^{(n)}+L^{2k}\left(\tilde{\mathcal{L}}^{(k+1)}-\tilde{R}^{\mu\nu}\hat Q_{\mu\nu}^{(k)}\right)+\sum_{n=k+2}^{\infty} L^{2(n-1)}\tilde{\mathcal{L}}'^{(n)}\right\}\, ,
\end{equation}
where all quantities are evaluated on $\tilde g_{\mu\nu}$, and\footnote{We have $\E^{\mu\nu}\tilde Q_{\mu\nu}^{(k)}=G^{\mu\nu}\tilde Q_{\mu\nu}^{(k)}=R^{\mu\nu}\hat Q_{\mu\nu}^{(k)}$.}
\begin{equation}
\hat Q_{\mu\nu}^{(k)}=\tilde Q_{\mu\nu}^{(k)}-\frac{1}{2}\tilde g_{\mu\nu}\tilde Q^{(k)}\, ,\quad  \tilde Q^{(k)}=\tilde g^{\alpha\beta}\tilde Q_{\alpha\beta}^{(k)}\, .
\end{equation}
Hence, all the terms containing up to $2k$ derivatives are unaffected, while the terms with $2(k+1)$ derivatives get the correction $-\tilde{R}^{\mu\nu}\hat Q_{\mu\nu}^{(k)}$. The higher-order terms will also get corrections that depend in a more complicated way on $\tilde Q_{\mu\nu}^{(k)}$, but if the starting action already contained all possible terms, the whole effect will be to change the couplings in the Lagrangian, and we denote this with the prime $\tilde{\mathcal{L}}'^{(n)}$. From this, it is clear that if we perform this type of field redefinition order by order, starting at $k=1$, we are able to remove all terms in the action that contain Ricci curvature --- except, of course, the Einstein-Hilbert term. In other words, any term containing Ricci curvature is meaningless from the point of view of EFT and we are free to add or remove all terms of that type.  Let us mention that this result is in agreement with the construction performed in \cite{Endlich:2017tqa}. 

Notice that in \req{eq:faction} we intentionally did not include a cosmological constant. When we add it, the effect of the redefinition \req{eq:metricredefk} is 
\begin{equation}\label{eq:factioncosmo}
\begin{aligned}
\tilde S=&\frac{1}{16\pi G}\int d^Dx\sqrt{|\tilde g|}\left\{-2\Lambda+\tilde R+\sum_{n=1}^{k-1} L^{2(n-1)}\tilde{\mathcal{L}}^{(n)}+ L^{2(k-1)}\left(\tilde{\mathcal{L}}^{(k)}+\frac{2(\Lambda L^2)}{D-2}\hat Q^{(k)}\right)\right.\\
&\left.+L^{2k}\left(\tilde{\mathcal{L}}^{(k+1)}-\tilde{R}^{\mu\nu}\hat Q_{\mu\nu}^{(k)}\right)+\sum_{n=k+2}^{\infty} L^{2n}\tilde{\mathcal{L}}'^{(n)}\right\}\, ,
\end{aligned}
\end{equation}
and now it changes the terms at order $2(k+1)$ and at order $2k$. This is a complication with respect to the case without cosmological constant because now we cannot remove the terms with Ricci curvature order by order. If we remove them at a given order, the field redefinition of the next order introduces the corrections $\frac{2(\Lambda L^2)}{D-2}\hat Q^{(k)}$, that will generically include again terms with Ricci curvature. Hence, the process cannot be carried out order by order because all steps are coupled. If one wants to remove all the terms with Ricci curvature up to order $2k$, it is necessary to consider the most general field redefinition up to that order, \ie including all the terms $\tilde Q_{\mu\nu}^{(m)}$ of order $m\le k$ at the same time. Nevertheless, note that this is just a technical complication: finding the precise field redefinition that removes the terms with Ricci curvature is more involved, but it can certainly be done. Thus, the conclusions are the same as in the case without cosmological constant.

\section{Mapping any theory to Generalized quasi-topological gravity}
After having reviewed the effect of field redefinitions on the gravitational action, we turn to the main question of this chapter:  can we map the most general effective theory of the form \req{eq:faction} to a Generalized quasi-topological gravity (GQG)? We recall these are the theories that satisfy the hypothesis of Theorem \ref{theo} in the previous chapter, and that have several remarkable properties. In particular, they allow for black hole solutions characterized by a single function (of the form \req{Fmetric}) and they only propagate a massless graviton on maximally symmetric backgrounds. Furthermore, we will see in the next chapters that for these theories we can obtain closed, exact expressions for the thermodynamic properties of black holes --- a fact that was collected in Conjecture \ref{conj}. 

\subsection{Explicit map up to sixth order}\label{sec:mapsix}
The lowest-order densities of this type (with four and six derivatives) were constructed in Sec. \ref{sec:construction}. Let us then see if we can re-express the general action \req{eq:faction}  using only those terms. For simplicity, we set the cosmological constant to $0$. 
First, it is useful to write the Lagrangians $\mathcal{L}^{(2)}$ and $\mathcal{L}^{(3)}$ in \req{eq:sixderiv2} using a different basis of invariants. In the case of the four-derivative action, the Riemann squared term can be traded by the Gauss-Bonnet density,
\begin{equation}
\mathcal{X}_{4}=R^2-4R_{\mu\nu}R^{\mu\nu}+R_{\mu\nu\rho\sigma}R^{\mu\nu\rho\sigma}\, ,
\end{equation}
that, as we know, is the only component of the GQG family at quadratic order. Thus, the most general four-derivative Lagrangian can be alternatively written as\footnote{The coefficients $\alpha_i$ are not the same as in \req{eq:sixderiv2}, but for clarity we do not introduce additional unnecessary notation.}

\begin{equation}
\mathcal{L}^{(2)}=\alpha_1R^2+\alpha_2 R_{\mu\nu}R^{\mu\nu}+\alpha_3 \mathcal{X}_4\, .
\end{equation}
On the other hand, we may rewrite the six-derivative action introducing the cubic Lovelock density $\mathcal{X}_6$ and the term $\mathcal{S}_D$ , both of them belonging to the GQG class. From their respective expressions, Eqs.~\req{eq:cubicLove2} and \req{SD}, we see clearly that they depend differently on the two pure-Riemann cubic combinations, hence we can trade those terms by $\mathcal{X}_6$ and $\mathcal{S}_D$. Then, we can rewrite $\mathcal{L}^{(3)}$ as
\begin{equation}
\begin{aligned}
\mathcal{L}^{(3)}=&\beta_1\mathcal{X}_{6}+\beta_2 \mathcal{S}_{D}+\beta_3 \tensor{R}{_{\mu \nu\rho\sigma}}\tensor{R}{^{\mu \nu\rho}_{\alpha}}R^{\sigma \alpha}+\beta_4\tensor{R}{_{\mu\nu\rho\sigma}}\tensor{R}{^{\mu\nu\rho\sigma}}R+\beta_5\tensor{R}{_{\mu\nu\rho\sigma}}\tensor{R}{^{\mu\rho}}\tensor{R}{^{\nu\sigma}}\\
&+\beta_6R_{\mu}^{\ \nu}R_{\nu}^{\ \rho}R_{\rho}^{\ \mu}+\beta_7R_{\mu\nu }R^{\mu\nu }R+\beta_8R^3+\beta_9 \nabla_{\sigma}R_{\mu\nu} \nabla^{\sigma}R^{\mu\nu}+\beta_{10}\nabla_{\mu}R\nabla^{\mu}R\, .
\end{aligned}
\end{equation}
Another possibility in $D\ge 5$ is to exchange $\mathcal{S}_{D}$ or $\mathcal{X}_6$ by the cubic Quasi-topological combination $\mathcal{Z}_{D}$. On the whole, we say that all these three densities belong to the family of generalized quasi-topological gravities, and the important observation is that we are able to trade the terms that only contain Riemann tensors by densities of this family.  The rest of terms contain Ricci curvature, and therefore they can be removed by using field redefinitions. Let us construct this redefinition explicitly for the four- and six-derivative terms. 
First, in order to remove the $R^2$ and $R_{\mu\nu}R^{\mu\nu}$ terms, we perform 
\begin{equation}
g_{\mu\nu}=\tilde g_{\mu\nu}+\alpha_2\ell^2 \tilde R_{\mu\nu}-\frac{\ell^2 \tilde R}{D-2}\tilde g_{\mu\nu}(2\alpha_1+\alpha_2)
\end{equation}
The effect of this transformation on the four-derivative terms is
\begin{equation}
\mathcal{L}^{(2)}\rightarrow \tilde{\mathcal{L}}^{(2)}=\alpha_3 \tilde{\mathcal{X}}_4\, .
\end{equation}
Now, this redefinition also affects the higher-order terms, but since we are starting from the most general theory, the only effect is to change the coefficients of these terms. In particular, for the six-derivative ones: $\beta_i\rightarrow\beta_{i}'$. 
Then, we may perform another redefinition of the metric,
\begin{equation}
\begin{aligned}
\tilde g_{\mu\nu}&=\tilde{\tilde{g}}_{\mu\nu}+\ell^4\left[\beta_3'\tensor{\tilde{\tilde R}}{_{\mu \alpha\rho\sigma}}\tensor{\tilde{\tilde R}}{_{\nu}^{\alpha\rho\sigma}}+\beta_5' \tilde{\tilde R}^{\alpha\beta}\tilde{\tilde R}_{\mu\alpha\nu\beta}+\beta_6' \tensor{\tilde{\tilde R}}{_{\mu}^{\alpha}}\tilde{\tilde R}_{\nu\alpha}+\beta_{7}'\tilde{\tilde R}\tilde{\tilde R}_{\mu\nu}-\beta_{9}'\tilde{\tilde{\nabla}}^2\tilde{\tilde R}_{\mu\nu}\right.\\
&\left. -\frac{1}{D-2}\tilde{\tilde{g}}_{\mu\nu}\left(\tensor{\tilde{\tilde R}}{_{\alpha\beta\sigma}}\tensor{\tilde{\tilde R}}{^{\alpha\beta\rho\sigma}}(\tilde\beta_3+2\tilde\beta_4)+\tilde{\tilde R}_{\alpha\beta}\tilde{\tilde R}^{\alpha\beta}(\beta_5'+\beta_6')+\tilde{\tilde R}^2(\beta_7'+2\beta_8')-\tilde{\tilde{\nabla}}^2 \tilde{\tilde R}(\beta_9'-2\beta_{10}')\right)
\right]\, 
\end{aligned}
\end{equation}
which leaves unaffected the four-derivative terms, while it cancels all the six-derivative terms that contain Ricci curvature:
\begin{equation}
\tilde{\mathcal{L}}^{(3)}\rightarrow \tilde{\tilde{\mathcal{L}}}^{(3)}=\beta_1'\tilde{\tilde{\mathcal{X}}}_{6}+\beta_2'\tilde{\tilde{\mathcal{S}}}_{D}\, .
\end{equation}
Hence, the most general action can be written, after all, as
\begin{equation}\label{eq:effectiveaction1}
\tilde{\tilde S}=\frac{1}{16\pi G}\int d^Dx \sqrt{|\tilde{\tilde g}|}\left\{\tilde{\tilde{R}}+\alpha_3 L^2\tilde{\tilde{\mathcal{X}}}_4+\beta_1'L^4\tilde{\tilde{\mathcal{X}}}_{6}+\beta_2'L^4\tilde{\tilde{\mathcal{S}}}_{D}+\mathcal{O}(L^6)\right\}\ .
\end{equation}

We can see that the reason why we can write the general theory \req{eq:faction} as a combination generalized quasi-topological terms is that there are more of these than pure Riemann invariants. 

\subsection{All $\mathcal{L}$(Riemann) gravities as GQGs}
Let us introduce some notation to formulate the problem in a clearer way. We will say that a given curvature invariant is reducible if it contains at least one factor of Ricci curvature or if it is equivalent, up to a total derivative, to another density that contains Ricci curvature. We can actually rephrase this in a better way.
\vskip0.2cm
\noindent
\textbf{Definition 1}: \emph{a curvature invariant is \emph{reducible} if it is a total derivative when evaluated on any Ricci-flat metric. The rest of them are called \emph{irreducible}.}
\vskip0.2cm
Observe this definition contains trivially the case in which the invariant vanishes on Ricci-flat metrics. Intuitively, the irreducible terms correspond to those formed purely from the Riemann tensor, without explicit factors of Ricci curvature. The point is that all reducible terms can be removed or introduced by using field redefinitions, but the irreducible ones cannot. Therefore, we may imagine that the most general higher-derivative gravity is obtained by including all the possible irreducible terms in the action and then we are free to add all the reducible terms we want, since this is just equivalent to choosing a frame. It is convenient to introduce here another concept.
\vskip0.2cm
\noindent
\textbf{Definition 2}: \emph{We say that a curvature invariant $\mathcal{L}$ is \emph{completable to a generalized quasi-topological gravity} if there exist a generalized quasi-topological density $\mathcal{Q}$ such that $\mathcal{L}-\mathcal{Q}$ is reducible.}
\vskip0.2cm
In other words, by adding reducible terms to $\mathcal{L}$ we are able to obtain a GQG term. Note that reducible terms are trivially completable to $0$.  Then, the question whether any higher-derivative gravity can be expressed as a sum of generalized quasi-topological terms is equivalent to the following question: are all irreducible densities completable to a GQG? We have just found that the answer is positive at least up to six-derivative terms. The reason is that there are more GQG terms than irreducible terms and we were able to ``complete'' all of them. In the case of the four derivative terms, the only irreducible density is the Riemann squared term, and this can be completed to the Gauss-Bonnet density. For the six-derivative terms, it turns out that all the terms containing derivatives of the Riemann are reducible, and the only irreducible terms are the two cubic Riemann terms. In general dimensions $D$ there are 3 GQGs that involve different combinations of these cubic terms, so they can always be completed. 

The problem of completing all irreducible invariants is of course different depending on the dimension, since in lower dimensions many of the densities we can construct are not linearly independent. Due to this, in lower dimensions the number of irreducible densities is significantly smaller, and therefore the problem of completing all the irreducible densities is simpler. Going back to the example of six-derivative terms, we find that the two cubic densities are independent when $D\ge 6$. In $D=4, 5$ only one of them is linearly independent, and in $D<4$ there is only Ricci curvature so all theories are reducible to Einstein gravity. On the other hand, in $D=4$ there are four GQGs, and in $D>4$ there are three of them. So, there are less irreducible terms in lower dimensions and there are more ways to complete them to a GQG theory. 

It is clear that we cannot make much progress by studying all the possible invariants order by order. The following result will be very useful in order to derive some general statements.

\begin{theorem}[Deser, Ryzhov, 2005 \cite{Deser:2005pc}]\label{th:th1} 
When evaluated on a general static and spherically symmetric ansatz,
\begin{equation}\label{SSS}
ds^2=-N(r)^2f(r)dt^2+\frac{dr^2}{f(r)}+r^2d\Omega_{(D-2)}^2\, ,
\end{equation}
all possible contractions of $n$ Weyl tensors\footnote{Recall that the Weyl tensor is defined as
\begin{equation}
   W_{\mu\nu\rho\sigma}=R_{\mu\nu\rho\sigma}-\frac{2}{(D-2)}\left(g_{\mu[c\rho}R_{\sigma]\nu}-g_{\nu[\rho}R_{\sigma]\mu}\right)+\frac{2}{(D-2)(D-1)}R g_{\mu[\rho}g_{\sigma]\nu}.
\end{equation}.} are proportional to each other. More precisely, let $(W^n)_i$ be one of the possible independent ways of contracting $n$ Weyl tensors, then for all $i$
\begin{equation}\label{Dser}
(W^n)_i|_{\rm SSS} = F(r)^n c_i  \, ,
\end{equation}
where $c_i$ is some constant which depends on the particular contraction, and $F(r)$ is an $i$-independent  function of $r$ given in terms of the functions appearing in the SSS ansatz \req{SSS}.  In other words, the ratio $[(W^n)_1/(W^n)_2]|_{\rm SSS}$ for any pair of contractions of $n$ Weyl tensors is a constant which does not depend on the radial coordinate $r$.
\end{theorem}

\noindent
\textbf{Proof.} 
When evaluated on \req{SSS} the Weyl tensor takes the form
\begin{equation}\label{fack}
\left. \tensor{ W}{^{\mu\nu}_{\alpha\beta}} \right|_{\rm SSS}=-2\chi(r) \frac{(D-3)}{(D-1)} \tensor{ w}{^{\mu\nu}_{\alpha\beta}}\, ,
\end{equation}
where 
\begin{equation}
\chi(r)=\frac{(-2+2f-2r f'+r^2f'')}{2r^2}+\frac{N'}{2r N}(-2f +3r f')+\frac{f N''}{N}
\end{equation}
is a function which contains the full dependence on the radial coordinate. On the other hand, $\tensor{ w}{^{\mu\nu}_{\alpha\beta}}$ is an $r$-independent tensorial structure which can be written as \cite{Deser:2005pc}
\begin{equation}\label{ww}
 \tensor{ w}{^{\mu \nu}_{\alpha \beta}}=2\tau^{[\mu}_{[\alpha} \rho^{\nu]}_{\beta]}-\frac{2}{(D-2)} \left(\tau^{[\mu}_{[\alpha} \sigma^{\nu]}_{\beta]}+\rho^{[\mu}_{[\alpha} \sigma^{\nu]}_{\beta]} \right)+\frac{2}{(D-2)(D-3)} \sigma^{[\mu}_{[\alpha} \sigma^{\nu]}_{\beta]}\, ,
\end{equation}
where $\tau$, $\rho$ and $\sigma$ are orthogonal projectors defined as\footnote{Namely, they satisfy $\tau \tau=\tau$, $\rho \rho=\rho$, $\sigma \sigma=\sigma$, $\tau \rho=\tau\sigma=\rho\sigma=0$. Also, their traces read ${\rm Tr} \tau={ \rm Tr}\rho=1$, ${\rm Tr}\sigma=D-2$.}
\begin{equation}
\tau_{\mu}^{\nu}= \delta^0_{\mu} \delta^{\nu}_0 \, , \quad \rho_{\mu}^{\nu}= \delta^1_{\mu} \delta^{\nu}_1 \, , \quad  \sigma_{\mu}^{\nu}=\sum_{m=2}^{D-1} \delta^m_{\mu} \delta^{\nu}_m \, .
\end{equation}
The important point is that any possible invariant $(W^n)_i$ constructed from the contraction of $n$ Weyl tensors will be given by
\begin{equation}\label{scs}
(W^n)_i|_{\rm SSS} =\left(-2 \chi(r) \frac{(D-3)}{(D-1)} \right)^n (w^n)_i\, ,
\end{equation}
where $ (w^n)_i$ stands for the constant resulting from the  contraction induced on the $w$ tensors defined in \req{ww} which we can identify with $c_i$ in  \req{Dser}. Therefore, $(W^n)_i|_{\rm SSS}$ takes the form \req{Dser} with $F(r)$ given by the function between brackets. QED.

So far we have not found examples of irreducible densities constructed from derivatives of the Riemann tensor. Somehow, the densities of fourth-order (that do not contain derivatives of the curvature) seem to be the most relevant ones. The following result will help us to argue that all theories of that family can be recasted in the form of a GQG.

\begin{theorem}\label{th:th2}
Let us consider the set of irreducible curvature invariants of a given order that are of the form $\mathcal{L}(g^{\mu\nu},R_{\mu\nu\rho\sigma})$. If one of these invariants is completable to a GQG and is non-vanishing when evaluated on a static, spherically symmetric ansatz, then all of them are completable.
\end{theorem}
\noindent
\textbf{Proof}: Let the order of these invariants be $2n$. Since they are irreducible and they do not contain derivatives of the curvature, they are formed from contractions of a product of $n$ Riemann tensors. We can write schematically $\mathcal{L}_{i}=\left(\text{Riem}^n\right)_i$, where the subscript $i$ denotes a specific way of contracting the indices. We can consider an alternative basis by replacing the Riemann tensor by the Weyl tensor in the expressions of these densities. Both ways of expressing these invariants are obviously equivalent since they differ in terms containing Ricci curvature. Thus, let us consider $\tilde{\mathcal{L}}_i=\left(W^n\right)_i$ and let us assume that some $\tilde{\mathcal{L}}_{i_0}$ is completable to a GQG.  Now, the condition that determines if a given density belongs to the generalized quasi-topological class only depends on the evaluation of the density on a static and spherically symmetric (SSS) metric ansatz --- see Theorem \ref{theo}. As we have just seen, for this type of metric, all the invariants formed from the Weyl tensor are proportional to the same quantity: they are given by \req{Dser}. Then, since by assumption $\tilde{\mathcal{L}}_{i_0}\big|_{SSS}\neq 0$, all the invariants $\tilde{\mathcal{L}}_i$ are proportional to $\tilde{\mathcal{L}}_{i_0}$ when evaluated on SSS metrics. Thus, if $\tilde{\mathcal{L}}_{i_0}$ can be completed to a GQG by adding a reducible piece $\mathcal{R}_{i_0}$, it is clear that any other term is completed to a GQG when we add the reducible piece $\frac{c_i}{c_{i_0}}\mathcal{R}_{i_0}$.  QED.
\vskip0.2cm

\noindent
This result can be rephrased in a more useful way as follows
\begin{corollary}\label{cor:cor2}
Let us consider the curvature invariants of a given order that are of the form $\mathcal{L}(g^{\mu\nu},R_{\mu\nu\rho\sigma})$, and let us assume that there exists one irreducible and non-trivial GQG density formed from these invariants. Then all of these invariants are completable to a GQG. 
\end{corollary}

We recall that irreducible means that the density does not vanish (modulo total derivatives) on Ricci-flat metrics and non-trivial means that it does not vanish for SSS metrics. We have compelling evidence that this type of generalized quasi-topological theories exist at any order and in any dimension. In particular, we provided explicit examples of these terms up to $n=10$ for the four-dimensional case in Sec.~\ref{sec:D4all}. Thus, Corollary \ref{cor:cor2} virtually proves that all the terms of the form $\mathcal{L}(g^{\mu\nu},R_{\mu\nu\rho\sigma})$ can be rewritten as a sum of GQG densities by using field redefinitions. In order to get a formal proof we should simply provide a closed expression for one of these densities at every order in curvature, or at least a systematic way to construct them. While this would be a tedious exercise, we are fully confident that such construction is possible. 

\subsection{Terms with covariant derivatives}\label{sec:covder}
On the other hand, the role of terms containing derivatives of the Riemann tensor is less clear. Let us note that, even if we start with some effective theory that does not include this type of term, they will appear when we perform field redefinitions. So we need to make sure as well that they can always be completed to a GQG. Now, the fact is that such terms have not been yet used to construct GQGs --- for what we know, these types of theories should exist as well.  One of the reasons is that, up to sixth-order in derivatives, all the invariants with derivatives of the curvature are actually reducible. Thus, in order to gain some insight about the behaviour of these terms let us consider in detail the case of the eight-derivative densities. 

The general eight-derivative Lagrangian can be constructed from the 92 terms that appear in the appendix of \cite{0264-9381-9-5-003} --- many of them will differ in total derivatives though.  
Let us first consider the densities that do not contain derivatives of the curvature. Looking at \cite{0264-9381-9-5-003}, we see that there is a basis of 26 of these invariants --- these are listed also in Table~\ref{tabla2} --- but, depending on the dimension, there are at most 7 independent, irreducible terms (this happens for $D>7$). Now, in \cite{Ahmed:2017jod} several non-trivial and irreducible generalized quasi-topological theories were constructed using those invariants --- and we provided a couple of examples in Sec.~\ref{sec:D4all} for the case $D=4$. Since, according to Corollary \ref{cor:cor2}, we only need one, this already proves that these 26 invariants can be rewritten as a sum of GQGs by means of field redefinitions.

Now we consider the terms with derivatives of the curvature.  Looking again at \cite{0264-9381-9-5-003}, we see, apparently, five irreducible terms

\begin{eqnarray}
\mathcal{L}_{1}&=&\tensor{R}{^{\mu\nu\rho\sigma}}\nabla_{\nu}\tensor{R}{^{\alpha\beta\lambda}_{\mu}}\nabla_{\sigma}\tensor{R}{_{\alpha\beta\lambda\rho}}\, ,\\
\mathcal{L}_{2}&=&\tensor{R}{^{\mu\nu\rho\sigma}}\nabla_{\rho}\tensor{R}{^{\alpha\beta\lambda}_{\mu}}\nabla_{\sigma}\tensor{R}{_{\alpha\beta\lambda\nu}}\, ,\\
\mathcal{L}_{3}&=&\tensor{R}{^{\mu\nu\rho\sigma}}\nabla^{\lambda}\tensor{R}{^{\alpha}_{\mu}^{\beta}_{\rho}}\nabla_{\lambda}\tensor{R}{_{\alpha\nu\beta\sigma}}\, ,\\
\mathcal{L}_{4}&=&\tensor{R}{^{\mu\nu\rho\sigma}}\tensor{R}{_{\mu}^{\alpha\beta\lambda}}\nabla_{\sigma}\nabla_{\lambda}\tensor{R}{_{\nu\alpha\rho\beta}}\, ,\\
\mathcal{L}_{5}&=&\nabla_{\alpha}\nabla_{\beta}\tensor{R}{_{\mu\nu\rho\sigma}}\nabla^{\alpha}\nabla^{\beta}\tensor{R}{^{\mu\nu\rho\sigma}}\, .
\end{eqnarray}

However, a careful analysis --- using commutation of covariant derivatives, the symmetries of the Riemann tensor and the Bianchi identity --- reveals that all of them can be decomposed as the sum of total derivative terms plus quartic curvature terms (without covariant derivatives) plus terms with Ricci curvature (hence reducible). This is, they can be expressed as
\begin{equation}\label{eq:Lred}
\mathcal{L}_{i}=\nabla_{\mu}J^{\mu}+\mathcal{Q}+R_{\mu\nu}F^{\mu\nu}\, ,
\end{equation}
for some tensors $J^{\mu}$ and $F^{\mu\nu}$ and some quartic density $\mathcal{Q}$.
In order to illustrate this, let us show how $\mathcal{L}_1$ is reduced to an expression of the form \req{eq:Lred}. First, we have
\begin{align}
\mathcal{L}_{1}&=\tensor{R}{^{\mu\nu\rho\sigma}}\nabla_{\nu}\tensor{R}{^{\alpha\beta\lambda}_{\mu}}\nabla_{\sigma}\tensor{R}{_{\alpha\beta\lambda\rho}}=\frac{1}{4}\tensor{R}{^{\mu\nu\rho\sigma}}\nabla^{\lambda}\tensor{R}{^{\alpha\beta}_{\mu\nu}}\nabla_{\lambda}\tensor{R}{_{\alpha\beta\rho\sigma}}\\
&=\frac{1}{4}\nabla_{\lambda}\left(\tensor{R}{^{\mu\nu\rho\sigma}}\nabla^{\lambda}\tensor{R}{^{\alpha\beta}_{\mu\nu}}\tensor{R}{_{\alpha\beta\rho\sigma}}\right)-\frac{1}{4}\tensor{R}{^{\mu\nu\rho\sigma}}\nabla^2\tensor{R}{^{\alpha\beta}_{\mu\nu}}\tensor{R}{_{\alpha\beta\rho\sigma}}-\frac{1}{4}\nabla_{\lambda}\tensor{R}{^{\mu\nu\rho\sigma}}\nabla^{\lambda}\tensor{R}{^{\alpha\beta}_{\mu\nu}}\tensor{R}{_{\alpha\beta\rho\sigma}}\, ,
\end{align}
where in the first equality we applied the differential Bianchi identity twice, and in the second we ``integrated by parts''.  Now we note that the last term in the second line is actually $-\mathcal{L}_{1}$, so we get
\begin{align}
\mathcal{L}_{1}&=\frac{1}{8}\nabla_{\lambda}\left(\tensor{R}{^{\mu\nu\rho\sigma}}\nabla^{\lambda}\tensor{R}{^{\alpha\beta}_{\mu\nu}}\tensor{R}{_{\alpha\beta\rho\sigma}}\right)-\frac{1}{8}\tensor{R}{^{\mu\nu\rho\sigma}}\nabla^2\tensor{R}{^{\alpha\beta}_{\mu\nu}}\tensor{R}{_{\alpha\beta\rho\sigma}}\, .
\end{align}
Then we are done, because the Laplacian of the Riemann tensor decomposes, using a schematic notation, as\footnote{Explicitly, one has \cite{Ricci}
\begin{align}
\nabla^{\alpha}\nabla_{\alpha}R_{\mu\nu\rho\sigma}=&+2\nabla_{[\mu|}\nabla_{\rho}R_{|\nu] \sigma}+2\nabla_{[\nu|}\nabla_{\sigma}R_{|\mu] \rho}-4\left[\tensor{R}{^\gamma_\mu^\lambda_\nu} R_{\gamma[\rho| \lambda |\sigma]}+ \tensor{R}{^\gamma_\mu^\lambda_{[\rho|}} R_{\gamma\nu\lambda|\sigma]}\right]\\ \notag &+g^{\gamma\lambda}\left[R_{\lambda\nu\rho\sigma}R_{\gamma\mu}+R_{\mu\lambda\rho\sigma}R_{\gamma\nu}\right]\, .
\end{align}
}   $\nabla^2\text{Riem}=\nabla\nabla \text{Ricci}+\text{Riem}^2$, so we can indeed express $\mathcal{L}_{1}$ as in \req{eq:Lred}. 
Proceeding similarly with the other terms we come to the same conclusion. 

Since total derivatives are irrelevant for the action, and since we can remove all the terms containing Ricci curvature by means of field redefinitions, the terms with covariant derivatives of the Riemann tensor only change the coefficients of the quartic terms, that are already present in the action. Hence, from the point of view of effective field theory, these densities are meaningless and can be removed. In addition, we conclude that all the eight-derivative terms can be recasted as a sum of GQGs by implementing field redefinitions. 

However, it is clear that we cannot repeat the analysis above order by order, since the number of basic invariants grows very quickly, and it is impossible to provide a list of all the invariants.  If we want to prove that all terms with covariant derivatives of the curvature can be mapped to GQG theories we need a different strategy. We propose here a possible way to do it, based again on Theorem~\ref{th:th1}, and we will illustrate it by considering terms containing two covariant derivatives. These come in two different types: those in which the two derivatives act on the same Riemann tensor, and those in which they act on two different Riemanns. We denote these types of invariants as $\mathcal{R}^{\{2\}}$ and $\mathcal{R}^{\{1,1\}}$. Now, all of the first kind can be transformed to a sum of terms of the second kind upon integration by parts, so we only need to consider the invariants of the form $\mathcal{R}^{\{1,1\}}$. The only densities of that type that are susceptible of being irreducible are those formed from the Weyl tensor, and they can be written schematically as
\begin{equation}\label{eq:Inv2}
\mathcal{R}^{\{1,1\}}=W^n \nabla W\nabla W\, ,
\end{equation}
for some value of $n$. We saw in \req{fack} that, when evaluated on a SSS metric the Weyl tensor has a very simple structure, so that any scalar formed from it is proportional to the same quantity. In the case of the covariant derivative of the Weyl tensor, we have
\begin{equation}\label{fack2}
\left. \nabla_{\sigma}\tensor{ W}{^{\mu\nu}_{\alpha\beta}} \right|_{\rm SSS}=-2\frac{(D-3)}{(D-1)}\left[\chi'(r)  \delta_{\sigma}^{1}\tensor{ w}{^{\mu\nu}_{\alpha\beta}}+\chi(r) \nabla_{\sigma}\tensor{ w}{^{\mu\nu}_{\alpha\beta}}\right]\, ,
\end{equation}
which now depends on two different tensorial structures. It is easy to check that, when those structures are contracted --- with each other or with themselves --- they only produce a number times $f(r)$. Thus, when evaluated on a SSS metric, all the invariants of the form \req{eq:Inv2} are given by 
\begin{equation}
\left.\mathcal{R}^{\{1,1\}} \right|_{\rm SSS}=f\chi^n\left(c_1\chi'^2+c_2\frac{\chi\chi'}{r}+c_3\frac{\chi^2}{r^2}\right)\, ,
\end{equation}
where $c_{1,2,3}$ are constants. Thus, there are at most three linearly independent terms of the form \req{eq:Inv2} when one considers SSS metrics. Hence, if we are able to find three independent terms of that kind which are completable to a GQG, this will imply that all densities containing two covariant derivatives are completable. 
Three possible terms of that type are 
\begin{align}
\mathcal{W}^{\{1,1\}}_1=&\nabla_{\nu}\left(\nabla^{\nu}\tensor{W}{_{\mu_1\mu_2}^{\mu_3\mu_4}}\tensor{W}{_{\mu_3\mu_4}^{\mu_5\mu_6}}\tensor{W}{_{\mu_5\mu_6}^{\mu_7\mu_8}}\ldots\tensor{W}{_{\mu_{2n+3}\mu_{2n+4}}^{\mu_1\mu_2}}\right)\\
&\notag-\nabla^2\tensor{W}{_{\mu_1\mu_2}^{\mu_3\mu_4}}\tensor{W}{_{\mu_3\mu_4}^{\mu_5\mu_6}}\tensor{W}{_{\mu_5\mu_6}^{\mu_7\mu_8}}\ldots\tensor{W}{_{\mu_{2n+3}\mu_{2n+4}}^{\mu_1\mu_2}}%\nabla_{\nu}\tensor{W}{_{\mu_1\mu_2}^{\mu_3\mu_4}}\nabla^{\nu}\tensor{W}{_{\mu_3\mu_4}^{\mu_5\mu_6}}\tensor{W}{_{\mu_5\mu_6}^{\mu_7\mu_8}}\ldots\tensor{W}{_{\mu_{2n+3}\mu_{2n+4}}^{\mu_1\mu_2}}\, ,\\
&&\nonumber\\
\mathcal{W}^{\{1,1\}}_2=&\nabla_{\nu}\tensor{W}{^{\nu\rho\delta}_{\mu_1}}\nabla_{\rho}\tensor{W}{_{\delta\mu_2}^{\mu_3\mu_4}}\tensor{W}{_{\mu_3\mu_4}^{\mu_5\mu_6}}\ldots\tensor{W}{_{\mu_{2n+1}\mu_{2n+2}}^{\mu_1\mu_2}}\, ,\\
&&\nonumber\\
\mathcal{W}^{\{1,1\}}_3=&\nabla_{\nu}\tensor{W}{^{\nu\rho\delta \gamma}}\nabla_{\sigma}\tensor{W}{^{\sigma}_{\rho\delta \gamma}}\tensor{W}{_{\mu_1\mu_2}^{\mu_3\mu_4}}\ldots\tensor{W}{_{\mu_{2n-1}\mu_{2n}}^{\mu_1\mu_2}} \, .
\end{align}
One can check that when evaluated on a general SSS metric they are linearly independent. For instance, Schwarzschild's metric is a ``double root'' of the third term and a simple root of the second one, while the first one is non-vanishing when evaluated on Schwarzschild's metric. This observation guarantees that they are independent, hence any other term $\mathcal{R}^{\{1,1\}}$ can be expressed as a linear combination of these terms when evaluated on \req{SSS}. This can be stated alternatively as 
\begin{equation}\label{eq:Inv2b}
\mathcal{R}^{\{1,1\}}=C_1 \mathcal{W}^{\{1,1\}}_1+C_2\mathcal{W}^{\{1,1\}}_2+C_3\mathcal{W}^{\{1,1\}}_3+\ldots \, ,
\end{equation}
where the ellipsis denote terms that vanish on SSS metrics --- which are trivially completable to a GQG --- and where $C_i$ are constants. Now, it is easy to check that, by means of field redefinitions, the densities $\mathcal{W}^{\{1,1\}}_{1,2,3}$ can be mapped to a sum of terms without covariant derivatives. Actually, both $\mathcal{W}^{\{1,1\}}_2$ and $\mathcal{W}^{\{1,1\}}_3$ are reducible because they are proportional to the divergence of Weyl's tensor, which depends only on Ricci curvature
\begin{equation}
\nabla_{\alpha}\tensor{W}{^{\alpha}_{\mu\nu\rho}}=\frac{2(D-3)}{D-2}\left[\nabla_{[\nu}R_{\rho]\mu}-\frac{1}{2(D-1)}g_{\mu[\rho}\nabla_{\nu]}R\right]\, .
\end{equation}
On the other hand, $\mathcal{W}^{\{1,1\}}_1$ is the sum of a total derivative plus a term that contains a Laplacian of Weyl's tensor. %can be decomposed in the following way. 
%\begin{equation}
%\begin{aligned}
%\mathcal{W}^{\{1,1\}}_1&=\frac{1}{n+1}\bigg[\nabla_{\nu}\left(\nabla^{\nu}\tensor{W}{_{\mu_1\mu_2}^{\mu_3\mu_4}}\tensor{W}{_{\mu_3\mu_4}^{\mu_5\mu_6}}\tensor{W}{_{\mu_5\mu_6}^{\mu_7\mu_8}}\ldots\tensor{W}{_{\mu_{2n+3}\mu_{2n+4}}^{\mu_1\mu_2}}\right)\\
%&-\nabla^2\tensor{W}{_{\mu_1\mu_2}^{\mu_3\mu_4}}\tensor{W}{_{\mu_3\mu_4}^{\mu_5\mu_6}}\tensor{W}{_{\mu_5\mu_6}^{\mu_7\mu_8}}\ldots\tensor{W}{_{\mu_{2n+3}\mu_{2n+4}}^{\mu_1\mu_2}}\bigg]\, ,
%\end{aligned}
%\end{equation}
Since the Laplacian can be expressed as $\nabla^2\text{Weyl}=\nabla\nabla \text{Ricci}+\text{Riem}^2$, we conclude that, by means of field redefinitions $\mathcal{W}^{\{1,1\}}_1$ can be reduced to a sum of terms without covariant derivatives. From the result in the previous section, we know that those terms are completable to a GQG, hence the density $\mathcal{W}^{\{1,1\}}_{3}$ is also completable. Therefore, we have expressed any term with two covariant derivatives as the sum of completable densities \req{eq:Inv2b}, and we conclude that all those terms are completable of a GQG. 
The result is actually stronger than that: since the densities $\mathcal{W}^{\{1,1\}}_{1,2,3}$ can be completed to a GQG without covariant derivatives of the Riemann tensor, this implies that any other term $\mathcal{R}^{\{1,1\}}$ can be completed to a GQG that, when evaluated on a SSS metric, is equivalent to a GQG without covariant derivatives. 

As we see, the main point in the preceding analysis is that, while there are many possible invariants formed from $W$ and $\nabla$, only few of them --- say $m$ of them --- are independent when evaluated on a SSS metric. Thus, we only need to show that there are $m$ invariants which are independent and completable to a GQG. We have found explicitly such invariants for the case of two covariant derivatives, where we only needed to find $m=3$ of them, but this analysis can presumably be extended to terms with an arbitrary number of covariant derivatives. Although finding examples of these completable terms is not straightforward, we see no reason to expect that this cannot be done.

\section{Discussion}
Based on the evidences provided in the last section, we are confident to state the following\footnote{Our conclusions also hold if one includes parity-breaking terms in the effective action, \ie those that involve the Levi-Civita symbol $\epsilon_{\mu_1\ldots \mu_D}$. Although we did not consider those terms explicitly, all such terms vanish for spherically symmetric configurations, hence all of them trivially belong to the GQG family.}
\begin{conjecture}\label{conj:3}
Any higher-derivative gravity Lagrangian can be mapped, order by order, to a sum of GQG terms by implementing redefinitions of the metric of the form \req{eq:metricredef}.
\end{conjecture}
Basically, there are many theories satisfying the GQG condition --- \ie the condition of the hypothesis of Theorem \ref{theo} in the previous chapter --- and the amount of terms we can modify in the action with field redefinitions is also very large. All in all, there so much freedom that field redefinitions can always bring the most general action \req{eq:faction} into a sum of GQG terms, order by order in the curvature. We have shown explicitly that this works at least up to order eight, and there is a priori no reason to expect that an exception will appear at higher orders.  

The main result of the Chapter is Theorem~\ref{th:th2}, which essentially tells us that if for a given order in curvature there is one GQG of the form $\mathcal{L}(g^{\alpha\beta}, R_{\mu\nu\rho\sigma})$, then all densities of that type and order are completable to a GQG. Since we know by experience that those terms exists, this result virtually proves that all $\mathcal{L}$(Riemann) terms can be mapped to a GQG.  We would have to provide an explicit construction of these terms in order to complete a formal proof. As we mentioned before, such systematic construction must be possible, but has not been carried out yet only because it is a highly laborious task. 

On the other hand, we have seen that the densities that contain explicit covariant derivatives of the Riemann tensor do not seem to play any role. In fact, we have checked that, up to eighth order, all the terms that contain derivatives of the Riemann tensor are irrelevant --- they can always be mapped to other terms that already appear in the action upon the application of field redefinitions.  
More generally, we have been able to prove that any term with two covariant derivatives can be completed to a GQG \emph{which is equivalent to a GQG of the form $\mathcal{L}(g^{\alpha\beta}, R_{\mu\nu\rho\sigma})$ when evaluated on a SSS metric.} Note that the last claim is slightly different from stating that the original term can be completed to a GQG of the form $\mathcal{L}(g^{\alpha\beta}, R_{\mu\nu\rho\sigma})$. It means that the GQG to which the original density is completed may, in principle, contain covariant derivatives of the curvature, but it is guaranteed that those terms vanish for a SSS metric. 
We argued that the previous conclusion can very likely be extended to densities with an arbitrary number of covariant derivatives, and it suggests a stronger conjecture:

\begin{conjecture}\label{conj:4}
Any higher-derivative gravity Lagrangian can be mapped, order by order, to a sum of GQG terms that, when evaluated on a SSS metric, are equivalent to GQGs of the type $\mathcal{L}(g^{\alpha\beta}, R_{\mu\nu\rho\sigma})$. 
\end{conjecture}
If correct, the second statement of this conjecture would imply that we can study the spherically symmetric black hole solutions of the most general higher-derivative gravity by analyzing only the solutions of the Generalized quasi-topological gravities of the form $\mathcal{L}(g^{\alpha\beta}, R_{\mu\nu\rho\sigma})$ --- like the ones we found in Section~\ref{sec:construction}.

The conclusion of this chapter is that the theories of the Generalized quasi-topological class, that we defined by the condition in Theorem \ref{theo}, are not just toy models with very interesting properties. According to our results, they capture, at the very least, a very large part of all possible effective theories of gravity, and very likely --- if Conjecture \ref{conj:3} is true --- they capture \emph{any theory at all}. From this point of view, we could think that GQG theories correspond to the most general EFT expressed in a frame in which the study of spherically symmetric black holes is particularly simple. While, in general, the profile of the
solutions will be different in every frame, we recall that black hole thermodynamics are invariant under the change of frame. An intriguing possibility, that we will elaborate  in Chapter~\ref{Chap:5}, is that, by studying the black hole thermodynamics in GQG theories, we could be able to obtain the thermodynamic quantities of black holes in any higher-derivative gravity.

\part{Asymptotically flat black holes}

\chapter{Black holes in Einsteinian cubic gravity}\label{Chap:4}
In the first part of this thesis we studied several aspects of general higher-order gravities that allowed us to classify them according to their properties. As a result, we identified some particular theories with very interesting features, such as the family of Generalized quasi-topological gravites introduced in Chapter \ref{Chap:2}. These theories have second-order linearized equations on constant curvature backgrounds, and their equations of motion in the presence of spherical symmetry have a very simple form. In addition, we argued in Chapter \ref{Chap:3} that, very likely, the theories of this class provide a ``basis'' to construct the most general EFT for gravity.

In this chapter and in the next one we study the black hole solutions of these theories. We focus on the four-dimensional case, which, for obvious reasons, is the most interesting one. Paradoxically, it is also the less explored. In fact, we are going to present the first examples of genuine, non-perturbative black hole solutions of higher-derivative gravity in four dimensions. 
Let us be more precise about the last statement. As we remarked in Section~\ref{sec:GoSs}, there are many examples of black hole solutions of higher-order gravity in the literature, but most of them correspond to somewhat unnatural situations.
Indeed, many of the solutions studied so far either (a) come from a theory that lacks a well-defined Einstein gravity limit \cite{Riegert:1984zz,Klemm:1998kf,Oliva:2010zd,Lu:2013hx,Banados:1993ur,Cai:2006pq,Cai:2009ac,Love}, (b) or are the same solutions of Einstein gravity ``embedded'' in some higher-order gravities \cite{delaCruzDombriz:2009et,Li:2017ncu}, (c) or are solutions that are not a smooth deformation of Einstein gravity solutions \cite{Lu:2015psa,Lu:2015cqa}. We consider these situations of reduced interest for our purposes. Case (a) is quite unrealistic because we are interested in corrections to Einstein's theory. Case (b) is in principle acceptable from the previous point of view, but it does not provide us with any new information about the effects of higher-curvature terms on black holes, since the theory is chosen in a way such that Einstein metrics are still solutions. Finally, the solutions of the type (c) are also somewhat unphysical since they do not exist in the Einstein gravity limit.

Let us, on the contrary specify the case that we consider to be interesting:
\begin{enumerate}
\item The theory has a well-defined Einstein-gravity limit, \ie it contains the Einstein-Hilbert term plus higher-derivative terms which are controlled by free couplings; when the couplings are set to zero one recovers Einstein's theory.  
\item The spherically symmetric black holes of the theory are a smooth (and non-trivial) deformation of Schwarzschild solution, in the sense that they reduce to it when the higher-order couplings are set to zero. 
\end{enumerate}
This is the situation we consider to be most natural: the theory contains corrections and these corrections modify the Einstein gravity solutions accordingly.
 %On the other hand, we can mention other possibilities that we consider to be much less interesting.  For instance, some theories, such as $f(R)$ gravity, do not modify vacuum Einstein gravity solutions --- in particular the Schwarzschild solution \cite{delaCruzDombriz:2009et}. In addition, there are theories, such as quadratic gravity in four dimensions, that possess the Schwarzschild metric as a solution but they also have disconnected branches of non-Schwarzschild solutions \cite{Lu:2015cqa}. However, in those cases one would consider the Schwarzschild metric as the relevant solution, since the properties of the other branches seem to be quite unphysical.
Focusing on this case, one finds that there are few examples of exact black hole solutions of that type in the literature, and all of them in higher dimensions $D\ge5$. Indeed, exact spherically symmetric black hole solutions have been constructed in Lovelock gravities \cite{Wheeler:1985nh,Wheeler:1985qd,Boulware:1985wk,Cai:2001dz,Dehghani:2009zzb,deBoer:2009gx}, and in Quasi-topological gravity  \cite{Quasi2,Quasi} and its higher-order extensions \cite{Dehghani:2011vu,Cisterna:2017umf}. The field equations of those theories in the presence of spherical symmetry are simple enough so that an analytic and explicit resolution is possible. For any other theory, the problem of solving the equations of motion for spherically symmetric and  static metrics is considerably more involved. Now, all Lovelock and Quasi-topological theories are either topological or trivial in $D=4$, which is the reason why exact solutions modifying the Schwarzschild black hole in a continuous way have not been constructed.  Nevertheless, in Chapter~\ref{Chap:2} we described a family of theories that generalize Lovelock and Quasi-topological gravities --- indeed, they are named Generalized quasi-topological gravities --- for which the spherically symmetric equations of motion are still simple enough. Fortunately, these theories do exist in four dimensions.

The first non-trivial example of these theories in $D=4$ is given by the Einsteinian cubic gravity (ECG) term, $\mathcal{P}$. This cubic density was first identified in Sec.~\ref{sec:ECG} by the special properties of its linearized equations, and then rediscovered in Sec.~\ref{cubic} for the remarkably simple form of its field equations in the presence of spherical symmetry. 
The aim of this chapter is to study spherically symmetric black hole solutions in four-dimensional Einsteinian cubic gravity.\footnote{When we refer to our theory as ``Einsteinian cubic gravity'' we imply that the Lagrangian contains the Einstein-Hilbert term plus the cubic correction $\mathcal{P}$.}  We will be able to solve the equations of motion non-perturbatively in the higher-order coupling and we will find that its black hole solutions have remarkable differences with respect to their Einstein gravity counterparts. 

As guaranteed by Theorem \ref{theo}, which is satisfied by Einsteinian cubic gravity in $D=4$, the problem of finding black hole solutions in ECG is reduced to solving a second-order ODE. We will find explicitly that equation and we will show that there is a unique asymptotically flat black hole solution for every value of the mass. The profile of these solutions has to be constructed numerically, but one of the most amazing properties of ECG is that the thermodynamic properties ---  entropy and temperature --- of these black holes can be computed exactly, without making use of any numerical methods or approximations.  We will see that for large masses, ECG black holes reduce to the usual Schwarzschild solution plus a small deviation --- interestingly, we will argue that this perturbation captures the leading-order correction to Schwarzschild's black hole in any higher-derivative gravity. 
On the other hand, we will see that for small masses, ECG black holes have radically different properties from those in Einstein gravity.

\section{Review of Einsteinian cubic gravity}
We consider the cubic-curvature extension of the Einstein-Hilbert action given by
\begin{equation}\label{eq:ECGc4}
S_{\rm ECG}=\frac{1}{16\pi G}\int d^4x\sqrt{|g|}\left\{-2\Lambda+R-\frac{\mu L^4}{8}\mathcal{P}\right\}\, ,
\end{equation}
where $\mathcal{P}$ is the Einsteinian cubic gravity density,
\begin{align}
\mathcal{P}=12 \tensor{R}{_{\mu}^{\rho}_{\nu}^{\sigma}}\tensor{R}{_{\rho}^{\alpha}_{\sigma}^{\beta}}\tensor{R}{_{\alpha}^{\mu}_{\beta}^{\nu}}+\tensor{R}{_{\mu\nu}^{\rho\sigma}}\tensor{R}{_{\rho\sigma}^{\alpha\beta}}\tensor{R}{_{\alpha\beta}^{\mu\nu}}-12R_{\mu\nu\rho\sigma}R^{\mu\rho}R^{\nu\sigma}+8\tensor{R}{_{\mu}^{\nu}}\tensor{R}{_{\nu}^{\rho}}\tensor{R}{_{\rho}^{\mu}}
\end{align}
Also,  $\mu$ is a dimensionless coupling,\footnote{The normalization factor of $-1/8$ is conventional.} while $L$ is a length scale that determines the distance at which the gravitational interaction is modified. Equivalently, it is related to the energy scale of new physics by $E\sim 1/L$. 
For arbitrary values of $\mu$, the cubic term corrects the Einstein field equations, which in this case read
\begin{align}\label{eq:fe}
\mathcal{E}_{\mu\nu}=G_{\mu\nu}+\Lambda g_{\mu\nu}
-\frac{\mu L^4}{8} \left(P_{\mu\sigma\rho\lambda}R_{\nu}\,^{\sigma \rho\lambda}-\frac{\mathcal{P}}{2}g_{\mu\nu}+2\nabla^{\alpha}\nabla^{\beta}P_{\mu\alpha\nu\beta}\right)
\end{align}
where
\begin{equation}
\begin{aligned}
\label{eq:PECG}
\tensor{P}{_{\mu\nu}^{\alpha\beta}}&=36\tensor{R}{_{[\mu|\sigma}^{[\alpha|}_{\rho}} \tensor{R}{_{|\nu]}^{\sigma|\beta]}^{\rho}}+3\tensor{R}{_{\mu\nu}^{\sigma\rho}}\tensor{R}{_{\sigma\rho}^{\alpha\beta}}-12 \tensor{R}{_{[\mu}^{[\alpha}}\tensor{R}{_{\nu]}^{\beta]}}\\ 
&-24R^{\sigma\rho}\tensor{R}{_{\sigma[\mu|\rho}^{[\alpha}}\tensor{\delta}{_{|\nu]}^{\beta]}}+24\tensor{R}{_{\sigma}^{[\alpha|}}\tensor{R}{^{\sigma}_{[\mu}}\tensor{\delta}{_{\nu]}^{|\beta]}}\, .
\end{aligned}
\end{equation}
Notice that these equations are in general of fourth order on account of the doble derivative of $P_{\mu\alpha\nu\beta}$. 
During the first part of this thesis we learned several interesting properties of Einsteinian cubic gravity. Let us collect here some of the facts we already know about this theory.
\begin{enumerate}
\item The linearized equations of \req{eq:ECGc4} around maximally symmetric backgrounds are of second order. Namely, they read
\begin{equation}
\E^{L}_{\mu\nu}=\frac{1}{16\pi G_{\rm eff}}G^{L}_{\mu\nu}\, ,
\end{equation}
where $G_{\rm eff}$ is the effective Newton's constant and $G^{L}_{\mu\nu}$ is the linearized cosmological Einstein tensor.
Hence, the theory only propagates a massless graviton on the vacuum. This property is actually satisfied by the Lagrangian of \req{eq:ECGc4}  in any spacetime dimension, not only in $D=4$, as we explained in Sec.~\ref{sec:ECG}. The effective cosmological constant of the previous theory is obtained by solving the equation
\begin{equation}
\Lambda_{\rm eff}-\frac{\mu}{9}L^4\Lambda_{\rm eff}^3=\Lambda\, ,
\end{equation}
while the effective Newton's constant is given by 
\begin{equation}
G_{\rm eff}=\frac{G}{1-\frac{\mu}{3}L^4\Lambda_{\rm eff}^2}\, .
\end{equation}

\item It satisfies Theorem \ref{theo} of Chapter~\ref{Chap:2}, so that we know it possesses static, spherically symmetric solutions of the form
\begin{equation}\label{eq:Fmetric2}
ds^2=-f(r)dt^2+\frac{dr^2}{f(r)}+r^2d\Omega_{(2)}^2\, ,
\end{equation}
\textit{i.e.}, characterized by a single function. In addition, since the field equations of \req{eq:ECGc4} are of fourth order, Theorem \ref{theo} tells us that the equation for $f(r)$ can be reduced to a second-order differential equation. Thus, the problem of finding spherically symmetric solutions is greatly simplified with respect to other theories. 
\item Finally, from the point of view of EFT, the ECG term $\mathcal{P}$ represents the leading-order parity-preserving correction to Einstein gravity. Since parity-breaking terms do not contribute to the field equations in the presence of spherical symmetry, this means that the theory above captures, modulo field redefinitions, the most general leading correction to Schwarzschild black hole in any higher-order gravity. 
\end{enumerate}

The third item justifies the search of black hole solutions in ECG, since, at least perturbatively in the coupling $\mu$, they represent the most general correction to Schwarzschild geometry. In this chapter, however, we aim to find fully non-perturbative solutions of the equations of motion of ECG, and this will be possible thanks to their simple form for SSS metrics, as remarked in the second item. 

\section{Asymptotically flat black holes}
\subsection{Spherically symmetric solutions of ECG}
Let us start by computing the equations of motion of \req{eq:ECGc4} for a SSS metric. We already know that ECG has spherically symmetric solutions of the form  \req{eq:Fmetric2}, so that we only need to obtain the equation that the function $f(r)$ satisfies. However, it is illustrative to show explicitly that the solutions have in fact the form \req{eq:Fmetric2}. Thus, we consider a general SSS metric ansatz of the form
\begin{equation}
ds^2_{N,f}=-N(r)^2f(r)dt^2+\frac{dr^2}{f(r)}+r^2d\Omega_{(2)}^2\,.
\end{equation}
As explained in Chapter~\ref{Chap:2}, we can obtain the equations of motion for $N$ and $f$ by evaluating the action in this metric ansatz and considering the reduced action as a functional of $N$ and $f$.  Integrating by parts several times, we find that we can write the reduced action functional $S[N,f]$ for ECG \req{eq:ECGc4} as
\begin{equation}
\label{Nfaction}
\begin{aligned}
S[N,f]=&\frac{1}{2G}\int dr \left\{N(r) \frac{d\E_f}{dr}+\mathcal{O}\left(N'^2/N\right)\right\}\, ,
\end{aligned}
\end{equation}
where $\mathcal{O}\left(N'^2/N\right)$ represents terms at least quadratic in derivatives of $N$, and
\begin{equation}
\E_f\equiv-\frac{1}{3}\Lambda r^3-(f-1)r-\frac{\mu L^4}{4} \left[f'^3+3\frac{f'^2}{r}-6f(f-1)\frac{f'}{r^2}-3ff''\left(f'-\frac{2(f-1)}{r}\right)\right]\, .
\end{equation}
Thus, when we take the variation with respect to $f$, we find that the equation $\delta S/\delta f=0$ is solved by $N'(r)=0$, in agreement with Theorem \ref{theo}. Thus, we fix $N(r)=1$. On the other hand, the variation with respect to $N$ evaluated on $N(r)=1$ yields
\begin{equation}\label{eq:feqq}
\frac{\delta S}{\delta N}\bigg|_{N=1}=\frac{1}{2G}\frac{d\E_f}{dr}=0\, ,
\end{equation}
and this is the only equation we have to solve. Equivalently, one may evaluate directly the field equations \req{eq:fe} on the single-function ansatz \req{eq:Fmetric2} and check that
\begin{equation}
\E_{tt}=-f^2 \E_{rr}\propto \frac{d\E_f}{dr}\, .
\end{equation}
Since both components are identical, it is consistent to set $N(r)=1$, and the equation we obtain for $f(r)$ is the same as the one we get from the reduced action.
Then, since the equation \req{eq:feqq} has the form of a total derivative, we can integrate it once to get $\E_f/(2G)=c$,
where $c$ is an integration constant. But we already determined in full generality in Sec.~\ref{bho} that this constant is always the ADM mass, $c=M$. Hence, we can finally write the equation for $f(r)$ as
\begin{equation}\label{eq:feqECG}
-(f-1)r-\frac{\mu L^4}{4} \left[f'^3+3\frac{f'^2}{r}-6f(f-1)\frac{f'}{r^2}-3ff''\left(f'-\frac{2(f-1)}{r}\right)\right]=\frac{1}{3}\Lambda r^3+2GM\, .
\end{equation}
In accordance with Theorem \ref{theo}, the problem is reduced to solving a second-order differential equation, where the ADM mass appears as an integration constant (we will check in a moment that $M$ is in fact the total mass). Note that, when the higher-order terms are set to zero, $\mu=0$, we obtain immediately the Schwarzschild-(A)dS solution,
\be
f_0(r)=1-\frac{2 GM}{r}-\frac{\Lambda r^2}{3}\, .
\ee
This is not a solution of \req{eq:feqECG} when $\mu\neq 0$, and in that case $f(r)$ will represent a smooth deformation of the previous function $f_0(r)$. The rest of the chapter is devoted to the study of the black hole solutions of Eq.~\req{eq:feqECG} and their properties.  Since our interest in this chapter is to study asymptotically flat solutions, from now on we set $\Lambda=0$. We will recover the cosmological constant in Chapter~\ref{Chap:7}, where we will study asymptotically AdS solutions. 

\subsection{Perturbative solution}
Given the complicated form of the differential equation \req{eq:feqECG}, finding an exact, analytic solution is a complicated task. A first possibility to simplify the problem is to treat the parameter $\mu$ --- or equivalently, the length scale $L$ --- as a small parameter, and to find the solution perturbatively in $\mu$.  At leading order in the coupling, the solution of \req{eq:feqECG} reads
\begin{equation}\label{eq:fpert}
f(r)=1-\frac{2GM}{r}-\frac{\mu L^4}{(GM)^4}\left(\frac{27 (GM)^6}{r^6}-\frac{46 (GM)^7}{r^7}\right)+\mathcal{O}(\mu^2L^8)\, .
\end{equation}
Actually, this is the approach that makes sense according to Effective Field Theory. In general, one would expect that, in the most general EFT for gravity --- as the one in Eq.~\req{eq:faction} --- one can expand the metric function $f(r)$ as a series in the length scale $L$:
\begin{equation}\label{eq:fexpand}
f(r)=f_0(r)+L^4f_4(r)+L^6 f_6(r)+L^8f_8(r)+\ldots
\end{equation}
Note that we are omitting the term $L^2$ because the quadratic curvature densities do not introduce corrections to the Schwarzschild solution, $f_0$. For this reason, the first correction in the most general EFT is related to the cubic terms, that are responsible for the term $L^4f_4(r)$ in \req{eq:fexpand}. We also saw, in Section~\ref{sec:mapsix}, that using field redefinitions all cubic terms can be mapped to the ECG density. Thus, the perturbative solution \req{eq:fpert} gives us the factor $f_4(r)$ in \req{eq:fexpand}, this is, the most general leading perturbative correction to the Schwarzschild solution in any higher-derivative gravity. 

It would be of course interesting to stop here to analyze the properties of this modified Schwarzschild solution \req{eq:fpert}. However, we do not do that because we will be able to study instead the properties of the exact, non-perturbative black hole solutions of ECG. Let us simply note that, at the level of the horizon $r_h\sim 2GM$, the corrections in \req{eq:fpert} are of the order of $L^4/(GM)^4\sim L^4/r_h^4$. Hence, $L$ is indeed related to the length scale at which the law of gravitation is corrected: black holes whose size is much larger than $L$ are barely modified, but those with radius of order $L$ or smaller have important corrections. Of course, when $r_h\sim L$, the perturbative expansion breaks down.  

Now, forgetting about EFT considerations and focusing on the theory \req{eq:ECGc4}, we may try to use a series expansion in $\mu L^4$ in order to solve the equation \req{eq:feqECG}. In fact, the expression \req{eq:fpert} only represents an accurate solution of Eq.~\req{eq:feqECG} when $\mu L^4/(GM)^4$ is very small, so we might expect that including higher powers of $\mu L^4$ we could get a solution valid even when $\mu L^4/(GM)^4$  is large.  With this purpose, we expand $f(r)$ as
\begin{equation}\label{eq:fser}
f(r)=\sum_{n=0}^{\infty}\left(\mu L^4\right)^nf_n(r)\, ,
\end{equation}
and plugging this into \req{eq:feqECG} we can solve order by order for the functions $f_n(r)$. However, when one analyzes the behaviour of $f_n$ for high-enough values of $n$, one concludes that the radius of convergence of the series \req{eq:fser} is zero.  This implies that, strictly speaking, the solution of equation \req{eq:feqECG} cannot be expanded as a perturbative series in $\mu$ around $\mu=0$. The reason must be that the solution is non-perturbative, namely, it must contain terms such as $e^{-\frac{g^2(r)}{\sqrt{\mu}}}$, which of course are invisible to the perturbative expansion --- we will provide evidence of this in the next subsection. Let us finally note that the fact that the series  \req{eq:fser} does not converge does not imply that the leading term \req{eq:fpert} is inaccurate: it does capture the leading correction providing that $\mu L^4/(GM)^4<<1$.

\subsection{Exact solution}\label{sec:exactECG}
We now address directly the problem of obtaining an exact solution of Eq.~\req{eq:feqECG}. Since it is a second-order differential equation, we need to fix two boundary conditions in order to find a solution. In the case at hand, we are looking for asymptotically flat black holes, and this naturally implies two conditions: asymptotic flatness and the existence of a regular horizon. Let us show how these conditions fix the solution by examining the behavior of Eq.~\req{eq:feqECG} in those limits.

\subsubsection{Asymptotic region}
The condition of asymptotic flatness implies that 
\begin{equation}\label{bdy}
\lim_{r\rightarrow\infty} f( r )=1\, .
\end{equation}
In order to determine whether this condition fixes an integration constant of equation \req{eq:feqECG}, let us solve it in the asymptotic region $r\rightarrow\infty$. A particular solution, $f_p$, can be found by assuming a $1/r$ expansion, and in that case we get
\begin{equation}\label{eq:fpart}
f_p(r)=1-\frac{2GM}{r}-\frac{27 \mu L^4(GM)^2}{r^6}+\mathcal{O}\left(\frac{1}{r^7}\right)\, .
\end{equation}
Observe that the $1/r$ coefficient is not modified by the corrections, hence the parameter $M$ is in fact the ADM mass. Now, there must be additional solutions since Eq.~\req{eq:feqECG} is a second-order differential equation. In order to find them, let us write the general solution in the asymptotic region as the sum of the particular solution $f_p$ plus another piece $f_h$,
\begin{equation}\label{eq:fdecompasymp}
f(r)=f_p(r)+f_h(r)\, .
\end{equation}
Since $\lim_{r\rightarrow\infty} f( r )=1$, $f_h$ must vanish asymptotically, and we can assume it is arbitrarily small. Then, inserting \req{eq:fdecompasymp} in Eq.~\req{eq:feqECG} and keeping only linear terms in $f_h$, we get a homogeneous equation for $f_h$
\begin{equation}\label{eq:homoeq}
A f_h''+Bf_h'+C f_h=0\, .
\end{equation}
Asymptotically, we find that the coefficients $A$, $B$ and $C$ behave as
\begin{equation}\label{eq:coeffhomo}
\begin{aligned}
A&=\frac{9 \mu  L^4 G M}{2 r^2}\left(1-\frac{2 GM}{r}\right)+\mathcal{O}\left(\frac{1}{r^4}\right)\, ,\\
B&=-\frac{9 \mu  L^4 G M}{r^3}\left(1-\frac{GM}{r}\right)+\mathcal{O}\left(\frac{1}{r^5}\right)\, ,\\
C&=-r+\frac{9\mu L^4 GM}{r^4}+\mathcal{O}\left(\frac{1}{r^5}\right)\, .
\end{aligned}
\end{equation}
Keeping only the leading term in each case, the general solution of equation \req{eq:homoeq} reads
\begin{equation}\label{eq:homosol}
f_h=c_1 r^{3/2} I_{3/5}\left(\frac{2 \sqrt{2} r^{5/2}}{15 L^2 \sqrt{\mu GM}}\right)+c_2 r^{3/2} K_{3/5}\left(\frac{2 \sqrt{2} r^{5/2}}{15 L^2 \sqrt{\mu GM}}\right)\, ,
\end{equation}
where $I_{\alpha}$ and $K_{\alpha}$ are modified Bessel functions of the first and second kind, respectively, and $c_{1,2}$ are two integration constants. Now let us examine the behaviour of this solution, for which we have to distinguish two cases: $\mu GM>0$ and $\mu GM<0$. Since we also assume that $GM>0$, these cases simply correspond to $\mu>0$ and $\mu<0$.  In the former case, the argument of the modified Bessel functions in \req{eq:homosol} is real and positive, and therefore they behave qualitatively as 
\begin{equation}\label{eq:homosol+}
f_h\sim c_1e^{\frac{2 \sqrt{2} r^{5/2}}{15 L^2 \sqrt{\mu GM}}}+c_2 e^{-\frac{2 \sqrt{2} r^{5/2}}{15 L^2 \sqrt{\mu GM}}} \quad\text{when}\quad r\rightarrow\infty\, .
\end{equation}
Hence, in order to get an asymptotically flat solution we must set $c_1=0$. Thus, the asymptotic flatness condition is fixing one integration constant of Eq.~\req{eq:feqECG}, leaving a one-parameter family of solutions. The remaining constant will be then fixed by the regular horizon condition, as we will see in a moment. 

Let us then consider the other possibility, $\mu<0$. In this case, the argument of the Bessel functions is a pure imaginary number, which implies that they acquire an oscillatory character and we get
\begin{equation}\label{eq:homosol-}
f_h\sim c_1 r^{1/4} \sin\left(\frac{2 \sqrt{2} r^{5/2}}{15 L^2 \sqrt{\mu GM}}\right)+c_2 r^{1/4}\cos\left(\frac{2 \sqrt{2} r^{5/2}}{15 L^2 \sqrt{\mu GM}}\right) \quad\text{when}\quad r\rightarrow\infty\, .
\end{equation}
These solutions are sick because they do not decay at infinity, but also because they oscillate wildly when $r\rightarrow \infty$. Thus, the only way of getting an asymptotically flat solution in this case is to set $c_1=c_2=0$. This means that there is a unique asymptotically flat solution to the equation \req{eq:feqECG} and we cannot impose further conditions. In particular, we will not be able to impose the regular horizon condition, and as a consequence, there are no asymptotically flat black holes in the case $\mu<0$.  For this reason, we will assume from now on that $\mu\ge0$.

Before passing to the next section, we can extract some interesting conclusions from the form of the asymptotic solution. As we remarked, the integration constant $c_2$ in \req{eq:homosol} will in general be non-vanishing. Note that the term multiplied by $c_2$ is non-analytic in $\mu$, in the sense that it cannot be expanded in a Taylor series around $\mu=0$. This explains why in the previous section we were unable to obtain a solution by performing a series expansion in the parameter $\mu$: the solutions of Eq.~\req{eq:feqECG} are typically non-perturbative in $\mu$. 
In addition, we see that this term decays at infinity faster than exponentially, and this has a nice interpretation in terms of the degrees of freedom that propagate in this theory. Let us recall that a massive mode of mass $m$ generically produces a contribution that decays as $\sim e^{-m r}$ --- we saw this explicitly in Sec.~\ref{sec:sollin}. Now, Einsteinian cubic gravity does not propagate additional modes in maximally symmetric backgrounds --- in the case at hand, Minkowski spacetime --- because they are infinitely massive.  However, as we discussed at the end of Chapter \ref{Chap:1}, there could be additional modes propagating in less symmetric backgrounds. Here, the fact that we get a contribution of the form $\sim e^{-\frac{2 \sqrt{2} r^{5/2}}{15 L^2 \sqrt{\mu GM}}}$ to the solution suggests that, in the background of a black hole, we have additional degrees of freedom, whose mass depends on the position and would be of the order
\begin{equation}\label{eq:madd}
m\sim \frac{2 \sqrt{2} r^{3/2}}{15 L^2 \sqrt{\mu GM}}\, .
\end{equation}
We observe that the mass of these hypothetical modes diverges for $r\rightarrow\infty$, so that they could never escape to infinity and they would be confined to a finite region in the vicinity of the black hole. This nice intuition could be confirmed by performing 
perturbation theory in the background of a black hole, but such analysis is beyond the scope of this thesis. 
Likewise, let us note that in the case $\mu<0$ the mass \req{eq:madd} would be imaginary, so that the additional modes would be tachyons. This is yet another reason to discard a negative coupling. 

%\begin{equation}
%f(r)=1-\frac{2GM}{r}-\frac{27 \mu L^4(GM)^2}{r^6}+\mathcal{O}\left(\frac{1}{r^7}\right)+c_2 r^{3/2} K_{3/5}\left(\frac{2 \sqrt{2} r^{5/2}}{15 L^2 \sqrt{\mu GM}}\right)\quad\text{when}\quad r\rightarrow\infty\, .
%\end{equation}

\subsubsection{Near-horizon region}
We have just seen that the asymptotic flatness condition fixes one of the integration constants of the equation \req{eq:feqECG}, leaving us with a one-parameter family of solutions. Thus, in order to determine a solution we still have to impose another condition. Since 
we intend to obtain black hole solutions, the remaining condition is nothing but the existence of an event horizon. In the metric \req{eq:Fmetric2}, a horizon is identified simply by the condition $f(r_h)=0$ for some $r_h>0$. However, in order for the horizon to be regular, we must also demand continuity and differentiability of the metric at that point. 
Thus, we assume the existence of some $r_h>0$ such that $f(r_h)=0$ and we demand that $f(r)$ can be expanded in a power series around $r=r_h$, as
\be\label{eq:fnh}
f(r)=\sum_{n=1}^{\infty} a_n (r-r_h)^n\, ,
\ee
for some coefficients $a_n$.
At first sight, it is quite puzzling how such a condition could impose any constraint in the solution, since we are just demanding analyticity at the point $r_h$, which, in addition, is undetermined. However, let us insert \req{eq:fnh} in the equation \req{eq:feqECG}. Expanding also the equation order by order in $(r-r_h)$, we find an expression of the form 
\begin{equation}
\sum_{n=0}^{\infty}C_n (r-r_h)^n=0\, ,
\end{equation}
which implies that every term must vanish independently, this is, $C_n=0$ $\forall\, n\ge0$. This provides us with a system of equations for the coefficients $a_n$ of the near-horizon expansion \req{eq:fnh}. The first few equations read
\begin{align}\label{eq:C0}
C_0= &-2 M+r_h-\mu L^4 \frac{ a_1^2}{4}\left(a_1+\frac{3}{r_h}\right)=0\, ,\\ \nonumber
C_1= &\, 1-a_1 r_h-\frac{3 a_1^2 \mu  L^4}{4 r_h^2}=0\, ,\\ \nonumber
C_2=&\, -a_1-a_2 r_h\\ \nonumber
&\, + \mu L^4 \left[a_3 \left(\frac{9 a_1^2}{4}+\frac{9 a_1}{2
   r_h}\right)+a_2 \left(-\frac{9 a_1}{2 r_h^2}-\frac{3 a_1^2}{r_h}+\frac{3}{2} a_1 a_2\right)+\frac{9
   a_1^2}{4 r_h^3}+\frac{3 a_1^3}{2 r_h^2}\right]=0\, .
\end{align}
The equation $C_1=0$ gives us a relation between $a_1$ and $r_h$ that allows us to obtain $a_1(r_h)$:
\begin{equation}\label{eq:a1sol}
a_{1}^{(\pm)}=\frac{2r_h \left(-r_h^2 \pm  \sqrt{r_h^4+3 \mu  L^4}\right)}{3 \mu  L^4}
\end{equation}
In order to describe a black hole solution we must have $a_1>0$, so that, out of the two possible roots $a_{1}^{(\pm)}$, we must choose $a_{1}^{(+)}$ (remember that we take $\mu>0$). Note also that, in the limit $\mu\rightarrow0$, this root reduces to the value of $a_1$ in the Schwarzschild solution: $\lim_{\mu\rightarrow 0} a_{1}^{(+)}=1/r_h$.
Now, using this relation in the equation $C_0=0$ we can get the mass as a function of $r_h$,
\begin{equation}\label{eq:MrhECG}
2GM=\frac{4r_h^3}{27 \mu ^2 L^8}\left(2 r_h^6+\left(-2 r_h^4 +3 \mu  L^4\right)\sqrt{r_h^4+3 \mu  L^4} \right)\, .
\end{equation}
Although it is not apparent written in this way, this expression also reduces to the Schwarzschild value $2GM=r_h$ when $\mu\rightarrow 0$. Let us remark that the expressions \req{eq:MrhECG} and \req{eq:a1sol} contain no approximations and are \emph{exact}. They appear as consistency conditions for the existence of a regular horizon, and any black hole solution must satisfy them. The fact that we can obtain explicit and analytic expressions for $M(r_h)$ and $a_1(r_h)$ is related to the very special form of the equations of motion of Einsteinian cubic gravity \req{eq:ECGc4}, and it is not possible to obtain an analogous result for other higher-order gravities. 

Now, one can check that the relation $M(r_h)$ is one-to-one for $M\ge 0$, $r_h\ge 0$, and this allows us to invert (implicitly) the relation \req{eq:MrhECG} in order to get $r_h(M)$. At the same time, we obtain the relation $a_1(M)$ using \req{eq:a1sol}. 
Thus, the two first equations in \req{eq:C0} fix the radius $r_h$ and the coefficient $a_1$ as functions of the mass. Let us new take a look at the next equations.  We see that from $C_2=0$ we can get $a_3$ as a function of $a_2$, $a_1$ and $r_h$, 
\begin{equation}
a_3= \left(\frac{9 a_1^2}{4}+\frac{9 a_1}{2r_h}\right)^{-1}\left[\frac{a_1+a_2 r_h}{\mu L^4}-a_2 \left(-\frac{9 a_1}{2 r_h^2}-\frac{3 a_1^2}{r_h}+\frac{3}{2} a_1 a_2\right)-\frac{9a_1^2}{4 r_h^3}-\frac{3 a_1^3}{2 r_h^2}\right]\, ,
\end{equation}
but since $a_1$ and $r_h$ both are fixed by the mass, this means that we get a relation of the form $a_3=a_3(a_2;M)$. Likewise, from the equation $C_3=0$ (that we do not show here) we obtain $a_4$ as a function of $a_3$, $a_2$, $a_1$ and $r_h$, but replacing the results of the previous equations, we end up with an expression of the form $a_4(a_2;M)$. Examining the following equations, one finds that the pattern is general: for a fixed mass, all the coefficients $a_{n\ge3}$ are determined by the value of $a_2$. In other words, the full near-horizon expansion \req{eq:fnh} is determined once $a_2$ is specified. Hence, despite Eq.~\req{eq:feqECG} being a second-order equation, we find only a one-parameter family of solutions possessing a regular horizon. This is telling us that imposing the existence of such horizon is actually fixing one of the integration constants of \req{eq:feqECG}. The reason must be that there are other solutions that possess a ``singular horizon'', \eg  this happens if the function $f(r)$ vanishes at some $r=r_h>0$, but is not differentiable there. In fact, the equation \req{eq:feqECG} is singular at the points in which $f(r)=0$, because the coefficient of $f''$ (which is the highest derivative) vanishes. Thus, on general grounds, the solution will not be differentiable at those points. What we have shown is that the solutions that are differentiable only contain one free parameter, which can be chosen to be $a_2\equiv f''(r_h)/2$.

\subsubsection{Complete solution}
In the two previous items we have learned that Eq.~\req{eq:feqECG} possesses a one-parameter family of asymptotically flat solutions, and another one-parameter family of solutions with a regular horizon. Thus, we expect that there exists at least \emph{one} solution satisfying both conditions; such solution would represent a black hole. In order to find it we must solve Eq.~\req{eq:feqECG} numerically. The resolution is a bit tricky because the equation is highly \emph{stiff} and also because the boundary conditions at the horizon are non-standard. We comment here on the basic idea behind the numeric resolution, and keep the details for Appendix~\ref{App:3}. 

First, we fix a mass $M$ for which we want to compute the solution. Then, we start the numerical method at the horizon. Using the equations  \req{eq:MrhECG} and \req{eq:a1sol} we determine the radius $r_h$ and the derivative of $f'(r_h)=a_1$. As we discussed earlier, this must be fixed if we want to produce a regular solution. Then, the initial condition to start the numerical method is the value of the second derivative, $f''(r_h)=2a_2$. We must choose this parameter so that the resulting solution is asymptotically flat. Since the asymptotic flatness condition imposes one constraint, we expect that there exists (at least) one value of $a_2$ that yields an asymptotically flat solution. That value can be found by using the ``shooting method'': we search for a value of $a_2$ such that the numerical solution, extended up to sufficiently large $r$, overlaps with the asymptotic expansion \req{eq:fpart} --- see Appendix~\ref{App:3}. The conclusion, after gathering large empirical evidence, is that the parameter $a_2$ yielding an asymptotically flat solution always exists and is unique for all $\mu\ge0$, $M>0$. Hence, Eq.~\req{eq:feqECG} possesses a unique solution corresponding to an asymptotically flat black hole.

Let us then examine the profile and properties of these solutions. In Fig.~\ref{fig1} we show the profile of $f(r)$ for a fixed mass and several values of the ECG coupling and we compare these solutions with the Schwarzschild one. We can see several differences. First, we note that, for large enough $r$, the solutions are almost indistinguishable from the Schwarzschild one, and the differences only become important as we move to lower values of $r$. Focusing on the horizon --- corresponding to the point where $f(r)$ vanishes --- we see that black holes in ECG are larger than its Einstein gravity counterparts, and the discrepancy grows when we increase the value of $\mu L^4/(GM)^4$. 
\begin{figure}[t!]
\centering 
\includegraphics[width=0.65\textwidth]{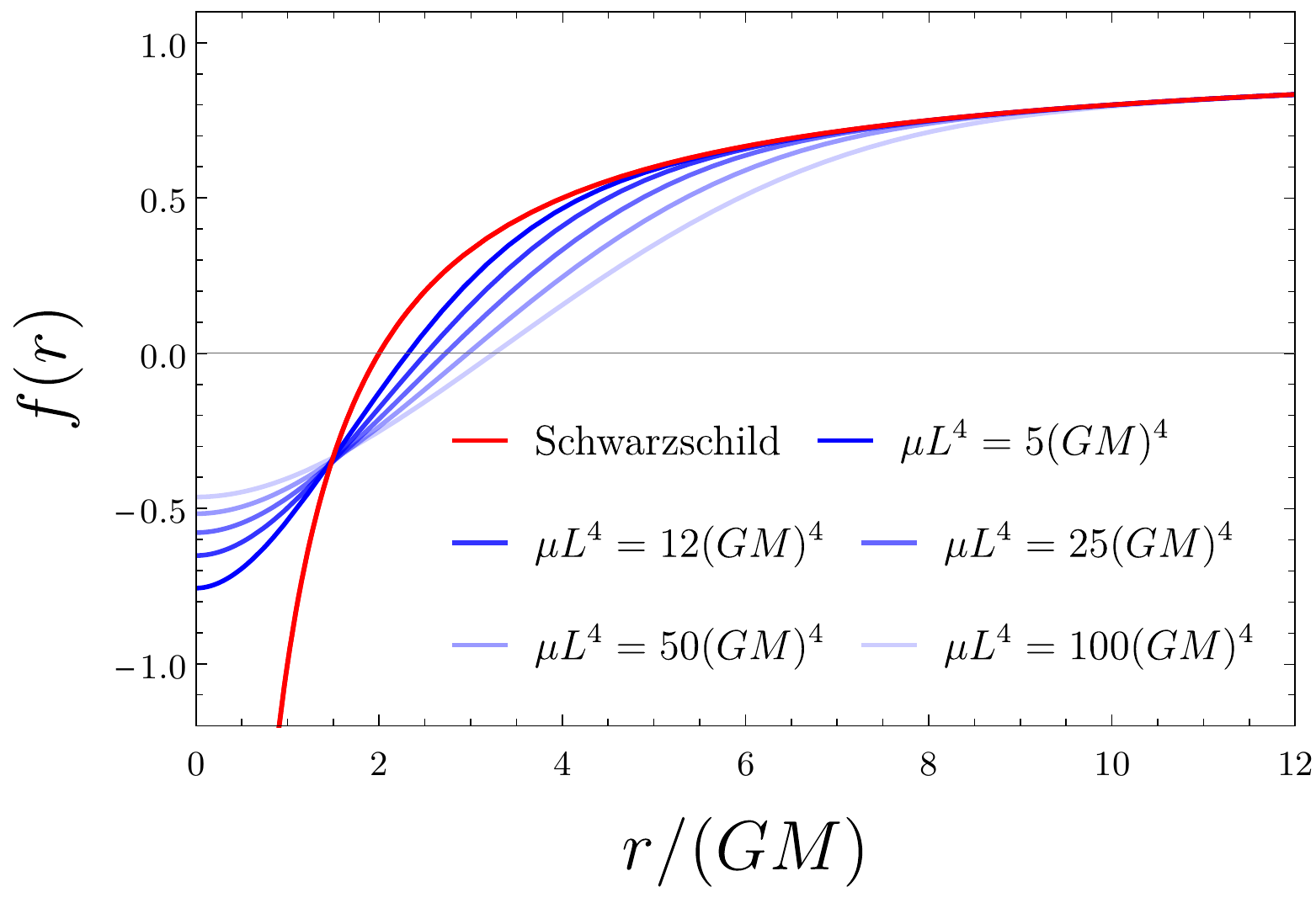}
\caption{Asymptotically flat black holes in Einsteinian cubic gravity: profile of $f(r)$ for a fixed mass $M$ and several values of the ECG coupling $\mu L^4$. The red line corresponds to the usual Schwarzschild blackening factor, $\mu=0$. } 
\labell{fig1}
\end{figure}
The relation between the mass $M$ and the radius $r_h$ of these black holes is exactly given by the expression \req{eq:MrhECG} that we found before. Using it, we plot in Fig.~\ref{fig2} the horizon radius as a function of the mass for ECG black holes. For convenience, we show $r_h$ and $GM$ in units of the natural length scale $\mu^{1/4}L$, which allows us to capture all the cases using a single curve. For large masses we recover the Schwarzschild prediction, but we observe that in the limit $M\rightarrow 0$ the behaviour is drastically different. Solving $r_h(M)$ from \req{eq:MrhECG} in that limit, we get
\begin{equation}
r_h=\frac{\sqrt{3}\mu^{1/6} L^{2/3}G^{1/3}}{2^{1/3}}M^{1/3} \quad \text{when}\quad M<<\frac{\mu^{1/4} L}{G}
\end{equation}
As an interesting observation, note this relation reminds the one of a usual matter distribution, $M\propto \rho\cdot r_h^3$, where the ``density'' in this case would be of the order of $\rho\sim 1/(G L^2)$. 
\begin{figure}[t!]
\centering 
\includegraphics[width=0.65\textwidth]{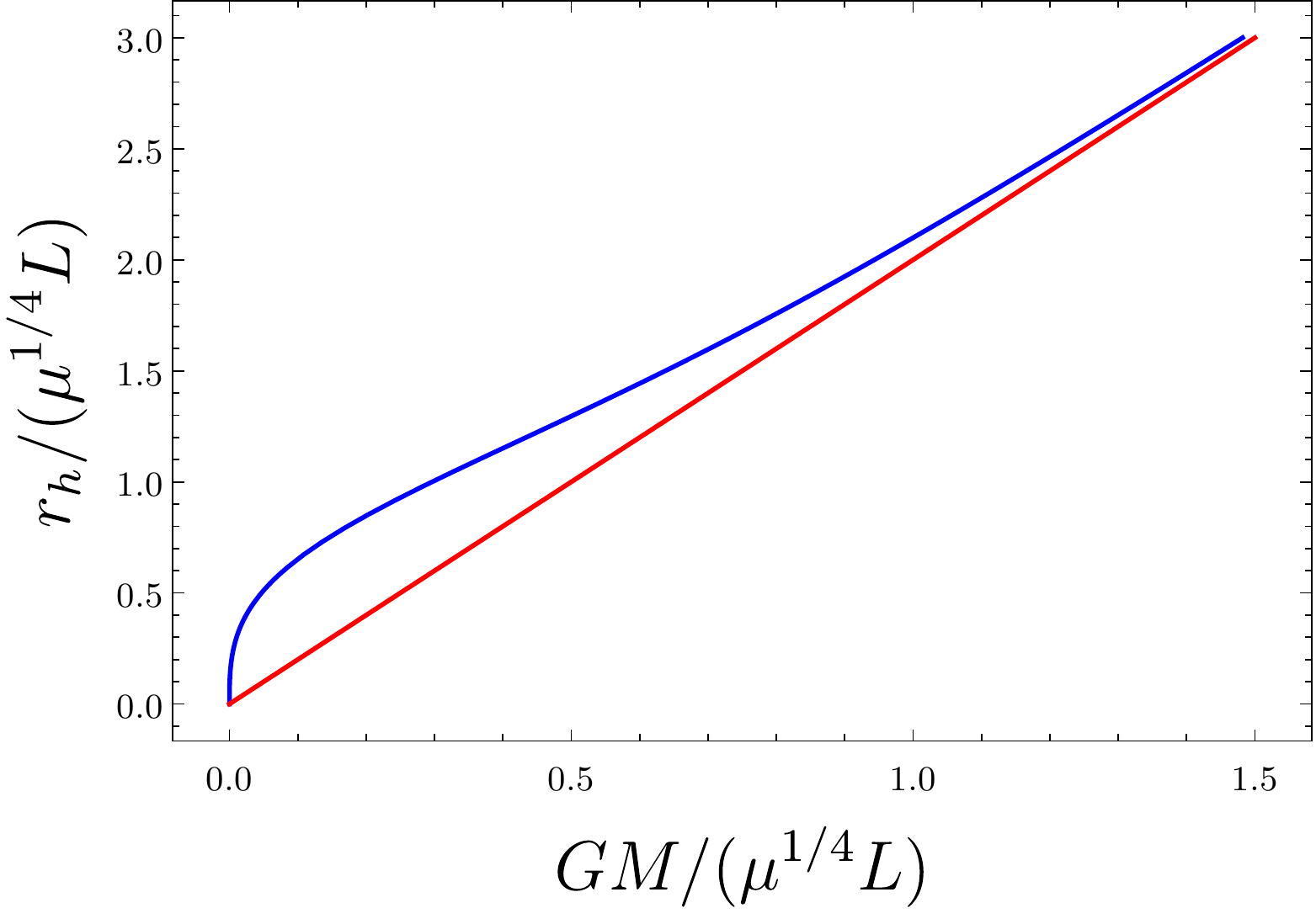}
\caption{We plot the horizon radius $r_h$ as a function of the mass for ECG black holes (blue line) and we compare it with the usual relation $r_h=2GM$ for the Schwarzschild black hole (red line). We have normalized $r_h$ and $GM$ by the natural length scale of the system, $\mu^{1/4}L$. For large masses, we recover the Schwarzschild value $r_h\approx 2GM$, but for small masses the relation is instead $r_h\propto M^{1/3}$.} 
\labell{fig2}
\end{figure}
But the most dramatic difference with respect to the Schwarzschild solution occurs in the interior of the black hole, which in practice can be described by the metric \req{eq:Fmetric2} using the negative part of $f(r)$, shown in Fig.~\ref{fig1} for different cases.\footnote{Of course, the coordinates we are using are not valid in order to probe the solution beyond the horizon, but since $f(r)$ is analytic, an analytic extension can be performed exactly as in the case of Schwarzschild geometry.}  The main difference is that, while $f(r)$ diverges at $r=0$ in the case of the Schwarzschild solution, the ECG black holes have a finite and smooth $f(r)$ everywhere. Indeed, looking at the behaviour of $f(r)$ near $r=0$ we see that it always takes the form
\begin{equation}
f(r)=-a+br^2 +\mathcal{O}(r^3) \quad \text{when}\quad r\rightarrow 0\, ,
\end{equation}
for some constants $a$ and $b$. However, despite having an apparent smooth character, the point $r=0$ is still a singularity, as can be checked by computing the Kretschmann invariant, 
\begin{equation}\label{krets}
R_{\mu\nu\rho\sigma}R^{\mu\nu\rho\sigma}=\frac{4(f(0)-1)^2}{r^4}+\mathcal{O}\left(\frac{1}{r^2}\right) \quad \text{when}\quad r\rightarrow 0\, .
\end{equation}
Nevertheless, the singularity problem is improved with respect to the Schwarzschild black hole, for which the Kretschmann invariant diverges as $\sim 1/r^6$. Furthermore, the character of the singularity in ECG black holes is also significantly different with respect to the one in a Schwarzschild black hole, in the following sense.  Physically, the singularity in the black hole interior is associated with the divergence of the tidal forces at some certain moment. In a Schwarzschild black hole, the tidal forces ``stretch'' in the radial direction and ``squash'' in the angular directions, and both effects diverge when the singularity is reached. However, in an ECG black hole, the singularity is entirely caused by the collapse of the angular directions, and only the tidal forces in the angular directions diverge.  In fact, one can check that, if one removes the angular components of the metric \req{eq:Fmetric2}, the resulting line element $ds_2^2=-f(r)dt^2+dr^2/f(r)$ is completely regular at $r=0$.  This implies, for instance, that radial geodesics of the metric \req{eq:Fmetric2} that fall inside the black hole do not find any singularity at $r=0$, and suggests that the metric  could actually be extended beyond $r=0$. This is a thrilling possibility that could be explored elsewhere. 

Let us conclude this section showing the solutions in Fig.~\ref{fig1} from a different perspective. In that figure, we plotted the profile of the solutions for a fixed mass and several values of the coupling, which was useful in order to compare the ECG solutions to the Schwarzschild one. But from a physical perspective, it is more illuminating to fix a value of the coupling and plot the solutions for different masses. We show the corresponding plot in Fig.~\ref{fig3}.
\begin{figure}[t!]
\centering 
\includegraphics[width=0.65\textwidth]{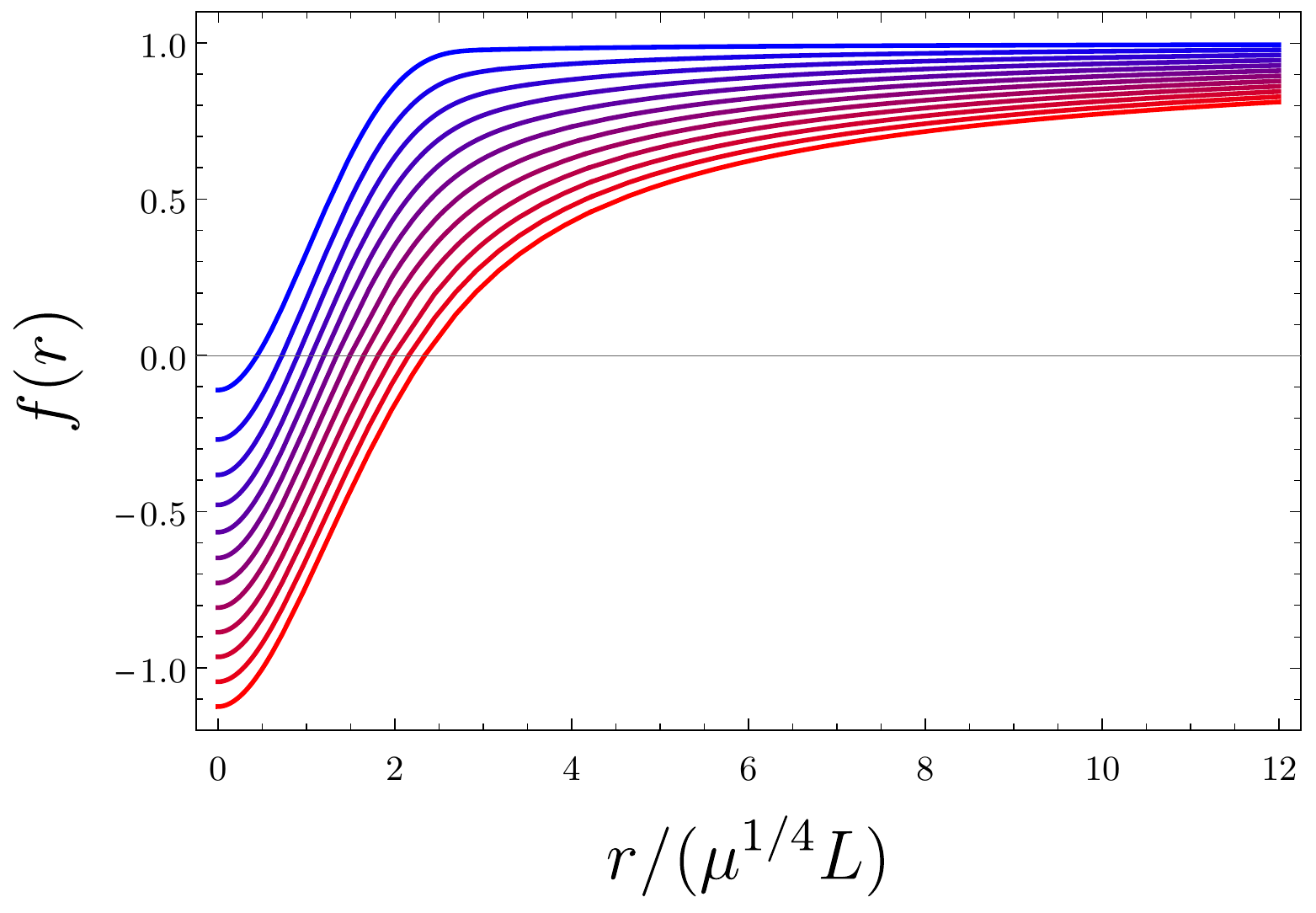}
\caption{Black holes of different masses for a fixed value of the ECG coupling $\mu L^4$. Red tones denote larger masses and blue tones smaller ones. The zero mass limit of these black holes is not flat space.} 
\labell{fig3}
\end{figure}
There, we see that, for large enough masses, the exterior solution is similar to the Schwarzschild one, but the interior solution is always drastically different. For smaller masses, the exterior solution also becomes very different, and we observe a fact that was unnoticed in Fig.~\ref{fig1}: the zero mass limit of these black holes is not flat space. In fact, as we decrease the mass, the horizon radius tends to zero, but we observe quite clearly that there is a ``potential well'' of length $\sim \mu^{1/4} L$ that is present for arbitrarily small masses. Thus, the limit $M\rightarrow 0$ of these solutions is quite exotic: it seems to correspond to a black hole of vanishing mass and area. In addition, the horizon, which is reduced to a point, is placed at an infinite distance,\footnote{This follows from the fact that the behaviour of $f(r)$ near $r=0$ for these solutions is $f(r)\propto r^2$.} which further characterizes this solution as an extremal black hole. We will learn more about this limit in the next section.

\section{Black hole thermodynamics}\label{Sec:thermoECG}
After constructing the spherically symmetric black hole solutions of Einsteinian cubic gravity, we turn our attention to the thermodynamic properties of these black holes. We already know that the parameter $M$ entering in the equation \req{eq:feqECG} is the total mass --- or energy --- of the system. We wish now to compute the temperature and entropy \cite{Hawking:1974sw,Bardeen:1973gs,Bekenstein:1973ur,Bekenstein:1974ax} of these black holes and to determine their dependence on $M$. 
As we show below, it is a formidable property of ECG that one can determine these quantities analytically and fully non-perturbatively in the higher-order coupling. 

\subsubsection{Temperature}
Let us first identify the Hawking temperature \cite{Hawking:1974sw} of ECG black holes. This can be done by considering the Euclidean version of the black hole metric \req{eq:Fmetric2} and finding the periodicity of the Euclidean time $\tau=i t$. Let us then write 
\begin{equation}
ds^2_{\rm E}=f(r)d\tau^2+\frac{dr^2}{f(r)}+r^2d\Omega_{(2)}^2\, ,
\end{equation}
and let us study the behaviour of this metric near $r=r_h$, which in the Lorentzian case corresponds to the horizon. Using the expansion \req{eq:fnh} and introducing the new radial coordinate $\rho=2\sqrt{\frac{r-rh}{a_1}}$, we get
\begin{equation}\label{eq:eunh}
ds^2_{\rm E}\approx \rho^2\left(\frac{a_1d\tau}{2}\right)^2+d\rho^2+r_h^2d\Omega_{(2)}^2\quad\text{when}\quad \rho\rightarrow 0\, .
\end{equation}
Now, observe that the metric of the plane can be written as $\rho^2d\phi^2+d\rho^2$, where the coordinate $\phi$ has period $2\pi$; otherwise, that metric would represent a cone, with a corresponding singularity at $\rho=0$. For the same reason, the Euclidean time $\tau$ in \req{eq:eunh} must have a period $\beta=4\pi/a_1$ in order to avoid a conical singularity. Since $\beta$ is the inverse of the Hawking temperature $T$, we simply get
\begin{equation}\label{eq:THdef}
T=\frac{a_1}{4\pi}\, .
\end{equation}
The same result can be found by using the identification $T=\frac{\kappa}{2\pi}$, where $\kappa$ is the surface gravity of the black hole. This result is general for any spherically symmetric black hole with metric \req{eq:Fmetric2}, but let us now apply it to ECG black holes. Remarkably enough, in the previous section we found an analytic expression for the parameter $a_1$ in terms of the horizon radius --- see \req{eq:a1sol}. Thus, the Hawking temperature of ECG black holes reads, exactly
\begin{equation}\label{eq:TECG}
T=\frac{r_h}{2\pi \left(r_h^2+ \sqrt{r_h^4+3 \mu  L^4}\right)}\, .
\end{equation}
This expression, together with the relation $M(r_h)$ given in \req{eq:MrhECG} allows us to obtain the thermodynamic relation $T(M)$. Alternatively, one can derive the following ``equation of state'' relating $T$ and $M$ by combining the equations in \req{eq:C0} and \req{eq:THdef},
\begin{equation}\label{eq:eqofstate}
\pi GMT-\mu (2\pi T L)^4-\left(2\pi GMT+\mu (2\pi T L)^4\right)^{3/2}=0\, .
\end{equation}
This equation can be solved explicitly for $M$, yielding a complicated expression. Note that for $\mu=0$ we recover the usual expression for the temperature of a Schwarzschild black hole, $T_{\rm Schw.}=(8\pi G M)^{-1}$. 
The relation $T(M)$ for ECG black holes is plotted in Fig.~\ref{fig4} and we observe remarkable differences with respect to the situation in Einstein gravity. 
\begin{figure}[t!]
        \centering
                \includegraphics[width=0.65\textwidth]{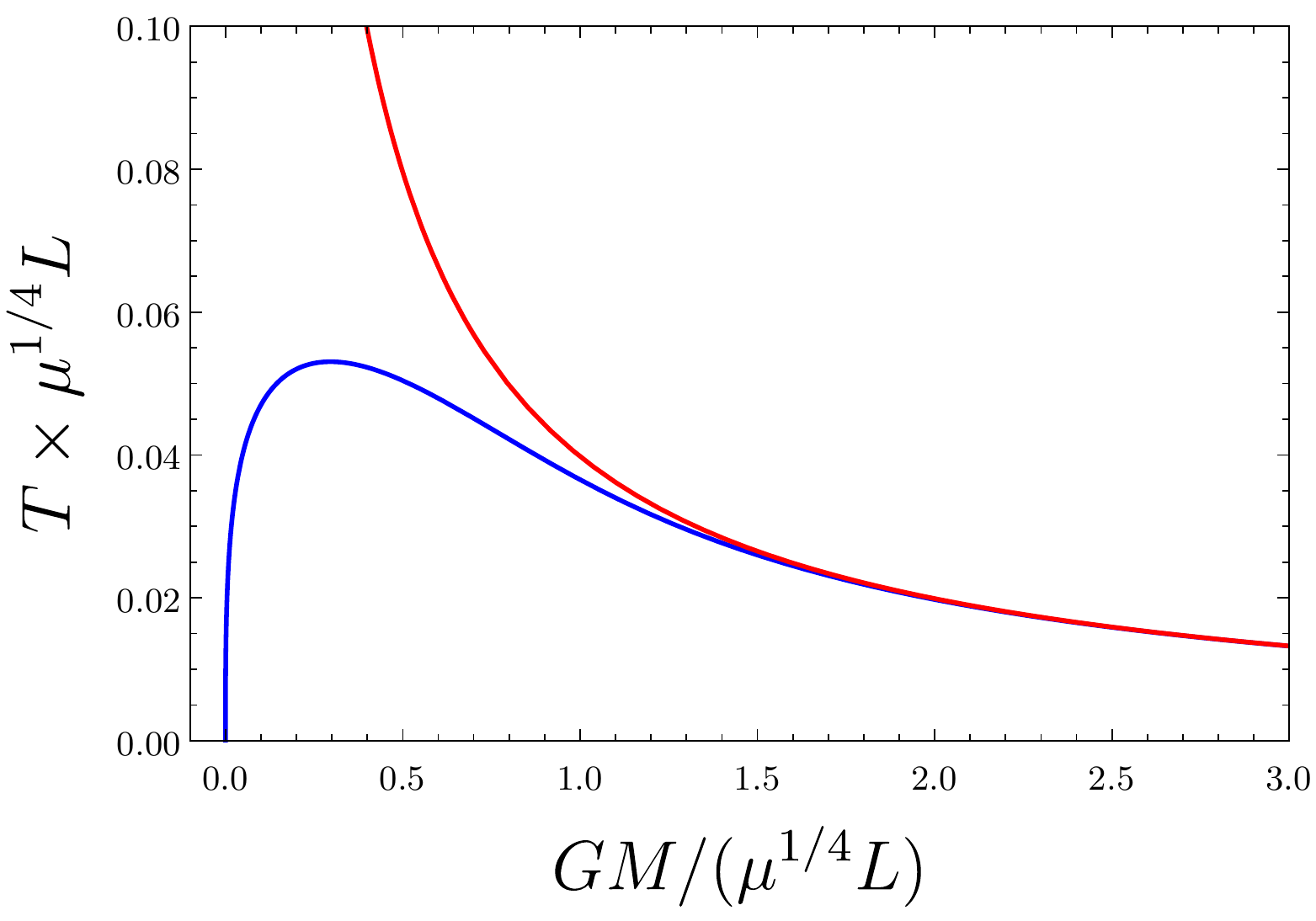}
        \caption{Hawking temperature as a function of the mass for Schwarzschild (red) and ECG (blue) black holes. We normalize both quantities using the natural scale in each case. We observe the existence of a maximum temperature in the case of ECG black holes ---see Eq.~\req{eq:Tmax}. Thus, for a given temperature $T<T_{\rm max}$ there are two different black hole solutions in ECG, a large one and a small one. For $M\rightarrow0$ the temperature of black holes in ECG vanishes, just the opposite as for Einstein gravity. }
\labell{fig4}
\end{figure}
As usual, for large masses we reproduce the EG prediction, but the behaviour for small masses is just the opposite: while the temperature of a Schwarzschild black hole diverges when $M\rightarrow 0$, the one of an ECG black hole vanishes in that limit. In fact, the relation $T(M)$ near $M=0$ reads
\begin{equation}
T=\frac{1}{2\pi}\left(\frac{GM}{2\mu L^4}\right)^{1/3} \quad\text{when}\quad M<<\frac{\mu^{1/4} L}{G}\, .
\end{equation}
Note also that in this limit the temperature is proportional to the horizon radius, $T\sim \frac{r_h}{2\pi\sqrt{3\mu}L^4}$. For $M=0$ we get $r_h=T=0$, so the corresponding solution in that case is an extremal black hole of vanishing area, as we advanced previously. 

Another interesting observation that we extract from Fig.~\ref{fig4} is that the temperature of black holes in Einsteinian cubic gravity is bounded from above: there is a maximum temperature $T_{\rm max}$ that is reached at a certain mass $M_{\rm max}$.  Extremizing $T$ with respect to $M$ in \req{eq:eqofstate}, we find
\begin{equation}\label{eq:Tmax}
T_{\rm max}=\frac{1}{6\pi \mu^{1/4}L}\, ,\qquad M_{\rm max}=\frac{8\mu^{1/4}L}{27 G}\, .
\end{equation}
The radius corresponding to these black holes of maximum temperature is simply $r_{h,{\rm max}}=\mu^{1/4} L$. 
Notice that these interesting properties are highly non-perturbative in the coupling and can never be seen in a perturbative approach.

\subsubsection{Entropy and first law}
Let us now compute the entropy of these black holes. The black hole entropy in higher-derivative gravity is no longer proportional to the area of the horizon; instead, it is given by Wald's formula \cite{Wald:1993nt,Iyer:1994ys,Jacobson:1993vj}, which reads
\begin{equation}\label{eq:Wald1}
S=-2\pi \int_{H} d^2x\sqrt{h} \frac{\delta \mathcal{L}}{\delta R_{\mu\nu\rho\sigma}}\epsilon_{\mu\nu}\epsilon_{\rho\sigma}\, ,
\end{equation}
where the integral is taken over the bifurcation surface of the horizon and $\frac{\delta \mathcal{L}}{\delta R_{\mu\nu\rho\sigma}}$ is the Euler-Lagrange derivative of the gravitational Lagrangian as if the Riemann tensor were an independent variable, this is,
\begin{equation}
\frac{\delta \mathcal{L}}{\delta R_{\mu\nu\rho\sigma}}=\frac{\partial \mathcal{L}}{\partial R_{\mu\nu\rho\sigma}}-\nabla_{\alpha}\left(\frac{\partial \mathcal{L}}{\partial \nabla_{\alpha}R_{\mu\nu\rho\sigma}}\right)+\ldots
\end{equation}
In addition, $h$ is the determinant of the induced metric on the horizon and $\epsilon_{\mu\nu}$ is the binormal of the horizon, normalized as $\epsilon_{\mu\nu}\epsilon^{\mu\nu}=-2$. 

Applying Wald's formula \req{eq:Wald1} to our theory \req{eq:ECGc4}, we get
\begin{equation}\label{eq:Wald2}
S=\frac{1}{4G} \int_{H} d^2x\sqrt{h}\left[1+\frac{\mu L^4}{16}P_{\mu\nu\alpha\beta}\epsilon^{\mu\nu}\epsilon^{\rho\sigma}\right]\, ,
\end{equation}
where $P_{\mu\nu\alpha\beta}$ is the tensor defined in \req{eq:PECG}. Now we have to apply it to the black hole metric \req{eq:Fmetric2}, evaluated at the horizon $r=r_h$. In the coordinates $r$ and $t$ we are using, the only non-vanishing components of the binormal are $\epsilon_{tr}=-\epsilon_{rt}=1$ (note that the sign is actually irrelevant). Thus, evaluating the expression \req{eq:Wald2} we obtain
\begin{equation}\label{eq:Wald33}
S=\frac{\pi r_h^2}{G} \left[1+\frac{\mu L^4}{4}P_{trtr}\right]\, ,
\end{equation}
where have already performed the integration on the horizon, which is possible because $P_{trtr}$ must be constant. In fact, evaluated at $r=r_h$ this quantity reads
\begin{equation}
P_{trtr}\Big|_{r=r_h}=-\frac{3 f'(r_h) \left(r_h f'(r_h)+4\right)}{r_h^3}\, .
\end{equation} 
Amazingly, this expression does not depend on second derivatives of $f$, and on the other hand we have $f'(r_h)=4\pi T$. In addition, we know the explicit relation between $T$ and $r_h$, which is given by \req{eq:TECG}. Putting everything together and implementing few simplifications, we obtain the following exact expression for the entropy written in terms of the radius
\begin{equation}\label{eq:SECG1}
S=\frac{2 \pi  \left(r_h^4-3 \mu  L^4\right) \left(\sqrt{3 \mu  L^4+r_h^4}-r_h^2\right)}{3G \mu  L^4}\, .
\end{equation}
Using then the Eqs.~\req{eq:MrhECG} and \req{eq:TECG}, we can write in a parametric way the thermodynamic relations $S(M)$ or $S(T)$. Again, it is a very non-trivial property of Einsteinian cubic gravity the fact that we can find a closed and exact expression for the entropy of black holes. 

Looking at the expression \req{eq:SECG1}, we observe that something quite funny happens with the entropy of these black holes: it vanishes for $r_h=(3\mu)^{1/4}L$ (which corresponds to a mass $GM=2 \left(2-\sqrt{2}\right) (\mu/27)^{1/4} L$). For even smaller values of the radius (or the mass), the entropy becomes negative, and for $r_h\rightarrow 0$ we get $S(r_h=0)=-2\pi\sqrt{3\mu} L^2/G$. Objects with negative entropy should be regarded as unphysical, and this is a possibility that we might consider; perhaps there is a minimum mass for which a black hole can be described semiclassically in an effective theory such as \req{eq:ECGc4}. However, all these black holes possess finite temperature and mass, and they are perfectly regular, so this conclusion seems to be unjustified. Let us recall that, unless an explicit definition of the entropy as a count of microstates is given, the thermodynamic entropy of a system is defined only up to the addition of a constant, so the negative entropies we are obtaining might not be meaningful.  In the case of gravity, this is manifest in the freedom to add topological terms in the action. In fact, there is no reason to discard a Gauss-Bonnet term in the action \req{eq:ECGc4}, since, as we saw in Section~\ref{sec:construction}, it is a member of the family of Generalized quasi-topological gravities, in which we are interested. Of course, we did not include it in \req{eq:ECGc4} because it has no effect on the equations of motion. However, it has an effect on black hole entropy --- see \eg \cite{Jacobson:1993xs,Chatterjee:2013daa}.
Thus, let us add a Gauss-Bonnet term to the gravitational Lagrangian in \req{eq:ECGc4} \ie $\mathcal{L} \rightarrow \mathcal{L}+ \frac{\alpha L^2}{16 \pi G} \mathcal{X}_4$, where  $\mathcal{X}_4=R^2-4R_{\mu\nu}R^{\mu\nu}+R_{\mu\nu\rho\sigma}R^{\mu\nu\rho\sigma}$, and where $\alpha$ is an arbitrary dimensionless coupling. 
The effect of that term is to add a universal constant contribution to the entropy of any horizon with spherical topology. Once this is taken into account, the entropy of ECG black holes reads 
\begin{equation}\label{eq:SECG2}
S=\frac{2 \pi  \left(r_h^4-3 \mu  L^4\right) \left(\sqrt{3 \mu  L^4+r_h^4}-r_h^2\right)}{3G \mu  L^4}+2\pi \alpha \frac{L^2}{G}\, ,
\end{equation}
and now its value in the zero-size limit is $S(r_h\rightarrow 0)=2\pi(\alpha-\sqrt{3\mu}) L^2/G$. Hence, taking $\alpha\ge\sqrt{3\mu}$ we avoid negative entropies. We recall at this point that the zero-mass limit of these solutions is not actually flat space, but rather a massless, extremal black hole. Essentially, the coupling $\alpha$ of the Gauss-Bonnet term determines the entropy of this exotic black hole. Such object could have in principle a finite entropy, but since it has vanishing temperature and mass, it seems reasonable that it has vanishing entropy too. Thus, from now on we set $\alpha=\sqrt{3\mu}$ so that $S(M=0)=0$. 

After the preceding discussion, let us us further explore the properties of the entropy \req{eq:SECG2}. First, using \req{eq:MrhECG} and \req{eq:TECG} it is possible to show that the First law of black hole mechanics
\begin{equation}
dM=T dS\, ,
\end{equation}
holds exactly. This is an extraordinary check of our calculations, since the three physical quantities appearing in this expression --- namely, the Abbott-Deser (or ADM) mass $M$, the Wald entropy $S$ and the Hawking temperature $T$ --- have been computed independently.

In Fig.~\req{fig5} we plot the entropy as a function of the mass and as a function of the temperature both for Einstein gravity and for ECG. It is interesting to look at the behaviour of the entropy in the limits of large and small masses. We get in those cases 
\begin{equation}
S=
\begin{cases}
&\displaystyle 4 \pi  G M^2+2\pi \sqrt{3\mu}\frac{L^2}{G}-\frac{7 \left(\pi  \mu  L^4\right)}{16 G^3 M^2}+\mathcal{O}\left(\frac{1}{M^4}\right) \quad\text{when}\quad M>>\frac{\mu^{1/4}L}{G}\, ,\\
&\displaystyle 6\left(\frac{\mu L^{4} M^2}{4G}\right)^{1/3}+\mathcal{O}(M^{4/3}) \quad\text{when}\quad M<<\frac{\mu^{1/4}L}{G}\, .
\end{cases}
\end{equation}
Even though the scaling with the mass is very different in both limits,  we observe that the entropy is $S\sim  A/(2G)$ for small masses --- for large masses we recover $S\sim A/(4 G)$. Therefore, the ``area law'' is respected in the limit $M\rightarrow 0$ although with a different numerical factor. 

On the other hand, when we consider the relation $S(T)$, we observe that it possesses two branches: for any value of $T<T_{\rm max}$ there are two different black hole solutions with that temperature, a ``large'' black hole and a ``small'' one. Another remarkable difference with respect the Einstein gravity case is that the limit of vanishing entropy for ECG black holes corresponds to $T\rightarrow 0$, while for EG $S=0$ is reached for $T\rightarrow\infty$.
\begin{figure}[t!]
        \centering   		
              \includegraphics[width=0.48\textwidth]{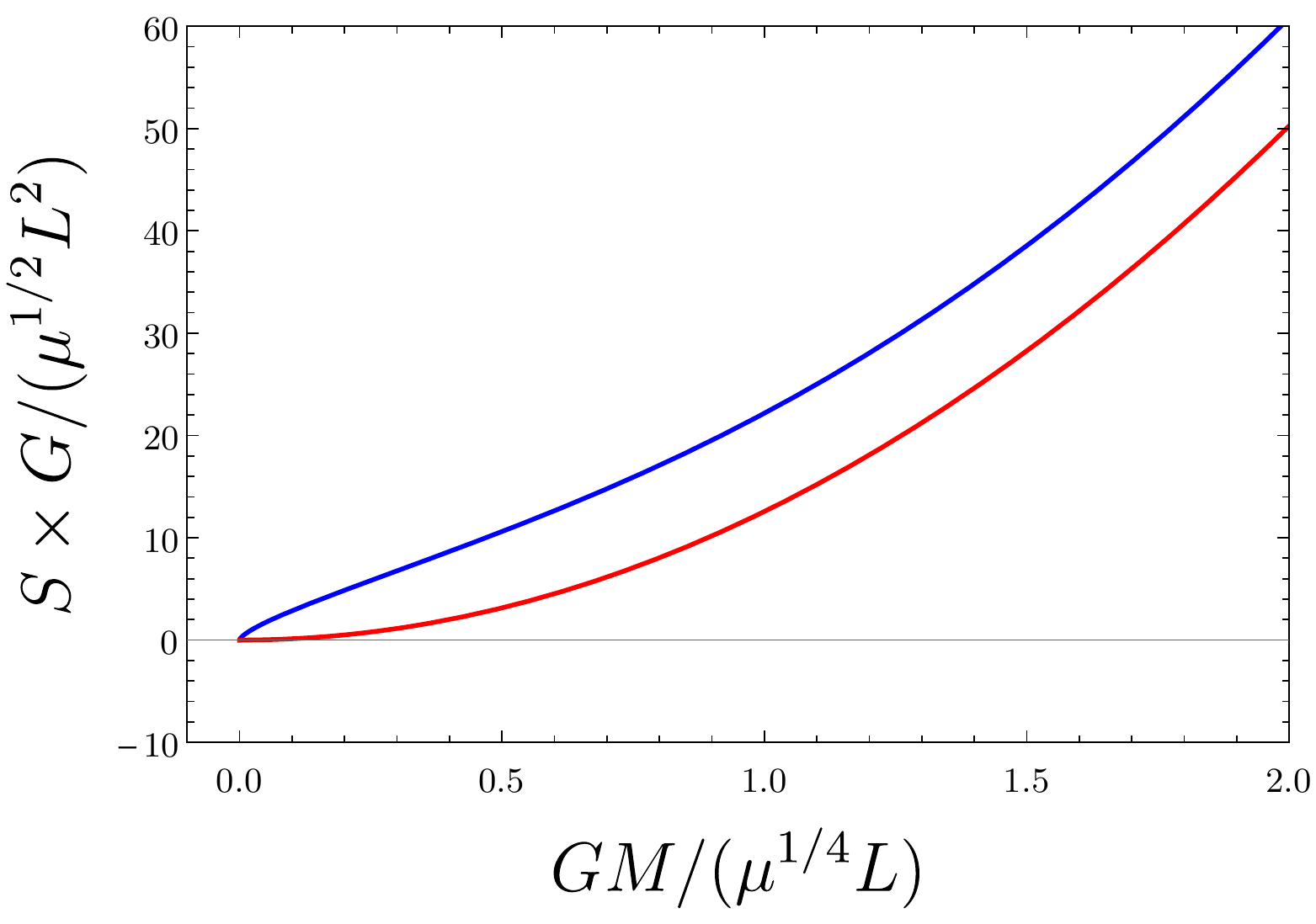}
              \includegraphics[width=0.48\textwidth]{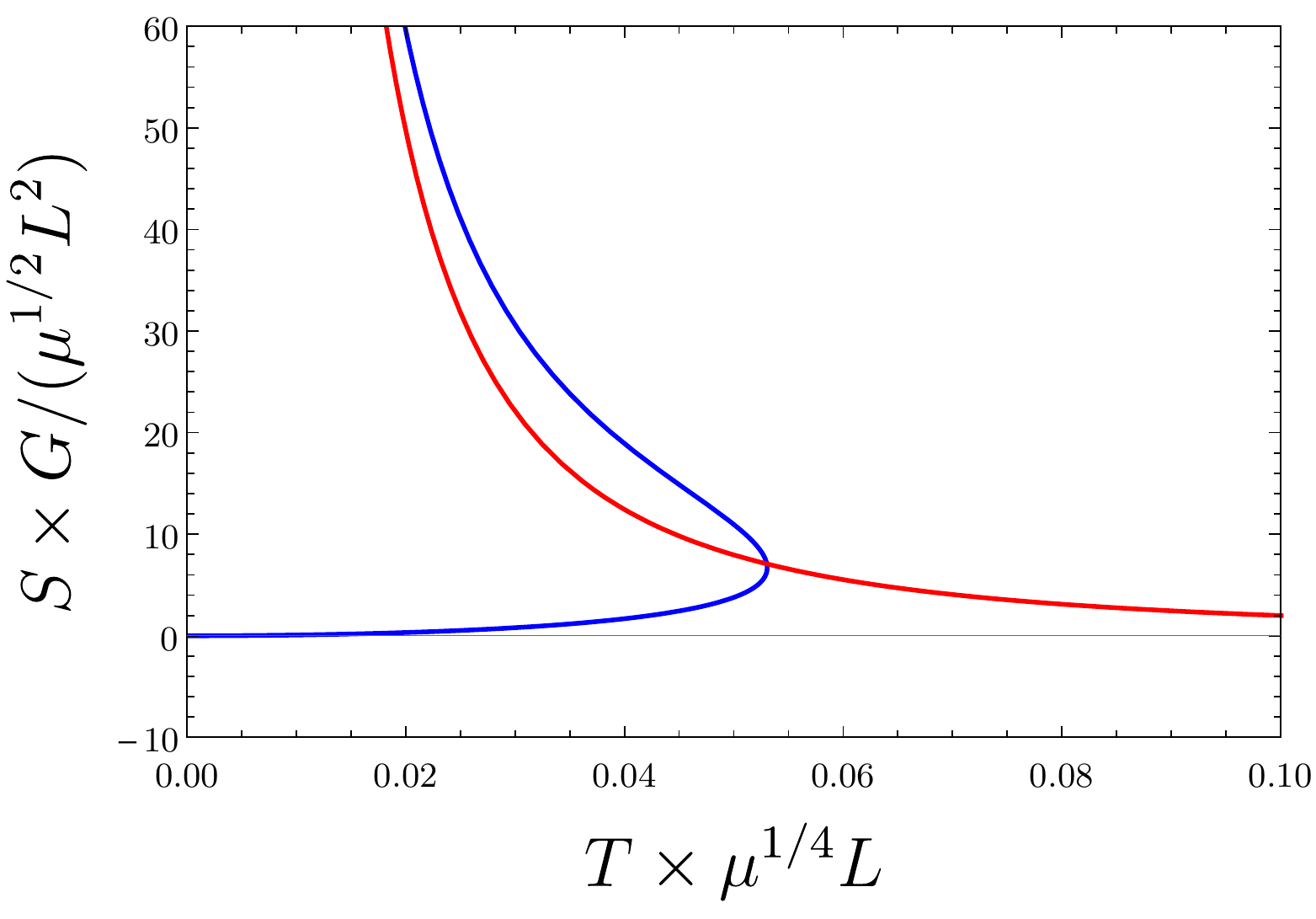}
        \caption{Left: we plot the entropy $S$ as a function of the mass $M$ for ECG black holes (blue) and for the usual Schwarzschild solution (red). Right: We show the relation $S(T)$, which in the case of ECG (blue) possesses two branches corresponding to ``large'' and ``small'' black holes. As usual, all quantities are expressed in appropriate natural units. }
\labell{fig5}
\end{figure}

\subsubsection{Stable black holes}
A very characteristic property of Schwarzschild  black holes is the fact they are thermodynamically \emph{unstable}. This means that, if one puts one of these black holes in a thermal background at some temperature $T_0$, the system always departs from equilibrium. In fact, if the initial temperature of the black hole is below $T_0$, then it absorbes radiation, hence gaining mass and decreasing its temperature even more according to the relation $T_{\rm Schw.}=(8\pi G M)^{-1}$. On the other hand, if the temperature of the black hole is above $T_0$, it will emit radiation, losing mass and getting hotter and hotter until eventually it ``explodes''. 

In more precise terms, the thermodynamic stability of a system is determined by the sign of the specific heat, defined as
\begin{equation}
C=T\left(\frac{\partial S}{\partial T}\right)_M.
\end{equation}
For Schwarzschild black holes, this quantity reads $C_{\rm Schw.}=-8\pi G M^2$, which is negative for arbitrary masses, implying that these black holes are unstable. %Now, large black holes in Einsteinian cubic gravity are also unstable, but remarkably enough, small black holes are stable. They correspond to the lower branch of the entropy $S(T)$ in the right plot of Fig.~\ref{fig5}.  
Now, in the case of ECG black holes, we can compute the specific heat using the relations \req{eq:SECG2} and \req{eq:TECG}. Parametrized in terms of $r_h$, it reads
\begin{equation}\label{cct}
C=\frac{4 \pi  r_h^2 \left(r_h^8+3 \mu ^2 L^8+ r_h^2\left(3 \mu  L^4-r_h^4\right) \sqrt{r_h^4+3 \mu  L^4}\right)}{3G\mu  L^4 \left(\mu  L^4-r_h^4\right)}\, .
\end{equation}
However, it is more meaningful to study the relation $C(T)$, which we plot in Fig.~\ref{fig6}. We observe that it has two branches: a negative one, which corresponds to large black holes, and a positive one, corresponding to small black holes. The solutions with positive specific heat are thermodynamically stable, and very different from the usual Schwarzschild solution, which has $C(T)<0$ for all $T$, as we also show in Fig. \ref{fig6}. Thus, these small black holes behave as a usual thermodynamic system: they get colder as they emit radiation and they tend to reach the equilibrium with the environment. The situation is reminiscent to the one observed in \cite{Myers:1988ze}, where certain odd-dimensional Lovelock black holes were shown to become stable for small enough masses.

\begin{figure}[t!]
\centering 
\includegraphics[width=0.65\textwidth]{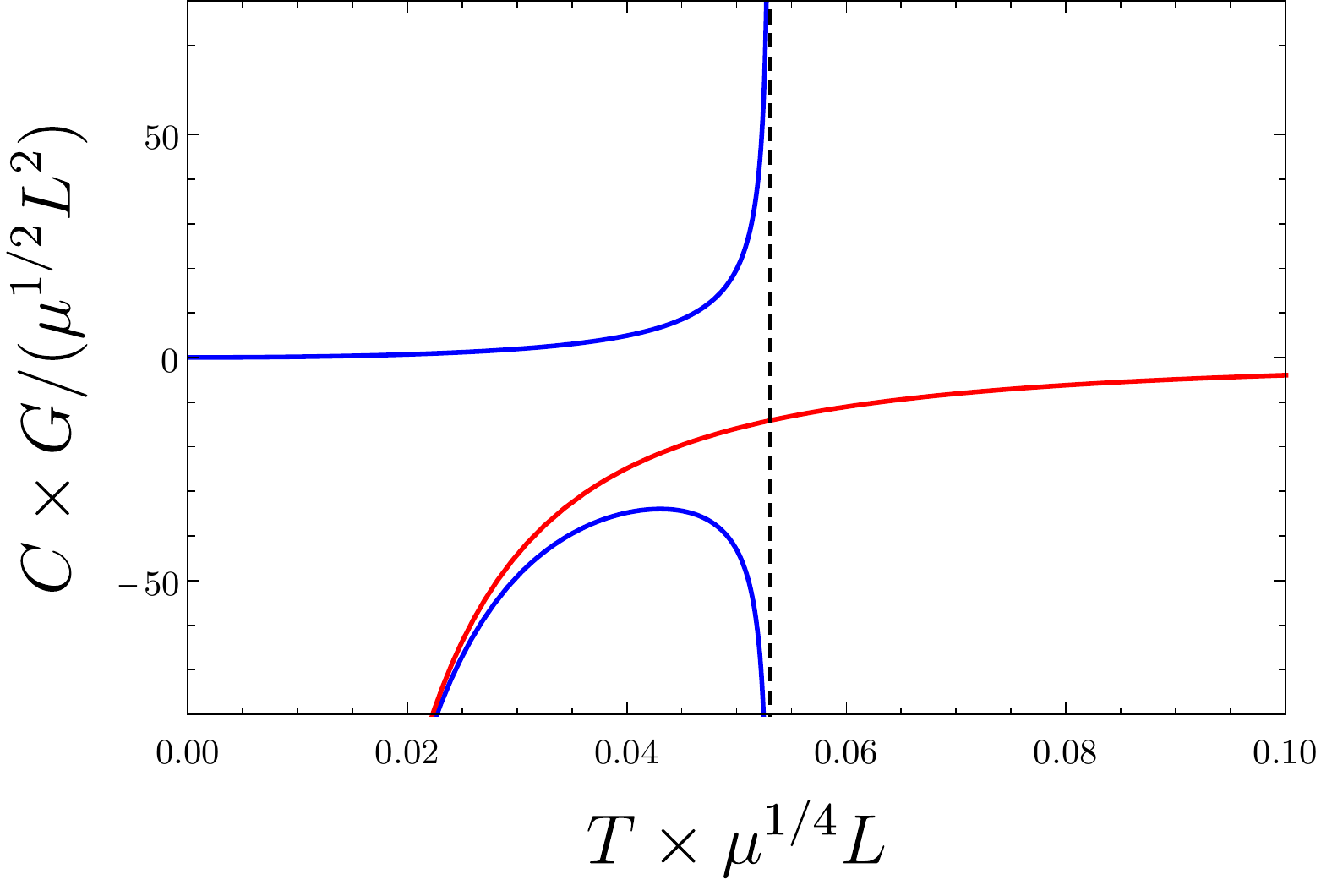}
\caption{We plot the specific heat $C$ as a function of the temperature for the ECG black holes (blue) and for the usual Schwarzschild solution (red).} 
\labell{fig6}
\end{figure}

Both branches of the specific heat diverge at $T=T_{\rm max}$, suggesting the presence of a phase transition, although the analogy is not completely accurate since there is no black hole solution for $T>T_{\rm max}$. In any case, at $T=T_{\rm max}$  the phases of stable and unstable black holes coalesce, and it is expected that some critical phenomenon takes place. The behaviour of the specific heat near $T=T_{\rm max}$ is
\begin{equation}
C\sim\pm\frac{4\pi \mu^{1/2}L^2}{\sqrt{3}G}\left(1-\frac{T}{T_{\rm max}}\right)^{-1/2}\, ,
\end{equation}
so that the critical exponent is $-1/2$.

%\section{Charged black holes}

\section{Discussion}
In this chapter we have constructed the asymptotically flat black hole solutions of Einsteinian cubic gravity \req{eq:ECGc4}. At the perturbative level, these solutions are relevant because they capture the leading correction to the Schwarzschild solution in any higher-derivative gravity. But more importantly, we have been able to compute the exact solution, which turns out to be much more interesting. These are the first examples of fully non-perturbative black hole solutions correcting the $D=4$ Schwarzschild solution in higher-order gravity, in the sense we explained in the introduction. 

Although the profile of the solutions has to be determined numerically, a very remarkable property of ECG is that some of the most interesting quantities of these black holes can be studied analytically. In this way, we obtained an exact relation between the mass $M$ and the radius of the horizon $r_h$, which modifies the standard Einstein gravity relation $r_h=2GM$. Likewise, we managed to obtain closed, analytic expressions for the thermodynamic quantities of these black holes --- this remarkable fact is captured in Conjecture \ref{conj}, which we presented in Chapter \ref{Chap:2}. Thanks to these relations being exact, we observed new non-perturbative phenomena that cannot be captured by perturbative methods.

While black holes in ECG have very similar properties to those in Einstein gravity when they have a large mass, they behave in a completely different way when their mass is below certain value. In particular, black holes in ECG have vanishing temperature in the zero-mass limit, just the opposite as for Einstein gravity black holes, whose temperature diverges in that limit. Related to this, we found that there is a maximum temperature a black hole can have in ECG. This implies that for a given value of the temperature below the maximum value there are  two types of black holes with very different thermodynamic properties. On the one hand, we have ``large'' black holes, which have negative specific heat and are unstable, just like Schwarzschild black holes. On the other hand, ``small'' black holes have a positive specific heat and are thermodynamically stable, and they represent a novelty not present in Einstein gravity. 
These properties are very appealing and they seem to draw a theory with an improved UV behaviour with respect to Einstein gravity. The drastic changes in small black holes could have important consequences for other areas of physics, but we will postpone this discussion for the next chapter. Of course, one could argue that, since we are only including a cubic term in the action, these nice properties could well be modified when one takes into account the effects of terms of higher-order. We will answer this question in the next chapter. 

Finally, let us mention that the analysis performed here can be extended in several ways. First, one can turn on a cosmological constant and study asymptotically (A)dS black hole solutions. We will study AdS solutions of ECG in Chapter~\ref{Chap:7}, and in that case one can also consider horizons with planar and hyperbolic geometries \cite{ECGholo}.
It is also possible to add a Maxwell term in the action in order to study charged black holes. In that case, the solutions are also given by the single-function metric \req{eq:Fmetric2} and the equations of motion can be solved similarly as we did here for neutral black holes --- in particular, it is possible to obtain again exact expressions for the thermodynamic quantities \cite{PabloPablo2,Hennigar:2017umz}. In addition, Ref.~\cite{Hennigar:2016gkm} studied the thermodynamic of black holes in ECG with a cosmological constant from the point of view of ``black hole chemistry'' \cite{Kubiznak:2014zwa,Kubiznak:2016qmn}. The same reference also studies black hole solutions in higher-dimensional ECG --- we remember that this theory was defined so that it possesses Einstein-like linearized equations in any dimension. However, when it comes to spherically symmetric solutions, all the nice properties that we found here do not extend to higher-dimensions (as we saw in Sec.~\ref{cubic}, the ECG density $\mathcal{P}$ only satisfies Theorem \ref{theo} in four dimensions), so the study of black hole solutions becomes much more involved.
In order to study higher-dimensional black holes with cubic curvature corrections it is more appropriate to consider the Quasi-topological \cite{Quasi,Quasi2} and Generalized quasi-topological \cite{Hennigar:2017ego,Mir:2019ecg} terms that we reviewed in Sec.~\ref{cubic}.

% which is determined by the scale of the higher-derivative terms and is of the order of $\sim L/G$.  

\chapter{Black hole thermodynamics at all orders in curvature}\label{Chap:5}
As proven by Hawking \cite{Hawking:1974sw}, a black hole with surface gravity $\kappa$ emits thermal radiation with a temperature $T= \kappa/(2\pi)$. In the prototypical case of a Schwarzschild black hole \cite{Schwarzschild:1916uq}, one gets the well-known relation
\begin{equation}
T=\frac{1}{8\pi G M}\, ,
\end{equation}
which implies that the temperature increases as the black hole radiates.  As a consequence, a Schwarzschild black hole will lose mass by means of radiation emission and it will evaporate in a time of the order of $\sim M_0^3/M_{\rm \ssc P}^4$, where $M_0$ is the initial black hole mass and $M_{\rm \ssc P}=G^{-1/2}$ is the Planck mass. 
%Strictly speaking, when the mass of the black hole is of the order of $M_{\rm \ssc P}$ the semiclassical approach used in the Hawking calculation stops making sense, and in that case the final product of the evaporation is an object of order-one entropy.
This suggests a violent ending for the evaporation process \cite{Hawking:1974rv}, and gives rise to the information paradox (see \cite{Harlow:2014yka,Chen:2014jwq} for recent reviews): a pure state collapsed to form a black hole would evolve into thermal radiation, in tension with the unitary evolution expected from quantum mechanics --- see \cite{Myers:1997qi}, though.

These considerations rely on the particular properties of the Schwarzschild black hole, which is the unique spherically symmetric, neutral black hole solution of Einstein gravity. However, as we have remarked previously in this thesis, the Einstein-Hilbert action is expected to be the first term of an effective theory containing an infinite series of higher-derivative corrections \cite{Nepomechie:1985us,Gross:1986iv,Gross:1986mw,tHooft:1974toh,Deser:1974cz}. One is then naturally led to wonder how these terms modify Schwarzschild's solution and how such modifications might affect its thermodynamic behavior, as well as the evaporation process  --- see \eg \cite{Myers:1988ze,Myers:1989kt,Callan:1988hs} for progress in this direction (mostly in higher-dimensions). The effect of the higher-derivative corrections becomes very important when the curvature is large enough, which means that small black holes will be specially affected. If the energy scale of the corrections is significantly below Planck's scale, the last stages of black hole evaporation could drastically depart from the prediction of Einstein gravity. 

Despite the interest of this problem, one of the main difficulties so far has been the lack of non-trivial extensions of Schwarzschild's solution in four-dimensional higher-order gravities.\footnote{We recall that our interest is on solutions that modify the Einstein gravity ones in a non-trivial way, and such that they reduce to the latter when the higher-derivative couplings are turned off.} In fact, as opposed to the $D\geq 5$ cases, all the higher-order Lovelock \cite{Lovelock1,Lovelock2} and Quasi-topological \cite{Quasi,Quasi2} invariants --- which due to their special properties do allow for such simple extensions, \eg \cite{Boulware:1985wk,Wheeler:1985qd,Myers:1988ze,Cai:2001dz,Dehghani:2009zzb,deBoer:2009gx} --- are  trivial in $D=4$.
Fortunately, the situation is about to change thanks to the new class of Generalized quasi-topological gravities (GQGs) that we described in Chapter \ref{Chap:2}. The first non-trivial member of this type of theories in $D=4$ is provided by Einsteinian cubic gravity, and in the last chapter we were able to construct the exact black hole solutions of this theory. As we saw, these solutions represent extensions of the Schwarzschild black hole in the sense that they approach this solution in the large mass limit. On the other hand, the properties of small black holes in ECG turned out to be drastically different from those of Schwarzschild's solution.
We found, in particular, that black holes become thermodynamically stable below certain mass. This would have a direct effect on the evaporation process, since one would expect in that case that the final ``Hawking explosion'' \cite{Hawking:1974rv} does not occur. However, one could argue that this conclusion cannot be trusted since we are only including a cubic term in an effective action which is supposed to contain an infinite number of terms. It could well happen that higher-order terms induce new effects with respect to the cubic correction. 

In order to determine whether this is the case, in this chapter we will study black hole solutions in all four-dimensional Generalized quasi-topological gravities.  For all of these theories, it is possible to construct the exact, non-perturbative black hole solutions, in the same way as we did for Einsteinian cubic gravity. More importantly, we will be able to study the thermodynamic properties of these black holes analytically, and we will see that some qualitative features are shared by all of these theories. In particular, we will show that the behaviour of small black holes is universal, provided there is at least one non-trivial correction to the Einstein-Hilbert action. We will then discuss the implications of the modified thermodynamic behaviour in the evaporation process of black holes, and will conclude that the new black holes have infinite lifetimes --- they never evaporate completely.

%In this paper, we will argue that these black holes are particular cases of an infinite family of four-dimensional solutions to a gravity theory involving terms of arbitrarily high order in curvature. As opposed to Schwarzschild's, the specific heat of the new black holes becomes positive below certain mass, which completely changes the evaporation process. 
%In particular, the new small black holes have infinite lifetimes. Besides, following the evaporation process till the point in which the semiclassical approximation breaks down, one finds that a naturally enormous time is required and that the resulting object has a huge entropy, in flagrant contrast with the Schwarzschild black hole case --- see Table \ref{tbl} below for a summary.  In the small mass regime, the new (asymptotically flat) black holes satisfy the Smarr-type relation: $M=\frac{2}{3}T S$, which, intriguingly, coincides with the result corresponding to a three-dimensional CFT at finite temperature.
%We argue that the above properties are universal for general values of the higher-order couplings.

%%%%%%%%%%%%%%%%%%%%%%%%%%%%%%%%%%%%%%%%%%%%%%%%%%%%%%%%%%%%%%%%%%%%%%%%%%%%%%%%%%%%%%%%%%%%%%%%%%%%%%%%%%%%%%%%%%%%%%%%%%%%%%%%

\section{Einstein-Hilbert action with an infinite number of corrections}\label{hoc}
Let us start with a brief summary of Generalized quasi-topological gravities. In Chapter~\ref{Chap:2} we defined this family of theories --- whose general cubic version was first proposed in \cite{Hennigar:2017ego} --- as those whose Lagrangians satisfy the hypothesis of Theorem \ref{theo}. 
Thus, according to that theorem, all of these theories have Einstein-like linearized equations on maximally symmetric backgrounds  --- see Chapter \ref{Chap:1} --- and they possess spherically symmetric solutions of the form
\begin{equation}\label{fmetric}
	ds_f^2=-f(r)dt^2+ \frac{dr^2}{f(r)} + r^2 d\Omega_{(2)}^2\, ,
\end{equation}
\ie characterized by a single function $f(r)$. In addition, $f(r)$ satisfies a differential equation which is reduced two orders with respect to the general order of the equations of motion. An even more remarkable property --- that we captured in Conjecture \ref{conj} --- is that the thermodynamic properties of black holes in these theories can be computed exactly and analytically.  We already observed this in the case of Einsteinian cubic gravity in the last chapter, and here we will show that it holds for all GQGs.
These properties make these theories ideal candidates to study spherically symmetric black hole solutions with higher-curvature corrections. 

Of course, one might worry that we are still not considering the most general higher-derivative gravity. However, we recall that according to the results of Chapter~\ref{Chap:3}, it is very likely that all higher-derivative Lagrangians --- understood as an EFT --- can be mapped to a sum of GQG terms using field redefinitions. Since black hole thermodynamics is invariant under redefinitions of the metric, we might actually capture the thermodynamic properties of black holes in any higher-derivative gravity using only GQG Lagrangians. We will come back to this intriguing point in the discussion section. 
By now, our problem will be to determine the black hole solutions only for Generalized quasi-topological gravities --- which on the other hand constitute already a very large family of theories. 

As we saw in Chapter~\ref{Chap:2}, the class of GQG theories contains and generalizes the family of Lovelock \cite{Lovelock1,Lovelock2} and Quasi-topological \cite{Quasi,Quasi2,Dehghani:2011vu,Cisterna:2017umf} gravities. For instance, all the GQG theories at cubic and quartic order in curvature in all dimensions where classified in \cite{Hennigar:2017ego} and \cite{Ahmed:2017jod}, respectively. 
But in this chapter we are interested in four-dimensional black holes, so let us recall the discussion in Section~\ref{sec:D4all} about the  structure of the GQG terms for $D=4$.
The first member (at the lowest order in the curvature) of the GQG family is the Gauss-Bonnet density $\mathcal{X}_{4}$, which is topological, so that it has no effect on the equations of motion.\footnote{However, it does have an effect on black hole entropy, as we saw in the last chapter. Hence, we will keep this term in the gravitational action.} At cubic order, we saw in Sec.~\ref{cubic} that there are two non-trivial densities of the GQG class: one of them is the Einsteinian cubic gravity density $\mathcal{P}$ and the other one was denoted by $\mathcal{C}$ --- see \req{eq:GQG4D}. However, when one analyzes the field equations for static and spherically symmetric metrics, one finds that only the ECG term $\mathcal{P}$ contributes in a non-trivial way --- and gives rise to the black solutions that we studied in the previous chapter.  Thus, when it comes to spherically symmetric black hole solutions, the density $\mathcal{C}$ is trivial\footnote{Note that this density does not vanish identically, so it will be non-trivial for other types of metrics.} and we only need to include the term $\mathcal{P}$ at cubic order.  In general, we observe that at a given curvature order, $n$, there are always several  GQG densities that we may denote by $\mathcal{R}_{(n)}^{i}$, where $i=1,\ldots, i_{\rm max}(n)$. Since we intend to study black hole solutions in all GQG theories, we should add a general linear combination of all of these densities in the gravitational Lagrangian. However, when we study the field equations of these terms for an arbitrary SSS metric, we come to the conclusion that, within a given order $n$, they all are proportional to each other. In other words, at a given order there is a unique way in which all GQGs modify the equation for $f(r)$ in \req{fmetric}. This further implies that performing a change of basis of the densities $\mathcal{R}_{(n)}^{i}$ we can always choose one of them (for instance, $\mathcal{R}_{(n)}^{1}$) as the one that contributes to the equations of motion for the metric \req{fmetric} while the rest of them are trivial in that case. Thus, the situation at cubic order with the terms $\mathcal{P}$ and $\mathcal{C}$ extends analogously to higher orders. 

The conclusion of the previous discussion is that, if we wish to study the spherically symmetric black hole solutions of the most general GQG, it is sufficient to include one non-trivial Generalized quasi-topological term at every order in curvature.
We shall denote this ``representative'' GQG term at order $n$ simply by  $\mathcal{R}_{(n)}$ (without superscript). Hence, we consider the action 
\begin{equation}\label{lal}
S= \frac{1}{16\pi G}\int d^4x\sqrt{|g|}\left[R+\sum_{n=2}^{\infty}L^{2n-2}\lambda_{n}\mathcal{R}_{(n)}\right]\, ,
\end{equation}
where $L$ is an overall length scale and $\lambda_n$ are arbitrary dimensionless couplings.  Of course, it is equivalent to consider instead of the density $\mathcal{R}_{(n)}$ another one proportional to it, because the whole effect is a rescaling of the coupling $\lambda_{n}$. Hence, we are free to choose a normalization for these densities, which can be done \eg by specifying their value on a constant-curvature background. We will assume a normalization such that
\begin{equation}\label{eq:densnorm}
\bar{\mathcal{R}}_{(n)}=-\frac{12}{n-2}\mathcal{K}^n\,\,\, \text{when evaluated on}\,\,\, \bar{R}_{\mu\nu\rho\sigma}=2\mathcal{K}g_{\mu[\rho}g_{\sigma]\nu}\, ,
\end{equation}
which is convenient because it simplifies the form of the vacuum embedding equation \req{Lambda-eq} --- see Appendix \ref{App:4}. The quadratic term cannot be chosen in this way, so we simply set $\mathcal{R}_{(2)}=\mathcal{X}_{4}$, which is of course the topological Gauss-Bonnet term. The $n=3$ term can be chosen to be the ECG density with an appropriate normalization, namely $\mathcal{R}_{(3)}=-\mathcal{P}/8$. As for the higher-order cases, we provided in Eqs. (\ref{r3})-(\ref{r10}) examples of these densities --- already normalized according to \req{eq:densnorm} ---  up to $n=10$. Of course, there is no reason to stop at that order because these densities must exist at every order, but we still lack a closed expression to generate these terms at arbitrary order. Despite this drawback, we are about to see that the structure of the field equations associated with these densities is much simpler than one could expect.

\subsection{Equations of motion}
We can now proceed to evaluate the field equations of \req{lal} on a static, spherically symmetric metric ansatz. As we did for the ECG case in Chapter~\ref{Chap:4}, this can be done by means of the reduced action method or by direct evaluation of the field equations on the single-function metric \req{fmetric} --- since the theory \req{lal} satisfies Theorem \ref{theo}, it is guaranteed that the equations can be solved with a metric of that form. In addition, the same theorem tells us that the only independent equation takes the form of a total derivative. Explicitly, the equations of motion evaluated on \req{fmetric} have the form
\begin{equation}
\E_{tt}\equiv-f^2\E_{rr}=\frac{f(r)}{16\pi G r^2}\frac{d\E_f}{dr}=0\, ,
\end{equation}
so that the quantity $\E_f=\E_f(r,f,f',f'')$ is constant. As we determined in Section~\ref{bho}, this constant is always proportional to the mass, the precise relation being $\E_f=2GM$ --- see \req{Cmass}. The left-hand-side of this equation will be a linear combination of the contributions coming from the different terms in the action \req{lal}, this is,
\begin{equation}\label{eq:allorders}
r(1-f)+\sum_{n=3}^{\infty}L^{2n-2}\lambda_{n}\E^{(n)}_f\left(r,f,f',f''\right)=2GM\, ,
\end{equation}
where we are already making explicit the contribution from the Einstein-Hilbert term. Then, we have to determine the contributions $\E^{(n)}_f$ order by order in the curvature. The few first terms, for $n=3,4,5$, yield

\begin{align}\label{eq:En345}
\E^{(3)}_f=&-\frac{3}{4} \left[\frac{f'^3}{3}+\frac{1}{r}f'^2-\frac{2}{r^2}f(f-1)f'  -\frac{1}{r}ff''\left(f'r-2(f-1)\right)\right]\, ,\\
\E^{(4)}_f=&-\frac{3}{2}\left(\frac{f'}{2r}\right)  \left[\frac{f'^3}{4}+\frac{f+2}{3r}f'^2-\frac{2}{r^2}f(f-1)f'  -\frac{1}{r}ff''\left(f'r-2(f-1)\right)\right]\, ,\\
\E^{(5)}_f=&-\frac{5}{2}\left(\frac{f'}{2r}\right)^{2}  \left[\frac{f'^3}{5}+\frac{f+1}{2r}f'^2-\frac{2}{r^2}f(f-1)f'  -\frac{1}{r}ff''\left(f'r-2(f-1)\right)\right]\, .
\end{align}
Remarkably, all of them have a very similar structure, and this allows us to guess a general pattern. In particular, we ``induce'' the following formula for the $n$-th order case
\begin{equation}\label{eq:Egeneraln}
\E^{(n)}_f=-\frac{n(n-1)}{8}\left(\frac{f'}{2r}\right)^{n-3}  \left[\frac{f'^3}{n}+\frac{(n-3)f+2}{(n-1)r}f'^2-\frac{2}{r^2}f(f-1)f'  -\frac{1}{r}ff''\left(f'r-2(f-1)\right)\right]\, ,
\end{equation}
which fits the cases $n=3,4,5$ in the equations above.
Now, using the densities in Eqs.~\req{r3}-\req{r10} we have checked that Eq.~\req{eq:Egeneraln} also holds for $n=6,7,8,9,10$, so we are confident to state that Eq.~\req{eq:Egeneraln} really provides the contribution from the density $\mathcal{R}_{(n)}$ for arbitrary $n$. Thus, even though we do not have a closed expression for these densities, we have been able to obtain a general formula for their contributions to the equations of motion. Equation~\req{eq:Egeneraln} represents the unique way in which one can modify the equation for the metric function $f(r)$ at every order in curvature within the family of GQG theories.

For $n=3$ and $n=4$, the above equation agrees with the ones found for Einsteinian cubic gravity in the previous chapter, Eq.~\req{eq:feqECG},  and the quartic version \cite{Ahmed:2017jod} of Generalized quasi-topological gravity \cite{Hennigar:2017ego}. 
It is possible to generalize Eq.~\req{eq:allorders} for arbitrary values of  the cosmological constant (here we have set $\Lambda=0$) and for general horizon geometries, \ie planar or hyperbolic besides spherical. This generalization, together with some other properties of the theory \req{lal}, are explored in Appendix~\ref{App:4}.

\section{Black hole solutions}\label{sec:bhs}
We are now ready to search for spherically symmetric black hole solutions of the theory \req{lal}. In order to do that, we have to solve Eq.~\req{eq:allorders} with the boundary conditions of asymptotic flatness and existence of an event horizon. The discussion is analogous to the one in Section~\ref{sec:exactECG} for the case of ECG, so let us be less detailed here. The idea is again to check that each condition fixes one integration constant of Eq.~\req{eq:allorders}, so that there will be one solution satisfying both conditions. 

%In the asymptotic region, the Einstein gravity contribution dominates $f(r)$, and the higher-order corrections can be assumed to be small. These considerations\footnote{Further details on this kind of analysis can be found in \cite{PabloPablo2,Hennigar:2017ego,Ahmed:2017jod}.} lead to the following asymptotic expansion:
In the asymptotic region, we can split the general solution to Eq.~\req{eq:allorders} as the sum of a particular solution plus another piece which will satisfy a homogeneous equation,
\begin{equation}\label{eq:asymptdecomp}
f(r)=f_p(r)+f_h(r)\, .
\end{equation}
The particular solution can be obtained assuming a $1/r$ expansion, and in that case the leading terms read
\begin{equation}
\begin{aligned}
f_p(r)=&1-\frac{2 G M}{r}+\frac{\lambda_3 L^4}{(GM)^4} \left(-\frac{27(GM)^6}{r^6}+\frac{46 (GM)^7}{r^7}\right)\\
&+\frac{\lambda_4L^6}{(GM)^6} \left(-\frac{54(GM)^9}{r^9}+\frac{97 (GM)^{10}}{r^{10}}\right)+\mathcal{O}(r^{-11})\, .
\end{aligned}
\end{equation}
Then, we plug the decomposition \req{eq:asymptdecomp} into the equation and we expand linearly in $f_h(r)$. Note that we can do it because $f_h$ must vanish asymptotically, hence we can assume it is arbitrarily small. The result is a second-order linear differential equation for $f_h$,
\begin{equation}\label{eq:homoeq2}
A f_h''+Bf_h'+C f_h=0\, .
\end{equation}
for some coefficients $A$, $B$ and $C$ that depend on $r$. Then, we have to determine the behaviour of the solutions of this equation in the limit $r\rightarrow\infty$. The discussion is different depending on which is the first non-vanishing coupling $\lambda_n$, so let us first assume that $\lambda_3\neq0$. In that case, the leading contributions to the coefficients $A$, $B$, $C$ come from the cubic term, and the discussion is the same as for Einsteinian cubic gravity in the previous chapter. In particular, these coefficients take the form in Eq.~\req{eq:coeffhomo} and the general solution for $f_h$ is given by \req{eq:homosol} (replacing $\mu\rightarrow\lambda_3$). We recall that this solution has a very different character depending on the sign of $\lambda_3$. For positive values of $\lambda_3$, the general solution contains an exponentially growing mode and an exponentially decaying one, so that we have to set the coefficient of the growing mode to zero, hence fixing one integration constant. On the other hand, for $\lambda_3<0$ all the solutions but the trivial one are singular, so that there is a unique asymptotically flat solution. Then, one is not able to impose the additional condition of existence of a regular horizon, and in that case we will not have asymptotically flat black hole solutions. 

Hence, providing that $\lambda_3>0$, the asymptotic flatness condition fixes one of the integration constants of Eq.~\req{eq:allorders}. Now, if $\lambda_3=0$ the leading contribution to the coefficients $A$, $B$, $C$ in Eq.~\req{eq:homoeq2} will come from a higher-order term and we have to repeat the analysis in that case.  Let us assume that $\lambda_4\neq0$ so that this is the leading contribution to these coefficients. We get
\begin{equation}\label{eq:coeffhomo2}
\begin{aligned}
A&=\frac{9\lambda_4 L^6  (GM)^2}{r^5}\left(1-\frac{2 GM}{r}\right)+\mathcal{O}\left(\frac{1}{r^7}\right)\, ,\\
B&=-\frac{45\lambda_4 L^6 (GM)^2}{r^6}\left(1-\frac{8 GM}{5r}\right)+\mathcal{O}\left(\frac{1}{r^8}\right)\, ,\\
C&=-r+\frac{18  \lambda_4 L^6 (GM)^2}{r^7}+\mathcal{O}\left(\frac{1}{r^8}\right)\, ,
\end{aligned}
\end{equation}
and keeping only the leading term in each case, the general solution to Eq.~\req{eq:homoeq2} reads
\begin{equation}
f_h(r)=c_1r^3 I_{\frac{3}{4}}\left(\frac{r^4}{12 GM L^3 \sqrt{\lambda_4}}\right)+c_2r^3 K_{\frac{3}{4}}\left(\frac{r^4}{12 GM L^3 \sqrt{\lambda_4}}\right)\, ,
\end{equation}
where $I_{\alpha}$ and $K_{\alpha}$ are modified Bessel functions of the first and second kind, respectively. Now, we can see that the discussion is the same as for the previous case. When $\lambda_4>0$, $I_{3/4}$ behaves as a growing mode so that we have to set $c_1=0$, while $K_{3/4}$ is exponentially decaying so we can keep it. On the other hand, for $\lambda_4<0$ both solutions oscillate wildly at infinity, hence both are singular and the only asymptotically flat solution is the one with $c_1=c_2=0$. Thus, there are no additional parameters that allow us to impose the existence of a regular horizon in that case. 

Now, if both $\lambda_3$ and $\lambda_4$ are zero, we would need to repeat the analysis for the first non-vanishing coupling $\lambda_{n_0}\neq0$. However, it is possible to check that the conclusion is always the same: the asymptotic flatness condition fixes one integration constant of Eq.~\req{eq:allorders} providing that the first non-vanishing coupling $\lambda_{n_0}$ is positive. We note that this is a necessary condition for the existence of asymptotically flat black hole solutions --- if $\lambda_{n_0}<0$ we are certain they do not exist --- but in all the cases in which we have explicitly constructed solutions, this condition seems to be also sufficient. We provide few examples below. 

Imposing the existence of a regular event horizon fixes the remaining integration constant of Eq.~\req{eq:allorders}. In order to see this, it is convenient to perform a Taylor expansion at the horizon,
\begin{equation}
f(r)=4\pi T (r-r_h)+\sum_{n=2}^{\infty}a_n (r-r_h)^n,
\label{series}
\end{equation} 
where $a_n=f^{(n)}(r_h)/n!$ and we have already taken into account that $f'(r_h)=4\pi T$, where $T$ is the black hole temperature. Solving \req{eq:allorders} for the first two orders in $(r-r_h)$ gives rise to the following relations:
 \begin{align} \label{massnn}
 &2GM=r_h-r_h\sum_{n=3}^{\infty}\lambda_n \left(\frac{2\pi TL^2}{r_h}\right)^{n-1}(n+(n-1)2\pi T r_h )\, , \\ 
 \label{temperaturenn}
 1=&4\pi Tr_h+\sum_{n=3}^{\infty}\lambda_n \left(\frac{2\pi TL^2}{r_h}\right)^{n-1}(n+(n-3)2\pi T r_h) .
 \end{align}
These equations fix $r_h$ and $T$ in terms of the black hole mass $M$. Note that these relations are exact, as they are necessary conditions for having a smooth near-horizon geometry.
Depending on the values of the $\lambda_n$ and the mass, these equations can have one, several, or even no solutions at all. If there are several solutions, it means that various possible black holes with the same mass can exist. However, only one of the solutions will smoothly reduce to Schwarzschild in the limit $\lambda_n\rightarrow 0$ for all $n$. This is the one which should be regarded as physical, and we will call it the ``Schwarzschild branch''. We will elaborate this point in the next section. 

For simplicity, we consider here the case in which there is a unique solution for any value of $M$, in whose case we can write  $T(M)$ and $r_h(M)$ without ambiguity. This happens, for example, if $\lambda_n\ge 0$ for all $n$, which, at the same time ensures the solution to be well-behaved asymptotically.

\begin{figure}[t!]
\begin{center}
\includegraphics[width=0.65\textwidth]{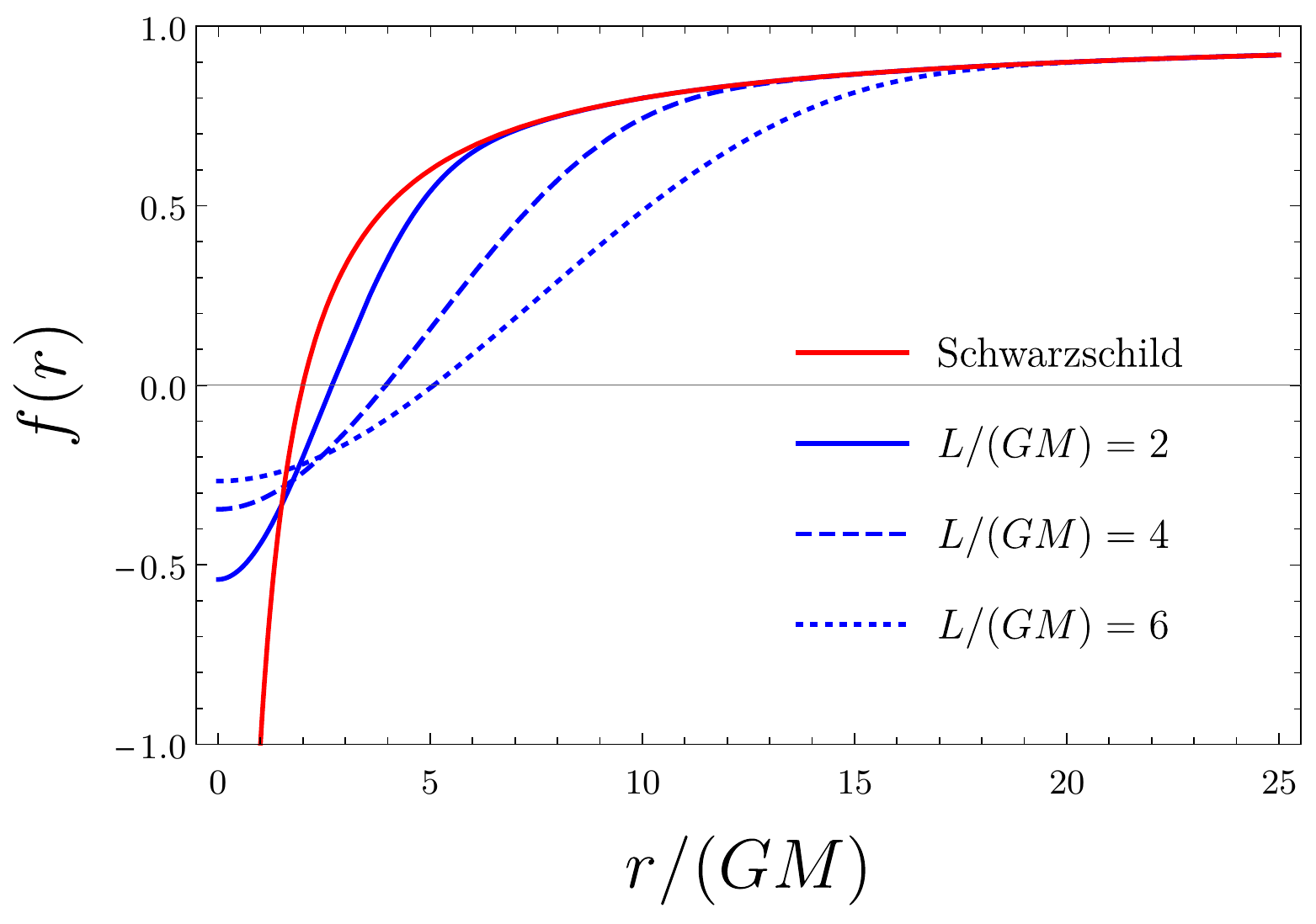}
\caption{Metric function $f(r)$ for Schwarzschild's solution (red) and for the new higher-order black holes (blue), with $\lambda_3=\lambda_4=\lambda_5=\lambda_6=1$, $\lambda_{n>6}=0$ and various values of the overall length scale $L$.}
\label{fig108}
\end{center}
\end{figure}

Once $T(M)$ and $r_h(M)$ have been determined, the $(r-r_h)^2$-order equation fixes $a_3$ as a function of $a_2=f''( r_h)/2$, the $(r-r_h)^3$ one fixes $a_4$ as a function of $a_3$ and $a_2$, and so on. In this process the only undetermined parameter is $a_2$, which means that the solution will be fully determined once we choose a value for it. Therefore, the family of solutions with a regular horizon is characterized by a single parameter, which must be carefully chosen so that the solution is asymptotically flat. 
% , {\rm i.e.}, we must glue a regular-horizoned solution with an asymptotically flat one. This can be done by choosing an appropriate value of $a_2$. Indeed, this value must be chosen with a high precision so that the growing exponential mode in \req{homo} is not excited. This is done by performing a numerical analysis, choosing a value for $a_2$ and computing the numerical solution. 
 In all cases studied, a numerical analysis shows that there is  a unique value of $a_2$ for which asymptotic flatness is achieved.  This means that there exists a unique asymptotically flat black hole, fully characterized by its mass $M$, which reduces to Schwarzschild's solution when the higher-order couplings are set to zero. %The results presented in \cite{PabloPablo3} imply that, just like for  Schwarzschild in the Einstein gravity case, this will also represent the exterior field of a generic spherically symmetric matter distribution. 

In Appendix \ref{App:3}, we provide a detailed discussion of the numerical construction of the solutions (which, unfortunately, do not seem accesible analytically). 
The resulting metric functions, $f(r)$, for a particular set of $\lambda_n$ and different values of $L$ are shown in  Fig. \ref{fig108}. Changing the $\lambda_n$ does not modify these curves qualitatively. In particular, we observe that the profile of these solutions is remarkably similar to the one of Einsteinian cubic gravity solutions. This already suggests that including higher-order terms may not change the interesting properties of black holes in ECG. We are going to see that this intuition is essentially correct.

\section{Thermodynamics}
We have just shown that Eq.~\req{eq:allorders} allows for black hole solutions and we have illustrated how to obtain them numerically. However, once we know that these solutions exist, we can compute many of their properties analytically, without using numerical methods. The aim of this section is to explore, in particular, the thermodynamic properties of these black holes. In fact, we have seen that the equations \req{massnn} and \req{temperaturenn} provide us implicit relations for the radius $r_h$ and the temperature $T$ as functions of the mass $M$.  Since we have an infinite number of higher-derivative couplings, one could expect that each of possible choice yields very different thermodynamic relations. However, we will show that, under some reasonable assumptions, the behaviour of the thermodynamic quantities is qualitatively the same for all the theories. 

\subsection{Thermodynamic relations and first law}
As we remarked earlier, the equations \req{massnn} and \req{temperaturenn} give us implicit, but otherwise exact, relations for $T(M)$ and $r_h(M)$. These equations are quite complicated, but we can massage them in order to bring them to a more useful form.  Replacing $T$ in terms of the variable $\chi=2\pi T L^2/r_h$ in Eqs.~\req{massnn} and \req{temperaturenn}, one realizes that these equations can be solved parametrically in terms of $\chi$, \ie we get explicit expressions for $r_h(\chi)$, $T(\chi)$, $M(\chi)$. In order to express the solution, it is useful to introduce the function $h$ defined by
\begin{equation}\label{eq:hfunc}
h(x):=x-\sum_{n=3}^{\infty}\lambda_{n} x^n\, .
\end{equation}
We assume that the coefficients $\lambda_{n}$ are well-behaved so that the series above converges at least for some finite range of $x$.   Then, the parametric solution of Eqs.~~\req{massnn} and \req{temperaturenn} reads
\begin{eqnarray}
\label{eq:rhchi}
r_h&=&L\left[\frac{h'(\chi)}{3h(\chi)-\chi h'(\chi)}\right]^{1/2}\, ,\\
\label{eq:Mchi}
GM&=&\frac{r_h^3}{L^2}h(\chi)=Lh(\chi)\left[\frac{h'(\chi)}{3h(\chi)-\chi h'(\chi)}\right]^{3/2}\, ,\\
\label{eq:Tchi}
T&=&\frac{\chi r_h}{2\pi L^2}=\frac{\chi}{2\pi L}\left[\frac{h'(\chi)}{3h(\chi)-\chi h'(\chi)}\right]^{1/2}\, .
\end{eqnarray}
Thus, by taking values of $\chi\ge0$ in the expressions above, we generate parametrically the different curves $r_h(M)$ or $T(M)$, which can contain several branches. Before taking a closer look to these relations, let us first compute the entropy of these black holes using Wald's formula, Eq.~\req{eq:Wald1}.
Evaluating this formula on the black hole metric \req{fmetric} for the theory \req{lal}, in the same way as we did in Section~\ref{Sec:thermoECG}, we find
\begin{equation}\label{eq:Wald3}
S=\frac{\pi r_h^2}{G} \left[1-2\sum_{n=2}^{\infty}L^{2n-2}\lambda_n P^{(n)}_{trtr}\right]\, ,
\end{equation}
where in each case $P^{(n)}_{trtr}$ is the component $\mu\nu\rho\sigma=trtr$ of the tensor $P^{(n)}_{\mu\nu\rho\sigma}=\frac{\partial \mathcal{R}_{(n)}}{\partial R^{\mu\nu\rho\sigma}}$ evaluated on the metric \req{fmetric} at the horizon $r=r_h$. Using our ``representative densities'' in Eqs.~\req{r3}-\req{r10}, we have computed these quantities for $n=2,3,...,10$. As happened in the case of the equations of motion \req{eq:Egeneraln}, we find very simple expressions for $P^{(n)}_{trtr}$, so that we are able to guess its form for arbitrary $n$. Thus, our final result for the entropy, including all the corrections, reads
\begin{align}\label{entropynn}
&S=\frac{\pi r_h^2}{G}\left[1-\sum_{n=3}^{\infty}n\lambda_n \left(\frac{2\pi TL^2}{r_h}\right)^{n-1} \left(\frac{n-1}{(n-2) 2\pi T r_h}+1\right)\right]+2\pi \lambda_2 \frac{L^2}{G}\, .%+\frac{4\pi}{GM_c^2}\sum_{n=3}^{\infty}\frac{\lambda_n \chi^{n-2}}{(n-2)}\, ,
\end{align}
We recall that the last, constant term is the contribution from the Gauss-Bonnet density, which is topological. Now, we have expressed the entropy as a function of $T$ and $r_h$ but these are not independent quantities. We can use the relations \req{eq:rhchi} and \req{eq:Tchi} in order to express the entropy in terms of $\chi$ as well. After some simplifications, it is also possible to write the final result in terms of the function $h$, and it  reads
\begin{align}\label{eq:Schi}
S=\frac{\pi L^2}{G}\left[\frac{h'^2(\chi)}{3h(\chi)-\chi h'(\chi)}+\int_{\chi_{2}}^{\chi} dx \frac{h''(x)}{x} \right]\, .%+\frac{4\pi}{GM_c^2}\sum_{n=3}^{\infty}\frac{\lambda_n \chi^{n-2}}{(n-2)}\, ,
\end{align}
In this expression we have absorbed the GB contribution in the inferior limit of integration, $\chi_2$, whose effect is to shift the entropy by a constant value.\footnote{The precise relation between the constant $\chi_2$ and the GB coupling $\lambda_2$ is $\int_{\chi_{2}}^{0} dxh''(x)/x=2\lambda_2$.}
Then, using \req{eq:Tchi} or \req{eq:Mchi} we generate the curves for the different thermodynamic relations $S(T)$ or $S(M)$, respectively. Of course, in order for these relations to be consistent, they must satisfy the first law of black hole mechanics.  Indeed, taking differentials in \req{eq:Schi} and in \req{eq:Mchi}, and using \req{eq:Tchi}, one can show that
\begin{equation}
dM=TdS\, ,
\end{equation}
\ie the first law holds. This is a very strong test for our computations, since the three quantities appearing in this expression have been computed independently, each one according to its own definition.

\subsection{The Schwarzschild branch}
Now that we have determined that these black holes satisfy consistent thermodynamic relations, let us explore in more detail their qualitative and quantitative features. One of the key aspects of the thermodynamic phase space of the theories in \req{lal} is the number of solutions of the equations \req{massnn} and \req{temperaturenn}, or equivalently, the branches of the relation $T(M)$, expressed parametrically by \req{eq:Mchi} and \req{eq:Tchi}.  For a given value of the mass, the number of such branches represents the number of different black hole solutions of that mass --- including the possibility that for some regions of the parameter space there are no solutions. Since we have an infinite number of parameters, a case-by-case analysis of black hole thermodynamics in these theories does not seem to be very practical. However, we are not interested in all possible black hole solutions, because some of them will not be physical. Note that for large enough masses, there exists always a unique solution that approaches the Schwarzschild one when $M\rightarrow\infty$. We will denote the family of solutions that are smoothly connected to this one when we decrease the mass as the ``Schwarzschild branch''. Our focus will be on the study of the thermodynamic properties of the black holes in this branch. Since we believe that large black holes are described approximately by the Schwarzschild solution, we expect that the solutions in the Schwarzschild branch are the physical ones, because they are obtained smoothly by removing mass from a large black hole. 

By definition, the Schwarzschild branch (SB) exists for arbitrarily large masses, but it may not extend to lower masses. Thus, we will distinguish two situations
\begin{enumerate}
\item{The Schwarzschild branch extends down to $M=0$.}
\item{The Schwarzschild branch stops at a minimum mass $M_{\rm min}>0$.}
\end{enumerate}
Every theory belongs to one or another category, and this distinction is all we will need in order to characterize the thermodynamic phase space of the higher-order gravities \req{lal}. Let us now analyze the thermodynamic relations of the Schwarzschild branch for theories of each type.

\subsubsection{Case 1}
Let us first consider the case in which the Schwarzschild branch connects black holes of arbitrarily large masses with solutions of vanishing mass. Then, we have to find a choice of couplings $\lambda_n$ for which this is the case. We remind that, according to our discussion in Section~\ref{sec:bhs}, there is a constraint in the higher-order couplings $\lambda_{n}$ in order for the corresponding theory to possess black hole solutions. Namely, the first non-vanishing coupling must be positive $\lambda_{n_0}>0$. Note that, for instance, if one considers a theory that only contains one higher-derivative correction, the coupling of that term must be positive. Thus, somehow the most natural choice is to take all the couplings to be non-negative $\lambda_{n}\ge0$ $\forall n\ge3$. It turns out that with this choice, the relation $T(M)$ defined  by Eqs.~\req{eq:Mchi} and \req{eq:Tchi} possesses a unique branch, and this branch extends down to $M=0$.\footnote{For instance, it is easy to check that for a fixed $r_h>0$ there is unique solution $T>0$ of Eq.~\req{temperaturenn}. Then, it is possible to show that \req{massnn} defines a bijective function $M(r_h)\ge0$ for $r_h\ge0$, so that there is a unique $r_h$ for every value of the mass.} Of course, there are more choices of couplings --- allowing some of them to be negative --- that still yield a unique branch of solutions, and even more general choices that produce a Schwarzschild branch connected to $M=0$, but we will restrict to positive couplings for the sake of simplicity. 

\begin{figure}[ht!]
\begin{center}
\includegraphics[width=0.65\textwidth]{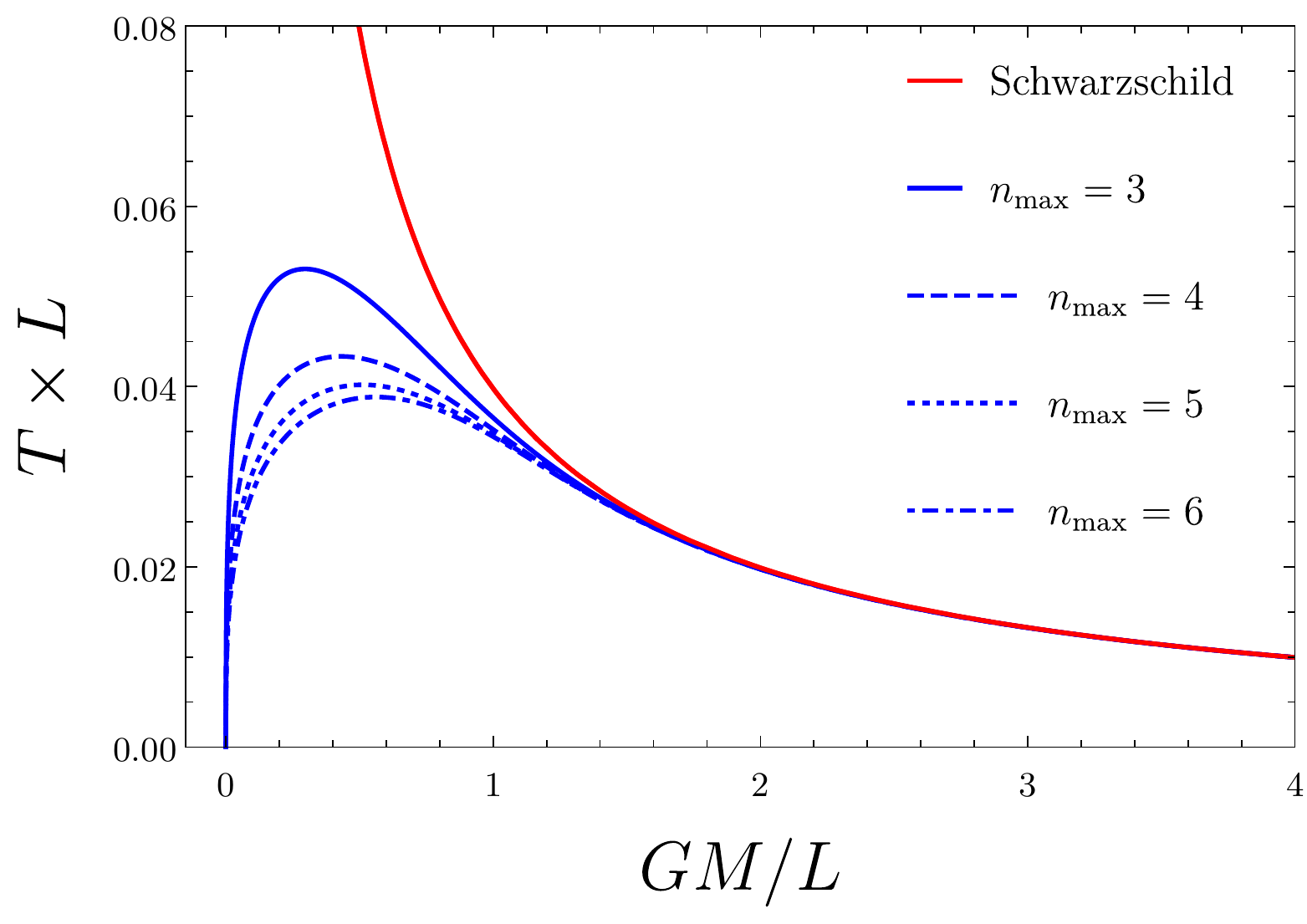}
\caption{Black hole temperature as a function of the mass for Schwarzschild's solution (red) and for the higher-order black holes with $\lambda_{n}=4^{n-3}$, $3\le n\le n_{\max}$, $\lambda_{n>n_{\rm max}}=0$. The shape of this curve is qualitatively the same for any other choice of couplings (except $\lambda_n=0$ for all $n$), provided it is smoothly connected to $M=0$.}
\label{fig1b}
\end{center}
\end{figure}

In Fig~\ref{fig1b} we plot the relation $T(M)$, obtained from Eqs.~\req{eq:Mchi} and \req{eq:Tchi}, for several choices of these couplings. In the different curves, we include all the terms in the action \req{lal} (with some particular weight) up to order $n_{\rm max}$. Thus, in the curve for $n_{\rm max}=3$ we only include the cubic correction (hence it corresponds to ECG), for $n_{\rm max}=4$ we include the cubic and quartic terms and so on. 
We observe that the qualitative profile of these curves is the same regardless the maximum order of the corrections. In fact, this is true even if we include an infinite number of corrections, something that can be done explicitly if we specify a simple value for all the couplings $\lambda_n$, so that the series in Eq.~\req{eq:hfunc} can be summed up. Furthermore, the qualitative features of the curves in Fig~\ref{fig1b} are shared by all the theories whose Schwarzschild branch extends down to $M=0$. In fact, in the next subsection we will show that the limit $M\rightarrow 0$ is universal in all theories; in particular the temperature always vanishes in that limit. This implies that the Schwarzschild branch must always have the profile in Fig~\ref{fig1b}: vanishing temperature in both limits $M\rightarrow \infty$ and $M\rightarrow 0$, and existence of a maximum temperature. 

\subsubsection{Case 2}
\begin{figure}[h!]
\begin{center}
\includegraphics[width=0.65\textwidth]{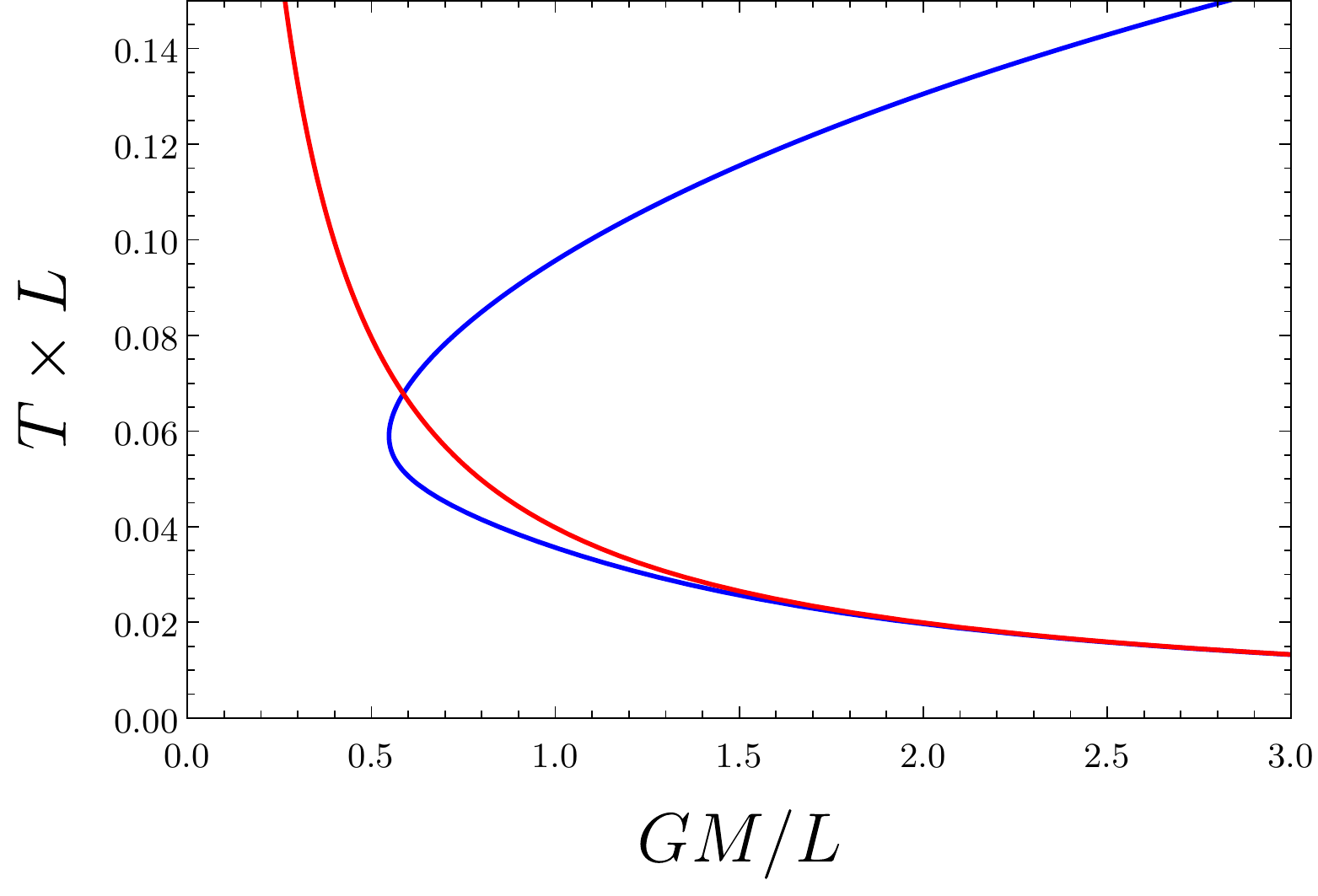}
\caption{Black hole temperature as a function of the mass for Schwarzschild's solution (red) and for a higher-order black hole with $\lambda_3=2$, $\lambda_4=-4$, $\lambda_{n>4}=0$ (blue). For large enough masses there are two different black hole solutions, but only one of them tends to the Schwarzschild solution when $M\rightarrow\infty$ (the Schwarzschild branch). For a certain mass both branches coalesce and there are no black hole solutions below that mass.}
\label{fig2b}
\end{center}
\end{figure}

Let us now provide an example of the opposite situation, namely, the Schwarzschild branch has an endpoint at a certain mass $M_{\rm min}>0$. Since for positive couplings it always extends to $M=0$, we must choose some couplings to be negative. On the other hand, at least the first non-vanishing coupling must be positive in order for the theory to possess black hole solutions. Thus, the most simple example we can consider is $\lambda_{3}>0$, $\lambda_{4}<0$ while the rest of couplings vanish. Even in this case, if $\lambda_{4}$ is not negative enough, the curve $T(M)$ has again the form of Fig~\ref{fig1b}. It is possible to show that the behaviour changes for $\lambda_{4}<-2(\lambda_3/3)^{3/2}$. In that case, one finds a diagram as the one shown in Fig~\ref{fig2b}.
There, we can see that for large enough masses there are two branches of black hole solutions: one with low temperature and one with high temperature. Only the first one reduces to the Schwarzschild solution for $M\rightarrow\infty$, hence it corresponds to the Schwarzschild branch. The other branch, which corresponds to a ``non-Schwarzschild'' solution, has a very exotic behaviour for large masses: it satisfies $r_h\propto T\propto M^{1/3}$, $S\propto M^{2/3}$. We observe that both branches coalesce at certain minimum mass and there are are no black hole solutions in this theory for lower masses. Thus, in particular the Schwarzschild branch does not extend down to $M=0$. 
The structure of the phase space becomes much more complicated as we turn on more couplings of arbitrary sign: we can have an increasing number of branches, perhaps some of them only existing for large masses, other only for small masses, etc. However, the Schwarzschild branch is unique, and in case it cannot be connected to $M=0$, it will always behave qualitatively as in Fig~\ref{fig2b}.

\subsection{Small black holes}
So far, we have provided examples of two different types of thermodynamic phase space for black holes: one (in Fig.~\ref{fig1b}) in which we find black hole solutions of arbitrarily small mass, and another one (in Fig.~\ref{fig2b}) in which there is minimum black hole mass. Focusing on the first possibility, we observe that the qualitative behaviour of the blue curves in Fig.~\ref{fig1b} in the limit $M\rightarrow 0$ seems to be very similar for all the higher-order theories. 
The aim of this section is to analyze in more detail the thermodynamic properties of these \emph{small} black holes. By small black hole we mean a solution whose mass is much smaller than $L/G$, so that the effects of the higher-order corrections are extreme. We are going to see that the properties of small black holes are universal for all the higher-order theories in \req{lal} --- the only exception being Einstein gravity. 

First, we need to identify the value of $\chi$ which corresponds to $M\rightarrow 0$ in the expression \req{eq:Mchi}. There, we see that $M=0$ is reached for a value $\chi_0>0$ such that
\begin{equation}\label{eq:chi0}
h'(\chi_0)=0\, ,
\end{equation}
and after a careful analysis we conclude that this is the only consistent way in which we can reach the zero mass limit.\footnote{With the only exception of EG, in which case it is reached for $\chi\rightarrow\infty$} Looking at \req{eq:rhchi} and \req{eq:Tchi}, we see that in this limit we also have $r_h=T=0$. Let us note that the equation \req{eq:chi0} may not 
have solutions, in which case it tells us that the theory does not possess black hole solutions below certain mass. 
Now, expanding the expressions \req{eq:rhchi}, \req{eq:Mchi}, \req{eq:Tchi} and \req{eq:Schi} around $\chi=\chi_0$ we find the thermodynamic quantities $T$, $S$ as well as the radius $r_h$ for a small, yet finite mass. The latter is related to the mass according to
\begin{equation}\label{eq:rhsmall}
r_h=\left(\frac{L^2GM}{h_0}\right)^{1/3}\, ,
\end{equation}
and we see that in order to avoid pathological situations we must in addition demand that $h_0\equiv h(\chi_0)>0$. Interestingly enough, the scaling is the same for all the theories, and it reminds us of the one of a usual matter distribution $M\propto r_h^3 \rho$, where the density would be $\rho\sim 1/(GL^2)$. However, it is not clear how seriously one can take this analogy.
 
As for the thermodynamic quantities, we observe that the entropy tends to a constant value in the zero-mass limit, namely $S\rightarrow \tfrac{\pi L^2}{G}\int_{\chi_{2}}^{\chi_{0}} dxh''(x)/x$. This is the same issue that we found in the case of ECG in the last chapter, and in that case we used the Gauss-Bonnet coupling $\lambda_2$ in order to impose the condition of vanishing entropy. We recall that this coupling is now absorbed in the integration limit $\chi_2$, so we are free to choose an appropriate value for it. In particular, for $\chi_2=\chi_0$ we get $S(M=0)=0$, and the thermodynamic relations for small masses read
\begin{eqnarray}
\label{eq:Ssmall}
S&=&\frac{3\pi L^2h_0^{1/3}}{G\chi_0}\left(\frac{GM}{L}\right)^{2/3}\, ,\\
\label{eq:Tsmall}
T&=&\frac{\chi_0}{h_0^{1/3}2\pi L}\left(\frac{GM}{L}\right)^{1/3}\, .
\end{eqnarray}
Remarkably enough, we find that  $S$ and $T$ scale with the mass in the same way for all the higher-derivative theories of the form \req{lal}, and the only theory-dependent parameter is the combination $\chi_0/h_0^{1/3}$. Furthermore, combining the expressions above we observe that all these solutions satisfy the following Smarr-like relation 
\begin{equation}\label{smmr}
M=\frac{2}{3}TS\, ,
\end{equation}
which contains no free parameters. Thus, this seems to be a universal property of small black holes in a vast number of higher-derivative gravities --- we have just shown it holds at least for all those belonging to the GQG family. Note that \req{smmr} differs from Schwarzschild's analogous relation, $M=2TS$, which is also satisfied for all the theories in \req{lal} in the large mass regime $M>> L/G$. 

The unusual exponents in Eqs.~\req{eq:Tsmall} and \req{eq:Ssmall} indicate that the properties of these small black holes are very different from those of Schwarzschild's one. In particular, as happened in the case of Einsteinian cubic gravity, they are thermodynamically stable, as can be checked by computing the specific heat, 
\begin{equation}
C(T)=\frac{24\pi^3h_0 L^4}{G \chi_0^3} T^2>0\, .
\end{equation}
This will have dramatic consequences for the evaporation process, as we explore in the next section.
Despite the unusual and interesting thermodynamic properties of these small black holes, the area law for the entropy still works. In fact, Eq.~\req{eq:Ssmall} can be rewritten as
\begin{equation}
S=\frac{3h_0}{\chi_0}\times \frac{A}{4G}\, ,
\end{equation}
so there seems to be a ``flow'' of the proportionality factor as we move from large black holes to small ones (it goes from $1/(4G)$ to $3h_0/(4\chi_0G)$) but on the whole the area law is respected.

As a conclusion, the findings of this section explain why the qualitative behaviour of all the curves in Fig.~\ref{fig1b} was the same: the limit $M\rightarrow 0$ of black holes is universal in all the higher-order gravities captured by the action \req{lal}. 

\section{Black hole evaporation}
%\commentt{Two cases: (a) Schwarzschild branch connected to $M=0$. (b) Schwarzschild branch has a minimum mass $M_0>0$}
%\begin{table*}[]
%	\centering
%	\begin{tabular}{c||c|c|c} 
%		&  Semiclassical approximation breakdown mass& Time till  $M\sim M_{\rm  \ssc min}$     & Entropy when $M\sim M_{\rm  \ssc min}$\\
%		\hline \hline
%		Schwarzschild  & $M_{\rm  \ssc min}\sim M_{\rm \ssc P}$ & $\Delta t \sim M_0^3/M_{\ssc \rm P}^4 $  & $S \sim 1$ \\ \hline
%		Higher-order BHs & $ M_{\rm  \ssc min}\sim  \sqrt{M_{\ssc \rm P}M_c}  $   & $\Delta t\sim M_{\rm \ssc P}^{7/2}/M_{c}^{9/2}$  & $S\sim M_{\rm \ssc P}/M_c$  %\\ 
%	\end{tabular}
%	\caption{We compare the result of the evaporation process for a Schwarzschild black hole to the one for the new higher-order solutions at the scale for which the semiclassical approximation stops making sense. When the minimum mass is reached, the Schwarzschild black hole has turned into a Planck-mass object of order-one entropy. For the new black holes, the entropy of the  resulting object is huge instead.  
%	}
%	\label{tbl}   
%\end{table*}   
Let us now explore how the evaporation process of black holes gets affected by the special thermodynamic behaviour of the new solutions. We can imagine that our initial black hole is large (with $GM>>L$), so that it is approximately described by the Schwarzschild solution. Then, as the black hole radiates it losses mass, and it moves in the phase diagram following the Schwarzschild branch, which we recall is the one that is smoothly connected to the Schwarzschild solution. In the previous section we have learned that there are essentially two qualitatively different possibilities for this branch. The first possibility is that it can be extended down to $M=0$, in which case it always has the form shown Fig.~\ref{fig1b}: there is a maximum temperature, reached at a certain mass of the order of $L/G$, and for smaller masses the thermodynamic relations are universally given by \req{eq:Tsmall} and \req{eq:Ssmall}. The other possibility is that there is an endpoint for the Schwarzschild branch at some finite mass, as illustrated in Fig.~\ref{fig2b}. In this case, an evaporating black hole will lose mass until it reaches the minimum value available in the phase diagram. However, at that point it still has a finite temperature, so it would continue radiating, hence losing even more mass. We can only speculate about what would happen in that case. For instance, in a situation as the one in Fig.~\ref{fig2b} there are no other black hole solutions below the minimum mass, so it seems that the only possibility would be for the black hole to turn into a singular solution (the phase diagram only shows solutions with a regular horizon). In other cases, there could be other branches of solutions and a phase transition could take place after the minimum mass is reached, or it could even happen that the black hole decays spontaneously into pure radiation. All these possibilities are mere speculations, but we can already agree that the presence of an endpoint in the Schwarzschild branch gives rise to unpleasant situations. Therefore, we will assume that the Schwarzschild branch is connected to $M=0$, in whose case the black hole will evaporate normally, although the time evolution may drastically differ from the standard Einstein gravity prediction.

Computing in a precise way the time evolution of the black hole mass due to Hawking radiation is a far-from-trivial task, since black holes do not radiate exactly as black bodies, and one has to take into account all of the standard model particles that can be radiated at a given temperature. A detailed computation in the case of Schwarzschild black holes was first performed by Page \cite{PhysRevD.13.198}.
Performing an analogous computation for our higher-order black holes would be beyond the scope of the present thesis. It will suffice for our purposes to assume that the power emitted by a black hole is given by the Stefan-Boltzmann law,
\begin{equation}
P=\gamma\frac{\pi^2}{60} A  T^4\, ,
\end{equation}
where $\gamma$ is a numerical factor that takes into account the number of species emitted as well as the fact that the black hole is not a perfect black body. Such factor would depend in principle on $T$, but we will make the approximation that it is a constant. This simple approximation will be enough in order to capture the correct scaling of the problem as well as other qualitative features. Then, taking into account that $P=-dM/dt$ and that $A=4\pi r_h^2$, we get the following equation for the mass-loss rate
\begin{equation}\label{eq:dMdt}
\frac{d M}{dt}=-\gamma\frac{\pi^3 r_h^2}{15} T^4\, .
\end{equation}
Using the existing relations between $r_h$, $T$ and $M$, one can integrate this equation in order to determine the time evolution of the mass. 

In the case of a Schwarzschild black hole, one gets $P\propto M^{-2}$, and the integration yields
\begin{equation}\label{eq:Mtschw}
M(t)=\left(M_0^3-\frac{\gamma\, t}{5120\pi G^2}\right)^{1/3}\, .
\end{equation}
Thus, after a time of the order of $M_0^3G^2=(M_0/M_{\rm \ssc P})^3 t_{\rm \ssc P}$, the black hole reaches zero mass, at which moment the temperature and the power diverge. The thermodynamic description actually breaks down slightly before that (because $S\sim 1$ when the mass of the black hole is of the order of $M_{\rm \ssc P}$), but everything points toward a violent ending of the evaporation process. 

\begin{figure}[h!]
\begin{center}
\includegraphics[width=0.55\textwidth]{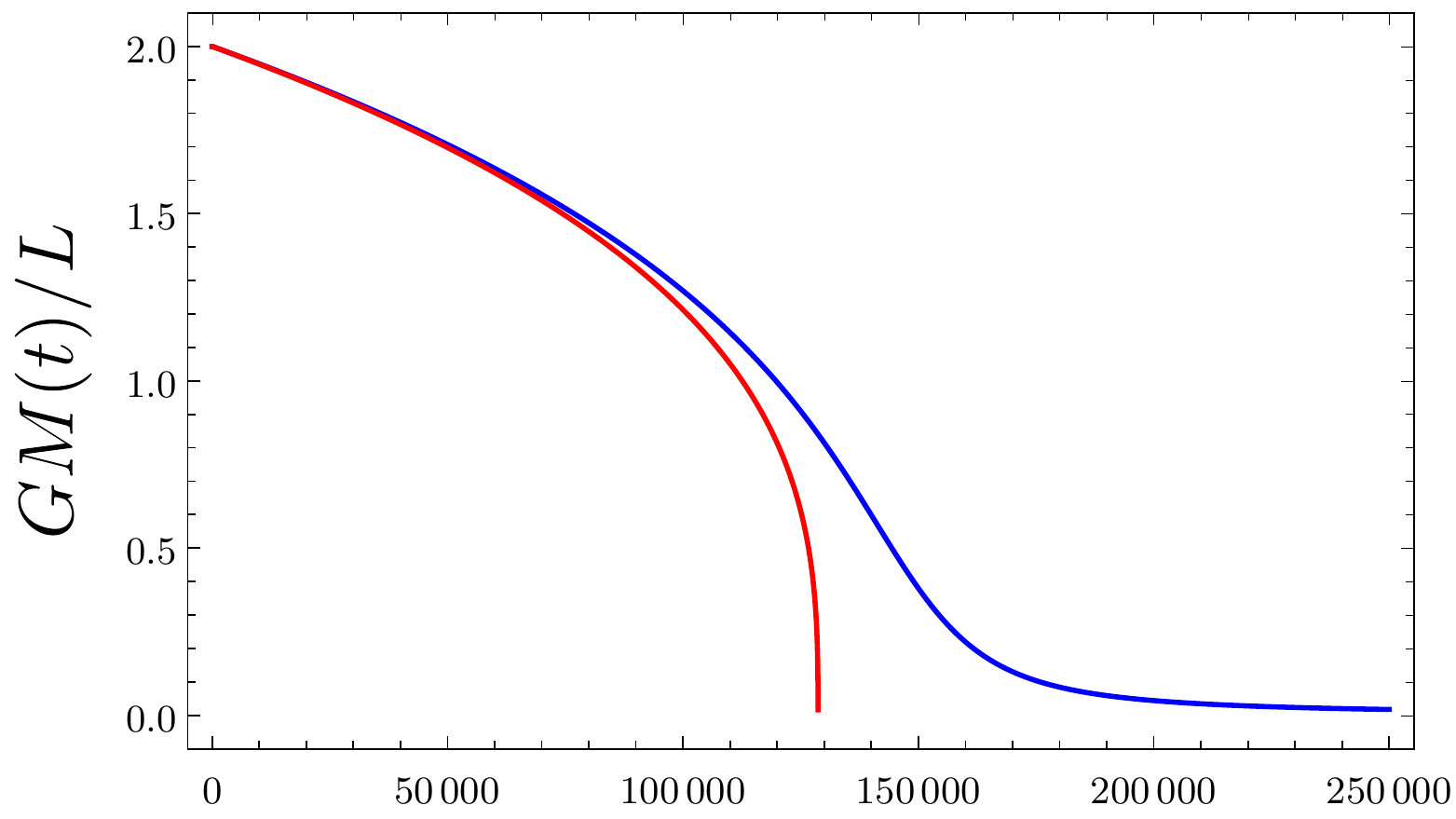}
\includegraphics[width=0.55\textwidth]{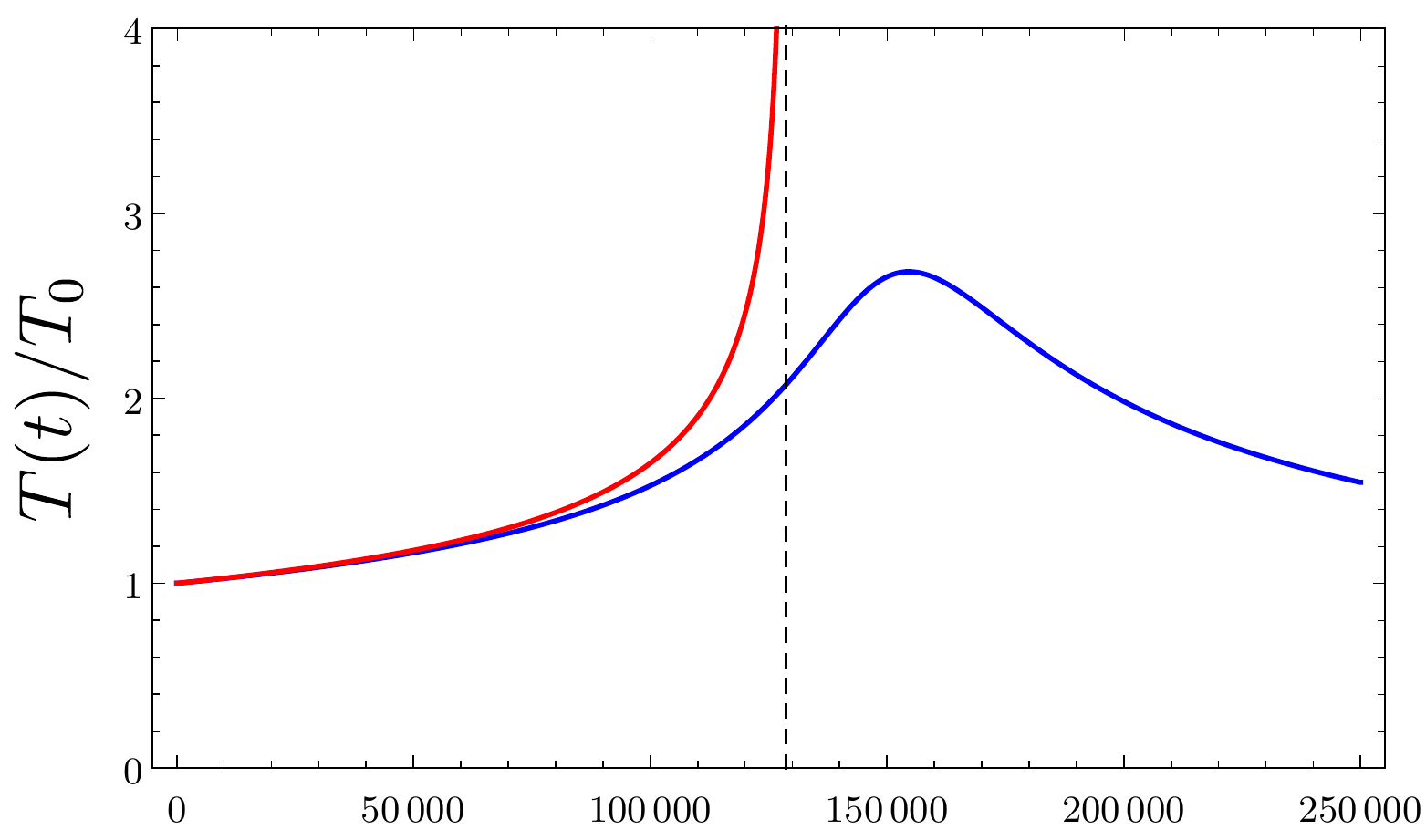}
\includegraphics[width=0.55\textwidth]{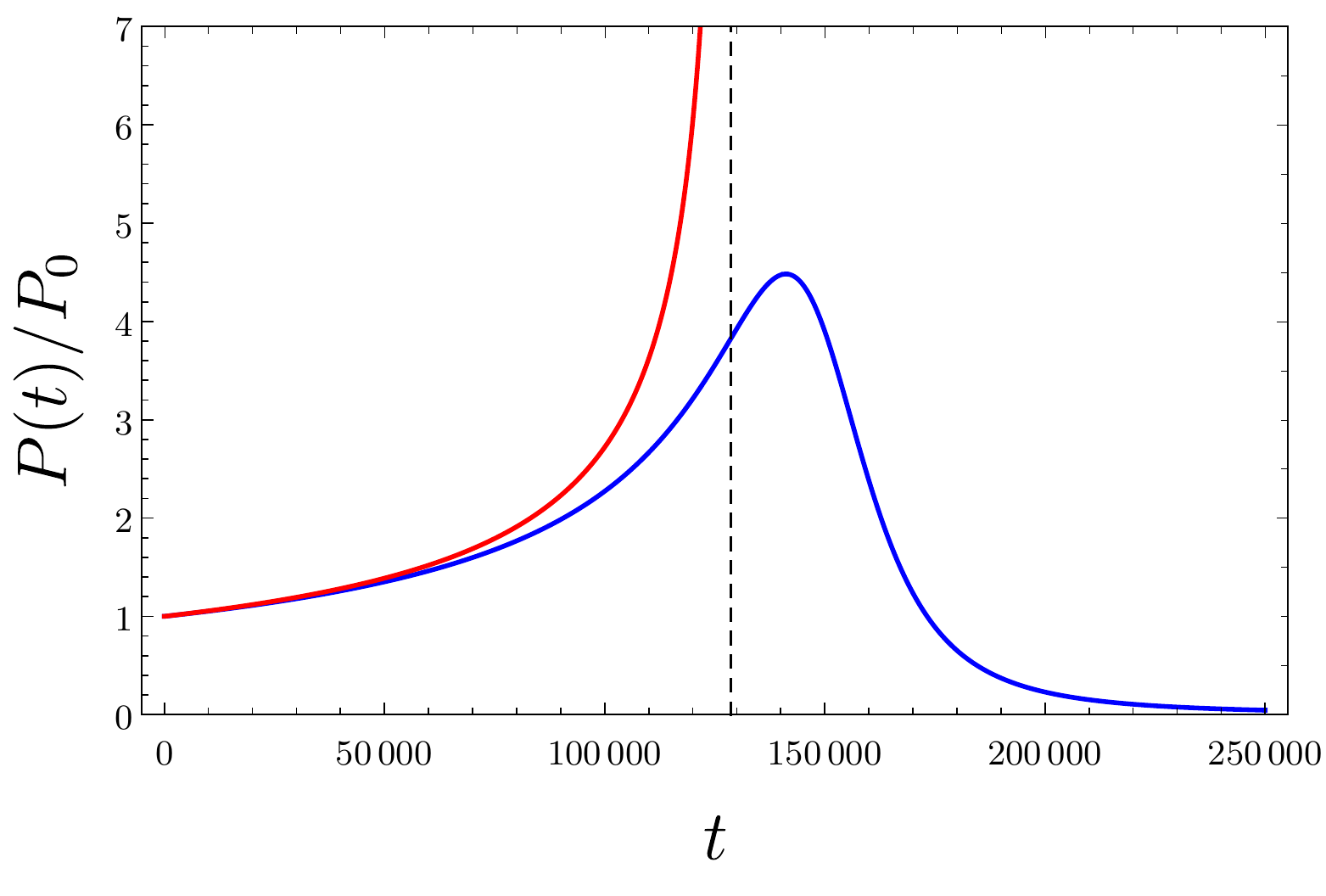}
\caption{Black hole evaporation process predicted by Einstein gravity (red curves) and for higher-order gravity (blue curves). We have taken Einsteinian cubic gravity as a working example, but the qualitative profiles are the same for any other theory whose Schwarzschild branch is connected to $M=0$. From top to bottom we represent the time evolution of the mass, the temperature and the power, respectively. }
\label{fig3b}
\end{center}
\end{figure}

However, the story is very different when higher-derivative corrections are taken into account --- a graphic comparison is shown in Fig.~\ref{fig3b}.  As long as the mass of the black hole is much larger than $L/G$, the time evolution is approximately given by \req{eq:Mtschw}, and during this period the temperature of the black hole increases. Then, after some finite time the black hole reaches a mass of the order of $L/G$, and at this point the corrections become important. As illustrated in Fig.~\ref{fig3b}, the black hole will reach a maximum temperature and after that moment the temperature starts decreasing as the black hole radiates. In the same way, the power reaches a maximum value and then decreases. When the mass is small enough, $M<<L/G$, the behaviour of these black holes is universally dictated by the formulas \req{eq:rhsmall}, \req{eq:Ssmall} and \req{eq:Tsmall}. Using those, we see that the emitted power scales as $P\propto M^2$, and integrating \req{eq:dMdt} in this regime we get
\begin{equation}\label{eq:Mstab1}
M(t)=M_0\left(1+\frac{\gamma \chi _0^4  G^2 M_0 t}{240 \pi  h_0^2 L^4}\right)^{-1}\, ,\quad \text{when}\,\, \, M<<\frac{L}{G}
\end{equation}
Thus, we observe that the mass never goes to zero in a finite time, hence these black holes have infinite lifetimes.\footnote{A similar behavior was previously observed, for example in: \cite{Myers:1988ze,Myers:1989kt} for certain higher-dimensional Lovelock black holes; \cite{Callan:1988hs} for $D(>4)$-dimensional black holes in dilaton gravity modified with a stringy Riemann-squared term;   and  in \cite{Easson:2002tg} for $D=2$ dilaton gravity black holes.}  The mass only tends to zero asymptotically,  as $t\rightarrow\infty$,  
\begin{equation}\label{eq:Mstab2}
M(t\rightarrow\infty)=\frac{240 \pi  h_0^2 L^4}{\gamma \chi _0^4  G^2}\times \frac{1}{t}\, ,
\end{equation}
and interestingly, its value in that limit is independent of the initial mass.
In addition, we recall that the zero mass limit of these solutions is not flat space: it is a black hole of vanishing mass, temperature, entropy and area, but which has a non-trivial gravitational field. Thus, an evaporating higher-order black hole never disappears completely: it leaves a ``remnant'' behind. This evaporation process is dramatically different to the one described before for Einstein gravity, which seemed to end with a violent explosion. 

\subsection{Limit of validity of the semiclassical description}
Even though these black holes never reach zero size in a finite time, their description as solutions of a classical theory stops making sense for small enough masses. The regime of validity of the semiclassical description is characterized by the fact that the horizon radius is much greater than the corresponding Compton wavelength, \ie $r_h\gg \lambda_{\rm \ssc Compton}\sim 1/M$.
Similarly, the thermodynamic description is valid whenever $S\gg 1$, and breaks down  as $S\sim 1$. A Schwarzschild black hole reaches the Planck mass in a time $\Delta t\sim M_0^3/ M_{\rm \ssc P}^4$, after which both conditions are violated. In contrast, for the new black holes, the condition on the entropy is never violated as long as we choose $L>>\ell_{\rm \ssc P}$. The semiclassical approximation stops making sense for a mass of the order of
\begin{equation}
	M_{\rm  \ssc min}\sim  \sqrt{M_{\ssc \rm P}M_{L}}\, ,
\end{equation}
where $M_{L}=1/L$ is the energy scale associated to the length scale $L$. 
Interestingly, when such mass is reached, the entropy is still large, namely $S_{\rm \ssc min} \sim L/\ell_{\rm \ssc P}$, so one could argue that the system still allows for a thermodynamic description according to the equations \req{eq:Ssmall}, \req{eq:Tsmall}.

Now, we want to remark two points about small black holes, that we recall are those with $M<< M_{\ssc \rm P}^2/M_{L} \equiv M_{\rm  \ssc max}$. First, given that $M_{L}<<M_{\ssc \rm P}$, there is a huge range of masses for which these objects can be described semiclassically, namely  $M_{\rm \ssc min } \ll M \ll M_{\rm  \ssc max}$. Second, for any black hole within this range of masses, the time needed to reach the semiclassical description breakdown is very large. In fact, the minimum mass is reached in a time 
$\Delta t \sim L^{9/2}/\ell_{\ssc \rm P}^{7/2}$ which, remarkably, does not depend on the initial mass $M_0$ as long as $M_0 \gg M_{\rm \ssc min } $.

It is illustrative to make a quantitative comparison with the Schwarzschild case. Let us assume that the higher-derivative corrections enter at a scale, say $M_L\sim 1$ GeV, which might seem to be too small, but it is really beyond any possibility of observation with the current experiments. In this case, the upper mass for small black holes is $\sim 10^{13}$ kg, and the semiclassical approximation breakdown mass would be $M_{\rm \ssc min}\sim 10^6$ TeV. At that point, the entropy would be of the order of $S_{\rm \ssc min}\sim 10^{19}$.  Now, consider one of the new higher-order black holes with the initial mass $M_0\sim 10^{6}$ kg (or, in fact, with any mass such that $10^6$ TeV $\ll M_0 \ll 10^{13}$ kg): it would need $\Delta t \sim 10^{25}$ times the age of the universe to reach $M_{\rm \ssc min}$.
In contrast, a Schwarzschild black hole of mass $M_0\sim 10^{6}$ kg, would evaporate down to the Planck mass in $\Delta t \sim 1$ minute, with a final entropy $S\sim 1$. 
Hence, in contrast to the Schwarzschild case, the lifetime of the new small black holes  (understood, in this case, as the period till the semiclassical approximation breakdown) is usually huge or, rather, infinite for all practical purposes.

\section{Conclusions}\label{conclusions}
In this chapter we have constructed exact black hole solutions in an extension of Einstein gravity containing an infinite number of higher-curvature corrections, corresponding to all of the Generalized quasi-topological gravities in four dimensions. We have seen that at every order in curvature there is a unique way in which these theories modify the equations of motion for spherically symmetric black holes, and we were able to find the general formula for the $n$-th order correction to this equation --- see Eqs.~\req{eq:allorders} and \req{eq:Egeneraln}. This allowed us to study in full generality the new black hole solutions at all orders in curvature. 

 We described how one can solve (numerically) the equations of motion in order to find the exact solutions, and we provided several examples in Fig.~\ref{fig108} --- see Appendix \ref{App:3} for a detailed discussion of the numerical construction.
 For a given value of mass, the modified equations of motion may allow for several black hole solutions (each one with different temperature, radius and entropy), but there is a unique solution that reduces to the Schwarzschild one in the limit $M\rightarrow \infty$. We focused our discussion on the family of solutions that is smoothly connected to this one, which we refer to as the Schwarzschild branch.

Even though the solutions cannot be constructed analytically, the relevant thermodynamic magnitudes can be accessed exactly and, in particular, the explicit relations between the horizon radius, the mass, the temperature and the entropy can be written in full generality for all the theories. These relations are given, parametrically, by Eqs.~\req{eq:rhchi}, \req{eq:Mchi}, \req{eq:Tchi} and \req{eq:Schi}. The theory-dependent quantity in these expressions is the function $h(x)$, defined in \req{eq:hfunc}. By analyzing those expressions, we have determined that there are essentially two qualitatively different cases with respect to the behaviour of the Schwarzschild branch. 
One possibility is that this branch has an endpoint at some minimum mass, so that it does not extend down to $M=0$ --- see Fig.~\ref{fig2b}. However, this case is somewhat odd because an evaporating black hole of the minimum mass still has a non-vanishing temperature but it cannot decay into other solution, which seems to be a pathological situation. 
The other possibility, that we consider most interesting, is that the Schwarzschild branch is connected to solutions of vanishing mass, as in Fig.~\ref{fig1b}. In this case, the specific heat of the black holes becomes positive below certain mass $M_{\rm \ssc max}\sim L/G$, and for lower masses, $M \ll M_{\rm \ssc max}$, the thermodynamic properties of these black holes are universal, and they are given by Eqs.~\req{eq:Ssmall} and \req{eq:Tsmall}. In this regime, the black holes satisfy the Smarr-like relation $M=2/3\cdot TS$ for arbitrary values of the new couplings --- the only exception to this is Einstein gravity, for which one has $M=2TS$ for every value of the mass. 
Thus, the characteristic profile of the relation $T(M)$ shown in Fig.~\ref{fig1b} is shared by all the higher-order theories whose Schwarzschild branch is connected to $M=0$. In particular, including only the cubic term in \req{lal} we obtain the same qualitative result as if we included an infinite number of terms. This answers one of the questions which we posed at the beginning of this chapter: the results obtained in Chapter \ref{Chap:4} in the case of Einsteinian cubic gravity are not substantially modified when one includes higher-derivative corrections --- as long as the theory allows for black hole solutions of arbitrarily small masses. Finally, we studied the evaporation process of black holes and we concluded that, unlike the case of Schwarzschild's solution, they new higher-order black holes never evaporate in a finite time. 

All these interesting results apply to the family of Generalized quasi-topological gravities, but, could they apply as well in more general cases? Indeed, in Chapter~\ref{Chap:3} we argued that, using field redefinitions, any higher-derivative gravity can be mapped to a sum of GQG terms. In particular, it is important to recall Conjecture \ref{conj:4}, which essentially tells us that, in order to capture the physics of static black holes in the most general higher-derivative gravity, it is enough to study the black hole solutions of the most general GQG theory of the form $\mathcal{L}(g^{\alpha\beta},R_{\mu\nu\rho\sigma})$. This is precisely what we have accomplished in the present chapter. 
Since black hole thermodynamics is invariant under a redefinition of the metric, this result would imply that the thermodynamic properties of black holes in \emph{any} higher-derivative gravity would coincide with those of the theories in \req{lal} for a \emph{certain} choice of the couplings $\lambda_n$.
In other words, if Conjecture \ref{conj:4} holds, then the relations \req{eq:Mchi}, \req{eq:Tchi} and \req{eq:Schi} represent the thermodynamic quantities of black holes in any higher-derivative gravity in four dimensions. The complicated part would be to determine in each case the values of the couplings $\lambda_n$ corresponding to a certain Lagrangian. But on the other hand, we have seen that the thermodynamic quantities exhibit a universal behavior, quite independent of the values of the couplings. For instance, the existence of stable black holes is something that we would expect on general grounds for arbitrary higher-derivative Lagrangians. 
Given the interest of this result, it would be of capital importance to explore the validity of Conjecture \ref{conj:4} or to attempt a proof of it in future works. In any event, it is already evident that the thermodynamic properties of black holes that we found here represent, at least, those of a very large portion of all possible higher-derivative gravity theories.

Given the presumable generality of our results, let us briefly discuss some of their possible implications before closing this chapter.

\subsubsection{Black hole remnants?}
 
The thermodynamic properties of the small higher-order black holes have dramatic consequences for the evaporation process. In particular, as opposed to the Schwarzschild case, the new black holes have infinite lifetimes, so one could speculate that they might act as \emph{eternal} information reservoirs evading possible unitarity violations. In fact, the zero-mass limit of these solutions is not empty space, because the gravitational field still has a non-trivial profile in that limit. This indicates that the final product of black hole evaporation is some sort of \emph{remnant}. However, it is difficult to see how such an object could act as an information reservoir, since, according to Eq.~\req{eq:Ssmall}, its entropy vanishes. %During the evaporation, the black holes require a (normally huge) time $\Delta t \sim M_{\rm \ssc P}^{7/2}/M_c^{9/2}$  to reach a mass $M_{\rm \ssc min}\sim \sqrt{M_{\rm \ssc P}M_c}$ for which the semiclassical description breaks down. At that point, the entropy of the resulting 
% \bl{\emph{remnant} is still huge, $S\sim M_{\rm \ssc P}/M_c$. This is in stark contrast with the Schwarzschild black hole case, which reaches the Planck mass with an order-one entropy, and hence fails to provide a reasonable possible account of the missing information. } 
%object is still huge, $S\sim M_{\rm \ssc P}/M_c$. This is in stark contrast with the Schwarzschild black hole case, which reaches the Planck mass with an order-one entropy.
%, and hence flagrantly fails to provide a reasonable possible account of the missing information. 
%One can only speculate on what would happen below $M_{\rm \ssc min}$. However, it is worth emphasizing that the tendency suggested by the evaporation process of the new black holes is very different from Schwarschild's: first, if we extrapolated the semiclassical approximation beyond $M_{\rm \ssc min}$, the Schwarzschild black hole would disappear soon after, while the new black holes would placidly keep on living forever; %and second, at the semiclassical breakdown point, the issue of being left with an extremely low entropy object gets notably softened. 
%\subsubsection{Implications for the information paradox?}

At this point, it is convenient to stress that attempts to resolve the information paradox involving remnant-like objects \cite{Aharonov:1987tp} have been usually argued to be very difficult to digest \cite{Preskill:1992tc,Harvey:1992xk,Giddings:1994qt,Susskind:1995da}. This is, in particular, because one can consider the evaporation process of arbitrarily large initial black holes, which would imply that an arbitrarily large amount of information would need to be stored in the remnant, forcing such finite-energy object to have an infinite number of internal states (or, in other words, infinitely many ``species'' of remnants would need to exist, one for each possible initial state
collapsing to a black hole).   This raises the question of how an arbitrary amount of information could be carried within a Planck volume (or, more generally, the volume associated to the semiclassical breakdown scale) and naturally leads to a problematic infinite production rate of remnant pairs --- see \eg \cite{Giddings:1994qt}. 
While these remain outstanding issues,\footnote{See \cite{Giddings:1992kn,Callan:1992rs,Giddings:1992hh} for some possible ways out.} our results illustrate that the evaporation process can drastically change  when higher-order terms are considered in the gravitational effective action, suggesting new perspectives for the final fate of black holes. 

\subsubsection{Microscopic stable black holes as dark matter}

Besides theoretical considerations, the existence of small (or microscopic), stable black holes in four dimensions could actually have consequences for our universe. In fact, it has been often argued that microscopic black holes could be responsible for dark matter --- see \eg \cite{Chen:2002tu,Easson:2002tg,Afshordi:2003zb,Nozari:2005ah,Mureika:2012na}. In the case of Einstein gravity there is a minimum mass such black holes can have, namely $M>M_{\rm ev}\sim 10^{14} g$ --- otherwise, they would have evaporated by now if they were formed in the early universe.  On the other hand, the new stable black holes have infinite lifetimes, so there is no minimum value for their masses. In fact, it is possible to derive a very simple estimation for the current mass of a primordial black hole of initial mass $M_0$. If $M_0>M_{\rm ev}$, then there is no difference with respect to Einstein gravity and we can approximate that the black hole keeps all its mass. On the contrary, if $M_0<M_{\rm ev}$ it will evaporate until it reaches the regime in which it is stable. In that case, using \req{eq:Mstab1} one can see that its current mass would be, approximately, $M_{\rm end}(M_0)=\min\{(L/\ell_{\rm\ssc P})^4 t_0^{-1}, M_0\}$, where $t_0$ is the age of the universe. Remarkably, we find that \emph{all} primordial black holes with initial masses between $M_{\rm ev}$ and $(L/\ell_{\rm\ssc P})^4 t_0^{-1}$ would have the same mass by now. This is a remarkably simple and powerful prediction which holds for an infinite number of higher-order gravities. In this scheme, the only important quantity is the length scale $L$, which determines the final mass of all primordial black holes within a huge range of masses. On the other hand, the order of magnitude of $L$ is almost unconstrained from known physics. 

Thus, these theories open a new window in the search for dark matter sources and it would be worth further exploring if these stable black holes are viable candidates. This possibility is even more tantalizing if we take into account that this family of higher-order gravities is also able to describe an inflationary phase in the early universe in a very natural way \cite{Arciniega:2018fxj,Cisterna:2018tgx,Arciniega:2018tnn}.

\part{Asymptotically anti-de Sitter solutions and Holography}\label{Part:III}

\chapter{Holographic aspects of Einsteinian cubic gravity}\label{Chap:6}

According to the AdS/CFT correspondence \cite{Maldacena,Gubser,Witten}, there exists an equivalence between a classical gravitational theory in a $D$-dimensional asymptotically AdS space and a conformal field theory that lives in a $(D-1)$-dimensional space, which can be identified with the boundary of AdS. The foundations of this duality are rooted in String Theory, which dictates the precise theories involved and the \emph{dictionary} between them. However, this correspondence has transcended its original confinements, and it is nowadays understood as a general principle, that is sometimes called \emph{gauge/gravity duality}. 
As we reviewed in the introduction (Section~\ref{sec:holotoymodels}), the AdS/CFT correspondence allows us to learn about CFTs by studying gravitational theories with a negative cosmological constant. Some of the aspects of a CFT that one can study by considering bulk gravity models include the $n$-point functions of the stress-energy tensor, trace anomalies, thermal properties, etc. 
%The holographic dictionary states that the metric perturbation induced in the boundary of AdS couples to the stress-energy tensor of the dual CFT. This allows us to determine any quantity of the dual CFT that only depends on the stress-energy tensor and its correlators --- we say in that case that the gravitational theory probes CFTs belonging to different universality classes. Likewise, one can explore the thermodynamic phase space of the CFT, taking into account the fundamental principle of the duality, \ie the equality of the partition functions, $\mathcal{Z}_{\rm grav}=\mathcal{Z}_{\rm CFT}$. Typically, the thermal states in the gravitational side are represented by black holes, and the gravitational partition function can be computed following the Euclidean path integral approach.
When the gravitational theory is Einstein gravity, one probes a very restricted set of CFTs, so it is interesting to consider modifications of EG that allow to explore a larger set of theories. % --- we elaborate this in Sec.\ref{Introduction-c6} below.  
We are interested in the case in which the modifications are given by higher-curvature interactions in the Einstein-Hilbert action.  However, one must take care when the gravitational theory is modified, because the holographic dictionary might be altered as well. In fact, in Chapter~\ref{Chap:1}, we saw that higher-derivative gravities usually introduce additional degrees of freedom with respect to Einstein gravity. Thus, the metric perturbation does not only contain the usual massless spin-2 graviton of Einstein gravity, but additional modes. Then, it is not clear that this perturbation couples only to the boundary stress-energy tensor and one would probably need to modify the holographic dictionary in some subtle way. In addition, among the new modes one finds a massive ghost-like graviton, which can be a cause of pathologies. 

Fortunately, these difficulties can be avoided if one chooses the higher-curvature Lagrangian in a suitable way. In Chapter~\ref{Chap:1} we presented a type of higher-order gravities whose linearized equations on maximally symmetric backgrounds coincide with those of Einstein gravity and, consequently, we named these theories Einstein-like. As we discussed at the end of Chapter~\ref{Chap:1}, these theories generally possess equations of motion of higher-order, but the only mode that can escape to infinity --- to the boundary of AdS --- is the Einstein graviton, because any other hypothetical mode becomes infinitely heavy as it tries to reach the boundary. Thus, for Einstein-like theories it is guaranteed that one can apply the holographic dictionary just like in Einstein gravity. However, the field-theoretic space of CFTs that one can explore with these theories is much larger than the one dual to Einstein gravity. 

We recall that the set of Einstein-like theories is still very large, so we can choose an appropriate subset of them. It is particularly interesting to consider the family of Generalized quasi-topological gravities, that we reviewed in Chapter~\ref{Chap:2}. These theories belong to the Einstein-like class, and additionally, possess simple black hole solutions for which we can determine the thermodynamic properties exactly --- this was illustrated in the previous two chapters for asymptotically flat black holes. Thus, GQGs are excellent candidates in order to probe the thermodynamic phase space of CFTs beyond Einstein gravity holography.
Additionally, as we argued in Chapter~\ref{Chap:3},  GQGs probably provide a basis to construct the most general EFT for gravity. In that case, they would also allow us to explore the AdS/CFT correspondence in the most general scenario.
The goal of the third part of this thesis is to provide a first study on the holographic aspects of these theories, focusing mainly in the case $D=4$, which has been so far the less explored one due to the lack of appropriate theories. 

In the present chapter we consider Einsteinian cubic gravity in $D=4$ with a negative cosmological constant and we identify the main entries in the holographic dictionary. In addition, by studying the AdS black holes of ECG, we will provide a detailed description of the thermodynamic phase space of the dual CFT, which possesses some important differences with respect to the Einstein gravity case. A more detailed summary of our findings in this chapter is provided in Sec.\ref{summ} below.
Then, our aim will be to study the \emph{squashed holography} of GQGs. As a first step, in Chapter~\ref{Chap:7} we will construct asymptotically AdS Euclidean-Taub-NUT solutions, and we will compute their partition functions analytically. In the case $D=4$, these solutions represent the first examples of modified Taub-NUT solutions in any higher-order gravity. Then, we will see in Chapter~\ref{Chap:8} that such solutions are dual to CFTs on squashed spheres, and that their gravitational on-shell Euclidean action computes the partition function of those CFTs. 
Combining this with the results of Chapters \ref{Chap:6} and \ref{Chap:7} we will derive new non-trivial relations for the free energy of general CFTs on squashed spheres.

\section{Introduction}\labell{Introduction-c6}
Higher-order gravities play an important role in AdS/CFT \cite{Maldacena,Gubser,Witten}. Perturbative corrections to the large-$N$ and strong-coupling limits of holographic CFTs are encoded, from the bulk perspective, in higher-curvature interactions which modify the semiclassical Einstein (super)gravity action --- see \eg \cite{Grisaru:1986px,Gross:1986iv,Gubser:1998nz,Buchel:2004di}.
%When understood as perturbative modifications of semiclassical Einstein (super)gravity, they encode corrections to the large-$N$ and strong-coupling limits of the corresponding CFT. 
The introduction of such terms, which is in principle fully controlled by String Theory, gives rise to
%In that framework, the structure of such terms is fully controlled and constrained by the underlying String Theory.
%From a different perspective, the addition of simple higher-curvature terms in the bulk action
 %can give access to 
 holographic theories belonging to universality classes different from the one defined by Einstein gravity \cite{Buchel:2008vz,Hofman:2008ar,Hofman:2009ug} --- \eg one can construct CFTs with $a\neq c$ in $d=4$ \cite{Nojiri:1999mh,Blau:1999vz}.  Some care must be taken, however. As shown in \cite{Camanho:2014apa}, higher-curvature terms making finite contributions to physical quantities in the dual CFT can become acausal unless new higher-spin ($J> 2$) modes appear at the scale controlling the couplings of such terms.
 
 %On the one hand, if one of these terms makes finite contributions to the dual theory, it means that the background scale has approached stringy or Planckian regimes, where infinitely many other higher-curvature terms would become relevant as well, suggesting the breakdown of the local field-theory description in the bulk. On the other, if we assume the new couplings to be controlled by scales much larger than 
 
In spite of this, a great deal of non-trivial information can be still obtained by considering particular higher-curvature interactions at finite coupling --- \ie beyond a perturbative approach. The idea is to select theories whose special properties make them amenable to calculations --- something highly nontrivial in general. The approach turns out to be very rewarding and, in some cases, it has led to the discovery of universal properties valid for completely general CFTs  \cite{Myers:2010tj,Myers:2010xs,Mezei:2014zla,Bueno1,Bueno2}. In other cases, higher-order gravities have served as a proof of concept, \eg providing counterexamples \cite{Buchel:2004di,Kats:2007mq,Brigante:2007nu,Myers:2008yi,Cai:2008ph,Ge:2008ni} to the Kovtun-Son-Starinets bound for the shear viscosity over entropy density ratio \cite{Kovtun:2004de} --- see discussion below. 
  Just like free-field theories, these holographic higher-order gravities should be regarded as toy models for which many calculations can be explicitly performed, hence providing important insights on physical quantities otherwise practically inaccesible for most CFTs --- see \eg \cite{HoloRen,Hung:2014npa,deBoer:2011wk,Bianchi:2016xvf} for additional examples.

A key property one usually demands from a putative holographic model of this kind is that it admits explicit AdS black-hole solutions. In $d\geq 4$, this canonically selects Gauss-Bonnet or, more generally, Lovelock gravities \cite{Lovelock1,Lovelock2}, for which numerous holographic studies have been performed in different contexts --- see \eg \cite{Camanho:2009vw,deBoer:2009pn,Buchel:2009sk,deBoer:2009gx,Camanho:2009hu,Grozdanov:2014kva,Grozdanov:2016fkt,Andrade:2016rln,Konoplya:2017zwo} and references therein. The next-to-simplest example in $d=4$ is Quasi-topological gravity (QTG) \cite{Quasi,Quasi2}, a theory which includes, in addition to the Einstein gravity and Gauss-Bonnet terms, an extra density, cubic in the Riemann tensor. Besides admitting simple generalizations of the Einstein gravity AdS black holes, and having second-order linearized equations of motion on maximally symmetric backgrounds, this theory contains three dimensionless parameters: the ratio of the cosmological constant scale over the Newton constant, $L^2/G$, and the new gravitational couplings, $\lambda$ and $\mu$. These can be translated into the three parameters characterizing the three-point function of the boundary stress tensor. As opposed to Lovelock theories, for which one of such parameters, customarily denoted $t_4$ \cite{Hofman:2008ar}, is always zero \cite{Buchel:2009sk,deBoer:2009gx,Camanho:2009hu,Camanho:2013pda}, the new QTG coupling gives rise to a nonvanishing $t_4$ \cite{Myers:2010jv}. For supersymmetric theories one also has $t_4=0$ \cite{Hofman:2008ar,Kulaxizi:2009pz}, so QTG provides a toy model of a non-supersymmetric CFT in four dimensions.

All studies performed so far involving finite higher-curvature couplings have been restricted to $d\geq 4$ --- observe that all theories mentioned in the previous paragraph reduce to Einstein gravity for $d=3$. Obviously, from the CFT side, there is no fundamental reason to exclude holographic three-dimensional theories. In fact, there exist many interesting CFTs in $d=3$ with known holographic duals, \eg \cite{Maldacena,Aharony:2008ug,Aharony:2008gk,Klebanov:2002ja,Leigh:2003gk,Aharony:2011jz}.
The actual reason for the absence of holographic studies involving higher-curvature terms in $d=3$ has been the lack of examples admitting generalizations of Einstein gravity black holes in four bulk dimensions. The situation has recently changed thanks to the discovery of Einsteinian cubic gravity (ECG) \cite{PabloPablo}, for which such generalizations are possible \cite{Hennigar:2016gkm,PabloPablo2} --- see Sec.~\ref{ECGG} for a detailed review. As we show here, ECG provides a holographic toy model of a nonsupersymmetric CFT in three dimensions, analogous to the one defined by QTG in four. The main purpose of this chapter is to study the behavior of several physical quantities in this new model. Just like it occurs for Lovelock and QTG in $d\geq 4$, all results can be obtained fully nonperturbatively in the new gravitational coupling, which provides a much better handle on the corresponding quantities than any possible perturbative calculation.

On a more general front, we propose a new method for computing Euclidean on-shell actions for asymptotically AdS$_{(d+1)}$ solutions of an important class of general higher-order gravities --- those for which the linearized equations become second-order on maximally symmetric backgrounds. Our generalized action represents a drastic simplification with respect to standard approaches, as it utilizes the same boundary term and counterterms as for Einstein gravity, but weighted by a universal quantity related to the entanglement entropy across a spherical region in the boundary theory. 

A more precise summary of our findings can be found next.

\subsection{Summary of results}\label{summ}
The chapter is somewhat divided into two main parts. In the first, which includes Sections \ref{ECGG}, \ref{BHs} and \ref{osa}, we develop some preliminary results and techniques which are necessary for the holographic computations which we perform in Sections \ref{tt} to \ref{shear}.
\begin{itemize}
\item{In Sec.~\ref{ECGG}, we start with a review of ECG and recent developments. Then, we characterize the AdS$_4$ vacua of the theory, and identify the range of (in principle) allowed values of the new coupling and its relation to the existence of a critical limit for which the effective Newton constant blows up.} 
\item{In Sec.~\ref{BHs}, we construct the AdS$_4$ black holes of ECG with general horizon topology.}
\item{In Sec.~\ref{osa}, we propose a new method for computing on-shell actions of asymptotically-AdS solutions of general higher-order gravities whose linearized spectrum on AdS$_{(d+1)}$ matches that of Einstein gravity. %We argue that the usual Gibbons-Hawking-York boundary term of general relativity can be still used in that case we argue should be valid for any theory with Einstein-like spectrum in an asymptotically AdS space. 
	We claim that the corresponding boundary term and counterterms can be chosen to be proportional to the usual Einstein gravity ones. Amusingly, we find that the proportionality factor is controlled by the charge $a^*$ characterizing the entanglement entropy across a spherical region $\mathbb{S}^{d-2}$ in the dual CFT. As a first consistency check of our proposal, we use our generalized action to prove the relation between $a^*$ and the on-shell gravitational Lagrangian $\mathcal{L}|_{\rm AdS}$ for odd-dimensional holographic CFTs with  higher-curvature duals. }
\item{In Sec.~\ref{tt}, we compute the charge $\ct$ controlling the correlator of the boundary stress-tensor from an explicit holographic computation and show that the result agrees with the (not so) naive expectation obtained from the effective Newton constant. We argue that the detailed cancellations between bulk and boundary contributions giving rise to the correct answer constitute a strong check of the generalized action proposed in the previous section.}
\item{In Sec.~\ref{therr}, we start with another check of our generalized action, consisting in an explicit calculation of the free energy of ECG AdS$_4$ black holes, which we show to agree with the one obtained using Wald's entropy approach. Then we compute the thermal entropy charge $\cs$, and we note that it presents notable differences with respect to previous results for other higher-curvature holographic models in $d\geq 4$. Then, we study the thermal phase space of holographic ECG with toroidal and spherical boundaries, respectively. In the latter case, we find that the standard Hawking-Page transition also occurs in ECG. However, the transition temperature increases with the ECG coupling, and actually diverges in the critical limit (for which thermal AdS always dominates). The phase diagram presents new phenomena, like the presence of `intermediate-size' black holes, a new phase of small and stable black holes, as well as the existence of a new critical point.}
\item{In Sec.~\ref{renyie}, we compute the Renyi entropy of disk regions in holographic ECG. In particular, we study the dependence of $S_q/S_1$ on the CFT-charges ratio $\ct/a^*$. Although the functional dependence is very complicated, we observe that the behavior is approximately linear for most values in the allowed range. We also obtain an exact result for the scaling dimension of twist operators, from which we are able to extract the value of the stress-tensor three-point function charge $t_4$, which is non-vanishing in general.}
\item{In Sec.~\ref{shear}, we compute the shear viscosity to entropy density ratio in ECG. Unlike all previous exact results ($d\geq 4$), the result turns out to be highly nonperturbative in the ECG coupling, as it involves a non-analytic function. Several approximations as well as a precise numerical evaluation are accesible. We find that violations of the KSS bound are strictly forbidden in ECG by the requirement that black holes have positive energy. On the other hand, we show that energy-flux bounds on $t_4$ impose a maximum value for the ratio, given by $(\eta/s)|_{\rm max.}\simeq 1.253/(4\pi)$.}
%\item{In section \ref{discu}, we make a quick summary of the different universal charges computed throughout the chapter and how they compare with the analogous ones for QTG in $d=4$. Here, we also speculate on the possible implications of the generalized on-shell action introduced in section \ref{osa} for holographic complexity.}
\item{In Appendix \ref{ttt}, we show that the scaling dimension of twist operators can be used to obtain the exact results for the stress-tensor three-point function parameters $t_2$ and $t_4$ for holographic theories in which explicit calculations of such quantities had been performed before. Appendix \ref{BTcheck} provides an additional check of our generalized action, in this case for a theory for which the generalized version of the Gibbons-Hawking-York term is explicitly known, namely, Gauss-Bonnet gravity. We show that our method gives rise to exactly the same divergent and finite terms as the standard prescription. Appendix \ref{2pbdy} contains some intermediate calculations omitted in Sec.~\ref{tt}. Finally, Appendix \ref{App:Gen} provides the generalization of some results of the chapter to higher orders in curvature. }
\end{itemize}

\subsubsection*{Note on conventions}
We set $c=\hbar=1$ throughout the chapter. $D$ stands for the number of spacetime dimensions of the bulk theory, and $d\equiv D-1$ for those of the boundary one. We use signature $(-,+,+,\dots)$, greek indices for bulk tensors, $\mu,\nu,\dots=0,1,\dots,D$, Latin indices from the beginning of the alphabet for boundary tensors, $a, b,\dots=0,1,\dots,d$ and $i,j,\dots=1,\dots,d$ for spatial indices on the boundary.
Our conventions for $\ctt$, $t_4$, $\cs$ and $a^*$ are the same as in \cite{Buchel:2009sk,Myers:2010jv,Myers:2010tj,Bueno2}. Superscripts `E' and `ECG' mean that the corresponding quantities are computed for Einstein and Einsteinian cubic gravities respectively, whereas we use the subscript `$E$' for Euclidean actions. $L$ is the cosmological constant length-scale ($-2\Lambda \equiv (D-1)(D-2)/L^2$) whereas $\tilde{L}$ stands for the AdS$_{D}$ radius. We often use $L$ for intermediate calculations (including on-shell actions, etc.), but normally present final results in terms of $\tilde{L}$. It is then important to keep in mind that, when expressing our results in terms of the ECG coupling $\mu$, there is some additional dependence hidden in $\tilde{L}=L/\sqrt{f_{\infty}}$, as $f_{\infty}$ is also a function of $\mu$ --- see Fig. \ref{ffffi} and \req{finfs}. 
 
 \section{Einsteinian cubic gravity}\label{ECGG}
Let us start with a quick review of four-dimensional Einsteinian cubic gravity (ECG) and its most relevant properties. The $D$-dimensional version of the theory was presented in \cite{PabloPablo}, where it was shown to be the most general diffeomorphism-invariant metric theory of gravity which, up to cubic order in curvature, shares the linearized spectrum of Einstein gravity on general maximally symmetric backgrounds in general dimensions\footnote{More concretely, the theory is selected by asking it to be the `same' for arbitrary $D$, in the sense that the coefficients relating the various cubic invariants entering its definition do not depend on $D$.}. This criterion selects the Lovelock densities --- cosmological constant, Einstein-Hilbert, Gauss-Bonnet and cubic Lovelock densities --- plus a new invariant, which reads
\begin{equation}
\mathcal{P}=12 \tensor{R}{_{\mu}^{\rho}_{\nu}^{\sigma}}\tensor{R}{_{\rho}^{\alpha}_{\sigma}^{\beta}}\tensor{R}{_{\alpha}^{\mu}_{\beta}^{\nu}}+\tensor{R}{_{\mu\nu}^{\rho\sigma}}\tensor{R}{_{\rho\sigma}^{\alpha\beta}}\tensor{R}{_{\alpha\beta}^{\mu\nu}}-12R_{\mu\nu\rho\sigma}R^{\mu\rho}R^{\nu\sigma}+8\tensor{R}{_{\mu}^{\nu}}\tensor{R}{_{\nu}^{\rho}}\tensor{R}{_{\rho}^{\mu}}\, .
\end{equation}
This invariant is neither trivial nor topological in $D=4$, so the action of the theory becomes
\begin{equation}\label{ECG}
I^{\rm ECG}=\frac{1}{16\pi G}\int d^4x \sqrt{|g|}\left[\frac{6}{L^2}+R-\frac{\mu L^4}{8} \mathcal{P} \right]\, ,
\end{equation}
in such a number of dimensions\footnote{From now on, we will always be referring to the four-dimensional version of the theory when referring to `ECG', unless otherwise stated.}. Here, $\mu$ is a dimensionless coupling. Note also that, for later convenience, in \req{ECG} we have chosen the cosmological constant to be negative, $-2\Lambda\equiv 6/L^2$, where $L$ is a length scale which will coincide with the corresponding AdS$_4$ radius for $\mu=0$. 

It was subsequently shown \cite{Hennigar:2016gkm,PabloPablo2} that \req{ECG} admits non-trivial generalizations of Einstein gravity's Schwarzschild black hole characterized by a single function $f(r)$ --- see next section. It was also observed \cite{Hennigar:2017ego,PabloPablo3,Ahmed:2017jod,PabloPablo4} that, in fact, ECG belongs to a broader class of theories --- coined \emph{Generalized quasi-topological gravities} in \cite{Hennigar:2017ego} --- which also includes Lovelock \cite{Lovelock1,Lovelock2} and Quasi-topological \cite{Quasi2,Quasi,Myers:2010jv,Dehghani:2011vu,Dehghani:2013ldu,Cisterna:2017umf} gravities as particular examples, and which are characterized by: having a well-defined Einstein gravity limit; sharing the linearized spectrum of Einstein gravity on general maximally symmetric backgrounds; admitting non-hairy single-function generalizations of Schwarzschild's black hole. If the action does not include derivatives of the Riemann tensor, the full non-linear equations of a given theory belonging to this class reduce, on a general static and spherically symmetric ansatz, to a single (at most second-order) differential or algebraic --- depending on the case \cite{PabloPablo2} --- equation for $f(r)$, which indeed can be seen to correspond to a unique non-hairy black hole whose thermodynamic properties can be exactly obtained by solving a system of algebraic equations without free parameters.
 
The thermodynamic properties of the asymptotically flat ECG black holes and its higher-curvature generalizations are very different from their Einstein gravity counterparts, as they become stable below a certain mass, which results in infinite evaporation times \cite{PabloPablo2,PabloPablo4}. The asymptotically-AdS black brane solutions of ECG, and generalizations above mentioned, have also been considered in \cite{PabloPablo3,Ahmed:2017jod,PabloPablo4} and, specially, in \cite{Hennigar:2017umz}. There, it was shown that, as opposed to  all previously considered higher-order gravities, the charged black brane solutions of the Generalized quasi-topological class in $D\geq 4$ generically present nontrivial thermodynamic phase spaces, containing phase transitions and critical points.
 
Another relevant development entailed the identification of a critical limit of ECG (for which the effective Newton constant diverges) \cite{Feng:2017tev}, corresponding to $\mu=4/27$. In that particular case, the black holes --- as well as other interesting solutions, such as bounce universes --- can be constructed analytically. 

More recently, some of the possible observational implications of the theory were studied in \cite{Hennigar:2018hza}. There, an observational bound on the ECG coupling was found using Shapiro time delay, and the effects of ECG on black-hole shadows were discussed, including possible measurable differences with respect to Einstein gravity predictions.  Comparisons between general relativity and other theories of gravity regarding black-hole observables are highly limited by the lack of explicit four-dimensional alternatives, which makes ECG particularly appealing for this purpose.

Finally, from the holographic front, let us mention that a study of R\'enyi entropies for spherical regions, similar to the one we perform in Sec.~\ref{renyie}, was carried out in \cite{Dey:2016pei} for ECG in $D=5$. However, it should be stressed that in dimensions greater than four, ECG does not belong to the Generalized quasi-topological class, in the sense that --- even though it shares the linearized spectrum of Einstein gravity --- simple black hole solutions satisfying the properties explained above do not exist for the theory and, as opposed to the $D=4$ case, one is restricted to perturbative calculations in the gravitational couplings, which makes them less interesting.

\subsection{AdS$_4$ vacua and linearized spectrum}
The AdS$_{4}$ vacua of \req{ECG} have a curvature scale $\tilde{L}$ related to the action length scale $L$ through
\begin{equation}\label{adsc}
\frac{1}{\tilde{L}^2}=\frac{f_{\infty}}{L^2}\, ,
\end{equation}
where $f_{\infty}$ is a solution to the algebraic equation
\begin{equation}\label{roo}
1-f_{\infty}+\mu f_{\infty}^3=0\, .
\end{equation}	
For negative values of $\mu$, two of the roots are imaginary, and one is  real and positive. For $0<\mu<4/27$, the three roots are real, one of them being negative and the other two positive. Finally, for $\mu>4/27$, two of the roots are imaginary, and the remaining one is negative. Hence, imposing $f_{\infty}>0$, constrains $\mu$ as
\begin{equation}\label{sis}
\mu<\frac{4}{27}\simeq 0.148\,.
\end{equation}
For larger values of $\mu$, no positive roots exist, which means that no AdS$_4$ vacuum exists in that case\footnote{This analysis is analogous to the one corresponding to QTG in $D\geq 5$ \cite{Quasi,Myers:2010jv}, with the difference that, in that case, the Gauss-Bonnet term is present, and the identification of the allowed stable vacua becomes more involved.}. However, not all real roots of \req{roo} satisfying \req{sis} give rise to stable vacua.

In order to see this, we can consider the linearized equations of motion of \req{ECG} on a general maximally symmetric background (in particular, one of these AdS$_4$), in the presence of minimally coupled fields. As already mentioned, these always reduce to the linearized equations of Einstein gravity, up to a normalization of the  Newton constant \cite{PabloPablo,Aspects}, namely 
\begin{equation}
G_{\mu\nu}^{ \ssc L}=8 \pi G_{\rm eff}^{\rm ECG} T_{\mu\nu}\, ,
\end{equation}
where $G_{\mu\nu}^{\ssc L}$ is the linearized Einstein tensor, $T_{\mu\nu}$ is the stress tensor of the extra fields, and $G_{\rm eff}^{\rm ECG}$ is the effective Newton's constant, which is given by
\begin{equation}\label{Geff}
G_{\rm eff}^{\rm ECG}=\frac{G}{1-3\mu f_{\infty}^2}\, .
\end{equation}
The sign of $G_{\rm eff}$ determines the sign of the graviton propagator. Whenever the denominator in the right-hand side --- which is nothing but (minus) the slope of \req{roo} --- is negative, the graviton becomes a ghost, and the corresponding vacuum is unstable. This imposes $\mu<0$ or $f_{\infty}^2< 1/(3\mu)$ for positive values of $\mu$. The condition kills one of the two positive roots of \req{roo} available for $0<\mu<4/27$, which would then correspond to unstable vacua. Hence, we conclude that, whenever \req{sis} is satisfied, there exists a single stable vacuum. No additional vacua exist for $\mu<0$, whereas an additional unstable vacuum exists for $0<\mu<4/27$. Special comment deserves the $f_{\infty}^2=1/(3\mu)$ case, corresponding to $\mu=4/27$, and for which $G_{\rm eff}\rightarrow +\infty$. This `critical' limit of the theory was identified in \cite{Feng:2017tev}, and gives rise to a considerable simplification of most calculations, as we further illustrate below.
\begin{figure}[t!]
	\centering 
	\includegraphics[width=0.65\textwidth]{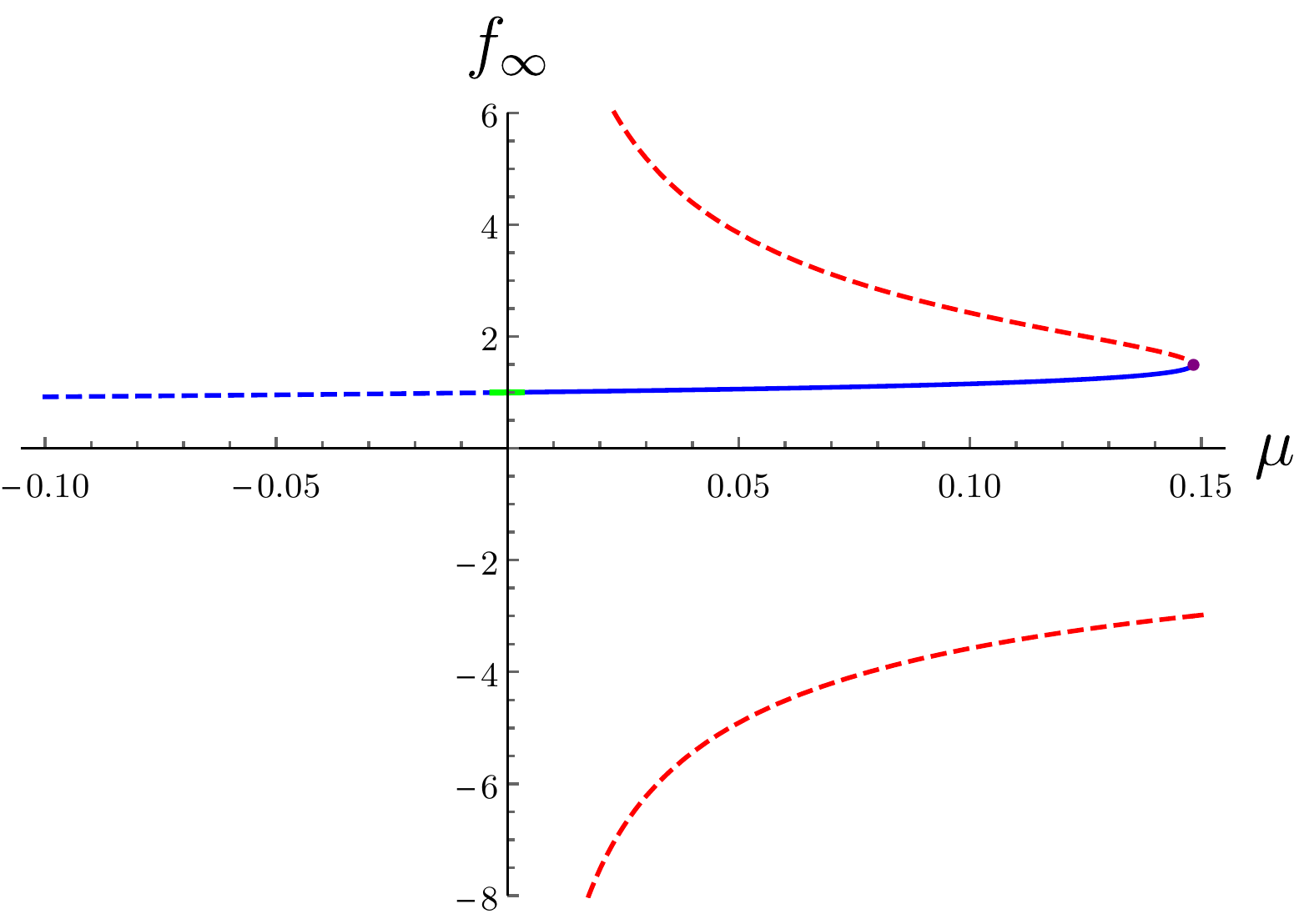}
	\caption{Real roots of \req{roo} for different values of $\mu$. The lower red dashed line corresponds to $f_{\infty}<0$, whereas the upper one corresponds to unstable vacua; the blue dashed line ($\mu<0$) corresponds to stable vacua which do not allow for black hole solutions --- see discussion under \req{hh}; the purple dot corresponds to the critical case, $\mu=4/27$; finally, the small green region corresponds to the set of parameters allowed by the positive-energy constraint $|t_4|\leq 4$ in \req{fifif}. }\label{ffffi}
\end{figure}

We summarize these observations in Fig. \ref{ffffi}, where we also include two additional constraints which we derive in Sections \ref{BHs} and \ref{t44}, respectively. The first comes from imposing the existence of black holes solutions, which restricts the allowed values to $0\leq \mu \leq 4/27$. The second follows from the positivity of energy fluxes at null infinity which, as we can see from the figure, produces the very stringent constraint, $-0.00322\leq \mu \leq 0.00312$.

Throughout this chapter, we will assume $\mu$ to lie in the range $0\leq \mu \leq 4/27$. From the two positive roots of \req{roo} in that range, we will be implicitly choosing the one corresponding to a stable vacuum, which is also the one connected to the Einstein gravity vacuum for $\mu\rightarrow 0$. While the positive-energy condition further limits this range, we find it convenient to also consider values close to $\mu=4/27$, for which many exact results can be obtained. Let us finally point out that the solution of \req{roo} corresponding to the relevant root (blue in Fig. \ref{ffffi}) can be written explicitly as
\begin{equation}\label{finfs}
f_{\infty}=\frac{2}{\sqrt{3\mu}}\sin \left[\frac{1}{3}\arcsin \left(\sqrt{\frac{27\mu}{4}} \right) \right]\, .
\end{equation}

%As we explain in the next subsection, additional bounds on $\mu$ can be obtained from 

\section{AdS$_4$ black holes}
\label{BHs}
ECG admits static asymptotically AdS$_4$ black holes of the form
\begin{equation}\label{bhss}
ds^2=-N^2 V_k(r) dt^2+\frac{dr^2}{V_k(r)}+\frac{r^2}{L^2} d\Sigma_k^2\, ,\quad \text{where} \quad d\Sigma_k^2= \begin{cases}
L^2 d\Omega^2_2\, ,\quad &\text{for}\quad k=+1\, , \\
d\vec{x}_2^2\, ,\quad &\text{for}\quad k=0\, ,\\
L^2 d\Xi^2\, ,\quad &\text{for}\quad k=-1\, ,\\
\end{cases}
\end{equation}
corresponding to spherical, planar and hyperbolic horizons, respectively, and where $V_k(r)$ is determined from the second-order differential equation
\begin{equation}\label{eqVs}
1-\frac{L^2 (V_k-k)}{r^2}-\frac{3L^6\mu}{4r^3} \left[\frac{V_k'^3}{3}+\frac{k V_k'^2}{r}-\frac{2 V_k (V_k-k)V'_k}{r^2}-\frac{V_k V_k''(rV'_k  -2 (V_k-k))}{r} \right]=\frac{\omega^3}{r^3}\,,
\end{equation}
where $\omega^3$ is an integration constant related to the ADM energy \cite{Abbott:1981ff,Deser:2002jk} of the solution --- see \req{adm}. Also, $N^2$ is a constant that we fix in different ways depending on the horizon geometry, \eg \cite{Quasi,Myers:2010jv,HoloRen}. In particular, we will choose $N^2=1$ for spherical horizons, $N^2=1/f_{\infty}$ for planar horizons, which sets the speed of light in the dual theory to one, and $N^2=L^2/(f_{\infty}R^2)$ for hyperbolic horizons, so that the boundary metric is conformally equivalent to that of $\mathbb{R}\times \mathbb{H}^2$, where $R$ is the curvature scale of the hyperbolic slices. 

The fact that ECG admits static solutions of the form \req{bhss}, characterized by a single function $V_k(r)$, such that the full nonlinear equations\footnote{These can be found explicitly in Eq.~\req{eq:fe}.}  of the theory reduce to a single third-order differential equation, which can in turn be integrated once to yield \req{eqVs}, is a highly non-trivial property of ECG \cite{Hennigar:2016gkm,PabloPablo2}. This property is shared by the higher-dimensional Lovelock \cite{Wheeler:1985nh,Wheeler:1985qd,Boulware:1985wk,Cai:2001dz,Dehghani:2009zzb}, QTG \cite{Quasi2,Quasi,Dehghani:2011vu,Cisterna:2017umf} (for these, the equation for $V_k(r)$ is algebraic instead) and Generalized quasi-topological \cite{Hennigar:2017ego} gravities, as well as by other higher-curvature theories of the same class, recently discovered and characterized \cite{PabloPablo4,Ahmed:2017jod}. As mentioned before, this property is related to the absence of extra modes in the linearized spectrum of the theory, and can be shown to lead to non-hairy black holes whose thermodynamic properties can be computed analytically on general grounds \cite{PabloPablo3}.

In \req{bhss}, it is customary to make the redefinition 
\begin{equation}
V_k(r)=k+\frac{r^2}{L^2}f(r)\, ,
\end{equation}
especially when dealing with the planar and hyperbolic cases. In terms of $f(r)$, \req{eqVs} reads
\begin{equation}\label{eqsf}
1-f+\mu \left[f^3+\frac{3}{2}r^2 f f'^2-\frac{r^3}{4}f' (f'^2-3f f'')+\frac{3}{4}k L^2 f' (rf''+3 f')\right]=\frac{\omega^3}{r^3}\, .
\end{equation}
Observe that this reduces to \req{roo} for constant $f(r)$ and $\omega^3=0$. In particular, asymptotically, we require $\lim_{r\rightarrow+\infty}f(r)=f_{\infty}$, which then makes \req{bhss} become the metric of pure AdS$_4$ with radius $\tilde{L}$ given by \req{adsc}, and a different boundary geometry for each value of $k$ \cite{Emparan:1999pm}.

%There are two cases in which we can solve \req{eqsf} analitically. These correspond to Einstein gravity ($\mu=0$), and to the critical theory ($\mu=4/27$). For those, one finds
%\begin{equation}
%f(r)=\begin{cases}
%1-\frac{k L^2\rh+\rh^3}{r^3}\quad {\rm \, \, if}\quad \mu=0\, ,\\
%\frac{3}{2}-\frac{k L^2+3/2\rh^2}{r^2}\quad {\rm if}\quad \mu=4/27\, .\\
%\end{cases}
%\end{equation}
%For intermediate values of $\mu$, the solutions can be thought of as interpolating between these two limits. 

\subsection{Asymptotic expansion}
For general values of $\mu$, finding analytic black hole solutions of \req{eqsf} looks extremely challenging (if not impossible). Let us then start by exploring the asymptotic and near horizon expansions, from which we can gain a lot of relevant information (and, in fact, argue that non-hairy black hole solutions do really exist for general values of $\mu$).

The first terms in the asymptotic expansion of $f(r)$ read
%In order to solve \req{eqsf}, we need to impose appropriate boundary conditions. Let us first consider the limit $r\rightarrow \infty$, where we impose that the solution is asymptotically AdS$_4$, this is, $\lim_{r\rightarrow\infty}f(r)=f_{\infty}$. In order to study this limit, we can work out an asymptotic expansion of the function $f$ in powers of $1/r$. For the first terms we get
\begin{equation}\label{Asympt}
f_{1/r}(r)=f_{\infty}-\frac{\omega^3}{(1-3\mu f_{\infty}^2)r^3}-\frac{21\mu f_{\infty}\omega^6}{2(1-3\mu f_{\infty}^2)^3r^6}+\mathcal{O}(r^{-8})\, .
\end{equation}
Note that \req{eqsf} is a second-order differential equation, which therefore possesses a two-parameter family of solutions. In order to capture the asymptotic behavior of the most general one, we write $f(r)=f_{1/r}(r)+h(r)$ and then expand \req{eqsf} linearly in $h$. Keeping only leading terms in $1/r$, we get the following equation for $h$\footnote{For instance, we assume that the term $h' L^2r^{-4}$ is negligible compared to $h'' r^{-1}$ when $r\rightarrow +\infty$.}:
\begin{equation}\label{heq}
h''(r)-\frac{4(1-3\mu f_{\infty}^2)^2}{9f_{\infty}\mu \omega^3} r h(r)=0\, .
\end{equation}
Leaving aside the limiting cases, corresponding to $\mu=0$ and $\mu=4/27$, we see that there are two possibilities, depending on the sign of $\mu \cdot \omega^3$.
If $\mu\cdot \omega^3>0$, \req{heq} has the following approximate solutions as $r\rightarrow+ \infty$\footnote{The exact solution of \req{heq} is given by the Airy functions,
\begin{equation*}
h(r)=A \textrm{AiryAi}\left[\left(\frac{4(1-3\mu f_{\infty}^2)^2}{9f_{\infty}\mu \omega^3} \right)^{1/3}r\right]+ B\textrm{AiryBi}\left[\left(\frac{4(1-3\mu f_{\infty}^2)^2}{9f_{\infty}\mu \omega^3} \right)^{1/3}r\right]\, ,
\end{equation*}
but we only need the asymptotic behavior for the discussion.}:
\begin{equation}\label{expA}
h(r)\sim A \exp\left[\frac{4|1-3\mu f_{\infty}^2|}{9\sqrt{f_{\infty}\mu\cdot \omega^3}}r^{3/2}\right]+B\exp\left[-\frac{4|1-3\mu f_{\infty}^2|}{9\sqrt{f_{\infty}\mu\cdot \omega^3}}r^{3/2}\right]\, .
\end{equation}
In order to obtain an asymptotically AdS$_4$ solution, we need to kill the growing mode, which forces us to set $A=0$. Therefore, this boundary condition fixes one of the integration constants required by \req{eqsf}. Now, even though the remaining exponentially decaying term is extremely subleading, in general we will have $B\neq 0$. In fact, this constant ends up being fixed by the horizon-regularity condition. In particular, this implies that the solutions show a strongly nonperturbative character, as   $\sim e^{-1/\sqrt{\mu}}$ terms generically appear. Indeed, it is possible to show that a series expansion of the full solution in powers of $\mu$ is always divergent.  

The second possibility corresponds to $\mu\cdot \omega^3<0$. An approximate solution of \req{heq} for large $r$ is then given by
\begin{equation}\label{hh}
h(r)\sim \frac{A}{r} \cos\left[\frac{4|1-3\mu f_{\infty}^2|}{9\sqrt{f_{\infty}|\mu\cdot \omega^3|}}r^{3/2}\right]+\frac{B}{r}\sin\left[\frac{4|1-3\mu f_{\infty}^2|}{9\sqrt{f_{\infty}|\mu\cdot \omega^3|}}r^{3/2}\right]\, .
\end{equation}
This solution is sick. Although $h(r)\rightarrow 0$ as $r\rightarrow +\infty$, the derivatives of $h$ diverge wildly in this limit, which would force us to set $A=B=0$ in order to get an asymptotically AdS$_4$ solution. However, this leaves us with no additional free parameters, and regularity at the (would-be) horizon cannot be imposed. Therefore, no regular black hole solution exists for $\mu\cdot \omega^3<0$: the solution is always sick, either at the horizon or at infinity. 

As shown later in \req{adm}, $\omega^3$ is proportional to the total energy $E$ (or mass) of the black hole, which leads us to impose $\mu\ge 0$. Hence, interestingly, the range of values of $\mu$ which allows for positive-energy solutions, forbids the negative-energy ones, which simply do not exist for $\mu\ge 0$.
%, in order for $f_{\infty}\mu \omega^3$ to be positive in \req{expA}. 
%\comment{This will imply that solutions with negative energy will not exist. It is interesting that in this theory negative energy configurations are forbidden by dynamics.} 

\subsection{Near-horizon expansion}
Let us now consider the near-horizon behavior. For that, we assume that there is a value $\rh$ of the radial coordinate for which the function $V_k$ vanishes and is analytic. Analyticity ensures that the solution can be maximally extended beyond the horizon using Kruskal-Szekeres-like coordinates. 

The derivative of $V_k$ at the horizon is related to the temperature through: $V_k'(\rh)=4\pi T/N$ so, in terms of $f$, the near-horizon expansion can be written as
\begin{equation}\label{nH}
k+\frac{r^2}{L^2}f(r)=\frac{4\pi T}{N}(r-\rh)+\sum_{n=2}^{\infty} a_n (r-\rh)^n\, ,
\end{equation}
where the relation between $f'(r)$ and the temperature reads in turn
\begin{equation}
T=\frac{N}{4\pi}\left[\frac{\rh^2}{L^2}f'(\rh) -\frac{2k}{\rh}\right]\,.
\end{equation}
Note also that $f(\rh)=-k L^2/\rh^2$. Now, if we plug \req{nH} into \req{eqsf} and we expand it in powers of $(r-\rh)$, we are led to the equation 
\begin{align}
0= 1+\frac{k L^2}{\rh^2}-\frac{\omega^3}{\rh^3}-\frac{4L^6\pi^2 T^2 \mu}{N^2 \rh^3}\left(\frac{3k}{\rh}+\frac{4\pi T}{N}\right)+\, \\
\left[-\frac{2kL^2}{\rh^2}+\frac{3\omega^3}{\rh^3}-\frac{4L^2\pi T}{N \rh}+\frac{24L^6\pi^2 T^2 \mu}{N^2 \rh^3}\left(\frac{k}{\rh}+\frac{2\pi T}{N}\right)\right](r-\rh)
+\mathcal{O}\left( (r-\rh)^2\right)&\, .
\end{align}
Since every coefficient must vanish independently, we get an infinite number of equations relating the parameters in the near-horizon expansion \req{nH}.
From the first two equations, we can obtain $T^{\rm ECG}$ and ${\omega}^{\rm ECG}$ as functions of $\rh$, the result being (in order to minimize the clutter, we often omit the `ECG' superscripts throughout the text)
\begin{align}\label{T}
T^{\rm ECG}&=\frac{N}{2\pi \rh} \left(k+\frac{3\rh^2}{L^2}\right)\left[1+\sqrt{1+\frac{3k L^4\mu}{\rh^4}\left(k +3\frac{\rh^2}{L^2}\right)} \right]^{-1} \, ,\\
{(\omega^{\rm ECG})}^3&=kL^2\rh+\rh^3-\frac{\mu L^6}{4}\left[\frac{3k}{\rh}\left(\frac{4\pi T^{\rm ECG}}{N}\right)^2+\left(\frac{4\pi T^{\rm ECG}}{N}\right)^3\right]\,.
\end{align}
These reduce to the usual Einstein gravity results for $\mu=0$, namely
\begin{align}
T^{\rm  E}=\frac{N}{4\pi \rh} \left(k+\frac{3\rh^2}{L^2}\right)\, ,\quad
{(\omega^{\rm  E})}^3=\rh^3+k \rh L^2\,.
\end{align}
The rest of equations, which we do not show here, fix all coefficients $a_{n>2}$ in terms of $a_2$. Hence, for a fixed $\rh$, the series \req{nH} contains a single free parameter, which is nothing but the value of $f''$ at the horizon. This  must be carefully chosen so that the solution has the appropriate asymptotic behavior, \ie so that $A=0$ in \req{expA}.

\subsection{Full solutions}\label{fullsol}
Equation \req{eqsf} can be solved analytically in two cases, namely: for Einstein gravity, $\mu=0$, and in the critical limit, $\mu=4/27$ \cite{Feng:2017tev}. For those, one finds\footnote{\label{BTZ}A curious property of the critical-theory solutions is that they look identical to three-dimensional BTZ black holes \cite{Banados:1992wn}, with an additional `angular' direction:
\begin{eqnarray}
ds^2_{\rm  ECG,\, crit}&=&-\frac{3(r^2-\rh^2)}{2L^2}dt^2-\frac{2 L^2 dr^2}{3(r^2-\rh^2)}+\frac{r^2}{L^2} d\Sigma_{(k)}^2\, ,\\
ds^2_{\rm BTZ}&=&-\frac{(r^2-\rh^2)}{L^2}dt^2-\frac{L^2 dr^2}{(r^2-\rh^2)}+r^2d\phi^2\, .
\end{eqnarray}
 We point out that an analogous behavior has been observed to occur for critical Gauss-Bonnet gravity ($\lambda_{\rm \ssc GB}=1/4$), see \eg \cite{Grozdanov:2016fkt} as well as for Einstein gravity coupled to an axionic field in a particular limit \cite{Davison:2014lua}. The connection of this phenomenon to other instances of background-symmetry enhancement --- \eg \cite{Compere:2012jk} --- deserves further attention.} 
\begin{equation}\label{limits}
f(r)=\begin{cases}
\displaystyle 1-\frac{\rh^3+k L^2\rh}{r^3}\quad {\rm\,\, if}\quad \mu=0\, ,\\
\displaystyle \frac{3}{2}-\frac{3\rh^2+2k L^2}{2r^2}\quad {\rm if}\quad \mu=4/27\, .\\
\end{cases}
\end{equation}
For intermediate values of $\mu$, the solutions can be constructed numerically. In order to do so, we solve \req{eqsf} setting the initial condition at the horizon, and then applying the shooting method to obtain the value of $a_2$ for which $f(r)\rightarrow f_{\infty}$. The differential equation \req{eqsf} is very stiff when $r$ is large but, by choosing $a_2$ accurately, it is always possible to extend the numerical solution well into the region in which the asymptotic expression \req{Asympt} applies. In all cases, there is a unique value of $a_2$ for which this happens. Hence, for each value of $\mu$ and each horizon geometry, there exists a unique regular  black fully characterized by $\rh$ (or, more physically, by $\omega^{\rm ECG}$).

\begin{figure}[t!]
	\centering 
	\includegraphics[width=0.55\textwidth]{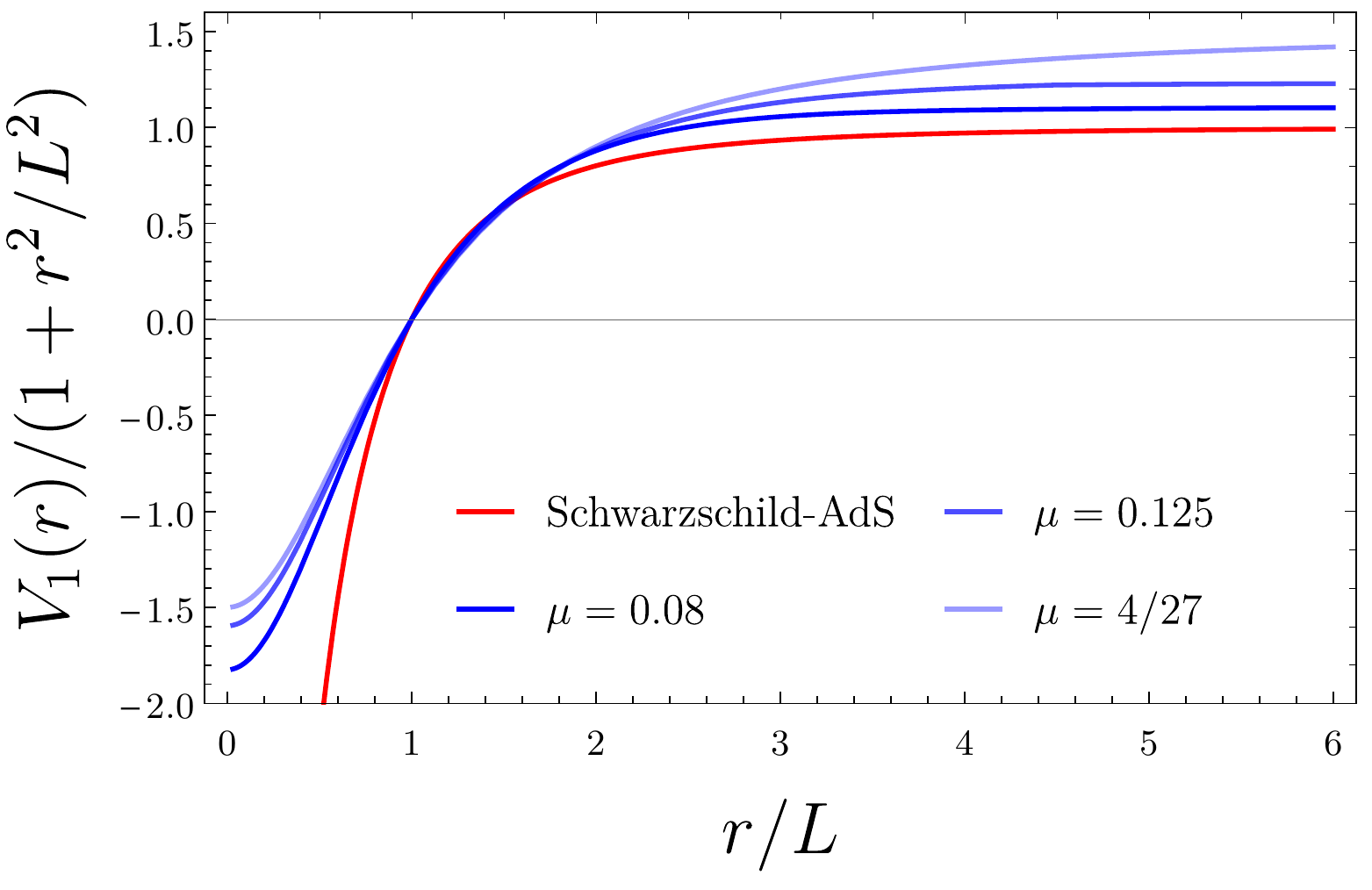}
	\includegraphics[width=0.55\textwidth]{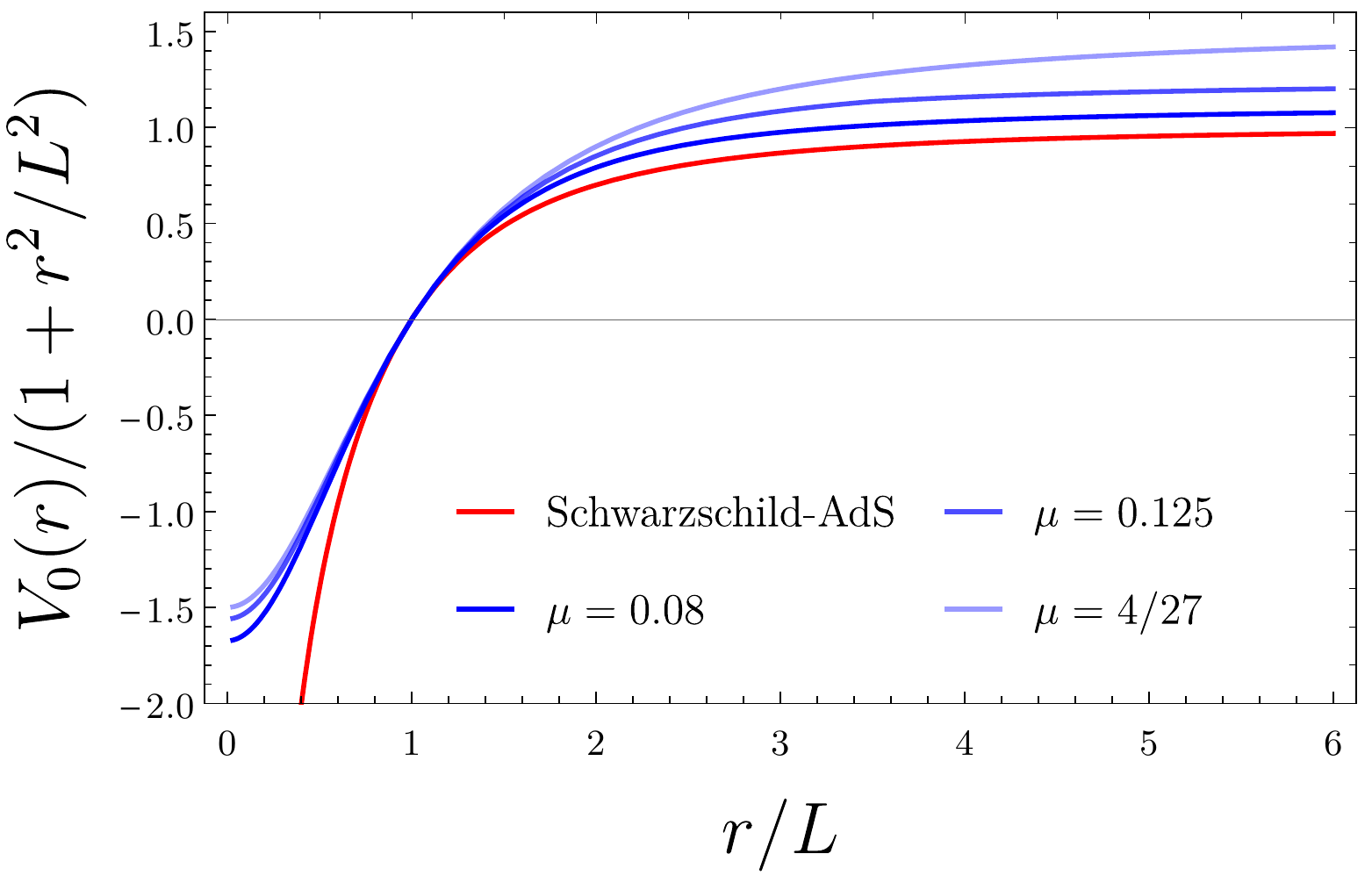}
	\includegraphics[width=0.55\textwidth]{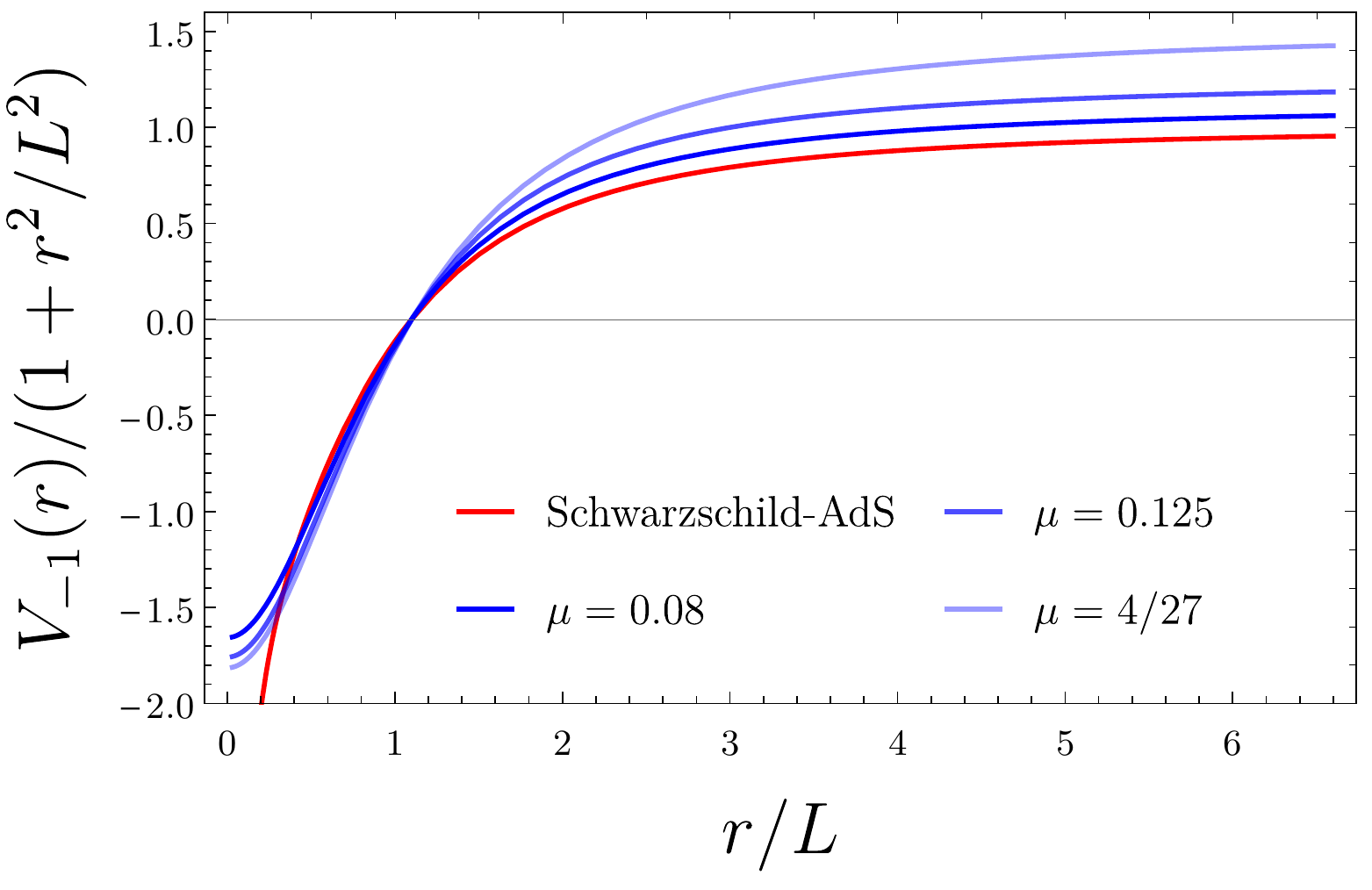}
	\caption{Black hole solutions for several values of $\mu$  including the Einstein gravity ($\mu=0$) and critical ($\mu=4/27$) cases. From top to bottom $k=1,0,-1$, and we have chosen $r_h=L$ for the spherical and planar cases and $r_h=1.1 L$ for the hyperbolic one. For the sake of clarity, we plot the ratio $V_k(r)/(1+r^2/L^2)$, which asymptotically tends to $f_{\infty}$ and at the same time allows to observe the behavior of $V_k(r)$ near $r=0$.}\label{fPlot1}
\end{figure}

In Fig. \ref{fPlot1} we show a couple of these numerical solutions
for a fixed value of the radius $r_h$. As we can see, the corresponding curves lie between the analytic limiting solutions in \req{limits}. For clarity reasons, we plot the quantity $V_k(r)/(1+r^2/L^2)=(k+r^2f(r)/L^2)/(1+r^2/L^2)$ instead of $f(r)$. Far from the horizon, this ratio behaves as $f(r)$ and it tends to the constant value $f_{\infty}$, which, as explained above, is different for each value of $\mu$ --- see Fig. \ref{ffffi}. Besides the exterior solutions, we also show plots of the black hole interior profiles, which present the curious feature of being regular, in the sense that the metric functions $V_k(r)$ are smooth. However, as observed in \cite{PabloPablo2} for the asymptotically flat case, curvature invariants still diverge. For example, in the critical case, one finds
\begin{equation}
R_{\mu\nu\rho\sigma}R^{\mu\nu\rho\sigma}=\frac{4k^2L^4+54r^4-6\rh^2 r^2+\rh^4-4kL^2(3r^2-\rh ^2)}{L^4r^4}\sim \mathcal{O}\left(r^{-4} \right)\, , %\quad R^2=\frac{(r_h^2+2kL^2-18r^2)^2}{L^4 r^4}\, ,
\end{equation}
 which is two powers of $r$ softer than in the usual Schwarzschild case. Such behavior is common to all solutions with $\mu\neq 0$. This singularity-softening phenomenon appears to be generic for higher-curvature generalizations of Einstein gravity black holes. For example, for the Gauss-Bonnet black hole \cite{Cai:2001dz}, one finds \cite{Ohta:2010ae} $R_{\mu\nu\rho\sigma}R^{\mu\nu\rho\sigma}\sim \mathcal{O}(r^{-{(D-1)}})$, which is in turn $(D-1)$ powers of $r$ softer than the Kretschmann invariant of the $D$-dimensional Schwarzschild black hole.

\section{Generalized action for higher-order gravities}\label{osa}

%$a^*$ The  charge appearing in the universal contribution to the entanglement entropy across a $S^{D-3}$ spherical region both in odd and even-dimensional theories\footnote{$a^*$ is an important quantity, as it reduces to the trace anomaly charge $a$ in even dimensions, and \comment{F theorem, bla, bla in odd $(D-1)$}. }. In particular, it was been argued \comment{it is proven for $(D-1)$ even and arguments for $(D-1)$ odd and more to come in the holoAspects of ECG paper} that for a general holographic higher-order gravity theory, this is given by\comment{check signs}
%\begin{equation}
%	a^*=\frac{\pi^{\frac{(D+1)}{2}} \tilde{L}^D}{2\pi \Gamma\left[\frac{D+1}{2}\right]}  \mathcal{L}|_{\rm AdS}\, .
%\end{equation}

When performing holographic calculations with higher-curvature bulk duals, one is faced with the challenge of identifying appropriate boundary terms which render the action differentiable, as well as counterterms which, along with those, give rise to finite and well-defined on-shell actions, when evaluated on stationary points of the functional. In this section, we propose a novel prescription for computing the on-shell action of arbitrary asymptotically AdS solutions of any $D$-dimensional higher-order gravity whose linearized spectrum on a maximally symmetric background matches that of Einstein gravity\footnote{\label{ELC}This property defines the `Einstein-like' class in the classification of \cite{Aspects}, and includes, in particular: Lovelock, QTG, ECG in general $D$ and, more generally, all theories of the Generalized QTG type. Additional examples of theories of this type can be found \eg in \cite{Karasu:2016ifk,Love,Li:2017ncu,Li:2017txk}.}. The procedure represents an important simplification with respect to previous methods, as it only makes use of the usual Gibbons-Hawking-York boundary term and the counterterms of Einstein gravity. As we argue here --- and illustrate throughout the rest of this chapter and the next one, and in Appendix \ref{BTcheck} with various non-trivial checks of the proposal --- such contributions can be also used to produce the correct on-shell actions for this class of higher-order theories. Interestingly, for those, the only modification with respect to the Einstein gravity case is that such contributions appear weighted by the Lagrangian of the corresponding theory evaluated on the AdS background, \ie $\mathcal{L}
|_{{\rm AdS}}$. This quantity has been argued to be proportional to the charge $a^*$ appearing in the universal contribution to the entanglement entropy of the dual theory across a $\mathbb{S}^{d-2}$, and our prescription can be used to actually prove such a connection explicitly for this class of theories, as we show below. %In the discussion section, we also speculate on the possible implications of our result in the context of the complexity$=$action paradigm. 
%\comment{Allows for calculations of Quasi-topological on-shell actions for non-planar horizons}

%In holographic calculations, one often needs to compute on-shell actions

% \comment{often needs to compute on-shell actions. Prescription for Einstein gravity includes  }

Let us start considering a general higher-curvature theory of the form
\begin{equation}\label{geng2}
I=\int_{\mathcal{M}}d^Dx\sqrt{|g|}\mathcal{L}(g^{\alpha\beta},R_{\mu\nu\rho\sigma})\, ,
\end{equation}
where the Lagrangian density $\mathcal{L}(g^{\alpha\beta},R_{\mu\nu\rho\sigma})$ is assumed to be constructed from arbitrary contractions of the Riemann and metric tensors. The variation of the action with respect to the metric yields
\begin{equation}
\delta I=\int_{\mathcal{M}}d^Dx\sqrt{|g|}\E_{\mu\nu}\delta g^{\mu\nu}+\epsilon \int_{\partial\mathcal{M}}d^{D-1}x\sqrt{|h|}n_{\mu}\delta v^{\mu}\, .
\label{var27}
\end{equation}
In this expression we defined
%The equations of motion of the theory take the form $\mathcal{E}_{ab}=0$, where
\begin{equation}
\mathcal{E}_{\mu\nu}\equiv P_{\mu}\,^{\alpha\beta\sigma}R_{\nu \alpha\beta\sigma}-\frac{1}{2}g_{\mu\nu}\mathcal{L}+2\nabla^{\alpha}\nabla^{\beta}P_{\mu \alpha\nu\beta}\, , %=\frac{1}{2}T_{ab}\, .
\label{fieldequations6}
\end{equation}
the equations of motion reading $\mathcal{E}_{\mu\nu}=0$, and
\begin{equation}
\delta v^{\mu}=2\tensor{P}{_{\alpha}^{\beta \mu \sigma}}\nabla^{\alpha}\delta g_{\beta\sigma}\, , \quad \text{where}\quad P^{\mu\nu\rho\sigma}\equiv \left[\frac{\partial \mathcal{L}}{\partial R_{\mu\nu\rho\sigma}}\right]_{g^{\alpha\beta}}\, .\label{Ptensor6}
\end{equation}
In addition, $n^{\mu}$ is the unit normal to $\partial \mathcal{M}$, normalized as $n^{\mu}n_{\mu}\equiv \epsilon=\pm 1$, and  $h_{\mu\nu}=g_{\mu\nu}-\epsilon n_{\mu}n_{\nu}$ is the induced metric. In order to have a well-posed variational problem, the action must be differentiable, in the sense that $\delta I\propto \delta g^{\mu\nu}$, so that $\delta I=0$ whenever the field equations --- and the boundary conditions --- are satisfied. %This requirement is essential if we want to define a path integral for quantum gravity, since the action must be stationary for solutions of the classical equations of motion. 
This is not the case of \req{var27}, due to the presence of the boundary contribution. In the case of Einstein gravity, $\mathcal{L}^{\rm E}=\left[R+(D-1)(D-2)/L^2\right]/(16\pi G)$, this problem is solved by the addition of the Gibbons-Hawking-York term \cite{York:1972sj,Gibbons:1976ue},
\begin{equation}
I_{\rm GHY}=\frac{\epsilon}{8\pi G}\int_{\partial\mathcal{M}}d^{D-1}x\sqrt{|h|}K\, ,
\end{equation}
where $K=K_{\mu\nu}g^{\mu\nu}$ is the trace of the second fundamental form of the boundary, $K_{\mu\nu}=h_{\mu}^{\ \alpha} \nabla_{\alpha} n_{\nu}$. When this term is included, the variation of the action, when we keep the induced metric $h_{ab}$ fixed at the boundary, reads\footnote{The indices $a,b$ are intrinsic boundary indices. Note that the conditions $\delta h_{ab}=0$ and $\delta h_{\mu\nu}=0$ are not equivalent \cite{Poisson:2009pwt}.}
\begin{equation}
\delta(I^{\rm E}+I_{\rm GHY})\Big|_{\delta h_{ab}|_{\partial\mathcal{M}}=0}=\frac{1}{16 \pi G}\int_{\mathcal{M}}d^4x\sqrt{|g|}\left[R_{\mu\nu}-\frac{1}{2}g_{\mu\nu}\mathcal{L}_{\rm EH}\right]\delta g^{\mu\nu}\, ,
\end{equation} 
 and so the action is stationary whenever the metric satisfies Einstein's field equations.

For higher-order gravities, the situation is much more involved in general.
 One of the main issues arises from the fact that these theories generally possess fourth-order equations of motion. This implies that the boundary-value problem is not fully specified by the induced metric on $\partial\mathcal{M}$, and one needs to impose additional boundary conditions on derivatives of the metric. Furthermore, even if we know which components of the metric and its derivatives to fix, determining what boundary term needs to be added to yield a differentiable action for such variations is a far from trivial task. Some notable examples for which differentiable actions have been constructed are: quadratic gravities (perturbatively in the couplings) \cite{Smolic:2013gz}, Lovelock gravities \cite{Teitelboim:1987zz,Myers:1987yn}, which are the most general theories with second-order covariantly-conserved field equations \cite{Lovelock1,Lovelock2} (and for which one only needs to fix $g_{\mu\nu}$ at the boundary), $f(R)$ \cite{Madsen:1989rz,Dyer:2008hb,Guarnizo:2010xr} and, more generally, $f($Lovelock$)$ gravities \cite{Love}. In these cases, it is also necessary to fix the value of some of the densities on the boundary --- \eg $\delta R\big|_{\partial\mathcal{M}}=0$ for $f(R)$ --- which is related to the fact that these theories propagate additional scalar modes. With the goal of providing a canonical formulation for arbitrary $f$(Riemann$)$ gravities, an interesting proposal for constructing satisfactory boundary terms for such general class of theories was presented in \cite{Deruelle:2009zk} --- see also \cite{Teimouri:2016ulk}.
 %In \cite{Deruelle:2009zk}, an interesting proposal for constructing satisfactory boundary terms for general $f$(Riemann$)$ gravity was presented --- see also \cite{Teimouri:2016ulk}. 
 Unfortunately, the procedure involves the introduction of auxiliary fields and it is quite implicit in general, which seems to limit its practical applicability in the holographic framework.
 % it is not clear how to understand it in a pure metric formalism.

The problem can be simplified if we specify the boundary structure in advance, \eg by restricting the analysis to spacetimes which are maximally symmetric asymptotically. %This approach was succesfully followed, \eg in \cite{} for \comment{bla bla Loevloek...}. 
%he problem of finding the boundary term which yields the action stationary is a difficult task, but we can relax the problem by specifying the boundary conditions on advance, and by finding a boundary term which yields the action stationary for those conditions. In particular, the most natural assumption is that the spacetime is asymptotically maximally symmetric, \textit{i.e.}, Minkowski, de Sitter or anti de Sitter. 
Let us, in particular, assume that the space is asymptotically AdS$_D$, so that the Riemann tensor behaves as $R_{\mu\nu\rho\sigma}\rightarrow -\tilde L^{-2} (g_{\mu\rho}g_{\nu\sigma}-g_{\mu\sigma}g_{\nu\rho})$ asymptotically. Then, on general grounds, the tensor $\tensor{P}{_{\mu\nu}^{\rho\sigma}}$ appearing in the boundary term in \req{var27} takes the simple form 
\begin{equation}\label{pp}
\tensor{P}{_{\mu\nu}^{\rho\sigma}}\rightarrow C(\tilde L^2)\delta_{[\mu}^{\ \rho}\delta_{\nu]}^{\ \sigma}+\text{subleading}\, ,
\end{equation}
where $C(\tilde L^2)$ is a constant which depends on the background curvature, and is in general given by\footnote{As shown in \cite{Aspects}, this quantity can be equivalently written as \begin{equation}
	C(\tilde L^2)=\frac{\tilde L^4}{D(D-1)}\frac{d\mathcal{L}
		|_{{\rm AdS}}}{d\tilde L^2}\, ,
	\end{equation}
	the relation between both expressions being nothing but the embedding equation of AdS$_D$ in the corresponding theory --- \eg \req{roo} for ECG.
	 } \cite{Aspects}
\begin{equation}\label{cdd}
C(\tilde L^2)=-\frac{\tilde L^2}{2(D-1)}\mathcal{L}
|_{{\rm AdS}}\, ,
\end{equation}
%\begin{equation}
%C(\tilde L^2)=\frac{\tilde L^4}{D(D-1)}\frac{d\mathcal{L}(\tilde L^2)}{d\tilde L^2}\, ,
%\end{equation}
%\comment{$C(\Lambda)$ es proporcional al coeficiente universal de la EE para una superficie esf'erica en cualquier n'umero de dimensiones (en dimensiones impares, este coincide con la free energy de la teor'ia puesta en una esfera, y en dimensiones pares coincide con el $a$-term de la Weyl anomaly)! Funciona para ECG y tambi'en para GB en dimensiones generales, en ambos casos, $C(\Lambda)$ y $F$ ambos $\propto (1-2(D-2)/(D-4))\lambda_{\ssc \rm GB} f_{\infty}$}
where $\mathcal{L}|_{{\rm AdS}}$ is the Lagrangian of the corresponding theory evaluated on the AdS$_D$ background with curvature scale $\tilde{L}$. 
%In \req{cdd}, the subleading terms will arise from 

For Einstein gravity, we simply have $C^{\rm E}=1/(16\pi G)$ and, in fact, there are no subleading terms in \req{pp} for any spacetime --- simply because $\tensor{P}{_{\mu\nu}^{\rho\sigma}}$ only involves products of deltas in that case. Now, asymptotically AdS$_D$ solutions of higher-order gravities will in general produce subleading contributions in \req{pp} as we move away from the asymptotic region. However, the leading term can still be canceled out by adding a generalized GHY term of the form
\begin{equation}\label{GGHY}
I_{\rm GGHY}=2 	C(\tilde L^2)\epsilon \int_{\partial\mathcal{M}}d^{D-1}x\sqrt{|h|}K\, .
\end{equation}
The question is, of course, whether or not the subleading terms for a given theory will give additional non-vanishing contributions asymptotically, forcing us to add extra terms. We expect this to be the case in general. In addition, one generally needs to specify extra boundary conditions, which is related to the metric propagating additional degrees of freedom. However, as we have mentioned, some theories --- see footnote \ref{ELC} ---
do not propagate additional modes on general maximally symmetric backgrounds. For those, the asymptotic dynamics is the same as for Einstein gravity, so it is reasonable to expect the only data that we need to fix on $\partial\mathcal{M}$ to be $g_{\mu\nu}$, and also that \req{GGHY} will be enough to make the action stationary for solutions of the field equations. 

% The dynamics of these theories is asymptotically the same as in Einstein gravity, up to rescaling of the gravitational constant, so we can expect that the only boundary data that we need to specify is $g_{\mu\nu}$ on $\partial\mathcal{M}$, and nothing else. In this case, we claim that this boundary term will be enough to make the action stationary for solutions of the equations of motion. \\
In order to obtain finite on-shell actions, one also needs to include counterterms, which only depend on the boundary induced metric. For asymptotically AdS$_D$ spacetimes, there is a generic way of finding them \cite{Emparan:1999pm}. Let us focus on Euclidean signature. In that case, we always have $\epsilon=+1$, and an additional global $(-)$ with respect to Lorentzian signature arises, \eg \cite{Myers:2010tj}, so we have
%
%We will work in Euclidean signature, so $\epsilon=+1$, but there is also a global change of sign in the action.
% Let us first recall the embedding equation for the AdS radius $\tilde L$ \cite{Aspects}:
%\begin{equation}
% \frac{d}{d\tilde L^2}\left(\tilde L^{D}\mathcal{L}(\tilde L^2)\right)=0
% \end{equation}
%Using this, we may rewrite the boundary term \ref{GGHY} as
%\begin{equation}
%I_{\rm GGHY}=\frac{\tilde L^2\mathcal{L}(\tilde L^2)}{(D-1)}\int_{\partial \mathcal{M}}d^{D-1}x\sqrt{h}K\, ,
%\end{equation}
%and the complete Euclidean action reads
\begin{equation}\label{SEcomplete}
I_E=-\int_{\mathcal{M}}d^Dx\sqrt{g}\mathcal{L}(g^{\alpha\beta},R_{\mu\nu\rho\sigma})-2 	C(\tilde L^2) \int_{\partial\mathcal{M}}d^{D-1}x\sqrt{|h|}K+I_{\rm GCT}\, ,
\end{equation}
where we seek to construct the generalized counterterms, $I_{\rm GCT}$. In order to identify all possible divergences, one possibility consists in evaluating the action on pure AdS$_D$ spaces with different boundary geometries \cite{Yale:2011dq}. Observe however that, whenever we evaluate the bulk term on pure AdS$_D$, this will produce an overall constant $\mathcal{L}
|_{{\rm AdS}}$, which is precisely proportional to $C(\tilde L^2)$. This already appears in front of the boundary term, and the result is that the combination of the bulk and boundary contributions reduce to those of Einstein gravity, up to a common overall $C(\tilde L^2)$. Hence, the divergences are exactly the same as for Einstein gravity, and we can use the same counterterms.
% But, anytime we evaluate the action on AdS$_D$ we get 
%\begin{equation}
%I_E=-C(\tilde L^2)\left[-\frac{2(D-1)}{\tilde L^2}\int_{\mathcal{M}}d^4x\sqrt{g}+2\int_{\partial \mathcal{M}}d^{D-1}x\sqrt{h}K\right]+I_{\rm GCT}\, .
%\end{equation}
%Now, the term between parenthesis is the same as in Einstein gravity when the AdS scale is $\tilde L$. 
%
%Therefore, the divergences of this action are the same as those of Einstein gravity times $C(\tilde L^2)$. Then, we can use the same counterterms as for EG, times this constant. 
For example, up to $D=5$ we find \cite{Emparan:1999pm,Yale:2011dq}
\begin{equation}\label{gass}
\begin{aligned}
I_{\rm GCT}&=-2C(\tilde L^2)\int_{\partial \mathcal{M}}d^{D-1}x\sqrt{h}\bigg[-\frac{D-2}{\tilde L}-\frac{\tilde L \Theta[D-4]}{2(D-3)}\mathcal{R}
%&-\frac{\tilde L^3\Theta[D-6]}{2(D-3)^2(D-5)}\left(\mathcal{R}_{ab}\mathcal{R}^{ab}-\frac{D-1}{4(D-2)}\mathcal{R}^2\right)
+\ldots\bigg]\, ,
\end{aligned}
\end{equation}
where $\Theta[x]=1$ if $x\geq 0$, and zero otherwise, and the dots refer to additional counterterms arising for $D\geq 6$. Combining \req{gass} with \req{SEcomplete}, we obtain the final form of the action.

Below, we show that \req{SEcomplete} successfully yields the right answers for ECG in various highly non-trivial situations in which the corresponding on-shell actions can be deduced from alternative considerations --- \eg it correctly computes the free energy of black holes, in agreement with the result obtained using Wald's entropy, as well as the holographic stress tensor two-point charge, $\ctt$, which can be alternatively deduced from the effective Newton's constant. Besides, in Appendix \ref{BTcheck} we consider arbitrary radial perturbations of AdS$_5$ in Gauss-Bonnet gravity, and show that \req{SEcomplete} produces exactly the same finite and divergent contributions as those obtained using the standard Gauss-Bonnet boundary term and counterterms, \eg \cite{Teitelboim:1987zz,Myers:1987yn,Emparan:1999pm, Mann:1999pc, Balasubramanian:1999re, Brihaye:2008xu,Astefanesei:2008wz}. 
%In the next subsection, we perform another check of \req{SEcomplete}.
%HEREEEE
%\comment{Y el GB de la induced metric? Clarificar qu'e pasa con los CT en general} \comment{counterterm related to conformal gravity}
%In appendix \ref{BTcheck} we compute the complete boundary contribution in \ref{SEcomplete} for Gauss-Bonnet gravity by using this method and we compare it to the actual Gauss-Bonnet boundary term. Of course, both terms coincide (by construction) when evaluated on AdS, so we consider radial perturbations and we find that both boundary terms give the same result for finite and divergent parts, meaning that they are equivalent, at least in the kind of situations considered. 

 %Boundary terms are introduced so that the variational principle is well defined 

\subsection{$a^*$ and generalized action}
Let us momentarily switch to $d\equiv D-1$ notation. As we have seen, both the boundary term and the counterterms appearing in \req{SEcomplete} have the property of being identical to those of Einstein gravity up to an overall constant $C(\tilde{L}^2)$ proportional to the Lagrangian of the corresponding theory evaluated on the AdS background \req{cdd}. Now, an interesting quantity that one would like to compute holographically is the charge $a^*$  appearing in the universal contribution to the entanglement entropy (EE) across a radius-$R$ spherical region $\mathbb{S}^{d-2}$, which, for a general CFT$_d$, is given by \cite{Myers:2010tj,Myers:2010xs,CHM}
\begin{equation}\label{asta}
S_{\rm \ssc EE\, univ.}=\begin{cases}
(-)^{\frac{d-2}{2}} 4a^* \log(R/\delta) \quad &\text{for even } d \, , \\
 (-)^{\frac{d-1}{2}}2\pi a^* \quad &\text{for odd } d\, .
\end{cases}
\end{equation}
$a^*$ coincides with the $a$-type trace-anomaly charge in even dimensional theories. In odd dimensions, $a^*$ is proportional to the free energy, $F=-\log Z$, of the corresponding theory evaluated on $\mathbb{S}^{d}$ \cite{CHM}, namely 
\begin{equation}\label{fffs}
F_{\mathbb{S}^{d}}= (-)^{\frac{d+1}{2}} 2\pi a^* \, , \quad \text{for odd } d\, .
\end{equation}
For even-dimensional holographic theories dual to any higher-order gravity of the form \req{geng2} in the bulk, $a^*$ is given by \cite{Imbimbo:1999bj,Schwimmer:2008yh}
\begin{equation}\label{astar}
a^*=-\frac{\pi^{d/2}\tilde{L}^{d+1}}{d \Gamma(d/2)}\mathcal{L}
|_{{\rm AdS}}\, ,
\end{equation}
\ie it is precisely proportional to the charge $C(\tilde L^2)$ defined in \req{cdd}, namely
\begin{equation}
C(\tilde{L}^2)=\frac{a^* }{ \Omega_{(d-1)}\tilde{L}^{d-1}} \, ,
\end{equation}
where $\Omega_{(d-1)}\equiv 2\pi^{d/2}/ \Gamma(d/2)$ is the area of the unit sphere $\mathbb{S}^{d-1}$.
For odd-dimensional theories, it was argued in \cite{Myers:2010tj,Myers:2010xs} that \req{astar} also yields the right $a^*$ for general cubic theories. We can readily extend this result to all theories for which \req{SEcomplete} and \req{gass} hold. From \req{fffs}, it follows that $(-)^{\frac{d+1}{2}} 2\pi a^*$ can be obtained from the on-shell action of pure Euclidean AdS$_{(d+1)}$ with boundary geometry $\mathbb{S}^d$. Since $C(\tilde{L}^2)$ appears as an overall factor in \req{SEcomplete} when evaluated in pure AdS, it follows that $F_{\mathbb{S}^{d}}$ matches the Einstein gravity result up to an overall factor $16 \pi G\cdot  C(\tilde L^2)$. Then, using the result for the free energy in Einstein gravity,
\begin{equation}
F^{\rm E}_{\mathbb{S}^{d}}= (-)^{\frac{d+1}{2}}\frac{\pi^{d/2}\tilde{L}^{d-1} }{4\Gamma(d/2)G}\, ,
\end{equation}
it follows immediately that for any theory of the form \req{geng2}, for which our generalized on-shell action can be used,
\begin{equation}
F_{\mathbb{S}^{d}}=16 \pi G\cdot  C(\tilde L^2)F^{\rm E}_{\mathbb{S}^{d}}= (-)^{\frac{d-1}{2}}\frac{2\pi^{d/2+1}\tilde{L}^{d+1}}{d \Gamma(d/2)}\mathcal{L}|_{{\rm AdS}}\, ,
\end{equation}
which takes the expected general form \req{fffs}, with $a^*$ precisely given by \req{astar}. Hence, we have obtained the expected form of the charge $a^*$ from an explicit holographic calculation of the free energy on $\mathbb{S}^{d}$ using our generalized action. The consistency between \req{SEcomplete} and \req{astar} provides support for both expressions.

Reversing the logic, we can rewrite our generalized action in terms of $a^*$, which is way more charismatic than $C(\tilde L^2)$. The result reads
\begin{equation}\label{SEcomplete2}
I_E=-\int_{\mathcal{M}}d^Dx\sqrt{g}\mathcal{L}(g^{\alpha\beta},R_{\mu\nu\rho\sigma})-\frac{2a^*}{\Omega_{(d-1)}\tilde{L}^{d-1}} \int_{\partial\mathcal{M}}d^{D-1}x\sqrt{|h|} \left[ K-\frac{d-1}{\tilde{L}}+\cdots \right]\, ,
\end{equation}
where we have omitted most of the counterterms in \req{gass}. The explicit appearance of $a^*$ in the boundary terms is rather suggestive, and somewhat striking.

\subsection{Generalized action for Quasi-topological gravity}
The QTG density in five bulk dimensions is given by \cite{Quasi2,Quasi}
\begin{equation}\label{eq:QuasiTopo5d}
\begin{aligned}
\mathcal{Z}_{5} =
&\tensor{R}{_{\mu}^{\rho}_{\nu}^{\sigma}}\tensor{R}{_{\rho}^{\alpha}_{\sigma}^{\beta}}\tensor{R}{_{\alpha}^{\mu}_{\beta}^{\nu}}
+ \frac{1}{56} \Big ( 
- 72 \tensor{R}{_{\mu \nu\rho\sigma}}\tensor{R}{^{\mu \nu\rho}_{\alpha}}R^{\sigma \alpha} 
+21 \tensor{R}{_{\mu\nu\rho\sigma}}\tensor{R}{^{\mu\nu\rho\sigma}}R
+120 \tensor{R}{_{\mu\nu\rho\sigma}}\tensor{R}{^{\mu\rho}}\tensor{R}{^{\nu\sigma}} \\
&+ 144 \tensor{R}{_{\mu}^{\nu}}\tensor{R}{_{\nu}^{\rho}}\tensor{R}{_{\rho}^{\mu}}
-132 R_{\mu\nu }R^{\mu\nu }R
+15R^3 \Big) \, .
\end{aligned}
\end{equation}

Just like ECG in $D=4$, the linearized equations of this theory on constant-curvature backgrounds are Einstein-like \cite{Quasi}. Hence, the method developed in the previous subsection should be valid for computing Euclidean on-shell actions of AdS$_5$ solutions of the theory. In this case, the full generalized action \req{SEcomplete2} is given by
\begin{equation}\label{QTBT}
\begin{aligned}
I_E^{\rm QTG}&=-\frac{1}{16\pi G}\int_{\mathcal{M}}d^5x\sqrt{g}\left[\frac{12}{L^2}+R+\frac{L^2\lambda}{2}\mathcal{X}_4+\frac{7\mu L^4}{4}  \mathcal{Z}_5\right]\\
&-\frac{1-6\lambda f_{\infty}+9\mu f_{\infty}^2}{8 \pi G}\int_{\partial \mathcal{M}}d^{4}x\sqrt{h}\bigg[K-\frac{3\sqrt{f_{\infty}}}{L}-\frac{L}{4\sqrt{f_{\infty}}}\mathcal{R}\Bigg]\, ,
\end{aligned}
\end{equation}
where we also included the Gauss-Bonnet density $\mathcal{X}_4=R^2-4\tensor{R}{_{\mu\nu}}\tensor{R}{^{\mu\nu}}+\tensor{R}{_{\mu\nu\rho\sigma}}\tensor{R}{^{\mu\nu\rho\sigma}}$. In this case, the charge $a^*$ reads \cite{Myers:2010jv}
\begin{equation}
a^{*{\rm QTG}}=\left(1-6\lambda f_{\infty}+9\mu f_{\infty}^2\right)\frac{\pi \tilde L^{3}}{8G}\, ,
\end{equation}
while $f_{\infty}$ is determined by the equation \cite{Quasi}
\begin{equation}
1-f_{\infty}+\lambda f_{\infty}^2+\mu f_{\infty}^3=0\, .
\end{equation}
A generalized boundary term for QTG was proposed in  \cite{Dehghani:2011hm}. It would be interesting to check whether \req{QTBT} provides the same results as those obtained using such term. As we mentioned above, in Appendix \ref{BTcheck} we perform an explicit check of that kind for Gauss-Bonnet gravity.
%It would be interesting to compare the results produced using \req{QTBT} with those obtained from the generalized boundary term constructed in  \cite{Dehghani:2011hm}.
%As we mentioned before, the exact generalization of the Gibbons-Hawking-York term is known for Gauss-Bonnet. In that case, we can compare both prescriptions, which we show to yield identical results for asymptotically AdS spaces in appendix \ref{BTcheck}. As for QTG, a boundary term was constructed in \cite{Dehghani:2011hm}
%On the other hand, no exact boundary term is known for QTG when $\mu\neq 0$, so \req{QTBT} should allow for new calculations.

\subsection{Generalized action for Einsteinian cubic gravity}
Let us now return to ECG. In that case, the full generalized Euclidean action \req{SEcomplete2} becomes
\begin{equation}\label{EuclideanECGc6}
\begin{aligned}
I_E^{\rm ECG}=&-\frac{1}{16\pi G}\int_{\mathcal{M}} d^4x \sqrt{|g|}\left[\frac{6}{L^2}+R-\frac{\mu L^4}{8} \mathcal{P} \right]\\
&-\frac{1+3\mu f_{\infty}^2}{8 \pi G}\int_{\partial \mathcal{M}}d^3x\sqrt{h}\left[K-\frac{2\sqrt{f_{\infty}}}{L}-\frac{L}{2\sqrt{f_{\infty}}}\mathcal{R}\right]\, ,
\end{aligned}
\end{equation}
where recall that $f_{\infty}$ can be obtained as a function of $\mu$ from \req{roo}. Observe also that the charge $a^*$ reads in this case
\begin{equation}\label{aae}
a^{* \rm ECG}=(1+3\mu f_{\infty}^2)\frac{\tilde L^2}{4 G}\, .
\end{equation}
We use \req{EuclideanECGc6} in several occasions in the remainder of the chapter, finding exact agreement with the expected results in all cases for which alternative methods can be used.

\section{Stress tensor two-point function charge $\ctt$}\label{tt}
In order to characterize the holographic dual of ECG, we must translate the two available dimensionless parameters in \req{ECG}, namely: $L^2/G$ and $\mu$, into universal defining quantities of the boundary theory. Since we are only considering the gravitational sector of the bulk theory, the most relevant `charges' to be identified in the CFT are those characterizing the boundary stress tensor. Conformal symmetry highly constrains the structure of stress-tensor two- and three-point functions \cite{Osborn:1993cr}. We will deal with the three-point function charges in Sec.~\ref{t44}. Let us start here with the stress-tensor correlator which, for an arbitrary CFT$_3$, is given by \cite{Osborn:1993cr}
\begin{equation}\label{2pf}
\braket{ T_{ab}(x)T_{cd}(x')} =\frac{\ctt}{|x-x'|^6}\mathcal{I}_{ab,cd}(x-x')\, ,
\end{equation}
where
\begin{equation}
\mathcal{I}_{ab,cd}(x)\equiv\frac{1}{2}\left(I_{ac}(x)I_{bd}(x)+I_{ad}(x)I_{bc}(x)\right)-\frac{1}{4}\delta_{ab}\delta_{cd}\, ,\quad \text{and} \quad I_{ab}(x)\equiv\delta_{ab}-2\frac{x_{a}x_{b}}{x^2}\ ,
\end{equation}
are fixed tensorial structures. This correlator is then fully characterized by a single theory-dependent parameter, customarily denoted $\ctt$. This quantity, which in even dimensions is proportional to the trace anomaly charge $c$, also plays a relevant role in three-dimensional CFTs --- see \eg \cite{Huh:2014eea,Diab:2016spb,Giombi:2016fct} for recent studies. As opposed to the $d=2$ case \cite{Zamolodchikov:1986gt}, $\ctt$ is not monotonous under general RG flows in three dimensional CFTs \cite{Nishioka:2013gza}. However, it universally shows up in various contexts, including relevant quantities in entanglement and R\'enyi entropies \cite{HoloRen,Hung:2014npa,Perlmutter:2013gua,Mezei:2014zla,Bueno1};  quantum critical transport --- see \eg \cite{Witczak-Krempa:2015pia,Lucas:2017dqa} and references therein; or partition functions on deformed curved manifolds \cite{Closset:2012ru,Bobev:2017asb,Fischetti:2017sut}.

In AdS/CFT, the dual of $T_{\mu\nu}(x)$ is the normalizable mode of the metric \cite{Witten,Gubser}. Hence, evaluating \req{2pf} entails determining the two-point  boundary  correlator  of  gravitons  in  the corresponding AdS  vacuum. For Einstein gravity in $d=3$, the result \cite{Buchel:2009sk,Liu:1998bu} reads
\begin{equation}
\ctte=\frac{3}{\pi^3}\frac{\tilde{L}^2}{G}\, .
\end{equation}
Naturally, the introduction of higher curvature terms in the bulk modifies this result, \eg \cite{Buchel:2009sk,Myers:2010jv,Bueno2}. In general, higher order gravities give rise to equations of motion involving more than two derivatives of the metric. In those cases, the metric generically contains additional degrees of freedom besides the usual massless graviton. From the holographic perspective, this means that the metric couples to additional operators which are typically nonunitary\footnote{See \eg \cite{Myers:2010tj,Bueno2} for more detailed discussions of this issue.}. This is not always the case, however. In fact, there exist families of higher order gravities whose linearized equations around maximally symmetric backgrounds are identical to those of Einstein gravity, up to a normalization of the Newton constant --- see footnote \ref{ELC} and  \eg \cite{Aspects} for details. For those, the only mode is the usual spin-2 graviton, the metric only couples to the stress tensor, and $\ctt$ can be straightforwardly extracted from the effective Newton constant. This generically reads $G_{\rm eff}=G/\alpha$, where $\alpha$ is a constant which depends on the new couplings. The appearance of $\alpha$ can be alternatively understood as changing the normalization of the graviton kinetic term which, holographically, gets translated into a modification of the stress-tensor correlator charge, which then becomes $\alpha \cdot \ctte$.

For ECG, using \req{Geff}, we find then
\begin{equation}\label{cttecg}
\ctt^{\rm ECG}= (1-3\mu f_{\infty}^2)\frac{3}{\pi^3}\frac{\tilde{L}^2}{G}\, .
\end{equation}
Observe that unitarity imposes $\ctt$ to be positive, which translates into $1-3\mu f_{\infty}^2 >0$. This is of course equivalent to asking the effective bulk gravitational constant to be positive.
It can be seen that this constraint is automatically satisfied whenever \req{sis} holds.

While we have been able to compute $\ctt$ for ECG using $G_{\rm eff}^{\rm ECG}$, it is instructive to obtain it from an explicit holographic calculation. This will also serve as a highly-nontrivial consistency check for the new on-shell action method introduced in the previous section.

Let us then consider a metric perturbation: $g_{\mu\nu}=\bar g_{\mu\nu}+h_{\mu\nu}$, on the Euclidean AdS$_4$ vacuum
\begin{equation}\label{EAdS}
ds^2=\frac{r^2}{L^2}\left[d\tau^2+dx^2+dy^2\right]+\frac{L^2}{r^2f_{\infty}}dr^2\, .
\end{equation}
Since all components of the two-point function will be determined by $\ctt$, computing one of them will be enough. It is then sufficient to consider a perturbation of the form $h_{xy}=\frac{r^2}{L^2}\phi(r,\tau)$. Plugging this into the Euclidean version of \req{ECG} and expanding up to quadratic order in $\phi$, we find
\begin{equation}\label{ttw}
I^{\rm ECG}_{E\, \rm \small Bulk}=\frac{(1-3\mu f_{\infty}^2)}{32\pi G}\int d^3xdr\left[\frac{1}{\sqrt{f_{\infty}}}(\partial_{\tau}\phi)^2+\sqrt{f_{\infty}}\frac{r^4}{L^4}(\partial_r\phi)^2\right]-\frac{1}{16\pi G}\int d^3x\, \Gamma_r\Big|_{r=r_{\infty}}\, ,
\end{equation}
where $\Gamma_r$ is a boundary term which appears after integration by parts --- see \req{gaga}. Recall also that, in these coordinates, the boundary corresponds to $\lim_{r\rightarrow \infty} r \equiv L^2/\delta$, where we introduce the UV cutoff $\delta\ll 1$. 
% Before considering the boundary terms in \req{EuclideanECGc6}, let us consider
%In addition to this, we have to take into account the generalized Gibbons-Hawking term as well as the counterterms, which are shown in \ref{EuclideanECGc6}. For now, let us consider only the bulk action. From it we can read 
The equation of motion for $\phi$ follows from \req{ttw}, and reads
\begin{equation}
\frac{\partial}{\partial r}\left(\frac{r^4}{L^4}\frac{\partial \phi}{\partial r}\right)+\frac{1}{f_{\infty}}\frac{\partial ^2\phi}{\partial \tau^2}=0\, .
\end{equation}
In order to solve it, we Fourier-transform the dependence on the coordinate $\tau$,
\begin{equation}
\phi(r,\tau)=\frac{1}{2\pi}\int dp \phi_0(p)e^{ip\tau}H_p(r)\, .
\end{equation}
$H_p$ satisfies the equation
\begin{equation}
\frac{d}{d r}\left(\frac{r^4}{L^4}\frac{d H_p}{d r}\right)-\frac{p^2}{f_{\infty}}H_p=0\, ,
\end{equation}
whose general solution reads
\begin{equation}
H_p(r)=c_1 e^{-\frac{L^2 |p|}{\sqrt{f_{\infty}}r}}\left(1+\frac{L^2 |p|}{\sqrt{f_{\infty}}r}\right)+c_2 e^{\frac{L^2 |p|}{\sqrt{f_{\infty}}r}}\left(1-\frac{L^2 |p|}{\sqrt{f_{\infty}}r}\right)\, .
\end{equation}
In order to get a regular solution, we set $c_2=0$, and we also fix $c_1=1$ so that $H_p(r \rightarrow L^2/\delta)=1$. With this solution, we evaluate the Lagrangian, which can be expressed as a total derivative. Further integrating over the $r$ coordinate and substituting the solution in Fourier space, we get
%, so the on-shell action reads
%\begin{equation}
%I_{\rm ECG}^{\rm \small Bulk}=\frac{(1-3\mu f_{\infty}^2)\sqrt{f_{\infty}}}{32\pi G}\int d^3xdr\partial_r\left(\frac{r^4}{L^4}\phi\partial_r\phi\right)-\frac{1}{16\pi G}\int d^3x %\Gamma_r\Big|_{r=r_0}\, .
%\end{equation}
%Then, we integrate over the $r$ coordinate and we substitute the solution in Fourier space. This way we get
\begin{equation}\label{kd}
I^{\rm ECG}_{E\, \rm \small Bulk}=\frac{\sqrt{f_{\infty}}V_{\mathbb{R}^2}}{64\pi^2 G_{\rm eff}^{\rm ECG} }\int dpdq\phi_0(p)\phi_0(q)\delta(p+q)\frac{L^4}{\delta^4}H_p\partial_r H_p\Big|_{r=L^2/\delta}-\frac{1}{16\pi G}\int d^3x \Gamma_r\Big|_{r =L^2/\delta}\, ,
\end{equation}
where $V_{\mathbb{R}^2}=\int dxdy$, and where we used $\int d\tau e^{i(p+\tau)}=2\pi \delta(p+q)$.

Let us now turn to the boundary contributions in the generalized action \req{EuclideanECGc6}. As we explain in Appendix \ref{2pbdy}, when these terms are added to \req{kd}, most divergences in $\Gamma_r\Big|_{r=L^2/\delta}$ disappear, and we are left with the following result for the full action:
%HEREEEE
 %Now we must consider the boundary contribution in \req{EuclideanECGc6}. When we include these terms we can see that they cancel most of the divergences which appear in $\Gamma_r$ \comment{Appendix with this} and at the end only one term survives. This way, the complete action $I_{\rm ECG}=I_{\rm ECG}^{\rm \small Bulk}+I_{\rm ECG}^{\rm \small Bdry}$ reads
\begin{align}\label{ICT}
I^{\rm ECG}_E&=I^{\rm ECG}_{E\, \rm \small Bulk}+I^{\rm ECG}_{E\, \rm \small GGHY+GCT}\\ \notag&=\frac{V_{\mathbb{R}^2}}{64\pi^2 G_{\rm eff}^{\rm ECG} \sqrt{f_{\infty}}}\int dpdq\phi_0(p)\phi_0(q)\delta(p+q)\left[f_{\infty}\frac{L^4}{\delta^4}H_p\partial_r H_p\Big|_{r=L^2/\delta }-\frac{L^2 p^2}{\delta} H_p^2\right]\, .
\end{align}
Observe that, even though $1/G_{\rm eff}^{\rm ECG}$ and $a^{* \rm ECG}$ have  a different dependence on $\mu$ --- see \req{Geff} and \req{aae} respectively --- and that it is $a^{* \rm ECG}$ the one  appearing as an overall constant in the generalized GHY term and the counterterms \req{EuclideanECGc6}, everything conspires to produce a single finite contribution which is instead proportional to $1/G_{\rm eff}^{\rm ECG}$, as it must.

If we take the limit $\delta \rightarrow 0$ explicitly in \req{ICT}, we get the simple result
\begin{equation}
I^{\rm ECG}_E[\phi_0]=-\frac{V_{\mathbb{R}^2}\tilde{L}^2}{64\pi^2 G_{\rm eff} }\int dpdq\phi_0(p)\phi_0(q)\delta(p+q)|p|^3\, .
\end{equation}
Using the holographic dictionary \cite{Witten}, we can compute one of the components of the boundary stress tensor two-point function in momentum space as
%
%Then we have expressed the on-shell action in terms of the boundary perturbations. Hence, according to the holographic dictionary we can compute the following component of the two-point function in momentum space:
\begin{equation}\label{TTECG}
\langle T_{xy}(0,0,p)T_{xy}(0,0,q)\rangle=-(2\pi)^2\frac{\delta^2 I^{\rm ECG}_E[\phi_0]}{\delta \phi_0(-p)\delta \phi_0(-q)}=\frac{\tilde{L}^2V_{\mathbb{R}^2}}{8 G_{\rm eff} } \delta(p+q)|p|^3\, .
\end{equation}
Now, from the CFT side, this is given by
\begin{equation}\label{df}
\langle T_{xy}(0,0,p)T_{xy}(0,0,q)\rangle=\int d^3x\int d^3x'e^{-i p\tau}e^{-iq\tau'}\langle T_{xy}(x)T_{xy}(x')\rangle \, ,
\end{equation}
where
\begin{equation}
\langle T_{xy}(x)T_{xy}(x')\rangle=\frac{\ctt}{2|x-x'|^6}\left[-1+2\frac{(\tau-\tau')^2}{|x-x'|^2}+8\frac{(x-x')^2(y-y')^2}{|x-x'|^4}\right]\, .
\end{equation}
The integration in \req{df} can be performed without further complications and we obtain the result
\begin{equation}
\langle T_{xy}(0,0,p)T_{xy}(0,0,q)\rangle=\frac{\pi^3 \ctt V_{\mathbb{R}^2}}{24}\delta(p+q)|p|^3\, .
\end{equation}
Comparing this expression with \req{TTECG}, we obtain the result for $\ctt$, which agrees with the one in \req{cttecg}, as it should. The fact that our generalized action \req{EuclideanECGc6} succeeds in providing the right answer for this quantity, including various non-trivial cancellations  between $I^{\rm ECG}_{E\, \rm \small Bulk}$ and $I^{\rm ECG}_{E\, \rm \small GGHY+GCT}$ --- see Appendix \ref{2pbdy} --- provides strong evidence for the validity of the method developed in Sec.~\ref{osa}. 
%Therefore, comparing with \ref{TTECG} we get the central charge $\ctt$:
%\begin{equation}
%\ctt=\frac{3(1-3\mu f_{\infty}^2)L^2}{\pi^3 G f_{\infty}}\, .
%\end{equation}

Note finally that, as explained at the beginning of this section, $\ctt$ provides information about many different physical quantities appearing in numerous contexts. Hence, by the same price we computed \req{cttecg}, we gain access to all such quantities for the CFT$_3$ dual to ECG.

\section{Thermodynamics}\label{therr}
In this section we study the thermodynamic properties of the ECG black holes constructed in Sec.~\ref{BHs}. First, we compute the Wald entropy, ADM energy and free energy of the solutions, and compare the result with the one obtained from an explicit on-shell action calculation, which serves as a further check of the method proposed in Sec.~\ref{osa}. Then, focusing on the flat boundary case, $k=0$, we identify the quantity $\cs$ which relates the thermal entropy density to the temperature, and show that, in contradistinction to Einstein gravity, it defines an independent charge with respect to $\ct$. In Subsections \ref{TT2} and \ref{SS2}, we study the phase space of holographic ECG on $\mathbb{S}^1_{\beta}\times \mathbb{T}^2$ and $\mathbb{S}^1_{\beta}\times \mathbb{S}^2$, respectively. In the first case, we show that the standard phase transition between the ECG AdS soliton and black brane keeps occurring at the same temperature as for Einstein gravity. In the second, we show that depending on the value of $\mu$, one, two or three black hole solutions can coexist at the same temperature. The dominating phases are still thermal AdS at small temperatures and large black holes at large temperatures, but the Hawking-Page-transition temperature becomes arbitrarily large as we approach the critical limit $\mu=4/27$. Besides, small black holes become thermodynamically stable for $\mu\neq0$, although their contribution to the partition function is always subleading with respect to thermal AdS. Nevertheless, we show that small black holes can become the dominant phase if a topological Gauss-Bonnet term is added to the action. In that case, we also observe the presence of a critical point in the phase diagram. 

\subsection{Entropy, energy and free energy}
Let us start by computing the Wald entropy of the solutions, which, for any covariant theory of gravity, is given by \cite{Wald:1993nt,Iyer:1994ys}
\begin{equation}
S=-2\pi \int_{\rm \ssc H}  d^{d-1}x\, \sqrt{h} \frac{\partial \mathcal{L}}{\partial R^{\mu\nu}\,_{\rho\sigma}}\varepsilon^{\mu\nu}\varepsilon_{\rho\sigma} \, ,
\end{equation}
where $\varepsilon_{\mu\nu}$ is the binormal to the horizon. Now, for metrics of the form \req{bhss}, the integration can be performed straightforwardly, yielding
\begin{equation}
S=-\frac{2\pi \rh^2}{L^2} V_{\Sigma} \left.\frac{\partial \mathcal{L}}{\partial R^{\mu\nu}\,_{\rho\sigma}}\varepsilon^{\mu\nu}\varepsilon_{\rho\sigma}  \right|_{r=\rh}\, ,
\end{equation}
where $V_{\Sigma}$ is the regularized volume of $\mathbb{S}^2$, $\mathbb{R}^2$ or $\mathbb{H}^2$ for $k=1,0,-1$ respectively. Explicitly, the final result for the ECG black holes reads
\begin{equation}\label{sth}
S^{\rm ECG}=\frac{\rh^2  V_{\Sigma}}{4G L^2}  \left[1-\frac{3 \mu L^4  \left(k+\frac{3\rh^2}{L^2}\right)\left[ \left(k+\frac{3\rh^2}{L^2}\right)+2k \left[1+\sqrt{1+\frac{3k L^4\mu}{\rh^4}\left(k +3\frac{\rh^2}{L^2}\right)} \right] \right]}{\rh^4 \left[1+\sqrt{1+\frac{3k L^4\mu}{\rh^4}\left(k +3\frac{\rh^2}{L^2}\right)} \right]^2}\right]   \, .
\end{equation}
Again, this reduces to the Einstein gravity result
\begin{equation}\label{E}
S^{\rm E}=\frac{\rh^2 V_{\Sigma} }{4GL^2}    \, ,
\end{equation}
when we set $\mu=0$.
Once we have $S(T)$ (defined implicitly by means of Eq.~\req{T}), we can use the first law, $dE=TdS$, to find the energy. The result is\begin{equation}\label{adm}
E^{\rm ECG}=\frac{{(\omega^{\rm ECG})}^3 V_{\Sigma} N}{8\pi G L^4}\, .
\end{equation}
As expected, this coincides with the result one would obtain for the generalized ADM energy from the asymptotic expansion \req{Asympt}.

The entropy of the solutions can be alternatively computed from the free energy as $S=-\partial F/\partial T$. Hence, we can perform an additional check of our generalized action \req{EuclideanECGc6}, which evaluated on the Euclidean version of the solutions --- for which we identify $t_E\sim t_E+\beta$ --- should yield the free energy as  $F^{\rm ECG}= I^{\rm ECG}_{E}/\beta$.
%Even though we have completely determined the thermodynamic properties of these black holes, it is important for the applications that we are going to consider, to motivate the thermodynamics from the point of view of Euclidean path integral. For that, we first need to define a well-posed action, in the sense that it is stationary for solutions of the equations of motion. In Appendix A we propose a way to achieve this in ECG by adding a generalized York-Gibbons-Hawking term. In Euclidean signature, we propose the following action functional
%\begin{equation}\label{EECG1}
%\begin{aligned}
%I^{\rm ECG}_{E}=&-\frac{1}{16\pi G}\int d^4x \sqrt{|g|}\left[\frac{6}{L^2}+R-\frac{\mu L^4}{8} \mathcal{P} \right]\\
%&-\frac{1+3\mu f_{\infty}^2}{8 \pi G}\int_{\partial \mathcal{M}}d^3x\sqrt{h}\left(K-\frac{2\sqrt{f_{\infty}}}{L}-\frac{L}{2\sqrt{f_{\infty}}}\mathcal{R}\right)\, ,
%\end{aligned}
%\end{equation}
%The free-energy of the solution should be obtained from the Euclidean on-shell action as $F= I^{\rm ECG}_{E}/\beta$. 
Plugging \req{bhss} in \req{EuclideanECGc6}, we find that the bulk term is a total derivative that can be integrated straightforwardly, namely
\begin{equation}
I^{\rm ECG}_{E\, {\rm Bulk}}=\frac{\beta N V_{\Sigma}}{16\pi G L^2 }\left[H(\rh)-H(L^2/\delta) \right]\, ,
\end{equation}
where
\begin{equation}
H(r)\equiv \frac{r^3}{L^2}\left[(2-4 f-r f')-\frac{\mu }{4}\left(2f+r f'\right)^2 \left(4f-r f'\right)\right]\, . 
\end{equation}
Using the asymptotic expansion \req{Asympt}, we get
\begin{equation}
H(L^2/ \delta)=\frac{2L^4}{\delta^3}(1-2f_{\infty}-2\mu f_{\infty}^3)+\frac{{(\omega^{\rm ECG})}^3}{L^2}\frac{(1+3\mu f_{\infty}^2)}{(1-3\mu f_{\infty}^2)}+\mathcal{O}(\delta)\, . 
\end{equation}
We can also evaluate the boundary contributions in \req{EuclideanECGc6}. For these, we use 
%the following results for the induced metric determinant, trace of 
%On the other hand we can also determine the boundary contribution by using again the asymptotic expansion. The extrinsic curvature and the the Ricci scalar of the boundary $r=L^2/\delta$ read
\begin{equation}
\begin{aligned}
d^3x\sqrt{h}&=N dt\wedge d\Sigma_{k} \left(\frac{\sqrt{f_{\infty}}L^3}{\delta^3}+\frac{k L}{2\delta\sqrt{f_{\infty}}}-\frac{{(\omega^{\rm ECG})}^3}{2\sqrt{f_{\infty}}L^3(1-3\mu f_{\infty}^2)}\right)+\mathcal{O}(\delta)\, ,\\
K&=\frac{3\sqrt{f_{\infty}}}{L}+\frac{k \delta^2}{2L^3\sqrt{f_{\infty}}}+\mathcal{O}(\delta^{4})\, ,\quad \mathcal{R}=\frac{2k\delta^2}{L^4}\, .
\end{aligned}
\end{equation}
Then, we find
\begin{equation}
I^{\rm ECG}_{E\, \rm \small GGHY+GCT}=-\frac{\beta N V_{\Sigma}(1+3\mu f_{\infty}^2)}{8\pi G L^4 }\left[\frac{L^6 f_{\infty}}{\delta^3} -\frac{{(\omega^{\rm ECG})}^3}{2(1-3\mu f_{\infty}^2)}\right]+\mathcal{O}(\delta)\, .
\end{equation}
Now, if we add up both contributions we obtain the finite result
\begin{equation}
I^{\rm ECG}_{E}=\frac{\beta N V_{\Sigma}}{16\pi G L^2 }H(\rh)\, ,
\end{equation}
where we made use of the AdS$_4$ embedding equation \req{roo}. Hence, all boundary contributions cancel out and the on-shell action is reduced to the evaluation of the function $H(r)$ at the horizon. Using the near-horizon expansion \req{nH}, we can finally write the free energy as
%HEEEEEREEEEE
%Therefore, after using the near-horizon expansion \req{nH} we get the free energy $F= I_{\rm ECG}^{E}/\beta$, which can be written as
\begin{equation}\label{fecg}
F^{\rm ECG}=\frac{N V_{\Sigma}}{8\pi G L^2 }\left[k\rh +\frac{\rh^3}{L^2}-\frac{2\pi T\rh^2}{N}+\mu L^4 \left(\frac{3k}{\rh}\left(\frac{2\pi T}{N}\right)^2+\left(\frac{2\pi T}{N}\right)^3\right)\right]\, .
\end{equation}
Note that this can be also written fully in terms of $\rh$ using \req{T}.
%By using \req{T} we can also write explicitly the free energy as a function of the radius $\rh$.
%\begin{equation}
%H(\rh)=2\rh^3 \left[ 1+\frac{L^2}{\rh^2}\left(k-P_k(\mu,r)\right)+\frac{\mu L^6}{\rh^6}\left(3k+P_k(\mu,r)\right) P_k(\mu,r)^2 \right]\, ,
%\end{equation}
%where
%\begin{equation}
%P_k(\mu,r)=\left(k+\frac{3\rh^2}{L^2} \right) \left[1+\sqrt{1+\frac{3k L^4\mu}{\rh^4}\left(k +3\frac{\rh^2}{L^2}\right)} \right]^{-1}\,.
%\end{equation}
%The final result for the free energy reads
%\begin{equation}\label{fecg}
%F_{\rm ECG}=\frac{N V_{\Sigma}}{16\pi G L^2 }H(\rh)\, .
%\end{equation}
When $\mu=0$, \req{fecg} reduces to the Einstein gravity result
\begin{equation}
F^{\rm E}=\frac{N V_{\Sigma} \rh}{16\pi G L^2 }\left(k-\frac{\rh^2}{L^2}\right)\, .
\end{equation}
Using \req{fecg} and the thermodynamic identity $S=-\partial F/\partial T$, we can recompute the entropy of the solutions. The result precisely matches \req{sth}, computed using Wald's formula, which provides another check for our generalized action.

\subsection{Thermal entropy charge $\cs$}\label{cssex}
%The planar case corresponds to $k=0$ and, for that, it is customary to make the redefinition
%\begin{equation}
%V_{k=0}(r)=\frac{r^2}{L^2}f(r)\, , 
%\end{equation}
When the boundary geometry is flat, $k=0$, it is convenient to set $N^2=1/f_{\infty}$, a choice which fixes the speed of light to one in the dual CFT \cite{Buchel:2009sk}. In that case, the thermodynamic expressions simplify considerably. In particular, we find
\begin{eqnarray}\label{Trh}
T=\frac{3 \rh}{4\pi L^2\sqrt{f_{\infty}}}\, , \quad {\omega}^3= \rh^3 \left(1-\frac{27}{4}\mu \right)\, , \\  \label{srh}
s=\frac{ \rh^2}{4GL^2}\left(1-\frac{27}{4}\mu\right)\, ,\quad \varepsilon=\frac{ \rh^3}{8\pi G L^4\sqrt{f_{\infty}}}\left(1-\frac{27}{4}\mu \right)\, ,
\end{eqnarray}
where we defined the entropy and energy densities $s\equiv S/V_{\mathbb{R}^2}$, $\varepsilon \equiv E/V_{\mathbb{R}^2}$. %Assuming $\omega^3>0$, the second equation also imposes the constraint \ref{sis}. 
%\comment{comparison with GB Quasitopo on the depend of the grav coupling}
We can explicitly write these quantities in terms of the temperature, the result being
%\comment{Is it possible to have black holes with $\omega^3<0$. I think the condition at asymptotic infinity reads $\omega^3\mu<0$, so if it is the case, one could also have black holes with positive values of $\mu$}
%It is convenient to define the entropy and energy densities as $s=S/V_2$ and $\rho=E/V_2$, respectively, where $V_2=\int dx_1dy_1$. From the general expressions \req{sth} and \req{E} we find the following result
\begin{equation}\label{entropy}
s=\frac{4\pi^2 \tilde{L}^2f_{\infty}^2}{9G}\left(1-\frac{27}{4}\mu\right) T^2\, ,\quad \varepsilon=\frac{8\pi^2 \tilde{L}^2f_{\infty}^2}{27G}\left(1-\frac{27}{4}\mu\right) T^3.
\end{equation}
%\begin{equation}
%s=\frac{16\pi^2 L^2 f_{\infty}}{9G}\left(1-\frac{27}{4}\mu \right)T^2
%\end{equation}
From \req{entropy}, it immediately follows that ECG black branes satisfy
%Note that the following relation holds,
\begin{equation}
\varepsilon=\frac{2}{3}T s\, ,
\end{equation}
as expected for a thermal plasma in a general three-dimensional CFT.

The dependence on the temperature of the thermal entropy density is also fixed for any CFT$_3$ to take the form
\begin{equation}
s=\cs T^2\, , 
\end{equation}
where $\cs$ is a theory-dependent quantity. From, \req{entropy}, it follows that
\begin{equation}\label{csss}
\cs^{\rm ECG}=\left(1-\frac{27}{4}\mu \right)f_{\infty}^2\, \cse\, ,\quad \text{where} \quad \cse=\frac{4\pi^2}{9}\frac{\tilde{L}^2}{G}\, ,
\end{equation}
is the Einstein gravity result --- see \eg \cite{Buchel:2009sk}. As we can see, in the holographic model defined by ECG, $\cs$ is no longer proportional to $\ctt$, and therefore defines an additional well-defined  independent `charge' which characterizes the theory\footnote{Observe that $\cs$ can be rewritten as $\cs^{\rm ECG}=f_{\infty}^2(1-3\mu f_{\infty}^2/4)(1-3\mu f_{\infty}^2)^2\cse$, %$=\textcolor{red}{\frac{(f_{\infty}+3)(3-2 f_{\infty})}{27f_{\infty}^2}\ct^{\rm ECG}}$ 
which makes it more obvious that this charge is not proportional to $\ct^{\rm ECG}$.}. For growing values of $\mu$, $\cs$ monotonously decreases with respect to the Einstein gravity value and, funnily, it vanishes for the critical case\footnote{This would seem to suggest that the black brane has a unique microstate in that case, but it is probably just another evidence of the problematic properties of the critical theory.} , $\mu=4/27$. 

The fact that $\cs$ vanishes for certain value of the gravitational coupling is quite unusual, and does not occur for QTG or Lovelock black holes (in the Einstein gravity branch) in any number of dimensions --- see \eg \cite{Buchel:2009sk,Quasi,Myers:2010jv,Dehghani:2009zzb,Camanho:2011rj}. In fact, in those cases, the only modification in $\cs$ with respect to Einstein gravity is an overall $f_{\infty}^{(d-1)}$ factor, \ie the result reads  $\cs^{\rm QTG/Lovelock}=f_{\infty}^{(d-1)} \cse$, where $\cse$ is the Einstein gravity result written in terms of $\tilde{L}$. In fact, in view of the results for those theories, one would have naively expected all `$(1-27/4\mu)$' factors in \req{Trh}-\req{csss} not to appear for ECG. This seems to be a simple manifestation of the fact that the theories belonging to the Generalized QTG class (including ECG) for which $f(r)$ is determined through a second-order differential equation possess rather different properties from those for which $f(r)$ is determined from an algebraic equation --- see below and \cite{Hennigar:2017ego,PabloPablo3,Ahmed:2017jod,Hennigar:2017umz} for more evidence in this direction.

%\subsection{Disk entanglement entropy universal term $a^*$}	
%\begin{equation}
%s_{\rm EE}=a_1 \frac{R}{\delta}-2\pi a^*\, , 
%\end{equation}
%where $a^*$ is a universal contribution given by
%\begin{equation}\label{aae}
%a^*=\left(4-\frac{3}{f_{\infty}}\right)a^*_{{\rm \ssc E}}\, , \quad \text{where} \quad a^*_{{\rm \ssc E}}=\frac{1 }{4}\frac{\tilde{L}^2}{G}
%\end{equation}	

\subsection{Toroidal boundary: black brane vs AdS$_4$ soliton}\label{TT2}
In this subsection we study the phase space of thermal configurations when the spatial dimensions of the boundary CFT form a torus $\mathbb{T}^2$. The first obvious saddle corresponds to Euclidean AdS$_4$ with toroidal boundary conditions, given by
\begin{equation}\label{TAdS}
ds^2=\frac{r^2}{ L^2}\left[d\tau ^2+dx_1^2+dx_2^2\right]+\frac{L^2}{r^2f_{\infty}}dr^2\, ,
\end{equation}
where the coordinates $x_1$ and $x_2$ are assumed to be periodic, $x_{1,2}\sim x_{1,2}+l_{1,2}$, where $l_{1,2}$ is the period of each coordinate. Without loss of generality we assume $l_1\le l_2$. As before, $\tau \sim \tau+\beta$. The next candidate is the Euclidean black brane
\begin{equation}\label{BB}
ds^2=\frac{r^2}{L^2}\left[\frac{f(r)}{f_{\infty}}d\tau^2+dx_1^2+dx_2^2\right]+\frac{L^2}{r^2f(r)}dr^2\, ,
\end{equation}
for which the temperature is fixed in terms of the horizon radius through \req{Trh}. Finally, it should be evident that moving the $f(r)/f_{\infty}$ factor from $g_{\tau \tau}$ to $g_{11}$ or $g_{22}$ should also give rise to solutions of ECG, \eg
\begin{equation}\label{soliton}
ds^2=\frac{r^2}{L^2}\left[d\tau^2+\frac{f(r)}{f_{\infty}}dx_1^2+dx_2^2\right]+\frac{L^2}{r^2f(r)}dr^2\, .
\end{equation}
These are the so called AdS$_4$ `solitons' \cite{Witten:1998zw,Horowitz:1998ha}. The crucial difference with respect to the black brane is that, for these, regularity no longer imposes a relation between the temperature and the horizon radius. Instead, it fixes the periodicity of $x_1$ (or $x_2$ if $f(r)/f_{\infty}$ appears in $g_{22}$ instead) in terms of $\rh$ as
\begin{equation}
l_{1,2}=\frac{4\pi L^2\sqrt{f_{\infty}}}{3 \rh}\, .
\end{equation}
Of course, $\tau$ is still periodic with period $\beta$, but, as opposed to the black-brane case, the temperature can be now arbitrary for a given value of $\rh$.

Now, the Euclidean action vanishes for pure Euclidean AdS$_4$, whereas for the black brane and the solitons we find, respectively 
\begin{eqnarray}\label{bbe}
I_E^{\rm bb}=-\frac{4\pi f_{\infty}L^2}{27 G}\left(1-\frac{27}{4}\mu\right)T^2 l_1l_2\, ,\quad 
I_E^{\rm soliton\, 1,2}=-\frac{4\pi f_{\infty}L^2}{27 G}\left(1-\frac{27}{4}\mu\right)\frac{l_1l_2}{Tl_{1,2}^3}\, .
\end{eqnarray}
The solution which dominates the partition function is the one with the smaller on-shell action (or free energy, $ \beta F\equiv I_E$). As we can see from \req{bbe}, for the set of values of $\mu$ for which the ECG solutions exist, the free energies of the black brane and the AdS solitons are always negative, just like for Einstein gravity, which implies that pure AdS$_4$ never dominates. 
We observe that for (arbitrarily) small temperatures, the partition function is dominated by the soliton with the shortest periodicity, the other one being always subleading. For large temperatures, the black brane dominates instead. At $T=1/l_1$, (recall we are assuming $l_1<l_2$), there is a first-order phase transition which connects both phases.
Hence, the phase-transition temperature is not modified with respect to Einstein gravity. The latent heat, computed as the difference between the energy densities of both configurations at $T=1/l_1$, does change and is given by
\begin{equation}
\delta Q=\frac{4\pi f_{\infty}L^2}{9 G}\left(1-\frac{27}{4}\mu\right)\frac{l_2}{l_1^2}\, .
\end{equation}
Again, something unusual happens in the critical limit. In that case, the free energy of both the black brane and the soliton --- which have a simple metric function given by $f(r)=\frac{3}{2}(r^2-\rh^2)/L^2$ --- vanishes. Then, for $\mu=4/27$, the black brane, the two solitons and pure AdS$_4$ are all equally probable configurations.
%When $\mu\rightarrow 4/27$ the latent heat vanishes. Indeed, the black brane and soliton solution exist in the critical limit, and they are simply given by $f(r)=\frac{3}{2}(r^2-\rh^2)/L^2$, with the corresponding value of $\rh$. However, in the critical limit the free energy of both configurations vanish. This means that the black brane, the solitons and pure AdS are equally probable configurations.

\subsection{Spherical boundary: Hawking-Page transitions}\label{SS2}
Let us now consider the boundary theory on $\mathbb{S}^1_{\beta}\times \mathbb{S}^2$. In that case, apart from Euclidean AdS$_4$ foliated by spheres, the other candidate saddle of the semiclassical action corresponds to the Euclidean spherically symmetric black hole
\begin{equation}\label{Spheric}
ds^2=\left[1+\frac{r^2}{L^2}f(r)\right]d\tau^2+\frac{dr^2}{\left[1+\frac{r^2}{L^2}f(r)\right]}+r^2d\Omega^2_{(2)}\, ,
\end{equation}
where we have chosen $N^2=1$. Also, note that the `volume' of the transverse space is, in this case, $V_{\mathbb{S}^2}=4\pi L^2$. As a function of the horizon radius, the temperature of these solutions is given by \req{T}
\begin{equation}\label{eq:temperatureBHsECG}
T(\rh)=\frac{1}{2\pi \rh }\left(1+3\frac{\rh^2}{L^2}\right)\left[1+\sqrt{1+\frac{3\mu L^4}{\rh^4}\left(1+3\frac{\rh^2}{L^2}\right)}\right]^ {-1}\ .
\end{equation}
The contribution coming from the cubic term in the action becomes less and less relevant as we make $\rh$ larger, but its effect is highly nonperturbative for small radius. For example, a non-vanishing value of $\mu$ makes the temperature vanish, instead of blowing up, as  $\rh\rightarrow 0$. More precisely, one finds $T\approx \rh/(2\pi\sqrt{3\mu}L^2)$ in that regime. This is no different from the behavior observed for the asymptotically flat ECG black holes \cite{Hennigar:2016gkm,PabloPablo2,PabloPablo4} --- small black holes do not care whether they are inside AdS$_4$ or flat space.

\begin{figure}[t!]
\centering
\includegraphics[width=0.65\textwidth]{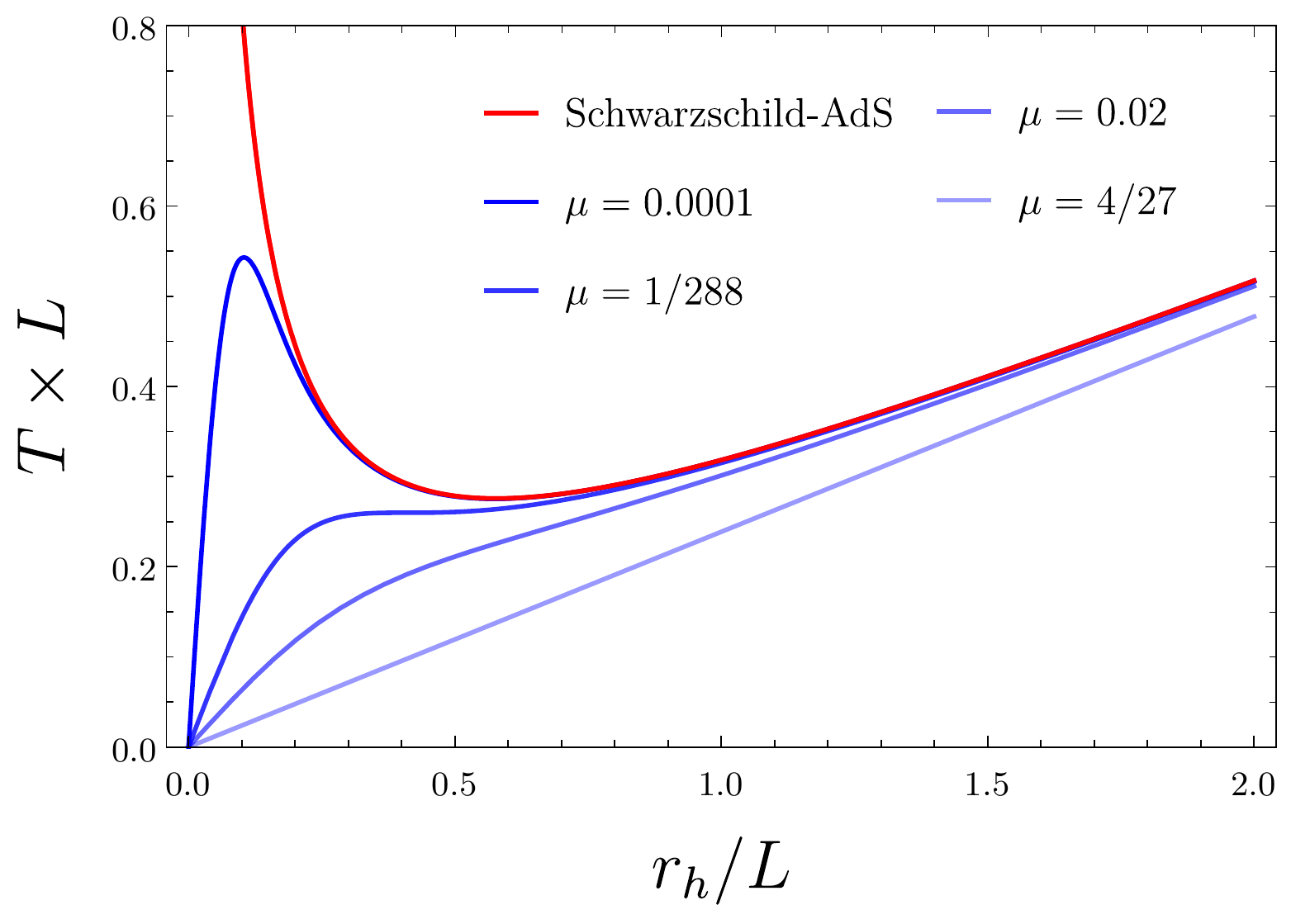}
\caption{Temperature as a function of the horizon radius for various values of $\mu \in [0,4/27]$. Depending on $\mu$, there exist one, two, or three black holes with the same temperature. }
\label{TempECG}
 \end{figure}

% For instance, one can see that, as a consequence of turning on the higher-order parameter $\mu$, the temperature does not diverge when $\rh\rightarrow 0$, even if $\mu$ is really small. Instead, one has $T\approx \frac{\rh}{2\pi\sqrt{3\mu}L^2}$, so that it vanishes when $\rh\rightarrow 0$.
Besides this, the introduction of the cubic term in the action leads to some additional differences with respect to Einstein gravity --- see Fig. \ref{TempECG}. For the usual Schwarzschild-AdS$_4$ Einstein gravity black hole, the temperature is always higher than a certain value,  $ T>T_{\rm min}\equiv\sqrt{3}/(2\pi L)$. In that case, for a given $T>T_{\rm min}$, there exist two black holes, one large, and one small. There are no solutions for which $T<T_{\rm min}$. For ECG the situation is quite different. On the one hand, one observes that there is no minimum temperature, this is, as long as $\mu\neq 0$, there always exists at least one black hole solution for a given $T$. We can distinguish two qualitatively different behaviors depending on $\mu$. For $0<\mu<\mu_{T}\equiv 1/288$, there is an interval %\footnote{The dependence of $T_{\rm min}$ and $T_{\rm max}$ on $\mu$ can be analytically obtained by finding the roots of the following cubic equation: $3 z^3 + (36 \mu- 1) z^2 + 12 \mu z + \mu =0$, which, in the interval under consideration, has three real roots, two of them positive and one negative. The minima and maxima are given by the square root of the two positive ones.} 
of temperatures $\left(T_{\rm min},T_{\rm max}\right)$ for which three black hole solutions with the same temperature exist. However, if $T\ge T_{\rm max}$ or $T<T_{\rm min}$. we just have one. On the other hand, if  $\mu>\mu_T$, there is always a single black hole solution for each temperature. In the critical limit, for which $f(r)=3(r^2-\rh^2)/(2L^2)$, the relation \req{eq:temperatureBHsECG} becomes linear \cite{Feng:2017tev}, and reads $T=3\rh/(4\pi L^2)$.

%\begin{itemize}
%\item When $0<\mu<\mu_{T}=1/288$, there is an interval\footnote{The dependence of $T_{\rm min}$ and $T_{\rm max}$ with $\mu$ can be analytically obtained by finding the roots of the following cubic equation: $3 z^3 + (36 \mu- 1) z^2 + 12 \mu z + \mu =0$, which, in the interval under consideration, has three real roots, two of them positive and one positive. The minima and the maxima are given by the square root of the two positive ones.}$\left(T_{\rm min},T_{\rm max}\right)$ where we have three black hole solutions with the same temperature. If $T\ge T_{max}$ or $T<T_{min}$ we just have one. 
%\item If $\mu>\mu_T$, we find that there is only one black hole solution at a given temperature. 
%\end{itemize}
%This is illustrated in fig. \ref{TempECG}.

% Besides, we see that for the critical vale $\mu=4/27$ the relation \req{eq:temperatureBHsECG} becomes linear: $T=\frac{3\rh}{4\pi L^2}$. In fact, it was found by Lu et al \comment{Cite here} that in the critical limit the solution \req{Spheric} becomes 
%\begin{equation}
%ds^2_{\rm cr}=-\frac{3(r^2-\rh^2)}{2L^2}dt^2+\frac{2L^2dr^2}{3(r^2-\rh^2)}+r^2d\Omega^2_{(2)}\, ,
%\end{equation}
%whose temperature precisely agrees with the previous value. 
 
In sum, at a fixed temperature $T$, we have several solutions with $\mathbb{S}^1_{\beta}\times \mathbb{S}^2$ boundary geometry: thermal AdS$_4$, and one or three black holes depending on the value of $\mu$. In order to identify which phase dominates the holographic partition function at each temperature, let us again compare the on-shell actions of the solutions. For thermal AdS$_4$, one finds a vanishing result, whereas for the black holes, the result can be obtained from \req{fecg}, from which we can obtain $I_E(T)$ implicitly using \req{eq:temperatureBHsECG}. 
\begin{figure}[t!]
	\centering 
	\includegraphics[width=0.47\textwidth]{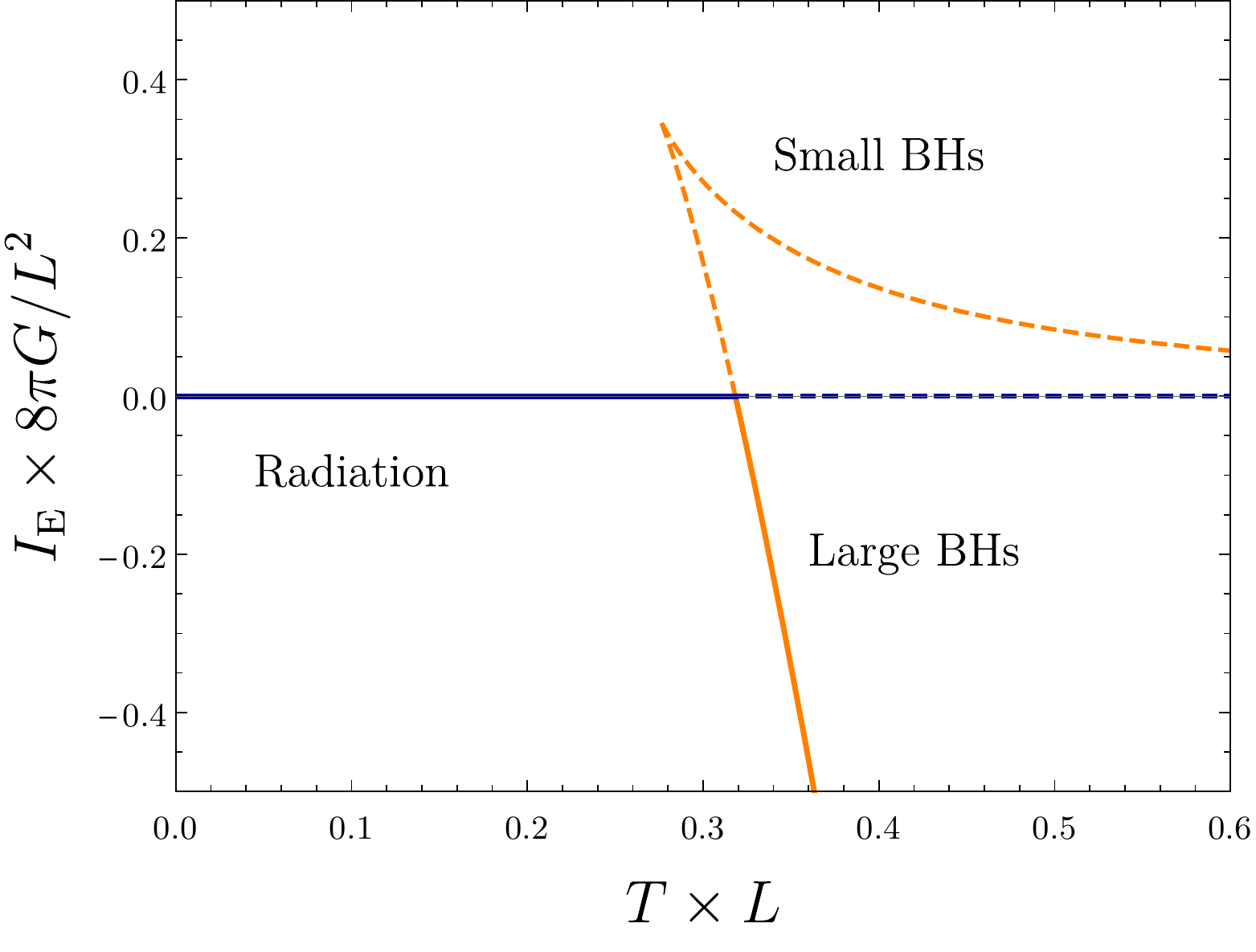}
	\includegraphics[width=0.47\textwidth]{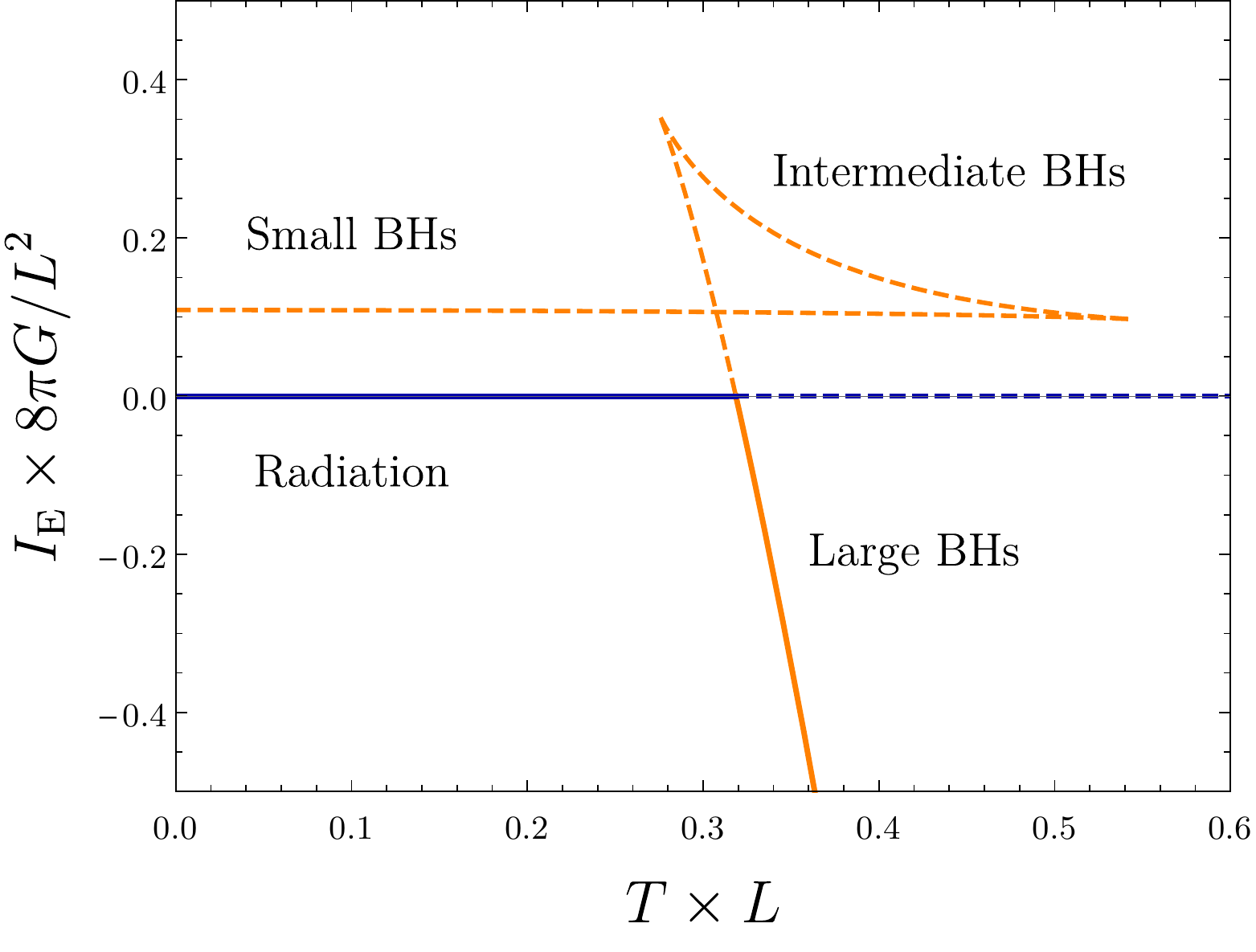}
	\includegraphics[width=0.47\textwidth]{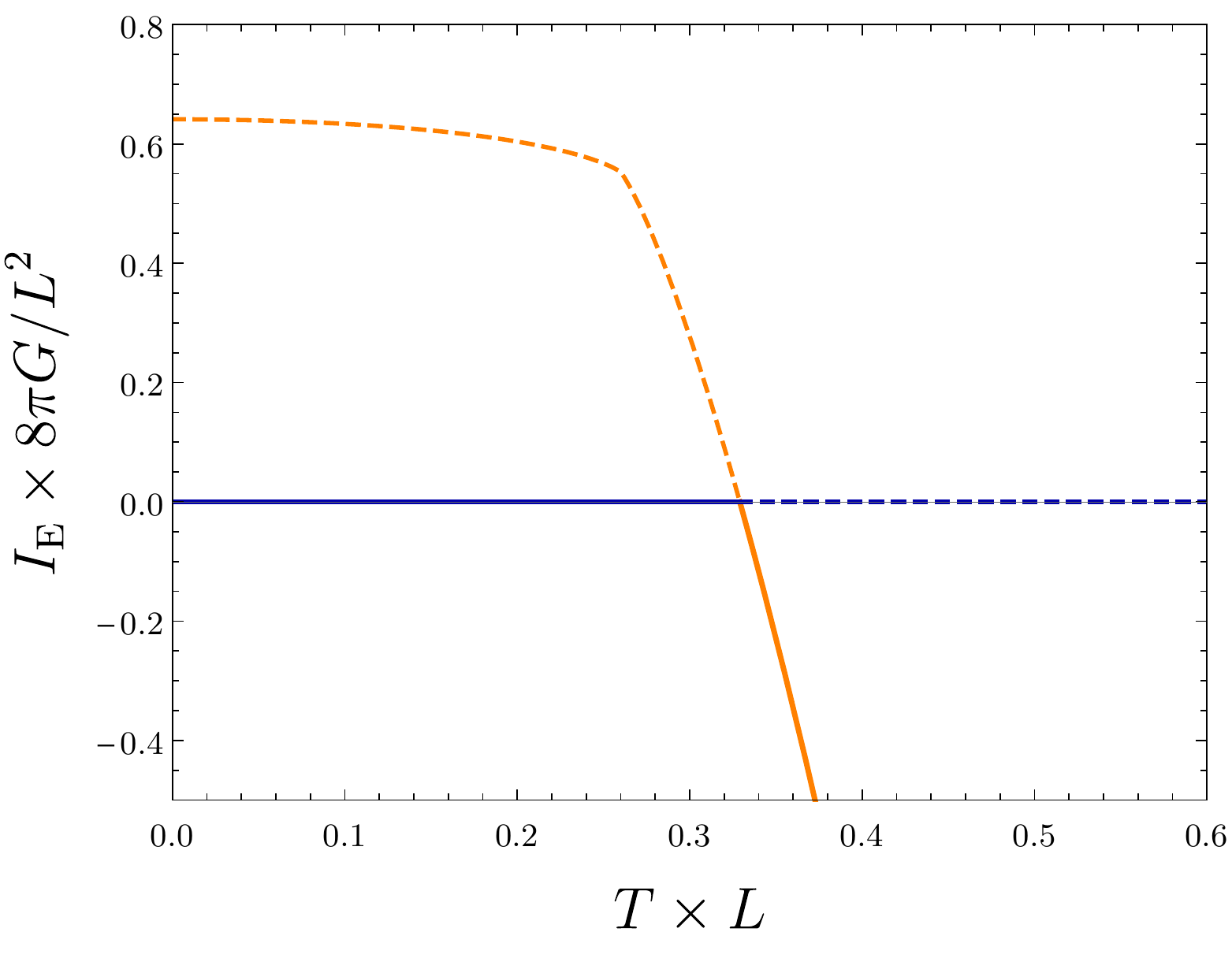}
	\includegraphics[width=0.46\textwidth]{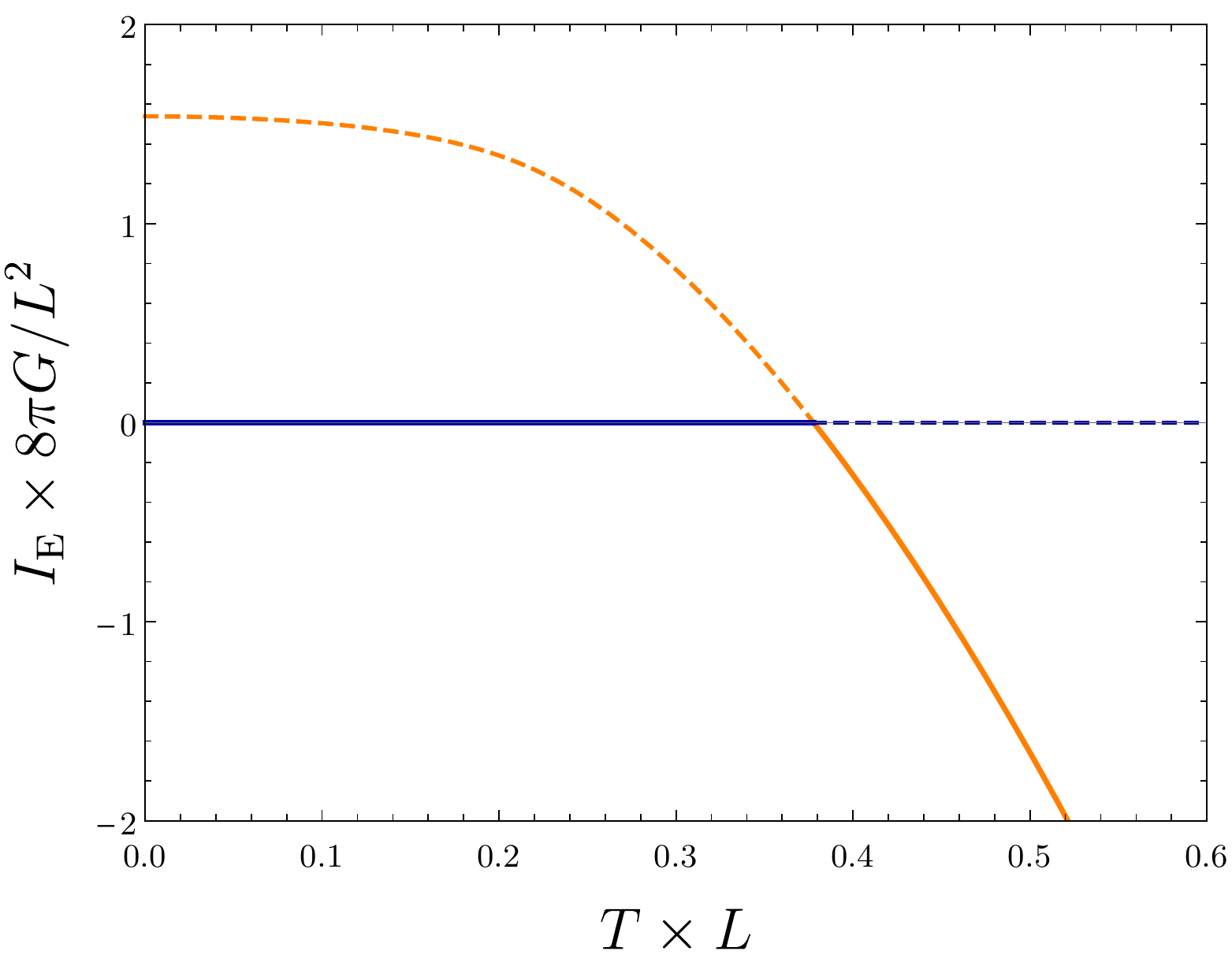}
	\caption{We plot $I_E$ as a function of the temperature for the different phases of holographic ECG in $\mathbb{S}^1_{\beta}\times \mathbb{S}^2$. Solid lines represent the dominant phase in each case. Blue lines correspond to thermal AdS$_4$, and orange lines to black holes. From left to right and top to bottom: $\mu=  0,\, 0.0001,\, 1/288,\, 0.02$. For $\mu=0$ we get the usual Einstein gravity result, with two orange branches corresponding to small and large black holes, and a Hawking-Page transition at $T_{\rm HP}=1/(\pi L)$. For $0<\mu<1/288$, there exist either one or three black-hole branches, depending on the temperature, while for $\mu>1/288$ there is a single black hole for every temperature. As $\mu$ approaches the critical value, the Hawking-Page transition temperature grows as $T_{\rm HP}\sim 3/(2\pi L \sqrt{1-27\mu/4})$. In the limit $\mu=4/27$, the on-shell action is constant (not shown in the figure), $I_E=4\pi L^2/(3G)$, so thermal AdS$_4$ always dominates, and there is no Hawking-Page transition.}\label{HP}
\end{figure}
In Fig.  \ref{HP}, we plot $I_E$ for various values of $\mu$. At a given temperature, we always have several possible phases: a pure thermal vacuum (radiation), and one or several black holes. The dominating phase (shown in solid line) is the one with smaller on-shell action. Regardless of the value of $\mu$, the qualitative behavior is always the same: for small temperatures, the partition function is dominated by radiation, while for large enough temperatures there is a Hawking-Page phase transition \cite{Hawking:1982dh,Witten:1998zw} to a large black hole. The temperature at which the transition occurs depends on $\mu$. For Einstein gravity, one finds $T_{\rm HP}=1/(\pi L)$, while for $\mu \ll 1$, this result gets corrected as 
\begin{equation}
T_{\rm HP}=\frac{1+10 \mu}{\pi L}+\mathcal{O}(\mu^2)\, .
\end{equation}
Hence, the introduction of the ECG density increases the temperature at which the transition occurs. The black-hole radius for which the phase transition takes place also grows if we turn on $\mu$, and is given by $\rh=L(1+26\mu +\mathcal O(\mu^2))$, and the same happens with the latent heat, $\delta Q=L/G\cdot \left(1+38\mu+\mathcal{O}(\mu^2)\right)$. As we increase $\mu$, the Hawking-Page transition temperature grows. In fact, it diverges in the critical limit $\mu=4/27$, which means that no transition at all occurs in that case. If we define $\epsilon\equiv 1-27/4\mu$, the transition temperature for $\epsilon \ll 1$ can be seen to be given by
\begin{equation}
T_{\rm HP}= \frac{3}{2\pi L \sqrt{\epsilon}}\left[1-\frac{\epsilon}{4}+\mathcal{O}(\epsilon^2)\right] \, ,\quad \text{which occurs for}\quad \rh= \frac{2L }{\sqrt{\epsilon}}\left[1-\frac{1}{4}\epsilon+\mathcal{O}(\epsilon^2)\right]\, .
\end{equation}
The reason for the disappearance of the transition is that the critical black holes have a temperature-independent on-shell action, namely\footnote{The fact that the on-shell action of black holes does not depend on the horizon size is yet another unusual property of the critical theory.}

\begin{equation}
I_E=\frac{4\pi L^2}{3G}\left[1-\frac{9+8\pi^2 T^2L^2}{18}\epsilon + \mathcal O(\epsilon^2)\right] \, ,
\end{equation}
which in the $\epsilon=0$ limit is a positive constant, therefore greater than the thermal AdS$_4$ value.\footnote{As $\epsilon \rightarrow 0$, the latent heat also diverges as $\delta Q=4 L/ (G\sqrt{\epsilon})\cdot \left(1-3\epsilon/4+\mathcal{O}(\epsilon^2)\right)$, although the entropy increase tends to a constant value, $\delta S=8L^2\pi/(3G)\cdot \left(1-\epsilon/2+\mathcal{O}(\epsilon^2)\right)$.}
Let us mention that the thermodynamic behavior of our black holes is qualitatively similar to the one observed for $D=5$ Gauss-Bonnet black holes \cite{Cai:2001dz}\footnote{See also \cite{Cho:2002hq} for the case of general quadratic gravity --- the analysis becomes perturbative in that case though.}. Just like for ECG, a new phase of small stable black holes appears also in that case, as a consequence of the Gauss-Bonnet term. Again, thermal AdS$_5$ is always globally preferred over such solutions. Observe also that the fact that there is no phase transition for critical ECG seems to be related to the fact that, in that case, the solutions become `very similar' to three-dimensional BTZ black holes (see footnote \ref{BTZ}), for which no Hawking-Page transition exists either \cite{Myung:2005ee}.

Although we have seen that only radiation and large black holes can dominate the partition function, the addition of a Gauss-Bonnet term in the action can drastically change the previous picture, as noted in Refs.~\cite{Mir:2019ecg,Mir:2019rik}. In fact, we already saw in Chapter~\ref{Chap:4} that the entropy of ECG black holes becomes negative in the zero-mass limit, but that this can be avoided if we introduce a topological GB term. Thus, by considering the Lagrangian $\mathcal{L}\rightarrow \mathcal{L}+\frac{\alpha L^2}{16\pi G}\mathcal{X}_4$, the entropy of spherical black holes is modified according to $S\rightarrow S+2\pi \alpha L^2/G$, while the on-shell Euclidean action is shifted in the opposite way $I_E\rightarrow I_{E}-2\pi \alpha L^2/G$. It is worth emphasizing that such constant contributions only appear in the presence of a spherical horizon, and in particular, the GB term does not change the on-shell action of pure radiation, which is still $I_E^{\rm radiation}=0$. Then, in the zero-size limit, the entropy and on-shell action of spherical black holes reads
\begin{equation}
\lim_{r_h\rightarrow 0}I_E^{\rm BHs}=-\lim_{r_h\rightarrow 0}S=(\sqrt{3\mu}-\alpha)\frac{2\pi L^2}{G}\, .
\end{equation}
Thus, it seems reasonable to choose $\alpha=\sqrt{3\mu}$, so that the entropy vanishes in that limit. The effect of that subtraction in the diagrams shown in Fig.~\ref{HP} is to shift the on-shell action of BHs so that for $T=0$ it vanishes. As a consequence, radiation never dominates the partition function, because for any positive temperature there is a phase of black holes with $I_{E}<0$. In that case, the phase space of the theory is qualitatively different depending on the value of $\mu$. 
For $\mu<\mu_{T}=1/288$ we would have a Hawking-Page transition between large and small black holes (this would correspond to the top right diagram in Fig.~\ref{HP}).
On the other hand, the phase space has a critical point (not to be confused with the critical limit of the theory) where the three black-hole phases in Fig. \ref{HP} (top right) stop existing separately. This occurs for $\mu=\mu_T$ which separates the cases for which there are three phases, from those for which there is only one. The phase transition is of second-order, and takes place at a temperature $T_{\rm c}=\frac{\sqrt{2/3}}{\pi L}$, corresponding to the non-smooth point on the dashed orange curve in Fig. \ref{HP} bottom left. The critical exponent of the specific heat at the transition turns out to be $-2/3$. More precisely, we find
	% As observed previously, this happens when $\mu\ge \mu_T$. For $\mu>\mu_T$ there are no different phases, but precisely at the critical point $\mu=1/288$ we find a second-order phase transition. This can be clearly observed as the non-smooth point in Fig. \ref{HP} bottom left. The phase transition occurs at a temperature $T_c=\frac{\sqrt{2/3}}{\pi L}$ and it can be shown that the critical exponent of the specific heat $(C=-T \partial^2F/\partial T^2)$ is $-2/3$. More exactly, we find
\begin{equation}
C\equiv -T\frac{\partial^2 F}{\partial T^2}=\frac{\pi 5^{4/3} L^2}{9\cdot 2^{7/3}G}\left|\frac{T}{T_{\rm c}}-1\right|^{-2/3}\, \quad \text{as} \quad T\rightarrow T_{\rm c}\, .
\end{equation}
Finally, for $\mu>\mu_{T}$ there is a single phase of black holes at every temperature and it always dominates the partition function.

In passing, let us point out that more sophisticated phase transitions connecting different AdS vacua have been identified for Lovelock gravities in various dimensions \cite{Camanho:2013uda}. It would be interesting to explore their possible existence in ECG or, more generally, for the class of theories introduced in  \cite{Ahmed:2017jod,PabloPablo4,Hennigar:2017ego}.

\section{R\'enyi entropy}\label{renyie}
R\'enyi entropies \cite{renyi1961,renyi1} are useful probes of the entanglement structure of quantum systems --- see \eg \cite{HoloRen,Klebanov:2011uf,Laflorencie:2015eck}, and references therein. Roughly speaking, given a state $\rho$ and some spatial subregion $V$ in a QFT, R\'enyi entropies characterize `the degree of entanglement' between the degrees of freedom in $V$ and those in its complement (when such a bi-partition of the Hilbert space is possible). More precisely, they are defined as
\begin{equation}\label{rr}
S_q(V)=\frac{1}{1-q}\log \Tr \rho_V^q \, , \quad q\geq 0\, ,
\end{equation}
where $\rho_V$ is the partial-trace density matrix obtained integrating over the degrees of freedom in the complement of the entangling region. Whenever \req{rr} can be analytically continued to $q\in\mathbb{R}$, the corresponding EE can be recovered as the $q\rightarrow 1$ limit of $S_q$.

In this section we use the methods developed in \cite{CHM,HoloRen} to compute the R\'enyi entropy for disk regions in the ground state of holographic ECG. In Subsection \ref{ere}, we study the dependence of $S_q/S_1$ on $\mu$, as well as on some of the charges characterizing the CFT. In Subsection \ref{twist}, we compute the conformal scaling dimension of twist-operators $h_q$ for ECG --- see below for definitions --- as an intermediate step to obtain in Subsection \ref{t44}, using the results in \cite{Chu:2016tps}, the charge $t_4$ characterizing the three-point function of the stress tensor.

\subsection{Holographic R\'enyi entropy}
\label{ere}
In \cite{CHM}, it was shown that the entanglement entropy across a radius-$R$ spherical region $\mathbb{S}^{d-2}$ for a generic $d$-dimensional CFT equals the thermal entropy of the theory at a temperature $T_0=1/(2\pi R)$ on the hyperbolic cylinder $\mathbb{R}\times \mathbb{H}^{d-1}$, where the curvature scale of the hyperbolic planes is given by $R$.
%\begin{equation}
%\see = S_{\rm thermal}
%\end{equation}
The result is particularly useful in the holographic context, where the latter can be computed as the Wald entropy of pure AdS$_{(d+1)}$ foliated by $\mathbb{R}\times \mathbb{H}^{d-1}$ slices\footnote{Observe that this means, in particular, that for odd-dimensional holographic CFTs, we can in principle access $a^*$ --- see \req{asta} --- in three different ways: 1) from an explicit EE calculation using the Ryu-Takayanagi functional \cite{Ryu:2006bv,Ryu:2006ef} or its generalizations, \eg \cite{Fursaev:2013fta,Dong:2013qoa,Camps:2013zua}, depending on the bulk theory; 2) from the Euclidean on-shell action of pure AdS$_{(d+1)}$ with $\mathbb{S}^{d}$ boundary \cite{CHM}; 3) from the Wald entropy of AdS$_{(d+1)}$ with $\mathbb{R}\times \mathbb{H}^{d-1}$ boundary \cite{CHM}.}. Later, in \cite{HoloRen}, it was argued that this result could be in fact extended to general R\'enyi entropies, the result being 
\begin{equation}\label{sqq}
S_q=\frac{q}{(1-q)T_0}\int_{T_0/q}^{T_0} S_{\rm \ssc thermal}(T) dT\, ,
\end{equation}
where $S_{\rm \ssc thermal}(T)$ is the corresponding thermal entropy on $\mathbb{R}\times \mathbb{H}^{d-1}$ at temperature $T$. While for $T=T_0$, general results for the EE across a spherical region can be obtained for arbitrary holographic higher-derivative theories as long as AdS$_{(d+1)}$ is a solution, the situation becomes more involved for general $q$. In that case, \req{sqq} requires that we know $S_{\rm \ssc thermal}(T)$ for arbitrary values of $T$. Holographically, the calculation can only be performed if the bulk theory admits hyperbolic black-hole solutions for which we are able to compute the corresponding thermal entropy. Examples of such theories for which R\'enyi entropies have been computed using this procedure include: Einstein gravity, Gauss-Bonnet, QTG \cite{HoloRen} and cubic Lovelock \cite{Puletti:2017gym}. Analogous studies for theories in which the corresponding black holes solutions were only accessible approximately --- typically at leading order in the corresponding gravitational couplings --- have also been performed, \eg in \cite{Galante:2013wta,Belin:2013dva,Dey:2016pei}. 
ECG allows us to perform the first exact calculation (fully nonperturbative in the gravitational couplings)  of the holographic R\'enyi entropy of a disk region in $d=3$ for a bulk theory different from Einstein gravity. 

Following \cite{HoloRen}, let us start by rewriting \req{sqq} as 
\begin{equation}\label{sex}
S_q=\frac{q}{(q-1)T_0}\left[\left. S(x)T(x) \right|^1_{x_q}-\int_{x_q}^1S'(x) T(x)dx \right]\, ,
\end{equation}
where we defined the variable $x\equiv \rh/\tilde{L}$, and where $S$ and $T$ stand for the thermal entropy and temperature of the hyperbolic AdS black hole of the corresponding theory. For $x=1$, one has, in general $T(1)=T_0$, whereas $x_q$ is defined as a solution to the equation $T(x_q)=T_0/q$. 
For ECG cubic gravity, the expressions for $S(x)$ and $T(x)$ can be extracted from \req{sth} and \req{T} respectively by setting $k=-1$,
\begin{align}\notag
S(x)&=\frac{x^2 \tilde{L}^2 V_{\mathbb{H}^2}}{4G}  \left[1-\frac{3 \mu f_{\infty}^2  \left(\frac{3x^2 }{f_{\infty}}-1\right)\left[  \left(\frac{3x^2 }{f_{\infty}}-1\right)-2 \left[1+\sqrt{1-\frac{3 f_{\infty}^2\mu}{x^4}\left(\frac{3x^2 }{f_{\infty}}-1\right)} \right] \right]}{x^4\left[1+\sqrt{1-\frac{3 f_{\infty}^2\mu}{x^4}\left(\frac{3x^2 }{f_{\infty}}-1\right)} \right] ^2}\right] \, , \\ \label{sx}
T(x)&=\frac{1}{2\pi R x} \left(\frac{3x^2 }{f_{\infty}}-1\right)\left[1+\sqrt{1-\frac{3 f_{\infty}^2 \mu}{x^4}\left(\frac{3x^2}{f_{\infty}}-1 \right)} \right]^{-1} \, ,
\end{align}
where, in addition, we have set $N^2=L^2/(f_{\infty}R^2)$. This makes the boundary metric conformally equivalent to 
\begin{equation}
ds^2_{\rm bdy}=-dt^2+R^2 d\Xi^2\, ,
\end{equation}
so that the boundary theory lives on $\mathbb{R}\times \mathbb{H}^{2}$, with the `radius' of the hyperbolic plane given by $R$, as required \cite{HoloRen}. From \req{sx}, it can be seen that $x_q$ corresponds to the real and positive solution of
%\begin{equation}\label{eq:x_q 1}
%\sqrt{1-\frac{3 f_{\infty}^2\mu}{x_q^4}\left(\frac{3x_q^2 }{f_{\infty}}-1\right)}=\frac{q}{x_q}\left(\frac{3x_q^2 }{f_{\infty}}-1\right)-1 \ ,
%\end{equation}
%which can also be written as 
%\begin{equation}\label{eq:x_q 2}
%x_q^2\left(\frac{3q^2x_q^2}{f_{\infty}}-q^2-2qx_q\right)=-3\mu f^2_{\infty}	\ ,
%\end{equation}
\begin{equation}\label{eq:x_q 2}
x_q^2\left(3q^2x_q^2-q^2-2qx_q\right)=3\mu f^2_{\infty}\left(q^2 x_q^4-1 \right)	\ ,
\end{equation}
which for Einstein gravity reduces to 
\begin{equation}
x^{\rm E}_{q}=\frac{1}{3q}\left(1+\sqrt{1+3q^2} \right)\ .
\end{equation}

Observe that we have not said anything yet about the divergent nature of $V_{\mathbb{H}^2}$. Of course, one expects the entanglement and R\'enyi entropies to contain (a particular set of) divergent terms, so one could have only expected some source of divergences to appear in the calculation. It is a remarkable feature of the procedure outlined above that all necessary divergent terms in the R\'enyi entropy  (and no others)  are produced by the volume of the hyperbolic plane. In the case of interest for us, corresponding to $d=3$, the regularized volume reads \cite{CHM}
\begin{equation}\label{reguV}
V_{\mathbb{H}^2}=2\pi \left[\frac{R}{\delta}-1 \right]\, ,
\end{equation}
	where we introduced a short-distance cut-off $\delta$. From this expression, we shall only retain the universal piece\footnote{As stressed in \cite{Casini:2015woa}, the universality of constant terms comes with a grain of salt. For example, in \req{reguV}, one could think of rescaling $R$ by an order-$\delta$ constant, which would pollute the constant term. In the case of EE, this issue was overcome in \cite{Casini:2015woa} using mutual information as a regulator. We will ignore this problem here.}, and hence we will replace $V_{\mathbb{H}^2} \rightarrow -2\pi$ from now on, keeping in mind that $S_q$  also contains a cut-off dependent `area' law piece.  
	Taking this into account, after some massaging, which includes using \req{roo}, we can check that 
\begin{equation}
T(1)=T_0\, \quad \text{and} \quad S(1)=-2\pi a^{*\rm ECG}\, ,
\end{equation}
where $a^{*\rm ECG}$ was defined in \req{aae}. Hence, we obtain the same result for the EE of a disk as the one found in Sec.~\ref{osa} from the free energy of holographic ECG on $\mathbb{S}^3$. This is another check of our proposed generalized action \req{EuclideanECGc6}.
%\begin{equation}
%S(x_q)=\frac{x_q^2 \tilde{L}^2}{4G}V_{\mathbb{H}^2}\left[1-\frac{3\mu f_{\infty}^2}{x_q^3 q^2}\left(x_q-2q\right)\right]
%\end{equation}

With all the above information together, we are ready to evaluate the R\'enyi entropy from \req{sex}. The result reads
\begin{equation}\label{reni}
S^{\rm ECG}_q=\frac{q}{(1-q)}\frac{\pi \tilde L^2}{2G}\left[1-x_q-\frac{x^2_q}{q}+x_q^3-\mu f_\infty^2 \left(\frac{3}{q^2 x_q}-3-\frac{1}{q^3}+x_q^3\right)\right]\, ,
\end{equation}
which reduces to the Einstein gravity one \cite{HoloRen} for $\mu=0$. In Fig. \ref{figr1}, we plot $S_q/S_1$ as a function of the R\'enyi index for various values of $\mu$.
\begin{figure}[t!]
	\centering 
	\includegraphics[width=0.65\textwidth]{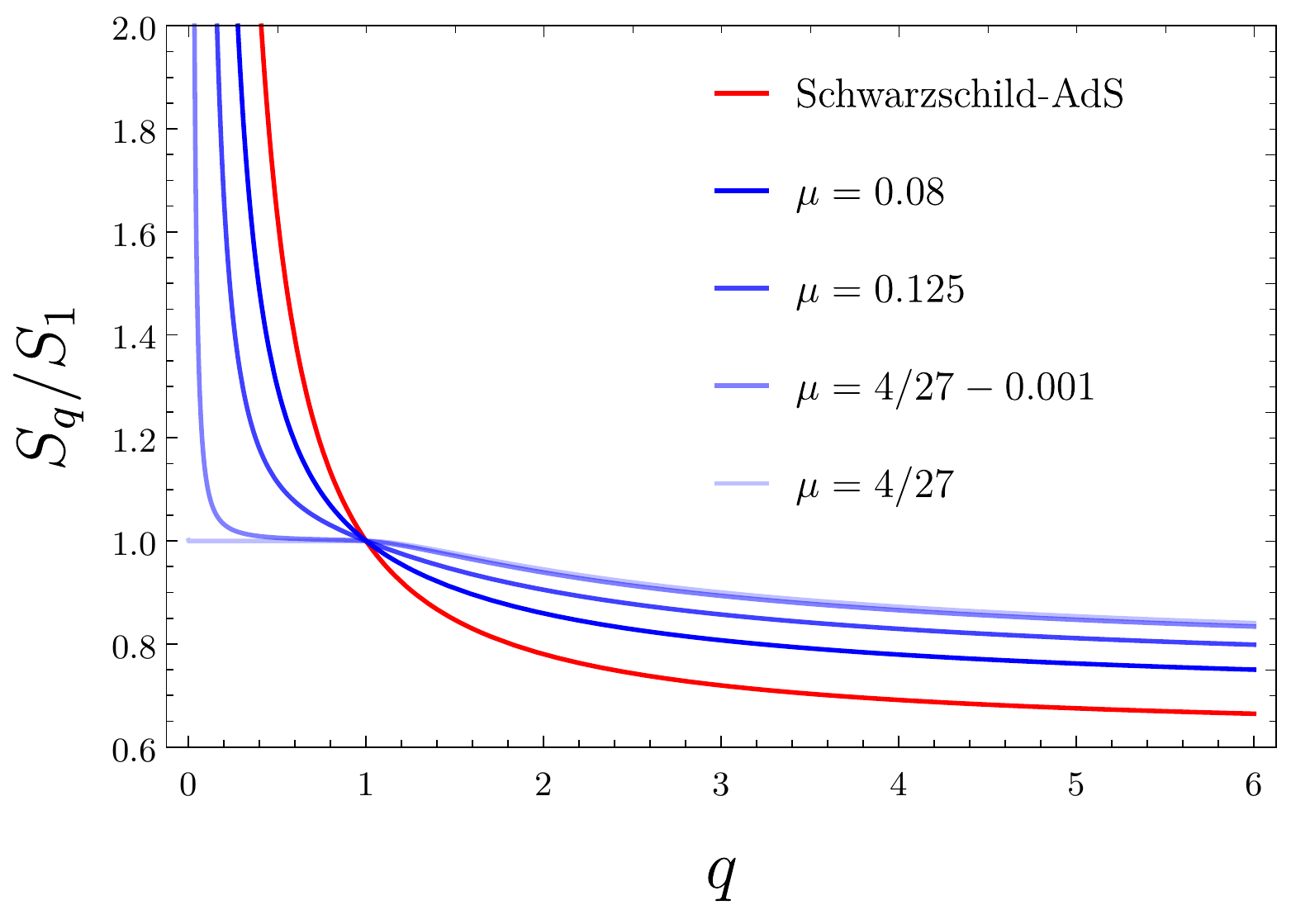}
	\caption{We plot the ratio of the R\'enyi entropy and the EE, $S_q/S_1$, as a function of the R\'enyi index $q$ for various values of the ECG coupling $\mu$.}
	\labell{figr1}
\end{figure}
As we increase $\mu$, $S_q/S_1$ becomes smaller in the $q<1$ region, but it remains larger than $1$ for all values of $\mu$. The opposite occurs for  $q>1$, where $S_q/S_1$ tends to grow as we increase $\mu$, but $S_q/S_1<1$ for all $\mu$.   In the critical limit, there is a jump, and $S_q/S_1$ no longer diverges near $q=0$. In fact, in that case, \req{reni} reduces to a $q$-independent constant for $q<1$, $S^{\rm crit.\, ECG}_q=-\pi \tilde{L}^2/G$. As we approach $\mu\rightarrow 4/27$, $S_{\infty}$ tends to another constant, $S_{\infty}\rightarrow -\pi \tilde{L}^2/G\times(1-1/(3\sqrt{2}))$. Note also that the curve is no longer concave for  $\mu \sim 0.135$ or larger.

 Explicit Taylor expansions of $S^{\rm ECG}_q$ around $q=\{0,1,\infty\}$ can be easily obtained. A few terms suffice in such expansions to provide excellent approximations to the exact curve for most values of $\mu$. At leading order we find, respectively,
\begin{eqnarray}
\lim_{q\rightarrow 1}S^{\rm ECG}_q&=&-2\pi a^{*\rm ECG}\ , \\ \label{peniss}
\lim_{q\rightarrow 0}S^{\rm ECG}_q&=&-\frac{1}{6\pi q^2}\cs^{\rm ECG}\ , \\
\lim_{q\rightarrow \infty}S^{\rm ECG}_q&=&-\frac{\pi \tilde L^2 }{2G}\left[1+3\mu f^2_\infty-\frac{2}{3\sqrt{3(1-\mu f_{\infty}^2)}}\right]\ .
\end{eqnarray}
The first result corresponds to the EE, and we have mentioned it already. As for the second, the appearance in the $q\rightarrow 0$ regime of the thermal entropy charge $\cs^{\rm ECG}$, identified in Sec.~\ref{cssex},  should not come as a surprise either. The reason is the following. As shown in \cite{HoloRen}, the R\'enyi entropy $S_q$ across a $\mathbb{S}^{d-2}$ in a general CFT$_d$ can be alternatively written as 
\begin{equation}\label{sqqs}
S_q=\frac{q}{(1-q)}\frac{R^{d-1}V_{\mathbb{H}^{d-1}}}{T_0}\left[\mathcal{F}(T_0)-\mathcal{F}(T_0/q)\right]\, ,
\end{equation}
where $\mathcal{F}(T)$ is the free energy density of the theory at temperature $T$ on $\mathbb{R}\times \mathbb{H}^{d-1}$. The point is that, as $q\rightarrow 0$, the second term in \req{sqqs} dominates over the first. Then, one can use the fact that, at high temperatures, the free energy density on $\mathbb{R}\times \mathbb{H}^{d-1}$ tends to the free energy density on $\mathbb{R}^d$ \cite{Swingle:2013hga}, $\mathcal{F}_{\mathbb{R}\times \mathbb{H}^{d-1}}(T)=\mathcal{F}_{\mathbb{R}^d}(T) \left[1+\mathcal{O}(1/(RT)^2)\right]$, since $1/R$ becomes irrelevant compared to $T$ in that regime. Using the general relation $\mathcal{F}_{\mathbb{R}^d}(T)=-\cs T^d/d$, valid for any CFT in flat space, it follows then that\footnote{See \cite{Bueno3,Galante:2013wta} for analogous arguments.}
\begin{equation}
\lim_{q\rightarrow 0}S_q=\frac{V_{\mathbb{H}^{d-1}}\cs }{d}\left(\frac{1}{2\pi q} \right)^{d-1} \, ,
\end{equation}
which should hold for any CFT$_d$ and, in particular, precisely agrees with \req{peniss} for ECG. Besides, we can readily check that
\begin{equation}
\left. \partial_q S^{\rm ECG}_q\right|_{q=1}=\frac{\pi^4}{12}\ctt^{\rm ECG}\, ,
\end{equation}
as expected from the general relation found in \cite{Perlmutter:2013gua}.

\begin{figure}[t!]
	\centering 
	\includegraphics[width=0.65\textwidth]{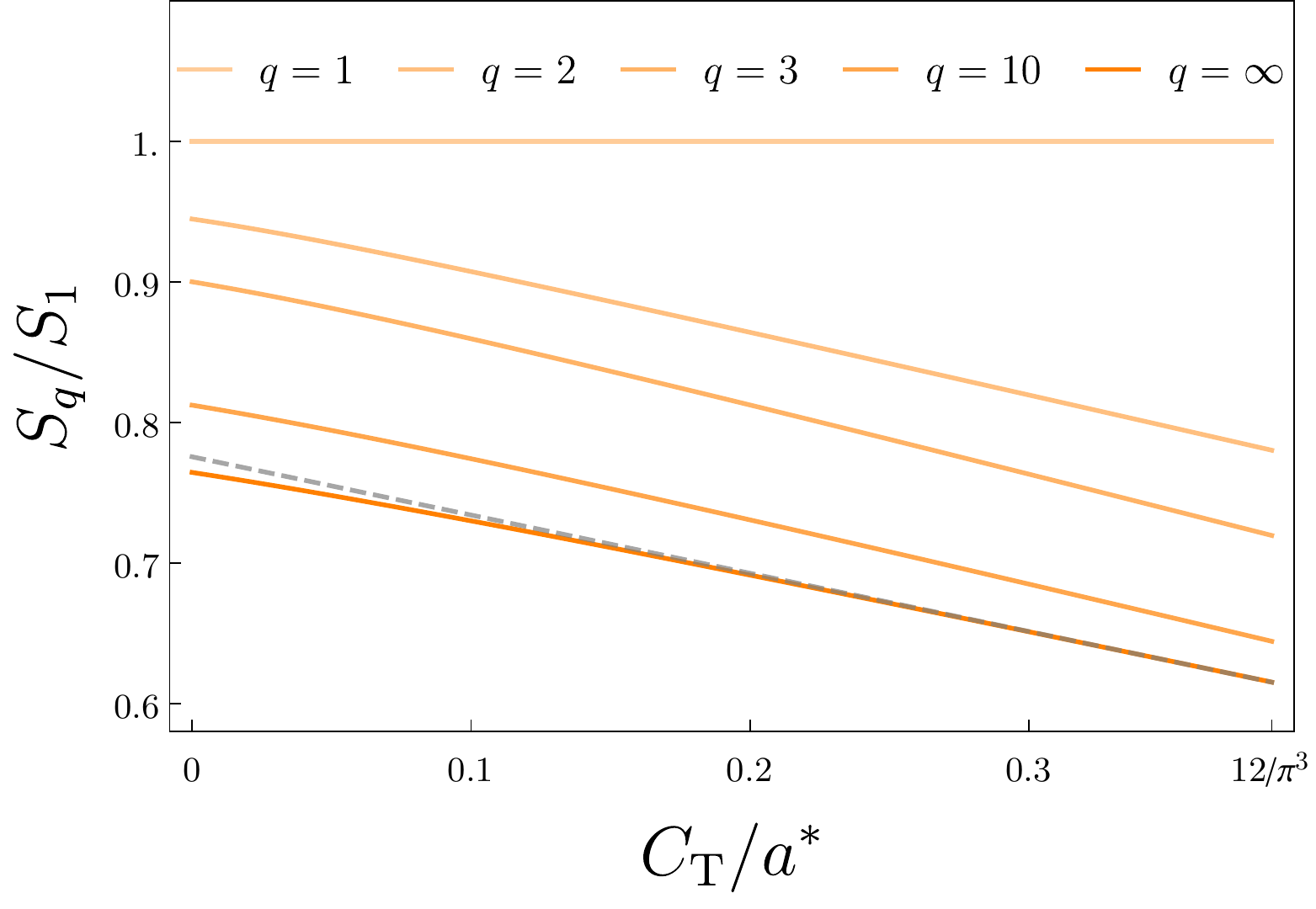}
	\caption{We plot the ratio of the R\'enyi entropy and the EE, $S_q/S_1$, as a function of the quotient $(\ctt/a^*)^{\rm ECG}$ for $q=\{1,2,3,10,\infty\}$. The limits of the range plotted correspond to the critical theory, $(\ctt/a^*)^{\rm ECG}=0$, and Einstein gravity $(\ctt/a^*)^{\rm ECG}=12/\pi^3$, respectively. The dashed line corresponds to the linear approximation to $S_{\infty}/S_1$ in \req{lin}.}
	\labell{figr2}
\end{figure}

Let us now gain some insight on the dependence of $S_q$ on quantities characterizing the CFT. In order to do that, we can use the relations
\begin{equation}
\frac{\tilde{L}^2}{G}= a^* \frac{\pi^3}{6}  \left[(\ctt/a^*)+\frac{12}{\pi^3} \right]\, ,\quad \mu f_{\infty}^2=-\frac{1}{3}\frac{\left[(\ctt/a^*)-\frac{12}{\pi^3} \right]}{\left[(\ctt/a^*)+\frac{12}{\pi^3}\right] }\, .
\end{equation}
It is then straightforward to substitute these in \req{eq:x_q 2} and \req{reni} to obtain $S_q$ as a function of $a^{*\rm ECG}$ and $(\ctt/a^*)^{\rm ECG}$.  Observe that $a^*$ appears as a global factor, so that $S_q/S_1$ is a function of $\ctt/a^*$ alone. We plot this ratio for several values of $q$ in Fig. \ref{figr2}. Observe that $\ctt/a^*$ takes values between $0$ and $12/\pi^3\simeq 0.3870$, corresponding to the critical value, $\mu=4/27$, and Einstein gravity respectively. Interestingly, even though the dependence of $S_q/S_1$ on  $\ctt/a^*$ is in principle highly non-linear, all curves seem to be approximately linear in the full range. In addition, we find that 
\begin{equation}
\left. \frac{S_q}{S_1}\right|_{({\ctt}_1/a_1^*)}<\left. \frac{S_q}{S_1}\right|_{({\ctt}_2/a_2^*)}\,  \quad \text{for} \quad ({\ctt}_1/a_1^*)> ({\ctt}_2/a_2^*)\, ,
\end{equation} 
\ie $S_q/S_1$ monotonously decreases as $\ctt/a^*$ grows, for all values of $q$. These features are very similar to the ones observed in \cite{HoloRen} for holographic Gauss-Bonnet in $d\geq 4$. 

We can gain some understanding on the approximately linear behavior of $S_q/S_1$ by expanding $S_{\infty}/S_1$ around the Einstein gravity value,  $(\ctt/a^*)^{\rm ECG}=12/\pi^3$. By doing so, we obtain
%\begin{align}\label{lin}
%\frac{S_{\infty}}{S_1}=1-\frac{1+\frac{\pi^3(\ctt/a^*)}{12}}{6\sqrt{\frac{\left[2(\ctt/a^*)+\frac{12}{\pi^3}\right]}{2\left[(\ctt/a^*)+\frac{12}{\pi^3}\right]}}} \simeq 
%\left[1-\frac{2}{3\sqrt{3}}\right]-\frac{5\pi^3 }{216\sqrt{3}} \left[(\ctt/a^*)-\frac{12}{\pi^3}\right]+\dots\, ,
%\end{align}
\begin{align}\label{lin}
	\frac{S_{\infty}}{S_1}=1-\frac{\pi^3  \left[(\ctt/a^*)+\frac{12}{\pi^3}\right]^{3/2}}{72  \left[(\ctt/a^*)+\frac{6}{\pi^3}\right]^{1/2}} \simeq 
	\left[1-\frac{2}{3\sqrt{3}}\right]-\frac{5\pi^3 }{216\sqrt{3}} \left[(\ctt/a^*)-\frac{12}{\pi^3}\right]+\dots\, ,
\end{align}
where the first omitted correction is quadratic in the expansion parameter. As it turns out, the linear approximation in \req{lin} fits the exact curve very well for most values of $\ctt/a^*$ --- see dashed line in Fig. \ref{figr2}. We suspect a similar phenomenon occurs for smaller values of $q$. 

In spite of this `pseudo-linearity', it seems clear that $S_q^{\rm ECG}$ does not have a simple dependence on universal CFT quantities. This fact, which agrees with the exact $d\geq 4$ results of \cite{HoloRen} for Gauss-Bonnet and QTG, was actually anticipated in that paper also for $d=3$, where $S_q$ was computed  at leading order in the gravitational coupling for a  bulk model consisting of Einstein gravity plus a  Weyl$^3$ correction.  

% If we restrict ourselves to the physical range \req{phiss}, the allowed values of $\ctt/a^*$ are
%\begin{equation}
%\frac{105}{107}\leq  \frac{\pi^3}{12}\frac{\ctt}{a^*} \leq 1\, .
%\end{equation}
%Then, we can rewrite \req{eq:x_q 2} and \req{reni}
%\begin{equation}\label{eq:x_q 22}
%x_q^2\left(3q^2x_q^2-q^2-2qx_q\right)=\frac{\left[(\ctt/a^*)-\frac{12}{\pi^3} \right]}{\left[(\ctt/a^*)+\frac{12}{\pi^3}\right] }\left(1-q^2 x_q^4 \right)	\ ,
%\end{equation}
%\begin{equation}\label{renii}
%S_q= a^* \frac{ V_{\mathbb{H}^2} \pi^3 q}{24(q-1)}  \left[(\ctt/a^*)+\frac{12}{\pi^3} \right]\left[1-x_q-\frac{x^2_q}{q}+x_q^3+\frac{1}{3}\frac{\left[(\ctt/a^*)-\frac{12}{\pi^3} %\right]}{\left[(\ctt/a^*)+\frac{12}{\pi^3}\right] } \left(\frac{3}{q^2 x_q}-3-\frac{1}{q^3}+x_q^3\right)\right]\ .
%\end{equation}

\subsection{Scaling dimension of twist operators}\label{twist}
Let us now turn to the scaling dimension of twist operators. In the context of computing R\'enyi entropies for some region $V$ using the replica trick, the boundary conditions which glue together the different copies of the replicated geometry at the entangling surface $\partial V$, can be alternatively implemented through the insertion of dimension-$(d-2)$ operators $\tau_q$ extending over $\partial V$ \cite{Calabrese:2004eu,HoloRen,Hung:2014npa,Swingle:2010jz}. The replicated-geometry construction is then replaced by a path integral over the symmetric product of $q$ copies of the theory on a single copy of the geometry, with the $\tau_q$ inserted. Given $V$, $\Tr \rho_V^q$ can be then obtained as the expectation value of these `twist operators', $\Tr \rho_V^q=\braket{\tau_q}_q$, computed in the $q$-fold symmetric product CFT. A natural notion of scaling dimension, $h_q$, can be defined for $\tau_q$ from the leading singularity appearing in the correlator $\braket{T_{ab}\tau_q}$, as the stress tensor is inserted close to $\partial V$. In particular \cite{HoloRen,Hung:2014npa},
\begin{equation}
\braket{T_{ab}\tau_q}_q=-\frac{h_q}{2\pi} \frac{b_{ab}}{y^d}\, ,
\end{equation}
where $b_{ab}$ is a fixed tensorial structure and $y$ is the separation between the stress-tensor insertion and $\partial V$.

 Our interest in the $h_q$ for ECG is mostly related to the use that we will make of them in the following subsection, so let us just reproduce the most relevant result needed to compute them for holographic CFTs \cite{HoloRen,Hung:2014npa}. This establishes that, given some higher-derivative bulk theory, $h_q$ can be obtained from the thermal entropy and temperature of the corresponding hyperbolic AdS black hole as 
\begin{equation}
h_q=\frac{2\pi R q}{(d-1)V_{\mathbb{H}^{d-1}}}\int_{x_q}^{1}T(x)S'(x)dx\ .
\end{equation}
%In order to evaluate this expression for ECG, it is convenient to use that 
%\begin{equation}
%\int_1^x T(x)S'(x)dx= I(x)\ ,
%\end{equation}
%where 
%\begin{equation}
%I(x)=-T_0 \frac{\tilde L^2 V_{\mathbb{H}^2}}{4G}\left[\frac{2x^3}{f_\infty}+\frac{2x\left(2-\frac{3x^2}{f_\infty}\right)\left(\frac{3x^2}{f_{\infty}}-1\right)}{3\left[1+\sqrt{1-\frac{3 f_{\infty}^2\mu}{x^4}\left(\frac{3x^2 }{f_{\infty}}-1\right)}\right]}+\frac{4x\left(\frac{3x^2 }{f_{\infty}}-1\right)^2}{3\left[1+\sqrt{1-\frac{3 f_{\infty}^2\mu}{x^4}\left(\frac{3x^2 }{f_{\infty}}-1\right)}\right]^2}\right] \ .
%\end{equation}
%We get
Then, using \req{sx}, we find, for the universal piece,
\begin{equation}\label{scalingdimension}
h^{\rm ECG}_q=-\frac{q\tilde L^2}{8G}\left[x_q^3-x_q-\mu f^2_{\infty}\left(x_q^3+\frac{2}{q^3}-\frac{3}{q^2x_q}\right)\right] \ ,
\end{equation}
which reduces to the Einstein gravity result \cite{HoloRen}
\begin{equation}
h_{q}^{\rm E}=\frac{q \tilde L^2}{8G}x_q\left(1-x_q^2\right)\, ,
\end{equation}
when $\mu=0$. It is easy to perform some checks of this result. In particular, we find
\begin{equation}
 \lim_{q\rightarrow 0}h^{\rm ECG}_q=-\frac{1}{12\pi^2 q^2}\cs^{\rm ECG}\, , \quad \partial_q h^{\rm ECG}_q|_{q=1} =\frac{\pi^3}{24}\ctt^{\rm ECG}\ ,
\end{equation}
as expected from the general identities found in \cite{Bueno3} and \cite{Hung:2014npa}, respectively.
Similarly, using \req{reni}, it is possible to verify that the general relations \cite{Bueno3}\footnote{For $j=1$, the second term is ignored.}
\begin{equation}
\left.\partial^j_q h_q\right|_{q=1}=\frac{1}{4\pi}\left[(j+1) \left. \partial^j_q S_q\right|_{q=1}+j^2 \left. \partial^{j-1}_q S_q\right|_{q=1} \right]\, ,
\end{equation}
hold for general $j$ and arbitrary values of $\mu$, as they should.

%Another check would be to take the limit $q\rightarrow 0$. For this, we use that 
%\begin{equation}
%x_q(q)=\frac{2f_\infty}{3 q}+\frac{1+\frac{27}{4}\mu}{2}q+\mathcal O(q^3)\ .
%\end{equation}
%Plugging this expansion into (\ref{scalingdimension}), we get 
%\begin{equation}\label{hq0}
%\lim_{q\rightarrow 0}h_q=-\frac{\tilde L^2}{27 G}\frac{f^2_{\infty}\left(1-\frac{27 }{4}\mu\right)}{q^2}=-\frac{\cs}{12\pi^2 q^2}\ ,
%\end{equation}
%where $\cs=\tfrac{4\pi^2\tilde L^2}{9G}f^2_{\infty}\left(1-\tfrac{27 }{4}\mu\right)$ is the thermal entropy charge.

\subsection{Stress tensor three-point function charge $t_4$}\label{t44}
For general CFTs in $d=3$, the stress tensor three-point function is a combination of fixed tensorial structures controlled by two theory-dependent quantities \cite{Osborn:1993cr}, which can be chosen to be $\ctt$ plus an additional parameter\footnote{In general, in $d=3$, there is also a parity-violating structure \cite{Giombi:2011rz,Maldacena:2011nz,Chowdhury:2017vel}, which is controlled by yet another parameter. Capturing this would require introducing another bulk density involving some contraction of curvature tensors with the Levi-Civita symbol --- see \eg \cite{Maldacena:2011nz}.}, $t_4$. The latter was originally introduced in  \cite{Hofman:2008ar}, where it was shown to appear in the general formula for the energy flux reaching null infinity in a given direction after inserting an operator of the form $\epsilon_{ij}T^{ij}$, where $\vec{\epsilon}$ is some symmetric polarization vector.  For any CFT$_3$, the result takes the general form
\begin{equation}\label{ees}
\braket{\mathcal{E}(\vec{n})}=\frac{E}{2\pi}\left[1+t_4 \left(\frac{|\epsilon_{ij}n^in^j|^2}{\epsilon_{ij}^*\epsilon_{ij}}-\frac{1}{8}\right)\right]\, ,
\end{equation}
where $E$ is the total energy, and $\vec{n}$ is the unit vector indicating the direction in which we are measuring the flux. Hence, the only theory-dependent quantity appearing in the above expression is $t_4$ which, along with $\ctt$, fully characterize $\braket{TTT}$ --- see \eg \cite{Buchel:2009sk,Bobev:2017asb} for the explicit connection.  For $d\geq 4$, there is an extra parity-preserving structure weighted by another theory-dependent constant, customarily denoted $t_2$.

Higher-dimensional versions of \req{ees} have been used to identify $t_4$ and $t_2$ for holographic theories dual to certain higher-order gravities in $d\geq 4$, such as Lovelock \cite{Buchel:2009sk,deBoer:2009gx} or QTG \cite{Myers:2010jv}. It is known that $t_4=0$ for general supersymmetric theories \cite{Hofman:2008ar,Kulaxizi:2009pz}, as well as for theories of the Lovelock class \cite{Buchel:2009sk,deBoer:2009gx,Camanho:2009hu,Camanho:2013pda}, including Einstein gravity in general dimensions. In fact, one of the original motivations for the construction of QTG in \cite{Quasi}, was to provide a nonperturbative holographic model with a non-vanishing $t_4$ in $d=4$. Here, we show that ECG provides an analogous model in $d=3$.

In order to determine $t_4$ for ECG, we will use the results in \cite{Chu:2016tps}, where it was shown that the scaling dimension of twist operators in holographic theories is related to the parameters controlling the stress-tensor three-point function. In particular, it was shown that the expression
\begin{equation}\label{minino}
\frac{h_q}{\ctt}=\frac{\pi^3}{24}(q-1)-\frac{\pi^3}{11520}(420+t_4)(q-1)^2+\mathcal{O}(q-1)^3\, ,
\end{equation}	
holds for general holographic higher-order gravities in $d=3$, at least at leading order in the couplings. Performing the corresponding expansion in the twist-operator scaling dimension \req{scalingdimension}, we find
\begin{equation}\label{t4ecg}
t^{\rm ECG}_4=\frac{-1260 \mu f_{\infty}^2}{(1-3\mu f_{\infty}^2)}\, ,
\end{equation}
which, as expected, vanishes for Einstein gravity. One may worry about the validity of \req{minino} beyond leading order, for which $t_4^{\rm ECG}= -1260 \mu +\mathcal{O}(\mu^2)$. However, we have good reasons to believe that \req{t4ecg} is correct for general values of $\mu$. First of all, observe that \req{t4ecg} singles out $\mu=4/27$ as a special value of the coupling, since $t^{\rm ECG}_4$ diverges in that case. Of course, this is nothing but the critical limit of the theory, for which some sort of bizarre behavior was to be expected. Secondly, in Appendix \ref{ttt}, we use the results found in \cite{HoloRen} for the twist-operator scaling dimensions in $d$-dimensional holographic Gauss-Bonnet and $d=4$ QTG, and show that the ($d$-dimensional versions of) \req{minino} provide expressions for $t_2$ and $t_4$ which exactly agree with the fully nonperturbative ones found in \cite{Buchel:2009sk} and \cite{Myers:2010jv}. These observations strongly suggest that \req{t4ecg} is an exact expression.
%At leading order in $\mu$, it reads
%\begin{equation}
%t_4= -1260 \mu +\mathcal{O}(\mu^2)\, .
%\end{equation}

Now, in $d=3$, imposing the positivity of energy fluxes in arbitrary directions gives rise to the constraint $-4\leq t_4 \leq 4$, which is valid for general CFTs \cite{Buchel:2009sk}, as long as the additional parity-odd structure is absent, as in the case of ECG\footnote{Observe, in particular, that for a CFT$_3$ consisting of $n_{\rm s}$ real conformal scalars and $n_{\rm f}/2$ Dirac fermions, $t_4=4(n_{\rm s}-n_{\rm f})/(n_{\rm s}+n_{\rm f})$, which therefore covers the full space of allowed values of $t_4$ \cite{Buchel:2009sk}, the limiting values corresponding to an arbitrary number of fermions, and to an arbitrary number of scalars, respectively.}. 
When written in terms of the gravitational coupling for ECG, this constraint translates into
\begin{equation}\label{fifif}
\frac{312}{313}\leq f_{\infty} \leq \frac{318}{317}\, ,
\end{equation}
which, together with the previous constraint $1\leq f_{\infty} \leq 3/2$ becomes
\begin{equation}
1\leq f_{\infty} \leq \frac{318}{317}\simeq 1.00315\, .
\end{equation}
This can in turn be explicitly written in terms of $\mu$ as
\begin{equation}\label{phiss}
0\leq \mu \leq \frac{100489}{32157432}\simeq 0.00312491\, .
\end{equation}
This reduces the range of allowed values of $\mu$ quite considerably.
Observe that for $f_{\infty}=318/317$, $t_4=-4$, which is precisely the value corresponding to a free fermion. The other limiting value, $t_4=4$, corresponding to a free scalar, would imply a negative value of $\mu$, and is therefore excluded. Observe also that the bound is maximally violated at the critical value $\mu=4/27$.

%%%%%%%%%%%%%%%%%%%%%%%%%%%%%%%%%%%%%%%%%%%%%%%%%%%%%%%%%%%%%%%%%%%
%%%%%%%%%%%%%%%%%%%%%%%%%%%%%%%%%%%%%%%%%%%%%%%%%%%%%%%%%%%%%%%%%%%
%%%%%%%%%%%%%%%%%%%%%%%%%%%%%%%%%%%%%%%%%%%%%%%%%%%%%%%%%%%%%%%%%%%

\section{Holographic hydrodynamics}\labell{shear}
One of the paradigmatic applications of higher-order gravities in the AdS/CFT context has been the construction of counterexamples to the famous Kovtun-Son-Starinets (KSS) bound for the shear viscosity over entropy density bound \cite{Kovtun:2004de}. The latter was originally conjectured to satisfy $\eta/s \geq \frac{1}{4\pi}$ (in natural units) for any fluid in any number of dimensions, the saturation occurring for holographic plasmas dual to Einstein gravity AdS$_{(d+1)}$ black branes.  Violations of the bound --- generically produced by finite-$N$ effects from the gauge-theory side --- were argued to occur for holographic plasmas dual to black branes in several higher-order theories --- see, \eg \cite{Buchel:2004di,Kats:2007mq,Brigante:2007nu,Myers:2008yi,Cai:2008ph,Ge:2008ni} for some of the earliest works and \cite{Cremonini:2011iq} for a review. A thorough study of various consistency conditions --- such as subluminal propagation of excitations, energy positivity  or unitarity --- on some of the holographic theories for which the corresponding branes could be actually constructed --- hence allowing for fully nonperturbative calculations in the higher-curvature couplings --- suggested that the bound can be lowered down to  $\eta/s \sim 0.4\cdot \frac{1}{4\pi}$ for $d=4$ \cite{Myers:2010jv}, and arbitrarily close to zero for large enough $d$ \cite{Camanho:2010ru}. 
These results give rise to three possibilities for finite-$d$: (i) the parameter space which would permit violations of the KSS bound is in fact not allowed by some other unidentified physical conditions --- see below --- and the KSS bound is true after all; (ii) there exists some lower bound, but it is lower than the KSS one; (iii) there is no bound at all. 
It was shown later \cite{Camanho:2014apa} that higher-derivative theories with nonperturbative couplings are in fact generally acausal unless the spectrum is supplemented by higher-spin modes. While it is still unclear under what circumstances such additional degrees of freedom play a relevant role --- specially given the success of holographic higher-curvature models in other holographic applications --- the reliability of the aforementioned conclusions regarding the fate of the bound was
put in suspense by this result. The current belief seems to be that some non-trivial bound, lower than the KSS one, does exist for general $d$ --- see \eg \cite{Fouxon:2008pz}.

In this section we compute the shear viscosity to entropy density ratio for ECG, providing the first calculation of such a quantity for a holographic higher-curvature gravity in $d=3$ which is fully nonperturbative in the gravitational coupling. We will proceed along the lines of \cite{Myers:2010jv,Paulos:2009yk}.  Let us start considering the ECG planar black hole in \req{bhss}, \ie we set $k=0$ and $N^2=1/f_{\infty}$, 
\begin{equation}\label{planarbh}
ds^2=\frac{r^2}{L^2}\left[-\frac{f(r)}{f_{\infty}}dt^2+dx_1^2+dx_2^2\right]+\frac{L^2}{r^2f(r)}dr^2\, .
\end{equation}
Now, it is convenient to perform the change of coordinates $z=1-\rh^2/r^2$, so that the horizon corresponds to $z=0$, the asymptotic boundary being at $z=1$. The metric reads then
\begin{equation}\label{mm}
ds^2=\frac{\rh^2}{L^2(1-z)}\left(-\frac{f(z)}{f_{\infty}}dt^2+dx_1^2+dx_2^2 \right)+\frac{L^2}{4f(z)(1-z)^2}dz^2\, .
\end{equation}
On the other hand, the cubic equation that determines $f(r)$, \req{eqsf}, reads, in terms of $z$
\begin{equation}\label{reee}
1-f(z)+\mu \left[f^3-3 (1-z)^2f f'^2 -2(1-z)^3 f' (f'^2-3f f'')\right]=\left(1-\frac{27}{4}\mu\right)(1-z)^{3/2}\, ,
\end{equation}
where now $f'\equiv df/dz$, and so on. In order to determine the shear viscosity, we will need the near-horizon behavior of $f$, so let us perform a Taylor expansion of the form
\begin{equation}\label{ftay}
f(z)=f_0' z +\frac{1}{2}f_0'' z^2 + \frac{1}{6}f_0'''(z)z^3+\dots,
\end{equation}
The coefficients in this expansion can be of course written in terms of those in the $r$-expansion series \req{nH}, but it is easier to work directly with the variable $z$. Inserting \req{ftay} in \req{reee} and imposing it to hold order by order in $z$, one finds
\begin{equation}\label{f0}
f_0'=\frac{3}{2}\, , \quad f_0'''=\frac{-144 \mu  f_0''^2+4 (135 \mu +4) f_0''-81 \mu +12}{216 \mu }\, ,\quad f_0^{(4)}=\ldots,
\end{equation}
etc. All the coefficients are determined by $f_0''$, whose value is fixed by the asymptotic condition $\lim_{z\rightarrow 1}f(z)=f_{\infty}$. Analogously to the discussion in Sec.~\ref{fullsol}, there is a unique value of this parameter for which the desired boundary condition is achieved. This defines $f_0''$ as a function of $\mu$, which we denote $f_0''(\mu)$. We can compute this numerically with arbitrary precision, but let us also try an analytic computation using the following logic. Observe that, for $\mu=0$, the solution is simply $f(z)=1-(1-z)^{3/2}$, from where we read all the derivatives 
\begin{equation}
f_0^{(n)}(0)=(-1)^{n+1}\frac{\Gamma(5/2)}{\Gamma(5/2-n)}.
\end{equation}
 Now, since the solution for general $\mu$ should reduce to the Einstein gravity one when $\mu\rightarrow 0$, the derivatives \req{f0} should coincide with the previous ones in that limit. It turns out that we can use this condition to determine the derivatives of $f_0''(\mu)$ with respect to $\mu$ at $\mu=0$. Let us see how this works. Obviously, we have $\lim_{\mu\rightarrow 0}f_0''(\mu)\equiv f_0''(0)=-3/4$. Then, we should also have $\lim_{\mu\rightarrow 0}f_0'''(\mu)\equiv f_0'''(0)=-3/8$. If we take this limit in the second equation of \req{f0}, we get the condition
 \begin{equation}
 \lim_{\mu\rightarrow 0}\left[\frac{-144 f_0''(\mu)^2+540f_0''(\mu) -81}{216}+\frac{2}{27}\frac{f_0''(\mu)+3/4}{\mu}\right]=-\frac{3}{8}\, .
 \end{equation}
 The limit of the first term is finite and we can simply substitute $f_0''(0)=-3/4$. However, in the second term we have
 \begin{equation}
\lim_{\mu\rightarrow 0} \frac{f_0''(\mu)+3/4}{\mu}=\lim_{\mu\rightarrow 0} \frac{f_0''(\mu)-f_0''(0)}{\mu}\equiv\frac{df_0''(\mu)}{d\mu}\bigg|_{\mu=0}\, .
 \end{equation}
 Therefore, this equation is actually giving us the value of the derivative of $f_0''(\mu)$ at $\mu=0$, the result being $243/8$. The same process can be repeated at every order and we can obtain all derivatives of this function at $\mu=0$. Up to second order, we have
\begin{equation}
f_0''(\mu=0)=-\frac{3}{4}\, ,\quad \frac{df_0''(\mu)}{d\mu}\bigg|_{\mu=0}=\frac{243}{8}\, ,\quad \frac{d^2f_0''(\mu)}{d\mu^2}\bigg|_{\mu=0}=-\frac{115911}{16}\, .
\end{equation}
Now, if the function $f_0''(\mu)$ were analytic, we could in principle construct it as
\begin{equation}\label{f0ser}
f_0''(\mu)=\sum_{n=0}^{\infty}\frac{1}{n!}\frac{d^nf_0''(\mu)}{d\mu^n}\bigg|_{\mu=0}\mu^n\, .
\end{equation}
However, a convergence analysis, including many terms in the expansion, reveals that this series is actually divergent for every $\mu\neq 0$ --- in other words, the radius of convergence is 0. The fact that the series diverges is telling us that the function does not allow for a Taylor expansion around $\mu=0$. This is an example of a $\mathcal{C}^{\infty}$ function which is not analytic\footnote{See \eg \cite{Pasini:2015zlx} for  another explicit example in a different context.}. 
Nevertheless, the series can be used to provide an approximate result for small enough $\mu$ if we truncate it at certain $n$. For example, to quadratic order we obtain
\begin{equation}\label{f0approx}
f_0''(\mu)\approx-\frac{3}{4}+\frac{243}{8}\mu-\frac{115911}{32}\mu^2\, ,
\end{equation}
but the approximation is only good for rather small values of the coupling, \eg for $\mu=0.003$, the error  is $\sim 3\%$ (with respect to the numerical value) and the precision is not increased by the addition of further terms. 
Observe also that in the  critical limit, $\mu=4/27$, we have $f_{\rm cr}(z)=\frac{3}{2}z$, and hence $f_0''(4/27)=0$ in that case. 

After this dissertation, which we will use to get a grasp on the small-$\mu$ behavior of $\eta$, let us now turn to the actual computation. In order to do so,  we perturb the black hole metric \req{mm} by shifting
\begin{equation}
dx_1\rightarrow dx_1+\varepsilon e^{-i\omega t} dx_2\,,
\end{equation}
where $\varepsilon$ is a small parameter. Then, the shear viscosity can be obtained as\footnote{See \cite{Fan:2018qnt} for a recent alternative method.} \cite{Paulos:2009yk}
\begin{equation}
\eta=-8\pi T \lim_{\omega,\epsilon\rightarrow 0}\frac{{\rm Res}_{z=0}\mathcal{L} }{\omega^2 \epsilon^2}\, ,
\end{equation}
where $\mathcal{L}$ is the corresponding full gravitational Lagrangian (including the $\sqrt{|g|}$ term) in \req{ECG} evaluated on the perturbed metric.  Using \req{ftay}, we can evaluate this quantity, and the result reads
\begin{equation}
\eta^{\rm ECG}=\frac{3\rh^2}{64\pi G L^2 f_0' } \left[2+(21f_0'^2+36f_0''^2-114f_0'f_0''+36f_0' f_0''')\mu \right]\, .
\end{equation}
Then, using the values of $f_0'$ and $f_0'''$ in \req{f0}, we find
\begin{equation}
\eta^{\rm ECG}=\frac{\rh^2}{32\pi G L^2} \left[5+27\mu+(4-36\mu)f_0''(\mu)\right]\,.
\end{equation}
Finally, from \req{entropy} it follows that the shear viscosity over entropy density ratio reads
\begin{equation}\label{eta/s}
\left[\frac{\eta}{s}\right]^{\rm ECG}=\frac{5+27\mu+(4-36\mu)f_0''(\mu)}{8\pi\left(1-\frac{27}{4} \mu\right)}\,.
\end{equation}
Some comments are in order. First, note that this expression is very different from the rest of nonperturbative results for $\eta/s$ available in the literature for $d\geq 4$ theories, corresponding to Lovelock \cite{Brigante:2007nu,Ge:2009eh,Brustein:2008cg,Shu:2009ax} and QTG \cite{Myers:2010jv}. In those cases, it is found that $\eta/s$ depends on the gravitational couplings in a polynomial way\footnote{Note however that, \eg for Gauss-Bonnet gravity, some of the remaining second-order coefficients have a nonpolynomial dependence on the corresponding coupling \cite{Grozdanov:2015asa,Grozdanov:2016fkt}. } --- see also \cite{Parvizi:2017boc}.
On the contrary, the ECG result has a very nonpolynomial character, for two reasons. First, the presence of the function $f_0''(\mu)$, which is non-analytic, implies that $\eta/s$ cannot be Taylor-expanded around $\mu=0$. And second, the denominator `$(1-27/(4 \mu))$' in \req{eta/s} is also a new feature, which gives rise to a divergence in the critical limit. The appearance of such contribution in the denominator is rooted in the different way in which ECG modifies the result for the thermal entropy charge $\cs$ with respect to the other theories mentioned above --- see discussion in Subsection  \ref{cssex}.

\begin{figure}[t!]
	\centering 
	\includegraphics[width=0.65\textwidth]{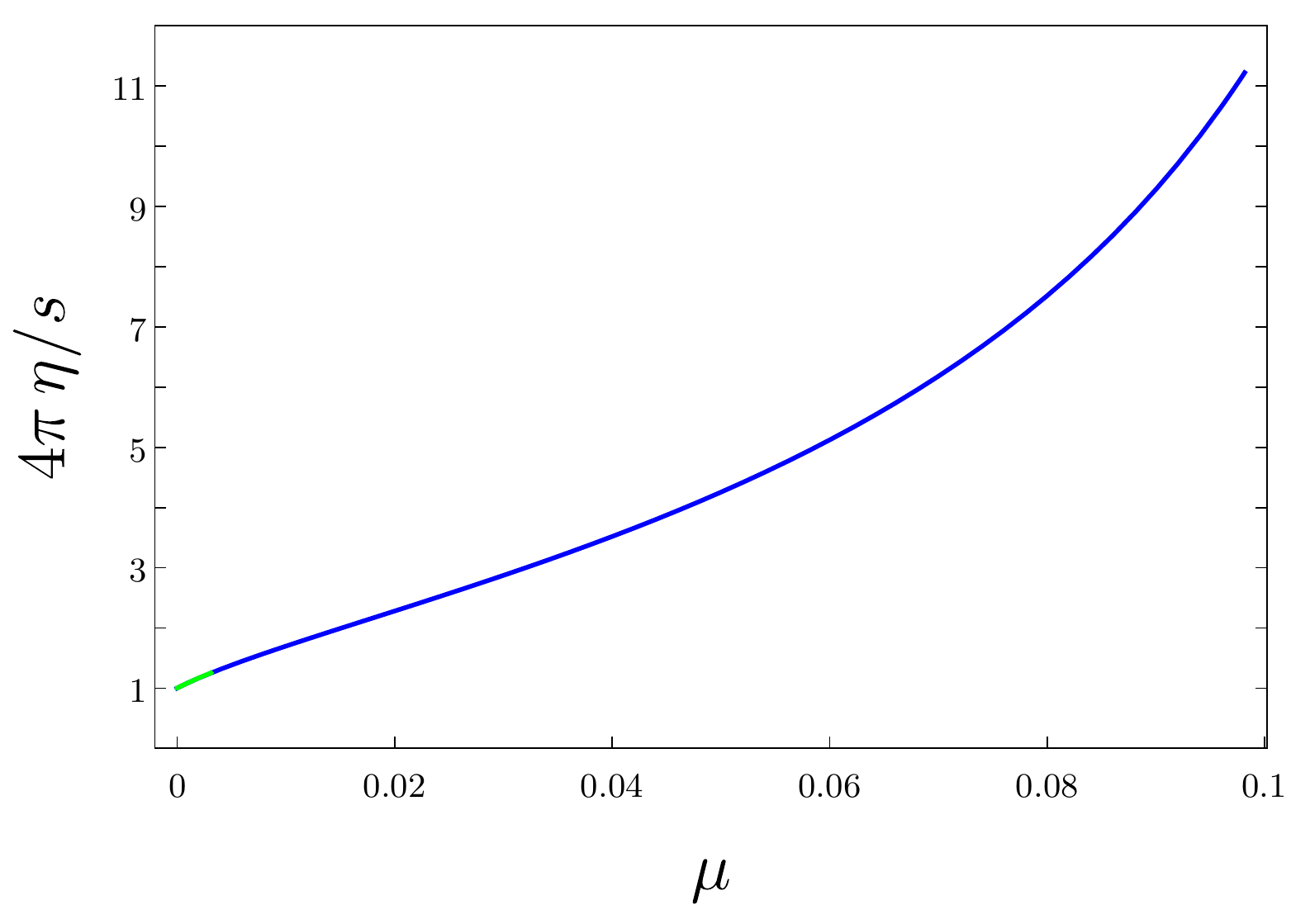}
	\caption{Shear viscosity to entropy density ratio as a function of $\mu$. The green line represents the region allowed by the constraint $t_4\ge -4$.}
	\label{ratio}
\end{figure}

Let us analyze the profile of $\eta/s$ as a function of $\mu$. When $\mu\ll 1$, we can use \req{f0approx} to obtain  
\begin{equation}
\left[\frac{\eta}{s}\right]^{\rm ECG}\approx \frac{1}{4\pi}\left(1+\frac{189 \mu }{2}-\frac{114453 \mu ^2}{16}\right)\, .
\end{equation}
Again, remember that, strictly speaking, this is not a Taylor expansion and it only provides a good approximation for very small $\mu$. In any case, note that the leading correction is positive, so $\eta/s$ is increasing with $\mu$.
On the other hand, in the critical limit, we have $f_0''(\mu\rightarrow 4/27)\rightarrow 0$, so the leading behavior of $\req{eta/s}$ can be captured analytically,
\begin{equation}
\left[\frac{\eta}{s}\right]^{\rm ECG}=\frac{9}{8\pi\left(1-\frac{27}{4} \mu\right)}+\mathcal{O}(1)\, ,\quad \text{for}\, \quad \mu\rightarrow\frac{4}{27}\, .
\end{equation}
Hence, this ratio takes arbitrarily high values as we approach the critical limit\footnote{Observe that, from this point of view, the critical limit of ECG is very different from that corresponding to its higher-dimensional cousins, such as Gauss-Bonnet. In that case, $\eta/s$ diverges for $\lambda_{\rm \ssc GB} \rightarrow -\infty$, while it stays finite for the critical value $\lambda_{\rm \ssc GB}=1/4$.}. The full profile of $\eta/s$ can be obtained with arbitrary precision from a numerical computation of $f_0''(\mu)$. The result is shown in Fig. \ref{ratio}.  
The curve is monotonically increasing, and blows up in the critical limit, $\mu=4/27$. Therefore, the KSS bound is not violated for any value of $\mu$ in the dynamically allowed region, $0\leq \mu\leq 4/27$, which is precisely a consequence of the nonexistence of $\mu<0$ solutions with positive energy. In that sense, as opposed to previously studied theories in higher dimensions, ECG simply does not allow for violations of the bound, not even in principle. It would be interesting to find out whether this phenomenon is common to the rest of $d=3$ theories constructed in \cite{PabloPablo4} and, more generally, to the new theories belonging to the Generalized QTG class \cite{Hennigar:2017ego,PabloPablo3,Ahmed:2017jod} in general dimensions.
% non-negative value of $\mu$ and in principle $\eta/s$ it could take any value above $1/(4\pi)$ . 

As we explained in Subsection \ref{t44}, imposing the positivity of energy fluxes in the CFT, gives rise to the constraint  $0\le \mu\le 0.00312$ --- see green region in Fig.  \ref{ratio}. This would imply a maximum possible value for $\eta/s $ in ECG, given by
\begin{equation}
\left[\frac{\eta}{s}\right]_{\rm max.}^{\rm ECG}\simeq 1.253\times \frac{1}{4\pi}\, .
\end{equation}
 %We find the maximum value of the shear viscosity to entropy density ratio compatible with this bound is $\max \eta/s \approx 1.253/(4\pi)$
%The fact that we cannot violate the KSS bound seems to be related to the fact that we are forced to choose $\mu\ge 0$, since for negative $\mu$ ECG does not allow for positive energy black holes or black branes. This is an intriguing connection that requires of further analysis. The computation performed here for the shear viscosity to entropy density ratio in ECG can be extended elsewhere for any (or all) of the theories introduced in \cite{PabloPablo4}. An interesting question is whether the addition of additional higher-curvature terms with arbitrary couplings makes possible the violation of the KSS bound, or if it is a common property of all these theories that they respect the bound.\\

From the results here, we can extract some general lessons regarding calculations of $\eta/s$ in higher-curvature holographic CFTs. First, we have seen that the ECG result is highly nonperturbative in the gravitational coupling. There is in principle no reason to expect this to be different for more general theories. The results found for Lovelock and QTG, polynomial in the gravitational couplings, are probably less generic --- for those, the metric function $f(r)$ is determined by an algebraic equation, which is a highly exceptional property \cite{PabloPablo3}.  Besides, as we have seen, there may be regions of the parameter space for which the corresponding black branes do not exist, even for arbitrarily small values of the couplings. None of this is seen when working perturbatively in the gravitational couplings, which means that the results obtained in that way must be taken carefully. This is a lesson which extends to most calculations in higher-curvature gravities.

\section{Conclusions}
In this chapter we have performed a thorough characterization of the holographic CFTs dual to Einsteinian cubic gravity. One of the improvements with respect to GR is that the 2- and the 3-point functions of the dual theory are now completely general, because we have two independent parameters, namely $L^2/G$ and $\mu$. In this sense, ECG provides us with a toy model of a nonsupersymmetric CFT ($t_4\neq 0$) in $d=3$, analogous to Quasi-topological gravity (QTG) in $d=4$. In addition, the parameter $\mu$ breaks the degeneracy of the charges, and this allows us to establish relations between several quantities, as we did for the case of R\'enyi entropies and the ratio $\ctt/a^*$ in Section~\ref{ere}. The values of the different charges and a comparison with the EG and QTG cases is shown in Table~\ref{diction}.

In Sec.~\ref{osa}, we proposed a new method for evaluating Euclidean on-shell actions for higher-order gravities whose linearized equations of motion on maximally symmetric backgrounds are of second order. Throughout this chapter, we have performed several successful and highly non-trivial checks of the proposal --- see Appendix \ref{BTcheck} as well. It would be interesting to perform further studies of our generalized action \req{SEcomplete} for other theories, such as higher-order Lovelock theories, Quasi-topological gravity and its higher-order generalizations and, more generally, for theories of the Generalized quasi-topological type. One of the most striking aspects of \req{SEcomplete} is that it avoids the --- usually very challenging --- problem of determining the correct generalization of the Gibbons-Hawking-York boundary term. At the same time, and somewhat surprisingly, it involves the universal charge $a^*$ controlling the EE of spherical regions in the corresponding dual CFT. This acts as a weight that changes from one theory to another. 

One of the most important facts about Einsteinian cubic gravity is that we are able to compute the thermodynamic properties of black holes analytically. We have shown that this can be done, equivalently, using Wald's entropy formula or the Euclidean action functional \req{SEcomplete}, which is the standard method required by holography. Many properties of the dual CFT, such as the the entropy density of a plasma, the temperature of the Hawking-Page transition, or Renyi entropies, can be studied exactly thanks to this.

Finally, we also studied the shear viscosity to entropy density ratio, unveiling a new phenomena: this quantity diverges for a certain value of the coupling $\mu$, which coincides with the critical limit of the theory. In addition, the KSS bound is not violated by any positive value of $\mu$, and is saturated only for $\mu=0$. The same behaviour has been recently observed for cubic and quartic Generalized quasi-topological gravities in higher dimensions \cite{Mir:2019ecg,Mir:2019rik}.

\begin{table*}[t] \hspace{-0.55cm}
	\begin{tabular}{|c|c|c|c|c|}
		\hline
		&   $\ctt$  & $\ctt\cdot t_4$   & $\cs $   & $a^* $\\
		\hline\hline
		Einstein &  $\frac{\Gamma(d+2)}{8(d-1)\Gamma(d/2)\pi^{(d+2)/2}}\frac{\tilde{L}^{d-1}}{G}$  & $0$ & $\frac{4^{d-2}\pi^{d-1}\tilde{L}^{d-1}}{d^{d-1}G}$ &  $\frac{\pi^{(d-2)/2}}{8\Gamma(d/2)}\frac{\tilde{L}^{d-1}}{G}$ \\ \hline
		ECG $(d=3)$	& $(1-3\mu f_{\infty}^2)\ctte$   & -$1260 \mu f_{\infty}^2 \ctte$ & $\left(1-\frac{27}{4}\mu \right)f_{\infty}^2 \cse $ & $\left(1+3\mu f_{\infty}^2\right)a^{*{\rm \ssc E}}$ \\ \hline
		QTG $(d=4)$	& $(1-3\mu f_{\infty}^2)\ctte $   & $3780 \mu f_{\infty}^2 \ctte$ & $f_{\infty}^3\cse $ & $\left(1+9\mu f_{\infty}^2\right)a^{*{\rm \ssc E}}$ \\
		\hline
	\end{tabular}
	\caption{From left to right: stress-tensor two- and three-point function charges $\ctt$ and $\ctt \cdot t_4$, thermal entropy charge $\cs$, and universal contribution to the entanglement entropy across a spherical region, $a^*$, for holographic theories dual to Einstein gravity in $d$ dimensions, ECG ($d=3$) and Quasi-topological gravity (with vanishing Gauss-Bonnet coupling) in $d=4$ \cite{Myers:2010jv}. }
	\label{diction}
\end{table*}

Before closing, let us mention some possible extensions of the work presented in this chapter. 
Many of the computations presented here can be repeated for the set of all GQGs that we considered in Chapter~\ref{Chap:5}, which share the nice property of ECG of having accessible black hole thermodynamics. We obtain some basic formulas for the holographic dictionary of those theories in Appendix~\ref{App:Gen}.
One could also consider the higher-dimensional versions of Generalized quasi-topological gravity --- certain holographic aspects of these theories at cubic and quartic order were analyzed in Refs. \cite{Mir:2019ecg,Mir:2019rik} --- which would allow to explore an even wider set of CFTs than those dual to Quasi-topological and Lovelock gravities \cite{Quasi2,Camanho:2009vw,deBoer:2009pn,Buchel:2009sk,deBoer:2009gx,Camanho:2009hu,Grozdanov:2014kva,Grozdanov:2016fkt,Andrade:2016rln,Konoplya:2017zwo}.
More generally, one could study the general set of Einstein-like theories \cite{Li:2019auk} --- which we recall contains the family of GQGs --- but in that case the analysis of black hole solutions becomes much more involved and only perturbative results can be accessed analytically.

\chapter{Euclidean AdS-Taub-NUT solutions}\label{Chap:7}

In this chapter we construct a plethora of new Euclidean AdS-Taub-NUT and bolt solutions of several four- and six-dimensional  higher-curvature theories of gravity with various base spaces $\mathcal{B}$. In $D=4$, we consider Einsteinian cubic gravity, for which we construct solutions with $\mathcal{B}=\mathbb{S}^2,\mathbb{T}^2$. These represent the first generalizations of the Einstein gravity Taub-NUT/bolt solutions for any higher-curvature theory in four dimensions. Then, we show that it is possible to extend this result to other higher-order theories by including an additional quartic curvature correction. 
In $D=6$ we consider quartic Quasi-topological and Generalized quasi-topological terms, for which we obtain new solutions with $\mathcal{B}=\mathbb{CP}^2$, although we show that the base spaces $\mathbb{S}^2\times\mathbb{S}^2,\mathbb{S}^2\times\mathbb{T}^2,\mathbb{T}^2\times\mathbb{T}^2$ are also allowed. In all cases, the solutions are characterized by a single metric function, and they reduce to the corresponding ones in Einstein gravity when the higher-curvature couplings are set to zero.  While the explicit profiles must be constructed numerically (except for a few cases), we  obtain fully analytic expressions for the thermodynamic properties of all solutions. The new solutions present important differences with respect to Einstein gravity, including regular bolts for arbitrary values of the NUT charge, critical points, and re-entrant phase transitions. 
We also find exotic analytic solutions in the critical limit of ECG that represent NUT-charged AdS wormholes or non-isotropic bouncing cosmologies depending on the sign of the cosmological constant. 

\section{Introduction}\labell{Introductionc7}
Higher curvature theories of gravity are proving to be increasingly useful in providing us with knowledge connected with fundamental questions in gravitational physics. Quadratic curvature theories are renormalizable
\cite{Stelle:1976gc}, and it has long been realized that corrections
to the Einstein-Hilbert action that go like powers in the curvature generically arise as low energy corrections from a UV complete theory of gravity, \eg String Theory~\cite{Zwiebach:1985uq}.  Furthermore, in the context of the AdS/CFT correspondence conjecture, higher curvature corrections correspond to $1/N$ corrections in the large $N$ limit of the dual CFT, allowing investigation of a much broader class of CFTs~\cite{Buchel:2008vz, Hofman:2009ug,Buchel:2009sk, Camanho:2009hu,deBoer:2009gx,Myers:2010jv}. 

Imposing the condition that higher-curvature gravity only propagates a massless graviton (even if only on a constant-curvature background) severely limits the number of sensible theories available, since they generically contain additional excitations. As we argued in the last chapter, such theories --- that are know to be of the ``Einstein-like'' class --- are the ones suited for holographic computations. The most well-known class of higher curvature theories that are Einstein-like is \textit{Lovelock gravity}~\cite{Lovelock1,Lovelock2}, in which the dimensionally extended Euler densities are included in the gravitational action. In $D \ge 2k+1$, the $k^{th}$ order Euler density is non-trivial;
in this sense Lovelock gravity is a natural extension of Einstein gravity, and is the unique higher curvature theory maintaining second order field equations for the metric.

However, other Einstein-like higher-curvature theories exist, and in this thesis we have described one class of particular interest known as Generalized quasi-topological gravity \cite{PabloPablo,Hennigar:2016gkm,PabloPablo2,Hennigar:2017ego,PabloPablo3,Ahmed:2017jod,PabloPablo4,Feng:2017tev}.
Let us once again refresh the properties of this family of theories.  The Lagrangian of the new family can be written schematically as
\begin{equation}
\mathcal{L}(g^{ab},R_{cdef})=\frac{1}{16\pi G}\left[\frac{(D-1)(D-2)}{L^2}+R+\sum_{n=2} \mu_n L^{2(n-1)}\mathfrak{R}_{(n)}\right]\, ,
\end{equation}
where $L$ is some length scale, $\mu_n$ are independent dimensionless couplings, and $\mathfrak{R}_{(n)}$ are certain  linear combinations of order-$n$ densities constructed from contractions of the metric and the Riemann tensor. The theories are characterized by the following properties:  i) they have second-order equations of motion when linearized around any maximally symmetric spacetime, \ie just like Einstein gravity, they only propagate a massless and traceless graviton on such backgrounds;
 ii) they possess a continuous and well-defined Einstein gravity limit corresponding to  $\mu_n\rightarrow 0$ for all $\mu_n$; iii) they admit generalizations of the Schwarzschild-(A)dS black hole --- so they reduce to it in the Einstein gravity limit --- characterized by a single function, %\ie of the form
\begin{equation}\label{FFbh}
ds^2=-f(r) dt^2+\frac{dr^2}{f(r)}+r^2 d\Sigma_{(D-2)}^2\, ,
% \quad d\Sigma_{k}^2=\begin{cases} L^2 d\Omega^2_{(D-2)} \quad &\text{for}\quad k=+1 \, ,\\ d\vec{x}_{(D-2)}^2 &\text{for}\quad k=0 \, ,\\  L^2d\Xi_{(D-2)}^2 &\text{for}\quad k=-1  \, , \end{cases}
\end{equation}
where $d\Sigma_{(D-2)}^2$ is the metric of the horizon cross sections (not necessarily spherical); iv) the function $f(r)$ is determined from (at most) a second-order differential equation which, for a fixed set of $\mu_{n}$, admits a unique black-hole solution completely characterized by its ADM energy and which, at least in the spherically symmetric case, describes the exterior field of matter distributions with that symmetry \cite{PabloPablo3}; v) the thermodynamic properties of such black holes can be obtained from a system of algebraic equations with no free parameters.

The above class of theories can be subdivided if one considers more restrictive criteria. In particular, replacing i) by the requirement that the full non-linear equations of the theory are second order selects the Lovelock family \cite{Lovelock1,Lovelock:1971yv,Concha:2017nca}. Keeping i) as it is, but replacing instead iv) by the requirement that $f(r)$ is determined by an algebraic equation, selects a more general class of theories, known as Quasi-topological gravities \cite{Quasi,Quasi2,Dehghani:2011vu,Cisterna:2017umf} (which of course include Lovelock as particular cases). More generally, the family of higher-curvature gravities satisfying i)-v) is larger than the Quasi-topological one. The missing theories possess black holes whose  metric function is determined by second-order differential equations, and the full set has been coined ``Generalized quasi-topological Gravity'' (GQG) in \cite{Hennigar:2017ego}. 

One intriguing feature of these new theories, in contradistinction to those belonging to the Quasi-topological subset (except for Einstein gravity itself) is that some of them are nontrivial in $D=4$. The simplest possible case of that kind, and the first to be identified, corresponds to a single additional cubic term and goes by the name of four-dimensional Einsteinian cubic gravity\footnote{The $D$-dimensional version of ECG was originally obtained as the most general cubic theory defined in a dimension-independent way --- \ie so that the relative coefficients of the cubic invariants involved do not depend on $D$ --- possessing second-order linearized equations on general maximally symmetric backgrounds \cite{PabloPablo}. However, it is only for $D=4$ that ECG additionally satisfies properties ii)-v) \cite{Hennigar:2016gkm,PabloPablo2}.} (ECG) \cite{PabloPablo}, whose action is given in \req{ECGAction} below. Many examples of GQG theories in general dimensions have now been constructed, and their respective black hole solutions studied and characterized \cite{Hennigar:2016gkm,PabloPablo2,Hennigar:2017ego,PabloPablo3,Ahmed:2017jod,PabloPablo4,Hennigar:2017umz,Feng:2017tev,Hennigar:2018hza,Colleaux:2017ibe,ECGholo,Peng:2018vbe,Dey:2016pei}.
 
A different class of exact static solutions of Einstein gravity is given by the Taub-NUT family. The Euclidean section of the corresponding metrics can be written as
\begin{equation}\label{FFnut}
ds^2=V_{\mathcal{B}}(r) (d\tau+ n A_{\mathcal{B}})^2+\frac{dr^2}{V_{\mathcal{B}}(r)}+(r^2-n^2)d\sigma_{\mathcal{B}}^2\, ,
\end{equation}
which, in even dimensions, can be understood as $U(1)$ fibrations over $(D-2)$-dimensional K\"ahler-Einstein base spaces $\mathcal{B}$ with metric $g_{\mathcal{B}}$. In \req{FFnut}, $\tau$ is a periodic coordinate parametrizing the $\mathbb{S}^1$, and $J=dA_{\mathcal{B}}$ is the K\"ahler form on ${\mathcal{B}}$. The non-triviality of the fibration is controlled by the presence of a non-zero parameter $n$, customarily called ``NUT charge''. Depending on the dimension of the set of fixed points of the $U(1)$ isometry --- namely those for which $V_{\mathcal{B}}(r)=0$ --- the solution is said to be a ``NUT'' or a ``bolt''. Taub-bolt solutions are characterized by $(D-2)$-dimensional fixed-point sets, whereas smaller dimensionalities give rise to Taub-NUT solutions.  

It has been known for some time that NUT-charged solutions exist
in Lovelock gravity \cite{Dehghani:2005zm,Dehghani:2006aa,Dotti:2007az,Hendi:2008wq}. A  broad understanding of their thermodynamics remains an ongoing subject of investigation \cite{Clarkson:2002uj,Mann:2004mi,Dehghani:2006dk,Ghezelbash:2007kw,Ghezelbash:2008zz,KhodamMohammadi:2008fh,Nashed:2012ud,Johnson:2014xza,Johnson:2014pwa,Lee:2014tma,Nashed:2015pga,Pradhan:2015jia,Lee:2015wua}
since it was realized that their contribution to the entropy does not obey the area law, even in Einstein gravity
\cite{Hawking:1998ct,Garfinkle:2000ms}.  A recent review of Taub-NUT spacetimes and their symmetries has appeared \cite{Frolov:2017kze}.

On general grounds, one expects two independent functions to be required to describe Taub-NUT solutions in general higher-curvature gravities.
The relevant observation for us is that both for Einstein gravity and Gauss-Bonnet,  all Taub-NUT solutions are characterized by a single function for each choice of base space.
 This is analogous to the situation encountered for static black-hole solutions.  One is then naturally led to wonder whether the rest of GQG theories also admit generalizations of the Einstein gravity Taub-NUT solutions characterized by a single function, $V_{\mathcal{B}}(r)$, just like they admit generalizations of the Schwarzschild black hole with that property. The answer turns out to be yes and, as we show here, a plethora of new Taub-NUT and Taub-bolt solutions of the form \req{FFnut} can be constructed in various dimensions and for different base spaces.

Each choice of $D$, base space and Taub-NUT class has its peculiarities, but some aspects of our construction can be explained in general. First of all, and in a similar fashion to what occurs with black holes, inserting ansatz \req{FFnut} in the equations of motion of the corresponding GQG theory we will observe that, whenever the corresponding theory admits  Taub-NUT solutions of that form\footnote{Not all GQG theories will admit all possible Taub-NUT solutions of the form \req{FFnut} for all possible base spaces.}, they reduce to a single  third-order equation for $V_{\mathcal{B}}(r)$. Interestingly, this equation always admits a simple integrable factor which allows us to integrate it once. Hence, in each case we are left with a single equation of the form
\begin{equation}\label{mastere}
\mathcal{E}_{\mathcal{B}}[V_{\mathcal{B}},V'_{\mathcal{B}},V''_{\mathcal{B}},r]=C\, ,
\end{equation}
where $C$ is an integration constant related to the ADM energy of the solution. 
Also, in analogy with the black-hole case, one of the integration constants of this second-order differential equation will always be fixed by imposing the solutions to be locally asymptotically AdS. With regards to the second, recall that the defining properties characterizing $V_{\mathcal{B}}(r)$ are, respectively,
%Throughout the paper, we will ask $V_{\mathcal{B}}(r)$ to satisfy
\begin{align}\label{smooth}
&V_{\mathcal{B}}(r)|_{r=n}=0\, , \quad V_{\mathcal{B}}'(r)|_{r=n}=4\pi/ \beta_{\tau}\,, \quad \text{for NUT}\, ,\\ \notag
&V_{\mathcal{B}}(r)|_{r=r_b}=0\, , \quad V_{\mathcal{B}}'(r)|_{r=r_b}=4\pi/ \beta_{\tau}\,, \quad \text{for bolt}\, ,
\end{align}
where $\beta_{\tau}$ is the period of $\tau$, and the bolt location satifies $r_b>n$. While the first condition determines whether we are considering a NUT or a bolt, the second ensures that the solutions are smooth at $r=n$ and $r=r_b$, respectively and, when possible, it will fix the other integration constant in \req{mastere} for our solutions. 
%. The other constant will in turn be fixed by imposing regularity at the fixed points of $\tau$. 

For concreteness, we shall restrict ourselves to four- and six-dimensional theories. In $D=4$, we will focus on the simplest possible modification to the Einstein-Hilbert action, namely, the ECG term. For this, we will construct solutions with base spaces $\mathcal{B}=\mathbb{S}^2$ and $\mathbb{T}^2$. These are, to the best of our knowledge, the first  higher-curvature generalizations of the Einstein gravity Taub-NUT solutions in four dimensions. Turning to $D=6$,  we find that no non-trivial solutions of the form \req{FFnut} can be constructed  at cubic order in the GQG family. They will exist, however, when quartic invariants are included, and we will restrict ourselves to that case. For those, we will construct solutions with
$\mathcal{B}=\mathbb{CP}^2,\,\mathbb{S}^2\times  \mathbb{S}^2,\, \mathbb{S}^2\times \mathbb{T}^2$ and $\mathbb{T}^2\times \mathbb{T}^2$.

Although we will not be able to solve \req{mastere} analytically for $V_{\mathcal{B}}(r)$ in general (except for the critical theories), the thermodynamic properties of the solutions will be accesible in a fully analytic fashion, again similar  to what happened for the black hole solutions constructed in \cite{Hennigar:2016gkm,PabloPablo2,Hennigar:2017ego,PabloPablo3,Ahmed:2017jod,PabloPablo4}. The relation between the ADM energy of the solutions, the NUT charge and $r_b$ (when present) will be accessible in each case from the asymptotic and near $r=n$ or $r=r_b$ expansions. On the other hand, in order to compute the free energy of the solutions, we will make use of the method introduced in \cite{ECGholo}. According to this, given some higher-curvature gravity with Lagrangian density $\mathcal{L}(g^{\alpha\beta},R_{\mu\nu\rho\sigma})$ whose linearized equations on pure AdS match those of Einstein gravity (up to a normalization of Newton's constant), the Euclidean on-shell action of any asymptotically AdS solution can be computed using the formula\footnote{The most remarkable aspect of \req{assd} is the fact that, for any theory of the kind explained above, the usual Gibbons-Hawking-York boundary term of Einstein gravity \cite{York:1972sj,Gibbons:1976ue} only appears modified through an overall factor proportional to $a^*$. This is a considerable simplification with respect to the standard approach of trying to construct the generalized version of $K$ which makes the corresponding gravitational action differentiable \cite{Myers:1987yn, Teitelboim:1987zz,Dehghani:2011hm}.}
\begin{equation}\label{assd}
I_E=-\int_{\mathcal{M}} d^D x \sqrt{g}\mathcal{L}(g^{\alpha\beta},R_{\mu\nu\rho\sigma})-\frac{2a^*}{\Omega_{(D-2)}\tilde{L}^{D-2}}\int_{\partial \mathcal{M}}\sqrt{h}\, \Big[ K+ \text{counterterms} \Big]\, ,
\end{equation}
where $\Omega_{D-2}\equiv 2\pi^{(D-1)/2}/\Gamma((D-1)/2)$ is the area of the unit sphere $\mathbb{S}^{D-2}$, $\tilde{L}$ is the AdS radius, and $a^*$ is the charge appearing in the universal contribution to the entanglement entropy across a spherical entangling surface $\mathbb{S}^{D-3}$ in the dual CFT. This quantity is related, for any higher-curvature theory of gravity, to the on-shell Lagrangian of the theory on pure AdS through \cite{Imbimbo:1999bj,Schwimmer:2008yh,Myers:2010tj,Myers:2010xs,ECGholo}
\begin{equation}
a^*=-\frac{\pi^{(D-1)/2}\tilde{L}^{D}}{(D-1)\Gamma\left[\frac{D-1}{2}\right]}\left. \mathcal{L}\right|_{\text{AdS}}\,.
\end{equation}
 In \cite{ECGholo}, it was also argued that the same counterterms required to produce finite on-shell actions for Einstein gravity solutions can also be used for higher-curvature gravities of this class if we weight them by the same overall coefficient. With minor modifications in the $D=6$ case --- see discussion below Eq. \req{full6} --- associated with the fact that the solutions are only locally asymptotically AdS, Eq. \req{assd}  satisfactorily removes all divergent terms in the corresponding on-shell actions, and yields thermodynamic masses that agree with the ADM ones in all cases.

The structure of this chapter is simple. In Section~\ref{ECGs} we construct Taub-NUT/bolt solutions of $D=4$ Einsteinian cubic gravity and in Section~\ref{quarticAp}, we repeat the analysis including an additional quartic density of the Generalized quasi-topological class. In Section~\ref{sIx} we construct the corresponding solutions for $D=6$ Quartic Generalized quasi-topological gravities. 
In each case, we compute the relevant thermodynamic quantities of the solutions, with special emphasis on the most standard cases $\mathcal{B}=\mathbb{S}^2$ and $\mathcal{B}=\mathbb{CP}^2$, for which we study the corresponding phase spaces finding interesting new phenomena. Subsection \ref{critic} is somewhat different from the rest. It is devoted to the critical limit of Einsteinian cubic gravity, for which the solutions can be constructed analytically. Some details regarding our numerical computations can be found in Appendix \ref{methods}.

\section{Four dimensions: Einsteinian cubic gravity} \label{ECGs}

The first theory we will consider is four-dimensional Einsteinian cubic gravity with a negative cosmological constant \cite{PabloPablo}.
 
Its Euclidean action reads %\comment{something about bhs and further developments?}
\begin{equation}\label{ECGAction}
I_E=-\frac{1}{16\pi G}\int d^4x \sqrt{g}\left[\frac{6}{L^2}+R-\frac{\mu L^4}{8} \mathcal{P} \right]\, ,
\end{equation}
where   the cubic density $\mathcal{P}$ is defined as
\begin{equation}
\mathcal{P}=12 \tensor{R}{_{\mu}^{\rho}_{\nu}^{\sigma}}\tensor{R}{_{\rho}^{\alpha}_{\sigma}^{\beta}}\tensor{R}{_{\alpha}^{\mu}_{\beta}^{\nu}}+\tensor{R}{_{\mu\nu}^{\rho\sigma}}\tensor{R}{_{\rho\sigma}^{\alpha\beta}}\tensor{R}{_{\alpha\beta}^{\mu\nu}}-12R_{\mu\nu\rho\sigma}R^{\mu\rho}R^{\nu\sigma}+8\tensor{R}{_{\mu}^{\nu}}\tensor{R}{_{\nu}^{\rho}}\tensor{R}{_{\rho}^{\mu}}\ .
\end{equation}
The explicit form of the field equations of \req{ECGAction} can be found in Eq.~\req{eq:fe}. 
The theory admits pure AdS$_4$ solutions of radius $\tilde{L}$ related to the action scale $L$ by $\tilde L^2=L^2/f_{\infty}$, where $f_{\infty}$ is determined through
\begin{equation}\label{finn}
1-f_{\infty}+\mu f_{\infty}^3=0\, .
\end{equation}
Throughout the chapter we will assume $0\leq \mu \leq 4/27$, for which a unique branch of stable AdS vacua reducing to the Einstein gravity one as $\mu\rightarrow 0$ exists. In general, stable vacua exist for $\mu<0$ as well, but these are eliminated by the requirement that black holes have positive energy \cite{Hennigar:2016gkm,PabloPablo2}. On the other hand, values of $\mu$ larger than $4/27$ always give rise to
unstable vacua. The ``critical'' limit of the theory \cite{Feng:2017tev}, corresponding to $\mu= 4/27$, warrants
special attention.  For that value of the coupling, the effective Newton constant diverges, and a number of simplifications take place, including the existence of analytic black hole solutions --- as well as various exotic results from the point of view of a putative CFT dual \cite{ECGholo}.

Let us consider a metric ansatz with NUT charge $n$ of the form \req{FFnut} where, initially, we choose the base spaces $\mathcal{B}=\mathbb{S}^2$, $\mathbb{T}^2$ and $\mathbb{H}^2$, although we shall only construct explicit solutions for the first two. The base-space metrics and $1$-forms appearing in \req{FFnut} can then be written, respectively, as 
\begin{equation}\label{base}
d\sigma^2_{\mathcal{B}}=\begin{cases}
d\theta^2+\sin^2\theta d\phi^2 \quad {\rm if}\quad {\mathcal{B}}=\mathbb{S}^2\, ,\\
\frac{1}{L^2}(d\eta^2+d\zeta^2)\quad {\rm if}\quad {\mathcal{B}}=\mathbb{T}^2\, ,\\
d\chi^2+\sinh^2\chi d\rho^2 \quad {\rm if}\quad {\mathcal{B}}=\mathbb{H}^2\, ,\\
\end{cases}\, \quad
A_{\mathcal{B}}=\begin{cases}
2\cos\theta d\phi \quad {\rm if}\quad {\mathcal{B}}=\mathbb{S}^2\, ,\\
\frac{2\eta d\zeta}{L^2} \quad {\rm if}\quad {\mathcal{B}}=\mathbb{T}^2\, ,\\
2\cosh \chi d\rho \quad {\rm if}\quad {\mathcal{B}}=\mathbb{H}^2\, .\\
\end{cases}
\end{equation}
%where we choose the periods of $\eta$ and $\zeta$ to be equal, $\beta_{\eta,\zeta}=l$.
We stress again that the most general ansatz for a Taub-NUT metric in a general higher-curvature gravity should involve an additional function --- for example, $g_{\tau\tau}=V_{\mathcal{B}}(r)N_{\mathcal{B}}(r)^2$ instead. It is a remarkable and highly nontrivial property of ECG that, when evaluated on \req{FFnut} with the above choice of base spaces, its field equations reduce to a single differential equation for $V_{\mathcal{B}}(r)$. This is given by (omitting the `$\mathcal{B}$' subscript to reduce the clutter)
\begin{equation}
	\begin{aligned}
		&-2 r V'+\frac{2 V \left(n^2+r^2\right)}{n^2-r^2}+\frac{2 k L^2-6 n^2+6 r^2}{L^2}+\mu L^4\Bigg[\frac{6 V^3 n^2 \left(n^4-16 n^2 r^2-45 r^4\right)}{\left(n^2-r^2\right)^5}\\
		&+\frac{3 V r^2 \left(V''\right)^2}{2 \left(r^2-n^2\right)}+\left(V'\right)^2 \left(\frac{3 V \left(n^4-37 n^2 r^2-2 r^4\right)}{\left(n^2-r^2\right)^3}-\frac{3 k \left(n^2+r^2\right)}{2 \left(n^2-r^2\right)^2}\right)-\frac{3 n^2 r \left(V'\right)^3}{\left(n^2-r^2\right)^2}\\
		&+V'' \left(\frac{6 V^2 \left(2 n^4-15 n^2 r^2-r^4\right)}{\left(n^2-r^2\right)^3}-\frac{6 V r V' \left(5 n^2+r^2\right)}{\left(n^2-r^2\right)^2}+\frac{3 V k \left(n^2-2 r^2\right)}{\left(n^2-r^2\right)^2}\right)\\
		&+V' \left(\frac{6 V k r^3}{\left(r^2-n^2\right)^3}-\frac{6 V^2 r \left(3 n^4+62 n^2 r^2+r^4\right)}{\left(n^2-r^2\right)^4}\right)\\
		&+V^{(3)} \left(-\frac{3 V^2 r \left(4 n^2+r^2\right)}{\left(n^2-r^2\right)^2}+\frac{3 V r^2 V'}{2 \left(r^2-n^2\right)}+\frac{3 V k r}{r^2-n^2}\right)\Bigg]=0\, ,
	\end{aligned}
\end{equation}
where $k=+1,0,-1$ for $\mathbb{S}^2$, $\mathbb{T}^2$  and $\mathbb{H}^2$, respectively.

Despite its challenging appearance, the above equation has two remarkable properties. First, it is of third order, instead of fourth, which is what one would have naively expected. Second, it allows for an integrable factor: after multiplying by $(1-n^2/r^2)$, the equation becomes a total derivative and it can be integrated once. By doing so, we are left with a second-order differential equation of the form \req{mastere}, namely
\begin{equation}
	\label{eqV}
	\begin{aligned}
		&V \left(\frac{2 n^2}{r}-2 r\right)+\frac{2 \left(k L^2 \left(n^2+r^2\right)-3 n^4-6 n^2 r^2+r^4\right)}{L^2 r}+\mu L^4 \Bigg[\frac{6 V^3 n^2 \left(n^2+9 r^2\right)}{r \left(n^2-r^2\right)^3}\\
		&+\left(V'\right)^2 \left(\frac{3 V n^2}{n^2 r-r^3}-\frac{3 k}{2 r}\right)-\frac{\left(V'\right)^3}{2}+V' \left(\frac{3 V^2 \left(17 n^2+r^2\right)}{\left(n^2-r^2\right)^2}+\frac{3 V k}{n^2-r^2}\right)+\\
		&V'' \left(-\frac{3 V^2 \left(4 n^2+r^2\right)}{r^3-n^2 r}+\frac{3 V V'}{2}+\frac{3 V k}{r}\right)\Bigg]=4C\, ,
	\end{aligned}
\end{equation}
where $C$ is an integration constant which will be related to the energy of the solution.  %Hence, in a very similar fashion to the case of black holes, the problem is reduced to a single second order differential equation for $V(r)$. 

We now require the metric \req{FFnut} to be locally asymptotically AdS; as a consequence we must demand $V(r)\rightarrow f_{\infty}\frac{r^2}{L^2}+\mathcal{O}(1)$ as $r\rightarrow +\infty$. Performing a $1/r$ expansion   we find
\begin{equation}
	V(r)=f_{\infty}\frac{r^2}{L^2}+k-5f_{\infty}\frac{n^2}{L^2}-\frac{2C}{r(1-3 f_{\infty}^2\mu)}+\mathcal{O}(r^{-2})
	= V_{p}(r)  \, .
	\label{rasymp}
\end{equation}
The effective Newton constant of the theory is given by
\beq
G_{\rm eff}=\frac{G}{1-3 f_{\infty}^2\mu}\, ,
\eeq
so, at least in the spherical case, we can identify the integration constant in \req{eqV} with the ADM mass of the solution as $C=GM$. In an abuse of notation, we will use this definition for all  base spaces. Now, note that since \req{eqV} is a second-order differential equation, it possesses a two-parameter family of solutions, of which \req{rasymp} corresponds to a particular one. In order to find the remaining asymptotic solutions, let us write $V(r)=V_{p}(r)+\frac{r^2}{L^2} g(r)$ and expand linearly in $g$. Taking into account only the leading terms when $r\rightarrow+\infty$, we find that $g$ satisfies the following equation
\begin{equation}\label{heq}
9 L^2 GM\mu f_{\infty} g''(r)-2r(1-3\mu f_{\infty}^2)^2g(r)=0\, .
\end{equation}
Leaving aside the limiting values $\mu=0,4/27$, the general solution is given by 
%\tcp{we recognize here the Airy equation, so that the general solution is given by
\begin{equation}\label{hSOL}
g(r)=A \textrm{AiryAi}\left[\left(\frac{2(1-3\mu f_{\infty}^2)^2}{9L^2{G M\mu f_{\infty}}}
\right)^{1/3}r\right]+B \textrm{AiryBi}\left[\left(\frac{2(1-3\mu f_{\infty}^2)^2}{9L^2{G M\mu f_{\infty}}}
\right)^{1/3}r\right]
\end{equation}
where $\textrm{AiryAi}[x]$ and $\textrm{AiryBi}[x]$ are the Airy functions of the first and second kind, respectively.
When $GM\mu>0$, the solution involving $\textrm{AiryBi}$ grows exponentially, while the one with $\textrm{AiryAi}$  decays.
Therefore, we must set $B=0$ in order for the solutions to be locally asymptotically AdS. Hence, we learn that the asymptotic boundary condition is fixing one of the integration constants in \req{eqV}. The remaining one will be fixed by the corresponding regularity conditions in the bulk, as we will show in the following sections. When $GM\mu<0$, the solutions  \req{hSOL} have an oscillatory character and they are all singular at infinity (except the trivial one, $g=0$). To remove this behaviour we would need to set both $A$ and $B$ to zero, which would fully specify the solution. This would leave us with no integration constants to impose regularity in the bulk. This behaviour is very similar to that found for the static black hole solutions of the theory \cite{Hennigar:2016gkm,PabloPablo2} --- see also \cite{PabloPablo4}, and leads us to choose $\mu\ge 0$, so that the solutions with $GM\mu>0$ have positive energy.  

\subsection*{Einstein gravity}
In the following subsections we will consider the base spaces $\mathbb{S}^2$ and $\mathbb{T}^2$ independently, and we will construct new Taub-NUT and bolt solutions for them for general values of $\mu$. It is illustrative however to start analyzing the Einstein gravity case, for which the analysis can be performed at the same time for all base spaces. Indeed, if we set $\mu=0$, \req{eqV} can be easily solved for $V_{\mathcal{B}}(r)$. Imposing the NUT condition $V_{\mathcal{B}}(r=n)=0$ first, one is left with
\begin{equation}\label{NUTE}
V_{\mathcal{B}}(r)=\frac{(r-n)\left[(r-n)(3n+r)+ kL^2\right]}{L^2(n+r)}\, ,
\end{equation}
where we already fixed the integration constant as
\begin{equation}\label{massE}
GM=k n - \frac{4n^3}{L^2}\, .
\end{equation}
The regularity condition \req{smooth} imposes
\begin{equation}\label{betaEin}
\beta_{\tau}=\frac{8\pi n}{k}\, ,
\end{equation}
which means that $\tau$ cannot be a compact coordinate for $\mathcal{B}=\mathbb{T}^2$ or, in other words, the solution is extremal, in the sense that the temperature $T\equiv 1/\beta_{\tau}$ is forced to vanish. Similarly, for $\mathcal{B}=\mathbb{H}^2$, one finds that the period of $\tau$ would need to be negative. This means that $V_{\mathbb{H}^2}(r)$ actually becomes negative for values of $r$ greater than $n$, which is forbidden by assumption. Hence, no regular Taub-NUT solution exists in that case for Einstein gravity. 

If we impose the bolt condition $V_{\mathcal{B}}(r=r_b)=0$ instead, we find
\begin{equation}\label{boltEi}
V_{\mathcal{B}}(r)=\frac{(r-r_b)\left[(6n^2 r r_b-3 n^4+k L^2 (n^2-r r_b)-r r_b (r^2+r r_b+ r_b^2)) \right]}{L^2(n^2-r^2)r_b}\, ,
\end{equation}
where the integration constant was fixed as
\begin{equation}
GM=\frac{k L^2(n^2+r_b^2)-3n^4-6n^2r_b^2+r_b^4}{2 L^2 r_b}\, .
\end{equation}
The regularity condition \req{smooth} fixes now the bolt radius as a function of $n$ and $\beta_{\tau}$, namely
\begin{equation}\label{boltrb}
r_b=\frac{2L^2 \pi}{3\beta_{\tau}} \left[ 1\pm \sqrt{ 1-\frac{3k \beta_{\tau}^2}{4L^2 \pi^2} +\frac{9n^2 \beta_{\tau}^2}{4L^4\pi^2}}\right]\, .
\end{equation}
In order for each solution to be allowed, it must be such that $r_b>n$.  Furthermore,
the quantity inside the square root must be positive, which restricts the allowed values of $n$ for which the corresponding solutions exists.

 On general grounds,	in order to remove the so-called Misner string \cite{Misner:1963fr}, an additional condition must be imposed on $\beta_{\tau}$ both for NUT and bolt solutions when $\mathcal{B}=\mathbb{S}^2$. As we explain in the next subsection, this reads $\beta_{\tau}=8\pi n$. It is a remarkable (and peculiar) fact that in Einstein gravity Eq.~\req{smooth} automatically implements this condition in the case of the NUT solution. In general, both conditions must be imposed separately.

\subsection{$\mathcal{B}=\mathbb{S}^2$}\label{S2cubic4d}
Let us now turn on the Einsteinian cubic gravity coupling. We begin by assuming the base space to be
%\comment{This needs to be fully rewritten and reorganized}
the one-dimensional complex projective space $\mathbb{CP}^1 $ or, equivalently, the two-dimensional round sphere, $\mathcal{B}=\mathbb{S}^2$. Then, the metric \req{FFnut} reads
\begin{equation}
	ds^2=V_{\mathbb{S}^2}(r)(d\tau+2 n \cos\theta d\phi)^2+\frac{dr^2}{V_{\mathbb{S}^2}(r)}+(r^2-n^2)\left(d\theta^2+\sin^2\theta d\phi^2\right)\, .
	\label{taub}
\end{equation}
This metric has ``wire singularities'' at $\theta=0, \pi$, for which it becomes noninvertible. As shown by Misner \cite{Misner:1963fr}, it can nevertheless be  made regular everywhere using two coordinate patches. The idea is to define new coordinates $\tau^{\pm}= \tau \pm 2n \phi$ covering the $\theta\geq \pi/2$ and $\theta \leq \pi/2$ regions respectively. In the overlap region, $\tau^+=\tau^-+4n\phi$, and since $\beta_{\phi} = 2\pi$, one is forced to impose the periods of $\tau^{\pm}$ to be $\beta_{\tau^{\pm}}=8 \pi n$.
%Note that since the 1-form $\cos\theta d\phi$ is singular at $\theta=0, \pi$, a proper definition of this metric involves two patches, in which we write $d\tau+2 n \cos\theta d\phi\rightarrow d\tau_{\pm}+2n(\cos\theta\pm 1)d\phi$, where the $+$ sign corresponds to the southern hemisphere, and the $-$ sign to the northern one. 
%In the overlap we have $\tau_--\tau_+=4 n \phi$, and one concludes that $\tau_+$ and $\tau_-$ must have a period of $8 n \pi$. 
For clarity reasons, in what follows we will work with the metric \req{taub} in a single patch, but taking into account  that the period of $\tau$ is related to the NUT charge through  $\beta_{\tau}=8\pi n$. Observe that this condition is a consequence of choosing $\mathcal{B}=\mathbb{S}^2$ and does not depend on the theory. When combined with the general regularity condition \req{smooth}, this gives rise to the conditions $V_{\mathbb{S}^2}'(r=n)=1/(2n)$ and $V_{\mathbb{S}^2}'(r=r_b)=1/(2n)$ respectively for NUTs and bolts.

The function $V_{\mathbb{S}^2}$ in \req{taub} is determined from \req{eqV} with $k=1$. Using the asymptotic expansion \req{rasymp}, we see that when $r\rightarrow +\infty$ the metric induced on a constant-$r$ hypersurface is given by
\begin{equation}
	\frac{^{(3)}ds^2}{r^2}=\frac{4 f_{\infty} n^2}{L^2}(d\psi+\cos\theta d\phi)^2+d\theta^2+\sin^2\theta d\phi^2+\mathcal{O}(r^{-2})\, ,
\end{equation}
where we have introduced the angle coordinate $\psi= \tau/(2n)$, whose period is $4\pi$. When $4 f_{\infty} n^2=L^2$, the previous metric is the one of a round $\mathbb{S}^3$. For any other value of $n$, it is the metric of a squashed sphere, and it is customary \cite{Chamblin:1998pz,Hartnoll:2005yc,Anninos:2012ft,Bobev:2016sap,Bobev:2017asb} to rewrite the NUT charge in terms of a `squashing parameter' $\alpha$ as $4 f_{\infty} n^2/L^2=1/(1+\alpha)$.
%\begin{equation}
%	\frac{4 f_{\infty} n^2}{L^2}=\frac{1}{1+\alpha}\, .
%\end{equation}
In order to specify the solution, we need to choose a boundary condition at some finite $r=r_b$.
Depending on whether we choose $r_b=n$, or $r_b>n$, we will be considering Taub-NUT or Taub-bolt solutions.
% When $\rh=n$, we will have a Taub-NUT solution, whereas Taub-bolt solutions will correspond to $\rh>n$.

%%%%%%%%%%%%%%%
\subsubsection{Taub-NUT solutions}\label{sec:TNECG}
\begin{figure}[t!]
	\centering 
	\includegraphics[width=0.65\textwidth]{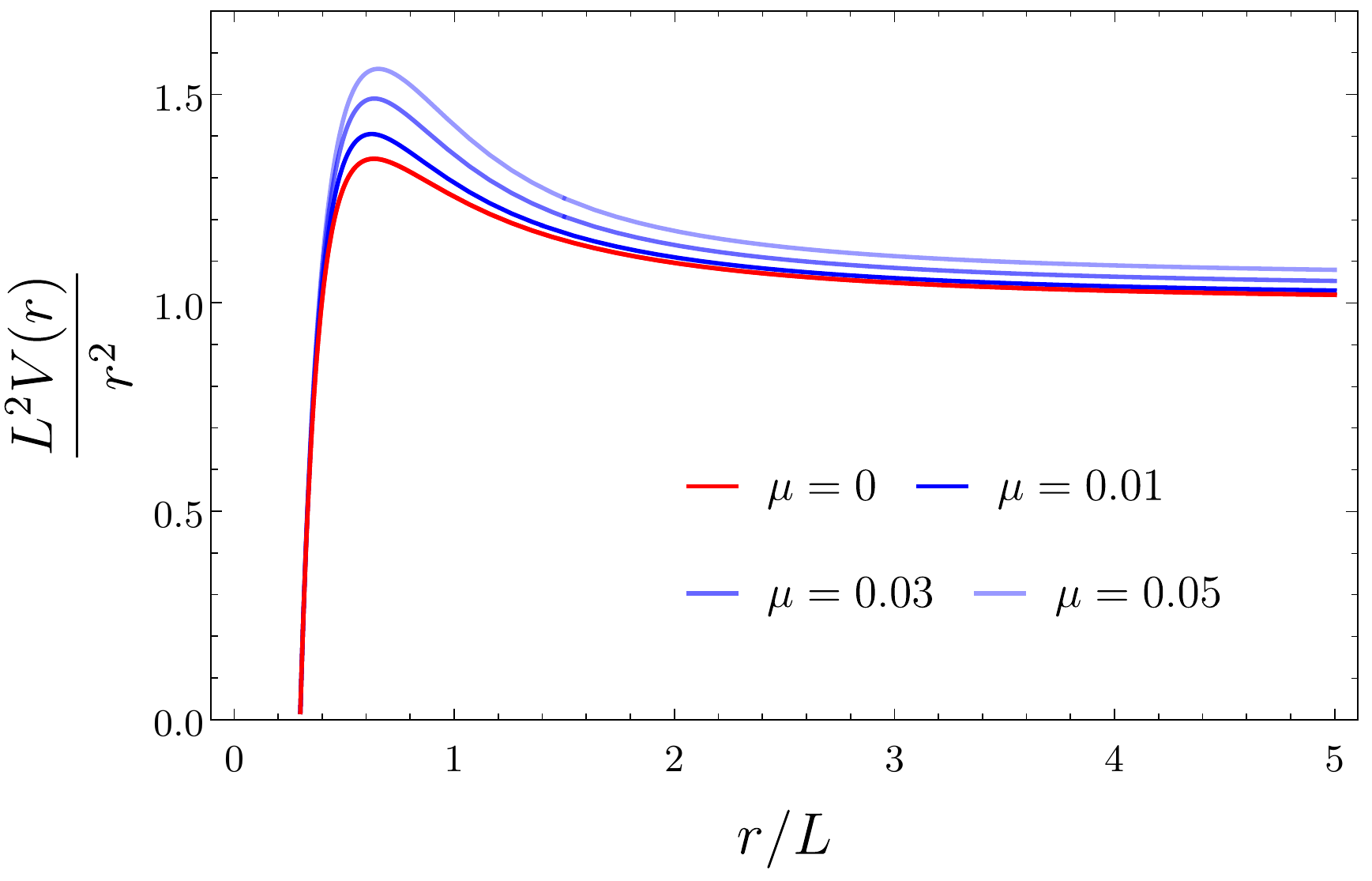}
	\caption{We plot the metric function $ V_{\mathbb{S}^2}(r)\cdot L^2/r^2$ corresponding to Taub-NUT solutions of ECG with $n/L=0.3$ for several values of $\mu$. The examples shown correspond to values of $\mu$ and $n$ which satisfy the positive-mass inequalities \req{condi}.  The red curve corresponds to the metric function for the Einstein gravity Taub-NUT solution given by \req{NUTE} with $k=1$. } 
	\label{NUTS2ECG}
\end{figure}

As we have explained, the Euclidean Taub-NUT metric is characterized by the conditions $V_{\mathbb{S}^2}(r=n)=0$ and $V_{\mathbb{S}^2}'(r=n)=1/(2n)$. Let us then expand $V_{\mathbb{S}^2}(r)$ around $r=n$ as 
\begin{equation}\label{near}
V_{\mathbb{S}^2}(r)=\frac{(r-n)}{2 n}+\sum_{i=2}^\infty (r-n)^i a_i\, ,
\end{equation}
for some $a_i$.
Plugging this expansion into \req{eqV}, we observe that the $\mathcal{O}\left(r-n\right)$ and $\mathcal{O}\left((r-n)^2\right)$ equations are automatically satisfied, whereas the $\mathcal{O}(1)$ one gives rise to the following relation between the mass and the NUT charge,
\begin{equation}\label{masss}
	GM=n-\frac{4 n^3}{L^2}-\frac{\mu L^4}{16 n^3}\, .
\end{equation}
Observe that this reduces to the Einstein gravity expression \req{massE} for $\mu=0$.
The following term in the expansion gives a relation between $a_3$ and $a_2$, which we can use to write the former as a function of the latter,
$a_3(a_2)$. Similarly, the following term allows us to obtain $a_4 (a_2 )$, and so on. Hence, as in the black hole case \cite{Hennigar:2016gkm,PabloPablo2}, the full series is determined
by a single free parameter $a_2$. This parameter must be chosen in a way such that $B = 0$ in \req{hSOL}, which ensures that the solution is locally asymptotically AdS. In practice, the shooting method can be used to identify $a_2$ for each value of $\mu$, so that the near $r=n$ expansion yields a good approximation to the exact solution that connects with the asymptotic expansion \req{rasymp}. There is a unique $a_2$ for each $\mu$ that does the job, corresponding to a unique Taub-NUT solution in each case. We plot the metric function $V_{\mathbb{S}^2}(r)$ for different values of $\mu$ in Fig. \ref{NUTS2ECG}. These solutions generalize the Einstein gravity Taub-NUT solution (the red curve in Fig. \ref{NUTS2ECG}), whose metric function is given by \req{NUTE} with $k=1$. As we can see, the qualitative behaviour of $V_{\mathbb{S}^2}(r)$ is very similar to that of Einstein gravity  for nonvanishing values of the ECG coupling.

One peculiarity of \req{masss} for nonvanishing $\mu$ is that the mass becomes negative for small values of $n$. In particular, the mass is non-negative only when 
\begin{equation}\label{condi}
\frac{L^2}{4f_{\infty}} \geq n^2 \geq \frac{L^2 \mu f_{\infty}^2}{8}\left[1+\sqrt{\frac{4-3\mu f_{\infty}^2}{\mu f_{\infty}^2}}\right]\, .
%\quad \text{or, equivalently, when} \quad	0\le\alpha\le-\frac{3}{2}+\frac{1}{2}\sqrt{\frac{f_{\infty}+3}{f_{\infty}-1}}\, .
\end{equation}
The existence of a finite lower bound for $n$
is a new feature, which does not occur for Einstein gravity. Indeed, in that case \req{condi} becomes $L^2/(4f_{\infty})\geq n^2 \geq 0$. 
%and $0\leq \alpha \leq +\infty$. 
For general values of the gravitational coupling, we cannot expect the solution to exist whenever $n$ lies outside the interval in \req{condi}, because the asymptotic behaviour for negative masses is pathological.  Indeed, as we explained in the discussion below Eq.~\req{hSOL}, negative mass solutions would be highly oscillating at infinity, and hence they are not asymptotically AdS. 
%The theory avoids dynamically solutions with negative mass. 
Solutions with zero mass occur when either the upper or the lower bounds are saturated. In terms of $\mu$, the $M=0$ condition reads
% whenever
\begin{equation}\label{num0}
\mu=\frac{16 n^4}{L^4}\left[1-\frac{4n^2}{L^2}\right]\, .
\end{equation}
In the case of Einstein gravity, the possibilities are $n^2=L^2/4$ and $n=0$, for which the solution reduces to pure Euclidean AdS$_4$ foliated by round $\mathbb{S}^3$ slices and $\mathbb{S}^1\times \mathbb{S}^2$ slices, respectively\footnote{Observe however that the $n\rightarrow 0$ limit is problematic, in the sense that the period of $\tau$ would vanish in that case. One can of course just set $n=0$ from the beginning, which makes the problem disappear.} \cite{Emparan:1999pm}. For any nonvanishing value of $\mu$, \req{num0} is satisfied identically for $n^2=L^2/(4f_{\infty})$, which can be straightforwardly checked using \req{finn}. In this case, the solution also reduces to pure AdS$_4$, the metric factor being simply given by $V_{\mathbb{S}^2}(r)=f_{\infty}r^2/L^2-1/4$. Besides this solution, there exists another one obtained by choosing $n$ to saturate the lower bound in \req{condi}, and which is analogous to the $n=0$ one in Einstein gravity. Interestingly, this solution no longer reduces to pure Euclidean AdS$_4$ for  $0< \mu < 4/27$ but, rather, it has a nontrivial profile. In the critical limit, $\mu=4/27$, the range allowed by \req{condi} collapses to a single possible value, corresponding to $n^2=L^2/6$. In that case, the solution does correspond to pure AdS$_4$. 
For other values of $n$, the critical solution can also be accessed analytically (see Section \ref{critic}), and the result for the metric function reads
\begin{equation}\label{cric}
V_{\mathbb{S}^2}^{\rm cr}(r)=\frac{3}{2L^2}(r^2-n^2)\, .
\end{equation}
This solution has a vanishing mass parameter $M=0$, namely, it only exists if we fix the integration constant $C$ to zero in \req{eqV}. Note that this solution has ${V_{\mathbb{S}^2}^{\rm cr}}'(n)=3n/L^2$;  it has a conical singularity at $r=n$ in all cases but one, corresponding to the value $n^2=L^2/6$, for which it becomes pure AdS$_4$, as mentioned above.

\begin{figure}[t!]
	\centering 
	\includegraphics[width=0.65\textwidth]{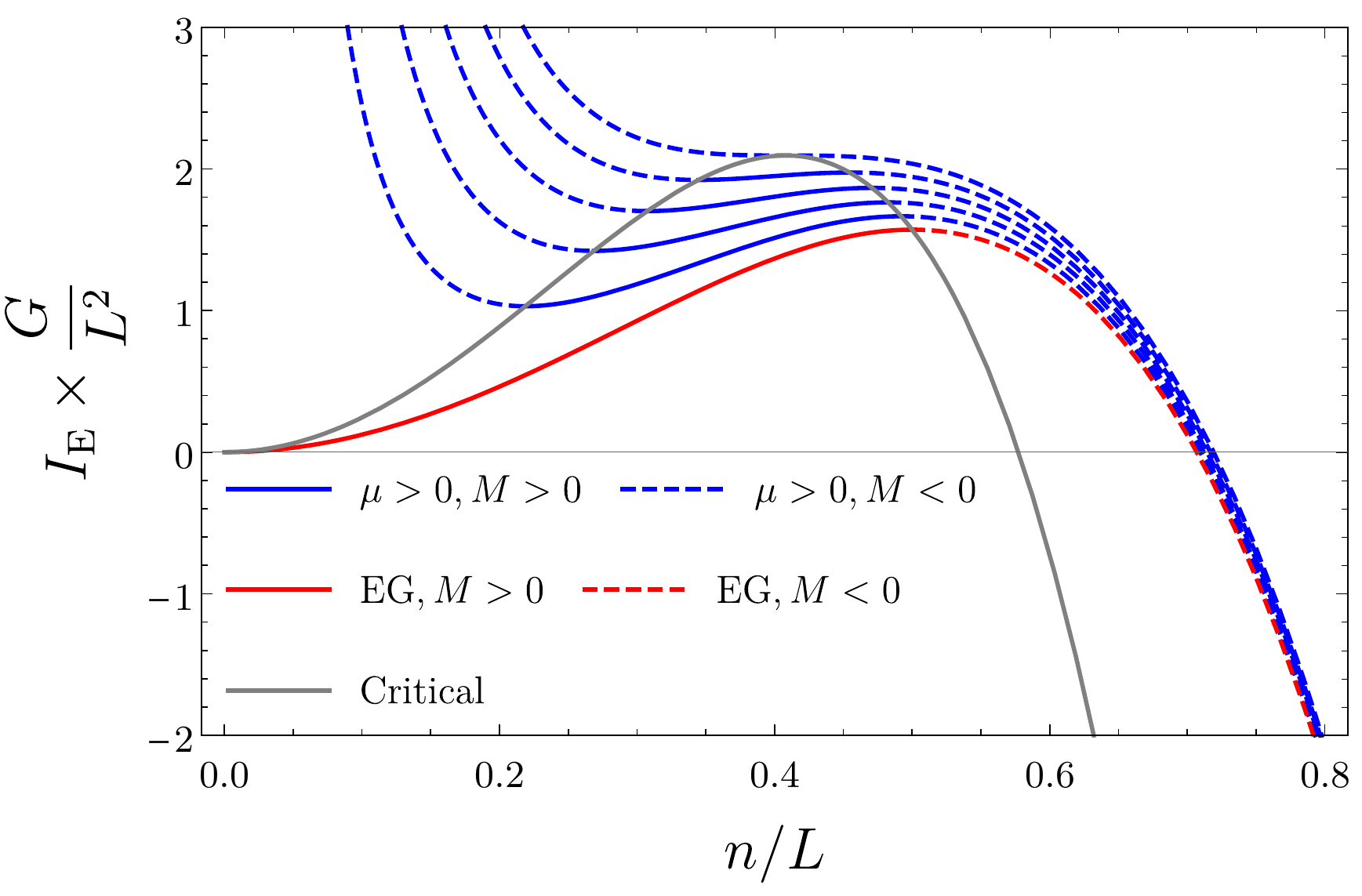}
	\caption{Free energy of the NUT solution with $\mathbb{S}^2$ base. In red we show the Einstein gravity result and in blue the ECG ones for $\mu=8/270,\, 16/270,\, 24/270,\, 32/270,\, 4/27$. The mass $M$ is proportional to the slope of each curve and solid lines represent $M>0$ while dashed lines represent $M<0$. For ECG only the solutions with $M>0$ exist. We also plot the free energy of the special critical solution \req{IEcr}, whose metric function is given in \req{cric}. Remarkably, this curve is the envelope of the free energies with $M>0$. }
	\label{solNUTS2}
\end{figure}

Let us now evaluate the on-shell action of the solutions. In order to do so, we make use of the generalized action \req{assd} where, for ECG, the charge $a^*$ is given by $a^*=(1+3\mu f_{\infty}^2)\tilde{L}^2/(4G)$.
Using this, the full ECG action takes the form
\begin{equation}\label{EuclideanECG}
\begin{aligned}
I_E=-\int  \frac{d^4x \sqrt{g}}{16\pi G}\left[\frac{6}{L^2}+R-\frac{\mu L^4}{8} \mathcal{P} \right]-\frac{(1+3\mu f_{\infty}^2)}{8 \pi G}\int_{\partial \mathcal{M}}d^3x\sqrt{h}\left[K-\frac{2\sqrt{f_{\infty}}}{L}-\frac{L}{2\sqrt{f_{\infty}}}\mathcal{R}\right]\, ,
\end{aligned}
\end{equation}
where $\mathcal{R}$ stands for the Ricci scalar of the induced metric on the boundary. The last two terms in the second line are the standard counterterms in $D=4$ which, as explained in the introduction, also appear weighted by $a^*$ without further modification according to the prescription in \cite{ECGholo}. An explicit evaluation of all the terms in the above expression for our new Taub-NUT solutions can be performed fully analytically using the near $r=n$ and asymptotic expansions. Let us present here the complete calculation in order to illustrate how it works. 
For the configuration \req{taub}, the Lagrangian is a total derivative, so the bulk part of the action can be integrated exactly,
\begin{equation}
I_{\rm bulk}=-\frac{4\pi \beta}{16 \pi G}\int_{n}^{L^2/\delta}dr(r^2-n^2)\mathcal{L}=-\frac{\beta}{4G}F(r)\Big|_{n}^{L^2/\delta}\, ,
\end{equation}
introducing a UV cutoff $\delta$, 
where $\beta=8\pi n$ is the periodicity of the Euclidean time and 
\begin{equation}
\begin{aligned}
F(r)&=\left(n^2-r^2\right) V'(r)-2 r V(r)+\frac{2 r \left(L^2-3 n^2+r^2\right)}{L^2}\\
&+\mu L^4 \Bigg[\left(-\frac{6 n^2 V(r)}{\left(n^2-r^2\right)^2}-\frac{21 n^2 \left(n^2+r^2\right) V(r)^2}{\left(n^2-r^2\right)^3}\right) V'(r)-\frac{\left(5 n^2+r^2\right) V'(r)^3}{4 n^2-4 r^2}\\
&-\frac{6 n^2 r \left(5 n^2+r^2\right) V(r)^3}{\left(n^2-r^2\right)^4}-\frac{6 n^2 r V(r)^2}{\left(n^2-r^2\right)^3}+\left(\frac{3 r}{2 \left(r^2-n^2\right)}-\frac{3 \left(9 n^2 r+r^3\right) V(r)}{2 \left(n^2-r^2\right)^2}\right) V'(r)^2\Bigg]\, .
\end{aligned}
\end{equation}
 Using \req{rasymp} to compute $F(L^2/\delta)$, we find
\begin{equation}
\begin{aligned}
I_{\rm bulk}=\frac{2\pi n}{G}\left[F(r\rightarrow n)-\left(\frac{2L^4}{\delta^3}-\frac{6 n^2}{\delta}\right)\left(1-2 f_{\infty}-2 f_{\infty}^3 \mu\right)-2G_{\rm eff}M\left(1+3 f_{\infty}^2 \mu\right)+\mathcal{O}(\delta/L^2)\right]\, .
\end{aligned}
\end{equation}
Now, for the boundary contributions, we use the trace of the extrinsic curvature at $r=L^2/\delta$, and the Ricci scalar of the induced metric, respectively given  by
\begin{equation}\label{K4D}
K=\frac{2(L^2/\delta)}{(L^2/\delta)^2-n^2}V(L^2/\delta)^{1/2}+\frac{1}{2}\frac{ V'(L^2/\delta)}{V(L^2/\delta)^{1/2}}\, , \quad 	\mathcal{R}=\frac{2(L^2/\delta)^2-2(1+V(L^2/\delta))n^2}{(n^2-(L^2/\delta)^2)^2}\, .
\end{equation}
%while the Ricci scalar of the induced metric reads
%\begin{equation}\label{R4D}
%	\mathcal{R}=\frac{2r_0^2-2(1+V(r_0))n^2}{(n^2-r_0^2)^2}\, .
%\end{equation}
Then, using the asymptotic expansion \req{rasymp} we find the boundary contribution
\begin{equation}
I_{\rm boundary}=\frac{2\pi n}{G}\left(1+3\mu f_{\infty}^2\right)\left[-f_{\infty}\left(\frac{2L^4}{\delta^3}-\frac{6 n^2}{\delta}\right)+ 2 G_{\rm eff}M\right]+\mathcal{O}(\delta/L^2)\, .
\end{equation}
Adding up bulk and boundary contributions, we find 
\begin{equation}
I_E=\frac{2\pi n}{G}\left[F(r\rightarrow n)-\left(\frac{2L^4}{\delta^3}-\frac{6 n^2}{\delta}\right)\left(1- f_{\infty}+ f_{\infty}^3 \mu\right)+\mathcal{O}(\delta/L^2)\right]=\frac{2\pi n}{G}F(r\rightarrow n)+\mathcal{O}(\delta/L^2)\, ,
\end{equation}
where in the last equality we used the defining equation of $f_{\infty}$, \req{finn}. Remarkably, all contributions coming from the boundary cancel out, including constant terms. Finally, taking the limit $\delta\rightarrow 0$ and using the expansion \req{near}, we are left with the simple result
\begin{equation}\label{freeee1}
	I_E=\frac{4\pi}{G}\left[n^2-\frac{2n^4}{L^2}+\frac{\mu L^4}{16 n^2}\right]\, .
\end{equation}
This reduces to the free energy of the corresponding Einstein gravity Taub-NUT solution when $\mu=0$, as it should. The energy and entropy can be easily obtained now from
$
	E=\partial I_E/\partial \beta$ and $S=\beta E-I_E\, .
$
Using this, we find that the energy precisely matches the result for the ADM mass obtained in \req{masss}, $E=M$, which is a highly nontrivial check of the calculation, whereas for the entropy we obtain
\begin{equation}\label{entR}
S=\frac{4\pi}{G}\left[n^2-\frac{6n^4}{L^2}-\frac{3\mu L^4}{16 n^2}\right] \, ,
\end{equation}
 which is not given by a simple area law due to contributions from the Misner string.

It is also possible to consider the thermodynamics of these NUT charged solutions from the perspective of extended phase space thermodynamics. Within this framework, one introduces potentials conjugate to the cosmological constant  --- interpreted as a pressure $P = - \Lambda/(8 \pi G)$ --- and any  higher-curvature couplings that appear in the action~\cite{Kastor:2009wy, Kastor:2010gq}.  These considerations are motivated by scaling arguments, since without these terms the Smarr relation fails to hold. In the case of Taub-NUT solutions in ECG, the extended first law reads
\begin{equation}
dE = TdS + VdP + \Upsilon^{\ssc \rm ECG} d ( \mu L^4 ) \, ,
\end{equation}
where we have restored the dimensions to the ECG coupling constant. The new potentials read
\begin{equation}
V = - \frac{8 \pi n^3}{3} \, , \quad \Upsilon^{\ssc \rm ECG} = \frac{1}{32 G n^3} \, .
\end{equation}
Interestingly, the thermodynamic volume here is precisely the same as for Taub-NUT solutions in Einstein gravity~\cite{Johnson:2014xza}. The same conclusion holds for the thermodynamic volume of black holes in  higher-curvature gravities that belong to the generalized quasi-topological class. With the thermodynamic quantities defined as above, the Smarr relation that follows directly from a scaling argument is found to hold:
\begin{equation}\label{smarrfS2}
E = 2 TS - 2 VP + 4 \mu L^4 \Upsilon^{\ssc \rm ECG} \, .
\end{equation}

In Fig. \ref{solNUTS2} we plot the Euclidean action \req{freeee1} for several values of $\mu$. Dashed lines correspond to negative values of the mass, whereas solid lines correspond to solutions with $M>0$. As we mentioned earlier, in principle we only expect solutions with positive mass $M>0$ to exist. A numerical analysis seems to confirm this, since we were not able to construct any solution with $M<0$.  This also constrains the validity of the thermodynamic expressions \req{freeee1} and \req{entR} to the interval defined by \req{condi}. This interval becomes smaller as $\mu$ grows, and it reduces to a single point, $n^2=L^2/6$, in the critical limit. Interestingly, we observe that the free energy of the critical theory solutions (solid gray curve) acts as an envelope of all possible solutions with positive mass and arbitrary values of $\mu$.  Observe that this free energy cannot be obtained from \req{freeee1} in the $\mu\rightarrow 4/27$ limit; the same applies to the mass, which cannot be obtained from \req{masss}. The correct result for the on-shell action associated to the critical solutions with metric function \req{cric} reads however
\begin{equation}\label{IEcr}
I_{ E}^{\rm cr}=\frac{8\pi n^2}{G}\left[1-\frac{3n^2}{L^2}\right]\, .
\end{equation}
As we mentioned above, all these solutions  except for the one with $n^2=L^2/6$ have conical singularities, so the result must be taken with care --- \eg the mass cannot derived from \req{IEcr} using standard thermodynamic identities. It is a remarkable and somewhat striking fact that the free energy of this singular solution, as given by \req{IEcr}, precisely separates the free energies of negative-mass solutions from those corresponding to completely regular positive-mass solutions for general values of $\mu$. The different nature of the critical solutions can also be seen from the fact that 
whenever $\mu \neq 4/27$, the solutions with $M=0$ correspond to values of $n$ for which $I_E(n)$ is locally extremized, whereas the whole  $\mu = 4/27$ curve has $M=0$.

\subsubsection{Taub-bolt solutions}
\begin{figure}[t!]
	\centering 
	\includegraphics[width=0.65\textwidth]{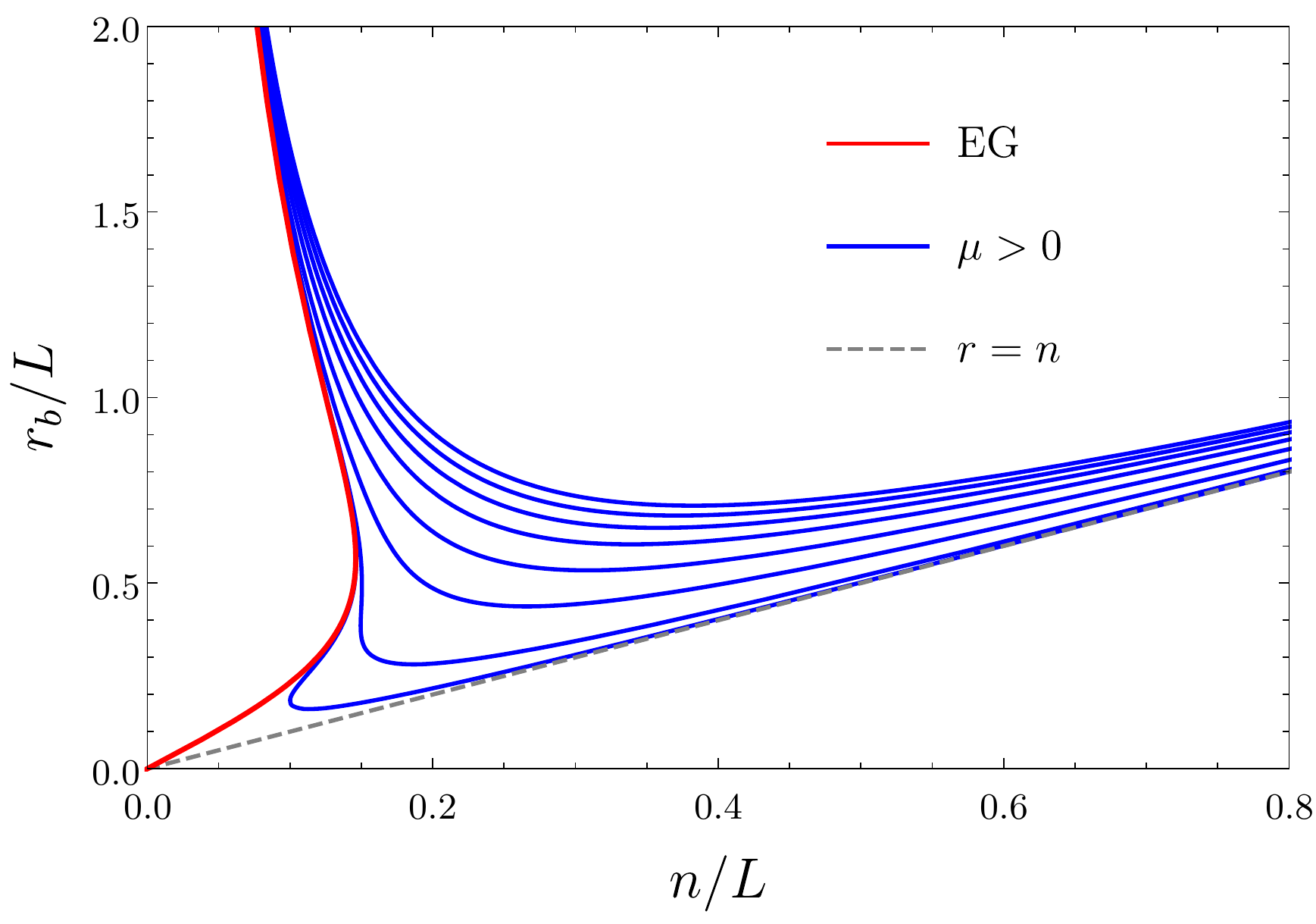}
	\caption{We show the bolt radius $r_b$ for several values of $\mu$. In red we show the Einstein gravity value $(\mu=0)$, and the blue lines correspond to $\mu=0.0001,\, 0.001126,\, 0.01,\, 8/270,\, 16/270,\, 24/270,\, 32/270,\, 4/27$. For any non-vanishing $\mu$ there is at least one solution for every value of $n$. For large $n$ there is a new solution which approaches $r_b\rightarrow n$ asymptotically. The gray dashed line corresponds to NUT solutions.}
	\label{rhbolt}
\end{figure}

Let us now turn to  bolt solutions. These are obtained by imposing $V_{\mathbb{S}^2}(r)$ to vanish for some $r_b>n$, \ie $V_{\mathbb{S}^2}(r_b)=0$, plus the regularity condition $V_{\mathbb{S}^2}'(r_b)=1/(2n)$. If we plug a Taylor expansion for $V_{\mathbb{S}^2}(r)$ around $r=r_b$ including these conditions in \req{eqV}, the equations corresponding to the first nontrivial orders give rise to two equations involving the mass of the solution $M$, the bolt radius $r_b$, and the NUT charge $n$. These read
\begin{eqnarray}\label{gmm3}
GM&=&\frac{n^2+r_b^2}{2r_b}+\frac{1}{L^2}\left[\frac{r_b^3}{2}-\frac{3 n^4}{2r_b}-3 r_b n^2\right]-\frac{\mu L^4}{64n^2}\frac{(6n+r_b)}{n r_b}\, ,\\ \label{rbb}
0&=&\frac{6}{L^2}(r_b^2-n^2)^2+(2-r_b/n)(r_b^2-n^2)-\frac{3\mu L^4}{8n^2}\frac{(r_b^2+n r_b+n^2)}{(r_b^2-n^2)}\, .
\label{Eqr}
\end{eqnarray}
The relation $r_b(n)$ has several remarkable differences with respect to the Einstein gravity case. Indeed, for $\mu=0$, \req{rbb} has two nontrivial roots, given by \req{boltrb}, namely
\begin{equation}\label{Einss}
r_b(\mu=0)=\frac{L^2}{12 n}\left[1\pm \sqrt{1-\frac{48 n^2}{L^2}+ \frac{144n^4}{L^4}}\right]\, .
\end{equation}
Since we want $r_b$ to be real and larger than $n$, this implies that $n/L<\left[(2-\sqrt{3})/12\right]^{1/2}\simeq 0.1494$. In particular, there is no bolt solution near the undeformed $\mathbb{S}^3$ case, corresponding to $n/L=1/2$. The situation is very different in ECG. Indeed, for any nonvanishing value of $\mu$ and for any value of $n$, there always exists at least one solution satisfying $r_b>n$. For small and large $n/L$, there is a unique solution in each case, while intermediate values of $n/L$ give rise to one or three possible solutions, depending on the value of $\mu$. For $\mu<0.001126$, there is a region of values of $n/L$ for which three solutions with $r_b>n$ exist. If $\mu$ is greater than this quantity, there is a two-to-one relation between $n$ and $r_b$ for all $n$. All this is shown in Fig. \ref{rhbolt}.

For the set of parameters for which a unique bolt solution exists, the profile of $V_{\mathbb{S}^2}(r)$ can be accessed numerically following exactly the same logic as for the NUT solutions. We plot the resulting metric functions for some values of $\mu$ in Fig. \ref{rhbolt}.
\begin{figure}[t!]
	\centering 
	\includegraphics[width=0.65\textwidth]{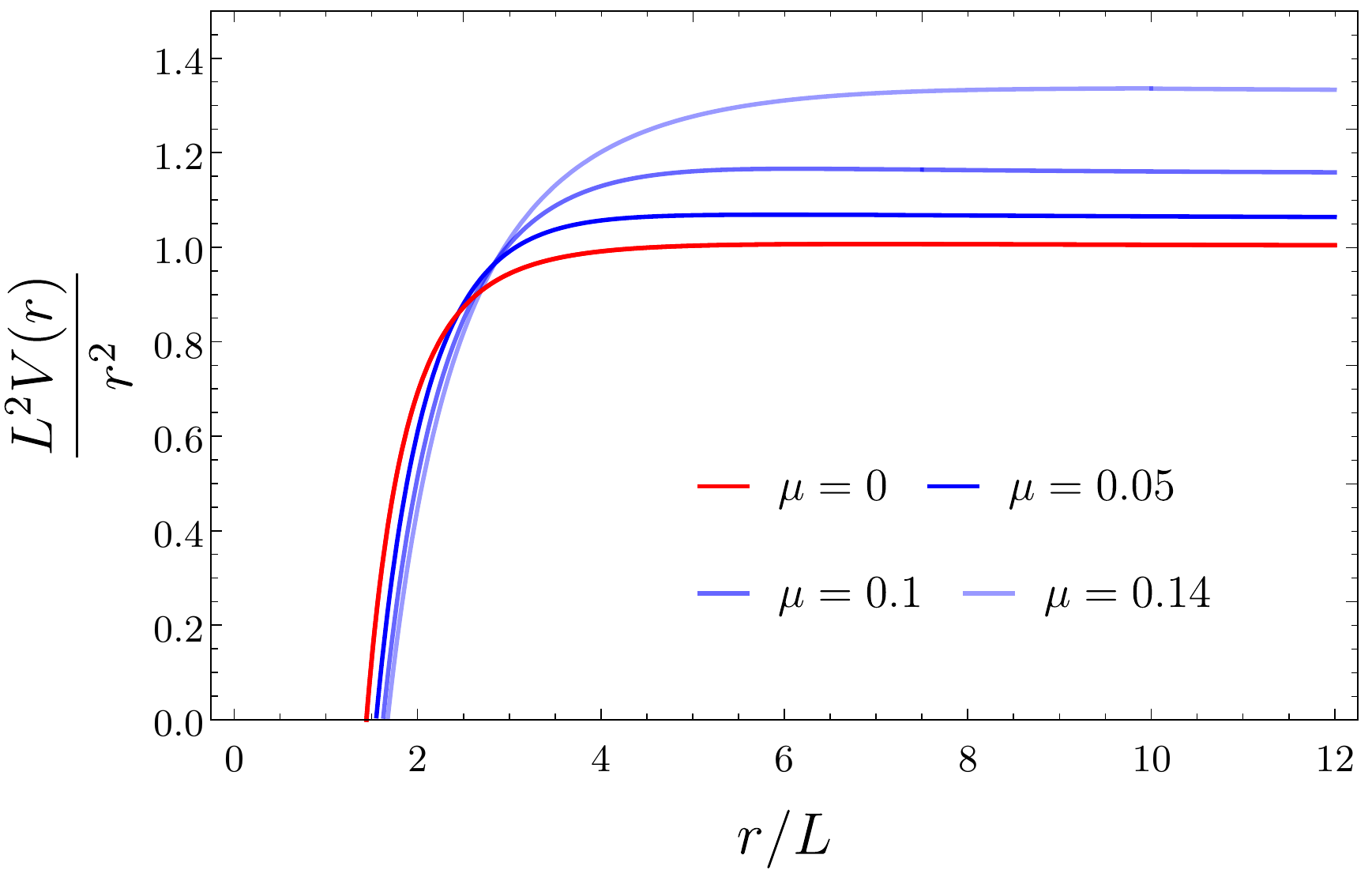}
	\caption{We plot the metric function $ V_{\mathbb{S}^2}(r)\cdot L^2/r^2$ corresponding to Taub-bolt solutions of ECG with $n/L=0.1$ for several values of $\mu$. We choose the largest bolt radius $r_b$ when there are several solutions for the fixed $n$. 
	 } 
	\label{boltS2ECG}
\end{figure}
We can also compute the on-shell action for the bolt solutions analogously to the NUT case.
%By using the equation \req{Eqr} in order to simplify the result, we get
The final result can be written as
\begin{equation}
I_E=\frac{\pi}{G}\left[n^2-r_b^2+4 n r_b+\frac{4n r_b}{L^2}(r_b^2-3n^2)+\mu L^4\frac{5 n^2+12 n r_b+r_b^2}{16n^2(r_b^2-n^2)}\right]\, .
\end{equation}
Using the chain rule and the relation \req{Eqr}, one can show again that $E\equiv \partial_{\beta} I_E= M$ as given in \req{gmm3}, which is a consistency check of the calculation.  In addition, the entropy, given by $S=\beta E-I_{\rm E}$, reads
\begin{equation}
S=\frac{\pi}{G r_b}\left[ -\frac{12 n^3}{L^2} \left(n^2+r_b^2\right)+4 n^3-n^2 r_b+r_b^3 +\frac{3 \mu  L^4 \left(4 n^3-n^2 r_b-8 n r_b^2-r_b^3\right)}{16 n^2(r_b^2-n^2) }\right]\,.
\end{equation}

 Just as in the case of the NUT solutions, we can also study the thermodynamics of the bolts in extended phase space. The extended first law has the same form as in the NUT case, but now the thermodynamic volume and coupling potential read
\begin{equation}
V = \frac{4 \pi \rh}{3}(\rh^2 - 3 n^2) \, , \quad \Upsilon^{\ssc \rm ECG} = \frac{\rh^2 +12 \rh n + 5 n^2}{128 G n^3 (\rh^2 - n^2)}\, ,
\end{equation}
and satisfy the Smarr formula that follows from scaling, which is of the same form as in the NUT case --- see \eqref{smarrfS2}. Note that, once again, the basic formula for the thermodynamic volume of the bolts is unaltered by the  higher-curvature terms~\cite{Johnson:2014xza}. However the thermodynamic volume is implicitly sensitive to the ECG coupling since the ECG term is important in determining the value of $\rh$ for a given $n$.

\begin{figure}[t!]
	\centering 
	\includegraphics[width=0.47\textwidth]{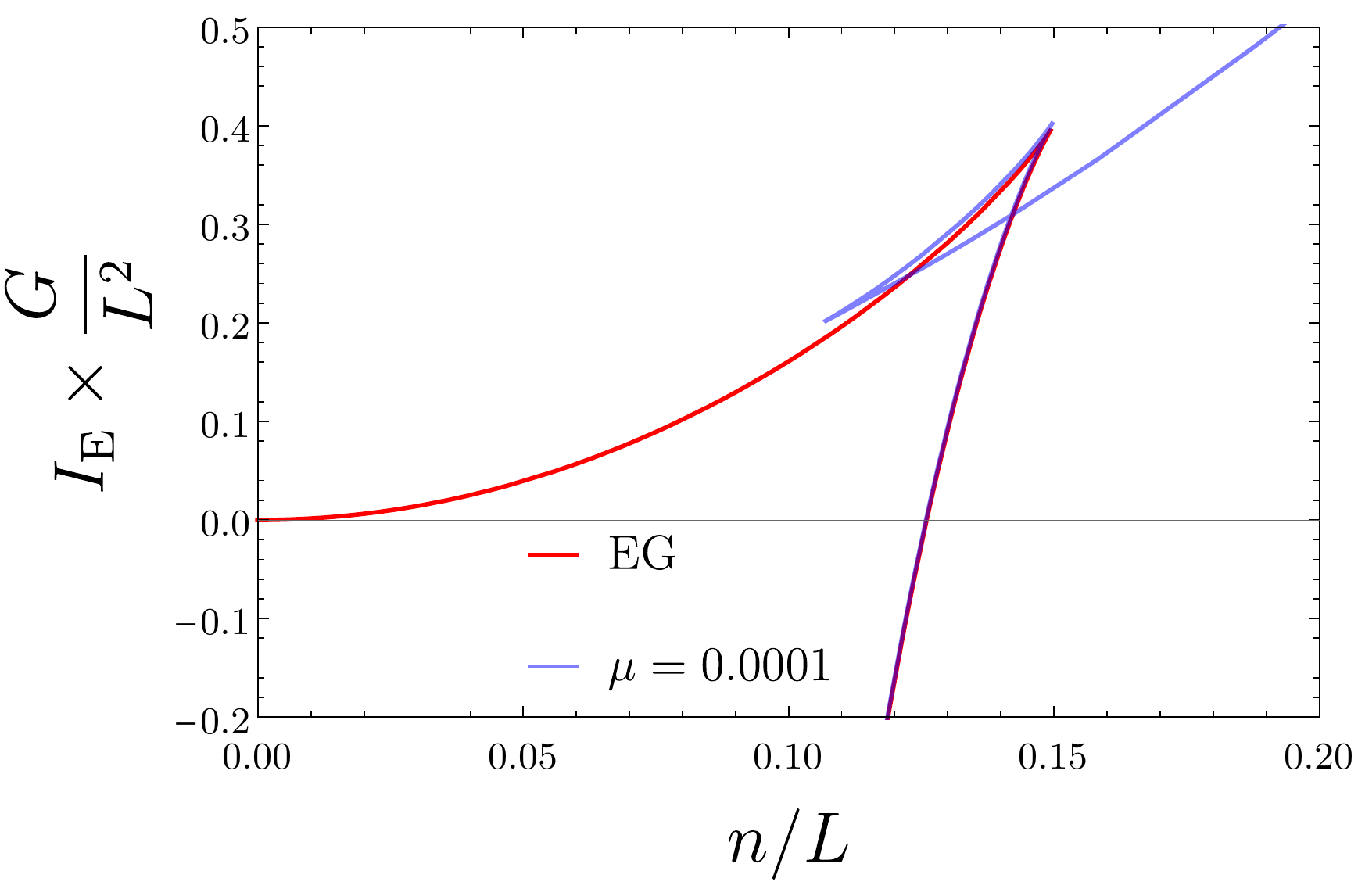}
	\includegraphics[width=0.47\textwidth]{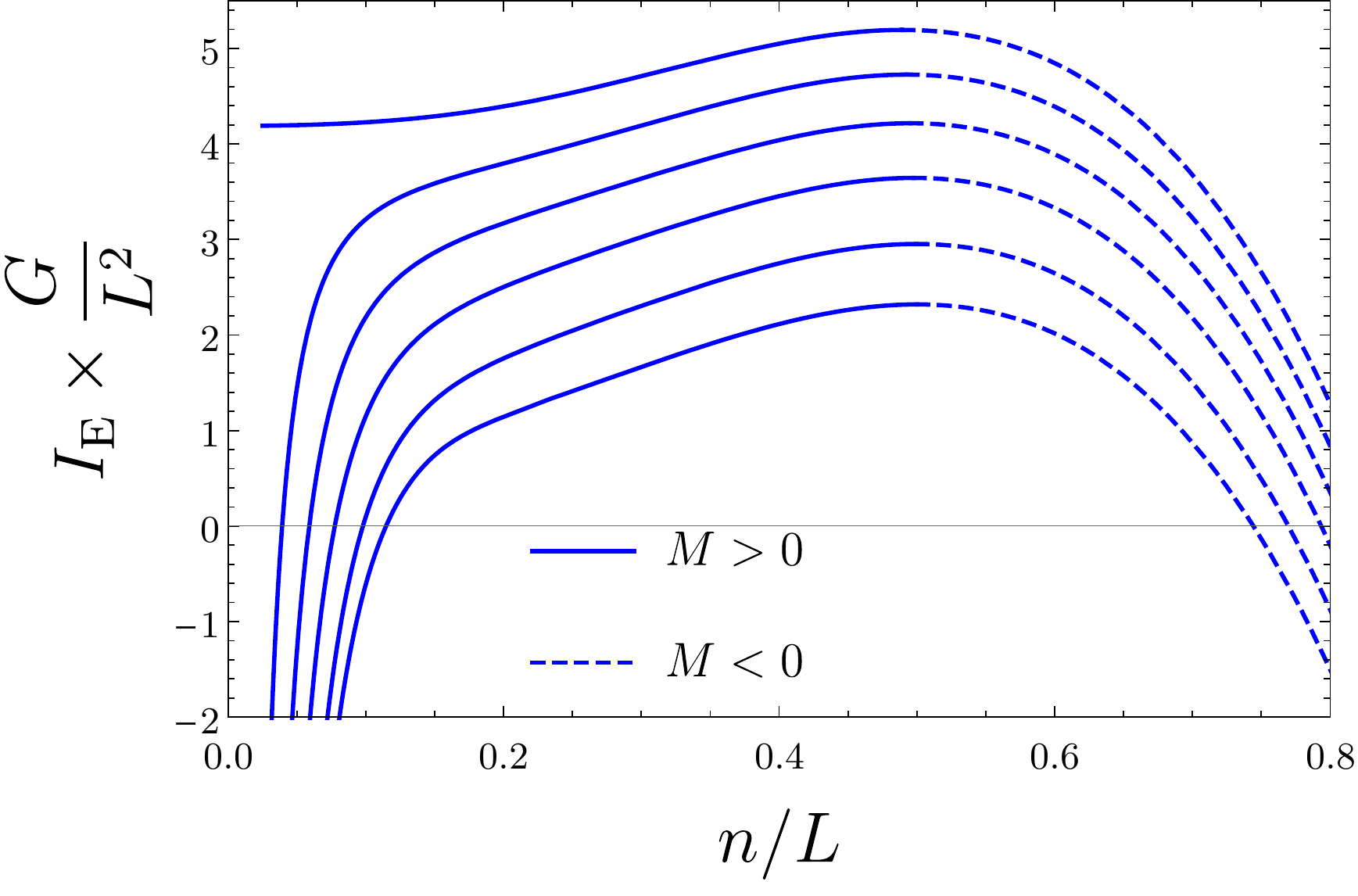}
	\caption{Euclidean on-shell action for $\mathcal{B}=\mathbb{S}^2$ bolt solutions in ECG. In the left panel we compare the Einstein gravity result with the ECG one with  $\mu=0.0001$, which contains three branches. For $\mu>0.001126$ there is only one branch, which is shown in the right panel for $\mu=0.01,\, 8/270,\, 16/270,\, 24/270,\, 32/270,\, 4/27$. The dashed lines correspond to $M<0$, so they should be excluded. As we can see, when $n\rightarrow0$ the free energy diverges to $-\infty$, except for the critical case $\mu=4/27$, which corresponds to the upper line.} 
	\label{IEboltS2ECG}
\end{figure}

Although we cannot solve \req{Eqr} exactly, we can study its behaviour in several limits. For example, let us consider the new branch of solutions for which $r_b$ is close to $n$ in the limit $\mu\ll 1$. We can expand $r_b$ in powers of $\mu^{1/2}$. To second order, we get
\begin{equation}
r_b=n+\frac{L^2}{n}\sqrt{\frac{3\mu}{8}}+\frac{3 \mu L^2}{16 n^3}\left(L^2-12 n^2\right)+\mathcal{O}(\mu^{3/2})\, .
\end{equation}
For Einstein gravity, we get $r_b=n$, and the solution reduces to the NUT one. However, for any given nonvanishing $\mu$, we have two inequivalent solutions:  the NUT constructed in the previous subsection, and this one. In particular, as opposed to the Einstein gravity case, a bolt solution does exist for $n^2=L^2/(4 f_{\infty})$, which corresponds to a nonsquashed spherical boundary geometry. 
% This is the only bolt solution for large enough $n$, including in particular the value $n^2=L^2/(4 f_{\infty})$. 
Expansions for the free energy and the mass of this branch of solutions can be easily obtained in the $\mu\ll 1$ limit, the results being
\begin{align}
I_E&=\frac{\pi}{G}\left[4 n^2-\frac{8 n^4}{L^2}+\frac{3L^2}{\sqrt{2}}\mu^{1/2}+\left(\frac{27 L^2}{8}-\frac{L^4}{8 n^2}\right)\mu\right]+\mathcal{O}(\mu^{3/2})\, ,\\
GM&=n-\frac{4 n^3}{L^2}+\frac{\mu L^4}{32 n^3}+\mathcal{O}(\mu^{3/2})\, .
\end{align}
Note that the mass is nonvanishing when the boundary geometry is that of a round $\mathbb{S}^3$, namely, $GM(n^2=L^2/(4f_{\infty}))=3\mu L/4+\mathcal{O}(\mu^{3/2})$, so the free energy is not extremized in that case. Instead, the maximum is reached for $n^2/L^2=1/4[1+\mu/2+\mathcal{O}(\mu^{3/2})]$, which is also the $M=0$ value. Greater values of $n$  would give rise to negative mass solutions, as illustrated in Fig. \ref{IEboltS2ECG}. Note also that $I_E(n^2=L^2/(4f_{\infty}))$  is greater than the one for the NUT solution, so that in the region near $n^2=L^2/(4f_{\infty})$, the NUT would dominate the corresponding holographic partition function.

We can also study the behaviour near $n= 0$, for which there is also a single bolt solution for each nonvanishing $\mu$. We get approximately
\begin{equation}
r_b=\frac{L^2}{6n}-2\left(1-\frac{27}{4}\mu\right)n+\mathcal{O}(n^2)\, ,
\end{equation}
and for the free energy,
\begin{equation}
I_E=\frac{\pi}{G}\left[-\frac{L^4}{108 n^2}\left(1-\frac{27}{4}\mu\right)+\frac{2L^2}{3}\left(1+\frac{27}{4}\mu\right)\right]+\mathcal{O}(n^2)\, .
\end{equation}
If we set $\mu=0$ in these expressions, we recover the small $n$ expansions for Einstein gravity bolts corresponding to the $(+)$ root in \req{Einss}. Observe that in the critical limit, $\mu=4/27$, the leading term disappears, and the on-shell action is finite for $n=0$.

We try to summarize the different possibilities in Fig. \ref{IEboltS2ECG}, where we plot $I_{\rm E}$ for ECG bolt solutions for several values of $\mu$. As we can see, the result is very different from that of Einstein gravity. There are two cases that we can distinguish: if $0<\mu<0.001126$, the diagram contains three branches, since there are three different bolt solutions; for $\mu>0.001126$ there is a single (elephant-shaped) branch. At $\mu\simeq 0.001126$, we expect to have a critical point which would represent a second-order phase transition if the bolt solution were dominant.  
 In all cases, the solutions exist for much larger values of $n$ than in Einstein gravity. However, there is an additional upper bound on $n$ coming from imposing $M>0$.

\subsubsection*{Free energy comparison}
\begin{figure}[t!]
	\centering 
	\includegraphics[width=0.47\textwidth]{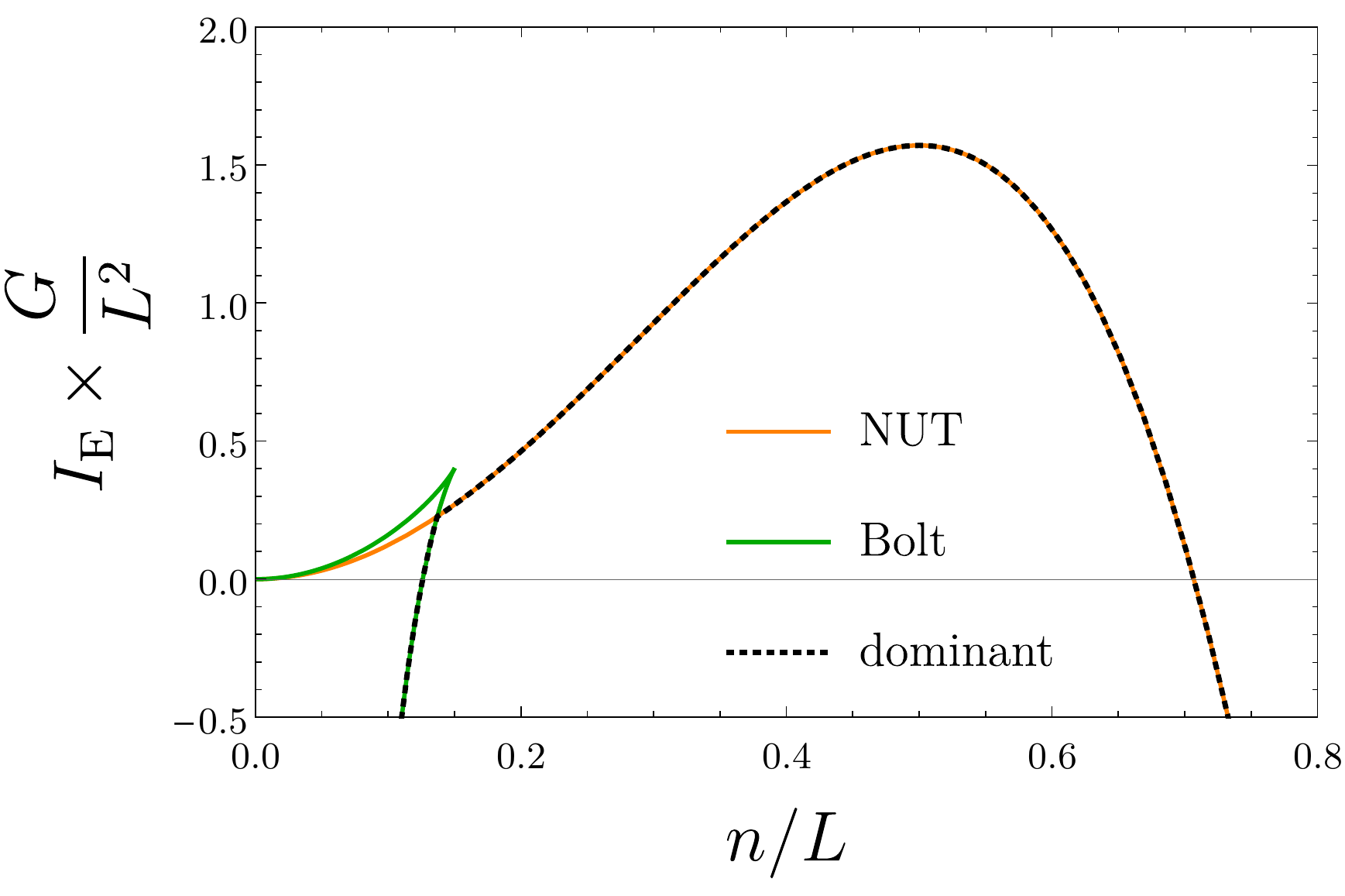}
	\includegraphics[width=0.46\textwidth]{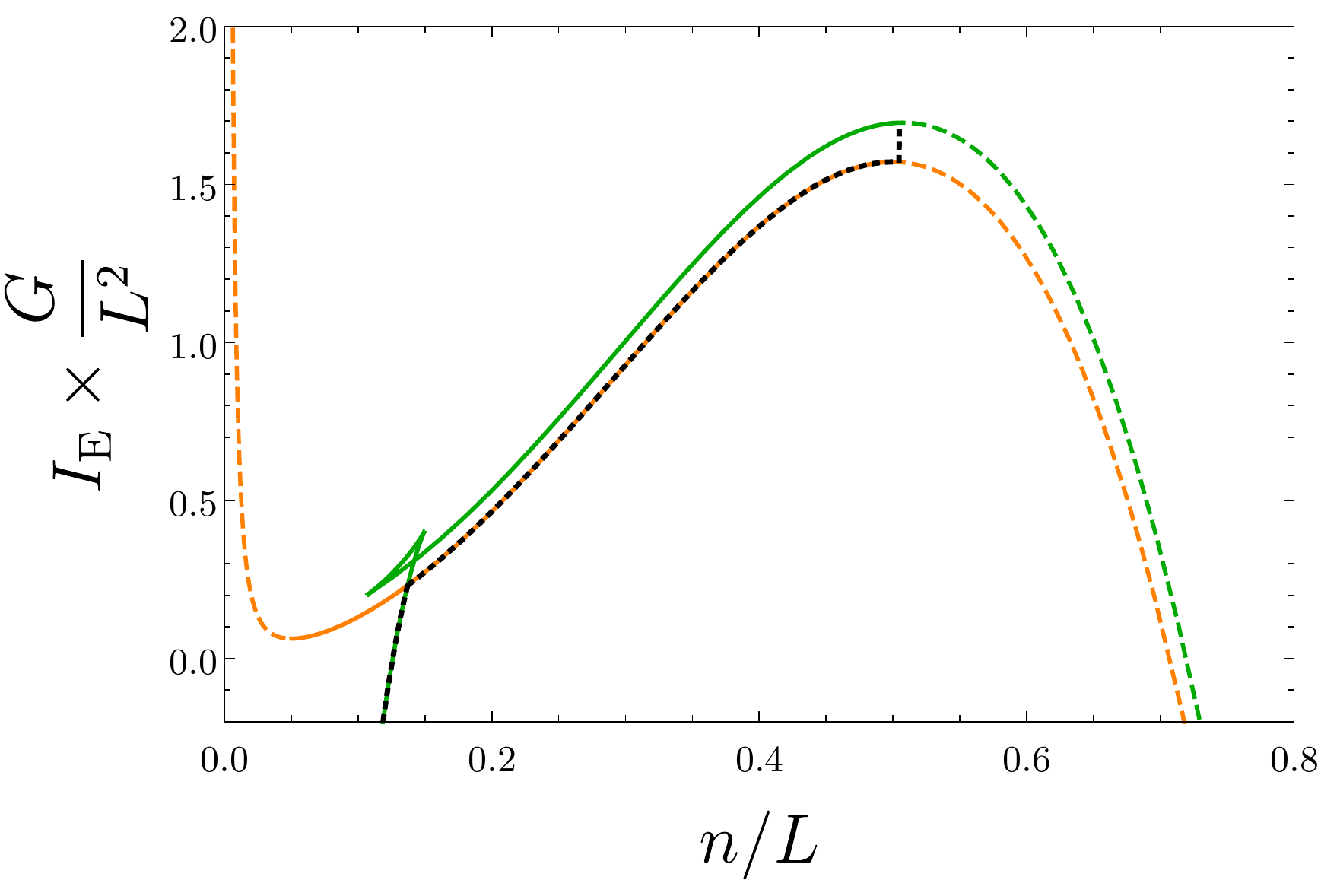}
	\includegraphics[width=0.47\textwidth]{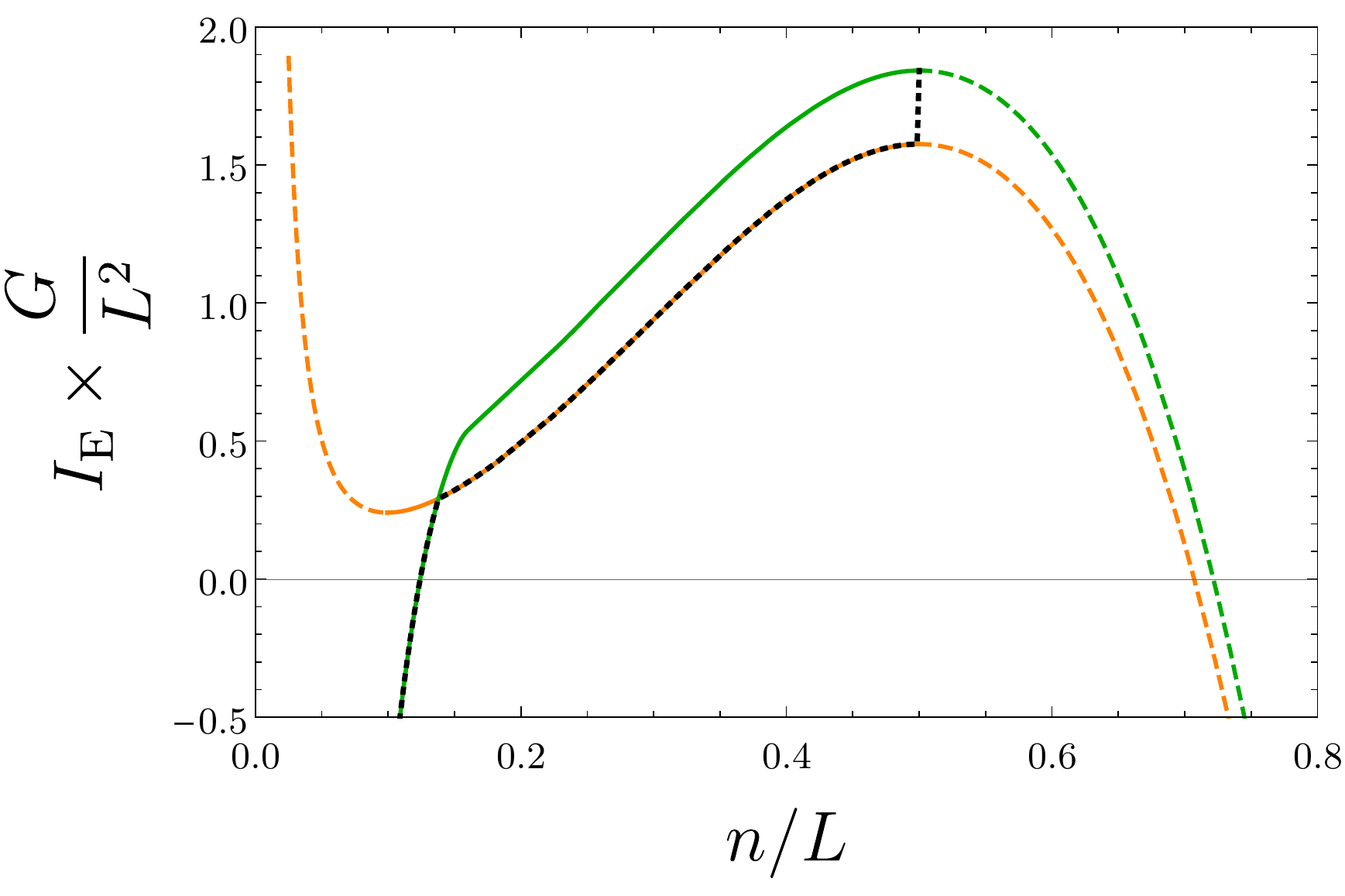}
	\includegraphics[width=0.46\textwidth]{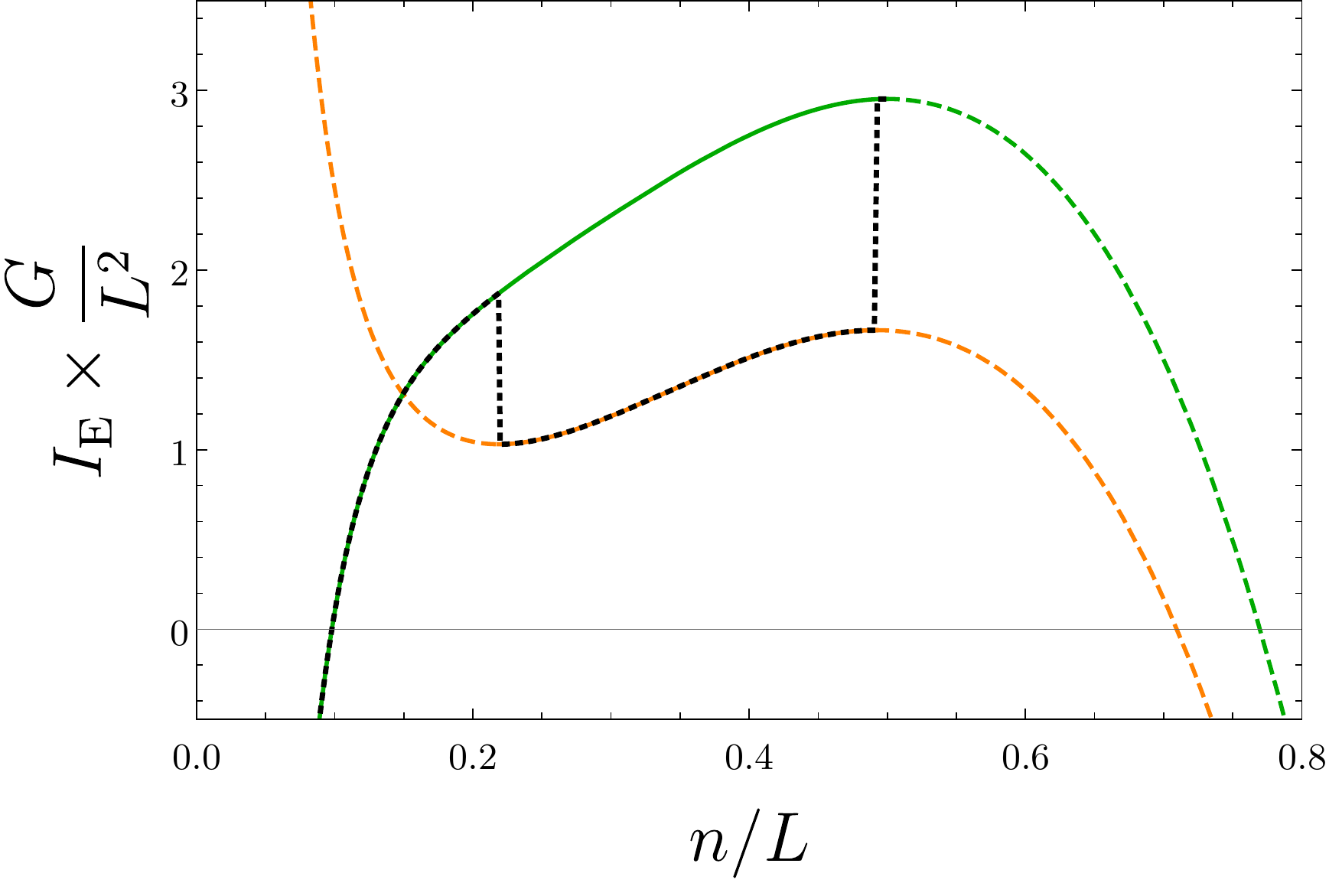}
	\caption{Comparison of Euclidean on-shell actions for NUT and bolt solutions for $\mathcal{B}=\mathbb{S}^2$ in ECG. Orange lines correspond to NUT solutions,  green ones correspond to bolt solutions, and the black dotted line represents the dominant contribution. Dashed lines represent configurations with $\mu M<0$, so that such solutions do not exist.
	Top, left: Einstein gravity result $(\mu=0)$. Top, right: $\mu=0.0001$. Bottom, left: $\mu=0.0015$. Bottom, right: $\mu=8/270\approx0.0296$. The vertical black dotted lines correspond to zeroth-order phase transitions in the points where the NUT solutions cease to exist.  } 
	\label{IEcomp4D}
\end{figure}

Finally, let us compare the Euclidean action of NUT and bolt solutions in order to determine which one dominates the partition function.  In Fig.~\ref{IEcomp4D} we compare the Euclidean actions for several values of $\mu$. The case for Einstein gravity is shown in the top-left panel, and we can see that the NUT solution dominates in all the region $n>L/6\sqrt{7-2\sqrt{10}}$, where a first order phase transition NUT/bolt takes place. In particular, there are no bolt solutions near the undeformed 3-sphere $n^2=L^2/4$. When we switch $\mu$ on, there are some drastic changes. Specially, we recall that for positive values of $\mu$ there are no solutions with negative mass. We plot with dashed lines the would-be Euclidean action of these solutions, but they do not actually exist. This has the effect of inducing zeroth-order phase transitions in the points where some solution ceases to exist. Another new feature is the existence of bolt solutions near the round 3-sphere $n^2=L^2/(4 f_{\infty})\equiv n_0^2$ for all values of $\mu>0$. In all the cases, we observe that for $n=n_0$ the NUT solution (corresponding to pure AdS) dominates, but for $n>n_0$ the NUT solution does not exist because it would have negative mass. However, for values of $n$ slightly larger than $n_0$, there is still a bolt solution of positive mass, and a zeroth-order phase transition from NUT to bolt must take place at $n_0$.  For larger values of $n$, the bolt solution also acquires a negative mass and there are no solutions. The behaviour is more interesting in the region $n<n_0$. In all the cases the NUT solution dominates until certain value $n=n_{\rm min}$, where there is a transition to a bolt solution. When $\mu<0.00569$, the transition is of first-order, as shown in top-right and bottom-left panels in Fig.~\ref{IEcomp4D}. When $\mu>0.00569$, the mass of the NUT solution vanishes before the value of the Euclidean action crosses that of the bolt solution, and a zeroth-order phase transition takes place, as shown in the bottom-right panel. After that phase transition the bolt solution exists and dominates for $0<n<n_{\rm min}$.

The appearance of zeroth-order transitions in Taub-NUT solutions is a new feature whose interpretation is not clear to us.  This seems to be a problem that only appears in four dimensions, since, as we will see, in six dimensions there is no restriction on the mass of the solutions.

\subsection{$\mathcal{B}=\mathbb{T}^2$}\label{SecT2}
Let us now consider a toroidal base space, so that the metric ansatz \req{FFnut} reads
\begin{equation}\label{k0metric}
ds^2=V_{\mathbb{T}^2}(r) \left(d\tau+ \frac{2n}{L^2} \eta d\zeta \right)^2+\frac{dr^2}{V_{\mathbb{T}^2}(r)}+\frac{(r^2-n^2)}{L^2}(d\eta^2+d\zeta^2)\, ,
\end{equation}
Here, the coordinates $(\eta,\zeta)$ parametrize a $\mathbb{T}^2$ with periods which we choose to be equal, $\beta_{\eta,\zeta}=l$.
%Here the coordinates $(\eta,\zeta)$ have a periodicity $l$, $(\eta,\zeta)\rightarrow (\eta+l,\zeta+l)$, and they parametrize a 2-torus. 
We note that, unlike the spherical case, the periodicity of the variable $\tau$, which we denote $\beta_{\tau}\equiv 1/T$, is not a priori fixed in terms of $n$  \cite{Dehghani:2005zm,Astefanesei:2004ji}. 
%Let us call this periodicity $\hat\beta$: $\tau\rightarrow\tau+\beta$, whose inverse we will call $\hat T=1/\hat\beta$. Let us then search for a regular, asymptotically AdS solution. 
The function $V_{\mathbb{T}^2}(r)$ satisfies \req{eqV} with $k=0$. From the general asymptotic expansion \req{rasymp}, we can obtain the metric of constant-$r$ hypersurfaces for $r\gg n$. This reads
\begin{equation}\label{bdryplanar}
\frac{^{(3)}ds^2_{\infty}}{r^2}= \left[\frac{f_{\infty}}{L^2}\left(d\tau+ \frac{2n}{L^2} \eta d\zeta \right)^2+\frac{(d\eta^2+d\zeta^2)}{L^2}\right]\, .
\end{equation}
If we define $z=\tau/(2n)$, $\eta/L=-x$, $\zeta/L=y$, this can be rewritten as
\begin{equation}\label{nil1}
\frac{^{(3)}ds^2_{\infty}}{r^2}= \left[\frac{4n^2 f_{\infty}}{L^2}\left(dz- x dy\right)^2+dx^2+dy^2\right]\, .
\end{equation}
Remarkably, when $n^2=L^2/(4f_{\infty})$ --- \ie for the same value $n$ for which in the $\mathcal{B}=\mathbb{S}^2$ case the corresponding boundary metric becomes that of a round $\mathbb{S}^3$ --- this reduces to the so-called Nil geometry\footnote{In fact, an additional change of variables can be used to rewrite \req{nil1} in the Nil form for any value of $n$, up to an overall factor. However, such coordinate change would involve making the periods of $\eta$ and $\zeta$ depend on $n$.} \cite{Thurston}. 
%\comment{has this been previously emphasized for Einstein gravity? Any further explanations about the interest of Nil metrics? }  \tcp{\bf [Not that I know of, except in the context of Thurston's classification of (2+1) geometries. SOL appears as a solution to BCEA theory (see gr-qc/9410021).]}
The appearance of this kind of geometry should not come as a surprise, as $\mathbb{T}^m$-bundles over tori $\mathbb{T}^n$ are always compact 2-step nilmanifolds (and vice versa) \cite{Palais} --- in our case above, $m=1$ and $n=2$. 
%http://vmm.math.uci.edu/PalaisPapers/TorusBndlesOverATorus.pdf

%This metric can be understood in different ways. First, if we make the change of coordinates $z=(2 f_{\infty} n/L^2)  \hat \tau$, $x=(-2n \sqrt{f_{\infty}}/L^2) \eta$, $y=(2n \sqrt{f_{\infty}}/L^2)\zeta$, it can be rewritten as
%\begin{equation}
%\frac{^{(3)}ds^2_{\infty}}{r^2}= \frac{L^2}{4n^2 f_{\infty}} \left[ (dz- x dy)^2+dx^2+dy^2  \right]\, .
%\end{equation}
%Up to an overall factor, this corresponds
%Remarkably, when $n^2=L^2/(4f_{\infty})$ --- \ie for the same value $n$ for which in the $\mathcal{B}=\mathbb{S}^2$ case the corresponding boundary metric becomes that of a round $\mathbb{S}^3$ --- this reduces 
%to the so-called Nil geometry \cite{}. 
%\comment{anything else}The appearance of this kind of metric should not come as a surprise, as $\mathbb{T}^m$-bundles over $\mathbb{T}^n$ are always compact 2-step nilmanifolds (and viceversa) \cite{} --- in our case above, $m=1$ and $n=2$. Observe also that the overall factor becomes one for the same value of $n$ for which, in the $\mathcal{B}=\mathbb{S}^2$ case, the corresponding boundary metric became that of a round $\mathbb{S}^3$, \ie for  $n^2=L^2/(4f_{\infty})$.
%http://vmm.math.uci.edu/PalaisPapers/TorusBndlesOverATorus.pdf

On the other hand, it is also natural to define $\hat \tau=\sqrt{f_{\infty}}\tau$, whose periodicity is $\beta_{\hat \tau}=\sqrt{f_{\infty}}\beta_{ \tau}\equiv 1/\hat T$. Then, \req{bdryplanar} reduces to the standard metric on $\mathbb{T}^3$ for $n=0$, 
\begin{equation}\label{hatT}
\frac{L^2}{r^2}\, \left. ^{(3)}ds^2_{\infty}\right|_{n=0}=d\hat \tau^2+d\eta^2+d\zeta^2\,,
\end{equation}
so we can also understand \req{bdryplanar} as a sort of twisted three-torus metric.% For later convenience, we reserve 

% Below, we will use  It turns out to be convenient to define 

% It is this periodicity that we will consider as the inverse of the temperature, $\beta=1/T$. 
\subsubsection{Taub-NUT solutions}
Let us start with the NUT solutions. Just like in the previous section, we assume that $V_{\mathbb{T}^2}(r=n)=0$, and we impose $V_{\mathbb{T}^2}'(r=n)=4\pi T$ in order to avoid a conical singularity at the NUT. Then, we can consider a Taylor expansion around $r=n$ of the form
\begin{equation}
V(r)=4\pi T (r-n)+\sum_{i=2}^{\infty}(r-n)^i a_ i \, .
\end{equation}
Plugging it into \req{eqV} and solving order by order in $(r-n)$, we obtain the following relations for the first terms
\begin{eqnarray}
GM&=&-\frac{4 n^3}{L^2}+\mu L^4 (4\pi  T)^3\, ,\\
0&=&\mu L^4 (4\pi T)^2\left(a_2-\frac{2\pi T}{n}\right)\, ,\\
0&=&8\pi T\left[-2+3\mu  L^4 \left(a_2^2 -\frac{3 \pi  a_2 T}{n}+\frac{2 \pi   T \left(5 \pi   T-2 a_3 n^2\right)}{n^2}\right)\right]\, .
\end{eqnarray}
The first equation fixes the ``mass'' $M$ in terms of $n$, $L$, $\mu$ and $ T$, while the rest give us relations between the coefficients of the expansion and the temperature. We can try to solve these relations in two inequivalent ways. 

The first possibility, which is the only one available for Einstein gravity, 
consists in setting $ T=0$ --- see equations \req{NUTE}, \req{massE} and \req{betaEin} with $k=0$. This solves the last two equations and, in fact, completely determines the series expansion for any value of $\mu$, \ie we can obtain $a_2$, $a_3$, etc., from the subsequent equations. The series is convergent in a vicinity of $r=n$. However, note that in that case, $GM=-4 n^3/L^2 <0$. Hence, according to the general discussion about the asymptotic behaviour, we expect this solution to be pathological at infinity unless some miraculous fine-tuning occurs. Unfortunately, this is not the case, and when we solve \req{eqV} starting from the near-horizon expansion with $ T=0$, the oscillatory character appears asymptotically. We are then led to conclude that regular extremal NUT solutions do not exist for any allowed value of $\mu$.
%For Einstein gravity, $\mu=0$, the only possibility is  $\hat T=0$. This choice solves the last two equations and, in fact, completely determines the series expansion
%We see that there are two possibilities. 
%HEEEREEE
%The first one,  and indeed the only one possible for $\mu=0$, is $\hat T=0$, so that the solution would be extremal. This choice solves the two last equations shown, and actually it completely determines the series expansion, this is, we can obtain $a_2$, $a_3$, etc. from the subsequent equations. This series expansion is convergent in a vicinity of $r=n$, so locally it is a solution. However, note that in this case $GM=-\frac{4 n^3}{L^2}<0$, since we are implicitly assuming $n>0$. Hence, according to the general discussion about the asymptotic behaviour, we expect that this solution will be sick at infinity unless some miraculous fine-tuning happens. Unfortunately this is not the case, and when we solve numerically \req{eqV} starting from this near-horizon expansion we observe that the oscillatory character appears asymptotically. Therefore, we are led to the conclusion that the extremal NUT solution does not exist for any positive value of $\mu$. Of course, it could exist for negative $\mu$ but we are not interested in that case here.\\

The second possibility is setting $a_2=2\pi T/n$, which solves the second equation. The following equations can be used to determine the remaining coefficients, which turn out to have a nonperturbative dependence on $\mu$, \eg 
\begin{equation}
a_3=\frac{2 \pi  T}{n^2}-\frac{1}{6 \pi \mu L^4  T}\, .
\end{equation}
Observe that these do not possess a finite limit when $\mu\rightarrow 0$. As a matter of fact, we have failed to construct these solutions for any nonvanishing value of $\mu$ different from the critical limit value $\mu=4/27$ --- see Section \ref{critic} --- so we strongly suspect that no regular NUT solution exists for $\mathcal{B}=\mathbb{T}^2$ for any $0<\mu<4/27$. 

%\comment{can we make a more conclusive statement here?} Hence, it seems that no NUT solution exist for $\mathbb{T}^2$ base space for any $0<\mu<4/27$. 

\subsubsection{Taub-bolt solutions}
\begin{figure}[t!]
	\centering 
	\includegraphics[width=0.65\textwidth]{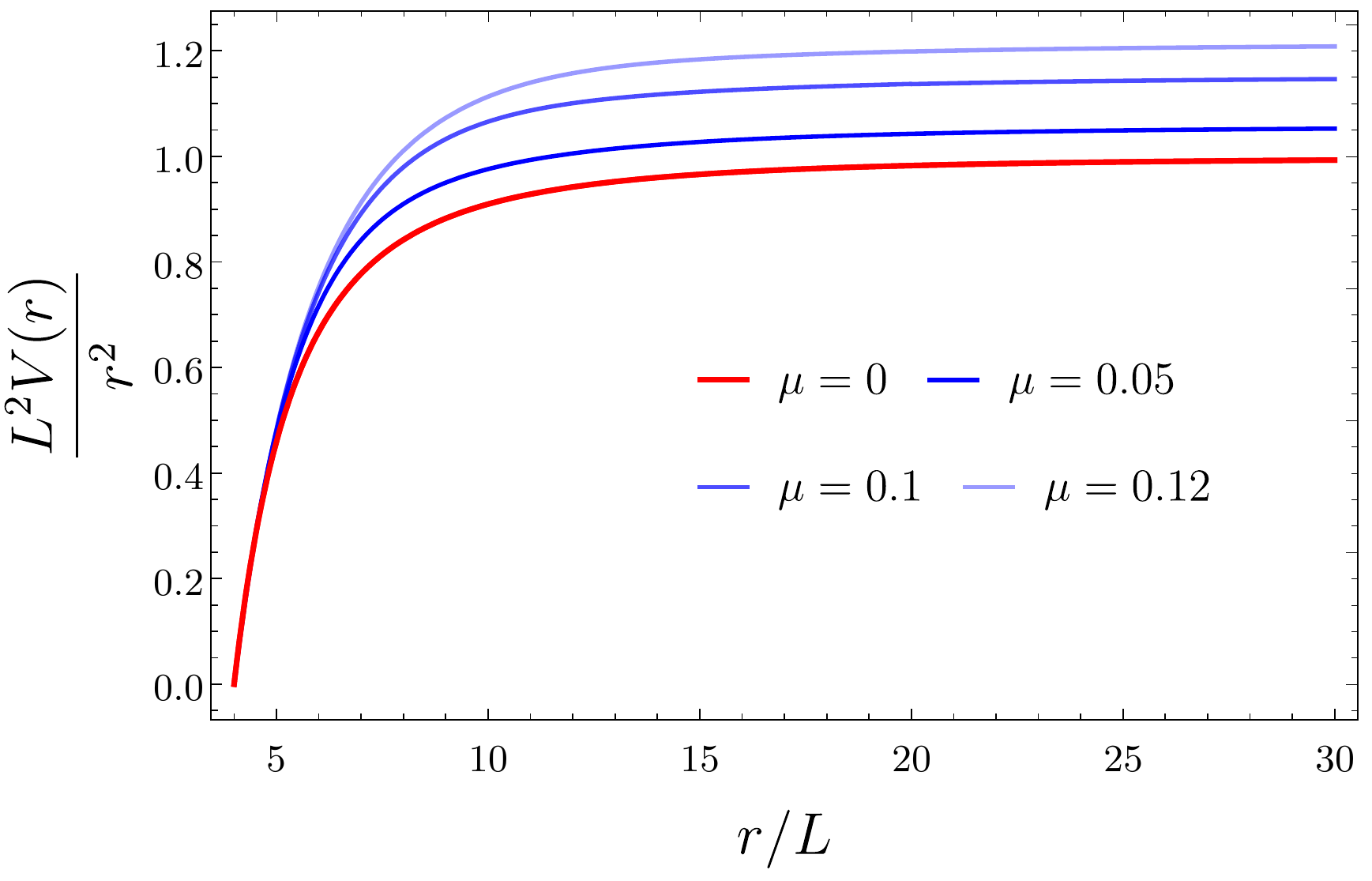}
	\caption{We plot the metric function $ V_{\mathbb{T}^2}(r)\cdot L^2/r^2$ corresponding to Taub-bolt solutions of ECG with $n/L=1$ and $r_b/L=4$ for several values of $\mu$.  	} 
	\label{Planarbolt}
\end{figure}

Fortunately, the situation is different for bolt solutions. In that case, we impose the existence of some $r_b>n$ such that near $r=r_b$,
\begin{equation}\label{PBexp}
V_{\mathbb{T}^2}(r)=4\pi  T (r-r_b)+\sum_{i=2}^{\infty}(r-r_b)^i   a_ i\, .
\end{equation}
This fixes $V_{\mathbb{T}^2}(r_b)=0$ and $V_{\mathbb{T}^2}'(r_b)=4\pi  T$. Again, we plug this expansion in \req{eqV}, and from the first two terms we get 
\begin{eqnarray}\label{MTplanar}
GM&=&\frac{ \left(-3 n^4-6 n^2 r_b^2+r_b^4\right)}{2L^2 r_b}-\frac{1}{8}\mu  L^4 (4\pi T)^3\, ,\\ \label{MTplanar2}
0&=&\frac{6 \left(r_b^2-n^2\right)^2}{L^2 r_b^2}-\frac{8 \pi   T \left(r_b^2-n^2\right)}{r_b}-\frac{3\mu L^4 n^2 (4\pi T)^3}{r_b(r_b^2-n^2)}\, .
\end{eqnarray}
As usual, the first equation fixes $M$, while the second relates $r_b$ to $n$ and $ T$. It turns out that for $\mu\ge 0$, there is a unique solution for $ T$ for every $n$ and $r_b>n$. The solution can be written explicitly as
\begin{equation}\label{Tplanar}
 T=\frac{\left(r_b^2-n^2\right) \left(2 r_b^{2/3}-\left(\sqrt{729 \mu  n^2+8 r_b^2}-27n \sqrt{\mu } \right)^{2/3}\right)}{12 \pi  L^2 n r_b^{1/3} \sqrt{\mu} \left(\sqrt{729 \mu  n^2+8 r_b^2}-27 n\sqrt{\mu}\right)^{1/3}}
\end{equation}
For a given $n$, this is a one-to-one relation between every $r_b>n$ and $ T>0$. In particular, we have $\lim_{r_b \rightarrow n} T=0$. However, in order to keep $GM$ positive, $r_b$ is bounded from below, $r_b\ge n \gamma(\mu)$, for some constant $\gamma(\mu)$. In particular, for the limiting cases $\mu=0$ and $\mu=4/27$, we have, respectively, $\gamma(0)=\sqrt{3+2\sqrt{3}}\simeq 2.5425$, and $\gamma(4/27)\simeq 4.0171$. In each case, for a given $n$, the radius $r_b$ and the ``mass'' $M$ are fixed by the periodicity of the coordinate $ \tau$. The remaining coefficients in the expansion \req{PBexp} are fully determined once we choose $a_2$, which is the only free parameter.  Once again, this  is fixed by demanding the solution to have the correct asymptotic behaviour. In all cases we find that there is one and only one value of $a_2$ for which this happens, and so the solutions are completely determined by $n$ and $ T$.  In Fig. \ref{Planarbolt} we show some of the metric functions corresponding to these solutions computed numerically.

Let us now study the thermodynamic properties of the solutions. The Euclidean action can be once again evaluated  using the generalized action \req{assd}, following the same steps as in the previous subsections.  The result is
%along the same lines as in \ref{S2cubic4d} and we find also that the computation is reduced to the evaluation of certain quantity at $r=\rh$. r_b
%The result for the free energy $F= T I_E$ reads
\begin{equation}\label{FEP4d}
I_E=
%\frac{l^2}{8 \pi G \sqrt{f_{\infty}} T} \left[\frac{r_b(r_b^2-3n^2)}{L^4}-\frac{(r_b^2-n^2)2 \pi T}{L^2}+\mu L^2\frac{(5n^2+r_b^2)(2\pi T)^3}{(r_b^2-n^2)}\right]\, ,
-\frac{l^2}{8 \pi GT }\left[\frac{r_b^4+3n^4}{2r_b L^4}-\mu L^2\frac{(11n^2+r_b^2)(2\pi T)^3}{(r_b^2-n^2)}\right]\, ,
\end{equation}
%\begin{equation}\label{FEP4d}
%\begin{aligned}
%F&=\frac{l^2}{8 \pi G \sqrt{f_{\infty}}}\left[\frac{r_b(r_b^2-3n^2)}{L^4}-\frac{(r_b^2-n^2)2 \pi \hat T}{L^2}+\mu L^2\frac{(5n^2+r_b^2)(2\pi\hat T)^3}{(r_b^2-n^2)}\right]\\
%&=\frac{l^2}{8 \pi G \sqrt{f_{\infty}}}\left[-\frac{r_b^4+3n^4}{2r_b L^4}+\mu L^2\frac{(11n^2+r_b^2)(2\pi\hat T)^3}{(r_b^2-n^2)}\right]\, ,
%\end{aligned}
%\end{equation}
where  we used \req{MTplanar2} to simplify the result. This on-shell action should be understood as a function of $T$ and $n$, which appear implicitly through $r_b$. In the case of Einstein gravity, for which the metric function can be obtained analytically --- see \req{boltEi} with $k=0$ --- the result for the on-shell action can be written explicitly as a function of $T$ and $n$. Using \req{boltrb} with $k=0$, one finds
%Since the general expressions are very involved, let us first study the case of Einstein gravity, which likely has not been much explored in the literature. When $\mu=0$ we can actually solve the field equations just by extracting $V(r)$ from \req{eqV}:
%\begin{equation}
%V_{\rm E}(r)=\frac{(r-r_b) \left(3 n^4+r r_b \left(r^2+r r_b+r_b^2-6 n^2\right)\right)}{L^2 r_b \left(r^2-n^2\right)}\, .
%\end{equation}
%We also obtain the explicit relation $r_b(n,T)$ from \req{MTplanar} (note that $\hat T=T$ in this case):
%\begin{equation}
%\rh=\frac{1}{3} \left(2 \pi  L^2  T+\sqrt{4 \pi ^2 L^4 T^2+9 n^2}\right)\, .
%\end{equation}
%Then, the free energy reads
\begin{equation}
I_{ E}=-\frac{l^2}{108 \pi G T L^4}\left[8 \pi^3 L^6  T^3+\left(4\pi^2 L^4 T^2+9 n^2\right)^{3/2}\right]\, .
\end{equation}
Now we must account for the fact that we have an extended thermodynamic phase space, since $n$ is in this case a free variable. However, $n$ cannot be the appropriate thermodynamic variable as it has units of length instead of energy. Hence, let us define $\theta\equiv 1/n$, which has the right units. Then, associated with $T$ and $\theta$, we have two potentials: the usual entropy $S$, and a new potential $\Psi$. In terms of the free energy $F\equiv T I_E$, these are given by
\begin{equation}\label{SPsi}
S=-\left(\frac{\partial F}{\partial T}\right)_{\theta}\, ,\quad \Psi=-\left(\frac{\partial F}{\partial \theta}\right)_{T}\, ,
\end{equation}
which explicitly read
\begin{eqnarray}
S&=&\frac{\pi l^2 T}{9 G} \left(2 \pi  L^2 T+\sqrt{4 \pi ^2 L^4 T^2+\frac{9}{\theta ^2}}\right)\, ,\quad \Psi=-\frac{l^2\sqrt{4 \pi ^2 L^4 T^2+\frac{9}{\theta ^2}}}{4 \pi  G L^4 \theta ^3 }\, .
\end{eqnarray}
Finally, the energy is defined as $E=F+TS+\theta\Psi$, so that, by construction it satisfies the first law
\begin{equation}
dE=T dS+\theta d\Psi\, .
\end{equation}
The energy is given by
\begin{equation}
E_{}=\frac{l^2}{27\pi G}\left[4 \pi^3 L^2 T^3+\left(2\pi^2T^2-\frac{9}{\theta^2 L^4}\right)\sqrt{4 \pi ^2 L^4 T^2+\frac{9}{\theta ^2}}\right]\, .
\end{equation}
Now, this is a thermodynamic energy, but the energy of the solution should be computed using the ADM formula, which in this case tells us that $E_{\rm ADM}=M l^2/(4\pi L^2)$. Using the expression for $M$ given in \req{MTplanar}, we have checked that both energies actually coincide $E_{\rm ADM}=E$. Hence, the introduction of the variable  $\theta$ is crucial for the first law of black hole mechanics to hold in this case.%\comment{if this hadn't been observed before for Einstein gravity, perhaps emphasize more}

\begin{figure}[t!]
	\centering 
	\includegraphics[width=0.47\textwidth]{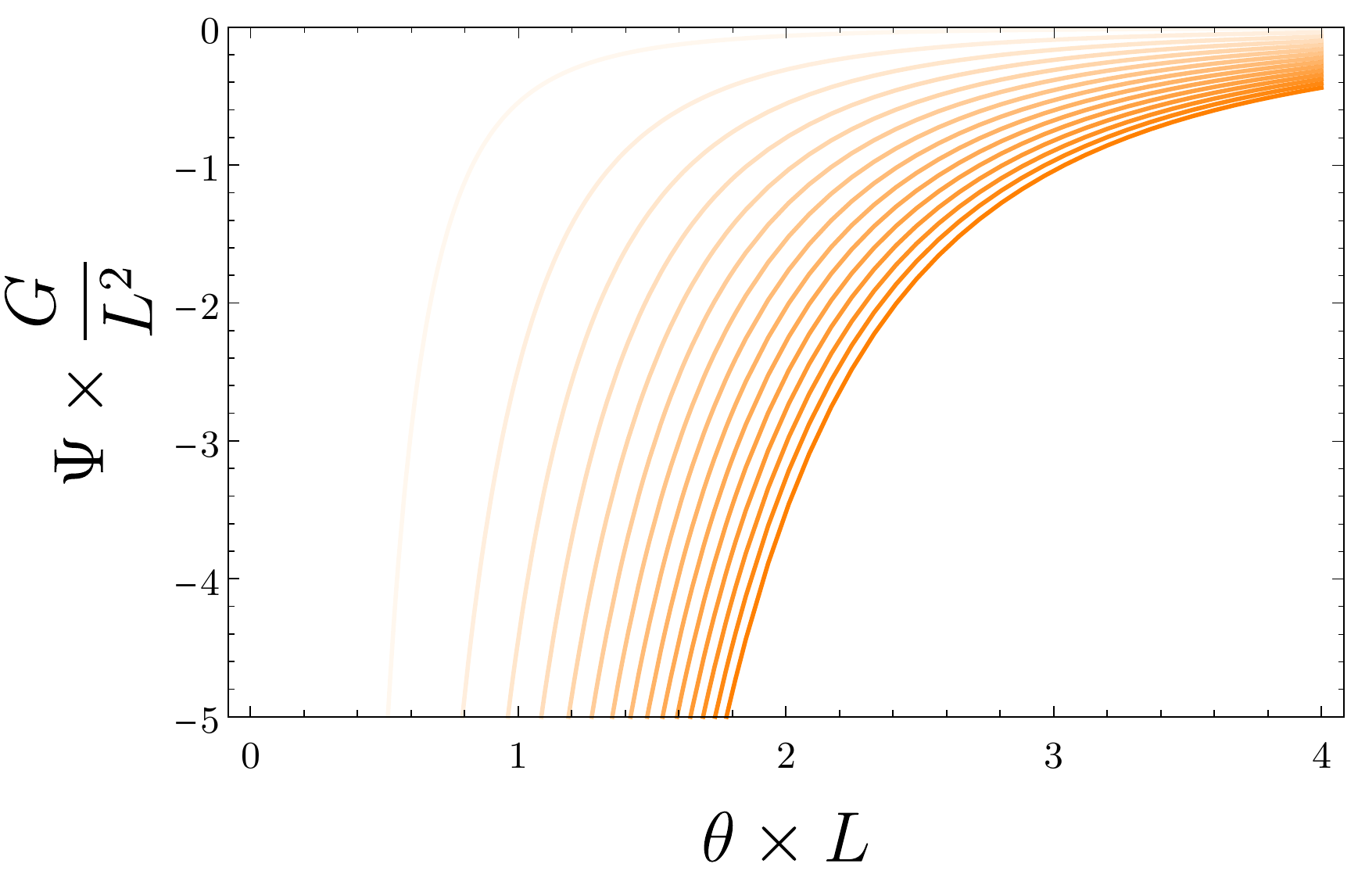}
	\includegraphics[width=0.47\textwidth]{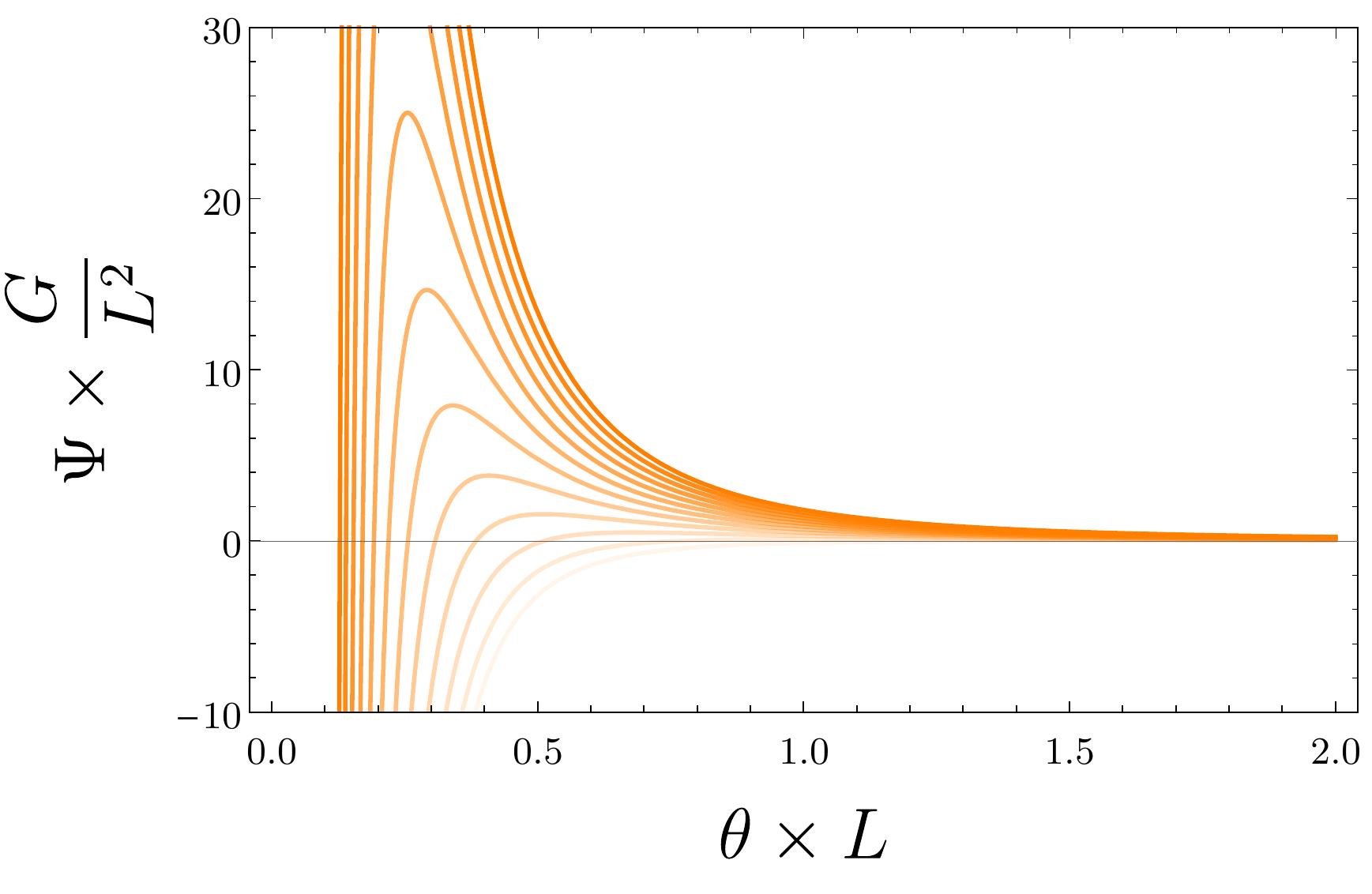}
	\caption{Isotherms in the $\Psi$-$\theta$ plane. Left: $\mu=0$. Right: $\mu=0.12$.} 
	\label{Psidiagram}
\end{figure}

This picture goes through nicely  when the ECG term is turned on. In that case, it is convenient to express the thermodynamic quantities in terms of the rescaled temperature $\hat T=T/\sqrt{f_{\infty}} $ introduced above equation \req{hatT}. In terms of this, we have the free energy $F(\hat T,\theta; \mu)=\sqrt{f_{\infty}}\hat T I_E$, which can be obtained from \req{FEP4d}, and the thermodynamic potentials $S(\hat T, \theta; \mu)$ and $\Psi(\hat T,\theta; \mu)$ defined as in \req{SPsi} (but with respect to $\hat T$ instead of $T$). We find that
\begin{equation}
E_{\rm ADM}=F+\hat TS+\theta\Psi\, ,
\end{equation}
where the ADM energy is now given by $E_{\rm ADM}=M l^2/(4\pi L^2\sqrt{f_{\infty}})$ and
\begin{eqnarray}
S&=&\frac{l^2}{4GL^2}\left[\frac{r_b^2\theta^2-1}{\theta^2}-12\mu L^4\frac{\pi^2f_{\infty} \hat T^2(5+r_b^2\theta^2)}{\rh^2\theta^2-1}\right]\, ,\\
\Psi&=&\frac{l^2}{8\pi G L^4\sqrt{f_{\infty}}\theta^3}\left[3r_b(\theta^2r_b^2-3)+4\pi L^2 \hat T \sqrt{f_{\infty}}(1-\theta^2r_b^2)\right]\,  .
\end{eqnarray}
It is interesting to study the isotherms on the diagram of $\Psi$-$\theta$. These are shown in Fig. \ref{Psidiagram} for $\mu=0$ and $\mu=0.12$. In the case of Einstein gravity --- the same happens for small values of $\mu$ --- the isotherms are monotonous. However, when $\mu$ is large enough, the diagram changes drastically. In that case, the isotherms develop a maximum, and the limit $\Psi \rightarrow 0$ corresponding to $\theta \rightarrow + \infty$ is approached from above instead of from below. However, the phase space seems to be free of critical points.

We have seen that to satisfy the quantum-statistical relation and first law, it is necessary to treat $\theta = 1/n$ as a thermodynamic variable. If we also wish to satisfy the Smarr formula, then once again we must consider both $\Lambda$ and $\mu$ to be thermodynamic parameters. The basic construction is identical to the $\mathcal{B} = \mathbb{S}^2$ case, but now we include $\theta$ as well. A simple computation then reveals the following for the thermodynamic volume and coupling potential
\begin{equation}
V = \frac{l^2 \rh}{3 L^2} \left(\rh^2 - \frac{3}{\theta^2} \right) \, , \quad \Upsilon^{\ssc \rm ECG} = \frac{\pi^2 l^2 T^3}{G L^2 } \left(\frac{\rh^2\theta^2 + 5}{\rh^2 \theta^2-1} \right)\, ,
\end{equation}
where in the second equation above we note that it is the un-normalized temperature that appears (i.e. $T$ rather than $\hat{T}$).  With these definitions, the extended first law and Smarr formula hold, with the latter being identical to Eq.~\eqref{smarrfS2} with the additional term $2\theta\Psi$ added.

Let us close this section by mentioning the possibility that the solutions considered in this subsection can be relevant holographically. In that context, and in analogy to the Taub-bolt solutions with $\mathcal{B}=\mathbb{S}^2$, we expect them to represent saddle points in the semiclassical partition function for boundary theories living on deformed tori with metric \req{bdryplanar}.
While in the spherical case the boundary is only characterized by $n$ --- or, equivalently, the squashing parameter $\alpha$ ---, the $\mathcal{B}=\mathbb{T}^2$ case is richer, given that  $n$ and $\hat T$ are independent parameters in that case.

\subsection{Exact Taub-NUT solutions in the critical limit}\label{critic}
In this subsection we study the Taub-NUT solutions of critical ECG, which can be constructed analytically. As we mentioned earlier, when $\mu=4/27$, the only AdS vacuum has a length scale $\tilde L^2=2 L^2/3$, and the linearized equations on that background vanish identically. %This is known as the critical limit of the theory.
 The field equations simplify considerably, which has allowed for the construction of analytic black hole solutions \cite{Feng:2017tev}.\footnote{The existence of special points in the parameter space in which the equations simplify has been previously reported in \cite{Dotti:2006cp} in the case of Lovelock gravity, where Lorentzian AdS wormholes similar to those in \cite{Feng:2017tev} were constructed.}  In the case of NUT-charged metrics, a similar simplification takes place, and we find the following family of exact Taub-NUT solutions,
\begin{equation}\label{tauc}
ds^2=(r^2-n^2)\left[\frac{3}{2L^2}\left(d\tau+n A_{\mathcal{B}}\right)^2+d\sigma_{\mathcal{B}}^2\right]+\frac{2 L^2dr^2 }{3(r^2-n^2)}\, .
\end{equation}
As we mentioned before, in the case of a spherical base space, $\mathcal{B}=\mathbb{S}^2$, this solution has a conical singularity at $r=n$, except for $n^2=L^2/6$, in whose case the solution is simply globally Euclidean AdS$_4$ --- also known as  $\mathbb{H}^4$. Hence, only the cases $\mathcal{B}=\mathbb{T}^2$, $\mathbb{H}^2$ are of interest in Euclidean signature. 

The solutions \req{tauc} can be analytically continued to Lorentzian signature in different ways, giving rise to very interesting metrics. For example if we make the replacement $n\rightarrow i n$ and $\tau=i t$ we get the following metric
\begin{equation}
ds^2=(r^2+n^2)\left[-\frac{3}{2L^2}\left(dt+n A_{\mathcal{B}}\right)^2+d\sigma_{\mathcal{B}}^2\right]+\frac{2 L^2dr^2 }{3(r^2+n^2)}\, .
\end{equation}
This metric is regular everywhere and, in fact, we can allow $r$ to take values in the whole real line. Hence, this solution usually represents a wormhole or wormbrane, depending on the topology, connecting two asymptotically AdS$_4$ regions. The cases $k=n=0$ and $k=-6n^2/L^2=-1$ are special as they correspond to pure AdS$_4$. Let us introduce a new radial coordinate $r=n \cosh\left(\rho/(\sqrt{2/3}L)\right)$, so that the metric reads
\begin{equation}\label{Lcrit}
ds^2=n^2\cosh^2\left(\frac{\rho}{\sqrt{2/3}L}\right)\left[-\frac{3}{2L^2}\left(dt+n A_{\mathcal{B}}\right)^2+d\sigma_{\mathcal{B}}^2\right]+d\rho^2\, ,
\end{equation}
which has an explicit wormhole character. In the spherical case $\mathcal{B}=\mathbb{S}^2$ the solution reads\footnote{Let us note that a very similar NUT-charged wormhole solution was reported in \cite{Ayon-Beato:2015eca} in the context of Einstein-Skyrme theory.}
\begin{equation}
ds^2=n^2\cosh^2\left(\frac{\rho}{\sqrt{2/3}L}\right)\left[-\frac{3}{2L^2}\left(dt+2n \cos\theta d\phi\right)^2+d\theta^2+\sin^2\theta d\phi^2\right]+d\rho^2\, .
\end{equation}
This solution has the problem that it suffers from closed time-like curves, because the time coordinate must be periodic $t\rightarrow t+8\pi n$. The $\mathcal{B}=\mathbb{T}^2$, $\mathbb{H}^2$ cases are free of them, because there is no periodicity condition on the time coordinate.
In particular, after some rescalings we can write the $\mathbb{T}^2$ solution as
\begin{equation}\label{whT2-1}
ds^2=\cosh^2\left(\frac{\rho}{\sqrt{2/3}L}\right)\left[-\left(dt+\frac{\sqrt{6}}{L}x dy \right)^2+dx^2+dy^2\right]+d\rho^2\, ,
\end{equation}
where the NUT charge has been absorbed in the period of the coordinates $x$, $y$. However, we can also allow $x$ and $y$ to be noncompact.
Interestingly, there is an inequivalent Lorentzian solution that can be obtained by rotating the coordinates as $(t,y)\rightarrow (i y, i t)$. This reads

\begin{equation}
ds^2=\cosh^2\left(\frac{\rho}{\sqrt{2/3}L}\right)\left[-dt^2+dx^2+\left(dy+\frac{\sqrt{6}}{L}x dt \right)^2\right]+d\rho^2\, .
\end{equation}

Going back to the general solution \req{Lcrit}, we can consider the following transformation: $t\rightarrow z$, $\rho \rightarrow i t$, $L\rightarrow i L$. Here we are changing the sign of $L^2$, which amounts to changing the sign of the cosmological constant in the ECG action \req{ECG}. Hence, the corresponding metric is a solution of the critical theory with a positive cosmological constant. The general solution reads
\begin{equation}
ds^2=-dt^2+n^2\cosh^2\left(\frac{t}{\sqrt{2/3}L}\right)\left[\frac{3}{2L^2}\left(dz+ n A_{\mathcal{B}}\right)^2+d\sigma_{\mathcal{B}}^2\right]\, .
\end{equation}
These represent bouncing cosmologies with different topologies for the spatial sections connecting two asymptotically (NUT charged) de Sitter spaces for $t\rightarrow\pm \infty$. The only exception is the case $k=1$, $n^2=L^2/6$, which is actually de Sitter space foliated by $\mathbb{S}^3$ spheres. Particularly relevant for cosmology is the flat case $k=0$, which after rescaling of the coordinates can be written as
\begin{equation}\label{bouncing}
ds^2=-dt^2+\cosh^2\left(\frac{t}{\sqrt{2/3}L}\right)\left[\left(dz+ \frac{\sqrt{6}}{L}x dy \right)^2+dx^2+dy^2\right]\, .
\end{equation}
The transverse geometry is again a Nil space. Interestingly, this solution represents a homogeneous but nonisotropic bouncing cosmology. Homogeneity follows from the fact that Nil space is a coset space and it possesses the isometries $(x,y,z)\rightarrow (x+a,y+b, z+c-a\sqrt{6}y/L)$, for arbitrary $(a, b, c)$. Let us also mention in passing that this solution seems to be disconnected from the isotropic and homogeneous bouncing solution found in \cite{Feng:2017tev}, since we do not recover it in any limit.

\section{Four dimensions: Quartic generalized quasi-topological term}\label{quarticAp}
As shown in \cite{PabloPablo4}, besides ECG, there are infinitely many theories involving terms of arbitrarily high order in curvature which allow for black hole solutions with $g_{tt}g_{rr}=-1$ in four dimensions --- this is also expected to be the case in higher dimensions. Hence, it is reasonable to expect that some of these theories will also possess NUT-charged solutions characterized by a single function $V_{\mathcal{B}}(r)$, \ie of the form \req{FFnut}. In this section we show this to be the case when we supplement the ECG action \req{ECGAction} with a particular quartic term belonging to the GQG class \cite{Ahmed:2017jod}. In particular, we study how the ECG Taub-NUT solutions with $\mathcal{B}=\mathbb{S}^2$ constructed in Sec.~\ref{S2cubic4d} are modified by the introduction of this term.
%Let us remark that this kind of theories are interesting not only because their solutions are way simpler to study --- the equations when two functions are needed are incredibly more involved  --- but also because they present \textit{integrability}. The field equations can be integrated once, and the thermodynamics of the system is algebraically determined even if we do not know how to solve the differential equation for $F$. In addition, as was shown in [on BHs] , any theory of this family only propagate a massless spin-2 graviton on maximally symmetric backgrounds --- in particular they do nor propagate ghosts at the linear level.
%Therefore, even though these theories are fine-tuned ---in the sense that they consist of very particular combinations of invariant ---  the interest in studying them is more than justified. On the one hand, they allow us to understand the kind of phenomena which happen when  higher-curvature corrections are taken into account. On the other hand, they are very rich toy models with a potential interest in holography. \\
Let us then consider the Euclidean action
\begin{equation}\label{QuarticT}
	I_{\rm E}=-\frac{1}{16\pi G}\int d^4x \sqrt{|g|}\left[\frac{6}{L^2}+R-\frac{\mu L^4}{8} \mathcal{P}-\frac{\xi L^6}{16}\mathcal{Q} \right]\, ,
\end{equation}
where 
\begin{equation}
	\begin{aligned}
		\mathcal{Q}=&-44R^{\mu\nu\rho\sigma }R_{\mu\nu }^{\ \ \alpha\beta }R_{\rho\ \alpha}^{\ \gamma \ \delta}R_{\sigma \gamma \beta \delta}-5 R^{\mu\nu\rho\sigma }R_{\mu\nu }^{\ \ \alpha\beta }R_{\rho\alpha}^{\ \ \gamma \delta}R_{\sigma \beta \gamma \delta}+5 R^{\mu\nu\rho\sigma }R_{\mu\nu\rho }^{\ \ \ \ \alpha}R_{\beta \gamma \delta \sigma}R^{\beta \gamma \delta}_{\ \ \ \ \alpha}\\&+24 R^{\mu\nu }R^{\rho\sigma \alpha\beta }R_{\rho\ \alpha \mu}^{\ \gamma}R_{\sigma \gamma \beta \nu}\, ,
	\end{aligned}
\end{equation}
is a particular GQG density.
%Here $\mathcal{Q}$ is a particular Generalized quasi-topological gravity term \comment{refs}.  The black hole solutions of this theory are the same as those studied in \comment{refs} as the field equations are equivalent for static, spherically symmetric metrics as those of theses references. 

\begin{figure}[t!]
	\centering 
	\includegraphics[width=0.65\textwidth]{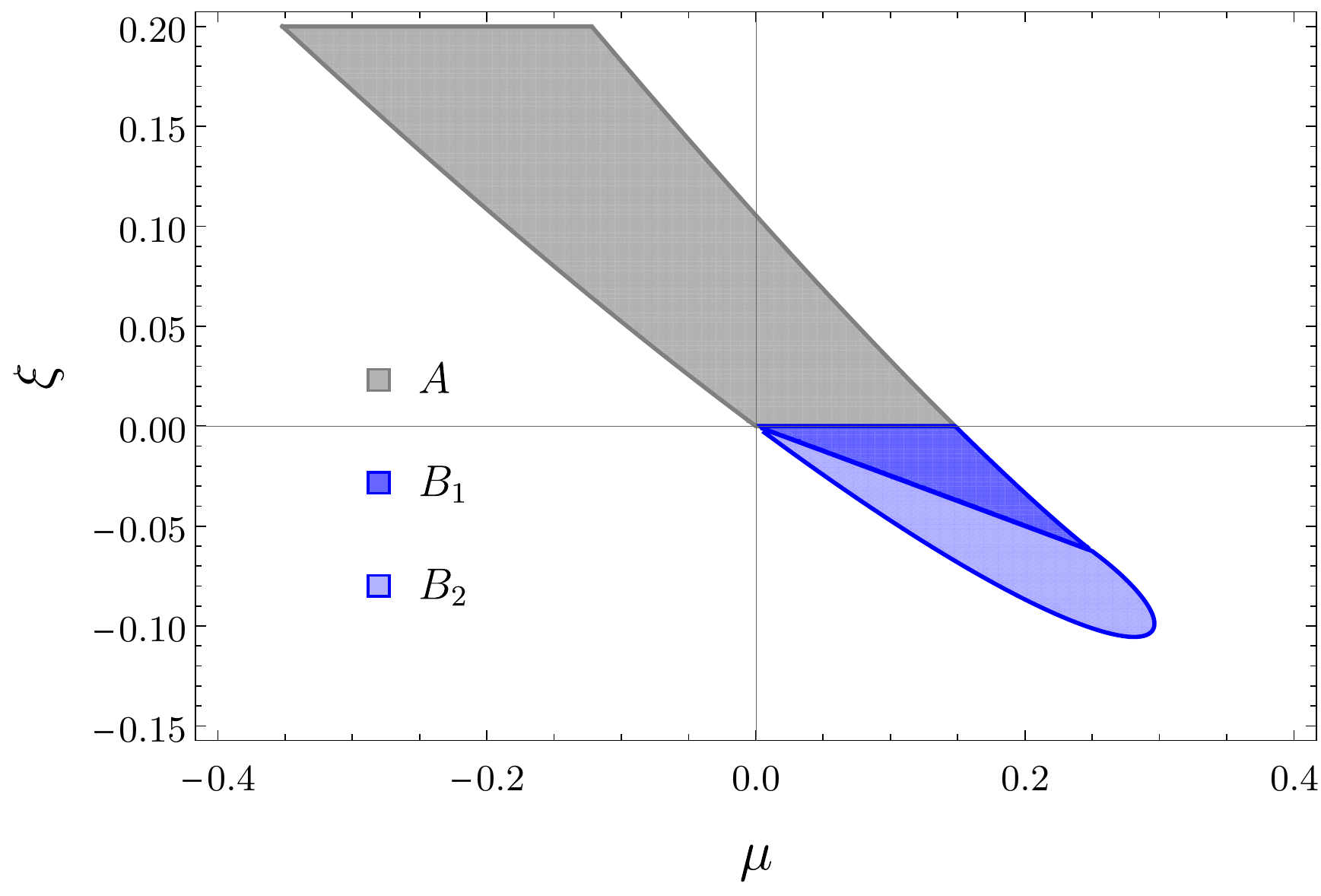}
	\caption{Region of the parameter space for which there is at least one physical AdS vacuum.}
	\label{VacuaQ4}
\end{figure}

Let us start by determining the AdS vacua of \req{QuarticT}. As usual, we write the relation between the action scale $L$ and the AdS radius $\tilde{L}$ as $\tilde L^2=L^2/f_{\infty}$. Then, the possible values of $f_{\infty}$ are determined by the positive roots of the polynomial
\begin{equation}\label{Qvacua}
	h(f_{\infty})\equiv1-f_{\infty}+\mu f_{\infty}^3+\xi f_{\infty}^4=0\, .
\end{equation}
For a given vacuum, the effective gravitational constant can be computed as $G_{\rm eff}=-G/h'(f_{\infty})$. Hence, in order to get a positive energy graviton, we must demand $h'(f_{\infty})<0$, the critical case corresponding to $h'(f_{\infty})=0$. Just like for ECG, there is an additional constraint coming from imposing the existence of positive-energy solutions. This reads $\mu+2f_{\infty}\xi\ge 0$ and, interestingly, it is equivalent to $h''(f_{\infty})\ge 0$ (assuming $f_{\infty}>0$). Therefore, we need to identify solutions to \req{Qvacua} satisfying $f_{\infty}>0$, $h'(f_{\infty})<0$ and $h''(f_{\infty})\ge 0$.  All these conditions bound the space of parameters $(\mu,\,\xi)$, and we can write the allowed set as
\begin{align}\label{cass}
\begin{rcases}
&\mu = \alpha^2(3+\beta)-4\alpha^3\\ 
&\xi = 3\alpha^4-(2+\beta)\alpha^3
\end{rcases}\quad  \text{where} \quad \alpha\ge 0,\, \beta\ge 0,\, 2\alpha-\beta\ge 1\, .
\end{align}
%\begin{equation}
%	(\mu,\, \xi)\in\left\{\left(\alpha^2(3+\beta)-4\alpha^3,3\alpha^4-(2+\beta)\alpha^3\right): \alpha\ge 0,\, \beta\ge 0,\, 2\alpha-\beta\ge 1\right\}\, .
%\end{equation}
If the parameters belong to this set, there exists at least one AdS vacuum satisfying all the aforementioned constraints with $f_{\infty}=1/\alpha$, $G_{\rm eff}=G/\beta$. Remarkably, we do not find any other allowed vacuum, so in this region of the parameter space the vacuum exists and it is unique. In Fig. \ref{VacuaQ4} we show the region defined by \req{cass}.
It is convenient to divide it into  three different zones. Region $A$ is the one with $\xi\ge 0$, and in this case the allowed AdS vacua is the second largest real root of $h$. The largest root has $h'(f_{\infty})>0$ so it is not allowed. Region $B_1$ corresponds to $\xi<0$. There a third root appears, which becomes the largest one. This one has $h'(f_{\infty})<0$ but $h''(f_{\infty})<0$, so it is not suitable. At this point the physical vacuum is the third largest root of $h$. If $\xi$ is negative enough, the two roots larger than the physical one disappear. They coalesce for $(\mu,\, \xi)=(3\alpha^2-4\alpha^3,3\alpha^4-2\alpha^3)$, $0\le\alpha\le 1/2$, in which case there appears a special critical point. This line is the one which separates regions $B_1$ and $B_2$ in Fig. \ref{VacuaQ4}.  Below that line, in region $B_2$, the physical vacuum is the largest root of $h$, which has interesting consequences for the Taub-NUT solutions.

There is a one-parameter family of critical theories, \ie for which $h'(f_{\infty})=0$. We can use $f_{\infty}$ to parametrize the value of the couplings in that case, namely
\begin{equation}
	\mu_{\rm cr}=\frac{3 f_{\infty}-4}{f_{\infty}^3}\, ,\quad \xi_{\rm cr}=\frac{3-2f_{\infty}}{f_{\infty}^4}\, .
\end{equation}
Of course, if we impose $\xi_{\rm cr}$ to be zero, we recover critical ECG, for which $f_{\infty}=3/2$ and $\mu_{\rm cr}=4/27$.
%In the case of Euclidean signature, we will work with the following regularized action functional \ref{GeneralSE}

%Let us propose again the NUT-charged ansatz
%\begin{equation}\label{4metric2}
%	ds^2=V(r) (d\tau+ n A_k)^2+\frac{dr^2}{V(r)}+(r^2-n^2)d\sigma_k^2\, ,
%\end{equation}
%whwre $A_k$ and $d\sigma_k^2$ are given in \ref{base}. 
When evaluated on the ansatz \req{FFnut} for $\mathcal{B}=\mathbb{S}^2,\mathbb{T}^2,\mathbb{H}^2$, we find again that the field equations of \req{QuarticT} reduce to a single equation for the function $V_{\mathcal{B}}$. As before, we find that this equation allows for an integrable factor $(1-n^2/r^2)$, and we can write it as in \req{mastere}, namely
%
%\begin{equation}%
	%\frac{d}{dr}\mathcal{E}_{\mathcal{B}}\left[V, V', V'', r\right]=0\, ,
%\end{equation}
\begin{equation}
	\label{eqV4}
	\begin{aligned}
		&V \left(\frac{2 n^2}{r}-2 r\right)+\frac{2 \left(k L^2 \left(n^2+r^2\right)-3 n^4-6 n^2 r^2+r^4\right)}{L^2 r}\\
		&+\mu L^4 \Bigg[\frac{6 V^3 n^2 \left(n^2+9 r^2\right)}{r \left(n^2-r^2\right)^3}
		+\left(V'\right)^2 \left(\frac{3 V n^2}{n^2 r-r^3}-\frac{3 k}{2 r}\right)-\frac{\left(V'\right)^3}{2}+V' \left(\frac{3 V^2 \left(17 n^2+r^2\right)}{\left(n^2-r^2\right)^2}+\frac{3 V k}{n^2-r^2}\right)\\
		&+V'' \left(-\frac{3 V^2 \left(4 n^2+r^2\right)}{r^3-n^2 r}+\frac{3 V V'}{2}+\frac{3 V k}{r}\right)\Bigg]+\xi L^6 \Bigg[\frac{V^4 \left(22 n^6+270 n^4 r^2+36 n^2 r^4\right)}{r \left(r^2-n^2\right)^5}\\
		&-\frac{4 V^3 k n^2 \left(n^2+9 r^2\right)}{r \left(n^2-r^2\right)^4}+\left(V'\right)^3 \left(\frac{k}{n^2-r^2}-\frac{V \left(15 n^2+r^2\right)}{2 (n-r)^2 (n+r)^2}\right)+\frac{\left(V'\right)^4 \left(9 n^2+3 r^2\right)}{8 n^2 r-8 r^3}\\
		&+\left(V'\right)^2 \left(\frac{3 V^2 \left(13 n^4+30 n^2 r^2+r^4\right)}{r \left(r^2-n^2\right)^3}-\frac{3 V k \left(n^2+r^2\right)}{r \left(n^2-r^2\right)^2}\right)+V'' \Bigg(\frac{24 V^3 n^2 \left(n^2+r^2\right)}{r \left(r^2-n^2\right)^3}\\
		&-\frac{6 V^2 k n^2}{r \left(n^2-r^2\right)^2}+\frac{3 V \left(V'\right)^2 \left(3 n^2+r^2\right)}{2 \left(r^3-n^2 r\right)}+V' \left(-\frac{3 V^2 \left(11 n^2+r^2\right)}{\left(n^2-r^2\right)^2}-\frac{3 V k}{n^2-r^2}\right)\Bigg)+\\
		&V' \left(-\frac{6 V^3 \left(43 n^4+21 n^2 r^2\right)}{\left(n^2-r^2\right)^4}-\frac{36 V^2 k n^2}{\left(n^2-r^2\right)^3}\right)\Bigg]=4C\, ,
	\end{aligned}
\end{equation}
where $k=+1,0,-1$ for $\mathcal{B}=\mathbb{S}^2$, $\mathbb{T}^2$  and $\mathbb{H}^2$, respectively. 
%Then, by integrating once we obtain
%\begin{equation}\label{eqV42}
%	\mathcal{E}_{\mathcal{B}}\left[V, V', V'', r\right]=4C
%\end{equation}

Let us start by determining the asymptotic behaviour in this case. As usual, we can separate the solution as the sum of a particular solution plus a homogeneous one. The particular solution can be obtained by performing a $1/r$ expansion, which yields
\begin{equation}\label{VpQ}
	V_p(r)=f_{\infty}\frac{r^2}{L^2}+k-5f_{\infty}\frac{n^2}{L^2}+\frac{2C}{h'(f_{\infty})r}+\mathcal{O}(r^{-2})\, ,
\end{equation}
where $h'(f_{\infty})=-1+3\mu f_{\infty}^2+4\xi f_{\infty}^3<0$, according to the unitarity constraint.  From this asymptotic expansion, and using the fact that $G_{\rm eff}=-G/h'(f_{\infty})$ \cite{Bueno:2018yzo},  we see that for a spherical base space, $C=GM$, where $M$ is the ADM mass \cite{Arnowitt:1960es,Arnowitt:1960zzc}, or more appropriately, the Abbott-Deser energy \cite{Abbott:1981ff, Deser:2002jk,Senturk:2012yi,Adami:2017phg}.   For the remaining topologies, $C$ is also proportional to the total energy, but the proportionality constant is different. If we now consider $V(r)=V_p(r)+\frac{r^2}{L^2}g(r)$ and expand linearly in $g$, we obtain the following differential equation keeping only the leading terms when $r\rightarrow\infty$\footnote{For example, we are neglecting a term $g'/r^3$ against $g''/r$.}
\begin{equation}
	-\frac{3L^2 C h''(f_{\infty})}{2h'(f_{\infty}) r}g''(r)+2 h'(f_{\infty}) g(r)=0\, .
\end{equation}
Just like for ECG, the solution is again given in terms of Airy functions,
\begin{equation}\label{asymptg}
	g(r)=A \textrm{AiryAi}\left[\left(\frac{4h'(f_{\infty})^2}{3L^2 C h''(f_{\infty})}
	\right)^{1/3}r\right]+B \textrm{AiryBi}\left[\left(\frac{4h'(f_{\infty})^2}{3L^2 C h''(f_{\infty})}
	\right)^{1/3}r\right]\, ,
\end{equation}
and the analysis is analogous. If $C h''(f_{\infty})>0$ there is a growing mode and a decaying one, so by eliminating the former we obtain an asymptotically AdS solution. If $C h''(f_{\infty})<0$ all solutions except the trivial one are pathological at infinity. Then, in order to ensure the existence of solutions of positive mass, $C>0$, we demand that $ h''(f_{\infty})>0$, which is the constraint anticipated before. 

\subsection{$\mathcal{B}=\mathbb{S}^2$}

From this point on, we focus on the case $\mathcal{B}=\mathbb{S}^2$. Then, the Taub-NUT metric takes the form \req{taub}, where $V_{\mathbb{S}^2}(r)$ satisfies \req{eqV4} with $k=1$. As usual, the period of $\tau$ is fixed to $\beta_{\tau}=8\pi n$, which removes the Dirac-Misner string.
%Let us now turn to study the solutions when the base space is $\mathbb{S}^2$
%\begin{equation}
%	ds^2=V(r)(d\tau+2 n \cos\theta d\phi)^2+\frac{dr^2}{V(r)}+(r^2-n^2)\left(d\theta^2+\sin^2\theta d\phi^2\right)\, ,
%	\label{taub2}
%\end{equation}
%where, we recall that $\tau$ has periodicity $8\pi n$. The function $V(r)$ satisfies \ref{eqV42} with $k=1$. Requiring that the solution is asymptotically AdS fixes one integration constant, as we have just seen, and the other constant is fixed by different regularity conditions.
\subsubsection*{Taub-NUT solutions}
Assuming $V_{\mathbb{S}^2}(n)=0$ and the regularity condition $V_{\mathbb{S}^2}'(n)=1/(2n)$, we can write an expansion around $r=n$ as
\begin{equation}
	V(r)=\frac{r-n}{2n}+\sum_{i=2}^{\infty} (r-n)^i a_i\, .
	\label{near2}
\end{equation}
If we introduce this expansion in \req{eqV4}, we obtain a series of relations that must be satisfied order by order in $(r-n)$. From the first one we read the mass of the solution, which is given by
%\begin{equation}
%	\left(-4 G M-\frac{16 n^3}{L^2}+4n-\frac{\mu  L^4}{4 n^3}-\frac{\xi L^6 }{16 n^5}\right)+\mathcal{O}\left((r-n)^3\right)=0\, ,
%\end{equation}
%from which we read the mass of the NUT solution,
\begin{equation}\label{mmas}
	GM=n-\frac{4 n^3}{ L^2}-\frac{\mu  L^4}{16 n^3}-\frac{\xi L^6 }{64 n^5}\, .
\end{equation}
Naturally, this generalizes the ECG result \req{masss} and reduces to it for $\xi=0$.
Also analogously to the ECG case, the following term in the expansion gives a relation between $a_3$ and $a_2$ from where we obtain $a_3(a_2)$, the next fixes $a_4(a_2)$, and so on. Therefore, once again, the complete series is determined by a single free parameter that must be chosen so that the condition $B=0$ in \req{asymptg} is met.
	
	% \comment{Are we going to plot numerical solutions? Perhaps unnecessary in this case. A comment should be made about them in any case.}\\
Let us now compute the Euclidean on-shell action of the solutions. For that, we use the generalized action \req{assd}, where the charge $a^*$ is given in this case by $a^*=(1+3\mu f_{\infty}^2+2\xi f_{\infty}^3)\tilde{L}^2/(4G)$. Then, we can write the full action as
\begin{equation}\label{EuclideanQ}
\begin{aligned}
I_E=-\int \frac{d^4x \sqrt{g}}{16\pi G} \left[\frac{6}{L^2}+R-\frac{\mu L^4}{8} \mathcal{P}-\frac{\xi L^6}{16}\mathcal{Q} \right]
-\frac{a^*}{2 \pi \tilde{L}^2}\int_{\partial \mathcal{M}}d^3x\sqrt{h}\left[K-\frac{2\sqrt{f_{\infty}}}{L}-\frac{L}{2\sqrt{f_{\infty}}}\mathcal{R}\right]\, ,
\end{aligned}
\end{equation}
The evaluation of all terms in \req{EuclideanQ} is analogous to the one performed in detail for ECG in Section~\ref{sec:TNECG}. We observe that the divergent terms coming from the various contribution cancel, and we are left with the following finite answer
\begin{equation}
	I_E=\frac{4\pi}{G}\left[n^2-\frac{2n^4}{L^2}+\frac{\mu L^4}{16 n^2}+\frac{\xi L^6}{128 n^4}\right]\, ,
\end{equation}
which generalizes the ECG result \req{freeee1}.
Taking into account that $\beta=8\pi n$, we can obtain the energy and the entropy $E=\partial I_E/\partial \beta$, $S=\beta E-I_E$. The first exactly coincides with the ADM mass in \req{mmas}, $E=M$, whereas for the entropy we find
\begin{equation}
	S=\frac{4\pi}{G}\left[n^2-\frac{6n^4}{L^2}-\frac{3\mu L^4}{16 n^2}-\frac{5\xi L^6 }{128 n^4}\right]\, ,
\end{equation}
which generalizes the ECG answer \req{entR}.

%%%%%%%%%%%%%%%%%%%%%%%%%%%%%%%%%%%
\subsubsection*{Taub-bolt solutions}
Let us now assume that $V_{\mathbb{S}^2}$ vanishes for some $r=r_b>n$. In order to avoid a conical singularity we demand again that $V_{\mathbb{S}^2}'(r_b)=1/(2n)$, so that $V_{\mathbb{S}^2}(r)$ should be Taylor-expanded as
\begin{equation}
	V_{\mathbb{S}^2}(r)=\frac{r-r_b}{2n}+\sum_{i=2}^{\infty} (r-r_b)^i a_i\, .
	\label{near3}
\end{equation}
Plugging this expansion into \req{eqV4},  we find that the mass of the bolt is given by
\begin{equation}
	GM=\frac{\left(n^2+r_b^2\right)}{2r_{b}}-\frac{3 n^4+6 n^2 r_{b}^2-r_{b}^4}{2L^2 r_{b}}-\frac{\mu  L^4 (6 n+r_{b})}{64n^3 r_{b}}-\frac{ \xi L^6 \left(9 n^2+16 n r_{b}+3 r_{b}^2\right)}{512n^4 r_{b} \left(r_{b}^2-n^2\right)}\, ,
\end{equation}
where $r_{b}$ is implicitly related to $n$ through 
\begin{equation}\label{bolt4Q}
	\begin{aligned}
		&\frac{6(r_{b}^2-n^2)^2}{L^2r_{b}^2}-\frac{(r_{b}^2-n^2)(r_{b}-2 n)}{ n r_{b}^2}-\frac{3 \mu  L^4 \left(n^2+n r_{b}+r_{b}^2\right)}{8 n^2 r_{b}^2(r_{b}^2-n^2) }\\
		&-\frac{L^6 \xi  \left(9 n^4+48 n^3 r_{b}+30 n^2 r_{b}^2+16 n r_{b}^3+r_{b}^4\right)}{128 n^4 r_{b}^2 \left(r_{b}^2-n^2\right)^2}=0\, .
	\end{aligned}
\end{equation}
Just as for ECG, the remaining equations fix the coefficients $a_{i>2}$ in terms of $a_2$, which must be chosen so that the solution is asymptotically AdS, a condition that selects a unique value of $a_2$.%\comment{again, not presenting plots of numerical solutions(?)}

\begin{figure}[t!]
	\centering 
	\includegraphics[width=0.47\textwidth]{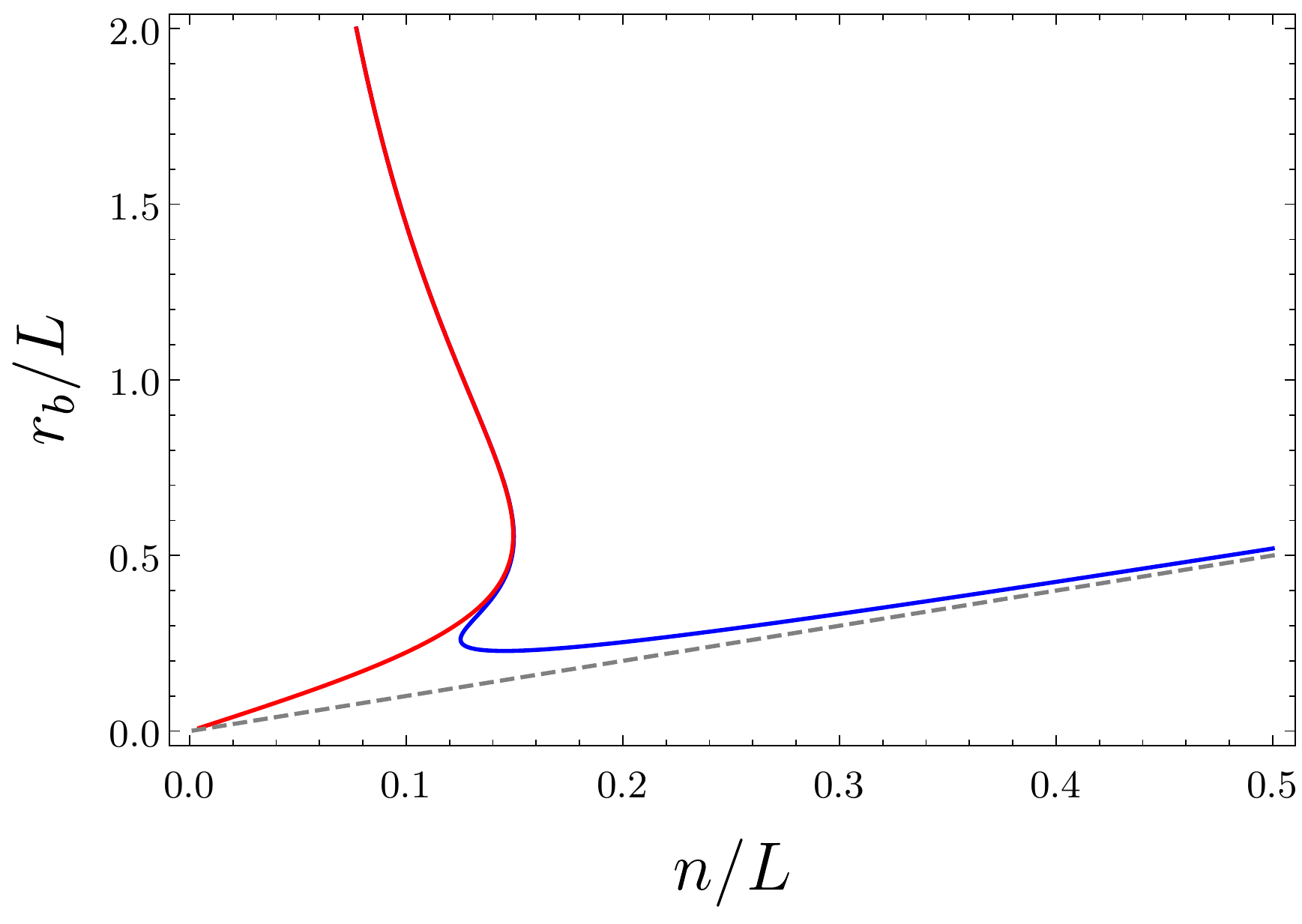}
	\includegraphics[width=0.47\textwidth]{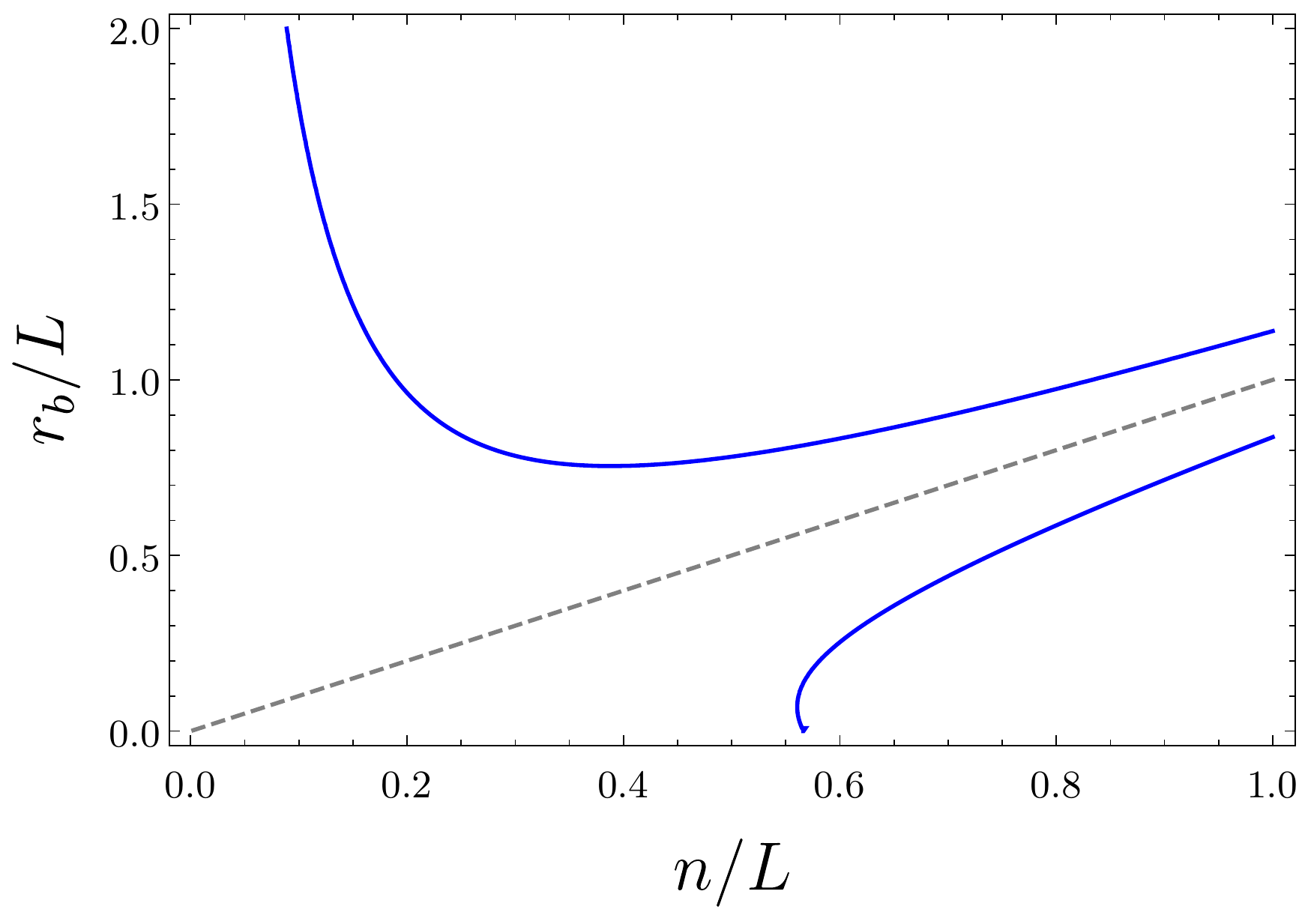}
	\includegraphics[width=0.47\textwidth]{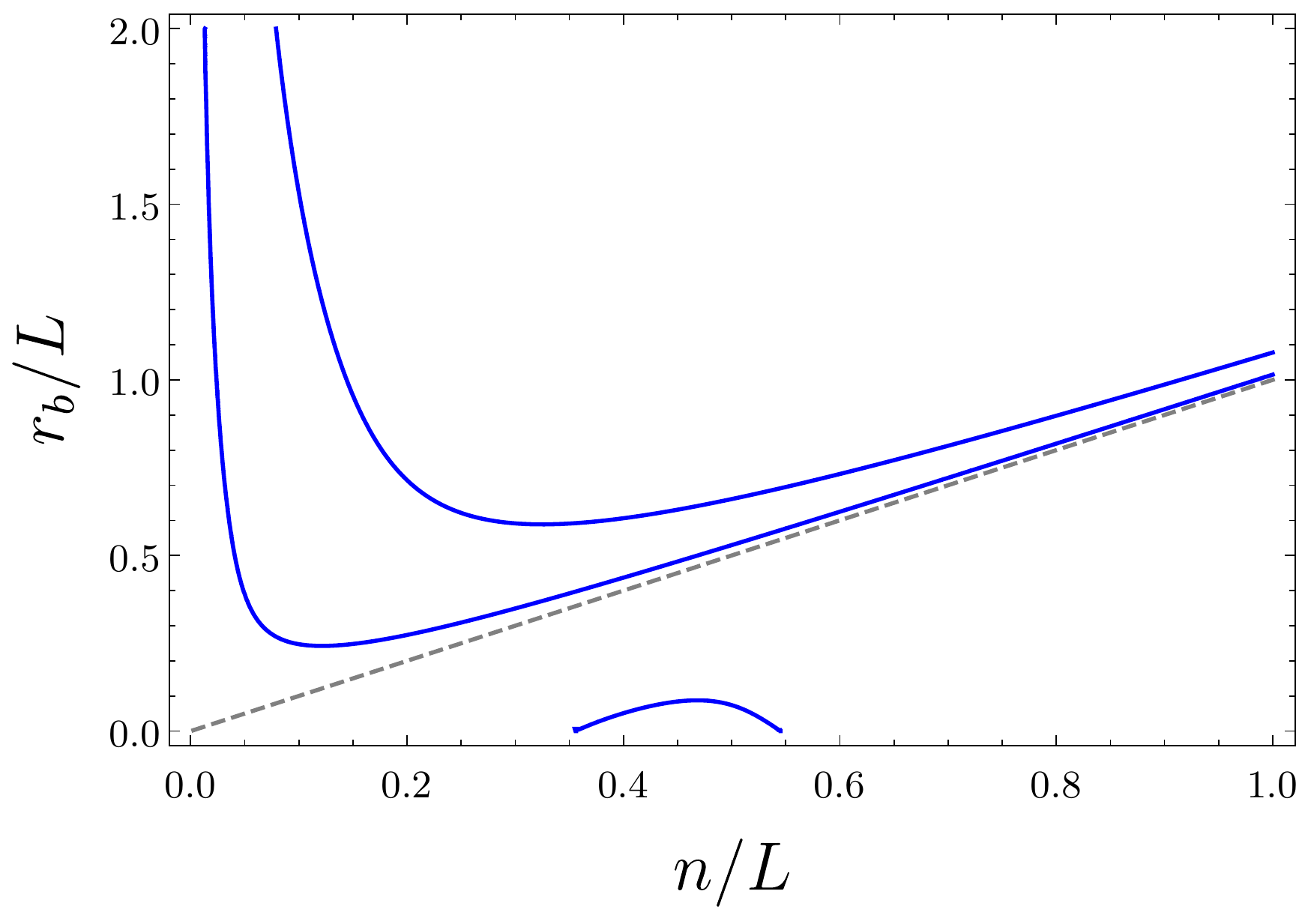}
	\includegraphics[width=0.47\textwidth]{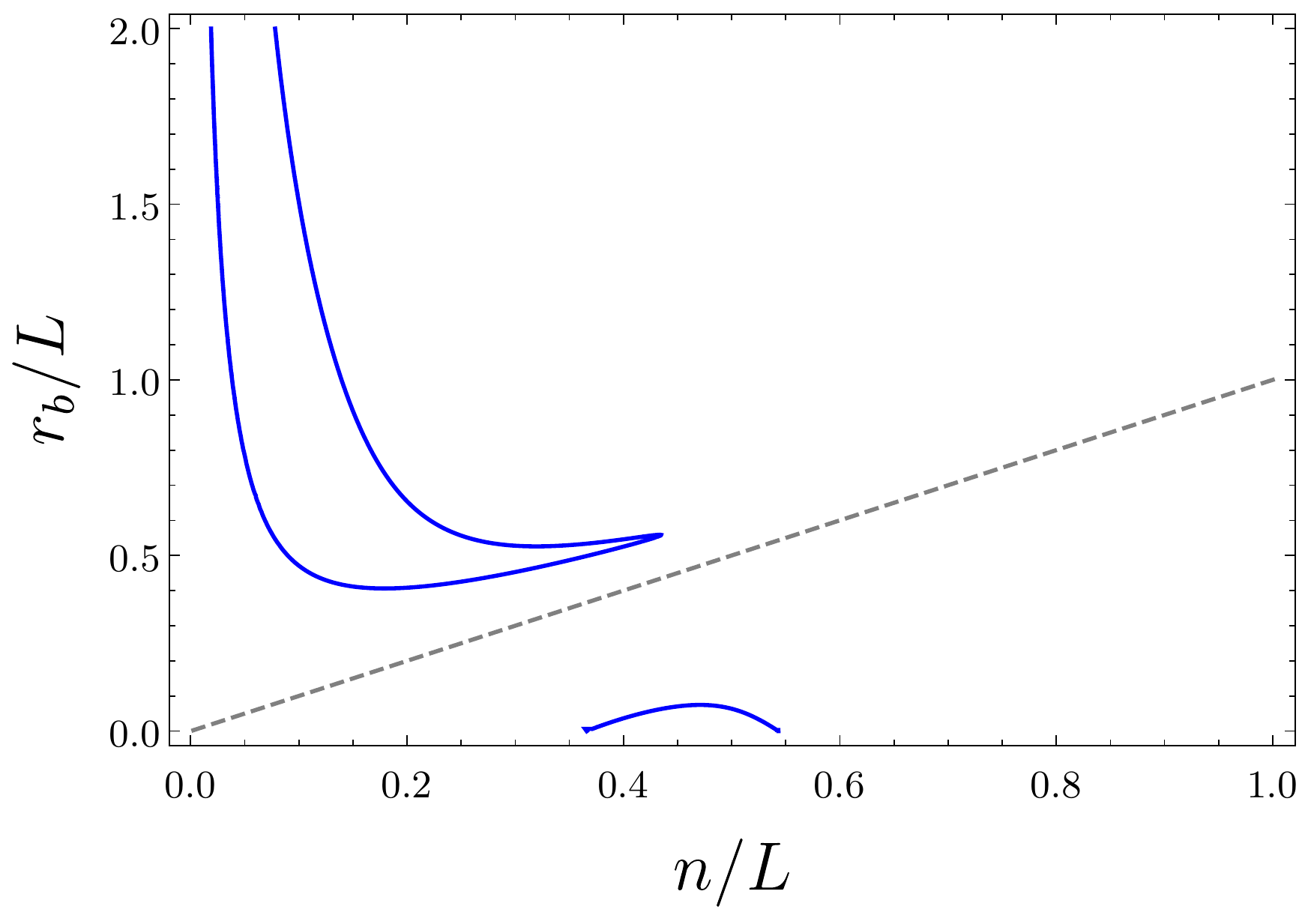}
	\caption{Roots of the equation \req{bolt4Q} for several values of the parameters. Only the roots above the reference dashed line $r_{b}=n$ admit the construction of bolt solutions. Upper left: behaviour in region $A$ when the parameters are very small ($\mu=\xi=10^{-5}$ in this case); there is a range of $n$ with three different bolt solutions. Upper right: $\mu=0.05$, $\xi=0.05$; this represents the typical case for region $A$. Lower left: region $B_1$ ($\mu=0.05$, $\xi=-0.002$); there are two roots for every value of $n$. Lower right: region $B_2$ ($\mu=0.05$, $\xi=-0.0052$); there are two roots if $n$ is smaller than certain value; both diverge as $n\rightarrow 0$. }
	\label{boltq4}
\end{figure}

The roots of \req{bolt4Q} behave in different ways depending on the values of the parameters. We can characterize several qualitative features depending on the region of the parameter space shown in Fig. \ref{VacuaQ4}. First, recall that in the case of Einstein gravity, $\mu=\xi=0$, there are two allowed roots  when $n/L<\left((2-\sqrt{3})/12 \right)^{1/2}$ --- see \req{Einss} --- and no solutions otherwise. One of the roots goes to zero for $n\rightarrow 0$ and the other one diverges. When $\mu\neq 0$ or $\xi\neq 0$ there is no root going to $0$ for $n\rightarrow 0$. In fact, in this limit we can expand $r_{b}$ as
\begin{equation}
	r_{b}=\frac{c_0L^2}{n}+c_1 n+\mathcal{O}(n^3)\, , \quad \text{with} \quad c_0^3(6c_0-1)=\frac{\xi}{128}\, ,\quad c_1=\frac{256 c_0^3-48c_0^2+\mu}{8 c_0^2(8c_0-1)}\, ,
\end{equation}
%with 
%\begin{equation}
%	c_0^3(6c_0-1)=\frac{\xi}{128}\, ,\quad c_1=\frac{256 c_0^3-48c_0^2+\mu}{8 c_0^2(8c_0-1)}\, ,
%\end{equation}
where we must demand $c_0>0$. The first equation gives us some information about the roots, depending on the region. If $\xi>0$, there is a unique value of $c_0$, so there is a single solution for $n\rightarrow 0$. Indeed, we observe that there is a unique branch in the diagram $(r_{b},n)$ if $\xi>0$ and that there is a solution for every value of $n$, including large values. When $-1/16<\xi<0$, there are two different roots $c_0$, so there are two different solutions for $n\rightarrow 0$. We see that if $\xi\in B_1$, then these solutions extend to every $n$, while for $\xi\in B_2$, the solutions only exist for $n$ smaller than certain value. Finally, if $\xi<-1/16$, we find that there are no bolt solutions. In Fig. \ref{boltq4} we summarize the different possibilities.

In all possible cases in which solutions exist, the mass when $n\rightarrow 0$ is given by
\begin{equation}
	GM=\frac{L^4(-256 c_0^3+48 c_0^2-\mu)}{64 n^3}+\mathcal{O}(n)\, ,
\end{equation}
which can be shown to be positive as long as the parameters lie in the allowed region.

Then, the Euclidean on-shell action can be computed along the same lines as in the NUT case using \req{EuclideanQ}. The final result reads
% and we find again that it may be expressed as $F=\frac{2\pi n}{G}H(\rh)$, where $H$  is given by \ref{Hq4}. When the expansion \ref{near3} is used, we obtain
\begin{equation}
		I_E=\frac{\pi}{G}\Bigg[\frac{ 4 n r_{b} \left(r_{b}^2-3 n^2\right)}{L^2}+n^2+4 n r_{b}-r_{b}^2+\frac{\mu  L^4 \left(5 n^2+12 n r_{b}+r_{b}^2\right)}{16n^2 (r_{b}^2-n^2)}
		+\frac{\xi L^6  \left(8 n^3+9 n^2 r_{b}+8 n r_{b}^2+r_{b}^3\right)}{64n^3 \left(r_{b}^2-n^2\right)^2}\Bigg]\, ,
\end{equation}
where $r_{b}$ is a function of $n$ given implicitly by \req{bolt4Q}. In the $n\rightarrow 0$ limit, we can write explicitly
\begin{equation}
	I_E=\frac{\pi L^2}{G}\left[\frac{(256 c_0^3-48 c_0^2+\mu)L^2}{16 n^2}+\frac{3\mu}{4 c_0}+12 c_0(8 c_0-1)+\mathcal{O}(n^2)\right]\, .
\end{equation}

\section{Six dimensions: Quartic generalized quasi-topological gravities}\label{sIx}

%\robiecomment{I've shuffled some constants around in the below expressions to, hopefully, make the presentation clearer. I will update notation in line with PB's suggestions in the near future.}

We now move on to consider theories in six dimensions. In this case the generalized quasi-topological gravity class \cite{Hennigar:2017ego,PabloPablo3,Ahmed:2017jod} includes additional densities beyond those in   $D=4$. In particular, the Gauss-Bonnet term $\mathcal{X}_4$ is no longer topological. Analytic generalizations of the Einstein gravity static black-hole \cite{Boulware:1985wk,Cai:2001dz} and Taub-NUT/bolt solutions \cite{Dehghani:2005zm,Dehghani:2006aa} have been constructed in the presence of this contribution. In particular, the Taub-NUT solutions of Gauss-Bonnet \cite{Dehghani:2005zm,Dehghani:2006aa}  
and 3rd-order Lovelock gravity \cite{Hendi:2008wq}  were, prior to these results, and to the best of our knowledge, the only known examples of solutions of that class for any  higher-curvature gravity theory.
 
 In principle, the six-dimensional GQG family includes two nontrivial terms at cubic order, corresponding to the usual Quasi-topological gravity density, plus an additional one. As observed in \cite{Quasi,Quasi2,Oliva:2010zd}, including the Quasi-topological gravity density in $D\geq 6$ is equivalent, from the point of view of static black-hole solutions, to including the cubic Lovelock interaction. In $D=6$, this is a topological term, and therefore no new nontrivial black holes exist in that case for Quasi-topological gravity. They do exist, however, when the additional GQG term is included \cite{Hennigar:2017ego}. Hence, following the same reasoning as for ECG in $D=4$, one would have expected that new Taub-NUT solutions of the form \req{FFnut} should also exist for GQG. Remarkably, we find that this is not the case, and that no cubic theory admits nontrivial generalizations of the Einstein gravity or Gauss-Bonnet Taub-NUT solutions characterized by a single function in six dimensions (independent of the base manifold considered).
% the black holes of  are equivalent to those of cubic Lovelock
% analytic black-hole and Taub-NUT/bolt solutions of the form \req{FFbh} and \req{FFnut} respectively have been constructed 
%Up to cubic order in curvature, the Lagrangian of the theory can be written as $\mathcal{L}=(D-1)(D-2)/L^2+R+\alpha \mathcal{X}_4+\mu \mathcal{Z}+\lambda \mathcal{S}$, where $\mathcal{X}_4$ is the usual Gauss-Bonnet term, and 
% these correspond to Gauss-Bonnet \cite{}, plus the cubic contributions corresponding to quasi-topological \cite{Quasi,Quasi2} and Generalized quasi-topological gravity.
%of theories admitting nontrivial static and spherically symmetric black holes characterized by a single function \req{FFbh} includes --- besides Einstein gravity --- Gauss-Bonnet at quadratic order, 
%Remarkably, we find that there are no non-trivial cubic theories in six dimensions, independent of the base manifold considered.
While there do exist cubic theories that satisfy the necessary constraints to admit solutions of the form~\eqref{FFnut}, it turns out that for any such theory the field equations on these spaces vanish identically.  This is analogous to the ${\cal C}^{(i)}_{D}$ terms discussed in~\cite{Ahmed:2017jod} for the case of static, spherically symmetric metrics.

Happily, at quartic order in curvature there exist non-trivial options of both Quasi-topological and Generalized quasi-topological type.\footnote{Recall that the distinction between both classes comes from considering the theories for static spherically symmetric spacetimes. While both admit solutions of the form \req{FFbh}, the field equations for the Quasi-topological theories reduce to algebraic polynomial equations for $f(r)$, while for those  of the generalized quasi-topological type, the metric function satisfies a non-linear second order differential equation in each case. } We will not present a detailed classification of such theories here --- see \cite{Dehghani:2011vu} and \cite{Ahmed:2017jod} --- but limit ourselves to some brief remarks. Beginning from a general action containing all 26 possible quartic invariants \cite{Aspects}, we constrain the action by imposing the conditions listed in Appendix A of~\cite{Ahmed:2017jod}. This selects theories admitting black hole solutions of the form \req{FFbh} and which, as a consequence \cite{PabloPablo3}, do not propagate ghosts on maximally symmetric backgrounds. Next, for each of the possible four dimensional base manifolds listed below, we generate additional constraints to ensure the corresponding theory admits solutions of the form~\eqref{FFnut}. Surprisingly, demanding the constraints to be simultaneously satisfied for all four base manifolds $\mathcal{B} = \mathbb{CP}^2, \mathbb{S}^2 \times \mathbb{S}^2$, $\mathbb{S}^2 \times \mathbb{T}^2$ and $\mathbb{T}^2 \times \mathbb{T}^2$ results in a family of theories that yield trivial field equations. When one relaxes this condition, considering only a subset of the base manifolds, then nontrivial options exist.    

The non-trivial theories can be classified into two groups: Quasi-topological and Generalized quasi-topological.
For the base manifold $\mathbb{S}^2 \times \mathbb{S}^2$, the only nontrivial theories are of the generalized quasi-topological type. The constraints can be satisfied simultaneously for $\mathcal{B} = \mathbb{CP}^2, \mathbb{S}^2 \times \mathbb{S}^2$ and $\mathbb{T}^2 \times \mathbb{T}^2$, resulting in a three parameter family of non-trivial theories, each making the same contribution to the field equations for a given base manifold.  We can deduce all of the relevant physics by considering only one member of this class, which we denote as $\mathcal{S}$ below.     

%\tcp{\bf [I think a bit more explanation is needed as to why some bases are admitted in some cases and others are not.]} \robiecomment{I don't think there is a good explanation here. It is just that, for some bases, there are not any theories (at the quartic level) that admit single function solutions. I think this is just a ``this is the way it is'' situation.}

Excluding the base manifold $\mathbb{S}^2 \times \mathbb{S}^2$, then Quasi-topological options exist for all remaining base manifolds. The constraints can be satisfied simultaneously, resulting in a three parameter family of non-trivial Quasi-topological theories. Again, each theory makes the same contribution to the field equations for a given base manifold, and we need only consider a single member of this family, which we denote as $\mathcal{Z}$ below. Thus, in six dimensions, we will consider the following two quartic Lagrangian densities:  
\begin{align}\label{eq:GQGTN1}
{\cal S} &=  992 R_{\mu}{}^{\rho} R^{\mu\nu} R_{\nu}{}^{\delta} R_{\rho\delta} + 28 R_{\mu\nu} R^{\mu\nu} R_{\rho\delta} R^{\rho\delta} - 192 R_{\mu}{}^{\rho} R^{\mu\nu} R_{\nu\rho} R - 108 R_{\mu\nu} R^{\mu\nu} R^2 
\nonumber\\
&+ 1008 R^{\mu\nu} R^{\rho\delta} R R_{\mu\rho\nu\delta} + 36 R^2 R_{\mu\nu\rho\delta} R^{\mu\nu\rho\delta} - 2752 R_{\mu}{}^{\rho} R^{\mu\nu} R^{\delta\tau} R_{\nu\delta\rho\tau} + 336 R R_{\mu}{}^{\tau}{}_{\rho}{}^{\gamma} R^{\mu\nu\rho\delta} R_{\nu\tau\delta\gamma} 
\nonumber\\
&- 168 R R_{\mu\nu}{}^{\tau\gamma} R^{\mu\nu\rho\delta} R_{\rho\delta\tau\gamma} - 1920 R^{\mu\nu} R_{\mu}{}^{\rho\delta\tau} R_{\nu}{}^{\gamma}{}_{\delta}{}^{\eta} R_{\rho\gamma\tau\eta} + 152 R_{\mu\nu} R^{\mu\nu} R_{\rho\delta\tau\gamma} R^{\rho\delta\tau\gamma} 
\nonumber\\
&+ 960 R^{\mu\nu} R_{\mu}{}^{\rho\delta\tau} R_{\nu\rho}{}^{\gamma\eta} R_{\delta\tau\gamma\eta} - 1504 R^{\mu\nu} R_{\mu}{}^{\rho}{}_{\nu}{}^{\delta} R_{\rho}{}^{\tau\gamma\eta} R_{\delta\tau\gamma\eta} + 352 R_{\mu\nu}{}^{\tau\gamma} R^{\mu\nu\rho\delta} R_{\rho\tau}{}^{\eta\sigma} R_{\delta\gamma\eta\sigma} 
\nonumber\\
&- 2384 R_{\mu}{}^{\tau}{}_{\rho}{}^{\gamma} R^{\mu\nu\rho\delta} R_{\nu}{}^{\eta}{}_{\tau}{}^{\sigma} R_{\delta\eta\gamma\sigma} + 4336 R_{\mu\nu}{}^{\tau\gamma} R^{\mu\nu\rho\delta} R_{\rho}{}^{\eta}{}_{\tau}{}^{\sigma} R_{\delta\eta\gamma\sigma} - 143 R_{\mu\nu}{}^{\tau\gamma} R^{\mu\nu\rho\delta} R_{\rho\delta}{}^{\eta\sigma} R_{\tau\gamma\eta\sigma} 
\nonumber\\
&- 436 R_{\mu\nu\rho}{}^{\tau} R^{\mu\nu\rho\delta} R_{\delta}{}^{\gamma\eta\sigma} R_{\tau\gamma\eta\sigma} + 2216 R_{\mu}{}^{\tau}{}_{\rho}{}^{\gamma} R^{\mu\nu\rho\delta} R_{\nu}{}^{\eta}{}_{\delta}{}^{\sigma} R_{\tau\eta\gamma\sigma} - 56 R_{\mu\nu\rho\delta} R^{\mu\nu\rho\delta} R_{\tau\gamma\eta\sigma} R^{\tau\gamma\eta\sigma} \, ,
\\
%%%%%%%%%
%%%%%%%%%
%%%%%%%%%
\label{eq:GQGTN2}
{\cal Z} &=  -112 R_{\mu}{}^{\rho} R^{\mu\nu} R_{\nu}{}^{\delta} R_{\rho\delta} - 36 R_{\mu\nu} R^{\mu\nu} R_{\rho\delta} R^{\rho\delta} + 18 R_{\mu\nu} R^{\mu\nu} R^2 - 144 R^{\mu\nu} R^{\rho\delta} R R_{\mu\rho\nu\delta} 
\nonumber\\
&- 9 R^2 R_{\mu\nu\rho\delta} R^{\mu\nu\rho\delta} + 72 R^{\mu\nu} R R_{\mu}{}^{\rho\delta\tau} R_{\nu\rho\delta\tau} + 576 R_{\mu}{}^{\rho} R^{\mu\nu} R^{\delta\tau} R_{\nu\delta\rho\tau} - 400 R^{\mu\nu} R^{\rho\delta} R_{\mu\rho}{}^{\tau\gamma} R_{\nu\delta\tau\gamma} 
\nonumber\\
&+ 48 R R_{\mu}{}^{\tau}{}_{\rho}{}^{\gamma} R^{\mu\nu\rho\delta} R_{\nu\tau\delta\gamma} + 160 R_{\mu}{}^{\rho} R^{\mu\nu} R_{\nu}{}^{\delta\tau\gamma} R_{\rho\delta\tau\gamma} - 992 R^{\mu\nu} R_{\mu}{}^{\rho\delta\tau} R_{\nu}{}^{\gamma}{}_{\delta}{}^{\eta} R_{\rho\gamma\tau\eta} 
\nonumber\\
&+ 18 R_{\mu\nu} R^{\mu\nu} R_{\rho\delta\tau\gamma} R^{\rho\delta\tau\gamma} - 8 R^{\mu\nu} R_{\mu}{}^{\rho\delta\tau} R_{\nu\rho}{}^{\gamma\eta} R_{\delta\tau\gamma\eta} + 238 R_{\mu\nu}{}^{\tau\gamma} R^{\mu\nu\rho\delta} R_{\rho\tau}{}^{\eta\sigma} R_{\delta\gamma\eta\sigma} 
\nonumber\\
&- 376 R_{\mu}{}^{\tau}{}_{\rho}{}^{\gamma} R^{\mu\nu\rho\delta} R_{\nu}{}^{\eta}{}_{\tau}{}^{\sigma} R_{\delta\eta\gamma\sigma} + 1792 R_{\mu\nu}{}^{\tau\gamma} R^{\mu\nu\rho\delta} R_{\rho}{}^{\eta}{}_{\tau}{}^{\sigma} R_{\delta\eta\gamma\sigma} - 4 R_{\mu\nu}{}^{\tau\gamma} R^{\mu\nu\rho\delta} R_{\rho\delta}{}^{\eta\sigma} R_{\tau\gamma\eta\sigma} 
\nonumber\\
&- 284 R_{\mu\nu\rho}{}^{\tau} R^{\mu\nu\rho\delta} R_{\delta}{}^{\gamma\eta\sigma} R_{\tau\gamma\eta\sigma} + 320 R_{\mu}{}^{\tau}{}_{\rho}{}^{\gamma} R^{\mu\nu\rho\delta} R_{\nu}{}^{\eta}{}_{\delta}{}^{\sigma} R_{\tau\eta\gamma\sigma} \, .
\end{align}
The generalized quasi-topological  term ${\cal S}$ is an appropriate choice for all base manifolds besides $\mathbb{T}^2 \times \mathbb{S}^2$, while the Quasi-topological term ${\cal Z}$ is an appropriate choice for all base manifolds besides $\mathbb{S}^2 \times \mathbb{S}^2$.

The complete action we consider is then
\be\label{eqn:quart_6d_act}
I_{\rm E} = - \frac{1}{16 \pi G} \int d^6x \sqrt{g} \left[\frac{20}{L^2} + R + \frac{\lambda_{\rm \ssc GB} L^2 }{6} {\cal X}_4 -\frac{ \xi L^6 }{216} \mathcal{S} - \frac{ \zeta L ^6}{144} \mathcal{Z}  \right] \, ,
\ee
where we have allowed for the possible contribution of the Gauss-Bonnet term. In this case, the AdS$_6$ vacua of the theory are characterized by being solutions to $h(f_\infty)=0$, where 
\be  \label{hfinf}
h(f_\infty) \equiv 1 - f_\infty + \lambda_{\rm \ssc GB} f_\infty^2 + \zeta f_\infty^4 + \xi f_\infty^4 \, ,
\ee
a definition that will turn out to be useful later on.

As anticipated, when we insert the single-function Taub-NUT ansatz \req{FFnut} in the equations of motion of this theory, we are left with a single independent equation for $V_{\mathcal{B}}$, which can be integrated once to leave it in the form \req{mastere},
%\be  \label{eee}
%\frac{d}{dr} {\cal E}_{\mathcal{B}}[V_{\mathcal{B}}, V_{\mathcal{B}}', V''] = 0
%\ee
where the function ${\cal E}_\mathcal{B}$ receives contributions from all terms in \req{eqn:quart_6d_act}, namely,
\be \label{eq6d}
 {\cal E}_\mathcal{B}^{\rm E} + \lambda_{\rm \ssc GB} L^2  {\cal E}_\mathcal{B}^{({\rm  GB})} + \xi L^6   {\cal E}_\mathcal{B}^{({\cal S})} + \zeta L^6  {\cal E}_\mathcal{B}^{(\mathcal{Z})}=% - \frac{4 G M}{9 \pi}
 C_{\mathcal{B}}\, ,
\ee 
%\robiecomment{Maybe we should denote the right-hand side here as just $C_{\mathcal{B}}$ and then start each subsection with a new definition for this parameter. The value of $-4 GM/(9\pi)$ is particular to the CP base.} 
where $ C_{\mathcal{B}}$ is an integration constant.
%where we have already expressed the integration constant in terms of $M$, which will be related to the ADM energy of the solutions below. 
The explicit form of the various terms appearing in the field equation is the following.
The Einstein gravity contributions to the field equation can be expressed in the form
\begin{align}
{\cal E}_\mathcal{B}^{\rm E} =& \frac{6 L^2(n-r)^2(n+r)^2 V - 6 r^6 + (30n^2 - 2 L^2)r^4 + (-90 n^4 + 12 L^2 n^2)r^2 - 30n^6 + 6 L^2 n^4}{3 L^2 r} \notag
\\
&-\frac{(3 n^4 + 6 n^2 r^2 - r^4)(1+\kappa)}{3r}\, ,
\end{align}
where $\kappa$ is defined by %\comment{shouldn't it be $-1$ for CP S2S2, $-2$ for S2T2 and $-3$ for T2T2 instead?}
\be \label{kappa6}
\kappa = \begin{cases} 
	-1 \quad &\text{for } \mathcal{B} = \mathbb{CP}^2 \text{ and }\mathbb{S}^2 \times \mathbb{S}^2\, ,
	\\
	0 \quad &\text{for } \mathcal{B} = \mathbb{S}^2 \times \mathbb{T}^2\, ,
	\\
	+1 \quad & \text{for } \mathcal{B} = \mathbb{T}^2 \times \mathbb{T}^2\, .
\end{cases}
\ee
Next are the Gauss-Bonnet contributions, which for the various base spaces read
\begin{align}
{\cal E}_{\mathbb{CP}^2}^{\rm  GB} =& -\frac{2n^2}{9 r} \left(9 V^2 + 6 V + 2 \right) -\frac{2r}{9 } \left(9 V^2 - 6 V + 2 \right) \, ,
\\
{\cal E}_{\mathbb{S}^2 \times \mathbb{S}^2}^{\rm  GB} =& -\frac{2n^2}{3 r} \left(3 V^2 + 2 V + 1 \right) -\frac{2r}{3 } \left(3 V^2 - 2 V + 1 \right) \, ,
\\
{\cal E}_{\mathbb{S}^2 \times \mathbb{T}^2}^{\rm GB} =& -\frac{2n^2}{3 r} V \left(3 V  + 1 \right) -\frac{2r}{3 } V \left(3 V - 1 \right) \, ,
\\
{\cal E}_{\mathbb{T}^2 \times \mathbb{T}^2}^{\rm GB} =& - 2 V^2 \left(\frac{n^2}{r} + r \right) \, .
\end{align}
The quartic contributions to the field equations are, of course, more complicated. The ones due to the Generalized quasi-topological term read
\begin{align}
{\cal E}_\mathcal{B}^{(\mathcal{S})} &= - \frac{16}{3} \Bigg[  \bigg(\frac{18n^4 + 37n^2 r^2 + 9r^4}{(n-r)^2(n+r)^2r} V^3  + \frac{19n^2 + 9 r^2}{(n-r)(n+r)} V^2 V' + \frac{n^2 + 9 r^2}{4 r} V (V')^2\bigg)V''
\nonumber\\
&- \frac{n^2 + 9 r^2}{16 r} (V')^4 + \frac{5n^2 - 3 r^2}{4(n-r)(n+r)} V (V')^3 + \frac{31n^4 + 98 n^2 r^2 + 9 r^4}{2(n-r)^2(n+r)^2r} V^2(V')^2   
\nonumber\\
&+ \frac{152n^4+143n^2r^2+9r^4}{(n-r)^3(n+r)^3}V^3V' +\frac{375n^6+1693n^4r^2+817n^2r^4+27r^6}{8 (n-r)^4(n+r)^4r} V^4 + \mathfrak{E}_\mathcal{B}^{(\mathcal{S})} \Bigg]  \, ,
\end{align}
where $\mathfrak{E}_\mathcal{B}$ is a base-dependent contribution, which takes the explicit form
\begin{align}
\mathfrak{E}_{\mathbb{CP}^2}^{(\mathcal{S})} &= \bigg(\frac{6(n^2+r^2)}{r(n-r)(n+r)} V + 3 V' + \frac{1}{2 r} \bigg) V V'' - (V')^3 + \bigg(\frac{n^2}{r(n-r)(n+r)}V - \frac{1}{4 r}\bigg)(V')^2 
\nonumber\\
&+ \bigg(\frac{2(14 n^2 + 3 r^2)}{(n-r)^2(n+r)^2} V^2 + \frac{V}{2(n-r)(n+r)} \bigg) V' + \frac{33n^4+86n^2r^2+9r^4}{2 (n-r)^3(n+r)^3r} V^3 
\nonumber\\
&+ \frac{3(n^2+r^2)}{2(n-r)^2(n+r)^2 r} V^2 \, ,
%%%%%%%%%%%%%
%%%%%%%%%%%%%
%%%%%%%%%%%%%
\\
\mathfrak{E}_{\mathbb{S}^2 \times \mathbb{S}^2}^{(\mathcal{S})} &= \bigg(\frac{6(n^2+r^2)}{(n-r)(n+r)r} V + 3 V' \bigg)V V'' - (V')^3 + \frac{n^2}{(n-r)(n+r)r} V (V')^2 
\nonumber\\	
&+ \frac{2(14 n^2 + 3 r^2)}{(n-r)^2(n+r)^2}V^2 V' 
+ \frac{33n^4+86n^2r^2+9r^4}{2 (n-r)^3(n+r)^3r} V^3 
- \frac{3(3n^2-r^2)}{4(n-r)^2(n+r)^2r} V^2 
\nonumber\\
&- \frac{V}{2(n-r)(n+r)r} - \frac{-r\log\left[\frac{r+n}{r-n} \right]+ 2 n}{16 n^3 r}\, ,
%%%%%%%%%%%%%
%%%%%%%%%%%%%
%%%%%%%%%%%%%
\\
\mathfrak{E}_{\mathbb{T}^2 \times \mathbb{T}^2}^{(\mathcal{S})} &=	0 \, .
\end{align}
On the other hand, the quartic Quasi-topological contributions yield
\begin{align}
{\cal E}_\mathcal{B}^{(\mathcal{Z})} =& \frac{2}{9} \bigg[ 40 \bigg(\frac{4 n^2 r}{(n-r)^2(n+r)^2} V^3 + 4 \frac{n^2}{(n+r)(n-r)} V' V^2 + \frac{n^2}{r} (V')^2 V  \bigg) V'' - \frac{10 n^2}{r} (V')^4 
\nonumber\\
&+ \frac{20 n^2}{(n-r)(n+r)} V (V')^3 + \frac{140n^2(n^2 + 2 r^2)}{r(n-r)^2(n+r)^2} V^2 (V')^2  + \frac{560n^2(n^2+r^2)}{(n-r)^3(n+r)^3} V^3 V'
\nonumber\\
&- \frac{(405 n^6 - 425 n^4 r^2 - 293 n^2 r^4 + 9 r^6) }{ r (n-r)^4(n+r)^4} V^4 + \mathfrak{E}_\mathcal{B}^{(\mathcal{Z})} \bigg]  \, ,
\end{align}
where now the base-dependent factors $\mathfrak{E}_\mathcal{B}^{(\mathcal{Z})}$ are
\begin{align}
\mathfrak{E}_{\mathbb{CP}^2}^{(\mathcal{Z})} =& \frac{10 n^2 }{r(n-r)(n+r)} V (V')^2 + \frac{40 n^2}{(n-r)^2(n+r)^2} V^2 V'  - \frac{4(45 n^4 + 8 n^2 r^2 + 3 r^4)}{r(n-r)^3(n+r)^3} V^3 
\nonumber\\
&- \frac{12(2n^2 + r^2)}{r(n-r)^2(n+r)^2} V^2 - \frac{2 V}{3r(n-r)(n+r)} \, ,
\\
\mathfrak{E}_{\mathbb{T}^2 \times \mathbb{S}^2}^{(\mathcal{Z})} &=	0  \, ,
\\
\mathfrak{E}_{\mathbb{T}^2 \times \mathbb{T}^2}^{(\mathcal{Z})} &=	0  \, .
\end{align}
 %are quite complicated and are presented in appendix~\ref{sec:app_feqs_6d}. 
%As already mentioned, ${\cal E}_\mathcal{B} $ in \req{eq6d} is the result of integrating once the equation for $V_{\mathcal{B}}(r)$ so, as before, we have
%\be \label{eq6D}
%{\cal E}_\mathcal{B} = - \frac{4 G M}{9 \pi}\, ,
%\ee
%where the integration constant is to be related to the ADM energy of the solutions below. 
It should be emphasized that while for static and spherically symmetric solutions the quartic Quasi-topological term yields algebraic field equations \cite{Dehghani:2011vu}, these become non-linear second-order differential equations for Taub-NUT metrics. This is an interesting difference with respect to the Gauss-Bonnet case, for which the equations determining the metric function are algebraic for both kinds of solutions.

\subsection*{Einstein gravity}
Just like in the four-dimensional case, in the following subsections we will be studying the different base spaces independently. As before, it is illuminating  to start with a quick study of the situation for Einstein gravity, for which the analysis can be performed at the same time for all base spaces. Indeed, if we set $\lambda_{\rm \ssc GB}=\xi=\zeta=0$, \req{eq6d} can be easily solved for $V_{\mathcal{B}}(r)$. Imposing the NUT condition $V_{\mathcal{B}}(r=n)=0$ first, one is left with
\begin{equation}\label{NUTE2}
V_{\mathcal{B}}(r)=\frac{(r-n)\left[6(r^3+3nr^2+n^2 r-5n^3)+(\kappa+3)(3n+r)L^2\right]}{6L^2(n+r)^2}\, ,
\end{equation}
 where we set the integration constant  
\begin{equation}\label{massE2}
%GM=6 \pi n^3 \left[\frac{12n^2}{L^2}+\kappa-1 \right]\, .
C_{\mathcal{B}}=-\frac{8n^3}{3}\left[\frac{12n^2}{L^2}+\kappa-1 \right]\, .
\end{equation}
The regularity condition \req{smooth} imposes
\begin{equation}\label{betaEin2}
\beta_{\tau}=\frac{24\pi n}{(1-\kappa)}\, .
\end{equation}
Hence, we find $\beta_{\tau}=12\pi n$ for $\mathcal{B}=\mathbb{CP}^2$ and $\mathcal{B}=\mathbb{S}^2\times\mathbb{S}^2$, $\beta_{\tau}=24\pi n$ for $\mathcal{B}=\mathbb{S}^2\times \mathbb{T}^2$, and $\beta_{\tau}=\infty$ for $\mathcal{B}=\mathbb{T}^2\times\mathbb{T}^2$, which forbids the existence of regular solutions with a compact $\mathbb{S}^1$ in that case. 
%So, as opposed to the four-dimensional case, base spaces involving factors of $\mathbb{T}^2$ can be used to construct regular solutions with finite periodicities for $\tau$. 

%which means that $\tau$ cannot be a compact coordinate for $\mathcal{B}=\mathbb{T}^2$ or, in other words, the solution is extremal, in the sense that the temperature $T\equiv 1/\beta_{\tau}$ is forced to vanish. Similarly, for $\mathcal{B}=\mathbb{H}^2$, one finds that the period of $\tau$ would need to be negative. Hence, no regular Taub-NUT solution exists in that case for Einstein gravity. This can also be seen from the fact that $V_{\mathbb{H}^2}(r)$ becomes negative for values of $r$ greater than $n$, which is forbidden by assumption.

If we impose the bolt condition $V_{\mathcal{B}}(r=r_b)=0$ instead, we find
\begin{align}\label{boltEi6d}
V_{\mathcal{B}}(r)=\frac{1}{6L^2(n^2-r^2)^2 r_b}&\left[6\left(r^6 r_b+15n^4r (r-r_b)r_b-r r_b^6-5n^6(r-r_b)-5n^2r r_b (r^3-r_b^3) \right) \right. \\ \notag & \left.-(\kappa-1)L^2 (r-r_b)\left(3n^4-6n^2 r r_b +r r_b (r^2+r r_b +r_b^2)\right)\right]\, ,
\end{align}
where in this case we related  $C_{\mathcal{B}}$  to $n$ and $r_b$ through
\begin{equation}
%GM=\frac{3\pi}{4r_b}\left[\frac{6(5n^6+15n^4r_b^2-5n^2r_b^4+r_b^6)}{L^2}+(\kappa-1)(3n^4+6n^2r_b^2-r_b^4)\right]\, .
C_{\mathcal{B}}=-\frac{1}{3r_b}\left[\frac{6(5n^6+15n^4r_b^2-5n^2r_b^4+r_b^6)}{L^2}+(\kappa-1)(3n^4+6n^2r_b^2-r_b^4)\right]\, .
\end{equation}
Finally, the regularity condition \req{smooth} produces the following relation between $r_b$, $n$ and the period of $\tau$,
\begin{equation}\label{rb6}
r_b=\frac{4L^2\pi}{10\beta_{\tau}}\left[1\pm \sqrt{1+\frac{5(\kappa-1)\beta_{\tau}^2}{8L^2\pi^2}+\frac{25n^2\beta_{\tau}^2}{4L^4\pi^2} }\right]\, .
\end{equation}
%\begin{equation}\label{boltEi}
%V_{\mathcal{B}}(r)=\frac{(r-r_b)\left[(6n^2 r r_b-3 n^4+k L^2 (n^2-r r_b)-r r_b (r^2+r r_b+ r_b^2)) \right]}{L^2(n^2-r^2)r_b}\, ,
%\end{equation}
%where the integration constant was fixed as
%\begin{equation}
%GM=\frac{k L^2(n^2+r_b^2)-3n^4-6n^2r_b^2+r_b^4}{2 L^2 r_b}\, .
%\end{equation}
%The regularity condition \req{smooth} fixes now the bolt radius as a function of $n$ and $\beta_{\tau}$, namely
%\begin{equation}\label{boltrb}
%r_b=\frac{2L^2 \pi}{3\beta_{\tau}} \left[ 1\pm \sqrt{ 1-\frac{3k \beta_{\tau}^2}{4L^2 \pi^2} +\frac{9n^2 \beta_{\tau}^2}{4L^4\pi^2}}\right]\, .
%\end{equation}
%In each case, one needs to choose the solution with $r_b>n$. Besides, the quantity inside the square root must me positive, which restricts the allowed values of $n$ for which the corresponding solution exists.
%\comment{more things here?}
Just like in $D=4$, we must require the quantity inside the square root to be positive and, of course, $r_b>0$, which in each case restricts the values of $n$ for which solutions exist.

 Besides the regularity condition \req{smooth}, additional constraints on $\beta_{\tau}$ arise both for NUTs and bolts when demanding the absence of Misner string singularities --- see \eg discussion in \cite{Mann:2005ra}. For example, for $\mathcal{B}=\mathbb{CP}^2$, we must demand $\beta_{\tau}=12\pi n$. Just like in $D=4$, for the Einstein gravity NUT this condition is automatically implemented by \req{smooth}. This is not the case in general, and the conditions must be imposed separately. 

\subsection{$\mathcal{B}=\mathbb{CP}^2$}
Let us now turn on again the  higher-curvature couplings in \req{eqn:quart_6d_act}. The first base space we consider is $\mathbb{CP}^2$. For this, we can write
\begin{align}
A_{\mathbb{CP}^2}&=6\sin^2 \xi_2 (d\psi_2+\sin^2\xi_1 d\psi_1)\, , \\
d\sigma_{\mathbb{CP}^2}^2&=6\left\{ d\xi_2^2+\sin^2\xi_2 \cos^2\xi_2(d\psi_2+\sin^2\xi_1 d\psi_1)^2 +\sin^2\xi_2 (d\xi_1^2+\sin^2\xi_1 \cos^2\xi_1 d\psi_1^2)\right\} \, ,
\end{align}
where the coordinate ranges\footnote{See, for example,~\cite{CPkbook} for a detailed discussion of $\mathbb{CP}^k$ in these octant coordinates.} are $0 \le \xi_{1,2} \le \pi/2$ and $0 \le \psi_{1,2} \le 2 \pi$. Now, we consider the metric asymptotically, making the rescalings $\tau \to 6  n\psi$ and $r \to r/\sqrt{6}$. This gives 
\be 
\frac{^{(5)}ds^2_{\rm bdry}}{r^2} = \frac{6 n^2 f_\infty}{L^2} \left(d\psi + \frac{A_{\mathbb{CP}^2}}{6} \right)^2 + \frac{1}{6} d\sigma^2_{\mathbb{CP}^2} 
\ee
at large $r$.
For the specific case of 
$
6 n^2 f_\infty/L^2 = 1 \, , 
$
this boundary metric is just that of a round $\mathbb{S}^5$ provided that the coordinate $\psi$ has period $2\pi$ to ensure regularity.  In all other cases, it is the metric of a squashed sphere \cite{Bobev:2017asb} and, in analogy with the $D=4$ case, it is customary to parametrize such squashing with the parameter $\alpha$, defined in terms of $n$ through
$
6 n^2 f_\infty/L^2= 1/(1 + \alpha)\, .
$

We begin our study of Taub-NUT/bolt solutions with this base space by considering the asymptotic behaviour of the metric. 
%For these considerations, it is convenient to introduce the polynomial,
%\be 
%h(f_\infty) = 1 - f_\infty + \lambda_{\rm \ssc GB} f_\infty^2 + \zeta f_\infty^4 + \xi f_\infty^4 \, ,
%\ee
%whose roots determine the vacua of the theory. 
%
%
The asymptotic solution for $V_{\mathbb{CP}^2}(r)$ consists of a particular and homogeneous solution. The particular solution is found by  expanding $V_{\mathbb{CP}^2}(r)$ in a $1/r$ series and solving the field equations to determine the constants order by order. The result is
\begin{align}\label{CP_asymp} 
V_p(r) %=%& f_\infty \frac{r^2}{L^2} + b_0 + \frac{b_2}{r^2} + \frac{b_3}{r^3} + {\cal O}(r^{-4})
%\nonumber\\
=&f_\infty \frac{r^2}{L^2} + \frac{1}{3} - \frac{3 \fin n^2}{L^2} - \frac{6 L^2 + 12 f_\infty n^2 - 6 L^2 f_\infty + 5 L^2 f_\infty^2 \lambda_{\rm \ssc GB} - 6 f_\infty^3 n^2 \lambda_{\rm \ssc GB}}{9 r^2 f_\infty (2 f_\infty^2 \lambda_{\rm \ssc GB} - 3 f_\infty + 4)} \left(1-\frac{6f_\infty n^2}{L^2} \right)	
\nonumber\\
& - \frac{C_{\mathbb{CP}^2}}{2 h'(f_\infty) r^3} + {\cal O} (r^{-4}) \, ,
\end{align}
where $h'(f_\infty)$ denotes the derivative of $h(f_\infty)$ --- see \req{hfinf} --- with respect to $f_\infty$. To obtain the form of the homogeneous equation, we again write $V(r) = V_p(r) + g(r)$, and work to linear order in $g(r)$. While both $\mathcal{S}$ and $\mathcal{Z}$ contribute in the same way to the particular solution, the contributions differ in the homogeneous equation.  The resulting equation, in the limit of large $r$, takes the form:
\be 
a(r) g''(r) + b(r) g'(r) + c(r) g(r) = 0\, ,
\ee 
where $a(r)$ and $b(r)$ are the leading terms in this expansion, taking the explicit forms
\begin{align}
a(r) =& \frac{8 f_\infty \xi L^2 r}{3 } \left(1-\frac{6f_\infty n^2}{L^2} \right)^2 + (1-\hat\xi)\frac{320 f_\infty \zeta n^2 L^4}{81 r} \left(1-\frac{6f_\infty n^2}{L^2} \right)^2 \, , 
\nonumber\\
b(r) =& -8 f_\infty \xi L^4 \left(1-\frac{6f_\infty n^2}{L^2} \right)^2 - (1-\hat\xi) \frac{1520 f_\infty \zeta n^2 L^4}{81 r^2} \left(1-\frac{6f_\infty n^2}{L^2} \right)^2 \, ,
\nonumber\\
c(r) = & - 2 r^3 h'(f_\infty) \, ,	
\end{align}
and we have defined 
\be
\hat\xi = \begin{cases}
	1 &\text{for } \xi \neq 0\, ,
	\\
	0 &\text{for } \xi = 0  \, ,
\end{cases}
\ee
to simplify the presentation of the terms above. We recognize that the contributions in parenthesis in $a(r)$ and $b(r)$ are directly related to the squashing parameter and vanish when the base is a round sphere; in that case, the solution reduces to just pure AdS. 

In the limit of large $r$, the homogeneous equation can be solved  in terms of special functions. First, when $\xi \neq 0$ the homogeneous solution reads
%\be 
%g(r) = C_1 r^{2 + 7(1-\hat\xi)/8} I_{1} \left(\sqrt{\frac{c(r)}{a(r)}} \right)  + C_2 r^{2 + 7(1-\hat\xi )/8}  K_{1} \left(\sqrt{\frac{c(r)}{a(r)}} \right) 
%\ee
\be 
g(r) = C_1 r^2 I_{1} \left(r\sqrt{-\frac{c(r)}{4 a(r)}} \right)  + C_2 r^{2}  K_{1} \left(r\sqrt{-\frac{c(r)}{4 a(r)}} \right)\, ,
\ee
while if $\xi = 0$ it takes the form:
\be 
g(r) = C_1 r^{23/8}  I_{\frac{23}{24}} \left(r \sqrt{-\frac{c(r)}{9 a(r)}} \right)  + C_2 r^{23/8}  K_{\frac{23}{24}} \left(r\sqrt{-\frac{c(r)}{9 a(r)}} \right)\, ,
\ee
where $I_\nu(x)$ and $K_\nu(x)$ are the modified Bessel functions of the first and second kinds, respectively. The explicit form of these solutions is not as important as what their asymptotic behaviour tells us: because $c(r) > 0$ by virtue of demanding the graviton is not a ghost, when $a(r) < 0$, in each case the homogeneous solution consists of a super-exponentially growing and super-exponentially decaying part. When $a(r) > 0$, both of the above  solutions oscillate more and more rapidly near infinity and are ultimately pathological. Therefore, to ensure that  the solutions are physically reasonable, we must demand that $a(r) < 0$,  while also requiring $h'(\fin) < 0$. The general solutions to these constraints with $\lambda_{\rm \ssc GB} \neq 0$ are a bit messy, and so we quote the result explicitly only in the case $\lambda_{\rm \ssc GB} = 0$. In that case it is a straight-forward matter to show that these conditions are satisfied --- independent of $n$ --- provided that
\begin{align} 
\label{eqn:good_vac_6dCP}
\xi &< {\rm min} \left\{0, \frac{27}{256} - \zeta \right\} \quad \text{if $\xi \neq 0$ or} \, ,
\nonumber\\
\zeta &< 0 \quad \text{ if $\xi = 0$ } .
\end{align}
Interestingly, in contrast to the four dimensional case, here the mass parameter $M$ does not enter into the constraints, with the result that there is no pathology associated with the negative mass solutions (see below). Furthermore, note that for $\zeta$ non-zero, simply demanding $\xi < 0$ is not enough since one must also require that $h'(\fin) < 0$ --- this is the origin of the more complicated constraint in that case. It can be shown that, for $\lambda_{\rm \ssc GB} = 0$, $\xi = 27/256 - \zeta$ corresponds to the critical limit of the theory, which has $f_\infty = 4/3$. 

A consequence of these bounds on the coupling is that, when one considers a theory that contains only a single one of the quartic terms, then it is not possible to reach the critical limit of the theory at physical coupling. This situation is similar to what happens to cubic GQG for spherically symmetric black hole solutions in $D \ge 6$.

\subsubsection{Taub-NUT solutions}

We now consider NUT solutions where $V_{\mathbb{CP}^2}(r=n) = 0$. Further restrictions on $V_{\mathbb{CP}^2}(r)$ arise due to regularity of the metric. Recall from the discussion above that the boundary is a squashed $\mathbb{S}^5$. Regularity of this boundary metric requires that $\psi := \tau/(6n)$ has period $2 \pi$, which in turn means $\tau \sim \tau + 12 \pi n$. A further constraint is imposed on the derivative of $V_{\mathbb{CP}^2}(r)$ near the NUT where the absence of conical singularities at a zero of $V_{\mathbb{CP}^2}$ requires that $\tau$ is periodic with period $\beta_\tau$ given by $\beta_\tau = 4 \pi /V_{\mathbb{CP}^2}'(r=n)$. Consistency of these two regularity conditions fixes $\beta_\tau = 12 \pi n$ and so we therefore have the following series expansion near the NUT:
\be 
V(r) =  \frac{(r-n)}{3n} + \sum_i^\infty (r-n)^i a_i  \, .
\ee 
Substituting this expression into the field equations, and expanding in $(r-n)$, we find
\begin{align}
\frac{16}{3} \frac{n^3(L^2 - 6 n^2)}{L^2} - \frac{8  n L^2 \lambda_{\rm \ssc GB}}{9}  - \frac{2 L^6 \left(\xi + \zeta \right)}{81 n^3} + \frac{4 G M }{9 \pi } + {\cal O} \left((r-n)^3 \right) = 0 \, ,
\end{align}
where we have conveniently redefined the integration constant $C_{\mathbb{CP}^2}=-4GM/(9\pi)$, where $M$ will correspond to the ADM mass of the solution.
The first condition (shown above explicitly) determines $M$ in terms of the couplings and the NUT charge, and the next two relations are automatically satisfied. The next non-trivial relation is linear in $a_3$, allowing one to solve for $a_3$ as a function of the free parameter $a_2$. This trend continues to higher order in the field equations, and thus there is a single free parameter that is left unfixed by the field equations and regularity conditions. This is  fully analogous to the $D=4$ case.

\begin{figure}[t!]
	\centering
	\includegraphics[width=0.47\textwidth]{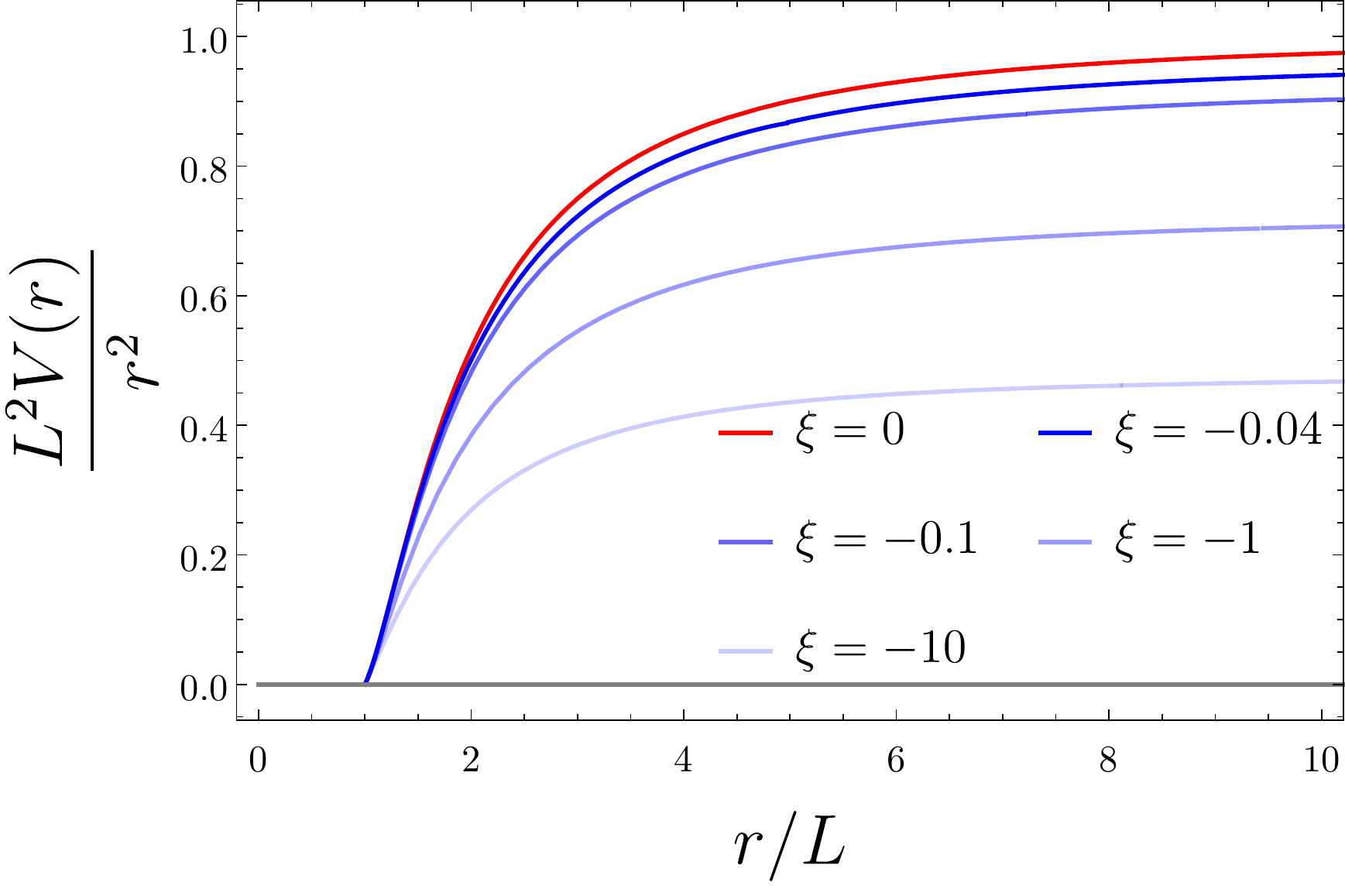}
	\includegraphics[width=0.47\textwidth]{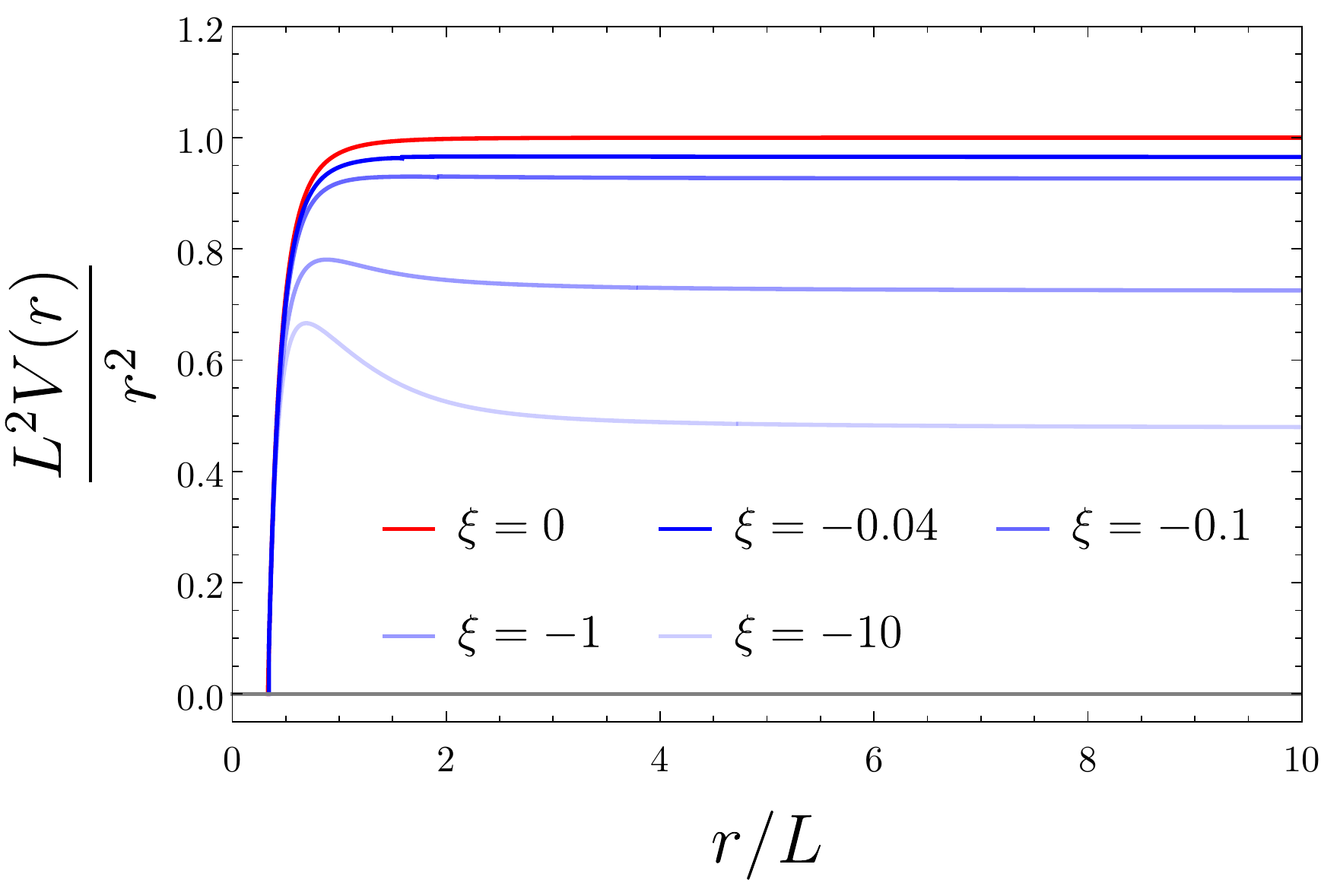}
	\includegraphics[width=0.47\textwidth]{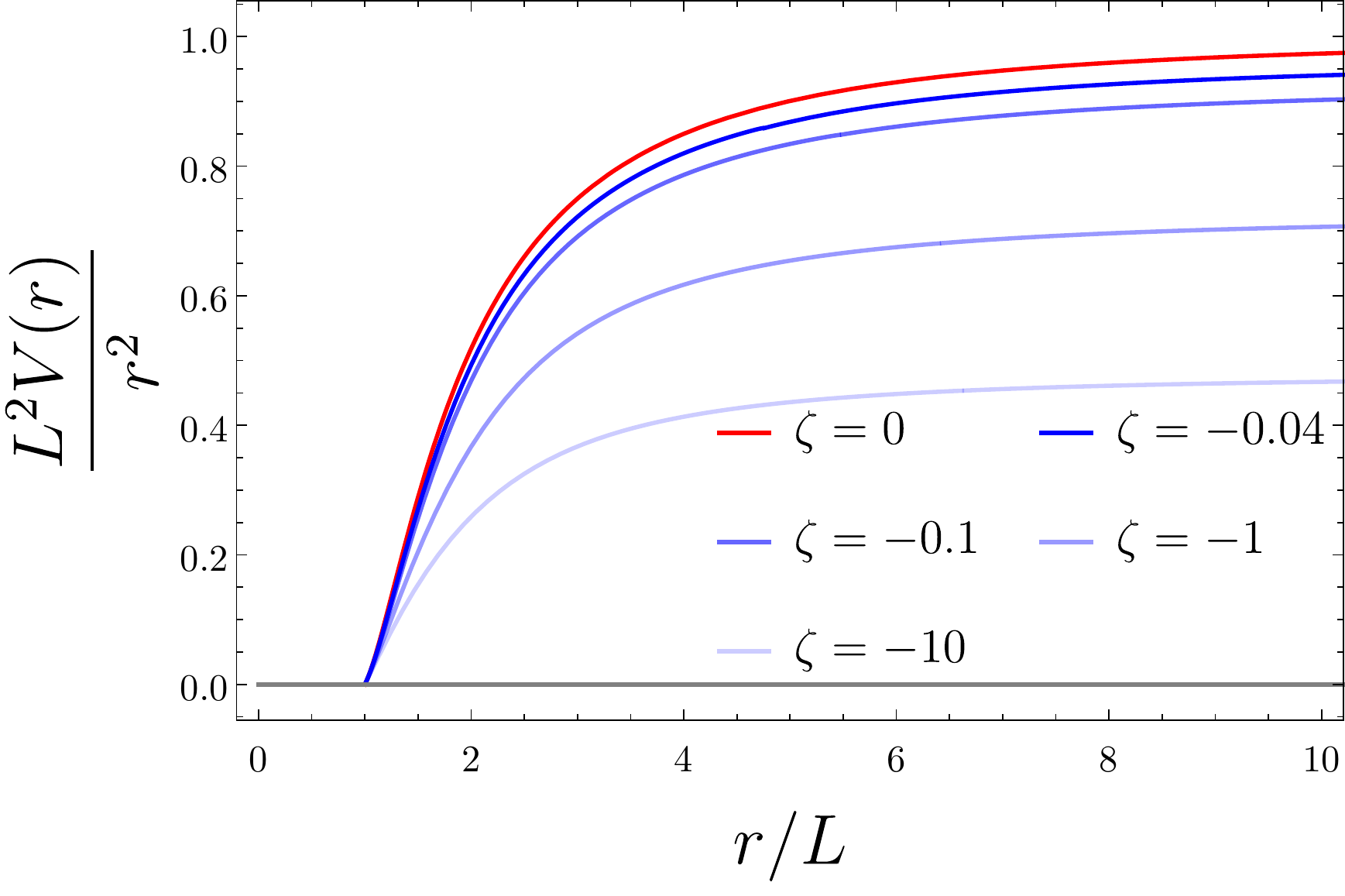}
	\includegraphics[width=0.47\textwidth]{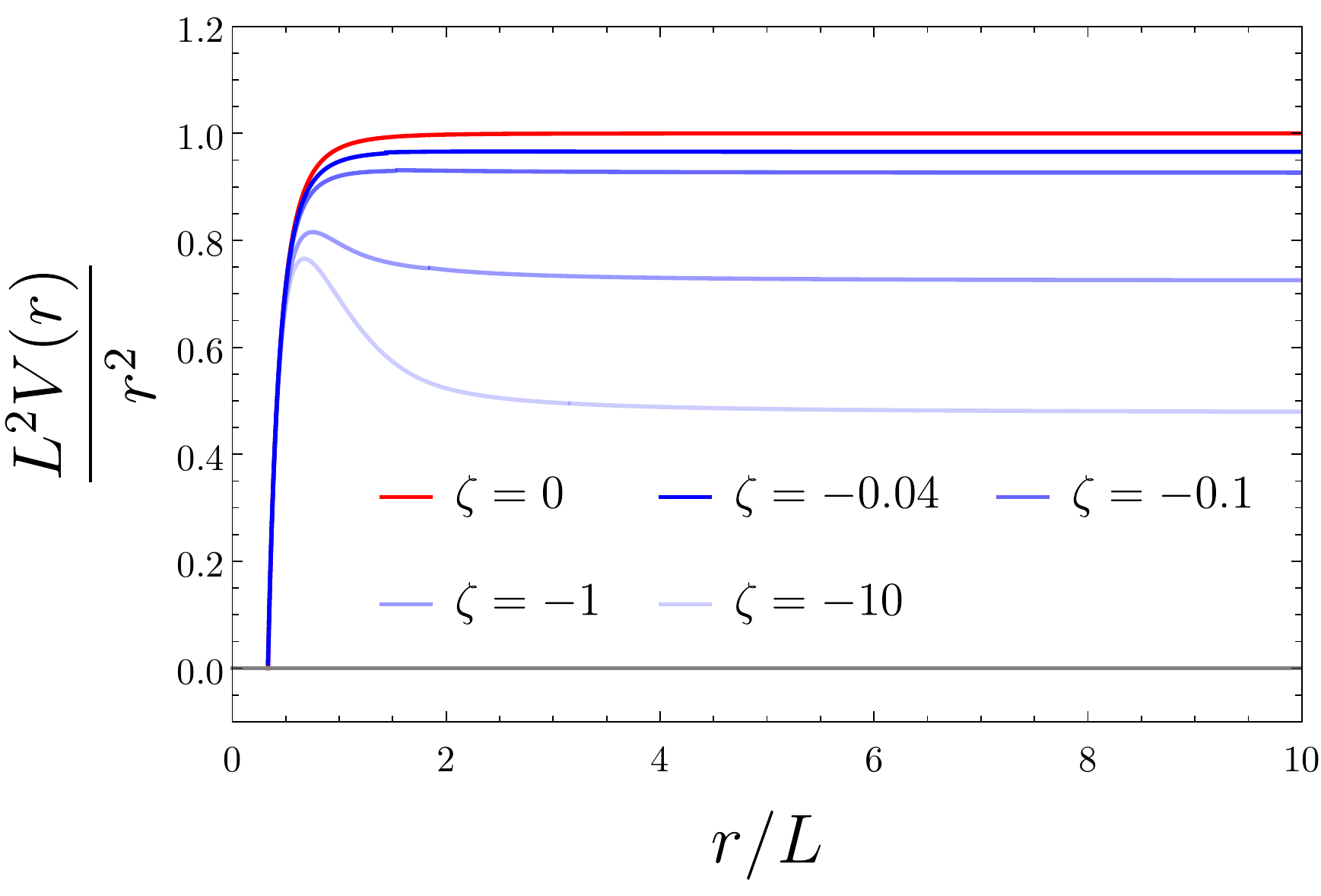}
	\caption{The metric function $L^2V_{\mathbb{CP}^2}(r)/r^2$ is plotted for NUT solutions of the quartic theories.  The top row depicts solutions of the quartic generalized quasi-topological theory with $n/L= 1$ (left) and $n/L = 1/3$ (right). The solutions with $n/L = 1$ all have positive mass, while those with $n/L = 1/3$ have negative mass. The bottom row depicts solutions of the quartic Quasi-topological theory  with $n/L= 1$ (left) and $n/L = 1/3$ (right). The solutions with $n/L = 1$ all have positive mass, while those with $n/L = 1/3$ have negative mass.} 
	\label{CP2-numeric_NUT}
\end{figure}

The near horizon solution can be joined to the asymptotic solution that was presented above by numerically integrating the field equations. The near horizon expansion is used as initial data, with the shooting method employed to determine the free parameter $a_2$. A careful choice of this parameter is required to ensure the growing modes present in the asymptotic solution are not excited. Ensuring this, we find a unique $a_2$ for which the solution can be integrated, with the result for several values of the coupling shown in Fig.~\ref{CP2-numeric_NUT}. For comparison, the Einstein gravity solution is shown in red and we see that the solutions to the higher curvature theories are qualitatively the same with the main difference being that they approach $\fin r^2/L^2$ with $\fin$ depending on the value of the couplings.  We also note that, while the top left and bottom left plots depict solutions with positive mass, the top right and bottom right plots depict solutions with negative mass. The fact that the negative mass solutions can be constructed and possess no inherent pathology is in contrast with the four dimensional case, where the negative mass solutions possessed pathological asymptotic structure.

Let us now turn to the free energy of the NUTs and compute the regularized on-shell action for these solutions. %The bulk contribution takes the form,
With minor modifications, the prescription \req{assd} introduced in \cite{ECGholo} can be used to eliminate the divergent terms in the on-shell action. The Euclidean action, completed with the generalized boundary term and counterterms is given by
\begin{align}\label{full6}
I_E &= - \int \frac{d^6 x \sqrt{g}}{16 \pi G} \left[ \frac{20}{L^2} + R + \frac{\lambda_{\rm \ssc GB} L^2 }{6} {\cal X}_4 -\frac{ \xi L^6 }{216} \mathcal{S} - \frac{ \zeta L ^6}{144} \mathcal{Z} \right] 
	\nonumber\\
&- \frac{1 - 4 \lambda_{\rm \ssc GB} \fin  + 8 (\xi + \zeta) \fin^3}{8 \pi G} \int d^5 x \sqrt{h} \left[K  - \frac{4 \sqrt{\fin}}{L} - \frac{L}{6 \sqrt{\fin}} {\cal R} - \frac{L^3}{18 \fin^{3/2}} \left( {\cal R}_{ab}{\cal R}^{ab} - \frac{5}{16} {\cal R}^2 \right)\right]
	\nonumber\\
&+ \frac{ \lambda_{\rm \ssc GB} \fin - 6 (\xi + \zeta) \fin^3}{8 \pi G} \frac{L^3}{18 \fin^{3/2}} \int d^5 x \sqrt{h} \left( 4 {\cal R}_{ij}{\cal R}^{ij} - \frac{5}{4} {\cal R}^2 + \frac{3}{2} {\cal X}^{(h)}_4 \right) \, .	
\end{align}
The evaluation is facilitated via the asymptotic expansion presented above and the expansions near $r = n$ in the NUT case or $r = r_{b}$ for the bolts. Near the boundary, the bulk action has several divergent components that are precisely canceled by the generalized boundary and counterterms. Note the addition of a new counterterm, nonproportional to $a^*$, on the last line above. This appears because, strictly speaking, the spacetime is not asymptotically AdS --- the boundary is not maximally symmetric except for the choice of NUT parameter that yields the undeformed five sphere. The additional counterterm was chosen since it vanishes identically when the boundary is maximally symmetric (and so could be dropped in those cases) but allows for the cancellation of the linear divergence in the case of $\mathcal{B} = \mathbb{CP}^2$ considered here.

Just like  the four-dimensional case discussed in detail in Section~\ref{sec:TNECG}, eliminating the divergent terms also removes all possible constant terms coming from boundary contributions, leaving us with the bulk action evaluated at $r=n$, and nothing else. The final result is
%the boundary makes no contribution to the on-shell action. We then have,
\be
\label{eqn:CP_nut_IE} 
I_E = \frac{36\pi^2}{ G}\left[n^4\left(\frac{4n^2}{L^2}-1\right) + \frac{L^2 n^2 \lambda_{\rm \ssc GB}}{3} -\frac{L^6(\xi + \zeta)}{108n^2} \right]  \, ,
\ee
from which the total energy and entropy can be found to be,
\begin{align} 
E &= \frac{12\pi}{G} \left[n^3\left(\frac{6n^2}{L^2}-1\right) + \frac{ n L^2 \lambda_{\rm \ssc GB}}{6} + \frac{L^6(\xi+\zeta)}{216 n^3} \right] = M \, , 
\nonumber\\
S &= \frac{36\pi^2}{G} \left[ n^4\left(\frac{20n^2}{L^2}-3\right)+  \frac{ n^2 L^2 \lambda_{\rm \ssc GB} }{3}+ \frac{L^6(\xi+\zeta)}{36n^2} \right]  \, ,
\end{align}
and the first law $dE = TdS$ is verified to hold.  

 Similar to the discussion for the $\mathbb{S}^2$ base in the case of ECG, here we can also enlarge the thermodynamic phase space and construct the extended first law. The expression is slightly more complicated reading
\begin{equation}\label{eflawCP}
dE = TdS + VdP + \Upsilon^{\ssc \rm GB} d(L^2\lambda_{\ssc \rm GB}) + \Upsilon^{\cal S} d (L^6\xi) + \Upsilon^{\cal Z} d(L^6\zeta) \, ,
\end{equation}
where we have again restored the dimensionality to the coupling constants. The potentials appearing in the extended first law read
\begin{equation}
V = \frac{48 \pi^3}{5} n^5 \, , \quad \Upsilon^{\ssc \rm GB} = \frac{1}{n G} \, ,\quad \Upsilon^{\cal S} = \Upsilon^{\cal Z} = -\frac{\pi}{36 G n^3} \, .
\end{equation}
The expression above for the thermodynamic volume holds also in Einstein gravity though there appear to be no previous computations of this quantity in the literature for higher dimensional Taub-NUT solutions. It is noteworthy that the thermodynamic volume here is positive, while the thermodynamic volume is negative in the $D = 4$ case. Of course, we find that the Smarr relation consistent with scaling is satisfied by the thermodynamic quantities defined above:
\begin{equation}
\label{smarrCP}
3 E = 4 TS - 2 PV + 2 \Upsilon^{\ssc \rm GB} L^2 \lambda_{\ssc \rm GB} + 6 \Upsilon^{\cal Z} (L^6 \zeta) + 6 \Upsilon^{\cal S} (L^6 \xi)   \, .
\end{equation}

Note that if we turn off the quartic couplings, then the result for the free energy reduces to that previously calculated for Einstein gravity and Gauss-Bonnet gravity in~\cite{Clarkson:2002uj, KhodamMohammadi:2008fh} up to an overall factor of $8/9$, the same discrepancy noted in~\cite{Bobev:2017asb}. We have carefully revisited the calculations in~\cite{Clarkson:2002uj, KhodamMohammadi:2008fh} and have traced the discrepancy to the ratio of volumes of  $\mathbb{S}^2 \times \mathbb{S}^2$ to $\mathbb{CP}^2$. In~\cite{Clarkson:2002uj} it is claimed that the thermodynamic quantities for both base spaces are identical. However, we have found this to be true only up to an overall ratio of the volumes of the base spaces. For $\mathbb{CP}^2$ normalized so that $R_{ab} = g_{ab}$ the volume is ${\rm Vol}\left(\mathbb{CP}^2 \right) = 18 \pi^2$, while the volume of $\mathbb{S}^2 \times \mathbb{S}^2$ is given by ${\rm Vol} \left(\mathbb{S}^2 \times \mathbb{S}^2 \right) = (4 \pi )^2$.  The ratio of these volumes is precisely $8/9$, which accounts for the observed discrepancy.\footnote{In~\cite{Clarkson:2002uj} a different set of coordinates is used, but the metric of $\mathbb{CP}^2$ is still normalized so that $R_{ab} = g_{ab}$. Using the fact that the coordinates in~\cite{Clarkson:2002uj} have ranges $0 \le u \le \infty$, $0 \le \theta \le \pi$, $0 \le \phi \le 2 \pi$ and $0 \le \psi \le 4 \pi$ we obtain the result presented here with the correct overall factor. In higher dimensions, ${\rm Vol}\left(\mathbb{CP}^k \right) = 2^k(k+1)^k \pi^k/k!$ and the volume of $k$ 2-spheres is ${\rm Vol} \left(\mathbb{S}^2 \times \cdots \times \mathbb{S}^2 \right) = (4 \pi )^k$. In higher dimensions, we find the ratio between thermodynamic quantities for the two bases is $2^k k!/(k+1)^k$.}

\begin{figure}[t!]
	\centering 
	\includegraphics[width=0.65\textwidth]{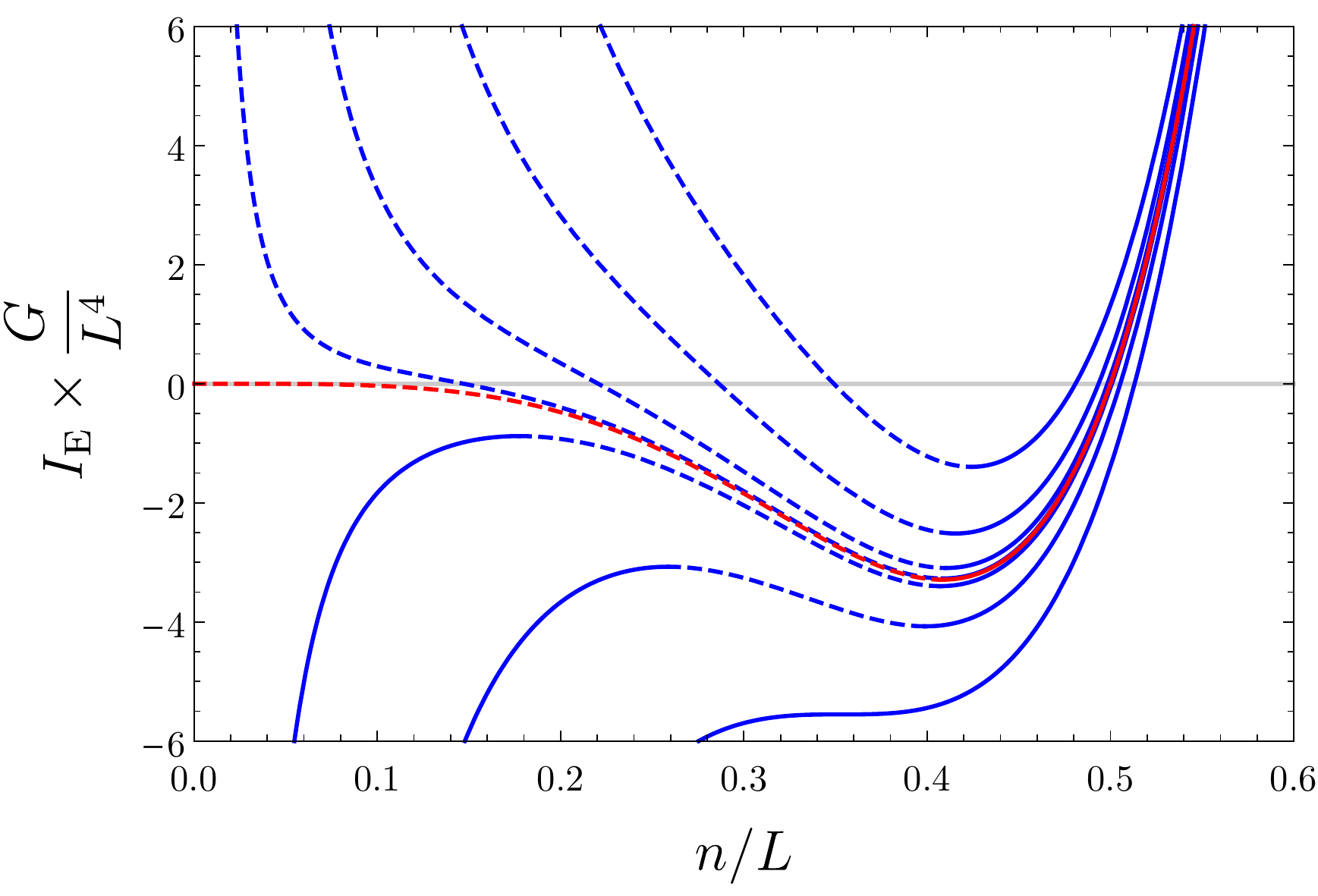}
	\caption{Euclidean on-shell action for $\mathcal{B} = \mathbb{CP}^2$ NUT solutions. The red curve corresponds to the Einstein gravity result, while the blue curves correspond to $\xi + \zeta = 27/256, 27/256 - 1/10, 27/256 - 10/95,  - 10^{-3}, - 10^{-2}, - 4  \times 10^{-2}$ and $-10^{-1}$ (bottom to top through a vertical slice). The dashed portions of the curves indicate solutions with negative mass, though there is no pathology associated with these solutions in this case.} 
	\label{CPNUTIE}
\end{figure}

In Fig.~\ref{CPNUTIE} we show plots of the Euclidean on-shell action for the NUT solutions with $\lambda_{\rm \ssc GB} = 0$. As is clear from Eq.~\eqref{eqn:CP_nut_IE}, this depends on the higher curvature couplings only through the combination $\xi + \zeta$. %, as we show several such possibilities. 
 In each case, there is only a single branch and, from the figure, we see that its qualitative structure depends on whether $\xi + \zeta$ is positive or negative. For consistency with the plots presented earlier in the document, we have indicated regions of negative mass with dashed curves. However, unlike the four dimensional case, there is no pathology associated with the negative mass solutions for the quartic theories in six dimensions. The region of negative mass solutions shrinks and eventually vanishes as $\xi + \zeta \to 27/256$, which corresponds to the critical limit of the theory.

\subsubsection{Taub-bolt solutions}

We now consider Taub-bolt solutions which satisfy $V_{\mathbb{CP}^2}(r_{b}) = 0$ for $r_{b} > n$.  In this section, we turn off the Gauss-Bonnet coupling to limit the size of the parameter space. Regularity demands that $V_{\mathbb{CP}^2}'(r_{b}) = 1/(3n)$, and therefore we write the near horizon expansion as
\be 
V(r) = \frac{(r-r_{b})}{3n} + \sum_{i=2}^\infty(r-r_{b})^i  a_i  \, .
\ee
Substituting this expansion into the field equations and solving order by order in $(r-r_{b})$, we find the first two relations fix the integration constant $C_{\mathbb{CP}^2}$ and the relationship between $r_{b}$ and $n$:
\begin{align}
\label{eqn:bolt_nh_cp2}
0 =&  \frac{4 G M}{9 \pi} + \frac{\xi L^6 ( 9 r_{b}^2 + 48 r_{b} n + 37 n^2)}{243n^4 r_{b}} - \frac{20 \zeta L^6}{729 n^2 r_{b}} 
\nonumber\\
&- \frac{2\left( L^2 r_{b}^4 - 6 L^2 r_{b}^2 n^2 - 3 L^2 n^4 + 3 r_{b}^6 - 15 r_{b}^4 n^2 + 45 r_{b}^2 n^4 + 15 n^6 \right)}{3 L^2 r_{b}} \, , 
\\ \label{eqn:bolt_nh_cp22}
0 =& \frac{2 (r_{b}^2 - n^2)^2(L^2 r_{b} - 3 L^2 n - 15 r_{b}^2 n + 15 n^3)}{3 n L^2 r_{b}^2} - \frac{\xi L^6 (3 r_{b}^4 - 46 r_{b}^2 n^2 - 48 r_{b} n^3 - 37 n^4)}{243 n^4r_{b}^2(r_{b}^2 - n^2)} 
\nonumber\\
&- \frac{20 \zeta L^6\left(r_{b}^2 + \frac{6}{5} n r_{b} + n^2 \right)}{729 n^2 r_{b}^2 (r_{b}^2 - n^2)} \, ,	
\end{align} 
where, just as in the NUT case, we have set $C_{\mathbb{CP}^2}=-4GM/(9\pi)$.

Let us now examine the second relation above in more detail. When the higher curvature terms are turned off, the bolt radius is given by \req{rb6} with $\kappa=-1$, \ie
\be 
r_{b}(\zeta, \xi=0) = \frac{L^2}{30 n} \left[1 \pm \sqrt{1 - 180 \frac{ n^2}{L^2} + 900 \frac{n^4}{L^4}} \right] \, .
\ee
Since $r_{b}$ must be real and larger than $n$, we must then have $n < \sqrt{15} (2-\sqrt{2})L/30$. As in the four-dimensional case, there is a maximum value of $n$ for bolts in Einstein gravity. In particular, this means that there does not exist a bolt solution near the undeformed five sphere, for which $n= L/\sqrt{6}$. Of course, the behaviour is different with higher curvature corrections, but there are some notable differences from what was observed in the four dimensional case. 

Depending on the relative size of $\xi$ and $\zeta$, the behaviour of $r_{b}$ as a function of $n$ can either resemble that of Einstein gravity (namely, there is a largest value of $n$ for which a bolt exists) or resemble that observed in the four dimensional cubic case discussed earlier in this chapter (bolts exist for arbitrarily large $n$). In fact, this classification is completely determined by the sign of the quantity $\zeta - 6\xi$. If this quantity is positive, there exists a maximum value of $n$; if this quantity is negative, bolts exist for arbitrarily large $n$.\footnote{This can be deduced in the following way. Take the numerator of Eq. \eqref{eqn:bolt_nh_cp22} and set $r_{b} = n + x$. Next, apply Descartes' rule of signs, treating $x$ as the independent variable, and notice that in the limit of large $n$ there will be a single sign flip provided that $\zeta - 6\xi < 0$. This guarantees a single positive root for $x$, which in turn guarantees the existence of a bolt with $r_{b} > n$. }

%\tcp{To classify the structure for intermediate values of $n$ is more difficult since the polynomial is of high order. We find (using the rule of signs) that for a given value of $n$, there will be either one or three values of $r_{b}$ provided that $\zeta - 6\xi < 0$ and $n < L/\sqrt{90}$. Whether there is precisely one or three will depend in a complicated manner on the particular values of $n$ and the couplings in this reduced parameter space.  For $n > L/\sqrt{90}$, there is a single bolt for a given $n$.  }

\begin{figure}[t!]
	\centering
	\includegraphics[width=0.65\textwidth]{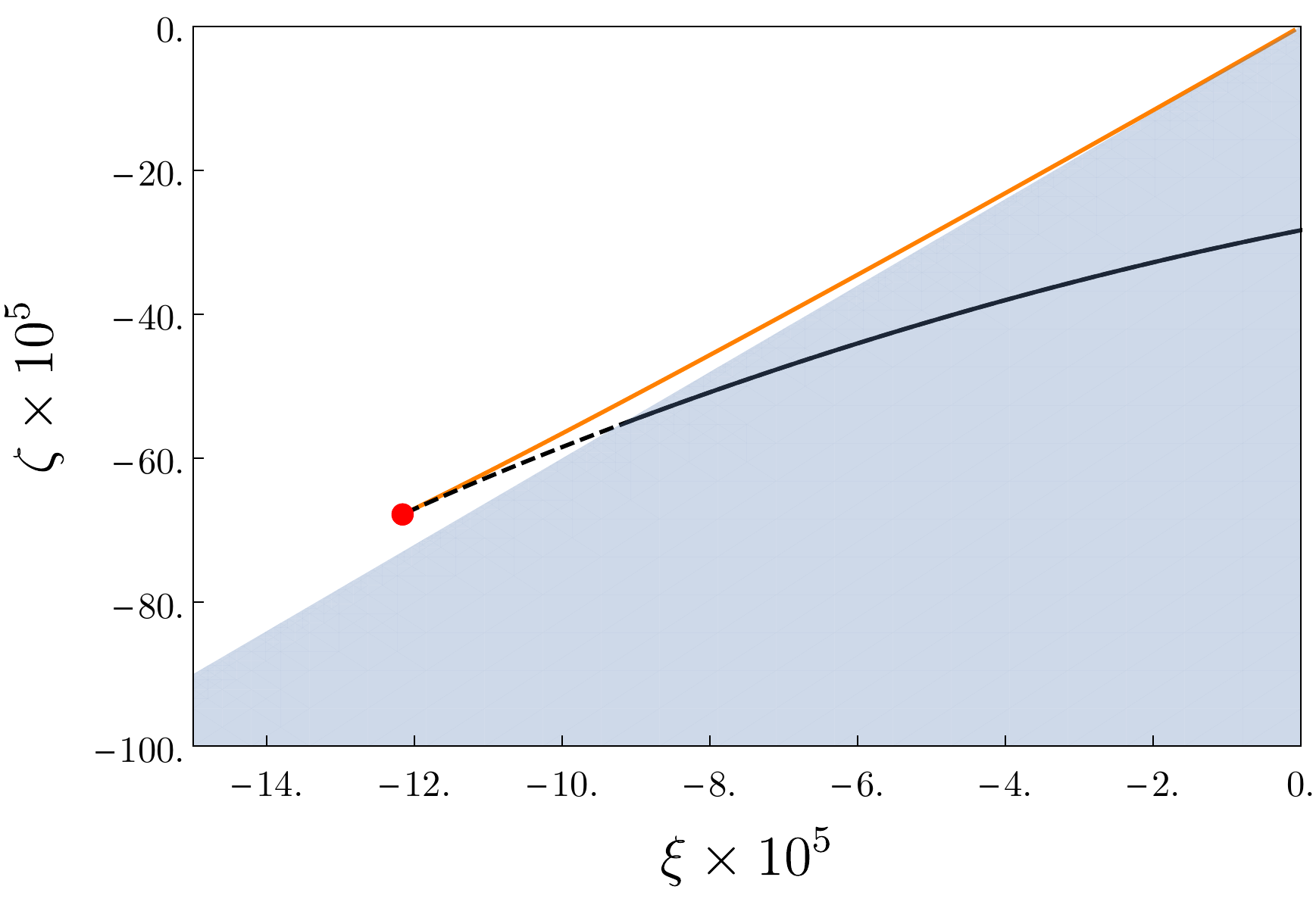}
	\caption{A breakdown of the coupling parameter space into useful regions for bolt solutions. Here the orange and black curves denote lines of `critical points', \ie for a given NUT charge, three solutions for $r_{b}$ coalesce. The red dot represents the single point in the physical parameter space where there is a coalescence of four roots. Within the blue shaded region, $\zeta < 6 \xi$ and there are bolts for arbitrarily large $n$. In the complement, $\zeta > 6 \xi$ and there is a largest value of $n$ for which bolts exist.  On the black locus of critical points, the critical point is always physical (i.e. of lowest free energy) within the blue region, otherwise (for the dashed portion of the curve) the situation can depend on which branch of the cusp minimizes the free energy, and also on whether or not there are re-entrant phase transitions as described in the text.  }
	\label{fig:bolt_params_CP2}
\end{figure}

From the perspective of the phase structure of the bolts, the most interesting scenario occurs when there are three values of $r_{b}$ for a particular $n$ --- one would expect these cases could yield swallowtail type behaviour and critical phenomena. We can constrain the regions of parameter space where three bolts exist by searching for `critical points'. More specifically, such a critical point would occur when $\partial n/\partial r_{b} = \partial^2 n /\partial r_{b}^2 = 0$, while respecting  Eq. \eqref{eqn:bolt_nh_cp22}. These points will mark transitions in the maximum number of bolts for given couplings. We were unable to solve the resulting constraints analytically, but it is straightforward to do so numerically. This results in the breakdown of parameter space shown in Fig.~\ref{fig:bolt_params_CP2}.  

\begin{figure}[t!]
	\centering
	\includegraphics[width=0.47\textwidth]{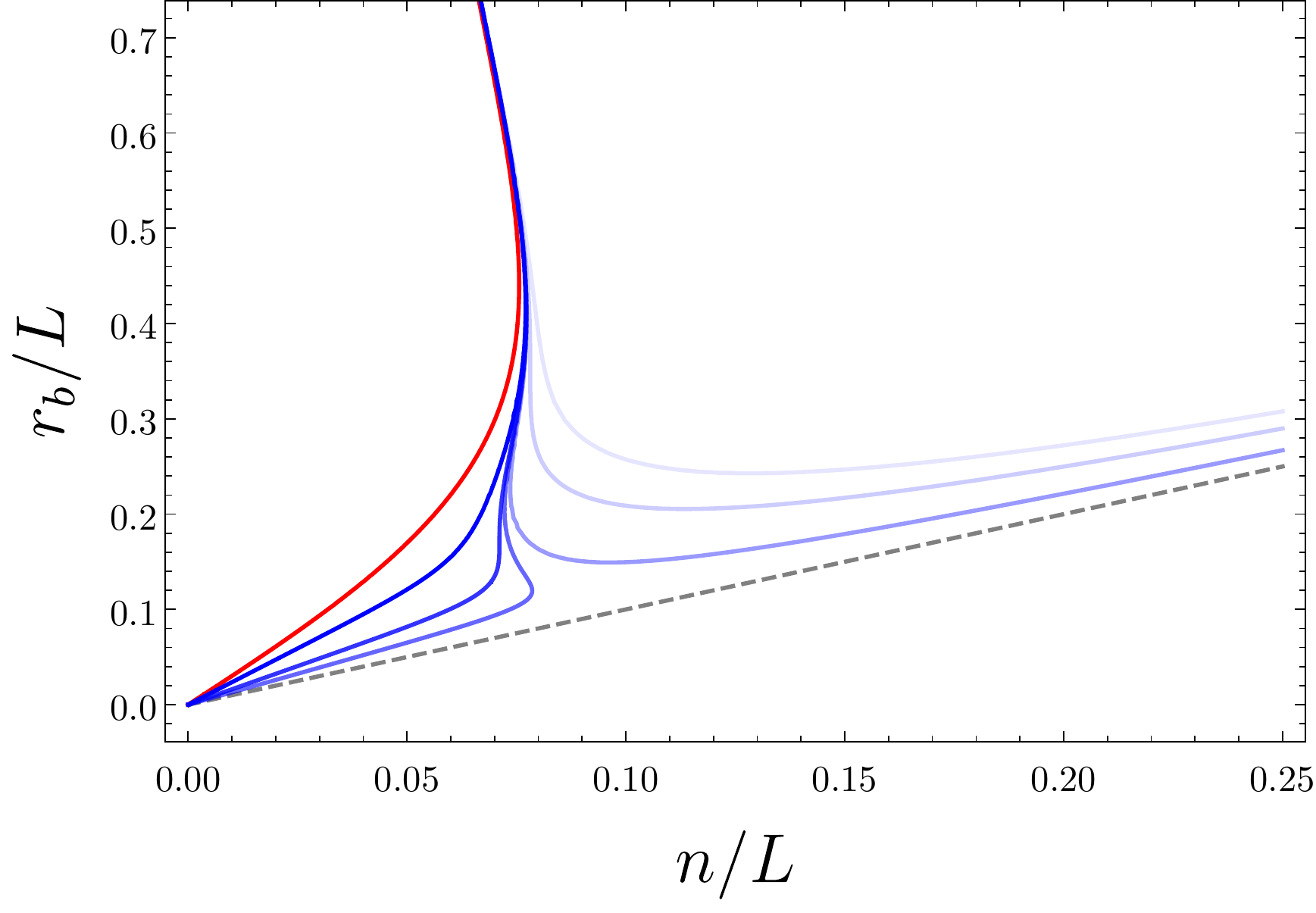}
	\includegraphics[width=0.47\textwidth]{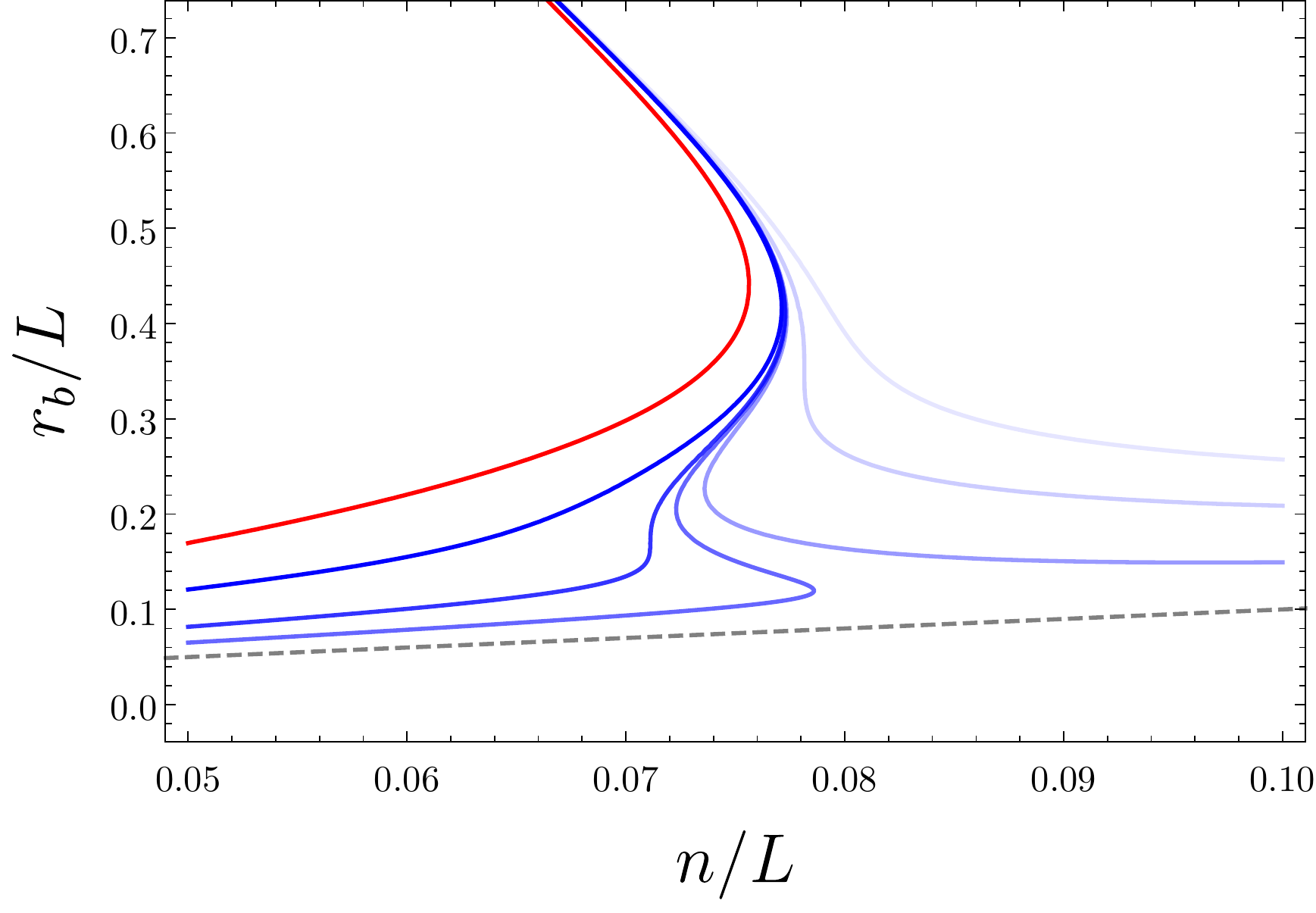}
	\includegraphics[width=0.47\textwidth]{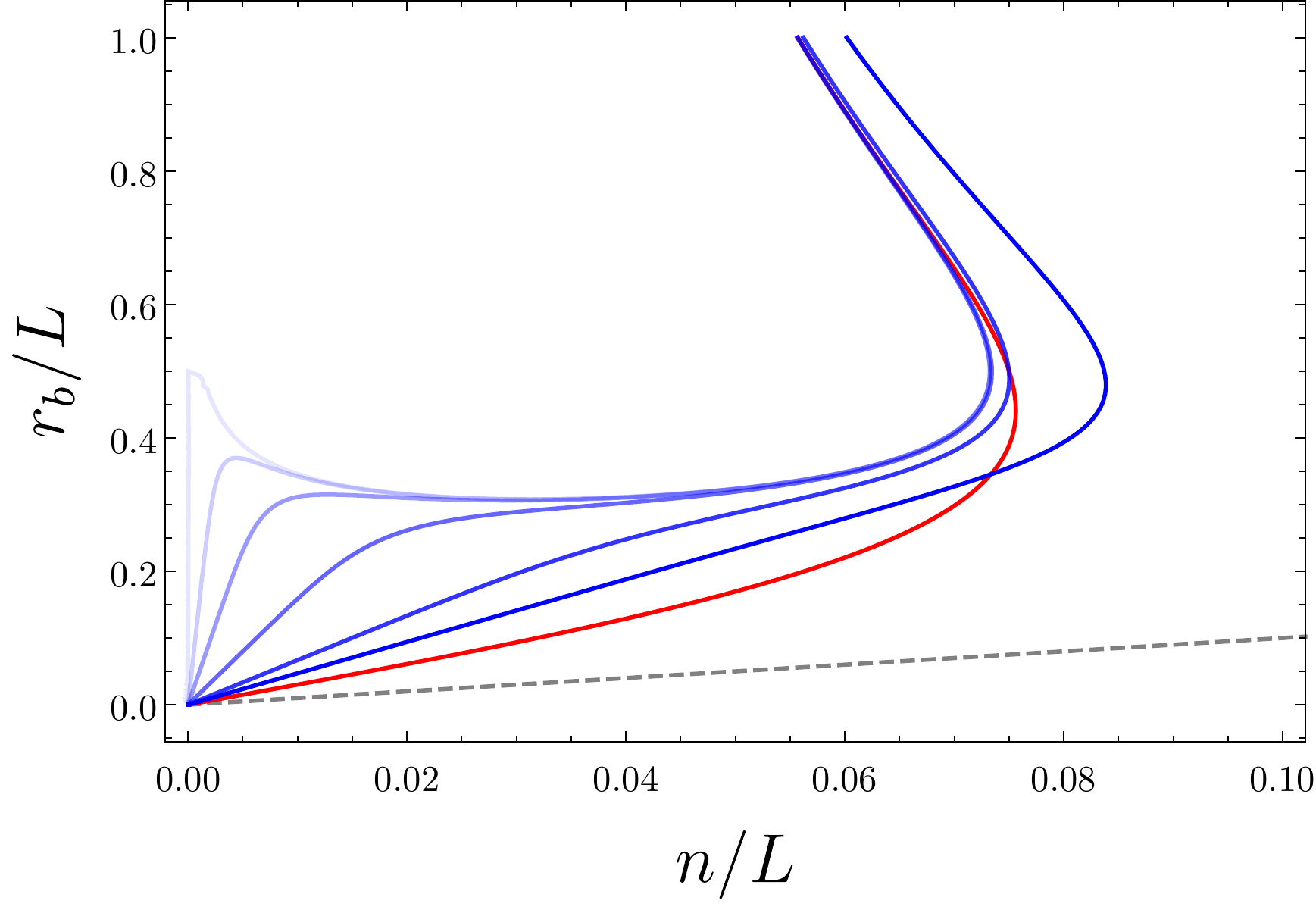}
	\includegraphics[width=0.47\textwidth]{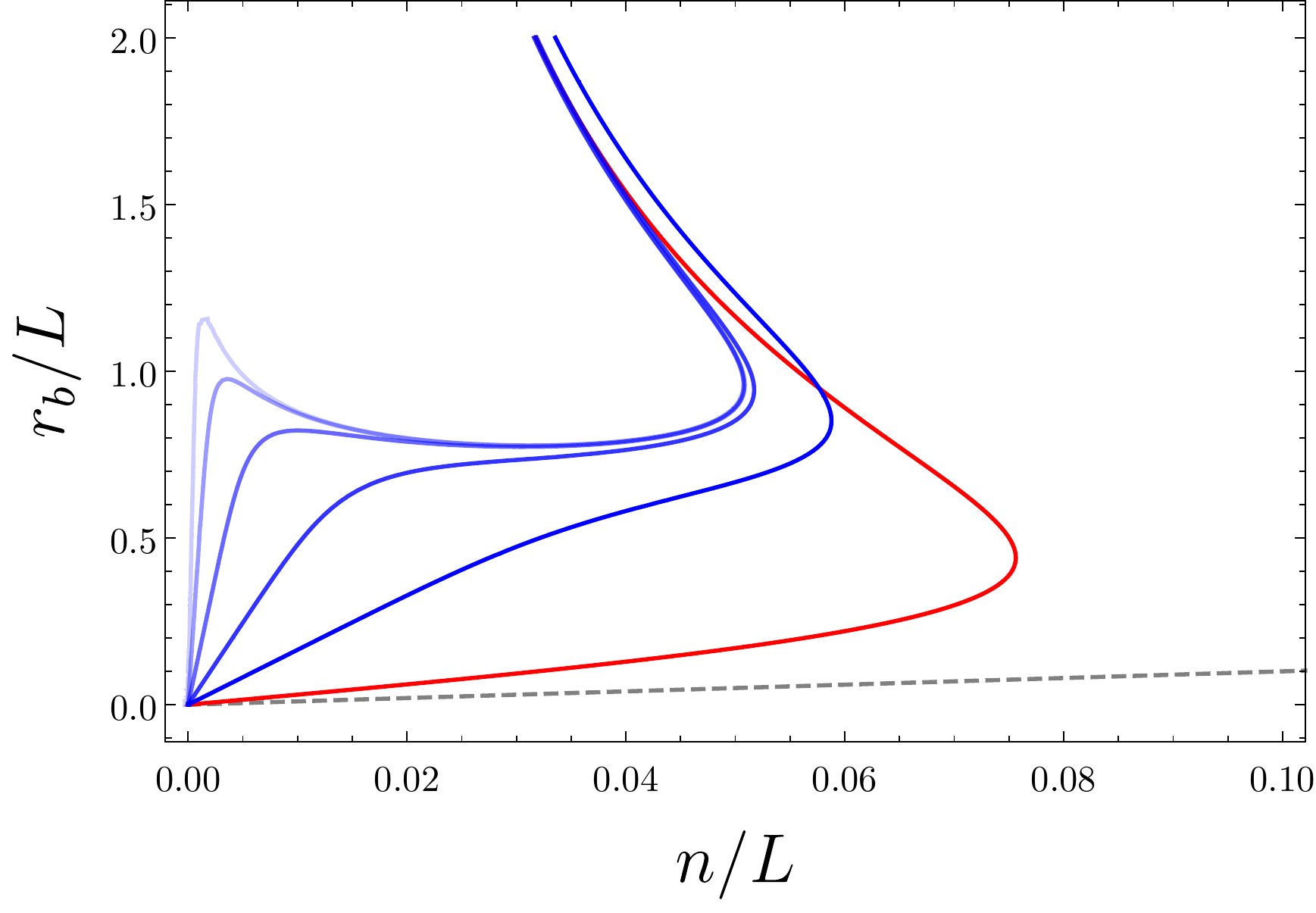}
	\caption{Top row: Plots of $r_{b}$ vs.~$n$ for fixed $\xi = -4 \times 10^{-5}$ with $\zeta \times 10^{5} = -20, -23.15, -23.8, -25, -38.03, -70$ (more to less opacity, or left to right for any horizontal slice through the plot). The right plot is a zoomed in version of the left, showing the interesting structure for bolt solutions. Bottom row: Plots of $r_{b}$ vs.~$n$ for positive $\zeta$. The left plot shows curves for $\zeta = 10^{-3}$ with $\xi = -10^{-3}, -10^{-4}, -10^{-5}, -10^{-6}, -10^{-7}, -10^{-10}$ (in order of decreasing opacity in the plot, or right to left along a horizontal slice through the plot). The right plot shows, for the same values of $\xi$, the result when $\zeta = 27/256 - \xi$, which corresponds to the critical limit. The behaviour when $\zeta > 0$ is all qualitatively similar.  In all plots, the red curve represents the Einstein gravity result, and the dashed, gray line represents the limiting circumstance of $r_{b} = n$.}
	\label{fig:quartic_CP_rh_vs_n}
\end{figure}

%While Fig.~\ref{fig:bolt_params_CP2} provides a useful breakdown of the parameter space, it is useful to understand the precise structure of the bolts in each region.

It is useful to understand the qualitative behaviour of the bolts in the various partitions of the parameter space shown in Fig.~\ref{fig:bolt_params_CP2}. We illustrate this in the top row of Fig.~\ref{fig:quartic_CP_rh_vs_n}, which represents a `vertical slice' through Fig.~\ref{fig:bolt_params_CP2} for $\xi = -4 \times 10^{-5}$. The plot on the right is a zoomed-in copy of the left, and the decreasing opacity of the blue curves (left to right) denotes $\zeta$ becoming more negative, while the red curve corresponds to the Einstein gravity result when both couplings vanish. We see that when $\xi$ and $\zeta$ are small (or, equivalently, when $r_{b}$ is large) the bolt radius reduces nicely to the Einstein gravity result. The interesting behaviour is observed for smaller bolt radius. The first curve corresponds to $\zeta = -20 \times 10^{-5}$ which is in the white region of Fig.~\ref{fig:bolt_params_CP2} and above the orange line. We see that in this case, the behaviour is similar to Einstein gravity, with two possible values for the bolt radius. As $\zeta$ is further decreased, the structure of the curve remains similar but a small `flattened' region begins to form, ultimately becoming vertical for $\zeta \approx  -23.16 \times 10^{-5}$ which corresponds to the point on the orange line of Fig.~\ref{fig:bolt_params_CP2}. Continuing to decrease $\zeta$ further, we see that a bump emerges, and as a result there are up to four values of $r_{b}$ for a given $n$. This behaviour continues until $\zeta < 6 \xi$, which corresponds to the blue shaded region of Fig.~\ref{fig:bolt_params_CP2}. At this point, the  structure of the curve changes drastically, and there are bolts for arbitrarily large $n$. Further, in the region where $\zeta < 6 \xi$ but remains above the region bounded by the black curve in Fig.~\ref{fig:bolt_params_CP2}, there are up to three bolts for a given value of $n$.  As $\zeta$ is further decreased we continue to see three bolts for a given $n$ until we reach the black curve of Fig.~\ref{fig:bolt_params_CP2}, which corresponds to $\zeta \approx      38.03 \times 10^{-5}$. At this point, the three bolts coalesce, and for values of $\zeta$ smaller than this there is only ever a single bolt for a given $n$.

% This continues until $\zeta \approx      38.03 \times 10^{-5}$, corresponding to the black curve of Fig.~\ref{fig:bolt_params_CP2}, after which there is only ever a single bolt for a given $n$.  

It is also possible for $\zeta$ to take on positive values, provided that $\xi < 0$ and $\zeta \le 27/256 - \xi$. The bottom row of plots in Fig.~\ref{fig:quartic_CP_rh_vs_n} shows representative behaviour in this case. The qualitative shape of the curve is controlled by the ratio $\xi/\zeta$. When $\xi/\zeta \to 0^{-}$, a peak forms at small $n$. The overall behaviour is similar to Einstein gravity: there is a maximum value of $n$ beyond which bolts cannot exist. For $n$ smaller than this value, there are two values of $r_{b}$  for any given $n$.

The above discussion highlights the general trend in this parameter space. The lines of `critical points' mark the boundaries where there is a change in the maximum number of bolts for a given NUT charge. For a fixed $\xi$, the structure is (referring to Fig.~\ref{fig:bolt_params_CP2}): two bolts and Einstein-like structure in the white region above the orange line; up to four bolts in the white region below the orange line; up to three bolts in the blue shaded region above the black line, and one bolt in the blue shaded region below the black line. When $\zeta$ takes on positive values, the structure remains the same as in the white region above the orange line, but a peak forms at small $n$ as $\xi/\zeta \to 0^{-}$.

\begin{figure}[t!]
\includegraphics[width=0.47\textwidth]{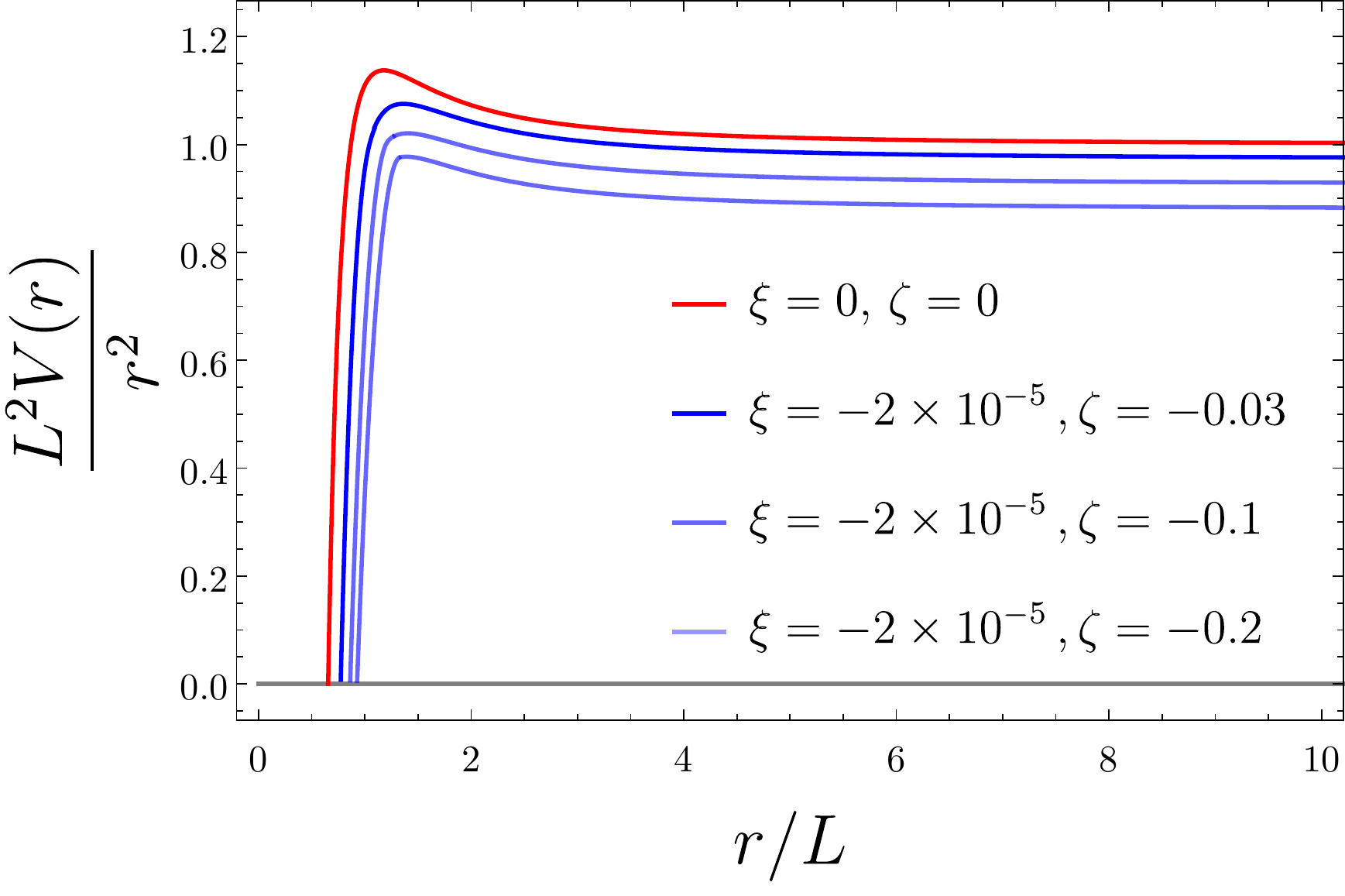}
\quad 
\includegraphics[width=0.47\textwidth]{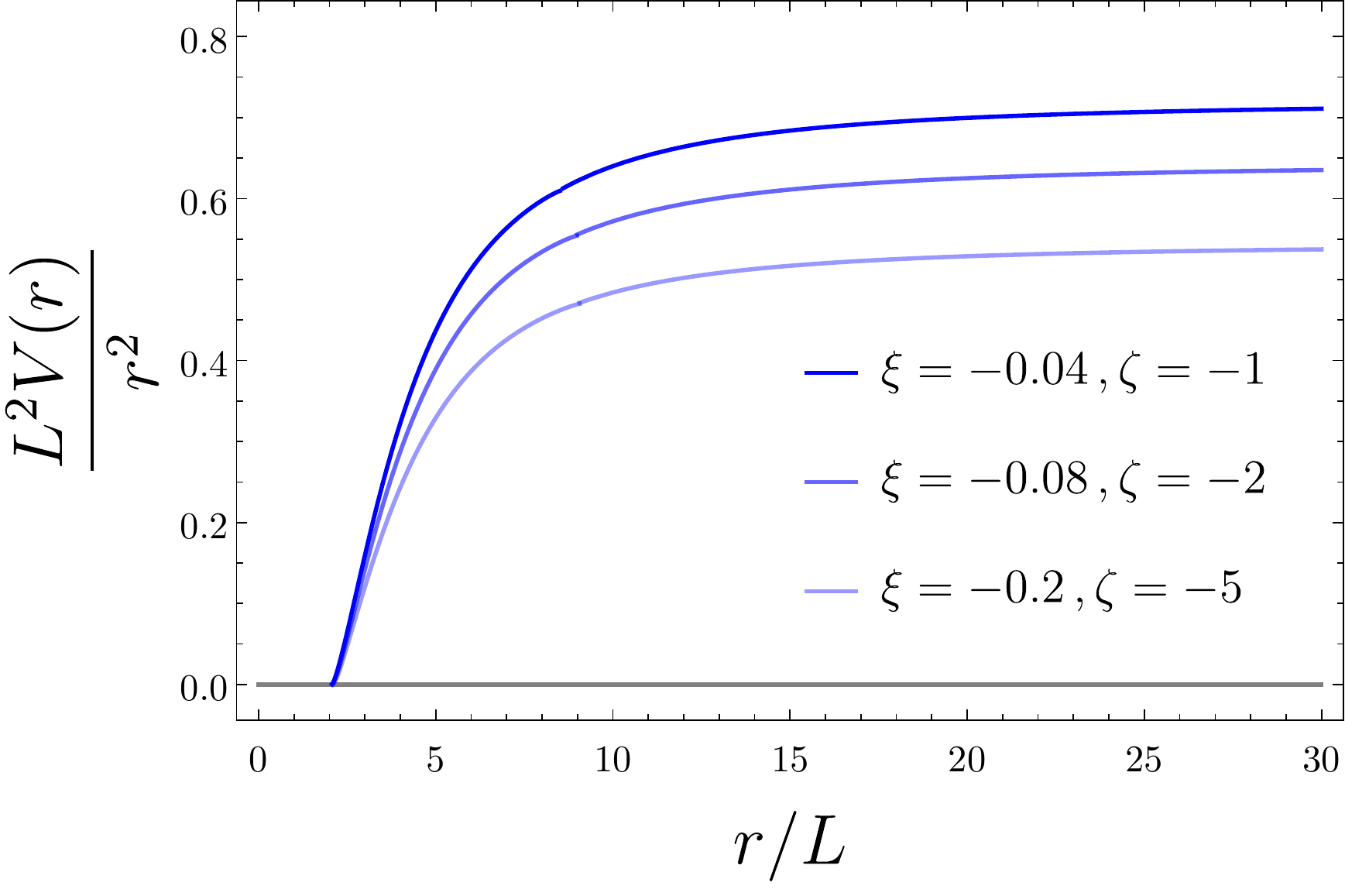}
\caption{The metric function $L^2 V_{\mathbb{CP}^2}(r)/r^2$ is plotted for bolt solutions of the quartic theories. The left plot is for different combinations of the quartic couplings with $n/L = 7/100$. For this value of the NUT parameter, the bolt solutions in the quartic theories can be compared to Einstein gravity solutions. In the right plot, the NUT parameter has been set to $n/L = 2$. For this value of the NUT parameter there are no bolt solutions in Einstein gravity and the existence of these solutions is purely because of the quartic curvature terms.}
\label{CP-bolt-numerics}
\end{figure}

So far, our study of the bolt solutions has focused on the properties of the near horizon solutions. It is important to verify that these near horizon solutions can be joined smoothly on to the asymptotic solution~\eqref{CP_asymp} that was presented at the beginning of this section. This can be shown by numerically solving the field equations, with some relevant examples shown in Fig.~\ref{CP-bolt-numerics}. The left plot shows example bolt solutions for $n/L = 7/100$. In this regime, both Einstein gravity and the quartic theories admit bolt solutions, and the two can be compared. The solutions are qualitatively similar but, of course, the solutions to the quartic solutions asymptote to $\fin r^2/L^2$ with $\fin \neq 1$. In the right plot, we show examples for $n/L = 2$ --- for this value of the NUT parameter, there are no bolt solutions in Einstein gravity.

 Finally, turning to the on-shell action, it can be computed using the same prescription as in the NUT case, but now evaluating for the bolt at $r = r_{b}$. In performing the calculation, we make use of the near horizon equation~\eqref{eqn:bolt_nh_cp2} to simplify the result. We find that,
\begin{align}
I_E &=  -\frac{ \pi^2}{54  L^2 G} \bigg[ 243 r_{b}^4 L^2 - 972 r_{b}^3(L^2 + 3 r_{b}^2) n -486 L^2 r_{b}^2 n^2  
\nonumber\\
&+ 972 r_{b}(3L^2 + 10 r_{b}^2)n^3 + 243 L^2 n^4 -14580 r_{b} n^5 + \frac{\zeta L^8}{n^3(r_{b}^2 - n^2)} (40 r_{b} n^2 + 24 n^3)
\nonumber\\
&- \frac{\xi L^8}{n^3(r_{b}^2-n^2)} (18 r_{b}^3 + 144 r_{b}^2 n  + 222 r_{b} n^2 )  \bigg] \, .
\end{align}
Making use of the chain rule and the second equation in~\eqref{eqn:bolt_nh_cp2}, we find that $E = \partial_\beta F = M$, justifying the terminology ``mass parameter'' used earlier.  The entropy is just given by $S = \beta E - I_E$ which reads
\begin{align}
S =& \frac{\pi^2}{54 r_{b} G} \bigg[243 r_{b}^5 - 486 r_{b}^3 n^2 - 2916 r_{b}^2 n^3 + 243 n^4 r_{b} - 2916 n^5 - \frac{4860(r_{b}^4 - 6 r_{b}^2 n^2 - 3 n^4) n^3}{L^2} 
	\nonumber\\ &+ \frac{8 \zeta L^6 (10 r_{b}^2 + 3 r_{b} n - 5 n^2)}{n (r_{b}^2 - n^2)} - \frac{6 \xi L^6(12 r_{b}^4 + 72 r_{b}^3 n + 65 r_{b}^2 n^2 - 48 r_{b} n^3 - 37 n^4)}{n^3(r_{b}^2 - n^2)} \bigg] \, .
\end{align}
We can study the extended thermodynamics of these bolts in the same manner as the NUTs. The extended first law has the same form as \eqref{eflawCP} but for the bolts the potentials are given by
\begin{equation}
\begin{aligned}
V &= \frac{6 \pi^2 r_{b}}{5} \left(3 r_{b}^4 - 10 r_{b}^2 n^2 + 15 n^4 \right) \, , \\
 \Upsilon^{\cal S} &= \frac{\pi r_{b}(3 r_{b}^2 + 24 r_{b} n + 37 n^2)}{108 n^4 G(r_{b}^2 - n^2)} \, ,\quad
 \Upsilon^{\cal Z} = - \frac{\pi (5r_{b} - 3 n)}{81 n^2 G ( r_{b}^2 - n^2)} \, ,
\end{aligned}
\end{equation}
and we recall that here we are working with $\lambda_{\ssc \rm GB} = 0$. These quantities also satisfy the Smarr relation that follows from scaling, which has the same form here as in~\eqref{smarrCP}.  Again, the formula for the thermodynamic volume is unaltered from its form in Einstein gravity. Though, since $r_{b}$ implicitly depends on the higher-curvature couplings, the numerical value of the thermodynamic volume for fixed $n$, $\xi$ and $\zeta$ will in general differ from the Einstein gravity value. Contrast this with the situation for the NUTs where the thermodynamic volume is completely insensitive to the theory of gravity, so long as the theory belongs to the generalized quasi-topological class.

\begin{figure}[t!]
	\centering 
	\includegraphics[width=0.47\textwidth]{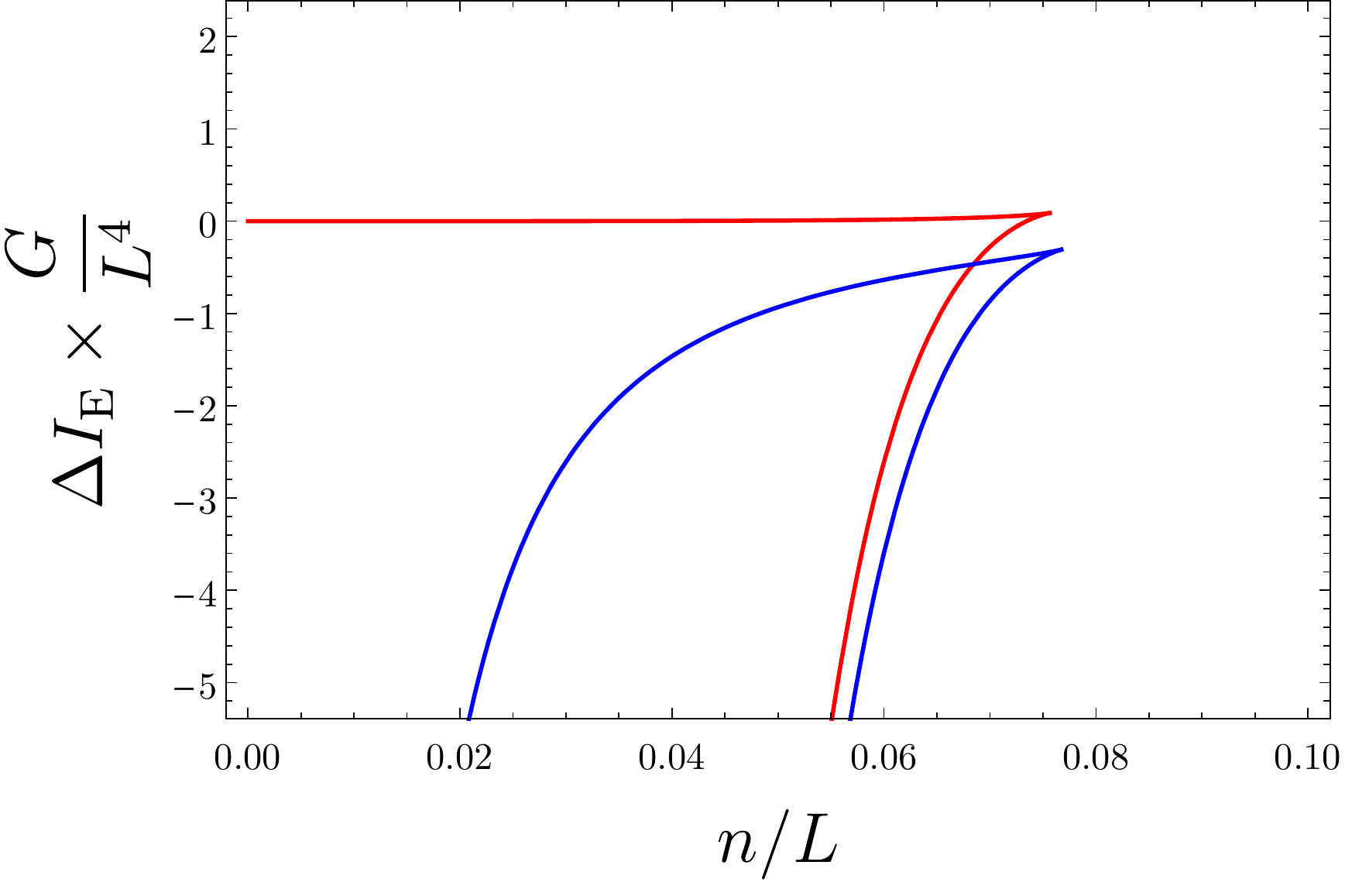}
	\includegraphics[width=0.47\textwidth]{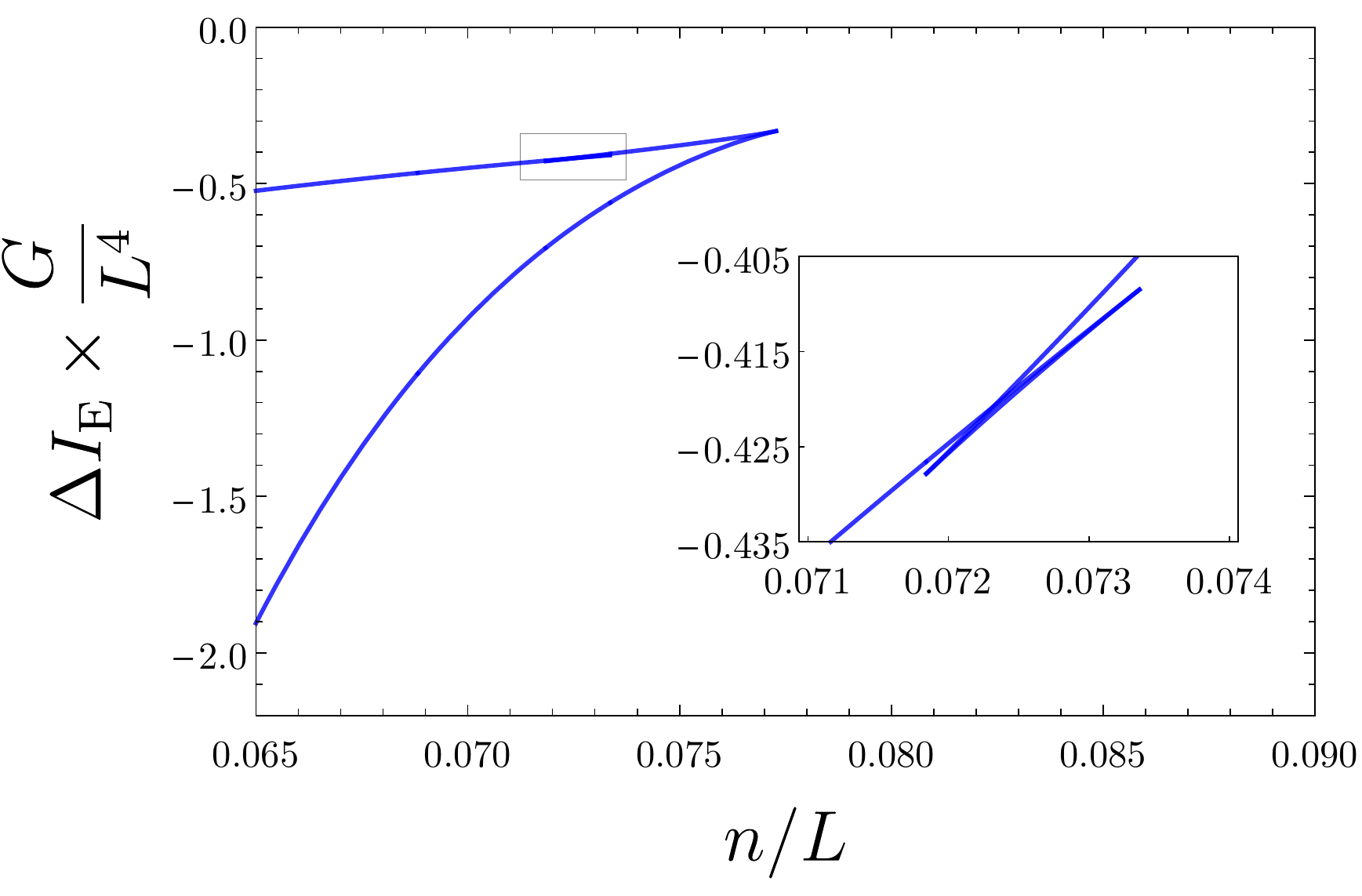}
	\includegraphics[width=0.47\textwidth]{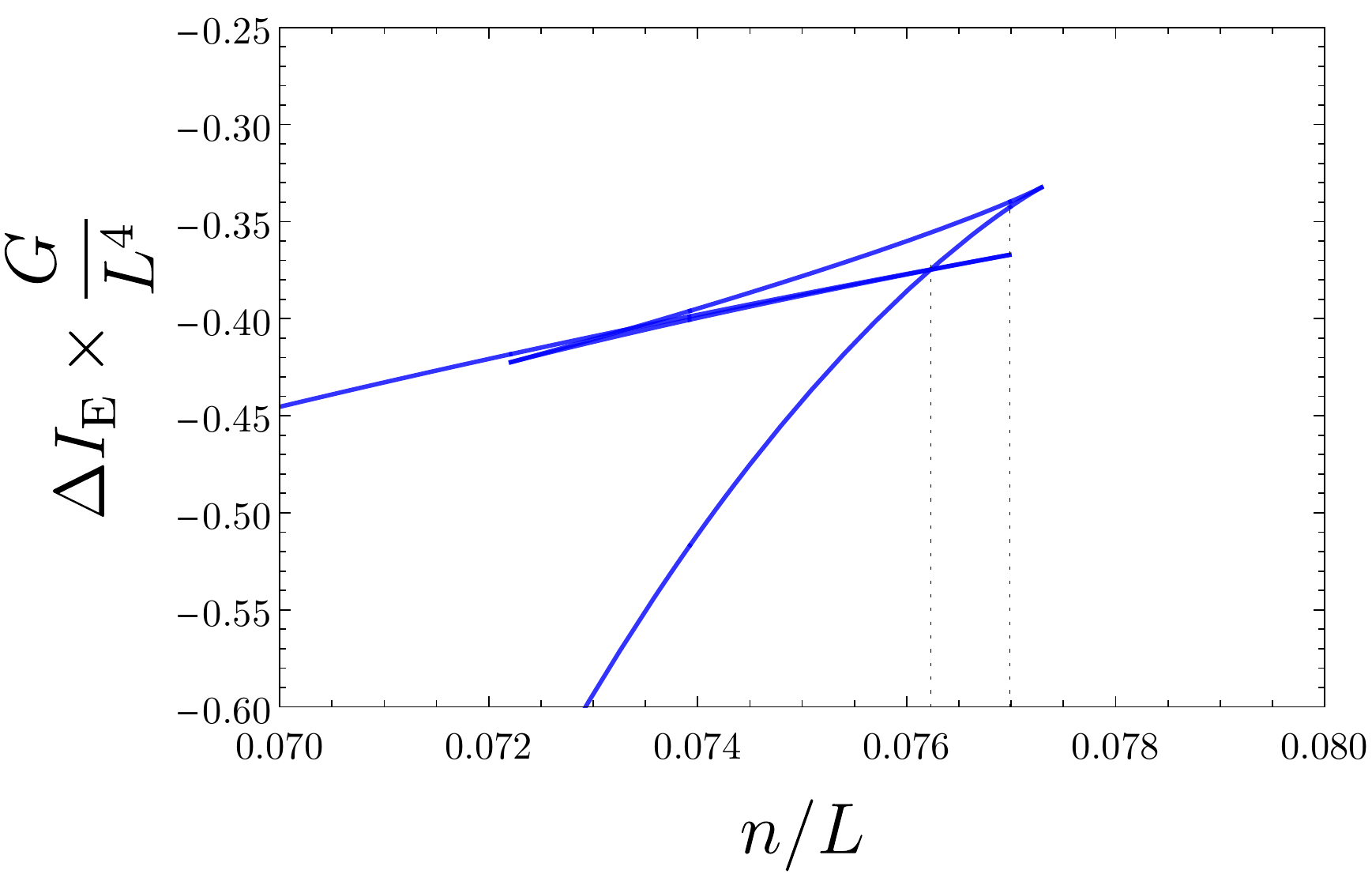}
	\includegraphics[width=0.47\textwidth]{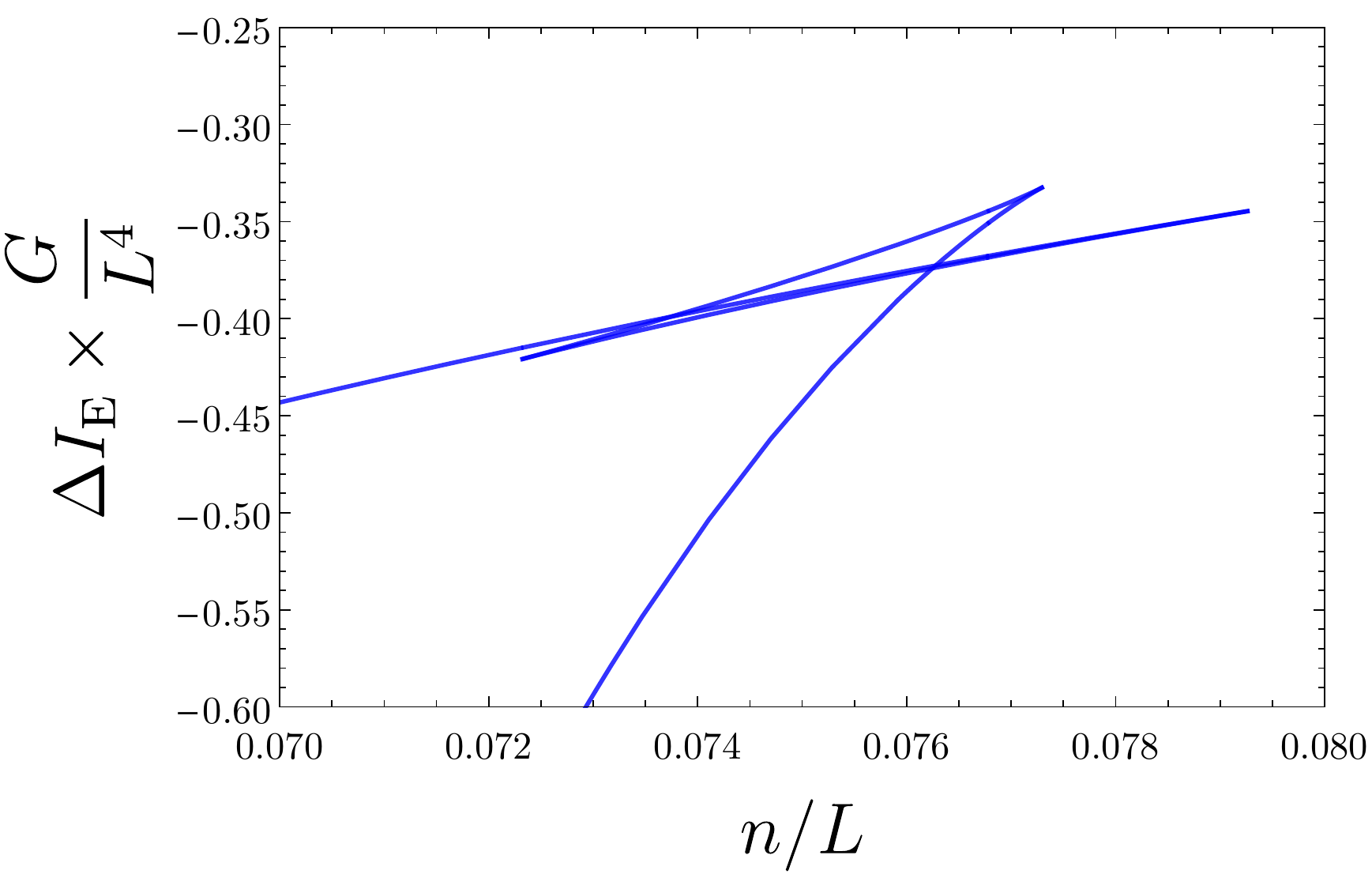}
	\includegraphics[width=0.47\textwidth]{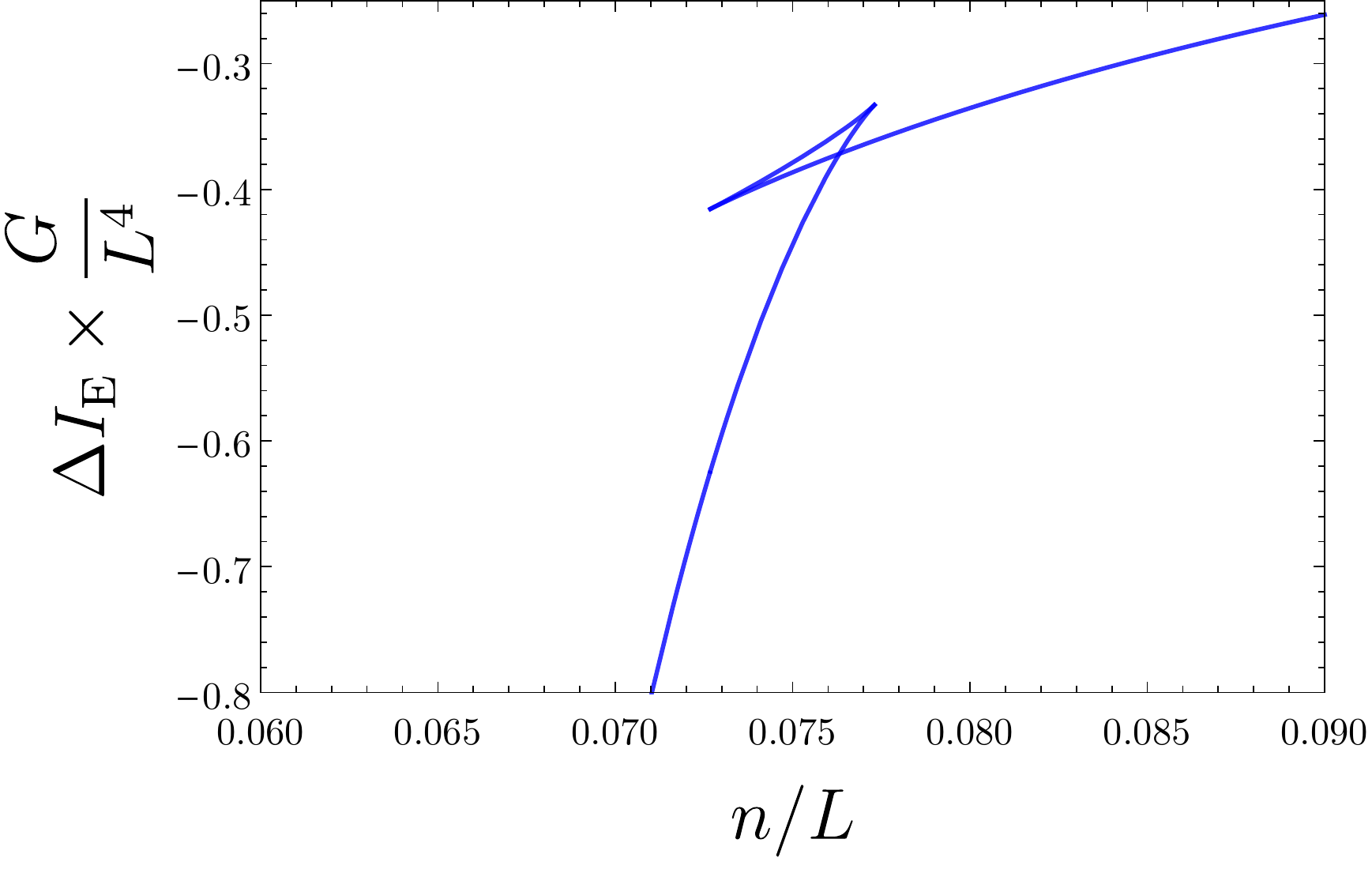}
	\includegraphics[width=0.47\textwidth]{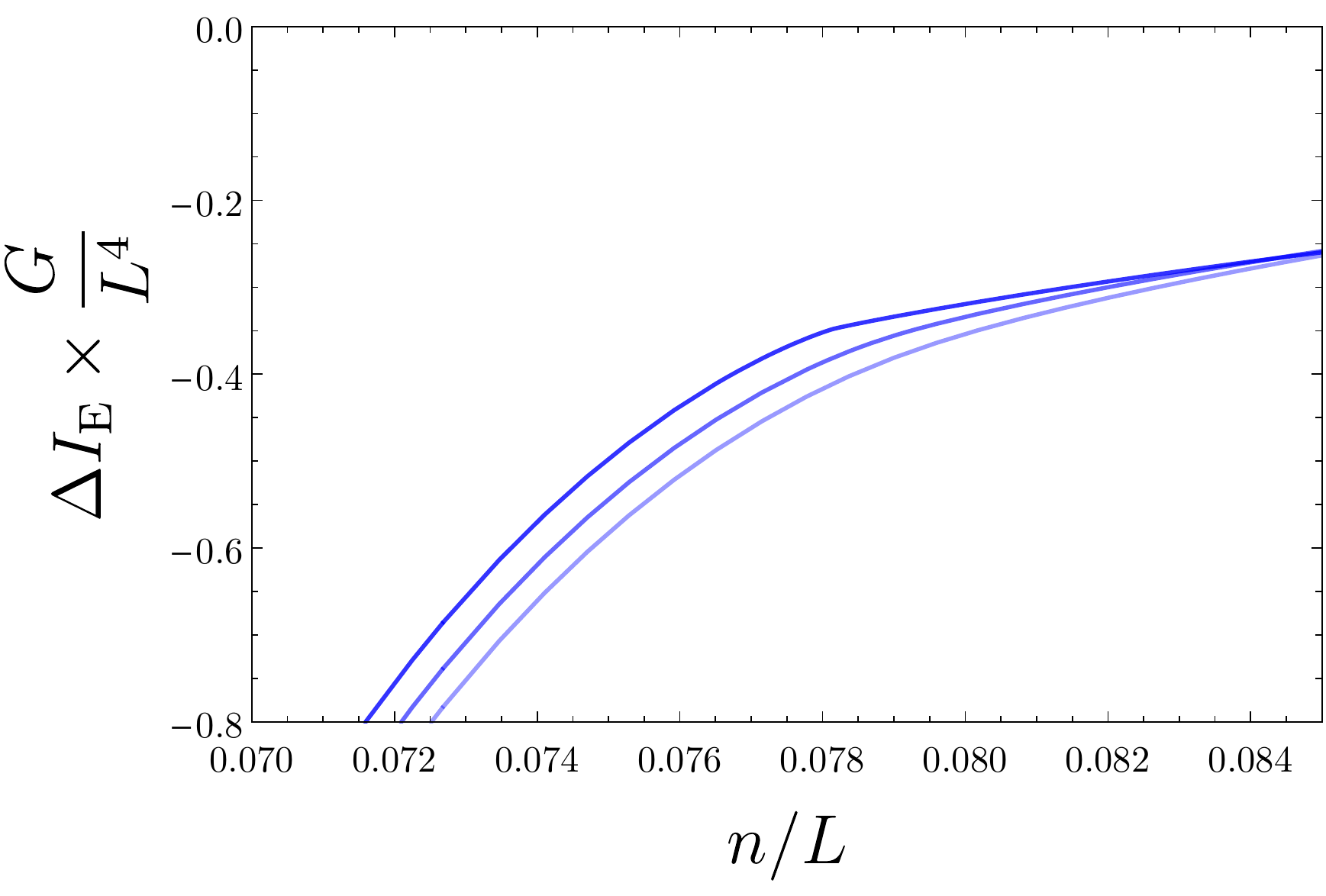}
	\caption{Euclidean on-shell action difference $\Delta I_{\rm E} = I_{\rm E}^{\rm bolt} - I_{\rm E}^{\rm NUT}$  for $\mathcal{B}=\mathbb{CP}^2$ solutions in the quartic theories. Red corresponds to Einstein gravity, while all blue curves have $\xi = -4 \times 10^{-5}$ for various values of $\zeta$. Top left: A comparison between Einstein gravity and the quartic theories with $\zeta = -10 \times 10^{-5}$, we see in both cases the action is a `cusp'.  Top right: Here $\zeta = -23.5 \times 10^{-5}$; the inset shows a zoomed-in plot of the boxed area, showing the swallowtail structure that has emerged. Center left: Here $\zeta = - 23.74 \times 10^{-5}$; the swallowtail now intersects the lower branch of the cusp. The vertical dotted lines correspond to a first order phase transition (leftmost line) and a zeroth-order phase transition (rightmost line). Center right: Here $\zeta = -23.82 \times 10^{-5}$. The swallowtail has elongated, and now extends past the cusp. Bottom left: Here $\zeta = -24.1 \times 10^{-5}$. Bolts now exist for all values of $n$, and there is a swallowtail structure present.  Bottom right: Here $\zeta = -38.026 \times 10^{-5} , -50 \times 10^{-5}$ and $-60 \times 10^{-5}$ (more to less opacity, respectively). Along the first curve, there is a critical point located at $n/L \approx 0.07815$, while the other two curves are smooth. The structure of the on-shell action is qualitatively similar to these last two curves for all $\zeta \lesssim -38.026 \times 10^{-5}$. } 
	\label{fig:CP_bolt_IE}
\end{figure}

The Euclidean on-shell action exhibits rich structure for the bolt solutions. In understanding the behaviour, it is helpful to once again refer to Fig.~\ref{fig:bolt_params_CP2}. As it turns out, this figure partitions the parameter space into regions where the behaviour is qualitatively similar.  Referring to Fig.~\ref{fig:bolt_params_CP2}, the most interesting changes in behaviour occur when the orange and black lines are crossed, which correspond to actual critical points in the thermal phase space marking the appearance/disappearance of swallowtail structures in the on-shell action.  Also when transitioning from the white-shaded to blue-shaded region, the action switches from terminating at a cusp at some finite $n$ to existing for all values of $n$. In the white region, in all cases but Einstein gravity there will be a zeroth order phase transition between bolt solutions and NUT solutions at the value of $n$ corresponding to the maximum value of $r_{b}$. In Einstein gravity there is also a phase transition at this point, but in that case it is first order. As an example that highlights the salient points pertaining to the bolts, let us once again consider the $\xi = -4 \times 10^{-5}$ slice through the parameter space for different values of $\zeta$ --- various relevant examples are shown in Fig.~\ref{fig:CP_bolt_IE}, where the Euclidean action of the NUT solutions has been subtracted off, $\Delta I_{\rm E} = I_{\rm E}^{\rm bolt} - I_{\rm E}^{\rm NUT}$. Though the discussion will make reference to numerical values in only this particular case, the qualitative features are general.

Particularizing now the discussion to $\xi = -4 \times 10^{-5}$, for positive $\zeta$ through to $\zeta  \approx  -23.1565 \times 10^{-5}$ (which corresponds to the orange line in Fig.~\ref{fig:bolt_params_CP2}), the behaviour is similar to Einstein gravity, with the on-shell action exhibiting two smooth branches that end at a cusp located at the maximum value of NUT charge. Precisely when $\zeta$ is chosen on the orange line shown in Fig.~\ref{fig:bolt_params_CP2} ($\zeta  \approx  -23.1565 \times 10^{-5}$ in this case),  the upper branch of $I_E$ develops a cusp, corresponding to a critical point in the system. As $\zeta$ is further decreased, a swallowtail emerges from the cusp on the upper branch, as shown in the second plot of Fig.~\ref{fig:CP_bolt_IE}. Further decreasing $\zeta$ elongates the swallowtail, and eventually it intersects the lower branch of $I_E$ --- for the particular case of $\xi = -4 \times 10^{-5}$, this intersection occurs for $\zeta \approx - 23.705 \times 10^{-5}$. This intersection then gives rise to a region where a re-entrant phase transition occurs as $n$ is increased, as shown in the  center-left plot of Fig.~\ref{fig:CP_bolt_IE}.   The two vertical black, dotted lines show the locations where these transitions occur. There is a first order phase transition from phase 1 to phase 2, followed by a zeroth order phase transition which returns the system back to the initial phase. It is in this sense that we have a re-entrant phase transition --- a monotonous variation of the NUT charge gives rise to two phase transitions with the final and initial phases coinciding. Let us note that re-entrant phase transitions were first observed in nicotine/water mixtures in~\cite{hudson1904mutual}. In the context of black hole physics, while somewhat exotic, they are well-established --- see~\cite{Altamirano:2013ane} for an example in a rotating black hole spacetime, and~\cite{Frassino:2014pha, Hennigar:2015esa} for black hole examples in higher curvature theories of gravity. We believe this is the first instance observed for NUT charged solutions.  As $\zeta$ is further decreased, the swallowtail continues to elongate, and for $\zeta \approx - 23.753 \times 10^{-5}$, the tip of the swallowtail extends past the cusp --- this ends the region of parameter space for which re-entrant phase transitions occur.  

There is a drastic change in structure at $\zeta = 6 \xi$. Corresponding to the boundary of the blue-shaded region in Fig.~\ref{fig:bolt_params_CP2}, this condition yields the largest $\zeta$ for which there is a maximum NUT parameter for which bolts exist. From the perspective of the on-shell action, essentially what happens is, at this point, the swallowtail has now elongated ``to infinity''. Between $\zeta = 6 \xi$ and $\zeta \approx - 38.026 \times 10^{-5}  $ the action displays a swallowtail structure that is associated with a first-order phase transition. The swallowtail vanishes at a critical point when $\zeta \approx  - 38.026 \times 10^{-5} $ (the black line in Fig.~\ref{fig:bolt_params_CP2}). For $\zeta \lesssim  - 38.026 \times 10^{-5} $, the on-shell action displays only a single branch for all values of $n$.

Lastly, let us make some remarks regarding the critical points that are present at some points in the parameter space. As mentioned above, the lines of critical points appearing in Fig.~\ref{fig:bolt_params_CP2} are bonafide critical points in the thermodynamic parameter space. When in the region of the parameter space corresponding to the white region of Fig.~\ref{fig:bolt_params_CP2}, the action exhibits a cusp structure qualitatively similar to that shown in the top left plot of Fig.~\ref{fig:CP_bolt_IE}. We find that one of the critical points always occurs on the upper branch of this cusp (those corresponding to the orange curve in Fig.~\ref{fig:bolt_params_CP2}). These critical points will, therefore, not be realized since they do not comprise the dominant contribution to the partition function. The critical points that correspond to the points on the black curve shown in Fig.~\ref{fig:bolt_params_CP2} belong to the lower branch of the cusp in the white region or are on the single physical branch in the blue shaded region. These critical points are physically realized.

At the critical point, certain physical quantities blow up in power law fashion. To get a sense of the critical exponents governing these divergences, we can study the behaviour of the specific heat,
\be 
C = - T \frac{\partial^2 F}{\partial T^2}  \propto \left(1 - \frac{T}{T_c} \right)^{\tilde\alpha}
\ee
where $F = T I_E$ and $\tilde\alpha$ is the critical exponent governing this divergence\footnote{We use the notation $\tilde\alpha$ to avoid confusion with much of the black hole chemistry literature, e.g.~\cite{Kubiznak:2016qmn}, where $\alpha$ is exclusively used in reference to the specific heat at constant volume. }. Due to the complexity of the equations relating the bolt radius to the NUT parameter, it is difficult to perform an analytic study near the critical point. Instead, to make progress, we plot 
\be 
\log \bigg| \frac{1}{T} - \frac{1}{T_c} \bigg| \quad \text{vs.} \quad \log \bigg| \frac{\partial F}{\partial T} (T) -  \frac{\partial F}{\partial T} (T_c)  \bigg|
\ee 
numerically and extract the slope of this line via a linear fit. As an example, we find in the case $\xi = -4 \times 10^{-5}$ the following fit:
\be 
\log \bigg| \frac{1}{T} - \frac{1}{T_c} \bigg| = 2.941  \log \bigg| \frac{\partial F}{\partial T} (T) -  \frac{\partial F}{\partial T} (T_c)  \bigg| + {\rm constant}
\ee
which after some simple algebra yields 
\be 
\tilde\alpha = 0.659
\ee
which is consistent with $\tilde\alpha = 2/3$ to within the numerical precision. This value for the critical exponent is often observed for the divergence of the specific heat at constant pressure in black hole systems --- see, e.g.,~\cite{Chamblin:1999hg}. In this sense, it is not surprising to find that the same critical exponent governs the behaviour near the critical point for the bolts. A numerical survey of many critical points for different values of the couplings shows that they are all consistent with this result.

The red dot shown in Fig.~\ref{fig:bolt_params_CP2} represents a special point in the parameter space where two critical points merge. Because of this, one might hope to see novel critical exponents similar to how the coalescence of multiple critical points leads to non-mean field theory critical exponents for Lovelock black holes~\cite{Dolan:2014vba}. However, unfortunately, this is not the case here. The reason is that as the red dot is approached, there is one critical point on the upper branch of the cusp and one on the lower branch. When these critical points merge, they also meet at the cusp which acts as a phase boundary --- no solutions exist beyond the tip of the cusp. To within the accuracy of our calculation, the critical exponent associated with each critical point as the cusp is approached remains consistent with $\tilde\alpha = 2/3$.

\vspace{.01cm}

\chapter{Free energy of CFTs on squashed spheres}\label{Chap:8}

Euclidean conformal field theories (CFTs) coupled to background fields can be used to learn important lessons about the dynamics of the theory in question.  A prototypical example corresponds to supersymmetric CFTs, where localization techniques have allowed for notable progress --- see \eg \cite{Pestun:2016zxk}. For non-supersymmetric theories, a natural possibility consists in coupling the theory to curved background metrics. This approach has produced some exact and universal results valid for general CFTs \cite{Bobev:2017asb,Fischetti:2017sut} and has found various applications, \eg in holographic cosmology \cite{Anninos:2012ft,Conti:2017pqc,Hertog:2017ymy,Hawking:2017wrd}. Particularly interesting is the case of spherical backgrounds, whose partition functions --- equivalently, free energies: $\mathcal{F}_{\mathbb{S}^d}=- \log |Z_{\mathbb{S}^d}|$ --- %in the absence of deformation 
have been conjectured to be renormalization-group monotones for general odd-dimensional QFTs \cite{Klebanov:2011gs,Casini:2012ei,Pufu:2016zxm}.  

In this chapter, we will consider CFTs on deformed spheres and study the effect that such deformations have on $\mathcal{F}$. The focus will be on a particular class of squashed spheres, $\mathbb{S}_{\varepsilon}^d$, which preserve a large subgroup of isometries of the round ones.\footnote{In particular, \req{squa} preserves a SU$(\frac{d+1}{2})\times$U$(1)$ subgroup of the usual SO$(d+1)$ preserved by the usual round-sphere metric in $d$-dimensions.} In particular, they are characterized by being Hopf fibrations over the complex projective space $\mathbb{CP}^{k}$ ($k\equiv (d-1)/2$), namely, $\mathbb{S}^1\hookrightarrow \mathbb{S}_{\varepsilon}^d\rightarrow \mathbb{CP}^k$. The metric on these  squashed-spheres is given by
\begin{equation}\label{squa}
ds^2_{\mathbb{S}_{\varepsilon}^d}=\frac{ds^2_{\mathbb{CP}^k}}{(d+1)}+(1+\varepsilon)\left(d\psi+\frac{A_{\mathbb{CP}^k}}{(d+1)}\right)^2\, ,
\end{equation}
where $\psi$ is a periodic coordinate which parametrizes the $\mathbb{S}^1$, $ds^2_{\mathbb{CP}^k}$ is the Einstein metric on $\mathbb{CP}^k$ normalized so that $R_{ij}=g_{ij}$, and $J=dA_{\mathbb{CP}^k}$ is the K\"ahler form on $\mathbb{CP}^k$. The parameter $\varepsilon$ measures the degree of squashing of the sphere and, in principle, it can take values in the domain $\varepsilon \in (-1,+\infty)$, the round-sphere corresponding to $\varepsilon=0$. In $d=3$, which is the simplest case, $\mathbb{CP}^1 \cong \mathbb{S}^2$, and we can write $ds^2_{\mathbb{S}^2}=d\theta^2+\sin^2\theta d\phi^2$, $A_{\mathbb{S}^2}=2\cos \theta d\phi$ in standard spherical coordinates.

This class of squashed spheres can be easily studied holographically \cite{Hawking:1998ct,Dowker:1998pi,Chamblin:1998pz,Emparan:1999pm,Hartnoll:2005yc,Bobev:2016sap}, as the relevant bulk geometries belong to the well-known AdS-Taub-NUT/bolt family. Our first main result --- see \req{fee0e} --- is a universal formula for the free-energy of a broad class of holographic CFTs on squashed-spheres. The formula is automatically regularized and, in fact, does not require knowing the corresponding NUT solutions explicitly. We will argue that it holds for an infinite number of higher-curvature bulk theories of the GQG type, and that it correctly reproduces all previous results available in the literature. Additionally it passes several consistency checks coming from field theory considerations.
Our second result --- see \req{3conj} --- is an expression for the subleading term in the small squashing-parameter expansion of $\mathcal{F}_{\mathbb{S}_{\varepsilon}^{3}}$ which, based on holographic and free field calculations we conjecture to be controlled by the stress-tensor three-point function coefficient $t_4$ for general CFTs. 
As an additional consequence of our results in the holographic context, we observe that, for the class of bulk theories just described, the function that determines the possible AdS vacua of the theory --- see \req{hgo} --- acts as a generating functional for the boundary stress-tensor, in the sense that we can easily characterize its correlators by taking ordinary derivatives of such function, drastically simplifying the standard holographic calculations  --- see \req{cte}, \req{tte}, \req{0p} and \req{conne}.

\section{Holography of Einstein-like higher-order gravities}
AdS/CFT \cite{Maldacena,Witten,Gubser} provides a powerful playground for exploring the physics of strongly coupled CFTs. In some cases, the possibility of mapping intractable field-theoretical calculations into manageable ones involving gravity techniques allows for the identification of universal properties valid for completely general CFTs.  In this context, higher-curvature gravities turn out to be very useful, as they define holographic toy models for which many explicit calculations --- otherwise practically inaccessible using field-theoretical techniques --- can be performed explicitly. The idea is that, if a certain property is valid for general theories, it should also hold for these models. This approach has been successfully used before, \eg in the identification of monotonicity theorems in various dimensions \cite{Myers:2010xs,Myers:2010tj}, or in the characterization of universal terms in the entanglement entropy \cite{Bueno1,Bueno2,Mezei:2014zla,Chu:2016tps}. Naturally, particular higher-curvature interactions generically appear as stringy corrections to the effective actions of top-down models admitting holographic duals \cite{Gross:1986mw}. For the purposes just described, however, it is more useful to consider bulk models which are particularly amenable to holographic calculations --- see \eg \cite{Camanho:2009hu,Buchel:2009sk,Myers:2010jv,deBoer:2009gx,Camanho:2013pda,ECGholo}. 
%After the results in \cite{}, a decent number of Taub solutions of the form \req{squa} are known in four and six bulk dimensions for various higher-curvature generalizations of Einstein gravity.

Let us start by considering a general higher-curvature Lagrangian, in $(d+1)$ bulk dimensions, that we can write as follows
\begin{equation}\label{hog}
\mathcal{L}=\frac{1}{16\pi G}\left[\frac{d(d-1)}{L^2}+R+\sum_{n=2} \mu_{n} L^{2(n-1)}\mathcal{R}_{(n)}\right]\, ,
\end{equation}
where $L$ is some length scale, $G$ is Newton's constant, the $\mu_{n}$ are dimensionless couplings, and the $\mathcal{R}_{(n)}$ stand for the higher-curvature terms, constructed from linear combinations of order-$n$ curvature invariants. In the previous Lagrangian we are explicitly including one higher-curvature term at every order, which suffices for our purposes, but we could add more terms. 
The first step in order to perform holographic computations for theories of the form \req{hog} is to determine their AdS vacuum. 
As we saw in Chapter~\ref{Chap:1}, the AdS vacua of these theories can be obtained by solving the equation\cite{Aspects}
\begin{equation}\label{hinf}
h(f_{\infty})\equiv\frac{16\pi G L^2}{d(d-1)}\left[\mathcal{L}(f_{\infty})-\frac{2f_{\infty}}{(d+1)}\mathcal{L}'(f_{\infty})\right]=0\, ,
\end{equation}
where $\mathcal{L}(f_{\infty})$ is the on-shell Lagrangian on pure AdS$_{(d+1)}$ with radius $L/\sqrt{f_{\infty}}$. This 
can be easily obtained evaluating all Riemann tensors in \req{hog} as $R_{\mu\nu\rho\sigma}=-f_{\infty}/L^2 (g_{\mu\rho}g_{\nu\sigma}-g_{\mu\sigma}g_{\nu\rho})$. Also, $\mathcal{L}'(f_{\infty})\equiv d\mathcal{L}(f_{\infty})/df_{\infty}$.  It is easy to see that the $\mathcal{R}_{(n)}$ can always be normalized so that the function $h(f_{\infty})$ in \req{hinf} takes the simple  form\footnote{The special case $n=(d+1)/2$ must be excluded from the sum, as no invariant of that order contributes to the vacuum equation.} 
\begin{equation}\label{hgo}
h(f_{\infty})=1-f_{\infty}+\sum_{n\neq (d+1)/2} \mu_n f_{\infty}^n\, .
\end{equation}
This function will play an important role in our discussion, as we will see.
Naturally, for Einstein gravity one just finds $f_{\infty}=1$, and the action scale $L$ coincides with the AdS radius. 

So far, the discussion applies to any higher-derivative gravity.  Let us further restrict \req{hog} to the particular subclass of theories whose linearized equations on maximally symmetric backgrounds are of second order. 
Namely, we restrict to those for which the linearized equations take the form $G_{\mu\nu}^{\rm L}=8\pi G_{\rm eff} T_{\mu\nu}$, where $G_{\mu\nu}^{\rm L}$ is the linearized Einstein tensor, $T_{\mu\nu}$ is some possible matter stress-tensor, and $G_{\rm eff}$ is the effective Newton constant. As we saw at the beginning of Chapter~\ref{Chap:6}, these theories  --- which we refer to as \emph{Einstein-like}  \cite{Aspects} ---  are very appealing for AdS/CFT, since the holographic dictionary of Einstein gravity can be applied straightforwardly to them as well. As a warm-up before presenting the main results of this chapter, it will be useful to derive some basic results for this family of theories.

%This subclass contains infinitely many theories and includes, among others: all Lovelock  \cite{Lovelock1,Lovelock2} and some $f($Lovelock$)$ theories \cite{Love}, Quasi-topological gravity \cite{Quasi2,Quasi} and its higher-curvature extensions \cite{Dehghani:2011vu,Cisterna:2017umf}, Einsteinian cubic gravity in general dimensions \cite{PabloPablo}, and Generalized Quasi-topological gravity \cite{Hennigar:2017ego}, among others \cite{Karasu:2016ifk,Li:2017ncu,Li:2017txk}. The vast majority of all known theories of the form \req{hog} admitting non-trivial black hole and Taub solutions belong to this class. 

\subsubsection*{Stress-tensor 2-point function from AdS equation}
It turns out that the function $h(f_{\infty})$ contains a surprisingly great deal of additional information for Einstein-like theories. 
Using the results in Chapter~\ref{Chap:1} --- in particular Eq.~\req{eq:kappaEL} ---  we can see that this function determines the effective gravitational constant through $G_{\rm  eff}=-G/h'(f_{\infty})$. From the dual CFT point of view, this translates into the following relation with the charge $\ctt$, which fully characterizes the CFT  stress-tensor two-point function%\footnote{Conformal invariance completely constrains the correlator $ \braket{ T_{\mu\nu}(x)\, T_{\lambda\rho}(0)} $ up to a theory-dependent quantity, customarily denoted $\ctt$, as
%$ \braket{ T_{\mu\nu}(x)\, T_{\lambda\rho}(0) } =\ctt\, \mathcal I_{\mu\nu,\lambda\rho}(x) / |x|^{2d}\,, $ where $\mathcal I_{\mu\nu,\lambda\rho}$ is a fixed dimensionless tensor structure \cite{Osborn:1993cr}.}
\footnote{Let us mention that \req{cte} was previously proven in the particular case of Lovelock theories in \cite{Camanho:2010ru,Camanho:2013pda}.} 
\begin{equation}\label{cte}
\ctt = -h'(f_{\infty})\ctte\, , 
\end{equation}
where $\ctte$ stands for the Einstein gravity result\footnote{Observe that our convention for $\ctt$ differs from that in \cite{Bobev:2017asb} by a factor $1/\mathbb{S}_d^2=\Gamma[d/2]^2/(4\pi^2)$. It agrees, however, with the convention in \cite{Myers:2010tj,Bueno2,Buchel:2009sk,Myers:2010jv}. Note also that it is customary to write Einstein gravity results in terms of $L/\sqrt{f_{\infty}}$, instead of $L$ alone. This is irrelevant for Einstein gravity itself, for which $f_{\infty}=1$, but needs to be kept in mind for higher-order theories.} 
\begin{equation}\label{cte2}
\ctte=\frac{\Gamma[d+2] (L/\sqrt{f_{\infty}})^{d-1}}{8\pi^{\frac{d+2}{2}}(d-1)\Gamma\left[\frac{d}{2} \right]G}\, .
\end{equation}
Let us remark that the general formula \req{cte} reproduces the result \req{cttecg} for Einsteinian cubic gravity in $d=3$, that was obtained applying a direct holographic procedure. 
The result above tells us that any holographic quantity that depends on $h'(f_{\infty})$ will be related to the central charge $\ctt$ --- a result that will be very useful for us.

\subsubsection*{Stress-tensor 3-point function in $d=3$}
We have been able to obtain a very simple and closed formula for the stress-energy tensor 2-point function of the CFT dual to any Einstein-like gravity, so we may wonder whether we can obtain an analogous result for for the 3-point function. This case is quite more challenging, so we will restrict to $d=3$ ($D=4$), which has been the must studied dimension throughout this thesis. In Chapter~\ref{Chap:6} we saw that, for parity-even three-dimensional CFTs, the 3-point function of the stress-energy tensor is characterized by two ``charges'', that can be chosen as $\ctt$ plus an additional dimensionless constant that is denoted by $t_4$ \cite{Osborn:1993cr,Hofman:2008ar}. In Section~\ref{t44} we obtained the value of $t_4$ for a CFT dual to ECG --- see Eq.~\req{t4ecg}. That result can be written in an appealing way in terms of the function $h(f_{\infty})$ --- which, for ECG reads $h(f_{\infty})=1-f_{\infty}+\mu f_{\infty}^3$ --- as follows,
\begin{equation}\label{t4d}
t_4=210 f_{\infty}\frac{h''( f_{\infty})}{h'( f_{\infty})}\, .
\end{equation}
Since this formula was derived using only one theory (ECG), it might not work in general, and the value of $t_4$ could be given by a different expression in other theories. However, we have shown in Appendix~\ref{App:Gen} that the formula above holds not only for ECG, but for \emph{all of the Generalized quasi-topological gravities in $D=4$}.  We remind that these are a subclass of the family of Einstein-like theories, but the fact that the result holds for all of them is good enough for our purposes. In addition, this strongly suggests that \req{t4d} might actually hold for all of the Einstein-like theories. Hence, one would be able to obtain the coefficient $t_4$ by taking a couple of derivatives of $h(f_\infty)$. This represents a dramatic simplification with respect to the standard holographic calculations involving energy fluxes --- see \eg \cite{Hofman:2008ar,Buchel:2009sk,Myers:2010jv}.

It is not clear whether a generalization of \req{t4d} exists for higher dimensions --- where now the 3-point function is controlled by two coefficients, $t_2$ and $t_4$. It would be interesting to determine if an expression of the form 
\begin{equation}
\label{tte}
a_{(d)} t_2 + b_{(d)} t_4 =f_{\infty} \frac{h''(f_{\infty})}{h'(f_{\infty})}\, ,
\end{equation}
holds for general Einstein-like theories in arbitrary dimensions, for some dimension-dependent constants $a_{(d)}$ and $b_{(d)}$. Using the available results for $t_2$ and $t_4$ in $d=4$ Quasi-topological gravity \cite{Myers:2010jv} and $d\geq 4$ Gauss-Bonnet gravity \cite{Buchel:2009sk}, it is straightforward to set: $b_{(4)}=-1/21$ and $a_{(d)}=-(d-2)(d-3)/[2d(d-1)]$. In fact, a formula equivalent to \req{tte} valid in the particular case of Lovelock theories --- for which $t_4=0$ ---  was shown to be true in \cite{Camanho:2010ru,Camanho:2013pda} for the same value of $a_{(d)}$. This provides additional support for the validity of \req{tte} for general Einstein-like theories. It would be interesting to test the validity for such additional theories in various dimensions and, if correct in general, to determine the value of $b_{(d\geq 5)}$.

\section{General formula for holographic free energy on squashed spheres}
Let us now come back to the main topic of this chapter: the holographic computation of the free energy of CFTs on squashed spheres. 
In AdS/CFT, the semiclassical partition function is exponentially dominated by the bulk geometry with the smallest on-shell action satisfying the appropriate boundary conditions. This means that the free energy of the holographic CFT can be accessed from the regularized on-shell action of the  bulk theory evaluated on the corresponding gravity solution \cite{Aharony:1999ti}.
 When the boundary geometry is a squashed-sphere of the form \req{squa}, the relevant bulk solutions are of the so-called Euclidean Taub-NUT/bolt class \cite{Hawking:1998ct,Chamblin:1998pz,Emparan:1999pm}. Such solutions are characterized by the NUT charge $n$ which, on general grounds, holography maps to the squashing parameter of the boundary geometry $\varepsilon$ through
\begin{equation}\label{n-epsilon}
\frac{n^2}{L^2} =\frac{(1+\varepsilon)}{(d+1) f_{\infty} }\, .
\end{equation}
%where $L$ is the cosmological constant action scale, and $f_{\infty}$ is defined so that $L/\sqrt{f_{\infty}}$ is the radius of the corresponding AdS$_{(d+1)}$ background.

Naturally, constructing Taub solutions is a more challenging task than classifying the vacua of the theory and, in fact, only a few examples of such solutions have been constructed for Einstein-like Lagrangians of the form \req{hog}. Fortunately, in the preceding chapter we have computed new Taub-NUT and Taub-bolt solutions in several higher-curvature gravities in various dimensions. 
The simplest instance in $d=3$ corresponds to Einsteinian cubic gravity, whose Taub-NUT/bolt solutions were obtained in Section~\ref{ECGs} \cite{NewTaub2}. % whose Lagrangian is given by \cite{PabloPablo}
%\begin{equation}\label{ecgg}
%\mathcal{L}^{\rm ECG}=\frac{1}{16\pi G}\left[\frac{6}{L^2}+R-\frac{\mu L^4}{8} \mathcal{P} \right]\, ,
%\end{equation}
%where $\mathcal{P}=12 R_{a\ b}^{\ c \ d}R_{c\ d}^{\ e \ f}R_{e\ f}^{\ a \ b}+R_{ab}^{cd}R_{cd}^{ef}R_{ef}^{ab}-12R_{abcd}R^{ac}R^{bd}+8R_{a}^{b}R_{b}^{c}R_{c}^{a}$ is a new cubic invariant and  $\mu$ is a dimensionless coupling. 
In $d\geq 5$, analytic Taub solutions have been constructed for Einstein \cite{Awad:2000gg} and Einstein-Gauss-Bonnet gravity \cite{Dehghani:2005zm,Dehghani:2006aa,Hendi:2008wq} and there have been a number of holographic applications of these solutions \cite{Astefanesei:2004kn,Clarkson:2004yp,Lee:2008yqa,Shaghoulian:2016gol}. In the previous chapter we generalized these results by constructing Taub-NUT/bolt solutions including quartic Quasi-topological and Generalized quasi-topological terms in $d=5$ \cite{NewTaub2}. 
In all these cases, the thermodynamic properties of the solutions can be accessed analytically. In particular, the computation of  regularized on-shell actions can be performed after the introduction of various boundary terms and counterterms which account for the various UV divergences \cite{Chamblin:1998pz,Balasubramanian:1999re, Brihaye:2008xu, Teitelboim:1987zz,Dehghani:2011hm,ECGholo}.  As long as the solution is the dominant saddle, the resulting on-shell action computes the free energy of the dual theory on a squashed sphere $\mathbb{S}^d_{\varepsilon}$. For sufficiently small $\varepsilon$, one finds that the relevant saddle is generically of the NUT type. For large enough $\varepsilon$, a NUT/bolt phase transition usually takes place, and even more exotic transitions between different bolt solutions are possible, as we studied in detail in Chapter~\ref{Chap:7}. 

We are interested in the behaviour near $\varepsilon=0$, where the NUT phase dominates. In order to illustrate these results, let us consider the case of ECG in $d=3$, whose action is given in \req{ECGAction}. The holographic free energy simply corresponds to the Euclidean action evaluated on the corresponding solution, and for Taub-NUT metrics the result is given in \req{freeee1}. Applying the relation \req{n-epsilon} between the NUT charge $n$ and the squashing parameter $\varepsilon$ we get
\begin{equation}\label{eq:FseECG2}
\mathcal{F}^{\rm ECG}_{\mathbb{S}_\varepsilon^{3}} = -\frac{\pi L^2 (1+\varepsilon)^2}{G f_\infty^2} \left[\frac{1}{2} - \frac{f_\infty}{(1+\varepsilon)} - \frac{\mu f_\infty^3}{(1+\varepsilon)^3} \right] \, ,
\end{equation}
where we remind that $f_{\infty}$ is a root of the equation $h(f_{\infty})=1-f_{\infty}+\mu f_{\infty}^3=0$. Additional examples are shown in Appendix~\ref{App:7}. 

Rather strikingly, we observe that the following simple pattern holds in all cases studied: the free energy of a holographic CFT on a squashed $\mathbb{S}^d_{\varepsilon}$ dual to a higher-order gravity theory can be obtained by evaluating the on-shell Lagrangian of the corresponding theory on an auxiliary pure AdS$_{(d+1)}$ space. The dependence on the squashing parameter appears encoded in the AdS radius of this auxiliary geometry, which is given by $L \sqrt{(1+\varepsilon)/f_{\infty}}$. Explicitly, we claim that the following formula holds 
\begin{equation}\label{fee0e}
\mathcal{F}_{\mathbb{S}_{\varepsilon}^{d}}=(-1)^{\frac{(d-1)}{2}}\frac{\pi^{\frac{(d+2)}{2}} }{\Gamma\left[\frac{d+2}{2}\right]}\frac{\mathcal{L}\left[f_{\infty}/(1+\varepsilon)\right] L^{d+1}}{[f_{\infty}/(1+\varepsilon)]^{\frac{(d+1)}{2}}}\, .
\end{equation}
Observe that this expression is automatically regularized, and it provides a direct entry of the holographic dictionary for at least certain higher-order gravities. In fact,  \req{fee0e}  cannot be valid for any theory since it is not invariant under field redefinitions --- at least for finite values of $\varepsilon$ --- so we have to determine for which theories it applies. We remark that the previous formula reproduces the result for the free energy of all the Taub-NUT solutions known in the literature --- including in particular the Einstein gravity result in arbitrary dimension. The point is that we only know how to compute the free energy of Taub-NUT solutions for a certain type of theories: the special Generalized quasi-topological gravities considered in Chapter~\ref{Chap:7} that allow for Taub-NUT solutions characterized by a single function \req{FFnut}. Thus, we claim that the formula \req{fee0e} computes exactly the free energy of CFTs dual these theories, which are a subset of the GQG family, but presumably an infinite one. 

However, the behaviour of \req{fee0e} near $\varepsilon=0$ turns out to be universal and provides us with relations that are valid for any Einstein-like theory. In fact, \req{fee0e} passes three non-trivial tests:

%\comment{comment here on how we realized this pattern, appendix?}
%%\begin{equation}\label{fee0e}
%F_{\mathbb{S}_{\varepsilon}^{d}}=-\frac{(2\pi)^{\frac{(d+1)}{2}} }{d!!}\frac{\mathcal{L}\left[f_{\infty}/(1+\varepsilon)\right] L^{d+1}}{[f_{\infty}/(1+\varepsilon)]^{\frac{(d+1)}{2}}}\, .
%\end{equation}

\begin{enumerate}
\item First, note that if we set $\varepsilon=0$, we must recover the result for the free energy of the theory on a round $\mathbb{S}^{d}$, which plays a crucial role in establishing monotonicity theorems, particularly in three-dimensions \cite{Klebanov:2011gs,Casini:2012ei,Pufu:2016zxm}. Evaluating \req{fee0e} at $\varepsilon=0$ we obtain
%For general odd-dimensional CFTs, this is proportional to the $a^*$ charge
\begin{equation}
\mathcal{F}_{\mathbb{S}^{d}}=-\frac{\pi^{\frac{(d+2)}{2}} \tilde L^{d-1}}{\Gamma\left[\frac{d+2}{2}\right]}  \mathcal{L}\Big|_{\rm AdS}\, ,
\end{equation}
where $\mathcal{L}|_{\rm AdS}$ stands for the Lagrangian of the corresponding theory evaluated on pure AdS$_D$ with radius $\tilde{L}=L/\sqrt{f_{\infty}}$.
Indeed, this quantity has been argued to satisfy  $\mathcal{F}_{\mathbb{S}^{d}} \propto \mathcal{L}|_{\rm AdS}$ for general higher-curvature bulk theories, with the proportionality coefficient precisely agreeing with the one predicted by \req{fee0e} --- see \eg \cite{Myers:2010tj,ECGholo}.
%find that for our odd-dimensional CFT$_{(D-1)}$'s ($D$ even), \comment{maybe use standard $d\equiv D-1$ notation here}
%\begin{equation}\label{ff}
%\mathcal{F}_{S^{(D-1)}}=-2\pi a^*\, .
%\end{equation}
%which is a first consistency check for \req{fee0e}. %Indeed, \req{ff} is nothing but the Casini-Huerta-Myers relation between the free energy of an odd-dimensional CFT on a  round $S^{D-1}$ and the entanglement entropy across a $S^{D-3}$, which holds for general theories. 

\item In addition, we know that the round sphere is a local extremum for the function $\mathcal{F}_{\mathbb{S}_{\varepsilon}^{d}}$ \cite{Bobev:2017asb}, namely, $d\mathcal{F}_{\mathbb{S}_{\varepsilon}^{d}}/d\varepsilon|_{\varepsilon=0}\equiv \mathcal{F}_{\mathbb{S}_{\varepsilon}^{d}}'(0)=0$ for general theories. This property is also nicely implemented in \req{fee0e}. Taking the first derivative in that expression and comparing with \req{hinf}, it is straightforward to show that
\begin{equation}\label{eq:Fprime}
\mathcal{F}_{\mathbb{S}_{\varepsilon}^{d}}'(\varepsilon)=(-1)^{\frac{(d-1)}{2}}\frac{\pi^{\frac{d}{2}} (d^2-1)L^{d-1}}{16\Gamma\left[\frac{d}{2}\right] Gf_{\infty}}\left(\frac{1+\varepsilon}{f_{\infty}}\right)^{(d-1)/2}h\left[f_{\infty}/(1+\varepsilon)\right]\, .
\end{equation}
Hence, evaluating at $\varepsilon=0$ we get $\mathcal{F}_{\mathbb{S}_{\varepsilon}^{d}}'(0)\propto h(f_{\infty})\, ,$
which of course vanishes by definition, as $h(f_{\infty})=0$ is nothing but the embedding condition of AdS$_{(d+1)}$ on the corresponding theory. It is remarkable how holography ties the CFT fact that round-spheres are local extrema of the free energy as a function of the squashing parameter, to the requirement that the AdS geometry solves the bulk field equations.

\item Furthermore, we know that $\mathcal{F}_{\mathbb{S}_{\varepsilon}^{d}}''(0)$ is fully determined by the stress tensor two-point function charge $\ctt$ for general odd-dimensional CFTs \cite{Bobev:2017asb}. In particular, for $d=3$ and $d=5$, it was found (in our conventions) that\footnote{Related expressions had been previously found in the context of three-dimensional $\mathcal{N}=2$ supersymmetric CFTs  \cite{Closset:2012ru} --- see also \cite{Closset:2012vp,Closset:2012vg,Hama:2011ea,Imamura:2011wg,Martelli:2011fu}. A detailed discussion of the connection can be found in section 5.1 of \cite{Bobev:2017asb}.}
\begin{equation}\label{sw2}
\mathcal{F}_{\mathbb{S}^3_{\varepsilon}}''(0)=-\frac{\pi^4}{3}\ctt\, , \quad \mathcal{F}_{\mathbb{S}^5_{\varepsilon}}''(0)=+\frac{\pi^6}{15}\ctt\, .
\end{equation}
%\begin{equation}\label{sw2}
%\mathcal{F}_{S^3_{\varepsilon}}''(0)=-\frac{\pi^2}{48}\ctt\, , \quad \mathcal{F}_{S^5_{\varepsilon}}''(0)=-\frac{3\pi^2}{320}\ctt\, .
%\end{equation}
Now, using \req{hinf}, \req{cte} and \req{fee0e}
%\begin{equation}
%\ctt = -\frac{\Gamma[D+1]\pi^{\frac{D-3}{2}} \tilde{L}^{D-2}}{2(D-2)\Gamma\left[\frac{D-1}{2} \right]^3 G} h'(f_{\infty})\, ,
%\end{equation}
%which should be valid at least for all theories sharing the linearized spectrum of Einstein gravity, together with \ref{hLD}
we find, after some manipulations,
\begin{equation}\label{genD2}
\mathcal{F}_{\mathbb{S}_{\varepsilon}^{d}}''(0)=\frac{(-1)^{\frac{(d-1)}{2}}\pi^{d+1} (d-1)^2 }{  2\,  d!}\ctt\, .
\end{equation}
%\begin{equation}\label{genD2}
%\mathcal{F}_{\mathbb{S}_{\varepsilon}^{d}}''(0)=-\frac{\pi^2 (d-1) (d-2)!!}{ 2^{(d+2)}\,  d (d-3)!! }\ctt\, .
%\end{equation}

This expression reduces to the general results in \req{sw2}, which is another highly non-trivial check of \req{fee0e}. Interestingly, it provides a generalization of the universal connection between $\mathcal{F}_{S^d_{\varepsilon}}''(0)$ and $\ctt$ which must hold for general odd-dimensional CFTs (holographic or not).

%Additional checks of formula \req{fee0e} for the available explicit examples are presented in appendix \ref{checks}.

\end{enumerate}

In passing, let us also note that from the boundary CFT point of view, \req{fee0e} tells us that the problem of computing $\mathcal{F}_{\mathbb{S}_{\varepsilon}^{d}}$ for a given theory, can actually be mapped to the one of evaluating the round $\mathbb{S}^d$ free energy for a different theory characterized by the same bulk Lagrangian, but different couplings $\tilde{\mu}_n$ such that $h(\tilde{f}_{\infty})=0$ is satisfied for $\tilde{f}_{\infty}\equiv f_{\infty}/(1+\varepsilon)$. 
Interestingly, a similar connection between $\mathfrak{F}_{\mathbb{S}_{\varepsilon}^{d}}$ and  $\mathfrak{F}_{\mathbb{S}^{d}}$ was found for $d=3$, $\mathcal{N}=2$ supersymmetric CFTs in  \cite{Hama:2011ea}.  Supersymmetry requires additional background fields to be turned on besides the metric, which makes the corresponding supersymmetric free energies  $\mathfrak{F}_{\mathbb{S}_{\varepsilon}^{d}}$ inequivalent from our $F_{\mathbb{S}_{\varepsilon}^{d}}$ \cite{Bobev:2017asb}. Nonetheless, the analogy suggests that similar mappings between the squashed and round sphere free energies may exist for higher-dimensional supersymmetric theories, or even for general CFTs.

\section{Universal expansion on the squashing parameter}

As we have seen, the leading term in the $\varepsilon\rightarrow 0$ expansion of $\mathcal{F}_{\mathbb{S}_{\varepsilon}^{d}}$ is quadratic in the deformation, and proportional to the stress-tensor two-point function charge $\ctt$ for general CFTs. A question left open in \cite{Bobev:2017asb} was the possibility that the subleading term, cubic in $\varepsilon$, could present an analogous universal behavior, in the sense of being fully characterized by the corresponding three-point function charges. 
Since $\varepsilon$ encodes a metric deformation, one expects $\mathcal{F}^{(n)}_{\mathbb{S}_{\varepsilon}^{d}}$ to involve integrated $n$-point functions of the stress tensor.
Focusing on the case of three-dimensional CFTs, the corresponding three-point function is completely fixed by conformal symmetry up to two theory-dependent quantities \cite{Osborn:1993cr}, which can be chosen to be $\ctt$ and the dimensionless parameter $t_4$ \cite{Hofman:2008ar}. Hence, we expect a linear combination of $\ctt$ and $\ctt t_4$ to appear in the $\mathcal{O}(\varepsilon^3)$ term.
The analysis in \cite{Bobev:2017asb} shows however that, besides these contributions, an additional correlator of the form $\braket{\frac{\delta T}{\sqrt{g}\delta g} T}$ --- which depends on additional details of the specific CFT --- appears at that order for general metric perturbations. The possibility that this term does not really contribute for certain metric perturbations, including our class of squashings, was left open.  

On the other hand, the available partial results --- numerical for a free scalar and a free fermion, and analytic for holographic Einstein gravity --- did not suffice to provide a conclusive answer, even in $d=3$. In particular, the exact result for the free energy in holographic Einstein gravity is a polynomial of order $2$ in $\varepsilon$, namely, $\mathcal{F}^{\rm E}_{\mathbb{S}_{\varepsilon}^{3}}=\pi L^2 (1-\varepsilon^2)/(2 G)$, which means that its Taylor expansion around $\varepsilon=0$ is trivial, and precisely ends with the quadratic piece --- which is of course controlled by $\ctt$ in agreement with \req{sw2}, as can be readily verified using \req{cte2}. 

Happily, the new Taub-NUT solutions constructed in the previous chapter for Einsteinian cubic gravity provide us with an additional family of holographic models for which we can access the cubic contribution, and explore its possible universality by testing it against the free-field numerics. Unlike the Einstein gravity case, the free energy of CFTs dual to ECG does contain a cubic term in the squashing parameter. We presented the exact result --- valid for finite values of $\varepsilon$ --- in \req{eq:FseECG2}, but expanding that formula up to cubic order we get
\begin{equation}\label{3conj1}
\mathcal{F}_{\mathbb{S}_{\varepsilon}^{3}}=(1+3\mu f_{\infty}^2)\frac{\pi L^2}{2f_{\infty}G}-\frac{\pi(1-3\mu f_{\infty}^2)L^2}{2 f_{\infty}G }\varepsilon^2\left[1+\frac{2\mu f_{\infty}^2}{1-3\mu f_{\infty}^2}\varepsilon + \mathcal{O}(\varepsilon^2) \right]\, .
\end{equation} 

Now, using the result obtained in Chapter~\ref{Chap:6} for the holographic mapping between boundary and bulk quantities of Einsteinian cubic gravity  \cite{ECGholo}, 
\begin{align}
\ctt^{\rm ECG}=(1-3\mu f_{\infty}^2)\frac{3L^2}{\pi^3 f_{\infty}G}\, ,\quad \ctt^{\rm ECG} t_4^{\rm ECG}=-3780\mu f_{\infty}\frac{L^2}{\pi^3 G}\, ,
\end{align}
we can express the squashed-sphere free energy of the corresponding dual theory for small values of $\varepsilon$ as 
\begin{equation}\label{3conj}
\mathcal{F}_{\mathbb{S}_{\varepsilon}^{3}}=\mathcal{F}_{\mathbb{S}_0^{3}}-\frac{\pi^4\ctt}{6}\varepsilon^2\left[1-\frac{t_4}{630}\varepsilon + \mathcal{O}(\varepsilon^2) \right]\, .
\end{equation} 

 As we can see, the leading correction to the round-sphere result agrees with the general result \req{sw2}, as it should. But now we have a nontrivial subleading piece, cubic in $\varepsilon$, and proportional to $\ctt t_4$. 
In principle, it is far from obvious that the cubic term should not depend on additional theory-dependent quantities. In order to test its generality, we can extend the calculation, done here for ECG, to other higher-order gravities. Let us take  our general formula for the free energy \req{fee0e} as the starting point. We have argued that this formula applies for an infinite number of theories ---at least for all the special Generalized quasi-topological gravities of the type studied in Chapter~\ref{Chap:7} --- and we have performed an explicit check for ECG and for a quartic GQG --- see Appendix \ref{App:7}. 
According to this formula, the third derivative of the free energy with respect to $\varepsilon$ reads
\begin{align}\notag
\mathcal{F}_{\mathbb{S}^d_{\varepsilon}}^{(3)}(0)=& \, -\frac{\pi L^{2}}{G}h''(f_{\infty})\, .
\end{align}
But now, we have seen that $h''(f_{\infty})$ is related to $t_4$ for a general GQG in $D=4$ according to \req{t4d}, this is, $t_4=210 f_{\infty} h''(f_{\infty})/h'(f_{\infty})$. Taking also into account the relation between $\ctt$ and $h'(f_{\infty})$ given by \req{cte}, we see that the third derivative $\mathcal{F}_{\mathbb{S}^d_{\varepsilon}}^{(3)}(0)$ precisely matches the prediction of \req{3conj}:
\begin{align}
\mathcal{F}_{\mathbb{S}^d_{\varepsilon}}^{(3)}(0)=& \, \frac{\pi^4 \ctt t_4}{630}\, .
\end{align}
Since both formulas used in the derivation of this result, \req{fee0e} and \req{t4d}, apply at the same time to an infinite number of higher-order gravities, it follows that for all these theories the expansion \req{3conj} holds. Thus, we are not able to find counterexamples to \req{3conj} within this family of theories. The robustness of the holographic result strongly suggests that this relation could actually be valid for general CFTs.

%For example, in Section~\ref{quarticAp} we computed the corresponding Taub-NUT solutions for a quartic GQG in $D=4$.  The free energy of a holographic CFT computed by evaluation of the Euclidean action on these solutions agrees with the general proposed formula \req{fee0e} --- see Appendix \ref{App:7} --- and there is no reason to expect that the situation does not extend to higher orders in the curvature. Therefore,  we can consider the formula \req{fee0e} applied to a Lagrangian containing GQG terms of arbitrarily high order. According to \req{fee0e}, the third derivative with respect to $\varepsilon$ reads
%\begin{align}\notag
%\mathcal{F}_{\mathbb{S}^d_{\varepsilon}}^{(3)}(0)=& \, -\frac{\pi L^{2}}{G}h''(f_{\infty})\, .
%\end{align}
%But we have seen that $h''(f_{\infty})$ is related to $t_4$ for a general GQG in $D=4$ according to \req{t4d}, this is, $t_4=210 f_{\infty} h''(f_{\infty})/h'(f_{\infty})$. Taking also into account the relation between $\ctt$ and $h'(f_{\infty})$ given by \req{cte}, we see that the third derivative $\mathcal{F}_{\mathbb{S}^d_{\varepsilon}}^{(3)}(0)$ precisely matches the prediction of \req{3conj}. Therefore, that formula applies at least to an infinite number of holographic higher-order gravities, which strongly suggests that it is actually valid for general CFTs.

\begin{figure}[t!]
\center
	\includegraphics[width=0.65\textwidth]{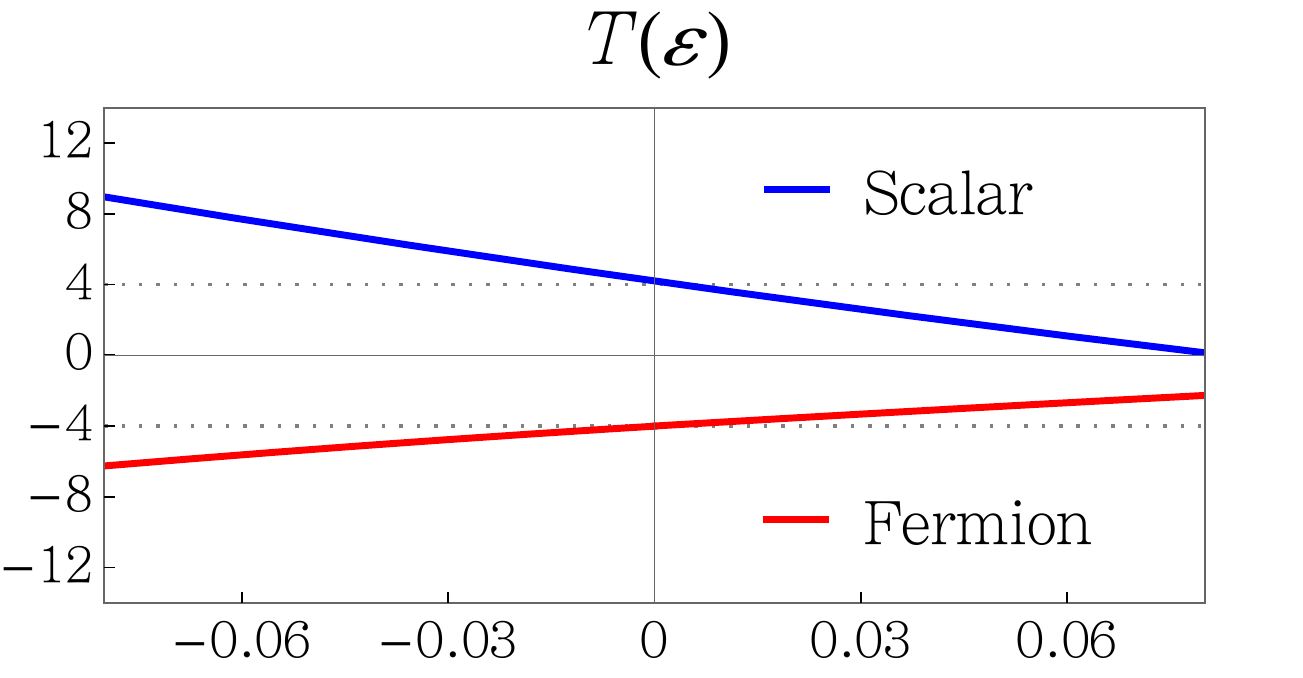}
	\caption{We plot the function $T(\varepsilon)$ defined in \req{tep} near $\varepsilon=0$ for a free scalar (blue) and a free fermion (red) using the numerical results for $\mathcal{F}^{\rm s,f}_{\mathbb{S}_{\varepsilon}^{3}}$ obtained in \cite{Bobev:2017asb}. We observe that $T(\epsilon=0)=t_4$ is satisfied in both cases with high accuracy, which provides strong evidence in favor of the conjectural general expression \req{3conj} suggested by the holographic Einsteinian cubic gravity result. Further details on the numerical calculations used to produce this plot can be found in Appendix~\ref{ffc}.}
	\label{fig.2}
\end{figure}

Finally, we can use the numerical free-field results in \cite{Bobev:2017asb} to perform two highly nontrivial tests of the possible validity of \req{3conj} beyond holography. In order to do so, we study the function
 \begin{equation}\label{tep}
T(\varepsilon)\equiv \frac{630}{\varepsilon}\left[ 1+\frac{6(\mathcal{F}_{\mathbb{S}_{\varepsilon}^{3}}-\mathcal{F}_{\mathbb{S}_0^{3}}) }{\pi^4 \ctt \varepsilon^2} \right]
 \end{equation}
 for the conformally-coupled scalar (s) and the free Dirac fermion (f) free energies near $\varepsilon=0$ --- details on the numerical method utilized in the computation of $\mathcal{F}^{\rm s}_{\mathbb{S}_{\varepsilon}^{3}}$ and $\mathcal{F}^{\rm f}_{\mathbb{S}_{\varepsilon}^{3}}$ can be found in Appendix~\ref{ffc}. Naturally, if \req{3conj} held for these theories, we should obtain $T(\varepsilon=0)=t_4$, which for the scalar and the fermion are respectively given by $t^{\rm s}_4=+4$ and $t^{\rm f}_4=-4$ \cite{Osborn:1993cr,Buchel:2009sk}. The result of this analysis is shown in Fig. \ref{fig.2}, where it is manifest that this is precisely satisfied in both cases. The extremely different nature of the theories and techniques used in deriving the holographic and free-field results make us think that this property extends to arbitrary CFTs.% More explicitly, we propose the following conjecture. 
% At this point, it is natural to wonder whether \req{3conj} may hold more generally than in this particular holographic model.

\begin{itemize}
\item \textit{Conjecture:} for general three-dimensional CFTs, the subleading term in the  squashing-parameter $\varepsilon$ expansion of the free energy $\mathcal{F}_{\mathbb{S}_{\varepsilon}^{3}}$ is universally controlled by the coefficient $t_4$ in the three-point function of the stress tensor. In particular, we conjecture that \req{3conj}
%\begin{equation}\label{3conj}
%\mathcal{F}_{\mathbb{S}_{\varepsilon}^{3}}=\mathcal{F}_{\mathbb{S}_0^{3}}-\frac{\pi^2\ctt}{96}\varepsilon^2\left[1-\frac{t_4}{630}\varepsilon + \mathcal{O}(\varepsilon^2) \right]\, ,
%\end{equation} 
holds for general theories. 
\end{itemize}
The level of evidence provided here in favor of \req{3conj} --- involving free-field and holographic higher-order gravity calculations ---  is very similar to the one initially presented in \cite{Bueno1,Bueno2} concerning the universal relation between the entanglement entropy of almost-smooth corner regions and the charge $\ctt$, which was eventually proven for general CFTs in \cite{Faulkner:2015csl}\footnote{In contrast to \req{3conj}, however, the subleading term in the smooth-limit expansion of the corner entanglement entropy (quartic in the deformation), was later shown not to be generically controlled by the stress tensor three-point function charges in \cite{Bueno:2015ofa}.}.

%The evidence comes from holographic theories (Einstein gravity and Einsteinian cubic gravity), plus scalar and fermion.
%\textbf{This holds (highly non-trivially) for the free scalar and the fermion!!}

One would expect that if our conjecture is true, an analogous expression should hold for the free energy of higher odd-dimensional squashed spheres. In that case, one would expect the $\mathcal{O}(\varepsilon^3)$ term to be controlled by some combination of $\ctt$, $t_4$ and the additional stress-tensor three-point function charge, $t_2$, which is nonvanishing for $d>3$.
In order to guess the exact relation, say, in $d=5$, one could compute $t_2$ and $t_4$ holographically for some of the six-dimensional bulk theories for which Taub-NUT solutions have recently been constructed  \cite{NewTaub2}, and follow the same steps taken here for Einsteinian cubic gravity. 

%As a related comment, let us mention that %plotting $\mathcal{F}_{\mathbb{S}_{\varepsilon}^{3}}$ for Einsteinian cubic gravity for general values of $\varepsilon$ --- see Appendix \ref{ffc} --- we observe 
%that 
%the sign of $t_4$ determines whether or not the round-sphere is a global maximum of the free energy as a function of $\varepsilon$ for Einsteinian cubic gravity, a feature which is also shared by the free fields \cite{Bobev:2017asb} --- see Appendix \ref{ffc}. Whenever $t_4\geq 0$, the round sphere is a global maximum, and whenever $t_4<0$, the free energy reaches a local minimum at some negative intermediate value of $\varepsilon$, and then grows past the round-sphere value as it approaches the limiting value $\varepsilon=-1$. It is tempting to speculate that this is also a general feature.

%provides an example of the power of higher-order gravities as holographic toy models from which universal information about general CFTs can be obtained

%\comment{blah blah it would be interesting to check this relation for additional theories, and perhaps try to prove it in general using field-theoretical techniques.}

\section{Discussion} 
In this chapter, we have presented a general formula for the free energy of odd-dimensional CFTs dual to a certain family of higher-curvature gravities with second-order linearized equations of motion. The formula \req{fee0e} is expected to hold in the region of parameter space for which Taub-NUT geometries dominate the corresponding semiclassical partition function, something that generically occurs for small enough values of $\varepsilon$. In our formula, UV regularization is automatically implemented as it only involves the evaluation of the Lagrangian of the corresponding theory on an auxiliary AdS geometry, which represents a drastic simplification with respect to the usual on-shell action approaches --- to the extent that it does not even require knowing the corresponding Taub-NUT bulk geometry. 
We have argued that \req{fee0e} satisfies various highly nontrivial properties expected from general CFT considerations \cite{Bobev:2017asb}, which AdS/CFT elegantly connects to bulk statements. Additionally, our formula is also satisfied in all known cases in which the corresponding holographic calculation involving the on-shell action of Taub-NUT geometries has been performed. Additional checks for other holographic theories or, preferably, a general proof of \req{fee0e} would be very desirable. 

Furthermore, we have conjectured that the subleading term in the free-energy squashing-parameter expansion is universally controlled by the stress-tensor three-point function coefficient $t_4$, as given in \req{3conj}, for general $(2+1)$-dimensional CFTs (holographic or not). In deriving \req{3conj}, we first have made use of the free energy result for holographic Einsteinian cubic gravity.  Then, we have seen that this result also applies when one takes into account the general formula \req{fee0e} together with the general expression \req{t4d} for the coefficient $t_4$ --- valid at least for all the GQGs in $D=4$ --- and finally we have cross-checked it with the numerical results corresponding to a conformally-coupled scalar and a free Dirac fermion, finding perfect agreement. 
These checks provide astonishing evidence of the validity of the conjecture \req{3conj}. In particular, we do not seem to be able to find a counterexample of the result within the holographic setup. Thus, the next natural step would be attempting a proof of \req{3conj} using field-theoretical techniques. However, as we mentioned earlier, the kind of integrals one would need to perform in that case look rather challenging \cite{Bobev:2017asb}.

It is not clear as of this moment whether the results presented for $d=3$ extend in some way to higher-dimensional CFTs. According to our general formula \req{fee0e} --- that we expect to be of applicability for an infinite set of higher-order gravities --- the value of the third derivative of the free energy reads
\begin{align} \label{sw3} 
\mathcal{F}_{\mathbb{S}^d_{\varepsilon}}^{(3)}(0)= \, \frac{(-1)^{\frac{(d+1)}{2}}\pi^{\frac{d}{2}}(d^2-1)L^{d-1}}{16\Gamma[\frac{d}{2}]f_{\infty}^{\frac{d-1}{2}} G}\left[(d-3)h'(f_{\infty})   -f_{\infty}h''(f_{\infty}) \right].
\end{align}
Therefore, the question whether $\mathcal{F}_{\mathbb{S}^d_{\varepsilon}}^{(3)}(0)$ depends only on the stress-energy tensor 3-point function coefficients is equivalent to whether $h''(f_{\infty})$ satisfies a relation as the one in \req{tte} for odd $d$. In order to test that, one would need to compute the energy fluxes \cite{Hofman:2008ar,Buchel:2009sk,Myers:2010jv} of several GQG theories of the type presented in Chapter \ref{Chap:7} --- such as \req{eq:GQGTN1} and \req{eq:GQGTN1} --- and check if the coefficients $t_2$, $t_4$ satisfy such relation. We have seen that in $d=3$ that relation exists --- see Eq.~\req{t4d} --- and consequently the result \req{3conj} seems to be universal.

In any event, we find that $h^{(n)}(f_\infty)$ appears to be related to the $(n+1)$-point function of the boundary stress tensor, therefore acting as some sort of generating functional. In fact, the 1-point function $h(f_{\infty})=0$ vanishes because it corresponds to the vacuum embedding equation. The 2-point function is generically determined by $h'(f_{\infty})$ according to \req{cte} for arbitrary Einstein-like theories, and, at least in $d=3$, the 3-point function is determined by $h''(f_{\infty})$. Interestingly,  the ``zero-point function'', corresponding to the regularized round-sphere free energy $\mathcal{F}_{\mathbb{S}^d}$,  also satisfies this pattern, as it can be extracted from an integral involving $h(f_{\infty})$, namely\footnote{In even-dimensional CFTs, this expression yields --- up to a  $2(-1)^{-1/2}/\pi$ factor --- the coefficient of the universal logarithmic contribution to the corresponding round-sphere free energy, given by $(-1)^{\frac{(d-2)}{2}}4a^*$, where $a^*$ is proportional to one of the trace-anomaly charges ($a$ in $d=4$), \eg \cite{Myers:2010tj}.  }
\begin{equation}\label{0p}
\mathcal{F}_{\mathbb{S}^d}=\frac{(-1)^{\frac{(d+1)}{2}} \pi^{\frac{d}{2}} (d+1)(d-1)L^{d-1}}{16 \Gamma\left[\frac{d}{2}\right]  G }\int^{f_{\infty}} \frac{ h(x)}{  x^{\frac{(d+3)}{2}}} dx\, ,
\end{equation}
which can be obtained by integrating formally both sides of \req{eq:Fprime}. In this expression we are supposed to evaluate the ``canonical'' primitive of the integrand on $f_{\infty}$, which, in the case $h(x)$ is a polynomial, is straightforward to obtain. 

Integrating by parts in this expression, and using \req{cte} and \req{cte2}, it is possible to find the suggestive relation
\begin{equation}\label{conne}
\ctt =\frac{(-1)^{\frac{(d-1)}{2}}  \Gamma[d+2] }{\pi^{d+1}(d-1)^2} \, f_{\infty}\left[ \frac{\partial \mathcal{F}_{\mathbb{S}^d}}{\partial f_{\infty}} \right]\, ,
\end{equation}
which is equivalent to the one recently found in \cite{Li:2018drw}, and which connects two seemingly unrelated quantities, such as $\ctt$ and  $\mathcal{F}_{\mathbb{S}^d}$.\footnote{In terms of $a^*$, the relation reads $\ctt= -2\Gamma[d+2]/[\pi^d (d-1)^2] \cdot f_{\infty} \left[\frac{\partial a^*}{\partial f_{\infty}}\right]$, which is valid in general (even and odd) dimensions.} Let us note that there is an ambiguity in this expression because there are many ways to express $\mathcal{F}_{\mathbb{S}^d}$ as a function of $f_{\infty}$ --- in fact, this quantity is supposed to be a constant fixed by $h(f_{\infty})=0$. There is a canonical form of $\mathcal{F}_{\mathbb{S}^d}$ in which \req{conne} works, and this form is $\mathcal{F}_{\mathbb{S}^d}=\mathcal{F}^{\rm E}_{\mathbb{S}^d}\left(1+\text{polynomial in }f_{\infty}\right)$, where $\mathcal{F}^{\rm E}_{\mathbb{S}^d}$ is the Einstein gravity result with the rescaled AdS scale $L/\sqrt{f_{\infty}}$. Expressed in this way, Eq.~\req{conne} remains the form of a flow equation for $\mathcal{F}_{\mathbb{S}^d}$ controlled by the parameter $f_{\infty}$.

%It is also natural to wonder about correlators involving four or more stress-tensors. The situation for those is a bit more complicated, as conformal symmetry no longer fixes them completely up to a few theory-dependent constants. %Hence, the connection between $\mathcal{F}_{S^3_{\varepsilon}}^{(n\geq 4)}(0)$, $h^{(n\geq 3)}(f_{\infty})$ and such correlators is less evident to us.

\chapter*{Conclusions\label{conclusions}}
\addcontentsline{toc}{chapter}{Conclusions}

%%%%%%%%%%%%%%%%%%%%%%%%%%%%%%%%%%%%%%%%%%%%%%%%%%%%%%%%%%%%%%%%%%%%%%
In this thesis, we have made an important progress in the classification of higher-curvature gravities according their linearized spectrum and a special family of theories possessing second-order linearized equations of motion has been identified. These theories have been called ``Einstein-like'' and they only propagate a massless graviton on constant curvature backgrounds. Furthermore, a special subset of these theories, known as Generalized quasi-topological gravities, turn out to be especially appropriate in order to study black hole solutions. In this way, we have been able to describe, non-perturbatively, four-dimensional black hole solutions with higher-curvature corrections. In particular, we managed to study the corresponding black hole solutions when the Einstein-Hilbert action is supplemented with an infinite number of higher-curvature terms. It should be noted that this is a quite remarkable achievement that does not have a precedent in the literature. Besides, we have observed that all these corrections have universal effects on the thermodynamic properties of black holes. In particular, the Hawking's temperature of the corrected black holes vanishes in the zero-mass limit --- an opposite behaviour to the one found in Einstein gravity, where the temperature diverges. As a consequence, small black holes are thermodynamically stable and do not evaporate in a finite time. In addition, since we argue that the theories considered could serve as a basis to construct the most general effective theory for gravity, there is a chance that these conclusions apply with generality beyond the concrete theories studied here. As an interesting consequence of the existence of small, stable black holes, we raise the possibility that these could be a constituent of dark matter. 

On the other hand, we have studied the holographic applications of Einsteinian cubic gravity in the context of the AdS$_4$/CFT$_3$ correspondence. The conformal field theory dual to Einsteinian cubic gravity turns out to belong to a different universality class from the one defined by holographic Einstein gravity, and this has allowed us to obtain new results. As a highlight, we have been able to establish that the partition function of a three-dimensional conformal field theory placed on a squashed sphere is determined, up to cubic order in the squashing parameter, by the two- and three-point functions of the stress-energy tensor.  Nevertheless, the main conclusion is that Einsteinian cubic gravity represents a very interesting holographic toy model that allows us to obtain consistent answers about conformal field theories. In this sense, we expect that all of the Generalized quasi-topological gravities --- which constitute a numerous family of theories --- will also give rise to consistent holographic duals which, on the other hand, could be easily studied. 

\appendix
%\include{appHEEHvLf}

%\appendix

\chapter{Linearized $\mathcal{L}($Riemann$)$ theories}

\section{Linearization procedure: examples}\label{lineapp}
In this appendix we apply the linearization procedure explained in section \ref{section2} to two instances. The first is a general quadratic theory in $D$-dimensions, for which we give details of all the steps involved in the linearization process. The second is a Born-Infeld gravity. Our goal in that case is to illustrate that our method can be easily applied to theories whose linearization would be difficult to achieve using different methods. 

\label{recipeex}
\subsubsection*{ Quadratic gravity}
Let us consider the most general quadratic gravity in general dimensions,
\begin{equation}
	S=\frac{1}{16\pi G}\int_{\mathcal{M}} d^Dx\sqrt{|g|}\left\{-2\Lambda+R + \alpha_1 R^2+\alpha_2 R_{\mu\nu}R^{\mu\nu}+\alpha_3 R_{\mu\nu\sigma\rho}R^{\mu\nu\sigma\rho}\right\}\, .
\end{equation}
In order to obtain $\mathcal{L}(\mathcal{K}, \alpha)$, we only have to substitute the Riemann tensors appearing in the above Lagrangian density by the expression (\ref{Riemalpha}) and use the algebraic properties of the auxiliary tensor $k_{\mu\nu}$ (\ref{k-def}) to compute all the contractions. We find
\begin{equation}
	\begin{aligned}
		R^2 \Big|_{(\mathcal{K},\alpha)} &=  \mathcal{K}^2 D^2(D-1)^2 + 2 \mathcal{K} \alpha D (D-1) \chi (\chi -1)+\alpha^2 \chi^2(\chi-1)^2\, , \\
		R_{\mu\nu} R^{\mu \nu} \Big|_{(\mathcal{K},\alpha)}&=  \mathcal{K}^2 D (D-1)^2 + 2 \mathcal{K} \alpha (D-1) \chi (\chi  - 1)   + \alpha^2 \chi (\chi - 1 )^2\, ,\\ 
		R_{\mu\nu\sigma\rho}R^{\mu\nu\sigma\rho}\Big|_{(\mathcal{K},\alpha)}
		%&=4(\mathcal{K} g_{\mu[\sigma}g_{\rho]\nu}+\alpha k_{\mu[\sigma}k_{\rho]\nu})(\mathcal{K} g^{\mu[\sigma}g^{\rho]\nu}+\alpha k^{\mu[\sigma}k^{\rho]\nu})\\
		%&=4\mathcal{K}^2 g_{\mu[\sigma}g_{\rho]\nu}g^{\mu[\sigma}g^{\rho]\nu}+8\mathcal{K}\alpha g_{\mu[\sigma}g_{\rho]\nu}k^{\mu[\sigma}k^{\rho]\nu}+4\alpha^2k_{\mu[\sigma}k_{\rho]\nu}k^{\mu[\sigma}k^{\rho]\nu}\\
		&=2D(D-1)\mathcal{K}^2+4 \mathcal{K} \alpha \chi(\chi-1)+2 \alpha^2 \chi(\chi-1)\, .
	\end{aligned}
\end{equation}
The final result for $\mathcal{L}(\mathcal{K}, \alpha)$ reads
	\begin{align}
		 \mathcal{L}(\mathcal{K},\alpha)
		=&\frac{1}{16 \pi G}\Big[-2\Lambda+\mathcal{K} D(D-1) + \alpha \chi (\chi-1)  \\ \notag
		&+\big(\mathcal{K}^2 D(D-1) + 2\mathcal{K}   \alpha \chi (\chi - 1) \big)    \big(D(D-1)\alpha_1 +(D-1)\alpha_2+2\alpha_3 \big)   \\ \notag
		&+ \alpha^2  \chi  (\chi-1) \big(\chi(\chi-1)\alpha_1+(\chi-1)\alpha_2+2\alpha_3 \big)\Big].
	\end{align}
Then, applying (\ref{tete}) we get
\begin{equation}
	e = \frac{1}{16\pi G}\left[\frac{1}{2}  +  \mathcal{K} \big(D(D-1)\alpha_1 +(D-1)\alpha_2+2\alpha_3 \big)\right]\, .
\end{equation}
The second derivative with respect to $\alpha$ yields
\begin{equation}
	\frac{\partial^2 \mathcal{L}}{\partial \alpha^2}= \frac{\chi (\chi - 1)}{8\pi G}\Big[\chi(\chi-1)\alpha_1+(\chi-1)\alpha_2+2\alpha_3\Big]\, .
\end{equation}
Hence, comparing with (\ref{teta}) we can easily obtain the values of $a$, $b$ and $c$. The result is
\begin{equation}
	a=\frac{ \alpha_3}{16 \pi G}\, , \quad   b=  \frac{\alpha_1}{32 \pi G }\, , \quad   c=  \frac{\alpha_2}{32 \pi G}\, .
\end{equation}
Inserting the values of $a,b,c$ and $e$ into (\ref{kafka})-(\ref{kafka2}) gives rise to equations \req{quadrit1}-\req{quadrit3} for $\kappa_{\rm eff}=8\pi G_{\rm eff}$, $m_s^2$ and $m_g^2$. 
%\begin{equation}
%	\begin{aligned}
%		\kappa_{\text{eff}} &= \frac{ \kappa }{  1+ 4 \mathcal{K} \kappa (\alpha_1 D(D-1)+\alpha_2 (D-1)-2 \alpha_3 (D-4))    }
%		\\
%		m_s^2 &= \frac{(D-2) + 4 (D-4)   \mathcal{K} \kappa   \left(  \alpha_1 D (D-1)   + \alpha_2 (D-1)  +  2
%			\alpha _3\right)}{2 \kappa 
%			\left(4 \alpha _1 (D-1)+\alpha _2 D  +   4 \alpha _3\right)}   \\
%		m_g^2 &=  \frac{-1  - 4 \mathcal{K} \kappa \left (   \alpha_1 D (D-1)    + \alpha_2 (D-1) 
%			-2      \alpha _3 (D-4)  \right)
%		}{2 \kappa \left(\alpha _2+4 \alpha _3\right)  }   .
%	\end{aligned}
%\end{equation}
%For example, for Gauss-Bonnet gravity ($\alpha_1 = \alpha_3=-\frac{1}{4}\alpha_2 = \alpha$) these quantities reduce, as expected, to
%\begin{equation}
%	\kappa_{\text{eff}} =\frac{\kappa }{1 + 4  \mathcal{K} \kappa  \alpha  (D-4) (D-3) 
%	}, \,\,
%	m_s  = + \infty,\,\,
%	m_g  = + \infty \quad \text{(Gauss-Bonnet)}.
%\end{equation}

Finally, from (\ref{Lambda-eq}) we see that the cosmological constant is related to the background scale $\mathcal{K}$ and the couplings of the theory through
\begin{equation}
	\Lambda = \frac{(D-1) (D-2) \mathcal{K}}{2} +  \mathcal{K}^2 (D-4) (D-1)  \big[ D (D-1) \alpha_1 + (D-1) \alpha_2 +2\alpha_3 \big]\,.
\end{equation}

\subsubsection*{Born-Infeld gravity}
Let us now consider the following theory, which has the form of a Born-Infeld model
\begin{equation}
	S=\frac{1}{\kappa^{\frac{D}{D-2}}  (1+\lambda)^{\frac{D-2}{2}}}   \int_\mathcal{M} d^Dx\left[\sqrt{ \big|g_{\mu\nu}(1+\lambda)+\kappa^{\frac{2}{D-2}} R_{\mu\nu} \big|}-\sqrt{|g_{\mu\nu}|}\right]\, ,
\end{equation}
where $|A_{\mu\nu}|$ stands for the absolute value of the determinant and $\lambda$ is a dimensionless parameter --- which we assume to be greater than $-1$. We work with $\kappa$ instead of $8\pi G$ in order to simplify the formulas. The normalization is chosen so that to leading order the action becomes Einstein-Hilbert
\begin{equation}
	S=\frac{1}{2\kappa}\int_\mathcal{M} d^Dx\sqrt{|g|}\Big[-2\Lambda+R+...\Big]\,,
\end{equation}
where $\Lambda=\big[(1+\lambda)^{1-D/2}-(1+\lambda)\big]\kappa^{\frac{2}{2-D}}$, and the ellipsis mean an infinite series of higher order terms in curvature.  Linearizing this theory can be a non-trivial task, due to the presence of the determinant and the square root. Using our method, it becomes quite easy though. First, extracting as common factor the square root of the metric determinant\footnote{We use that $|A_{\mu\nu}|=|g_{\mu\alpha}A^{\alpha}_{\ \nu}|=|g_{\mu\nu}||A^{\alpha}_{\ \beta}|$.}, we find the Lagrangian density
\begin{equation}
	\kappa^{\frac{D}{D-2}}  (1+\lambda)^{\frac{D-2}{2}}     \mathcal{L}=\sqrt{|(1+\lambda)\delta^{\mu}_{\ \nu}+\kappa^{\frac{2}{D-2}} R^{\mu}_{\ \nu}|}-1.
\end{equation}
Now, we follow our recipe and  substitute the ``Riemann tensor'' \req{Riemalpha} in this expression
\begin{equation}
	\kappa^{\frac{D}{D-2}}  (1+\lambda)^{\frac{D-2}{2}}    \mathcal{L}(\mathcal{K},\alpha)=\sqrt{ \Big| \left(1+\lambda+\kappa^{\frac{2}{D-2}} \mathcal{K}(D-1) \right)\delta^{\mu}_{\ \nu}+\alpha\kappa^{\frac{2}{D-2}} (\chi-1) k^{\mu}_{\ \nu} \Big|}-1\, .
\end{equation}
The determinant can be computed using (\ref{k-def}) and the identity
\begin{equation}
	|A|=e^{\operatorname{tr}(\log A)}\, .
\end{equation}
The result is 
\begin{equation}
	\kappa^{\frac{D}{D-2}}  (1+\lambda)^{\frac{D-2}{2}}     \mathcal{L}(\mathcal{K},\alpha)= \big ( 1+\lambda+\kappa^{\frac{2}{D-2}}\mathcal{K}(D-1) \big)^{D/2}\Big(1+\frac{\alpha\kappa^{\frac{2}{D-2}}(\chi-1)}{1+\lambda+\kappa^{\frac{2}{D-2}}\mathcal{K}(D-1)}\Big)^{\chi/2}-1\, .
\end{equation}
This ``prepotential'' contains all the information about the linearized theory. Let us begin by determining $\mathcal{K}$. The equation for the background curvature (\ref{Lambda-eq}) becomes
\begin{equation}
	\big[1+\lambda+\kappa^{\frac{2}{D-2}}\mathcal{K}(D-1)\big]^{D/2}-1=\kappa^{\frac{2}{D-2}} \mathcal{K}(D-1)\big[1+\lambda+\kappa^{\frac{2}{D-2}} \mathcal{K}(D-1)\big]^{D/2-1}\, .
\end{equation}
A simple algebraic manipulation yields
\begin{equation}
	1=(1+\lambda)\big[1+\lambda+\kappa^{\frac{2}{D-2}}\mathcal{K}(D-1)\big]^{D/2-1}\, .
\end{equation}
Thus, since we have assumed $\lambda>-1$, this equation has always one solution:
\begin{equation}
	\mathcal{K}=\frac{1}{\kappa^{\frac{2}{D-2}}(D-1)}\big[(1+\lambda)^{-2/(D-2)}-(1+\lambda)\big]\, .
	\label{backgroundLambda}
\end{equation}
Now we can compute the parameters $a,b,c$ and $e$. From (\ref{tete}) we get
\begin{equation}
	e=\frac{1}{4\kappa}(1+\lambda)^{-D/2}\,,
\end{equation}
where we already evaluated the expression on the background.
On the other hand, the second derivative of $\mathcal{L}(\mathcal{K},\alpha)$ with respect to $\alpha$ evaluated at $\alpha=0$ yields
\begin{equation}
	\frac{1}{4\chi(\chi-1)}\frac{\partial^2\mathcal{L}}{\partial\alpha^2}\Big|_{\alpha=0}=\frac{1}{16} \kappa ^{\frac{4-D}{D-2}}(\chi-1)(\chi-2)(1+\lambda)^{-\frac{D^2-2D-4}{2(D-2)}}\, ,
\end{equation}
where we have also made use of (\ref{backgroundLambda}). Now, comparing this expression with (\ref{teta}), we find the values of the parameters, namely
\begin{equation}
	a=0, \quad b=\frac{1}{16} \kappa ^{\frac{4-D}{D-2}} (1+\lambda)^{-\frac{D^2-2D-4}{2(D-2)}}, \quad c=-\frac{1}{8} \kappa ^{\frac{4-D}{D-2}} (1+\lambda)^{-\frac{D^2-2D-4}{2(D-2)}}\, .
\end{equation}
Finally, using  (\ref{kafka})-(\ref{kafka2}) we can compute the physical parameters $\kappa_{{\text{eff}}}$, $m_s$ and $m_g$
\begin{equation}
	\kappa_{{\text{eff}}}=\kappa(1+\lambda)^{D/2}\, , \quad m_s^2=2(1+\lambda) \kappa^{\frac{2}{2-D}} \, , \quad m_g^2= 2(1+\lambda)^{-2/(D-2)} \kappa^{\frac{2}{2-D}} \,.
\end{equation}
Therefore, we have completely characterized the linearized spectrum of this Born-Infeld model. Since we assumed that $\lambda>-1$, all quantities are finite and real, and everything is well-defined. For $D>2$, the background (\ref{backgroundLambda}) is dS ($\mathcal{K}>0$) when $\lambda<0$, AdS ($\mathcal{K}<0$) when $\lambda>0$ and flat when $\lambda=0$. In all cases we have, apart from the massless graviton, a massive scalar and a massive spin-2 graviton. The masses squared and the effective gravitational constant are always positive.

\section{Classification of theories: examples}\labell{Classificationexamples}
In this appendix we provide numerous examples of the different classes of theories characterized in section \ref{Classification}.

\subsection*{Theories without massive graviton}
In this appendix we will study general $f($Lovelock$)$ theories, which are a paradigmatic example of theories which only propagate the usual massless graviton plus the scalar at the linearized level \cite{Love}.

\subsubsection*{$f($Lovelock$)$ gravities}
The most general $f($Lovelock$)$ action can be written as
\begin{equation}
S= \int_{\mathcal M} d^D x \sqrt{|g|}  f (\mathcal{X}_2,\mathcal{X}_4, \dots, \mathcal X_{2\lfloor D/2 \rfloor})\, ,
\end{equation} 
where $f$ is some differentiable function of the dimensionally-extended Euler densities\footnote{Namely,  $\mathcal L_k$ becomes the Euler density when evaluated for a $2k$-dimensional manifold.}
\begin{equation}
\mathcal{X}_{2k}  \equiv  \frac{1}{2^k} \delta_{\alpha_1 \beta_1 \dots \alpha_k \beta_k}^{\mu_1 \nu_1 \dots \mu_k \nu_k} {R_{\mu_1 \nu_1}}^{\alpha_1 \beta_1} \cdots {R_{\mu_k \nu_k}}^{\alpha_k \beta_k}\, ,
\end{equation}
where the generalized Kronecker symbol is defined as $\delta^{\mu_1 \nu_1 \dots   \mu_k \nu_k}_{\alpha_1 \beta_1 \dots \alpha_k \beta_k} \equiv (2k)! \delta^{[\mu_1}_{\alpha_1} \delta^{\nu_1}_{\beta_1} \cdots \delta^{\mu_k}_{\alpha_k} \delta^{\nu_k]}_{\beta_k}$.
Note that the first two densities are nothing but the Einstein-Hilbert term, $\mathcal{X}_2=R$ and Gauss-Bonnet gravity, $\mathcal{X}_4=R^2  - 4 R_{\mu\nu}R^{\mu\nu} + R_{\mu\nu\rho\sigma} R^{\mu\nu\rho\sigma}$.
A corollary from the results presented in section \ref{fscalars} is that $f($Lovelock$)$ theories inherit the property of Lovelock gravities of not propagating the massive graviton\footnote{In appendix \ref{fscalarsapp} we show how the linearized equations of $f(R)$ can be obtained from those of Einstein gravity. The procedure can be naturally applied as well to $f($Lovelock$)$ theories starting from Lovelock, and the results will match the ones presented in this appendix.}. 
This means that the linearized equations  of motion for $f($Lovelock$)$ gravities should not involve the $\bar{\square}G^L_{\mu\nu}$ term. This is indeed the case. In particular, they read \cite{Love}
\begin{equation}
\mathcal E_{\mu\nu}^L = \alpha \, G_{\mu\nu}^L + \mathcal{K} \, \beta  \, \bar g_{\mu\nu} R^L + \frac{\beta}{D-1} \left ( \bar g_{\mu\nu} \bar \Box - \bar \nabla_\nu \bar \nabla_\mu \right) R^L = 0\,,
\end{equation}
where $\alpha$ and $\beta$ are the following constants\footnote{Note that $\lfloor D/2 \rfloor$ stands for the largest integer smaller or equal to  $D/2$.}
\begin{align}\labell{alpha}
\alpha &\equiv  \sum_{k=1}^{\lfloor D/2 \rfloor} \partial_k f (  \mathcal{\bar X}) \frac{k (D-3)!}{(D-2k - 1)!}  \mathcal{K}^{k-1}\,, \\ \labell{alphabeta}
\beta &\equiv \sum_{k,l=1}^{\lfloor D/2 \rfloor} \partial_k \partial_l f (  \mathcal{\bar X})  \frac{k l (D-2)! (D-1)!}{(D-2k)!(D-2l)!} \mathcal{K}^{k+l -2}\, .
\end{align}
Here $\partial_lf(\bar{\mathcal{X}})$ means that we should take a formal derivative of $f$ with respect to the corresponding dimensionally-extended Euler density, and then evaluate the result in the background.
Comparing with the linearized equations (\ref{lineareq33s}), we see that $\alpha$ determines the effective Einstein constant $G_{\rm eff} $  and $\beta$ is related to the mass of the scalar field
\begin{equation}\labell{msss}
G_{\rm eff} = \frac{1}{16\pi \alpha} \,, \quad \quad m_s^2 = \frac{D-2 -2  \beta  D  \mathcal{K}}{2 \beta } \,.%, \quad m_g  = + \infty.
\end{equation}
Note that for $\beta=0$ the   scalar mode   is also absent, and  the only physical field is the massless graviton. This applies \eg to pure Lovelock gravities, but also to other non-trivial theories \cite{Love} --- some of which we review in the last epigraph of this section.  The parameters $a,b,c$ and $e$ are given by
\begin{equation}
\begin{aligned}\labell{lovelove}
a&= -\frac{1}{2}c = - \frac{   \alpha - 2e}{4 (D-3) \mathcal{K} } \,, \quad b= \frac{\beta }{4
   (D-1)}-\frac{\alpha -2 e}{8 (D-3) \mathcal{K} } \,, 
   %\quad c= \frac{2 e-\alpha }{6 \mathcal{K} -2 D \mathcal{K} }  \\
    \quad e= \frac{ f( \mathcal{\bar X})}{4 \mathcal{K} (D-1)}\,,
\end{aligned}
\end{equation}
 and the background embedding equation \req{Lambda-eq} reads in turn
 \begin{equation}
 f(\mathcal{ \bar X})=\sum_{k=1}^{\lfloor D/2 \rfloor} \frac{2 k (D-1)!}{(D-2k )!}  \mathcal{K}^{k}  \partial_k f(  \mathcal{\bar X}) \, .
 \end{equation}
 An interesting subclass we shall not consider here is that of Lovelock-Chern-Simons theory \cite{Chamseddine:1990gk,Banados:1993ur}, which is a particular case of the Lovelock theory. This is most naturally defined in general dimensions in terms of the tetrad and the spin connection. Their corresponding equations are first order, and when the torsion is set to zero, the metric field equations become second order, and the theory is a particular case of the Lovelock action considered in this thesis, \ie with a metric-compatible connection. In the latter case, the degrees of freedom propagated by the theory on a msb are of course the $D(D-3)/2$ of the usual massless graviton. Interestingly, if the torsionless condition is relaxed, the number of dynamical degrees of freedom is in fact greater --- see \eg \cite{Banados:1996yj}.

\subsection*{Theories without dynamical scalar}
  \subsubsection*{Conformal gravity}
In the case of quadratic gravity, the most general  theory which does not propagate a scalar field is \cite{Hassan:2013pca}
\begin{equation}\label{nodynscalar}
S=\frac{1}{16\pi G}\int_{\mathcal{M}}d^Dx\sqrt{|g|}\left\{ -2\Lambda+R +  \beta \Big (R^2- \frac{4(D-1)}{D} R_{\mu\nu}R^{\mu\nu} \Big)+\gamma \mathcal X_4\right\} \,,
\end{equation}
where $\mathcal X_4$ is again the Gauss-Bonnet term and $\beta$ and $\gamma$ are dimensionful constants. Observe that for $D = 3$, this action   is equivalent to \emph{new massive gravity}  \cite{Bergshoeff:2009hq}.
%, which is a ghost-free higher curvature theory that only propagates a  single massive spin-2 state with two helicities. However, in $D\ge 4$ the ghost problem persists.
There are two different interesting ways of writing this theory in terms of other well-known curvature tensors.  Firstly, it was observed in \cite{Kan:2013moa} that      the contraction of the Einstein tensor $G_{\mu\nu}$ with the Schouten tensor\footnote{The Schouten tensor is defined as   $S_{\mu\nu} \equiv \frac{1}{D-2} \left (  R_{\mu\nu} - \frac{1}{2(D-1)} R g_{\mu\nu}\right)$.}  $S_{\mu\nu}$ is proportional to the curvature invariant in (\ref{nodynscalar}) that multiplies $\beta$
\begin{equation}
G_{\mu\nu} S^{\mu\nu}= - \frac{D}{4(D-2)(D-1)} \left ( R^2 - \frac{4(D-1)}{D}  R_{\mu\nu} R^{\mu\nu} \right) \,.
\end{equation}
Therefore, by rescaling $\beta$ we see that the theory   is equivalent to
\begin{equation}
S= \frac{1}{16 \pi G}\int_{\mathcal{M}}d^Dx\sqrt{|g|}\left\{ -2\Lambda+R + \bar   \beta G_{\mu\nu} S^{\mu\nu} +   \gamma \mathcal X_4 \right\} \,.
\end{equation}
Secondly, it turns out that the quadratic part of (\ref{nodynscalar}) is equivalent to   the higher dimensional version of conformal gravity, consisting of the square of the Weyl tensor, together with a Gauss-Bonnet term. The square of the Weyl tensor   is   in fact  equal to\footnote{The Weyl tensor is defined as $C_{\mu\nu\rho\sigma} \equiv R_{\mu\nu\rho\sigma} - \frac{2}{D-2} \left ( g_{\mu[\rho} R_{\sigma]\nu} - g_{\nu [\rho} R_{\sigma]\mu}\right) + \frac{2}{(D-1)(D-2)} R g_{\mu[\rho} g_{\sigma] \nu}$.}
\begin{equation}
\begin{aligned}
C_{\mu\nu\rho\sigma} C^{\mu\nu\rho\sigma}  %&=  R_{\mu\nu\rho\sigma} R^{\mu\nu\rho\sigma} - \frac{4}{D-2} R_{\mu\nu} R^{\mu\nu}  + \frac{2}{(D-2)(D-1)} R^2  \\
&=    \mathcal X_4  - \frac{D(D-3)}{(D-2)(D-1)} \left ( R^2 - \frac{4(D-1)}{D} R_{\mu\nu}R^{\mu\nu}   \right) \,.
\end{aligned}
\end{equation}
By using this relation and redefining the couplings,  the theory can   be written as
\begin{equation}\label{higherconformalgravity}
S=\frac{1}{16 \pi G}\int_{\mathcal{M}}d^Dx\sqrt{|g|}\left\{ -2\Lambda+R+\tilde \beta C_{\mu\nu\rho\sigma} C^{\mu\nu\rho\sigma} + \tilde \gamma \mathcal X_4 \right\} \,.
\end{equation}
Thus, we observe that conformal gravity in any dimension is free of the scalar mode, and only propagates the two gravitons. Finally,   for this theory the effective gravitational constant   and the mass of the extra graviton read  respectively
\begin{align}
G_{\rm eff}
&=\frac{G }{1- 2 \mathcal{K} (D-3) (2 \tilde \beta -   \tilde \gamma (D-4)  )  } \,, \\
   m_g^2 &=\frac{2-D+2\mathcal{K} (D-3) (D-2) (2 \tilde \beta -    \tilde  \gamma (D-4)
) }{4 \tilde \beta  (D-3)  } \,.
   \label{mgquadratictheorieswithoutms}
\end{align}
If the numerator of (\ref{mgquadratictheorieswithoutms}) becomes zero, then the extra graviton is massless. This particular case will be analyzed  in the epigraph on critical gravities. Note finally that in $D=3$ both the Weyl tensor and the Gauss-Bonnet term vanish identically, so the theory reduces to  Einstein gravity plus cosmological constant.

\subsection*{Theories with two massless gravitons}

The following is an example of a theory propagating two massless gravitons in addition to the scalar field,
\begin{equation}\label{massless2}
S=\frac{1}{16\pi G}\int_{\mathcal{M}}d^Dx\sqrt{|g|}\left\{  -2\Lambda+R +\alpha R^2 - D \Big ( \alpha + \frac{1}{8 \Lambda } \Big) R_{\mu\nu}R^{\mu\nu}  \right\}\, .
\end{equation}
Note that the $m_g^2=0$ condition has the unpleasant feature of mixing the couplings of terms of different order in curvature. In this case, we see that the $R_{\mu\nu}R^{\mu\nu}$ coupling depends on $\Lambda$. 
For this theory, the background scale is related to the cosmological constant by
\begin{equation}
\mathcal{K} = \frac{4 \Lambda}{D(D-1) } \,.
\end{equation}
In addition, the effective gravitational constant and the mass of the scalar field read
    \begin{equation}
     \hat G_{\rm eff} = \frac{2 (D-1) G \mathcal{K} }{1+ 2 \mathcal{K} \alpha  D (D-1) } \,, \qquad m_s^2 = -\frac{ 4 (D-1) \mathcal{K} }{D+ 2 \mathcal{K}\alpha  (D-1) (D-2)^2 } \,.
    \end{equation}
As far as we know, this theory has not been considered before.

\subsection*{Critical gravities}
\emph{Critical gravity} was   introduced in \cite{Lu} as the four-dimensional quadratic theory for  which   the extra graviton is massless and the scalar mode is absent. Hence, it  is a  special case of the theories considered in the last two epigraphs --- (\ref{nodynscalar}) and (\ref{massless2}) --- in the particular case of $D=4$. The following action is   a generalization of critical gravity to general dimensions   \cite{Kan:2013moa} 
\begin{equation}
S= \frac{1}{16\pi G}\int_{\mathcal{M}}d^Dx\sqrt{|g|}\left\{ -2\Lambda+R - \frac{ D^2}{8\Lambda (D-2)^2 } \Big  ( R^2  - \frac{4 (D-1)}{D} R_{\mu\nu}R^{\mu\nu}   \Big   )    \right\} \,.
\end{equation}
It can be obtained by setting    $\beta =  - D^2/  (8 \Lambda (D-2)^2 )$ and $  \gamma=0$ in (\ref{nodynscalar}) or, alternatively, from (\ref{massless2}) if we put $\alpha = - D^2/  (8 \Lambda (D-2)^2 )$ there. In $D=4$, this is the critical theory considered by  \cite{Lu}, and for $D=3$, it is equivalent to   \emph{critical new massive gravity}  with a cosmological constant \cite{Liu:2009bk}. 
Furthermore, the effective gravitational constant of this theory is
\begin{equation}
\hat G_{\rm eff} =-\frac{1}{2} (D-2)^2 G  \mathcal{K} \,,
\end{equation}
which, assuming $G>0$, is only positive for $\mathcal{K}<0$.

\subsection*{Einstein-like theories}
In section \ref{EQG1} we already constructed examples of \emph{Einstein-like} theories in the sense defined in section \ref{Classification}, \ie theories which only propagate a massless graviton on a msb. However, the theories considered in that section had the additional property of being defined in a dimension-independent manner and we coined them \emph{Einsteinian}. In this appendix we would like to present some more examples of \emph{Einstein-like} theories whose definition does however depend on the spacetime dimension.

\subsubsection*{Quasi-topological gravity}
The first example is \emph{Quasi-topological gravity} \cite{Quasi, Quasi2,Myers:2010jv}. This is a cubic theory which has the nice property of admitting analytic black hole solutions --- which generalize Schwarzschild-AdS and its Gauss-Bonnet generalization \cite{Cai:2003gr}. It consists of a combination of all Lovelock gravities up to cubic order
plus an additional ``Quasi-topological"   term:
\begin{equation}\label{qtopo}
S =\frac{1}{16\pi G}  \int_{\mathcal M} d^D x \sqrt{|g|} \left \{  -2 \Lambda + R    
+ \alpha \mathcal X_4 +  \beta \mathcal X_6 + \gamma \mathcal Z_{D}  \right\} \,.
\end{equation}
%\begin{equation}
% \begin{aligned}
%\beta_1 &= 1 \, ,    \qquad \qquad \qquad \qquad \quad \,\,\,\,\,
%\beta_2 = 0 \, , \\
%\beta_3 &= - \frac{3(D-2)}{(2D-3)(D-4)} \, , \qquad \,
%\beta_4 = \frac{3(3D-8)}{8(2D-3)(D-4)} \, ,\\
%\beta_5 &= \frac{3D}{(2D-3)(D-4)} \, , \qquad \quad
%\beta_6 = \frac{6(D-2)}{(2D-3)(D-4)}\,, \\
%\beta_7 &= - \frac{3(3D-4)}{2(2D-3)(D-4)}\,, \quad  \,\,\,\,
%\beta_8 = \frac{3D}{8(2D-3)(D-4)}\,.
%\end{aligned}
%\end{equation}
Here the cubic Lovelock term is given by  
\begin{equation}
\begin{aligned}\label{xx6}
\mathcal X_6 \equiv 
&-8 R_{\mu\ \nu}^{\ \rho \ \sigma}R_{\rho\ \sigma}^{\ \delta \ \gamma}R_{\delta\ \gamma}^{\ \mu \ \nu} 
+4 R_{\mu\nu }^{\ \ \rho\sigma }R_{\rho\sigma }^{\ \ \delta\gamma }R_{\delta\gamma }^{\ \ \mu\nu }
- 24 R_{\mu\nu\rho\sigma }R^{\mu\nu\rho }_{\ \ \ \delta}R^{\sigma \delta} \\
& + 3 R_{\mu\nu\rho\sigma }R^{\mu\nu\rho\sigma }R
+24  R_{\mu\nu\rho\sigma }R^{\mu\rho}R^{\nu\sigma} 
+ 16 R_{\mu}^{\ \nu}R_{\nu}^{\ \rho}R_{\rho}^{\ \mu}
 -12 R_{\mu\nu }R^{\mu\nu }R
+  R^3 \, ,
\end{aligned}
\end{equation}
and  the Quasi-topological one in general dimensions reads in turn \cite{Quasi,Quasi2}
\begin{equation}
\begin{aligned}
\mathcal Z_{D} \equiv 
&R_{\mu\ \nu}^{\ \rho \ \sigma}R_{\rho\ \sigma}^{\ \delta \ \gamma}R_{\delta\ \gamma}^{\ \mu \ \nu} 
+ \frac{1}{(2D-3)(D-4)} \Big ( 
- 3(D-2) R_{\mu\nu\rho\sigma }R^{\mu\nu\rho }_{\ \ \ \delta}R^{\sigma \delta} \\
&+ \frac{3 (3D-8)}{8}  R_{\mu\nu\rho\sigma }R^{\mu\nu\rho\sigma }R
+3D R_{\mu\nu\rho\sigma }R^{\mu\rho}R^{\nu\sigma} \\
&+ 6(D-2) R_{\mu}^{\ \nu}R_{\nu}^{\ \rho}R_{\rho}^{\ \mu}
- \frac{3(3D-4)}{2}  R_{\mu\nu }R^{\mu\nu }R
+ \frac{3D}{8} R^3 \Big) \, .
\end{aligned}
\end{equation}
The physical quantities for \req{qtopo} read
 \begin{equation}
 G_{\rm eff} = \frac{G}{f(\alpha, \beta, \gamma, \mathcal{K})} \,, \quad m_{s} = + \infty \,, \quad m_g = + \infty \,,
 \end{equation}
 where
 \begin{equation*}
 \begin{aligned}
  f(\alpha, \beta, \gamma, \mathcal{K}) \equiv&+ 1 + 2 \mathcal{K} \alpha  (D-4) (D-3) \\ &
   + 3   \mathcal{K} ^2  \beta  (D-6)   (D-5 )     (D-4) (D-3)\\ &  + \frac{3 (D-6) (D-3)}{8 (2D-3)} \mathcal{K} ^2 \gamma (16   +  3    D (D-5)) \, .
\end{aligned}
\end{equation*}
Hence, as explained in \cite{Quasi}, this theory shares the linearized spectrum of Einstein gravity. Let us close this section by mentioning that a quartic version of quartictopological gravity was constructed in \cite{Dehghani:2011vu}. It would be interesting to use our results in section \ref{quartic} to check that such theory also presents an Einstein-like spectrum. More recently, a quintic version was constructed \cite{Cisterna:2017umf}, and in that case the linearization method presented in Sec.~\ref{chap:2sec:1} was used to show that the theory has Einstein-like spectrum. 

  \subsubsection*{Special $f($Lovelock$)$ theories}
 The second example we would like to consider corresponds to a particular family of $f($Lovelock$)$ gravities. As we explained before, all $f($Lovelock$)$ theories are free of the massive graviton, but do in general propagate the extra scalar. However, as pointed out in \cite{Love} it is possible to construct non-trivial theories --- \ie different from the pure Lovelock case --- which are also free of the extra scalar and hence share the linearized spectrum of Einstein gravity. 
 
 Indeed, whenever $\beta$, as defined in \req{alphabeta}, vanishes, the mass of the scalar diverges --- which is obvious from \req{msss}. This is achieved whenever $\partial_k \partial_l f(\bar{\mathcal{L}})=0$ for all $k, l$, which leaves us with nothing but the usual Lovelock theory or, alternatively, if
 \begin{equation}
 \sum_{k,l=1}^{\lfloor D/2 \rfloor} \partial_k \partial_l f (  \mathcal{\bar L})  \frac{k l (D-2)! (D-1)!}{(D-2k)!(D-2l)!} \mathcal{K}^{k+l -2}=0\, , \quad  \partial_k \partial_l f (  \mathcal{\bar L})  \neq 0\, ,
 \end{equation}
for some $k, l$. This equation is \eg satisfied by all theories of the form \cite{Love}
\begin{equation}
S=\frac{1}{16\pi G}\int_{\mathcal{M}}d^Dx\sqrt{|g|}\left\{-2\Lambda+R+\lambda  \left(R^u \mathcal{L}_2^s-\gamma R^{2s+u}\right)\right\}\, ,
\label{parafernalia}
\end{equation}
where $\gamma$ is the dimension-dependent constant
\begin{equation}
\gamma\equiv \frac{u^2+4(s-1)s+u(4s-1)}{(u+2s)(u+2s-1)}\frac{(D-2)^s(D-3)^s}{D^s(D-1)^s}\, ,
\end{equation}
for any $u,s\geq0$. In particular, for $s=u=1$ one finds the cubic class of theories
\begin{equation}
S=\frac{1}{16\pi G}\int_{\mathcal{M}}d^Dx\sqrt{|g|}\left\{-2\Lambda+R+ \lambda  \left[R\mathcal{L}_2-\left(\frac{2(D-2)(D-3)}{3D(D-1)} \right) R^{3}\right]\right\}\, .
\label{parafernalia2}
\end{equation}
The $D=4$ case of \req{parafernalia2} was also considered in \cite{Karasu:2016ifk} in a slightly different context.
%\begin{equation}
%S=\int_{\mathcal{M}}d^4x\sqrt{|g|}\left\{\frac{1}{2\kappa}(-2\Lambda+R)+\lambda  \left(R \mathcal{L}_2-\frac{1}{9} R^{3}\right)\right\}\, .
%\label{parafernalia3}
%\end{equation}
The effective gravitational constant of \req{parafernalia2} reads
\begin{equation}
G_{\rm eff}=\frac{G}{1+  (D-6)(D-3)(D-1)D \lambda \mathcal{K}^2 }\, .
\end{equation}

\section{$f($scalars$)$ theories: examples}\label{fscalarsapp}
Let us now illustrate how the expressions obtained in section \ref{fscalars} can be used to easily compute the values of $a,b,c$ and $e$ for theories consisting of functions of invariants, as long as we know the values of those parameters for the invariants themselves.
\subsubsection*{ $f(R)$ gravity}
Let us first consider $f(R)$ gravity, whose Lagrangian in general dimensions we write as
\begin{equation}
	S=\int_\mathcal{M} d^Dx\sqrt{|g|}f( R )\, .
\end{equation} 
According to   table \ref{tabla2}, for $R$ we have $a=b=c=0$, $e=\frac{1}{2}$ and $\bar R=D(D-1)\mathcal{K}$. Therefore, using the transformation rules (\ref{transfrules0}) for the theory above we have
\begin{equation}
	a=c=0\, , \quad b=\frac{1}{4}f''(\bar R)\, , \quad e=\frac{1}{2}f'(\bar R)\, .
\end{equation}
Note that these expressions can also be easily obtained from the general $f($Lovelock$)$ ones \req{lovelove}. Also, according to (\ref{Lambda-eq}) the background curvature $\mathcal{K}$ is determined by the equation
\begin{equation}\label{f(R)embedding}
	f(\bar R)=2(D-1) \mathcal{K} f'(\bar R)\, .
\end{equation}
If $f''(\bar R)\neq 0$, we have a scalar mode with mass
\begin{equation}\label{frscalar}
	m_s^2=\frac{(D-2)f'(\bar R)-2\bar R f''(\bar R)}{2(D-1)f''(\bar R)}\, .
\end{equation}
The effective gravitational constant is in turn given by
\begin{equation}
	G_{\rm eff}=\frac{G}{f'(\bar R)}\, .
\end{equation}

\subsubsection*{$f(R, R_{\mu\nu}^2, R_{\mu\nu\rho\sigma}^2)$ gravity}
Let us now study all theories that can be constructed as functions of invariants up to quadratic order \cite{Carroll:2004de}.  The independent scalars are $R$, $Q\equiv R_{\mu\nu}R^{\mu\nu}$, and $K\equiv R_{\mu\nu\rho\sigma}R^{\mu\nu\rho\sigma}$, so let us consider an action of the form
\begin{equation}
	S=\int_\mathcal{M} d^Dx\sqrt{|g|}f(R,Q,K)\, .
\end{equation}
%We work in four dimensions in order to simplify the formulas. 
This theory includes, as particular cases, $f( R )$ and general quadratic gravities. In order to simplify the following expressions, let us write $\mathcal{R}\equiv(R,Q,K)$. Evaluated on the background, the invariants read
\begin{equation}
	\bar{\mathcal{R}}=\big (  D (D-1) \mathcal{K} ,  D (D-1)^2 \mathcal{K}^2, 2 D (D-1) \mathcal{K}^2  \big )\, .
\end{equation}
%\begin{equation}
%\bar{\mathcal{R}}=(12\mathcal{K}, 36\mathcal{K}^2, 24\mathcal{K}^2).
%\end{equation}
Then, the background embedding equation (\ref{Lambda-eq}) can be written in terms of these background   scalars $\bar R, \bar Q, \bar K$ as
\begin{equation}
	\bar R \partial_R f(\bar{\mathcal{R}})  + 2  \bar Q \partial_Q f(\bar{\mathcal{R}}) + 2 \bar K \partial_K f(\bar{\mathcal{R}})  = \frac{D}{2} f(\bar{\mathcal{R}})\, ,
	%\mathcal{K}      \partial_R f(\bar{\mathcal{R}})  + 2 \mathcal{K}^2 \left (    (D-1)  \partial_Q f(\bar{\mathcal{R}}) + 2   \partial_K f(\bar{\mathcal{R}})  \right) = \frac{1}{2(D-1)}f(\bar{\mathcal{R}})
\end{equation}
 which, in particular, generalizes (\ref{f(R)embedding}) for this theory.
Finally, the parameters $a,b,c$ and $e$ are given by
\begin{equation}
	\begin{aligned}
		a&=\partial_K f(\bar{\mathcal{R}}),\\
		b&=\Big[\frac{1}{4}\partial_R\partial_R f(\bar{\mathcal{R}})+ (D-1)\mathcal{K} \partial_R\partial_Q  f(\bar{\mathcal{R}})+2\mathcal{K}\partial_R\partial_K f(\bar{\mathcal{R}})\\
		&+(D-1)^2\mathcal{K}^2\partial_Q\partial_Q f(\bar{\mathcal{R}})+4 (D-1) \mathcal{K}^2 \partial_Q\partial_K  f(\bar{\mathcal{R}})+4\mathcal{K}^2\partial_K\partial_K f(\bar{\mathcal{R}})\Big],\\
		c&=\frac{1}{2}\partial_Qf(\bar{\mathcal{R}}),\\
		e&=\frac{1}{2}\left[\partial_Rf(\bar{\mathcal{R}})+2 (D-1) \mathcal{K} \partial_Qf(\bar{\mathcal{R}})+4\mathcal{K}\partial_Kf(\bar{\mathcal{R}})\right]\, ,
	\end{aligned}
\end{equation}
from which one can easily obtain the values of $G_{\rm eff}$, $m_s^2$, $m_g^2$.

\section{Einsteinian quartic gravities}\label{appEQG}
%%%%%%%%%%%%%%%%%
Here we provide the explicit expressions for the conditions $F_g^{(2)}(\alpha_i)=F_s^{(2)}(\alpha_i)=F_g^{(3)}(\beta_i,D)=F_s^{(3)}(\beta_i,D)=F_g^{(4)}(\gamma_i,D)=F_s^{(4)}(\gamma_i,D)=0$ appearing in section \ref{EQG1}. These read:
\begin{align}
F_g^{(2)}(\alpha_i)\equiv&+\frac{1}{2}\alpha_2+2\alpha_3=0\, ,\\  F_s^{(2)}(\alpha_i)\equiv &+2\alpha_1+\frac{1}{2}\alpha_2=0\, ,\\
F_g^{(3)}(\beta_i,D)\equiv&-\frac{3}{2}\beta_1+12\beta_2+2D\beta_3+2D(D-1)\beta_4
+\left(D-\frac{3}{2}\right)\beta_5\\ \notag&+\frac{3}{2}(D-1)\beta_6+\frac{1}{2}D(D-1)\beta_7=0\, , \\
F_s^{(3)}(\beta_i,D)\equiv &
+\frac{3}{2}\beta_1+2\beta_3+8\beta_4+\left(D+\frac{1}{2}\right)\beta_5+\frac{3}{2}(D-1)\beta_6
\\ \notag&+(D-1)\left(\frac{D}{2}+4\right)\beta_7+6D(D-1)\beta_8=0\, , 
\label{betaS}
\end{align}

\begin{align}
F_g^{(4)}(\gamma_i,D)\equiv&+(4D-9)\gamma_{1}+2(D+3)\gamma_{2}+(2D-9)\gamma_{3}+24\gamma_4+48\gamma_{5}+8\gamma_{6}\\ \notag&+8D(D-1)\gamma_{7}-\frac{1}{2}(D+3)\gamma_{8}+6(2D-1)\gamma_{9}+(2D^2-D-3)\gamma_{10}\\ \notag&-\frac{3}{2}D(D-1)\gamma_{11}+12D(D-1)\gamma_{12}+\left(2D^2+\frac{1}{2}D-3\right)\gamma_{13}\\ \notag&+\frac{1}{2}(3D^2-8D+6)\gamma_{14}
+(2D^2-3)\gamma_{15}+(2D^2+D-3)\gamma_{16}\\ \notag&+D(D-1)(2D-1)\gamma_{17}+2D^2(D-1)\gamma_{18}+2D^2(D-1)^2\gamma_{19}\\ \notag&+(D-1)(2D-3)\gamma_{20}+\frac{1}{2}D(D-1)(2D-3)\gamma_{21}+3(D-1)^2\gamma_{22}\\
&\notag+D(D-1)^2\gamma_{23}+\frac{3}{2}D(D-1)^2\gamma_{24}+\frac{1}{2}D^2(D-1)^2\gamma_{25}=0\, .
\end{align}
\begin{align}
F_s^{(4)}(\gamma_i,D)\equiv&+7\gamma_{1}+2(D-1)\gamma_{2}+5\gamma_{3}+8\gamma_{6}+32\gamma_{7}+\frac{5}{2}(D-1)\gamma_{8}\\ \notag&+6\gamma_{9}+3(D-1)\gamma_{10}+\frac{3}{2}(D^2+3D-8)\gamma_{11}+24\gamma_{12}+\frac{3}{2}(3D-2)\gamma_{13}\\ \notag&+\frac{1}{2}(3D^2+4D-10)\gamma_{14}+(4D-1)\gamma_{15}+5(D-1)\gamma_{16}\\ \notag&+(D+16)(D-1)\gamma_{17}+2(D+6)(D-1)\gamma_{18}+20D(D-1)\gamma_{19}\\
\notag&+(D-1)(2D+1)\gamma_{20}+\frac{1}{2}(D-1)(2D^2+13D-12)\gamma_{21}\\ \notag&+3(D-1)^2\gamma_{22}+(D-1)^2(D+8)\gamma_{23}+\frac{3}{2}(D-1)^2(D+4)\gamma_{24}\\ \notag&+\frac{1}{2}D(D-1)^2(D+20)\gamma_{25}+12D^2(D-1)^2\gamma_{26}=0\, .
\end{align}
Solving the last two equations order by order in $D$ gives rise to the constraints which characterize Einsteinian quartic gravities \req{ggf}.

\chapter{Redefining the metric}\label{App:2}

Implementing a differential change of variables directly in the action can be problematic if one is not careful enough. In order to see this, let us consider the equations of motion of $\tilde g_{\mu\nu}$ --- related to $g_{\mu\nu}$ according to $g_{\mu\nu}=\tilde g_{\mu\nu}+\tilde Q_{\mu\nu}$ --- by computing the variation of the new action $\tilde S[\tilde g_{\mu\nu}]=S[g_{\mu\nu}]$:\footnote{Note that in the second term we used the chain law for the functional derivative, that in general is given by
\be
\frac{\delta S}{\delta \phi}\frac{\delta \phi}{\delta \psi}=\frac{\delta S}{\delta \phi}\frac{\partial \phi}{\partial \psi}-\partial_{\mu}\left(\frac{\delta S}{\delta \phi}\frac{\partial \phi}{\partial_{\mu}\psi}\right)+\ldots
\ee
}

\begin{equation}\label{eq:badeq}
\frac{\delta \tilde S}{\delta \tilde g_{\mu\nu}}=\frac{\delta S}{\delta g_{\mu\nu}}+\frac{\delta S}{\delta g_{\alpha\beta}}\frac{\delta \tilde Q_{\alpha\beta}}{\delta \tilde g_{\mu\nu}} \Bigg|_{g_{\mu\nu}=\tilde g_{\mu\nu}+\tilde Q_{\mu\nu}}
\end{equation}
Now, it is clear that we can always solve these equations if
\begin{equation}\label{eq:goodeq}
\frac{\delta S}{\delta g_{\mu\nu}} \Bigg|_{g_{\mu\nu}=\tilde g_{\mu\nu}+\tilde Q_{\mu\nu}}=0\, .
\end{equation}
In other words, implementing the change of variables directly in the equations of the original theory produces an equation that solves the equations of $\tilde S$. However, the equations of $\tilde S$ contain more solutions. These additional solutions are spurious and appear as a consequence of increasing the number of derivatives in the action, so they should not be considered. 
A possible way to formalize this intuitive argument consists in introducing auxiliary field so that the redefinition of the metric becomes algebraic. Let us consider the following action
\begin{equation}
\begin{aligned}
S_{\chi}=&\int d^D\sqrt{|g|}\Big\{\mathcal{L}\left(g^{\mu\nu}, \chi_{\mu\nu\rho\sigma}, \chi_{\alpha_1,\mu\nu\rho\sigma},\chi_{\alpha_1\alpha_2,\mu\nu\rho\sigma},\ldots\right)+\frac{\partial \mathcal{L}}{\partial \chi_{\mu\nu\rho\sigma}}\left(R_{\mu\nu\rho\sigma}-\chi_{\mu\nu\rho\sigma}\right)\\&+\frac{\partial \mathcal{L}}{\partial \chi_{\alpha_1,\mu\nu\rho\sigma}}\left(\nabla_{\alpha_1}R_{\mu\nu\rho\sigma}-\chi_{\alpha_1,\mu\nu\rho\sigma}\right)+\frac{\partial \mathcal{L}}{\partial \chi_{\alpha_1\alpha_2,\mu\nu\rho\sigma}}\left(\nabla_{\alpha_1}\nabla_{\alpha_2}R_{\mu\nu\rho\sigma}-\chi_{\alpha_1\alpha_2,\mu\nu\rho\sigma}\right)+\ldots\Big\}\, ,
\end{aligned}
\end{equation}
where we have introduced some auxiliary fields $\chi_{\mu\nu\rho\sigma}$, $\chi_{\alpha_{1},\mu\nu\rho\sigma}$, \ldots $\chi_{\alpha_{1}\ldots\alpha_{n},\mu\nu\rho\sigma}$. Let us convince ourselves that this action is equivalent to \req{eq:generalhdg}. When we take the variation with respect to $\chi_{\alpha_{1}\ldots\alpha_{i},\mu\nu\rho\sigma}$, we get
\begin{equation}
\sum_{j=0}\frac{\partial^2 \mathcal{L}}{\partial\chi_{\alpha_1\ldots \alpha_i,\mu\nu\rho\sigma} \partial \chi_{\beta_1\ldots\beta_j,\lambda\eta\gamma\chi}}\left(\nabla_{\beta_1}\ldots\nabla_{\beta_j}R_{\lambda\eta\gamma\chi}-\chi_{\beta_1\ldots\beta_j,\lambda\eta\gamma\chi}\right)=0\, .
\end{equation}
In this way, we get a system of algebraic equations for the variables $\chi_{\alpha_{1}\ldots\alpha_{i},\mu\nu\rho\sigma}$ that always has the following solution

\begin{eqnarray}
\label{eq:solchi}
\chi_{\mu\nu\rho\sigma}&=&R_{\mu\nu\rho\sigma}\, ,\\
\chi_{\alpha_1,\mu\nu\rho\sigma}&=&\nabla_{\alpha_1}R_{\mu\nu\rho\sigma}\, ,\\
\chi_{\alpha_1\alpha_2,\mu\nu\rho\sigma}&=&\nabla_{\alpha_1}\nabla_{\alpha_2}R_{\mu\nu\rho\sigma}\, ,\\
&\ldots&
\end{eqnarray}
If $\mathcal{L}$ is a non-degenerate function of the auxiliary variables, this is also the unique solution, and we will assume that this is the case. When we plug this solution back in the action we recover \req{eq:generalhdg}, so that both formulations are equivalent. 

Now let us perform the following redefinition of the metric in $S_{\chi}$:
\begin{equation}
g_{\mu\nu}= \tilde g_{\mu\nu}+\tilde Q_{\mu\nu}\, ,\quad\text{where}\,\,\, \tilde Q_{\mu\nu}=\tilde Q_{\mu\nu}\left(\tilde g^{\alpha\beta}, \chi_{\alpha\beta\rho\sigma}, \chi_{\alpha_1,\alpha\beta\rho\sigma},\ldots\right)\, ,
\end{equation}
this is, $\tilde Q_{\mu\nu}$ is a symmetric tensor formed from contractions of the $\chi$ variables and the metric, but it contains no derivatives of any field. In this way, the change of variables is algebraic and can be directly implemented in the action. Therefore, we get
\begin{equation}
\tilde{S}_{\chi}\left[\tilde{g}_{\mu\nu}, \chi\right]=S_{\chi}\left[\tilde{g}_{\mu\nu}+\tilde Q_{\mu\nu}, \chi\right]\, ,
\end{equation}
where we are collectively denoting $\chi$ to all the auxiliary variables for simplicity.  Now, both actions are equivalent, and so are the field equations:
\begin{eqnarray}
\label{eq:eqmetric}
\frac{\delta \tilde{S}_{\chi}}{\delta\tilde g_{\mu\nu}}&=&\frac{\delta S_{\chi}}{\delta g_{\alpha\beta}}\left(\tensor{\delta}{^{\mu}_{\alpha}}\tensor{\delta}{^{\nu}_{\beta}}+\frac{\partial \tilde Q_{\alpha\beta}}{\partial \tilde g_{\mu\nu}}\right)\bigg|_{g_{\mu\nu}= \tilde g_{\mu\nu}+\tilde Q_{\mu\nu}}\, ,\\
\label{eq:eqchi}
\frac{\delta \tilde{S}_{\chi}}{\delta\chi}&=&\frac{\delta {S}_{\chi}}{\delta\chi}+\frac{\delta {S}_{\chi}}{\delta g_{\mu\nu}}\frac{\partial \tilde Q_{\mu\nu}}{\partial\chi}\bigg|_{g_{\mu\nu}= \tilde g_{\mu\nu}+\tilde Q_{\mu\nu}}
\end{eqnarray}
The first equation implies that $\frac{\delta S_{\chi}}{\delta g_{\mu\nu}}=0$, and using this into the second one, we see that the equations for the auxiliary variables become $\delta S_{\chi}/\delta\chi=0$, that of course have the same solution as before \req{eq:solchi}. When we take that into account, $\tilde Q_{\mu\nu}$ becomes a tensor constructed from the curvature of the original metric $g_{\mu\nu}$, so that we get
\begin{equation}
g_{\mu\nu}= \tilde g_{\mu\nu}+\tilde Q_{\mu\nu}\left(\tilde g^{\alpha\beta}, R_{\alpha\beta\rho\sigma}, \nabla_{\alpha_1}R_{\alpha\beta\rho\sigma},\ldots\right)\, .
\end{equation}
Note that this is slightly different to \req{eq:metricredef}, because $\tilde Q_{\mu\nu}$ is formed with the curvature tensor of $g_{\mu\nu}$, but the indices are contracted with $\tilde g^{\mu\nu}$. Nevertheless, if we work perturbatively, we can replace $R_{\alpha\beta\rho\sigma}$ by $\tilde R_{\alpha\beta\rho\sigma}$, and the expression is the same as \req{eq:metricredef}.

Now, according to Eq.~\req{eq:eqmetric}, the equation for the metric $\tilde g_{\mu\nu}$ is simply obtained from the equation of $g_{\mu\nu}$ by substituting the change of variables:
\begin{equation}\label{eq:consistenteq}
\frac{\delta S_{\chi}}{\delta g_{\mu\nu}}\bigg|_{g_{\mu\nu}= \tilde g_{\mu\nu}+\tilde Q_{\mu\nu}}=0
\end{equation}

However, note that this is not the same as substituting \req{eq:solchi} in the action and taking the variation, which would yield
\begin{align}\label{eq:fake}
\frac{\delta \tilde{S}_{\chi}\left[\tilde{g}_{\mu\nu}, \chi(\tilde g_{\mu\nu})\right]}{\delta\tilde g_{\mu\nu}}&=\frac{\delta \tilde S_{\chi}}{\delta \tilde g_{\mu\nu}}+\frac{\delta \tilde{S}_{\chi}}{\delta\chi}\frac{\delta\chi}{\delta\tilde g_{\mu\nu}}\\
&=\frac{\delta S_{\chi}}{\delta g_{\alpha\beta}}\left(\tensor{\delta}{^{\mu}_{\alpha}}\tensor{\delta}{^{\nu}_{\beta}}+\frac{\partial \tilde Q_{\alpha\beta}}{\partial \tilde g_{\mu\nu}}\right)\bigg|_{g_{\mu\nu}= \tilde g_{\mu\nu}+\tilde Q_{\mu\nu}}-\frac{\delta {S}_{\chi}}{\delta g_{\alpha\beta}}\frac{\partial \tilde Q_{\alpha\beta}}{\partial\chi}\frac{\delta\chi}{\delta\tilde g_{\mu\nu}}\bigg|_{g_{\mu\nu}= \tilde g_{\mu\nu}+\tilde Q_{\mu\nu}}\, .\nonumber
\end{align}
This equation is formally different to \req{eq:eqmetric} due to the second term, and it is equivalent to \req{eq:badeq}. The second term appears because the auxiliary variables $\chi(\tilde g_{\mu\nu})$ do not solve the equation $\frac{\delta \tilde{S}_{\chi}}{\delta\chi}=0$, but $\frac{\delta {S}_{\chi}}{\delta\chi}=0$. However, we must solve $\frac{\delta \tilde{S}_{\chi}}{\delta\chi}=0$ in order to get a solution of $\tilde{S}_{\chi}\left[\tilde{g}_{\mu\nu}, \chi\right]$, and according to \req{eq:eqchi} this would only happen if $\frac{\delta {S}_{\chi}}{\delta g_{\mu\nu}}\frac{\partial \tilde Q_{\mu\nu}}{\partial\chi}=0$, so that the only consistent solutions of \req{eq:fake} are those that satisfy \req{eq:consistenteq}. 
This explains why the only solutions of \req{eq:badeq} that we should consider are the ones that satisfy \req{eq:goodeq}.

\chapter{Numerical construction of the solutions}\label{App:3}
In this appendix we explain the numerical procedure which allows us to construct the black hole solutions analyzed in the main text. The same procedure has been previously used \eg in \cite{PabloPablo2,Hennigar:2017ego,PabloPablo3,Ahmed:2017jod}.
First, we observe that \req{eq:feqECG} is a \emph{stiff} differential equation. This is because the terms involving derivatives of $f(r)$ appear as corrections, and, in particular, the coefficient that multiplies $f''$ is usually small. Thus, the numerical resolution is problematic in terms of stability and we need to use, at least, \emph{A-stable} methods. In our calculations, we used implicit Runge-Kutta methods, but, of course, other methods with larger stability regions can be used as well. Once we have an appropriate numerical method, \req{eq:feqECG} can be solved by imposing the boundary conditions explained in the text.

In principle, we should start the solution at the horizon $r=r_h$, where we know $r_h$ and $a_1=f'(r_h)$ in terms of the mass and the solution is specified once we choose a value of $a_2=f''(r_h)/2$, as explained in the main text. In practice, the numerical method cannot be started at $r_h$, since at that point the equation is singular. Instead, we start the solution for some other value, $r_h+\epsilon$, very close to the horizon ($\epsilon \ll 1$). Then, the second-order Taylor polynomial of $f$ around the horizon can be used to compute $f(r_h+\epsilon)$ and $f'(r_h+\epsilon)$. This yields $f(r_h+\epsilon)=a_1\epsilon+a_2\epsilon^2$,  $f'(r_h+\epsilon)=a_1+2\epsilon a_2$. The numerical resolution can then be started at $r_h+\epsilon$ by using these initial conditions, where the only free parameter is $a_2$. %Of course, $\epsilon$ has to be as small as possible in order to obtain a reliable solution. 
Then, $a_2$ is chosen by imposing the solution to be asymptotically flat. From the asymptotic expansion analysis in section \ref{sec:exactECG}, we know that there exists a family of solutions which are exponentially growing when $r\rightarrow\infty$. Almost any choice of $a_2$ will excite this mode and the solution will not be asymptotically flat. Indeed, we expect that there is a unique choice of $a_2$ for which the solution is asymptotically flat. 

In order to find $a_2$, we use the shooting method to glue the numeric solution with the asymptotic expansion \req{eq:fpart}. The idea is the following: we first fix a value $r_{\infty}$ sufficiently large, for which the asymptotic expansion $f_{\rm asympt.}(r_{\infty})$ is a good approximation. Then we choose a value for $a_2$ and we compute numerically the solution up to $r_{\infty}$, which would yield a value $f_{\rm numeric}(r_{\infty}; a_2)$. The appropriate value of $a_2$ is such that it glues both solutions: $f_{\rm numeric}(r_{\infty}; a_2)=f_{\rm asympt.}(r_{\infty})$. In all the cases analyzed, there is  a unique value of $a_2$ for which this happens, and it must be chosen with great precision.  In practice, it is difficult to extend the numeric solution to very large $r$, since the equation becomes more and more stiff, but it is always easy to compute the numeric solution up to a value for which it overlaps with the asymptotic expansion.  Once $a_2$ has been determined, the interior solution $r<r_h$ can also be computed by starting the numeric method at $r_h-\epsilon$. In this case, the numerical resolution offers no problems.

Once a value of $\mu L^{1/4}$ has been chosen, we can compute $r_h$ and $a_1$ as functions of the mass using \req{eq:a1sol} and \req{eq:MrhECG}. Equivalently, $a_2$ will also be a function of the mass, although we do not know it explicitly.  Applying the previous method to various values of $M$, we can find the corresponding values of $a_2$. In Fig. \ref{figa2}, we show $a_2(M)$ constructed from the interpolation of discrete values. As we can see, in the limit $M>> L/G$, we recover the Schwarzschild value $a_2=-(2 G M)^{-2}$. On the other hand, its is possible to prove that
\begin{equation}
\lim_{M\rightarrow 0} a_2=\lim_{M\rightarrow 0}\frac{a_1}{2 r_h}=\frac{1}{\sqrt{3\mu}L^2}\, ,
\end{equation}
which is in fact valid for arbitrary values of $\mu$.
Finding an explicit expression for $a_2(M)$ (or at least a more efficient way of computing it for different values of the couplings) would be of interest, in particular in order to perform additional studies of the solutions presented here \eg in the contexts of holography or gravitational phenomenology, \eg along the lines of \cite{Brigante:2007nu,deBoer:2009pn,Camanho:2009vw,Buchel:2009tt,Cai:2009zv,Camanho:2009hu,Buchel:2009sk,Myers:2010jv,Quasi} and \cite{Holdom:2016nek}, respectively. 
In this respect, let us mention that an improvement with respect to the numerical scheme that we used here was presented in \cite{Hennigar:2018hza}, where ECG solutions were constructed using a continuous fraction approach \cite{Rezzolla:2014mua,Kokkotas:2017zwt}.

\begin{figure}[ht!]
	\centering 
	\includegraphics[scale=0.56]{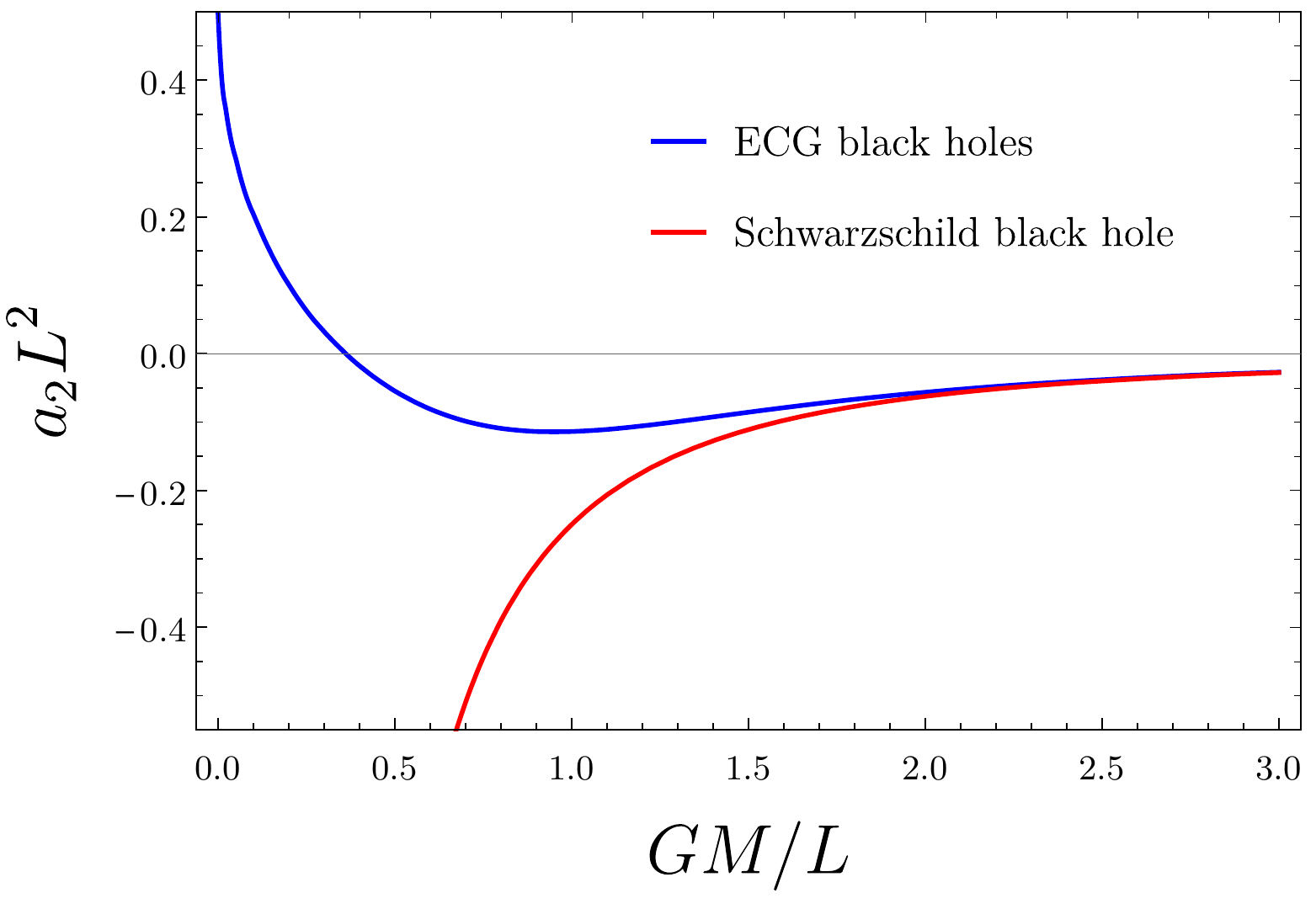}
	\caption{We plot $a_2$ as a function of the mass (interpolation) for  $\mu=4/3$  and for the Schwarzschild solution, $a_2=-(2 G M)^{-2}$.} 
	\labell{figa2}
\end{figure}

\chapter{General formulas for $D=4$ GQG theories}\label{App:4}

\section{(A)dS asymptotes and general horizon geometries}\label{mastereq}
In the main text we have focused on static, spherically symmetric and asymptotically flat solutions. Our results can be easily extended to (A)dS asymptotes as well as to planar or hyperbolic horizon geometries. In particular \req{lal} can be generalized to
\begin{equation}
S= \frac{1}{16\pi G}\int d^4x\sqrt{|g|}\left[R-2\Lambda+\sum_{n=2}^{\infty}L^{2n-2}\lambda_n\mathcal{R}_{(n)}\right]\, ,
\end{equation}
We can search for solutions of the form
\begin{equation}
ds^2=-f(r)dt^2+\frac{dr^2}{f(r)}+r^2d\Sigma_{(k)}^2\, ,
\label{kmetric2}
\end{equation}
where $d\Sigma_{(k)}^2$ is the metric of a 2-dimensional maximally symmetric space of curvature $k=1,0,-1$, \ie, the metric of the unit sphere, flat space or hyperbolic space. The  generalized version of the equation which determines the metric function $f(r)$ then reads (note that $M$ is no longer the mass in the planar and hyperbolic cases):
\begin{align}
&-(f-k)r-L^4\sum_{n=3}^{\infty}\frac{n(n-1)\lambda_n}{8}\left(\frac{L^2f'}{2r}\right)^{n-3}\Bigg[\frac{f'^3}{n}+\frac{(n-3)f+2k}{(n-1)r}f'^2-\frac{2}{r^2}f(f-k)f'\\
&-\frac{1}{r}f f''\left(f'r-2(f-k)\right)\Bigg]=2GM+\frac{1}{3}\Lambda r^3\, .
\label{kfequationn2}
\end{align}
The generalized versions of equations \req{massnn} and \req{temperaturenn} read
%By performing a Taylor expansion around the horizon we can obtain the relations among $r_h$, $T$ and $M$:

\begin{align} 
 2GM=&-\frac{1}{3}\Lambda r_h^3+kr_h-r_h\sum_{n=3}^{\infty}\lambda_n \left(\frac{2\pi TL^2}{r_h}\right)^{n-1}(kn+(n-1)2\pi T r_h )\, , \\ 
 k-r_h^2\Lambda=&4\pi Tr_h+\sum_{n=3}^{\infty}\lambda_n \left(\frac{2\pi TL^2}{r_h}\right)^{n-1}(kn+(n-3)2\pi T r_h) .
 \end{align}
 
Wald's entropy reads in turn

\begin{align}
&S=\frac{V_k r_h^2}{4G}\left[1-\sum_{n=3}^{\infty}n\lambda_n \left(\frac{2\pi TL^2}{r_h}\right)^{n-1} \left(\frac{k(n-1)}{(n-2) 2\pi T r_h}+1\right)\right]+2\pi k \lambda_2 \frac{L^2}{G}\, ,
\end{align}
where $V_k$ is the volume of the transverse space with metric $d\Sigma_{(k)}^2$. We note that $M$ is not actually the mass if $k\neq 1$, but the energy density in every case is $\rho=E/V_k=M/(4\pi)$. In the same way, we define the entropy density $s=S/V_k$. Then, the equations above can be solved parametrically in terms of $\chi=2\pi T L^2/r_h$ as follows

\begin{eqnarray}
r_h&=&L\left[\frac{k h'(\chi)}{3h(\chi)-\chi h'(\chi)}\right]^{1/2}\, ,\\
T&=&\frac{\chi r_h}{2\pi L^2}=\frac{\chi}{2\pi L}\left[\frac{k h'(\chi)}{3h(\chi)-\chi h'(\chi)}\right]^{1/2}\, ,\\
\rho&=&\frac{r_h^3}{4\pi G L^2}h(\chi)=\frac{Lh(\chi)}{4\pi G}\left[\frac{k h'(\chi)}{3h(\chi)-\chi h'(\chi)}\right]^{3/2}\, ,\\
s&=&\frac{1}{4G}\left[h'(\chi)r_h^2+k L^2\int_{\chi_{2}}^{\chi} dx \frac{h''(x)}{x} \right]\\\nonumber
&=&\frac{k L^2}{4G}\left[\frac{h'^2(\chi)}{3h(\chi)-\chi h'(\chi)}+\int_{\chi_{2}}^{\chi} dx \frac{h''(x)}{x} \right]\, ,
\end{eqnarray}
where
\begin{equation}
h(x):=\frac{L^2\Lambda}{3}+x-\sum_{n=3}^{\infty} \lambda_n x^n\, .
\end{equation}

%We can compute explicitly the thermodynamics in the limit $M<<M_{\rm max}$, corresponding to small stable black holes. We obtain the following relations, 
%\begin{equation}
%r_h=\left[\frac{M}{\zeta \chi ^3 M_c^2 M_{\rm \ssc P}^2} \right]^{1/3},\quad T=\frac{1}{4\pi}\left[\frac{M M_c^4}{\zeta  M_{\rm \ssc P}^2} \right]^{1/3},\quad \mathsf{S}=6\pi \left[\frac{\zeta^{1/2}  M M_{\rm \ssc P} }{ M_c^2} \right]^{2/3}+\frac{2\pi M_p^2}{M_c^2}\left(\lambda_2-2\sum_{n=3}^{\infty}\frac{\lambda_n}{n-2}\chi^{n-2}\right).
%\end{equation}
%In these expressions $\chi$ and $\zeta$ are constants which are determined through the equations
%\begin{equation}
%\sum_{n=3}^{\infty}\frac{2\lambda_n}{n-1}\chi^{n-1}=1,\quad \zeta=\frac{\Lambda_0}{3M_c^2\chi^3}+\sum_{n=3}^{\infty}\frac{\lambda_n}{n}\chi^{n-3}.
%\end{equation}

\section{Linearized equations}\label{linearized}
We can obtain the embedding equation of a maximally symmetric background of curvature $\mathcal{K}$ by plugging $f(r)=k-\mathcal{K}r^2$ in \req{kfequationn2} with $M=0$. Then we get the equation
\begin{equation}
h\left(-L^2\mathcal{K}\right)=0
%\Lambda+M_c^2\sum_{n=3}^{\infty}\frac{\lambda_n (-1)^n}{n(n-1)}\left(\frac{2\Lambda}{M_c^2}\right)^n=\frac{1}{3}\Lambda_0\, ,
\label{Lambda}
\end{equation}
which determines the possible vacua of the theory. For any of them, we know that the theory only propagates a massless graviton, due to the results in Chapter~\ref{Chap:2} \cite{PabloPablo3}.\footnote{We have verified this explicitly for all densities in Eqs.~\req{r3}-\req{r10}.} Thus, the linearized equations satisfied by a metric perturbation $h_{\mu\nu}$ over a maximally symmetric background $\bar{g}_{\mu\nu}$, read simply
\begin{equation}
G_{\mu\nu}^L=8\pi G_{\rm eff} T_{\mu\nu}\, ,
\end{equation}
where $G_{\mu\nu}^L$ is the linearized Einstein tensor, given in Eq.~\req{eq:GL}, and where the effective gravitational constant $G_{\rm eff}$ reads
\begin{equation}
G_{\rm eff}=\frac{G}{h'\left(-L^2\mathcal{K}\right)}\, .
\end{equation}
Observe that, for theories with Einstein-like spectrum, this can be generically obtained by using the fact that $G/G_{\rm eff}$ is the slope of Eq.~\req{Lambda} evaluated on the background \cite{PabloPablo,Aspects}.
Finally, in the transverse gauge, $\bar \nabla_{\mu}h^{\mu\nu} =\bar \nabla^{\nu}h$, we can write
\begin{equation}
-\bar\Box h_{\mu\nu}=16\pi G_{\rm eff} T_{\mu\nu}\, .
\end{equation}

\chapter{Holographic studies of ECG}\label{App:5}

\section{$\braket{TTT}$ parameters from $h_q$}\label{ttt}
In this appendix we show that the formulas in \cite{Chu:2016tps} for the twist operator scaling dimensions $h_q$ around $q=1$ can be used to obtain the exact values of the parameters $t_2$ and $t_4$ for holographic Gauss-Bonnet in general dimensions, and for QTG in $d=4$. The general-$d$ version of \req{minino} reads  \cite{Chu:2016tps}
\begin{equation}\label{Miaoformula}
\frac{h_q}{\ctt}=2\pi^{\frac{d}{2}+1}\frac{\Gamma(d/2)}{\Gamma(d+2)}(q-1)+\frac{h''_q(1)}{2\ctt}(q-1)^2+\mathcal O(q-1)^3\ ,
\end{equation}
where 
\begin{equation}\label{Miaoformula2}
\begin{aligned}
\frac{h''_q(1)}{\ctt}=&-\frac{2\pi^{1+d/2}\Gamma(d/2)}{(d-1)^3d(d+1)\Gamma(d+3)}\Bigg[d\left(2d^5-9d^3+2d^2+7d-2\right)\\
&+(d-2)(d-3)(d+1)(d+2)(2d-1)t_2+(d-2)(7d^3-19d^2-8d+8)t _4\Bigg]\ .
\end{aligned}
\end{equation}
This expression is valid for general holographic higher-order gravities, at least at leading order in the gravitational couplings. 
%We want to show that, for some theories, this expression also holds  to all order in the couplings. In order to do so, we consider Gauss-Bonnet in arbitrary dimensions and quasi-topological gravity in $4+1$ dimensions.
\subsection{Gauss-Bonnet in arbitrary dimensions}
In this case, the expression for the scaling dimension of twist operators is given by \cite{HoloRen} 
\begin{equation}\label{eq:scaling_dimension_GB}
\frac{h_q}{\ctt}=\frac{\Gamma(d/2)}{4\Gamma(d+2)}\pi^{1+d/2} qx_q^{d-4}(x_q^2-1)\left[d-3-(d+1)x_q^2+(d-3)\frac{1-2\frac{d-1}{d-3}\lambda f_\infty}{1-2\lambda f_\infty}(x_q^2-1)\right]\, ,
\end{equation}
where $x_q$ satisfies the following quartic equation 
\begin{equation}
x_q^4 d-\frac{2}{q}x_q^3-(d-2)x_q^2+\lambda f_\infty \left[4\frac{x_q}{q}-x_q^4 d+d-4\right]=0\, .
\end{equation}
A Taylor expansion around $q=1$ gives 
\begin{equation}
x_q=1+\frac{1}{1-d}(q-1)+\frac{d}{(d-1)^3}\frac{-2d+3+\lambda f_\infty (4d-10)}{-2+4\lambda f_\infty}(q-1)^2+\mathcal O(q-1)^3 \, .
\end{equation}
Plugging this expansion into \req{eq:scaling_dimension_GB}, we find 
\begin{align}
&\frac{h_q}{\ctt}=\frac{2\Gamma(d/2)\pi^{1+d/2}}{\Gamma(d+2)}(q-1)\\ \notag
&-\frac{(d-1)\Gamma(d/2)\pi^{1+d/2}}{\Gamma(d+2)}\left[-1+4d-2d^2+\lambda f_\infty(6-16d+4d^2)\right](q-1)^2+\mathcal O(q-1)^3\, .
\end{align}
Comparing this with (\ref{Miaoformula}), we find that $t_2$ and $t_4$ should be given by
\begin{equation}\label{eq:t2_GB}
t_2=\frac{4d(d-1) \lambda f_\infty}{(d-2)(d-3) (1-2\lambda f_\infty)}\, , \quad t_4=0\, ,
\end{equation}
which matches the exact nonperturbative result \cite{Buchel:2009sk}.

\subsection{Quasi-topological gravity}
In this case, the scaling dimension $h_q$ was obtained in \cite{HoloRen} in terms of the charges $a$, $c$ and $t_4$ of the theory as
\begin{equation}\label{eq:hq_quasitopo}
h_q=\frac{aq}{4\pi x_q^2}(x_q^2-1)\left[x_q^4\left(1-5\frac{c}{a}-10\frac{c}{a}t_4\right)-x_q^2\left(1-\frac{c}{a}-8\frac{c}{a}t_4\right)+2\frac{c}{a}t_4\right]\ ,
\end{equation}
where 
\begin{eqnarray}\label{centralchargesquasitopological}
c&=&\pi^2\frac{\tilde L^3}{8\pi G}\left(1-2\lambda f_\infty -3\mu f_\infty^2\right)\ , \\
a&=&\pi^2\frac{\tilde L^3}{8\pi G}\left(1-6\lambda f_\infty +9\mu f_\infty^2\right)\ , \\
t_4&=&\frac{3780\mu f_\infty^2}{1-2\lambda f_\infty-3\mu f_\infty^2}\ ,
\end{eqnarray}
 and where $x_q$ satisfies the following quartic equation
\begin{equation}\label{eq:xq-quasitopo}
2x_q^6-\frac{x_q^5}{q}-x_q^4+2\lambda f_\infty x_q^3\left(\frac{1}{q}-x_q^3\right)+\mu f_\infty^2\left(-1+\frac{3x_q}{q}-2x_q^ 6\right)=0\ .
\end{equation}
Moreover, we have \cite{Myers:2010jv}
\begin{equation}
t_2=\frac{24f_\infty\left(\lambda-87\mu f_\infty\right)}{1-2\lambda f_\infty-3\mu f_\infty^2}\ ,
\end{equation}
which properly reduces to the Gauss-Bonnet formula (\ref{eq:t2_GB}) for $\mu=0$ and $d=4$. Before computing the Taylor expansion of $h_q$ around $q=1$, we invert (\ref{centralchargesquasitopological}) and find\footnote{There seems to be a small typo in eq. (2.58) of \cite{HoloRen}. Note also that our convention for $t_4$ differs by a factor of $1890$ with respect to that in  \cite{HoloRen}, but agrees with the one in \cite{Myers:2010jv}.}
\begin{eqnarray}
\frac{\tilde L^3}{8\pi G}&=&\frac{a}{2\pi^2}\left(3\frac{c}{a}\left(1+\frac{3t_4}{1890}\right)-1\right)\, , \\
\lambda f_\infty&=&\frac{1}{2}\frac{\frac{c}{a}\left(1+\frac{6t_4}{1890}\right)-1}{3\frac{c}{a}(1+\frac{3t_4}{1890})-1}\, , 
\label{eq:lambda}
\\
\mu f_\infty^2&=&\frac{\frac{c}{a}\frac{t_4}{1890}}{3\frac{c}{a}\left(1+\frac{3t_4}{1890}\right)-1}\, .
\end{eqnarray}
and rewrite (\ref{eq:xq-quasitopo}) in terms of $c/a$ and $t_4$. We get 
\begin{equation}
x_q(q)=1-\frac{q-1}{3}+\frac{4+\frac{8t_4}{1890}-\frac{2}{3}\frac{a}{c}}{9}(q-1)^2+\mathcal O(q-1)^3 \, ,
\end{equation}
and plugging it into (\ref{eq:hq_quasitopo}), we find 
\begin{equation}
\frac{h_q}{c}=\frac{2}{3\pi}(q-1)+ \frac{7 \frac{a}{c}-24-\frac{84 t_4}{1890}}{27\pi}(q-1)^2+\mathcal O(q-1)^3 \, .
\end{equation}
Comparing the leading term, we notice that $\ctt$ should be related to $c$ via $\tfrac{\ctt}{c}=\tfrac{40} {\pi^4}$, which is correct. Finally, using
\begin{equation}
\frac{a}{c}=1-\frac{t_2}{6}-\frac{4}{45}t_4\ ,
\end{equation}
we find
\begin{equation}
\frac{h''_q(1)}{2\ctt }=-\pi^3\frac{102+7t_2+4t_4}{6480}\ ,
\end{equation}
which exactly agrees with the general formula \req{Miaoformula2} when we particularize it to $d=4$.

\section{Generalized action for Gauss-Bonnet gravity}\label{BTcheck}
In this appendix we perform an additional check of the generalized action introduced in section \ref{osa}. In particular, we apply it here to a theory for which the exact generalization of the GHY term is known, namely, $D$-dimensional Gauss-Bonnet gravity \cite{Teitelboim:1987zz,Myers:1987yn}. The full Euclidean action of the theory reads
\begin{equation}
I_E^{\rm GB}=-\frac{1}{16\pi G}\int_{\mathcal{M}}d^Dx\sqrt{g}\left[\frac{(D-1)(D-2)}{L^2}+R+\frac{L^2\lambda}{(D-3)(D-4)}\mathcal{X}_4\right]+I^{\rm GB}_{\rm GHY}+I^{\rm GB}_{\rm CT}\, ,
\end{equation}
where the generalization of the GHY term reads
\begin{equation}
\begin{aligned}
I^{\rm GB}_{\rm GHY}&=-\frac{1}{8\pi G}\int_{\partial \mathcal{M}}d^{D-1}x\sqrt{h}\Bigg\{K+\frac{L^2\lambda}{(D-3)(D-4)}\delta^{a_1a_2a_3}_{b_1b_2b_3}K^{b_1}_{a_1}\left(\mathcal{R}^{b_2b_3}_{a_2a_3}-\frac{2}{3}K^{b_2}_{a_2}K^{b_3}_{a_3}\right)\Bigg\}\, ,
\end{aligned}
\end{equation}
and the counterterms can be chosen as \cite{Emparan:1999pm, Mann:1999pc, Balasubramanian:1999re, Brihaye:2008xu,Astefanesei:2008wz}
\begin{align}\notag
I^{\rm GB}_{\rm CT}&=\frac{1}{8\pi G}\int_{\partial \mathcal{M}}d^{D-1}x\sqrt{h}\Bigg\{\frac{(D-2)(f_{\infty}+2)}{3f_{\infty}^{1/2}L}+\frac{ L(3f_{\infty}-2)}{2f_{\infty}^{3/2}(D-3)}\mathcal{R}+\frac{ L^3\Theta[D-6]}{2f_{\infty}^{5/2}(D-3)^2(D-5)}\\
&\times\left[(2-f_{\infty})\left(\mathcal{R}_{ab}\mathcal{R}^{ab}-\frac{D-1}{4(D-2)}\mathcal{R}^2\right)-\frac{(D-3)(1-f_{\infty})}{D-4}\mathcal{X}_4(h)\right]+\ldots\Bigg\}\, .
\end{align}
With these boundary contributions, the Gauss-Bonnet action functional is differentiable and finite in AdS spaces. Since Gauss-Bonnet gravity has an Einstein-like spectrum in pure AdS$_D$ (in fact, in any background), our generalized GHY term should also be applicable to GB gravity, as long as the boundary consists only of asymptotically AdS pieces. The prescription in \req{SEcomplete} gives the following result when applied to the Gauss-Bonnet Lagrangian,
\begin{equation}
\begin{aligned}
I_{\rm GGHY}^{\rm GB}+I_{\rm GCT}^{\rm GB}&=-\frac{1-2\lambda f_{\infty}\frac{D-2}{D-4}}{8 \pi G}\int_{\partial \mathcal{M}}d^{D-1}x\sqrt{h}\bigg[K-\frac{D-2}{\tilde L}-\frac{\tilde L}{2(D-3)}\mathcal{R}\\
&-\frac{\tilde L^3\Theta[D-6]}{2(D-3)^2(D-5)}\left(\mathcal{R}_{ab}\mathcal{R}^{ab}-\frac{D-1}{4(D-2)}\mathcal{R}^2\right)+\ldots\bigg]\, ,
\end{aligned}
\end{equation}
where we included a set of counterterms valid up to $D=7$ and where $\tilde L=L/\sqrt{f_{\infty}}$. Recall that the coefficient in front of the integral is proportional to the universal constant $a^*$ appearing in the EE across a spherical region, which for GB gravity reads
\begin{equation}
a^*=\left(1-2\lambda f_{\infty}\frac{d-1}{d-3}\right)\frac{\tilde L^{d-1}\Omega_{(d-1)}}{16 \pi G}\, .
\end{equation}
In order to compare both boundary terms, let us consider a metric of the form
\begin{equation}
ds^2=f(r)d\tau^2+\frac{dr^2}{f(r)}+r^2d\Sigma_{k}^2\, ,
\end{equation}
where $d\Sigma_{k}^2$ is the metric of a maximally symmetric space of curvature $k=-1,0,1$ and $\tau$ has period $\beta$. For $f(r)=f_{\infty} r^2/L^2+k$, the previous metric reduces to pure Euclidean AdS$_D$, with the boundary at $r=+\infty$, which we regulate as $r\rightarrow L^2/\delta$. Let us now switch on arbitrary radial perturbations
\begin{equation}
f(r)=f_{\infty}\frac{r^2}{L^2}+k+\frac{f_1}{r}+\frac{f_2}{r^2}+\frac{f_3}{r^3}+\ldots\, .
\end{equation}
Evaluated at  $r\rightarrow L^2/\delta$, the boundary terms coming from both prescriptions yield, respectively
\begin{eqnarray*}
I_{\rm GHY+CT}^{\rm GB}&=&\frac{\beta \pi}{4 G}\left[\frac{(5 f_{\infty}-6) L^6}{\delta^4}
+\frac{f_1 (5 f_{\infty}-6) L^2}{f_{\infty}\delta}+\frac{(5 f_{\infty}-6) \left(4 f_2 f_{\infty}-3 L^2 k^2\right)}{8 f_{\infty}^2}+\mathcal{O}(\delta^2)\right]\, ,\\
I_{\rm GGHY+GCT}^{\rm GB}&=&\frac{\beta \pi}{4G}\left[\frac{(5 f_{\infty}-6) L^6}{\delta^4}
+\frac{f_1 (5 f_{\infty}-6) L^2}{f_{\infty}\delta}+\frac{(5 f_{\infty}-6) \left(4 f_2 f_{\infty}-3 L^2 k^2\right)}{8 f_{\infty}^2}+\mathcal{O}'(\delta^2)\right]\, .
\end{eqnarray*}
This is, all divergent and finite terms are equal! The difference only appears in the decaying terms, which of course give no contribution to the action. For the sake of simplicity, we evaluated the above expressions for $D=5$, but it is straightforward to check that the same phenomenon happens in higher dimensions (with the expressions above we have checked $D=6,7$). Therefore, at least from a practical point of view, our generalized boundary term is as good as the Gauss-Bonnet one when applied to asymptotically AdS spaces. We expect our method to work also for general Lovelock gravities, as well as for QTG, and the rest of theories belonging to the Einstein-like class in the classification of \cite{Aspects}.
%The advantage of the new boundary term is that it can be applied to arbitrary higher-order theories, providing that their linearized equations are Einstein-like. 

\section{Boundary terms in the two-point function}\label{2pbdy}
In this appendix we evaluate explicitly the boundary contribution in \req{EuclideanECGc6} for the metric perturbation considered in section \ref{tt}. The sum of all boundary contributions appearing in \req{ICT}, which includes the one coming from $I^{\rm ECG}_{E\, \rm Bulk}$ in \req{ttw}, as well as the generalized GHY term and the counterterms in  \req{EuclideanECGc6}, reads
\begin{equation}
I_{E\, \rm bdry}^{\rm ECG}=-\frac{1}{8\pi G}\int d^3x\left[\frac{1}{2}\Gamma_r+(1+3\mu f_{\infty}^2)\sqrt{h}\left(K-\frac{2\sqrt{f_{\infty}}}{L}-\frac{L}{2\sqrt{f_{\infty}}}\mathcal{R}\right)\right]\, ,
\end{equation}
where $\Gamma_r$ comes from integration by parts in the bulk action, and is given by
\begin{align}\label{gaga}
\Gamma_r=\frac{1}{\sqrt{f_{\infty}}}\bigg[&-\frac{2(4f_{\infty}-3)r^3}{L^4}+\frac{4f_{\infty}-3}{L^4}\left(2r^4\phi\partial_r\phi+r^3\phi^2\right)+\frac{(f_{\infty}-1)r^5}{L^4}(\partial_r\phi)^2\\
&+6\mu f_{\infty}^2\left(-\frac{r}{2}(\partial_{\tau}\phi)^2+r^2\partial_r\phi\partial_{\tau}^2\phi\right)\bigg]\, .
\end{align}
The rest are the boundary terms in the action \req{EuclideanECGc6}. The induced metric on a hypersurface of fixed $r$ is
\begin{equation}
^{(3)}ds^2=\frac{r^2}{L^2}\left(d{\tau}^2+dx^2+dy^2+2dxdy\phi(r,{\tau})\right)\, .
\end{equation}
At quadratic order in $\phi$ we have
\begin{equation}
\sqrt{h}=\frac{r^3}{L^3}\left(1-\frac{1}{2}\phi^2\right)\, ,\quad \mathcal{R}=\frac{L^2}{2 r^2}\left(3(\partial_{\tau}\phi)^2+4\phi\partial_{\tau}^2\phi\right)\, , \quad K=\frac{3\sqrt{f_{\infty}}}{L}-\frac{r\sqrt{f_{\infty}}}{L}\phi\partial_r\phi\, .
\end{equation}
Then, we obtain at that order
\begin{equation}
\begin{aligned}
I_{E\, \rm bdry}^{\rm ECG}=\frac{1}{8\pi G}\int d^3x\bigg[&\frac{3r^3}{L^4\sqrt{f_{\infty}}}(1-f_{\infty}+\mu f_{\infty}^3)\left(-1+\frac{\phi^2}{2}+r\phi\partial_r\phi\right)-\frac{3(f_{\infty}-1)r^5}{2\sqrt{f_{\infty}}L^4}(\partial_r\phi)^2\\
&+\frac{r}{\sqrt{f_{\infty}}}\left((1+3\mu f_{\infty}^2)\left(\frac{3}{4}(\partial_{\tau}\phi)^2+\phi\partial_{\tau}^2\phi\right)+\frac{3\mu}{2}f_{\infty}^2(\partial_{\tau}\phi)^2\right)\\
&-3\mu f_{\infty}^{3/2}r^2\partial_r\phi\partial_{\tau}^2\phi\bigg]\, .
\end{aligned}
\end{equation}
The first term vanishes  because $1-f_{\infty}+\mu f_{\infty}^3=0$ is precisely the AdS$_4$ embedding equation \req{roo}. Now it proves useful to perform the Fourier transformation of $\phi$:
\begin{equation}
\phi(r,{\tau})=\frac{1}{2\pi}\int dp \phi_0(p)e^{ip {\tau}}H_p(r)\, ,
\end{equation}
with $H_p(r)=e^{-\frac{L^2 |p|}{\sqrt{f_{\infty}}r}}\left(1+\frac{L^2 |p|}{\sqrt{f_{\infty}}r}\right)$. Then,
\begin{equation}
\begin{aligned}
I_{E\, \rm bdry}^{\rm ECG}=&\frac{V_{\mathbb{R}^2}}{16 \pi^2 G}\int  dpdq \delta(q+p)\phi_0(p)\phi_0(q)\bigg[-\frac{3(f_{\infty}-1)r^5}{2\sqrt{f_{\infty}}L^4}(\partial_r H_p)^2\\
&+\frac{r H_p^2}{\sqrt{f_{\infty}}}\left((1+3\mu f_{\infty}^2)\left(-\frac{3}{4}p q-q^2\right)-\frac{3\mu}{2}f_{\infty}^2pq\right)+3\mu q^2 f_{\infty}^{3/2}r^2H_q\partial_rH_p \bigg]\, .
\end{aligned}
\end{equation}
Now, since $\partial_rH_p\sim 1/r^3$ , the first and last terms vanish for $r\rightarrow\infty$. Then, we are left with the final result
\begin{equation}
\begin{aligned}
I_{E\, \rm bdry}^{\rm ECG}=-\frac{V_{\mathbb{R}^2}(1-3\mu f_{\infty}^2)}{64\pi^2 G\sqrt{f_{\infty}}}\int dpdq \delta(q+p)\phi_0(p)\phi_0(q) p^2 rH_p^2(r)\, ,
\end{aligned}
\end{equation}
which appears in the main text.

\section{Generalization to higher orders}\label{App:Gen}
The results in Chapter~\ref{Chap:6} can be generalized for all the theories in \req{lal}, 
\begin{equation}\label{lal2}
\mathcal{L}(g^{\mu\nu},R_{\mu\nu\rho\sigma})= \frac{1}{16\pi G}\left[\frac{6}{L^2}+R+\sum_{n=2}^{\infty}L^{2n-2}\lambda_{n}\mathcal{R}_{(n)}\right]\, .
\end{equation}
Let us briefly study several holographic aspects of this theory --- they will be of vital importance in Chapter~\ref{Chap:8}.  First, let us introduce the function
\begin{equation}\label{hinf0}
h(x)\equiv\frac{16\pi G L^2}{6}\left[\mathcal{L}(x)-\frac{x}{2}\mathcal{L}'(x)\right]=0\, ,
\end{equation}
where $\mathcal{L}(x)$ is the on-shell Lagrangian on pure AdS$_{4}$ with radius $L/\sqrt{x}$. For the Lagrangian above, this function reads explicitly
\begin{equation}
h(x)=1-x+\sum_{n=3}^{\infty} \lambda_n x^n\, .
\end{equation}

Then, according to the results of Chapter~\ref{Chap:1} --- see equation \req{Lambda-eq} ---, the possible AdS vacua of the theory have a radius $\tilde L=L/\sqrt{f_{\infty}}$, where $f_{\infty}$ is a positive solution of the equation $h(f_{\infty})=0$. In this way, the roots of $h(x)$ represent the AdS vacua, and this is a general property of any higher-derivative gravity, not only of \req{lal2}. 
Now, once a vacuum is chosen we can compute the effective Newton's constant. For any Einstein-like theory we know that $G_{\rm eff}$ can be again computed in terms of $h$ (see \req{eq:kappaEL}),\footnote{Of course, we must demand $h'(f_{\infty})<0$ in order for the vacuum to be physical.}
\begin{equation}
G_{\rm eff}=-\frac{G}{h'(f_{\infty})}\, .
\end{equation}
But we have learned in this chapter that the effective Newton's constant is essentially what determines the central charge of the 2-point function, $\ctt$.
\begin{equation}\label{cttall}
\ctt = -h'(f_{\infty})\frac{3}{\pi^3}\frac{\tilde{L}^2}{G}\, .
\end{equation}
This result applies, in particular, to the theories \req{lal2}, since they are Einstein-like. Furthermore, they have the additional advantage of allowing for simple black hole solutions whose thermodynamic properties can be studied analytically. 
In fact, we already studied in Appendix~\ref{App:4} the black hole solutions of \req{lal2} for any value of the cosmological constant. Adapting those results in order to match the conventions of the different metrics in \req{eqVs}, we obtain the general expressions for the radius $r_h$, the temperature $T$, the total energy $E$ and the entropy $S$,
\begin{eqnarray}
r_h&=&L\left[\frac{k h'(\chi)}{3h(\chi)-\chi h'(\chi)}\right]^{1/2}\, ,\\
T&=&\frac{N\chi}{2\pi L}\left[\frac{k h'(\chi)}{3h(\chi)-\chi h'(\chi)}\right]^{1/2}\, ,\\
E&=&-\frac{V_{\Sigma}  N h(\chi)}{4\pi G L}\left[\frac{k h'(\chi)}{3h(\chi)-\chi h'(\chi)}\right]^{3/2}\, ,\\
S&=&-\frac{k V_{\Sigma} }{4G}\left[\frac{h'^2(\chi)}{3h(\chi)-\chi h'(\chi)}+\int_{\chi_{2}}^{\chi} dx \frac{h''(x)}{x} \right]\, ,
\end{eqnarray}
and from these one can also get the free energy as $F=E-TS$. These quantities are expressed parametrically in terms of $\chi$, and by taking values of that variable we generate the different curves such as $S(E)$, $r_h(E)$, $S(T)$, etc. We do not intend to perform here a detailed characterization of these relations, but they can be used in order to obtain a general formula for the 3-point function charge $t_4$, using the formulas \req{Miaoformula} and \req{Miaoformula2} found in \cite{Chu:2016tps}. 
Applying the previous expressions to the case $k=-1$ (for which $N=\tilde L/R$ and $V_{\Sigma}=-2\pi L^2$), we can write a parametric expression for the scaling dimension of twist operators,
\begin{eqnarray}
h_q&=&\frac{L^2h(\chi)h'(\chi)}{4G\chi(\chi h'(\chi)-3h(\chi))}\, ,\\
q&=&\frac{\sqrt{f_{\infty}}}{\chi}\left[\frac{h'(\chi)}{\chi h'(\chi)-3h(\chi)}\right]^{-1/2}
\end{eqnarray}
As a check, for $\chi=f_{\infty}$ we get $q=1$, $h_1=0$. In the same way, all the derivates of $h_q$ at $q=1$ are going to be related to derivatives of $h(\chi)$ at $\chi=f_{\infty}$. For the first and second derivatives we obtain 
\begin{eqnarray}
h_q'(1)&=&-\frac{h'(f_{\infty})}{8 f_{\infty}G}\, ,\\
h_q''(1)&=&\frac{14 h'(f_{\infty})+7f_{\infty}h''(f_{\infty})}{64 f_{\infty} G}\, .
\end{eqnarray}
We check that $h_q'(1)$ precisely coincides with \req{Miaoformula} if we take into account \req{cttall}, while $h_q''(1)$ together with \req{Miaoformula2} yields the following formula for $t_4$
\begin{equation}
t_4=210 f_{\infty}\frac{h''( f_{\infty})}{h'( f_{\infty})}\, .
\end{equation}
Since this expression only involves the evaluation of certain quantity on the vacuum, it is probably correct not only for all of the GQGs in \req{lal2}, but for all of the Einstein-like theories.

\chapter{Numerical methods for Taub-NUT solutions}\label{App:6}\label{methods}

We have presented in our investigation a number of numerical solutions for the NUT and bolts. Here we provide some details on how these solutions were obtained.  The differential equations solved here are in general stiff, which results in difficulties in the numerical scheme. All numerical solutions presented in this work were obtained using Mathematica, utilizing the \emph{ImplicitRungeKutta} method of \emph{NDSolve}.  This method satisfies A-stability, making it a suitable method for stiff differential equations. High \emph{WorkingPrecision} was used in the numerical solver, ranging between 20 and 50 on a case by case basis.

\begin{figure}[t] 
\centering
\includegraphics[scale=0.47]{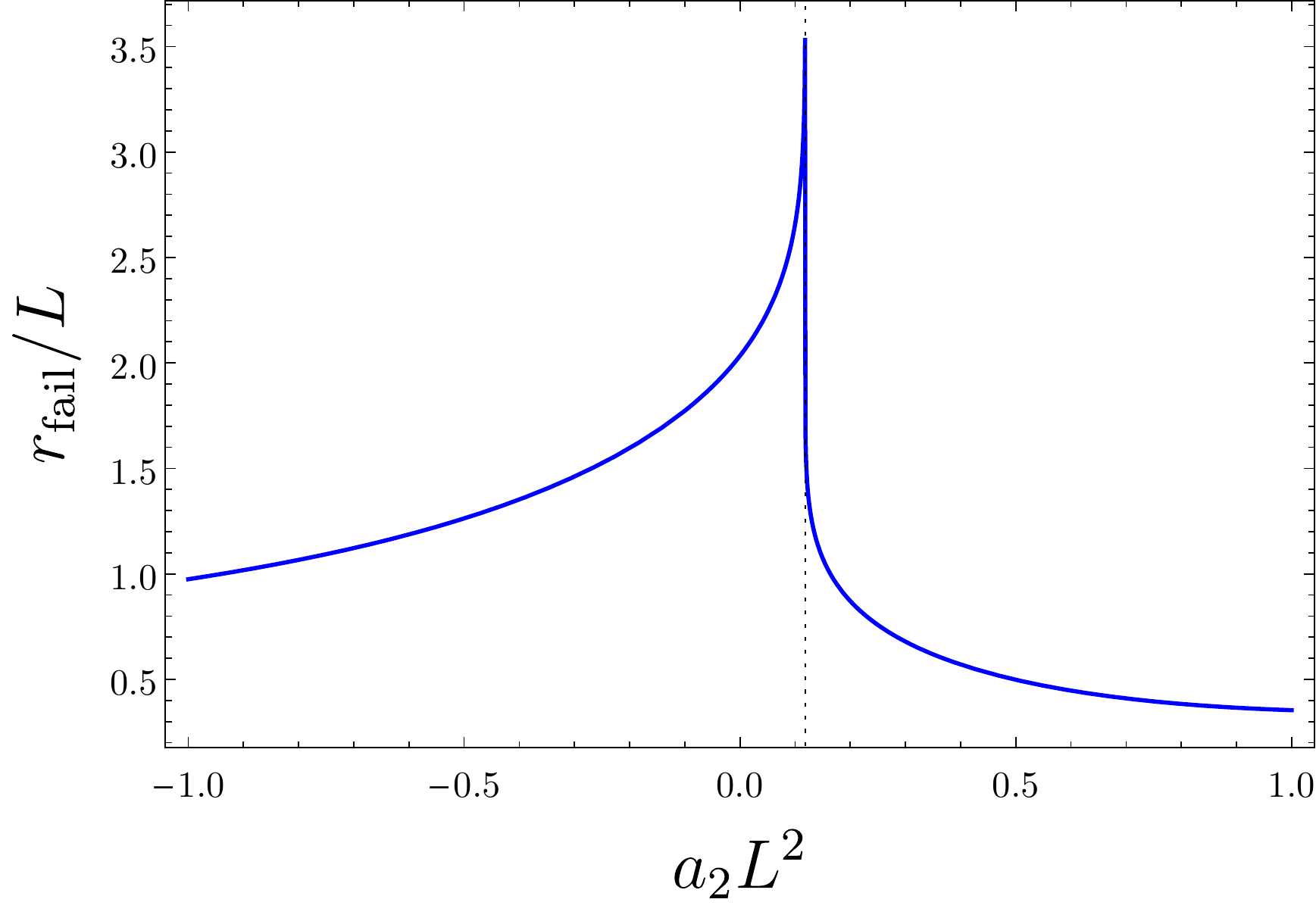}
\quad
\includegraphics[scale=0.47]{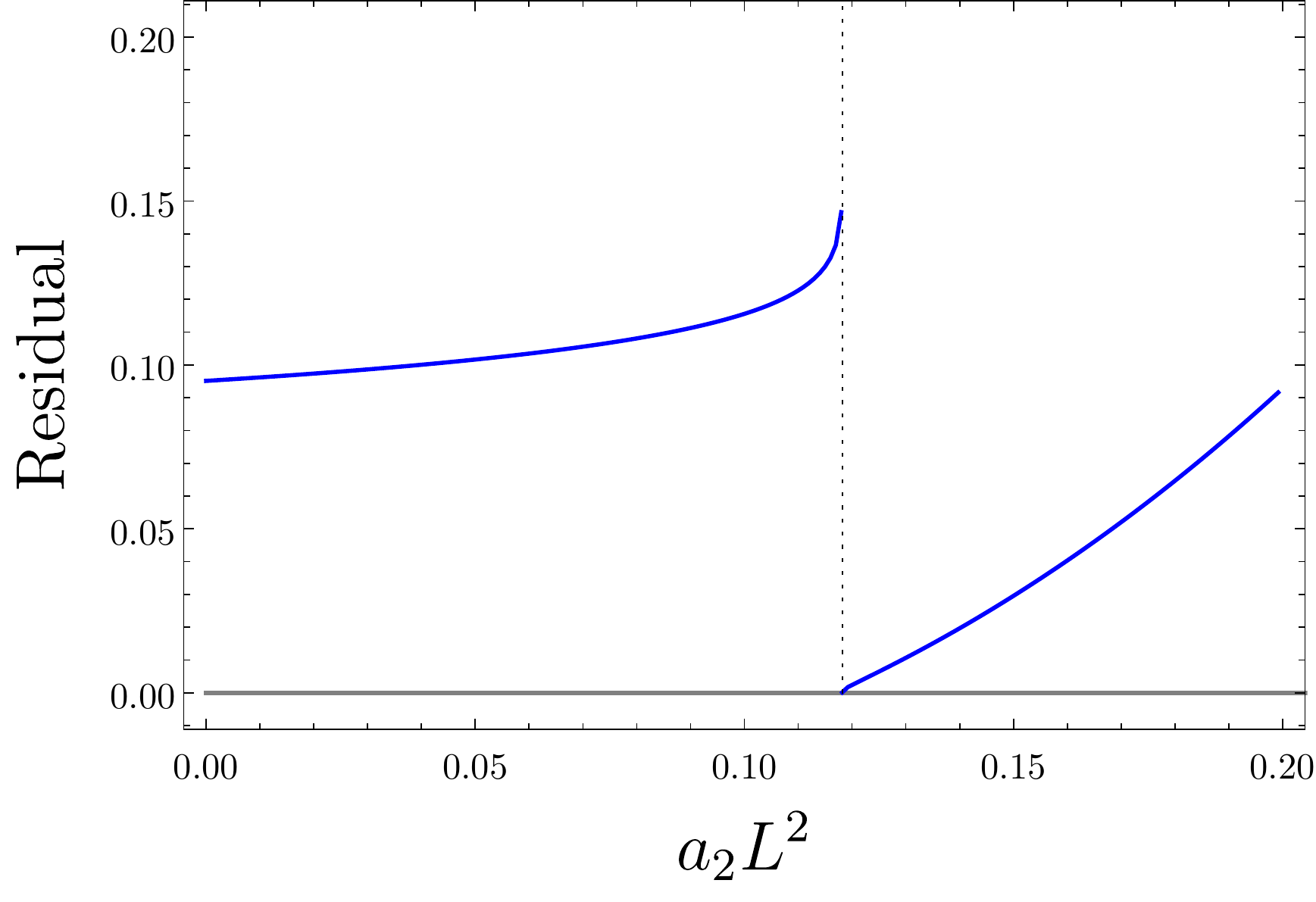} 
\caption{Left: A plot of $r_{\rm fail}$ the $r$ value at which the numerical scheme fails vs.~the shooting parameter $a_2$. Right: A plot of the residual as a function of the shooting parameter $a_2$. In both plots, the dotted line corresponds to $a_2 L^2 = 0.1181855186708097$. Both plots are for the case of NUT solutions in the quartic generalized quasi-topological theory for $\mathcal{B} = \mathbb{CP}^2$ with $\xi = -10$, $n/L = 1/3$ and $\epsilon = 10^{-2} L$.  A working precision of 40 was used in producing these particular plots. }
 \label{numerics-a2}
\end{figure}

Let us make some remarks on the details of the numerical scheme, focusing on the $\mathcal{B} = \mathbb{CP}^k$ bases. The metric function $V_{\mathcal{B}}(r)$ was expanded near a NUT or a bolt as 
\be\label{numericV}
V_{\mathbb{CP}^k}(\epsilon) = \frac{\epsilon}{(k+1)n} + a_2 \epsilon^2 \, ,
\ee
where $\epsilon = (r-n)$ for the NUTs or $\epsilon = (r-\rh)$ for the bolts is taken to be some small, positive quantity --- typically $10^{-2}L-10^{-3}L$ in this work. The parameter $a_2$ is not fixed by the near horizon solution, and must be determined via the shooting method. Specifically, for a given choice of $a_2$, Eq.~\eqref{numericV} is used to generate initial data for the differential equation, namely $V_{\mathbb{CP}^k}(\epsilon)$ and ${V'}_{\mathbb{CP}^k}(\epsilon)$.  Finding a numerical solution then reduces to finding a sensible value of $a_2$. 

\begin{figure}[t]
\centering 
\includegraphics[scale=0.65]{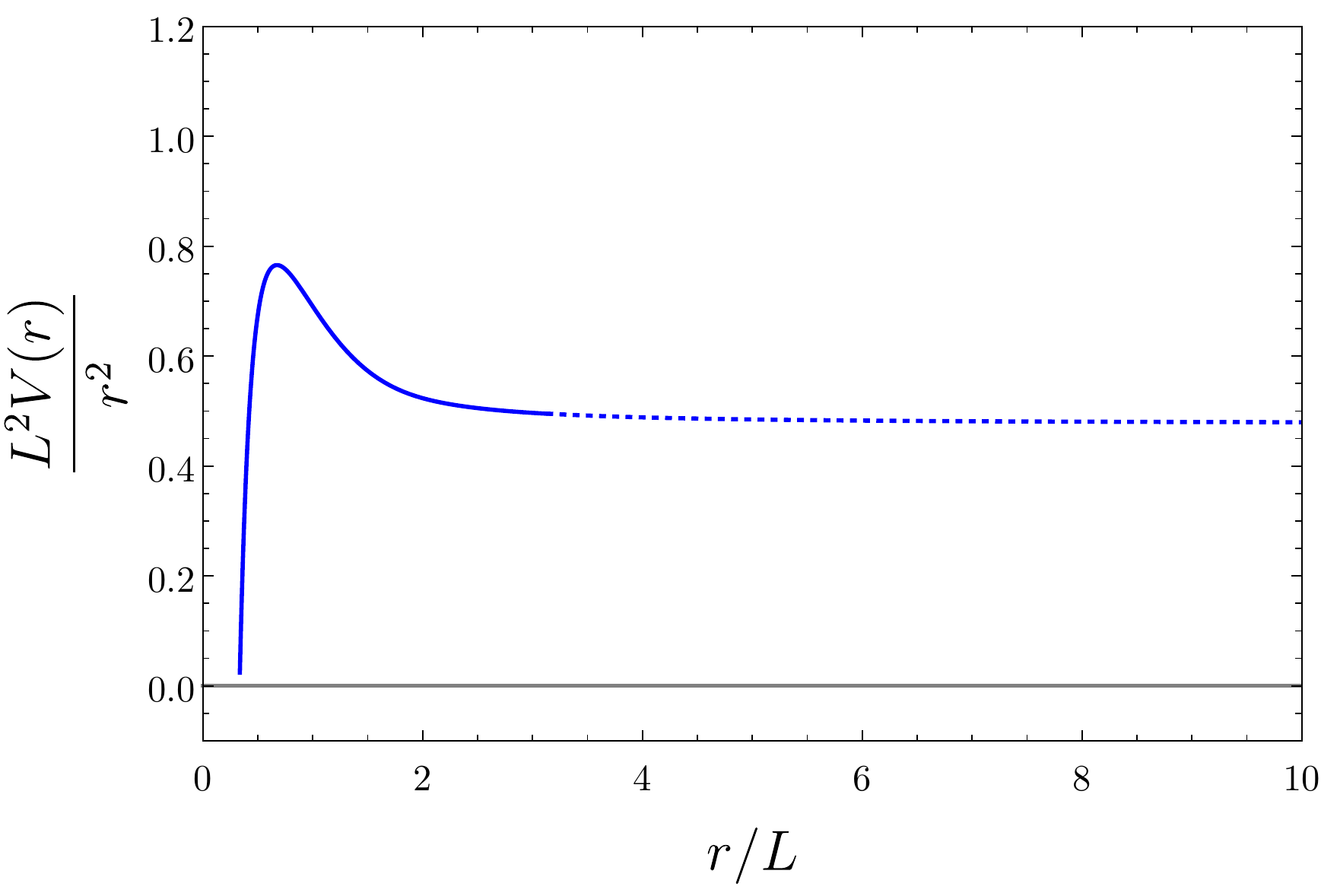} 
\caption{An example of a numeric solution. Here the solid blue curve corresponds to the result of the numeric integration, while the dotted curve corresponds to the asymptotic expansion. The plot is for the case of NUT solutions in the quartic  Quasi-topological theory for $\mathcal{B} = \mathbb{CP}^2$ with $\zeta = -10$, $n/L = 1/3$, $a_2 L^2 = 0.1181855186708097$ and $\epsilon = 10^{-2} L$.  A working precision of 40 was used in producing this plot. }
\label{numericEx}
\end{figure}

A generic choice of $a_2$ will lead to the excitation of the growing modes that appear in the asymptotic expansion of the metric function. The correct choice of $a_2$ will result in a numerical solution that approaches the $1/r$ part of the asymptotic expansion at sufficiently large $r$.  Regardless of the choice of $a_2$, the numerical scheme will eventually breakdown because of the accumulation of errors due to finite working precision. It is useful to study the point at which the numerical solution fails as a function of the shooting parameter $a_2$ --- an example of this is shown in the left plot of Fig.~\ref{numerics-a2}. This figure makes clear that there is a special value of $a_2$ that allows the solution to be integrated the furthest. It also appears that this is the unique value of $a_2$ that joins the numerical solution smoothly onto the asymptotic expansion --- see Fig.~\ref{numericEx}.

While with the proper choice of $a_2$ the solution can be visually seen to join onto the asymptotic expansion smoothly, it is nice to have quantitative confirmation of this. In the right plot of Fig.~\ref{numerics-a2} we show a residual that measures how closely the numerical solution matches the asymptotic expansion in a region where they overlap. The residual shown was calculated according to
\be 
{\rm Residual} = \int_{0.9 r_{\rm fail}}^{r_{\rm fail}} \frac{L \big|  V_{\rm numeric}(r) - V_{1/r}(r) \big|}{r^2}  dr \, ,
\ee
where again $r_{\rm fail}$ is the point at which the numerical solution breaks down. In performing this calculation, terms up to ${\cal O}(r^{-3})$ where included in the asymptotic expansion. The plot shows that the error blows up $a_2 L^2 = 0.1181855186708097$ is approached from the left, while it goes to zero when approached from the right. This confirms that the numerical solution is indeed becoming arbitrarily close to the asymptotic solution, and the asymptotic solution can be used to continue the solution to infinity.

On the contrary, there are some regions in the parameter space (for example, when the mass is negative in $D=4$) for which we argued that the solutions do not exist due to a bad asymptotic behaviour. In those cases we are not able to match the numerical solution with the asymptotic expansion, and this confirms that those solutions do not exist. 

Several strategies may be used in order to improve the precision of the numerical methods. For example, instead of working with the function $V(r)$ one may work with $f(r)=L^2 V(r)/r^2$, which should approach the constant $f_{\infty}$ at infinity. Also, more terms can be included in the expansion (\ref{numericV}), so that one does not need to choose a very small $\epsilon$ (we recall that the full expansion depends only on $a_2$).  Let us close by mentioning that the numerical problem is considerably more stiff in $D=6$ than in $D=4$. In the latter case we do not require to increase substantially the \emph{WorkingPrecision} and the parameter $\epsilon$ can be chosen as small as $10^{-3}L$.  The numerical integration in $D=6$ is less stable and requires a larger value of $\epsilon$ and higher values of \emph{WorkingPrecision}.

\chapter{Free energy of CFTs on squashed spheres}\label{App:7}

\section{Explicit checks of formula \req{fee0e}}\label{checks}
%\comment{present GB in general $d$,Quartic GQTG in $d=5$, ECG in $d=3$ and the quartic examples as highly non trivial checks  }

We have verified that our conjectured formula \req{fee0e} correctly reproduces the free energies of all Taub-NUT solutions known in the literature, computed using the standard on-shell action approach. This includes Einstein gravity and Gauss-Bonnet in general dimensions as well as the recently constructed solutions of Einsteinian cubic gravity and Quartic Generalized Quasi-topological gravities in $d=3$ and $d=5$ respectively.

In the case of $(d+1)$-dimensional Gauss-Bonnet, the complete Euclidean action, including the generalized Gibbons-Hawking boundary term \cite{Gibbons:1976ue, Myers:1987yn, Teitelboim:1987zz} and counterterms \cite{Balasubramanian:1999re, Brihaye:2008xu} reads
\begin{align} 
I_E^{\rm GB} &= - \int \frac{d^{d+1} x \sqrt{g}}{16 \pi G} \left[\frac{d(d-1)}{L^2} + R + \frac{\lambda_{\ssc \rm GB} L^2 {\cal X}_4}{(d-2)(d-3)}  \right] - \frac{1}{8 \pi G} \int_{\partial} d^d y \sqrt{h} \bigg[K +\\
& \frac{2 L^2 \lambda_{\ssc \rm GB}}{(d-2)(d-3)} \left[{\cal J} - 2 {\cal G}_{ij}K^{ij} \right] \bigg]
\nonumber
- \frac{1}{8  \pi G} \int_\partial d^d y\sqrt{h} \bigg\{- \frac{(d-1)(\fin + 2)}{3 L \fin^{1/2}} \\&\nonumber- \frac{L(3\fin - 2) \Theta[d-3]}{2 \fin^{3/2} (d-2)} {\cal R} 
- \frac{L^3 \Theta[d-5]}{2 \fin^{5/2}(d-2)^2(d-4)}\left[(2-\fin) \left({\cal R}_{ij}{\cal R}^{ij} - \frac{d}{4(d-1)}{\cal R}^2 \right)\right.\\&\left.\nonumber - \frac{(d-2)(1-\fin)}{d-3} {\cal X}_4^{(h)} \right] + \cdots \bigg\} \, ,
\end{align}
where ${\cal X}_4 = R_{\mu\nu\rho\sigma}R^{\mu\nu\rho\sigma} - 4 R_{\mu\nu}R^{\rho\sigma} + R^2$ is the Gauss-Bonnet density, $K_{ij}$ is the extrinsic curvature of the boundary with $K = h^{ij}K_{ij}$ its trace, ${\cal J} = h^{ij}{\cal J}_{ij}$ with
\begin{equation} 
{\cal J}_{ij} = \frac{1}{3} \left(2 K K_{ik}K^k_j + K_{kl}K^{kl}K_{ij} - 2 K_{ik}K^{kl}K_{lj} - K^2 K_{ij} \right)\, ,
\end{equation}
and ${\cal G}_{ij}$ is the Einstein tensor of the boundary metric $h_{ij}$.  We have also explicitly included the counterterms that ensure a finite on-shell action for  $d < 7$. The dots stand for additional contributions that are required in higher-dimensions.  Computing the on-shell action of Taub-NUT solutions in this theory yields 
\begin{align}
\mathcal{F}^{\rm EGB}_{\mathbb{S}_{\varepsilon}^{d}}=\frac{(-1)^{\frac{(d-1)}{2}}\pi^{\frac{d}{2}}(1+\varepsilon)^{\frac{(d+1)}{2}}d(d-1)L^{d-1}}{16\Gamma\left[\frac{d+2}{2}\right]f_{\infty}^{\frac{(d+1)}{2}} G}\left[1-\frac{f_{\infty}(d+1)}{(d-1)(1+\varepsilon)}+\frac{(f_{\infty}-1)(d+1)}{(d-3)(1+\varepsilon)^2}\right]\, ,
\end{align}
which is in precise agreement with the result obtained using the conjectured relationship~\req{fee0e}.

Our next example is ECG plus a quartic generalized quasi-topological term in $d = 3$. The Euclidean action with generalized boundary and counterterms reads~\footnote{Note that here we use the simple method for generating generalized boundary and counterterms introduced in~\cite{ECGholo}. There it was found that for Einstein-like higher-order gravities a finite on-shell action for asymptotically AdS spaces is obtained by using the Gibbons-Hawking-York boundary term along with the counterterms for Einstein gravity all weighted by $a^*$ --- c.f. Eq.~(4.19) of that work.}
\begin{equation}
\begin{aligned}
I_E=&-\int \frac{d^4x \sqrt{g}}{16\pi G} \left[\frac{6}{L^2}+R-\frac{\mu L^4}{8} \mathcal{P}-\frac{\xi L^6}{16}\mathcal{Q} \right]\\
&
-  \frac{(1 + 3 \mu \fin^2 + 2 \xi \fin^3)}{8 \pi G}\int_\partial d^3x\sqrt{h}\left[K-\frac{2\sqrt{f_{\infty}}}{L}-\frac{L}{2\sqrt{f_{\infty}}}\mathcal{R}\right]\, ,
\end{aligned}
\end{equation}
where 
\begin{align}
\mathcal{P}= &12 \tensor{R}{_{\mu}^{\rho}_{\nu}^{\sigma}}\tensor{R}{_{\rho}^{\alpha}_{\sigma}^{\beta}}\tensor{R}{_{\alpha}^{\mu}_{\beta}^{\nu}}+\tensor{R}{_{\mu\nu}^{\rho\sigma}}\tensor{R}{_{\rho\sigma}^{\alpha\beta}}\tensor{R}{_{\alpha\beta}^{\mu\nu}}-12R_{\mu\nu\rho\sigma}R^{\mu\rho}R^{\nu\sigma}+8\tensor{R}{_{\mu}^{\nu}}\tensor{R}{_{\nu}^{\rho}}\tensor{R}{_{\rho}^{\mu}}\, ,
\\\nonumber
\mathcal{Q}=&-44R^{\mu\nu\rho\sigma }R_{\mu\nu }^{\ \ \alpha\beta }R_{\rho\ \alpha}^{\ \gamma \ \delta}R_{\sigma \gamma \beta \delta}-5 R^{\mu\nu\rho\sigma }R_{\mu\nu }^{\ \ \alpha\beta }R_{\rho\alpha}^{\ \ \gamma \delta}R_{\sigma \beta \gamma \delta}+5 R^{\mu\nu\rho\sigma }R_{\mu\nu\rho }^{\ \ \ \ \alpha}R_{\beta \gamma \delta \sigma}R^{\beta \gamma \delta}_{\ \ \ \ \alpha}\\
\nonumber&+24 R^{\mu\nu }R^{\rho\sigma \alpha\beta }R_{\rho\ \alpha \mu}^{\ \gamma}R_{\sigma \gamma \beta \nu}\, .
\end{align}
Evaluating the on-shell action for Taub-NUT solutions we find ~\cite{NewTaub2}
\begin{equation}
\mathcal{F}_{\mathbb{S}_\varepsilon^{3}} = -\frac{\pi L^2 (1+\varepsilon)^2}{G f_\infty^2} \left[\frac{1}{2} - \frac{f_\infty}{(1+\varepsilon)} - \frac{\mu f_\infty^3}{(1+\varepsilon)^3} - \frac{\xi f_\infty^4}{(1+\varepsilon)^4} \right] \, ,
\end{equation}
which matches precisely the results from the conjectured formula~\req{fee0e}.

As our last example, the Euclidean action with generalized boundary terms for the quartic generalized quasi-topological theories in $d=5$ is given by~\cite{NewTaub2}
\begin{align}\label{full6}
I_E =& - \int \frac{d^6 x \sqrt{g}}{16 \pi G} \left[ \frac{20}{L^2} + R + \frac{\lambda_{\rm \ssc GB} L^2 }{6} {\cal X}_4 -\frac{ \xi L^6 }{216} \mathcal{S} - \frac{ \zeta L ^6}{144} \mathcal{Z} \right] 
	\nonumber\\
&- \frac{1 - 4 \lambda_{\ssc \rm GB} \fin  + 8 (\xi + \zeta) \fin^3}{8 \pi G} \int_\partial d^5 x \sqrt{h} \left[K  - \frac{4 \sqrt{\fin}}{L} - \frac{L}{6 \sqrt{\fin}} {\cal R} - \frac{L^3}{18 \fin^{3/2}} \left( {\cal R}_{ij}{\cal R}^{ij} - \frac{5}{16} {\cal R}^2 \right)\right]
	\nonumber\\
&+ \frac{ \lambda_{\ssc \rm GB} \fin - 6 (\xi + \zeta) \fin^3}{8 \pi G} \frac{L^3}{18 \fin^{3/2}} \int_\partial d^5 x \sqrt{h} \left( 4 {\cal R}_{ij}{\cal R}^{ij} - \frac{5}{4} {\cal R}^2 + \frac{3}{2} {\cal X}^{(h)}_4 \right) \, ,	
\end{align}
where ${\cal X}_4$ is the Gauss-Bonnet density and 
\begin{align}
{\cal S} &=   992 R_{\mu}{}^{\rho} R^{\mu\nu} R_{\nu}{}^{\delta} R_{\rho\delta} + 28 R_{\mu\nu} R^{\mu\nu} R_{\rho\delta} R^{\rho\delta} - 192 R_{\mu}{}^{\rho} R^{\mu\nu} R_{\nu\rho} R - 108 R_{\mu\nu} R^{\mu\nu} R^2 
\nonumber\\
&+ 1008 R^{\mu\nu} R^{\rho\delta} R R_{\mu\rho\nu\delta} + 36 R^2 R_{\mu\nu\rho\delta} R^{\mu\nu\rho\delta} - 2752 R_{\mu}{}^{\rho} R^{\mu\nu} R^{\delta\tau} R_{\nu\delta\rho\tau} + 336 R R_{\mu}{}^{\tau}{}_{\rho}{}^{\gamma} R^{\mu\nu\rho\delta} R_{\nu\tau\delta\gamma} 
\nonumber\\
&- 168 R R_{\mu\nu}{}^{\tau\gamma} R^{\mu\nu\rho\delta} R_{\rho\delta\tau\gamma} - 1920 R^{\mu\nu} R_{\mu}{}^{\rho\delta\tau} R_{\nu}{}^{\gamma}{}_{\delta}{}^{\eta} R_{\rho\gamma\tau\eta} + 152 R_{\mu\nu} R^{\mu\nu} R_{\rho\delta\tau\gamma} R^{\rho\delta\tau\gamma} 
\nonumber\\
&+ 960 R^{\mu\nu} R_{\mu}{}^{\rho\delta\tau} R_{\nu\rho}{}^{\gamma\eta} R_{\delta\tau\gamma\eta} - 1504 R^{\mu\nu} R_{\mu}{}^{\rho}{}_{\nu}{}^{\delta} R_{\rho}{}^{\tau\gamma\eta} R_{\delta\tau\gamma\eta} + 352 R_{\mu\nu}{}^{\tau\gamma} R^{\mu\nu\rho\delta} R_{\rho\tau}{}^{\eta\sigma} R_{\delta\gamma\eta\sigma} 
\nonumber\\
&- 2384 R_{\mu}{}^{\tau}{}_{\rho}{}^{\gamma} R^{\mu\nu\rho\delta} R_{\nu}{}^{\eta}{}_{\tau}{}^{\sigma} R_{\delta\eta\gamma\sigma} + 4336 R_{\mu\nu}{}^{\tau\gamma} R^{\mu\nu\rho\delta} R_{\rho}{}^{\eta}{}_{\tau}{}^{\sigma} R_{\delta\eta\gamma\sigma} - 143 R_{\mu\nu}{}^{\tau\gamma} R^{\mu\nu\rho\delta} R_{\rho\delta}{}^{\eta\sigma} R_{\tau\gamma\eta\sigma} 
\nonumber\\
&- 436 R_{\mu\nu\rho}{}^{\tau} R^{\mu\nu\rho\delta} R_{\delta}{}^{\gamma\eta\sigma} R_{\tau\gamma\eta\sigma} + 2216 R_{\mu}{}^{\tau}{}_{\rho}{}^{\gamma} R^{\mu\nu\rho\delta} R_{\nu}{}^{\eta}{}_{\delta}{}^{\sigma} R_{\tau\eta\gamma\sigma} - 56 R_{\mu\nu\rho\delta} R^{\mu\nu\rho\delta} R_{\tau\gamma\eta\sigma} R^{\tau\gamma\eta\sigma} \, , 
\\
%%%%%%%%%
%%%%%%%%%
%%%%%%%%%
{\cal Z} &=  -112 R_{\mu}{}^{\rho} R^{\mu\nu} R_{\nu}{}^{\delta} R_{\rho\delta} - 36 R_{\mu\nu} R^{\mu\nu} R_{\rho\delta} R^{\rho\delta} + 18 R_{\mu\nu} R^{\mu\nu} R^2 - 144 R^{\mu\nu} R^{\rho\delta} R R_{\mu\rho\nu\delta} 
\nonumber\\
&- 9 R^2 R_{\mu\nu\rho\delta} R^{\mu\nu\rho\delta} + 72 R^{\mu\nu} R R_{\mu}{}^{\rho\delta\tau} R_{\nu\rho\delta\tau} + 576 R_{\mu}{}^{\rho} R^{\mu\nu} R^{\delta\tau} R_{\nu\delta\rho\tau} - 400 R^{\mu\nu} R^{\rho\delta} R_{\mu\rho}{}^{\tau\gamma} R_{\nu\delta\tau\gamma} 
\nonumber\\
&+ 48 R R_{\mu}{}^{\tau}{}_{\rho}{}^{\gamma} R^{\mu\nu\rho\delta} R_{\nu\tau\delta\gamma} + 160 R_{\mu}{}^{\rho} R^{\mu\nu} R_{\nu}{}^{\delta\tau\gamma} R_{\rho\delta\tau\gamma} - 992 R^{\mu\nu} R_{\mu}{}^{\rho\delta\tau} R_{\nu}{}^{\gamma}{}_{\delta}{}^{\eta} R_{\rho\gamma\tau\eta} 
\nonumber\\
&+ 18 R_{\mu\nu} R^{\mu\nu} R_{\rho\delta\tau\gamma} R^{\rho\delta\tau\gamma} - 8 R^{\mu\nu} R_{\mu}{}^{\rho\delta\tau} R_{\nu\rho}{}^{\gamma\eta} R_{\delta\tau\gamma\eta} + 238 R_{\mu\nu}{}^{\tau\gamma} R^{\mu\nu\rho\delta} R_{\rho\tau}{}^{\eta\sigma} R_{\delta\gamma\eta\sigma} 
\nonumber\\
&- 376 R_{\mu}{}^{\tau}{}_{\rho}{}^{\gamma} R^{\mu\nu\rho\delta} R_{\nu}{}^{\eta}{}_{\tau}{}^{\sigma} R_{\delta\eta\gamma\sigma} + 1792 R_{\mu\nu}{}^{\tau\gamma} R^{\mu\nu\rho\delta} R_{\rho}{}^{\eta}{}_{\tau}{}^{\sigma} R_{\delta\eta\gamma\sigma} - 4 R_{\mu\nu}{}^{\tau\gamma} R^{\mu\nu\rho\delta} R_{\rho\delta}{}^{\eta\sigma} R_{\tau\gamma\eta\sigma} 
\nonumber\\
&- 284 R_{\mu\nu\rho}{}^{\tau} R^{\mu\nu\rho\delta} R_{\delta}{}^{\gamma\eta\sigma} R_{\tau\gamma\eta\sigma} + 320 R_{\mu}{}^{\tau}{}_{\rho}{}^{\gamma} R^{\mu\nu\rho\delta} R_{\nu}{}^{\eta}{}_{\delta}{}^{\sigma} R_{\tau\eta\gamma\sigma} \, .
\end{align}
are two densities belonging to the quartic generalized quasi-topological family of theories~\cite{Ahmed:2017jod}.  Computing the on-shell Euclidean action for Taub-NUT solutions of this theory yields
\begin{equation}
\mathcal{F}_{\mathbb{S}_\varepsilon^{5}} = \frac{\pi^2 L^4 (1+\varepsilon)^3}{ G f_\infty^3 } \left[\frac{2}{3} - \frac{f_\infty}{1+\varepsilon} + \frac{2 \lambda_{\ssc \rm GB} f_\infty^2}{(1+\varepsilon)^2} - \frac{2 (\xi + \zeta) f_\infty^4}{(1+\varepsilon)^4} \right] 
\end{equation}
which, again, precisely matches the result obtained using the conjectured relationship~\req{fee0e}.

Finally, let us note that on-shell Euclidean actions for Taub solutions in Einstein gravity and Gauss-Bonnet gravity have been previously computed in~\cite{Clarkson:2002uj, KhodamMohammadi:2008fh}. Our results agree with those calculations up to an overall factor of $8/9$ in $d=5$, a factor of $3/4$ in $d = 7$, and more generally by a factor of
\begin{equation}
\frac{2^k k!}{(k+1)^k}
\end{equation} 
in $d = 2k+1$ dimensions. These factors are precisely the ratio of the volume of a product of $k$ 2-spheres to the volume of $\mathbb{CP}^k$. This discrepancy was observed in~\cite{Bobev:2017asb} in the case $d = 5$. Both there and in the present work properly accounting for these factors is important for matching the general results expected from field theory considerations, \eg the proportionality factor between $\mathcal{F}_{\mathbb{S}_\varepsilon^{5}}''(0)$ and $\ctt$. This, combined with our careful analysis of the computations in~\cite{Clarkson:2002uj, KhodamMohammadi:2008fh}, gives us confidence that the results presented here are correct.

\section{Free-field calculations}\label{ffc}
The numerical results for a free (conformally-coupled) scalar field and a free Dirac fermion used in the main text were presented in \cite{Bobev:2017asb}. We quickly summarize them here, along with some further details on the manipulations we performed to produce the curves in Fig. \ref{fig.2}. %We also perform some additional comparisons with the Einsteinian cubic gravity result for finite values of the squashing parameter.

In each case, the corresponding partition functions are given, for a generic background metric, by
\begin{equation}
Z^{\rm s}=\int \mathcal{D}\phi \, e^{-\frac{1}{2}\int d^3 x \sqrt{g} \left[(\partial \phi)^2+\frac{R \phi^2}{8} \right]}\, ,\quad 
Z^{\rm f}=\int \mathcal{D}\psi \, e^{-\int d^3 x \sqrt{g} \left[\psi^{\dagger} \left(i \slashed{D} \right)\psi  \right]}\, ,
\end{equation}
where $\phi$ and $\psi$ stand, respectively, for a bosonic scalar field and a Dirac fermion. Also, $R$ stands for the Ricci scalar of $g_{\mu\nu}$ and $\slashed{D}$ is the Dirac operator on the corresponding background. For a given background geometry $\mathcal{M}$, the free energy of these fields can be written in a unified way as
\begin{equation}
\mathcal{F}_{\mathcal{M}}=\frac{(-1)^f}{2^{(f-1)}}\log \det \left[\mathfrak{D}_{\mathcal{M}}/\Lambda^f \right]\, .
\end{equation}
Here,  $\Lambda$ is an energy cutoff, $\mathfrak{D}_{\mathcal{M}}$ stands for the conformal Laplace operator or the Dirac operator in each case, and $f=1,2$ for the fermion and the scalar respectively. Using a heat-kernel regulator \cite{Anninos:2012ft,Vassilevich:2003xt}, one can write the above expression as  
\begin{equation}
\log \det \left[\mathfrak{D}_{\mathcal{M}}/\Lambda^f \right]=\sum_i \int_{1/\Lambda^2}^{\infty}\frac{dt}{t}e^{-t\lambda_i^{3-f}}\, ,
\end{equation}
where $\lambda_i$ are the eigenvalues of $\mathfrak{D}_{\mathcal{M}}$. This expression can  in turn be split into two parts, containing UV and IR modes, respectively. Once the UV divergences are conveniently identified and regularized introducing appropriate counter-terms --- something that can be done numerically in a systematic way, as explained in~\cite{Bobev:2017asb} --- one is left with a finite and unambiguous answer for the free energy, in each case. 

\begin{figure}[t]
	\centering
	 	\includegraphics[scale=0.7]{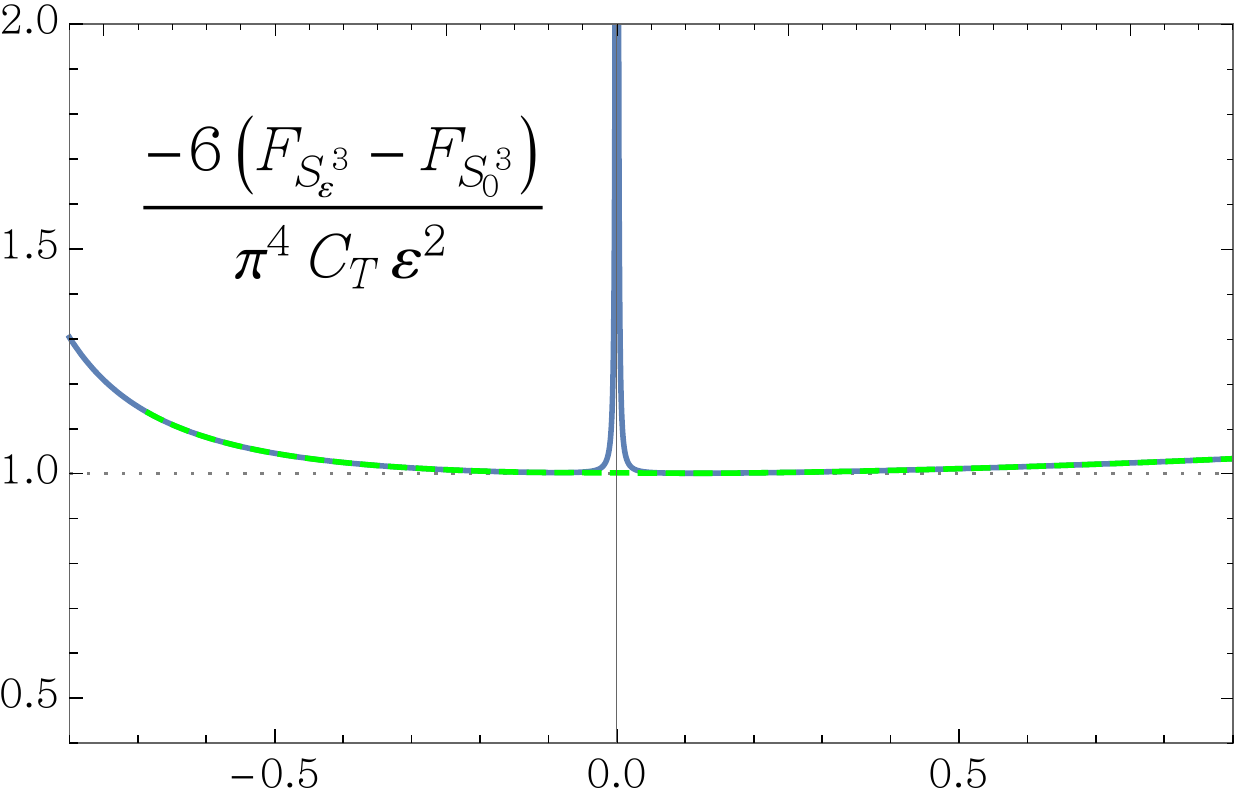}
	\caption{We plot  $-6(\mathcal{F}_{\mathbb{S}_{\varepsilon}^{3}}-\mathcal{F}_{\mathbb{S}_{0}^{3}})/(\pi^4\ctt \varepsilon^2)$ as a function of $\varepsilon$ using the numerical results in \cite{Bobev:2017asb}. In general, this function must cross the $\varepsilon=0$ axis at $1$, which requires a small treatment (resulting in the dashed green line) of the numerical data in a small neighborhood of $\epsilon=0$.}
	\label{fign}
\end{figure}

Plots of the numerical results obtained for $\mathcal{F}^{\rm f}_{\mathbb{S}_{\varepsilon}^{3}}$ and $\mathcal{F}^{\rm s}_{\mathbb{S}_{\varepsilon}^{3}}$ as functions of the squashing parameter $\varepsilon$ can be found in~\cite{Bobev:2017asb}.
Here we would like to make a technical comment about the procedure followed to obtain the curves appearing in Fig. \ref{fig.2}. Naturally, the idea is to plug the numerical results for $\mathcal{F}^{\rm f}_{\mathbb{S}_{\varepsilon}^{3}}$ and $\mathcal{F}^{\rm s}_{\mathbb{S}_{\varepsilon}^{3}}$ into the function $T(\varepsilon)$ and identify the value $T(0)$. In practice, this procedure requires a small treatment of the data near $\varepsilon=0$. The issue comes from the fact that $T(\varepsilon)$ involves dividing numerical expressions by $\varepsilon$, which produces divergences very close to $\varepsilon=0$. For example, we know that, for any CFT, the $\varepsilon=0$ limit of $-6(\mathcal{F}_{\mathbb{S}_{\varepsilon}^{3}}-\mathcal{F}_{\mathbb{S}_{0}^{3}})/(\pi^4\ctt \varepsilon^2)$ must be equal to one. However, the interpolating curves we obtain from the numerical data do not exactly account for the divergent piece in the denominator, which produces a spurious behavior very close to $\varepsilon=0$. This is illustrated in Fig. \ref{fign} in the case of the scalar. Luckily, the issue appears only within a very small neighborhood of $\varepsilon=0$, and we can safely correct these numerical values with a simple interpolation without losing any physical information about the tendency of the function in that region. Taking this into account, the functions $T(\varepsilon)$ can be safely constructed without polluting the numerical data, producing compelling evidence in favor of our general conjecture in \req{3conj}.

\chapter{Resumen\label{resumen}}

Las teorías de gravedad de orden superior en curvatura son extensiones de la relatividad general que aparecen como descripciones efectivas de teor\'ias de gravedad cu\'antica, tales como la teor\'ia de cuerdas. Mientras que a bajas energ\'ias el comportamiento del campo gravitacional en las teorías de curvatura superior es pr\'acticamente indistinguible del predicho por relatividad general, las diferencias pueden ser dram\'aticas en situaciones de gravedad extrema, como en el caso de los agujeros negros. Por lo tanto, es una tarea apasionante el estudiar c\'omo las geometr\'ias de agujero negro son modificadas por las correcciones de curvatura, con la esperanza de que algunas caracter\'isticas problem\'aticas de los agujeros negros observadas en relatividad general sean mejoradas, indicando as\'i los efectos de una teor\'ia de gravedad cu\'antica subyacente. Sin embargo, hay algunas dificultades asociadas a las teorías de orden superior en derivadas, como la existencia de inestabilidades, la propagación de ``fantasmas'', o simplemente la extremada complejidad de las ecuaciones diferenciales que rigen la din\'amica del campo gravitacional. En esta tesis, identificamos una nueva familia de gravedades de orden superior en curvatura que son capaces de evitar algunos de estos problemas.  Conocidas como \emph{gravedades cuasi-topol\'ogicas generalizadas} (Generalized quasi-topological gravity, GQG), tales teor\'ias representan extensiones de relatividad general que est\'an ausentes de inestabilidades y ``fantasmas'' a nivel lineal, y cuyas ecuaciones para m\'etricas est\'aticas y esf\'ericamente sim\'etricas adquieren una forma suficientemente sencilla que permite el estudio no perturbativo de soluciones de agujero negro. El miembro no trivial m\'as sencillo de esta clase de teor\'ias (que fue el primero en ser descubierto), es conocido como \emph{Einsteinian cubic gravity} (ECG), y jugar\'a un papel protagonista en esta tesis.  Adem\'as de sus interesantes propiedades, argumentaremos que las teor\'ias del tipo GQG capturan la correcci\'on de curvatura superior m\'as general cuando se tiene en cuenta la posibilidad de realizar redefiniciones de la m\'etrica.  Entonces usaremos estas teor\'ias para estudiar de un modo no perturbativo las correcciones al agujero negro de Schwarzschild en cuatro dimensiones, prestando especial atenci\'on a la modificaci\'on de las relaciones termodin\'amicas. La predicci\'on m\'as notable de estas teor\'ias es que la temperatura de Hawking de los agujeros negros neutros y est\'aticos tiende a cero en el l\'imite de masa peque\~na, en lugar de divergir como predice la relatividad general. Como consecuencia, los agujeros negros peque\~nos son termodin\'amicamente estables y su proceso de evaporaci\'on conlleva un tiempo infinito.  Adem\'as, las gravedades de orden superior en curvatura encuentran aplicaciones muy interesantes en la correspondencia anti-de Sitter/teor\'ia de campos conforme (AdS/CFT en ingl\'es), una dualidad que relaciona una teor\'ia cl\'asica de gravedad en un espacio AdS con una teor\'ia cu\'antica de campos que vive en la frontera de este espacio. En este contexto, las gravedades de curvatura superior se pueden considerar ``modelos de juguete'' hologr\'aficos que permiten, por ejemplo, obtener lecciones universales acerca de la din\'amica de las teor\'ias conformes de campos, o poner a prueba la generalidad de los resultados predichos por las aplicaciones hologr\'aficas de relatividad general. En esta tesis, exploramos varios  aspectos hologr\'aficos de Einsteinian cubic gravity, la cual proporciona un modelo de juguete para una teor\'ia conforme de campos no supersim\'etrica en tres dimensiones. Adem\'as, construimos nuevas soluciones eucl\'ideas tipo AdS-Taub-NUT, que permiten estudiar teor\'ias conformes de campos definidas sobre esferas aplastadas. Usando estos resultados, deducimos una expresi\'on universal para la expansi\'on de la energ\'ia libre de estas teor\'ias en tres dimensiones hasta orden c\'ubico en el par\'ametro de deformaci\'on.

\chapter{Conclusiones\label{conclusiones}}

%%%%%%%%%%%%%%%%%%%%%%%%%%%%%%%%%%%%%%%%%%%%%%%%%%%%%%%%%%%%%%%%%%%%%%
En esta tesis se ha producido un gran avance en la clasificaci\'on de teor\'ias de orden superior en curvatura en funci\'on de su espectro de modos a nivel lineal, y se ha identificado una familia especial de teor\'ias cuyas ecuaciones linealizadas son de segundo orden. Estas teor\'ias han sido llamadas ``Einstein-like'' y s\'olo propagan un gravit\'on sin masa sobre espacios maximalmente sim\'etricos. Algunas de dichas teor\'ias, llamadas gravedades cuasi-topol\'ogicas generalizadas, resultan adem\'as especialmente apropiadas para estudiar soluciones de agujero negro. De esta manera hemos podido describir de forma no perturbativa agujeros negros en cuatro dimensiones cuando se incluyen correcciones de curvatura superior. En particular, hemos sido capaces de estudiar los agujeros negros resultantes cuando la teor\'ia de Einstein es corregida con un n\'umero infinito de t\'erminos de curvatura. Ha de ser notado que esta es una haza\~na notable y que no tiene precedente en la literatura. Adem\'as hemos observado que todas estas correcciones tienen unos efectos universales sobre las propiedades termodinámicas de los agujeros negros. En particular, la temperatura de Hawking de los agujeros negros corregidos tiende a cero cuando la masa es peque\~na, un comportamiento opuesto al observado en relatividad general, donde la temperatura diverge. 
Como consecuencia, los agujeros negros peque\~nos son estables y no se evaporan en un tiempo finito. Adem\'as, dado que argumentamos que las teor\'ias consideradas podr\'ian servir como una base para construir la teor\'ia efectiva de gravedad m\'as general posible, cabe la posibilidad de que estas conclusiones se apliquen de hecho con casi total generalidad m\'as all\'a de las teor\'ias estudiadas. 
Como una consecuencia interesante de la existencia de agujeros negros peque\~nos estables, planteamos la posibilidad de que estos pudieran ser un constituyente de la materia oscura. 

Por otra parte, hemos estudiado las aplicaciones hologr\'aficas de ``Einsteinian cubic gravity'' en el contexto de la correspondencia AdS$_4$/CFT$_3$. La teor\'ia de campos conforme dual a Einsteinian cubic gravity pertenece a una clase de universalidad diferente a la definida por la relatividad general, y esto nos ha permitido obtener nuevos resultados.  Por ejemplo, hemos podido establecer que la funci\'on de partici\'on de una teor\'ia conforme de campos tridimensional en una esfera aplastada viene determinada, hasta orden c\'ubico en el par\'ametro de deformaci\'on, por las funciones de dos y tres puntos del tensor de energ\'ia-momento de la teor\'ia.  
Sin embargo, la principal conclusi\'on es que Einsteinian cubic gravity representa un modelo de juguete hologr\'afico muy interesante que nos permite obtener respuestas consistentes acerca de teor\'ias de campos conformes. En este sentido, esperamos que todas las gravedades cuasi-topol\'ogicas generalizadas, las cuales constituyen una familia numerosa, dar\'an lugar tambi\'en a duales hologr\'aficos consistentes que, no obstante, pueden ser estudiados f\'acilmente.

%\include{app3b}
%\include{app4}

%
%\newpage
%\thispagestyle{empty}
%\phantom{lala}

\renewcommand{\leftmark}{\MakeUppercase{Bibliography}}
\phantomsection
\bibliographystyle{JHEP-2}
\bibliography{Gravities}
\label{biblio}

%%%%%%%%%%%%%%%%%%%%%%%%%%%%%%%%%%%%%%%%%%%%%%%%%%%%%%%%%%%%%%%%%%%%%%%%%%%%%%%%%%%

%%%%% That's all Folks!    %%%%%%%%%%%%%%%%%%%%%%%%%%%%

\newpage
\thispagestyle{empty}
\phantom{lala}

\end{document}